\pdfoutput=1
\documentclass[a4paper,11pt,twoside,openright]{UGthesis}
\usepackage{fancyhdr}
\usepackage{graphicx}
\usepackage{array}
\usepackage{amsmath}
\usepackage{amssymb}
\usepackage{amsthm}
\usepackage{tabularx}
\usepackage{multibox}
\usepackage{lscape}
\usepackage{longtable}
\usepackage{multibox}
\usepackage{cite}
\usepackage{amsbsy}
\newcommand\ffam{\sffamily}
\newcommand\fser{\bfseries}
\newcommand\fsh{\upshape}

\numberwithin{equation}{section}

\newcommand\rctr{\renewcommand{\theenumi}{\roman{enumi}}}

\newcommand{\beq}{\begin{equation}}
\newcommand{\eeq}{\end{equation}}
\newcommand{\eqnlab}[1]{\label{eqn:#1}}
\newcommand{\Eqnref}[1]{Eq.~(\ref{eqn:#1})}
\newcommand{\beqa}{\begin{eqnarray}}
\newcommand{\eqa}{\end{eqnarray}}
\newcommand{\hs}{\hspace{0.1 cm}}

\newcommand{\al}{\alpha}
\newcommand{\be}{\beta}
\newcommand{\ga}{\gamma}
\newcommand{\de}{\delta}
\newcommand{\si}{\sigma}

\newcommand{\f}{\frac}
\newcommand{\mc}{\mathcal}
\newcommand{\la}{\lambda}
\newcommand{\G}{\Gamma}
\newcommand{\om}{\omega}
\newcommand{\pa}{\partial}
\newcommand{\nn}{\nonumber}

\newcommand{\g}{\mathrm{g}}
\newcommand{\tr}{\mathrm{tr}}

\newcommand{\ti}{\tilde}

\newcommand{\mf}{\mathfrak}

\newcommand{\ph}{\phantom}

\newcommand{\p}{\prime}

\newcommand{\mbb}{\mathbb}

\newcommand{\tm}{\tilde{m}}
\newcommand{\tn}{\tilde{n}}
\newcommand{\bas}{\backslash}

\newcommand{\xit}{\tilde{\xi}}
\newcommand{\xib}{\bar{\xi}}
\newcommand{\xitb}{\bar{\xit}}
\newcommand{\pp}{\prime\prime}

\newcommand{\mfVV}{\pa_{\mu}\mathrm{V}(x)\mathrm{V}(x)^{-1}}
\newcommand{\tV}{\mathrm{V}}
\newcommand{\mcBe}{\mc{B}_{\mc{M}_{\be}}}

\newcommand{\mgb}{\bar{\mf{g}}}
\newcommand{\mh}{\mf{h}}

\newcommand{\cG}{{\mathfrak{g}}}
\newcommand{\cH}{{\mathfrak{h}}}
\newcommand{\cK}{{\mathfrak{k}}}

\newcommand{\Zn}{{\mathbb Z}}
\newcommand{\pg}{SU(2,1;\mbb{Z}[i])}
\newcommand{\noi}{\noindent}
\def\ch{{\cal H}}
\def\cn{{\cal N}}
\def\ca{{\cal A}}
\def\ce{{\cal E}}
\def\cf{{\cal F}}

\newcommand{\x}{{\bf x}}

\newcommand{\A}{\mathrm{A}}
\newcommand{\B}{\mathrm{B}}
\newcommand{\QK}{\mathrm{QK}}
\newcommand{\SK}{\mathrm{SK}}

\newcommand{\cx}{{\mathbb C}}
\newcommand{\cxh}{\mathbb{CH}}
\newcommand{\cF}{{\mathcal F}}
\newcommand{\cZ}{{\mathcal Z}}

\newcommand{\cW}{{\mathcal{W}}}
\newcommand{\cosm}{{\mathcal{K}}}
\newcommand{\tcosm}{{\tilde{\mathcal{K}}}}
\newcommand{\tchi}{\tilde{\chi}}

\newcommand{\vb}[2]{e_{#1}^{\phantom{#1}{#2}}}
\newcommand{\hvb}[2]{\hat{e}_{#1}^{\phantom{#1}{#2}}}
\newcommand{\tvb}[2]{{\tilde{e}}_{#1}^{\phantom{#1}{#2}}}
\newcommand{\ivb}[2]{u_{#1}^{\phantom{#1}{#2}}}
\newcommand{\tvbint}{\tilde{e}_{\mathrm{int}}}

\newcommand{\tF}{\tilde{F}}
\newcommand{\tP}{\tilde{P}}

\newcommand{\tH}{\tilde{H}}
\newcommand{\vscalar}[2]{\vec{#1}\cdot\vec{#2}}
\newcommand{\gphi}{\vec{g}\cdot\vec{\phi}}
\newcommand{\fphi}[1]{\vec{f}_{#1}\cdot\vec{\phi}}
\newcommand{\Pc}{\mathcal{P}}
\newcommand{\tHHHH}{\tH_{a\gamma}\tH_b^{\phantom{b}\gamma}
\tH^a_{\phantom{a}\delta}\tH^{b\delta}}
\newcommand{\tHPPHa}{\tH^{c\alpha}\tP_{\alpha cd} \tP^{\beta
de}\tH_{e \beta}}
\newcommand{\tHPPHb}{\tH^{c\alpha}\tP_{\beta cd} \tP^{\beta
de}\tH_{e \alpha}}
\newcommand{\tHPHa}{\tH_{c\alpha}\tP^{\alpha cd}
\tH_d^{\phantom{d}\beta}}
\newcommand{\tHPHb}{\tH_{c\alpha}\tP^{\beta cd}
\tH_d^{\phantom{d}\alpha}}
\newcommand{\vg}{\vec{g}}
\newcommand{\vp}{\vec{\phi}}
\newcommand{\gp}{\vec{g}\cdot\vec{\phi}}
\newcommand{\vf}{\vec{f}}
\newcommand{\fip}{\vec{f}_i\cdot\vec{\phi}}
\newcommand{\ve}{\vec{e}}
\newcommand{\vo}{\vec{\omega}}
\newcommand{\val}{\vec{\al}}
\newcommand{\vL}{\vec{\Lambda}}
\newcommand{\vl}{\vec{\lambda}}

\newcommand{\lb}{\left[}
\newcommand{\rb}{\right]}
\newcommand{\cV}{{\cal V}}
\newcommand{\cQ}{{\cal Q}}
\newcommand{\cP}{{\cal P}}
\newcommand{\cD}{D^{(0)}}
\newcommand{\eps}{\epsilon}

\newcommand{\vet}{\varepsilon}
\newcommand{\bvet}{\bar\varepsilon}
\newcommand{\pt}{\psi}
\newcommand{\bpt}{\bar\psi}
\newcommand{\lt}{\lambda}
\newcommand{\blt}{\bar\lambda}

\newcommand{\venew}{\tilde{\varepsilon}}
\newcommand{\psinew}{\tilde{\psi}}
\newcommand{\lambdanew}{\tilde{\lambda}}

\newcommand{\tten}{\tilde{10}}

\DeclareMathOperator{\diag}{diag}
\DeclareMathOperator{\Exp}{Exp}
\DeclareMathOperator{\spn}{span}
\DeclareMathOperator{\rank}{rank}
\DeclareMathOperator{\ad}{ad}

\DeclareMathOperator{\mult}{mult}
\DeclareMathOperator{\htx}{ht}

\newenvironment{theorem}
{\vspace{1 em} \noindent{\bf Theorem:}}
{\vspace{1 em}}

\RequirePackage{hyperref}

\begin{document}

%%%%%%%%%%%%%%%%%%%%%%%%%%%%%%%%%%%%
%
% Cover
%
%%%%%%%%%%%%%%%%%%%%%%%%%%%%%%%%%%%%

%\include{cover}

%\blankpage

\setcounter{page}{1}
\pagenumbering{arabic}
%\pagenumbering{roman}
%\setcounter{page}{1}

%%%%%%%%%%%%%%%%%%%%%%%%%%%%%%%%%%%%%
%
% Titelblad
%
%%%%%%%%%%%%%%%%%%%%%%%%%%%%%%%%%%%%%
\pagestyle{empty}
\begin{center}
{\upshape\sffamily\bfseries\huge 
\noindent
Arithmetic and Hyperbolic Structures\\[2mm] 
 in String Theory }
 \\[4mm]
%and Supergravity} \\%[4mm]
%{\upshape\sffamily\bfseries\large PRELIMINARY VERSION}\\[4mm]
\end{center}

%\vspace*{1mm}
\begin{center}
        \rule{110mm}{2pt}
\end{center}

\vspace*{2mm}
\begin{center}
  {\fsh\ffam\fser\Large Daniel Persson}\\
\end{center}
%\vfill

%\vspace*{4mm}
\begin{center}
  {\it Institut f\"ur Theoretische Physik\\
  ETH Z\"urich\\
  CH-8093 Z\"urich, Switzerland}\\
\end{center}
%\vfill

%\vspace*{4mm}
\begin{center}
  {\tt daniel.persson@itp.phys.ethz.ch}\\
\end{center}
%\vfill

\vspace*{1cm}

\centerline{\ffam\fser Abstract}
\medskip
\normalsize
%       \noindent\input{abstract}
\noindent This monograph is an updated and extended version of the author's PhD thesis. It consists of an introductory text followed by two separate parts which are loosely related but may be read independently of each other. In {\bf Part~I} we analyze certain hyperbolic structures arising when studying gravity in the vicinity of a spacelike singularity (the ``BKL-limit''). In this limit, spatial points decouple and the dynamics exhibits ultralocal behaviour which may be mapped to an auxiliary problem  given in terms of a (possibly chaotic) hyperbolic billiard. In all supergravities arising as low-energy limits of string theory or M-theory, the billiard dynamics takes place within the fundamental Weyl chambers of certain hyperbolic Kac-Moody algebras, suggesting that these algebras generate hidden infinite-dimensional symmetries of the theory. We develop in detail the relevant mathematics of Lorentzian Kac-Moody algebras and hyperbolic Coxeter groups, and explain with many examples how these structures are intimately connected with gravity. We also construct a geodesic sigma model invariant under the hyperbolic Kac-Moody group $E_{10}$, and analyze to what extent its dynamics reproduces the dynamics of type II and eleven-dimensional supergravity. 
%We investigate the modification of the billiard dynamics when the original gravitational theory is formulated on a compact spatial manifold of arbitrary topology, revealing fascinating mathematical structures known as \emph{galleries}. We further use the conjectured hyperbolic symmetry $\mc{E}_{10}$ to generate and classify certain cosmological ($S$-brane) solutions in eleven-dimensional supergravity. Finally, we show in detail that eleven-dimensional supergravity and massive type IIA supergravity are dynamically unified within the framework of a geodesic sigma model for a particle moving on the infinite-dimensional coset space $K(\mc{E}_{10})\bas \mc{E}_{10}$. 

 {\bf Part~II} of the thesis is devoted to a study of how (U-)dualities in string theory provide powerful constraints on perturbative and non-perturbative quantum corrections. These dualities are described by certain arithmetic groups $G(\mbb{Z})$ which are conjectured to be preserved in the effective action. The exact couplings are generically given by automorphic forms on the double quotient $G(\mbb{Z})\bas G /K$, where $K$ is the maximal compact subgroup of the Lie group $G$. We discuss in detail various methods of constructing automorphic forms, with particular emphasis on the special class of non-holomorphic Eisenstein series. We provide detailed examples for the physically relevant cases of $SL(2,\mbb{Z})$ and $SL(3,\mbb{Z})$, for which we construct their respective Eisenstein series and compute their (non-abelian) Fourier expansions. We also show how these techniques can be applied to hypermultiplet moduli spaces in type II Calabi-Yau compactifications, and we provide a detailed analysis for the universal hypermultiplet. In this case, we propose that the quantum theory is invariant under the Picard modular group $SU(2,1;\mbb{Z}[i])$, and we construct an $SU(2,1;\mbb{Z}[i])$-invariant Eisenstein series whose non-abelian Fourier expansion exhibits the expected contributions from D2- and NS5-brane instantons.

\newpage

\thispagestyle{empty}

%\vspace*{-1cm}
%\vspace*{4mm}

\begin{center}
  {Thesis for the degree of D\small{OCTOR} \small{OF} P\small{HILOSOPHY} in Physics at}\\
  {C\small{HALMERS} U\small{NIVERSITY} \small{OF} T\small{ECHNOLOGY}}\\
  { \small{AND}}\\
  { D\small{OCTOR} \small{OF} S\small{CIENCE} in Theoretical Physics and Mathematics at}\\
  { U\small{NIVERSITE} L\small{IBRE} \small{DE} B\small{RUXELLES}}
\end{center}

\vspace*{0.5cm}

\begin{center}
{\upshape\sffamily\bfseries\huge 
\noindent
Arithmetic and Hyperbolic Structures\\ in String Theory 
}
 \\[4mm]
%and Supergravity} \\%[4mm]
%{\upshape\sffamily\bfseries\large PRELIMINARY VERSION}\\[4mm]
\end{center}

\vspace*{2mm}
\begin{center}
        \rule{110mm}{2pt}
\end{center}

\vspace*{4mm}
\begin{center}
  {\fsh\ffam\fser\Large Daniel Persson}\\
\end{center}
%\vfill
\vspace*{0.5cm}

%
%\vspace*{-40mm}
%%\begin{center}
%%  {\fsh\ffam\fser PRELIMINARY VERSION}
%%\end{center}
%\vfill
%  
%  
%  
%\vspace*{-40mm}
%\begin{center}
%  {\fsh\ffam\fser %avh.tex $-$ 
%  %\today
%  }
%\end{center}
%%\vfill

\begin{center}
\begin{minipage}[t]{40mm}
\begin{center}
\includegraphics[width=4cm]{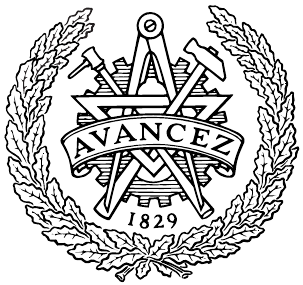}
%\vspace{-10mm}
%\caption{$s_1$ and $s_2$ generate a Coxeter group of order 6.}
\end{center}
\end{minipage}
\hspace{1cm}
\begin{minipage}[t]{35mm}
\begin{center}
\includegraphics[width=3.5cm]{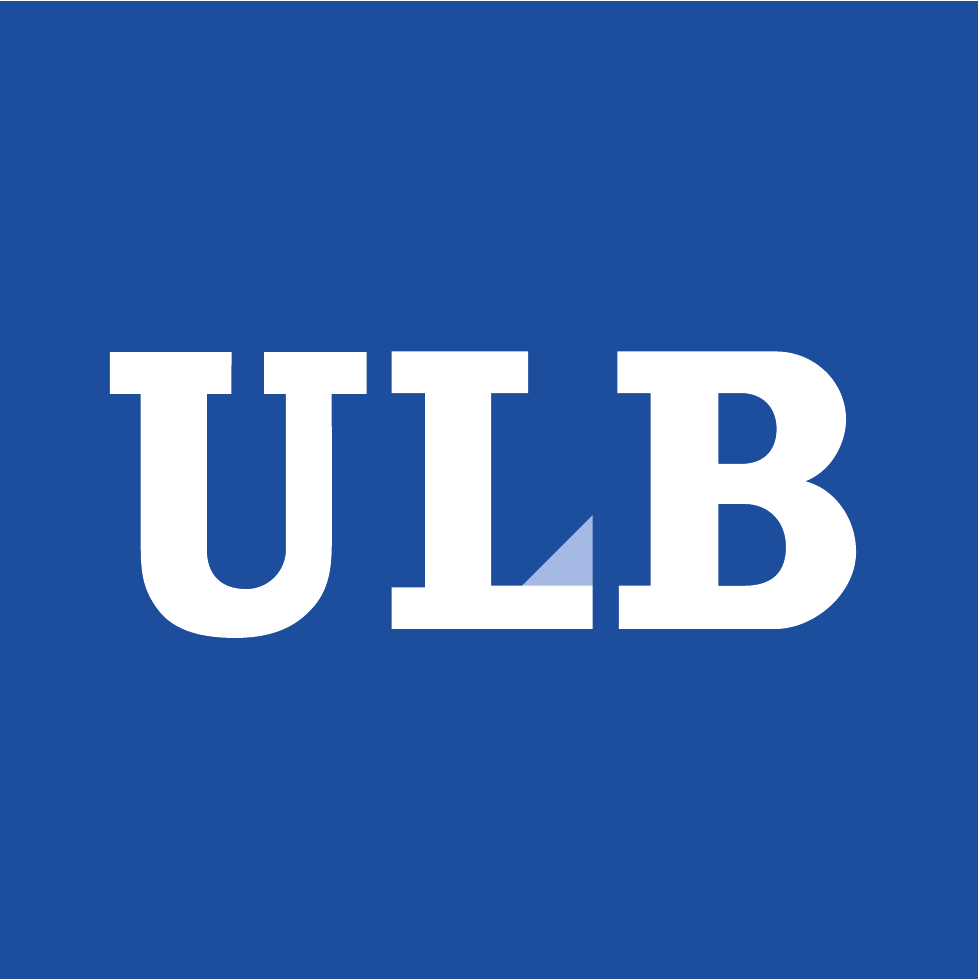}
%\vspace{4.1mm}
%\caption{$s_1$ and $s_2$ generate a Coxeter group of order 8.}
\end{center}
\end{minipage}
\end{center}

\vspace*{1cm}

{ \noindent \emph{This doctoral thesis is the result of joint supervision between Chalmers University of Technology and Universit\'e Libre de Bruxelles under the framework of the double doctoral degree agreement signed by both universities on the 8th of September 2008.}}

\vfill
\begin{center}
        { 
Fundamental Physics\\*[1mm]
        C\small{HALMERS} U\small{NIVERSITY} \small{OF} T\small{ECHNOLOGY} \\*[1mm]
        G\"oteborg, Sweden\\*[1mm]
        and\\*[1mm]
        Physique Th\'eorique et Math\'emathique\\*[1mm]
        U\small{NIVERSITE} L\small{IBRE} \small{DE} B\small{RUXELLES} \& I\small{NTERNATIONAL} S\small{OLVAY} I\small{NSTITUTES}\\*[1mm]
        Bruxelles, Belgium\\*[1mm]
        2009}
\end{center}

\newpage
%%%%%%%%%%%%%%%%%%%%%%%%%%%%%%%%%%%%
%
% Tryckort-sida
%
%%%%%%%%%%%%%%%%%%%%%%%%%%%%%%%%%%%%

%\mbox{}
%\thispagestyle{empty}
%\newpage
%\pagestyle{empty}

\vspace*{3cm}

\noindent 
Arithmetic and Hyperbolic Structures in String Theory \\
Daniel Persson\\
%ISBN 978-91-7385-301-9

%\vspace{0.5cm}
%\noindent \copyright\, Daniel Persson, 2009.

%\vspace{0.5cm}
%\noindent Doktorsavhandlingar vid Chalmers tekniska h\"ogskola\\
%Ny serie nr 2982\\
%ISSN 0346-718X 

%\vspace{0.5cm}
%\noindent Fundamental Physics\\
%Chalmers University of Technology\\
%SE-412 96 G\"oteborg, %\\
%Sweden%\\
%%Telephone +46 (0)31-772 10 00
%\vspace{0.5cm}

%

%%done at the\\\\
%\noindent
%Physique Th\'eorique et Math\'ematique\\
%Universit\'e Libre de Bruxelles \& International Solvay Institutes\\
%Campus Plaine CP 231\\
%1050 Bruxelles, Belgium\\

\vspace{1cm}

{ \noindent \emph{The research presented in this thesis has been funded by The International Solvay  Institutes, by a grant from PAI, Belgium, by IISN-Belgium
(conventions 4.4505.86, 4.4514.08, 4.4511.06 and 4.4514.08), by the Belgian National Lottery, by the European Commission FP6 RTN
programme MRTN-CT-2004-005104 and by the Belgian Federal Science Policy Office
through the Interuniversity Attraction Poles P5/27 and P6/11.\\ I am very grateful for all the generous support.}}

\vspace{.4cm}

\newpage
%\blankpage
%%%%%%%%%%%%%%%%%%%%%%%%%%%%%%%%%%%%%%%
%
% Sammanlaggnings-sida
%
%%%%%%%%%%%%%%%%%%%%%%%%%%%%%%%%%%%%%%%

%\thispagestyle{empty}

\noindent This thesis is based on the following eight papers, henceforth referred to as {\sffamily\bfseries Paper I--VIII}:

\bigskip

\begin{enumerate} 
\def\theenumi{\Roman{enumi}}

\item[{\sffamily\bfseries I}]
	M.~Henneaux, M.~Leston, D.~Persson and Ph.~Spindel, \cite{Henneaux:2006gp}\\
	{\it Geometric Configurations, Regular Subalgebras of $E_{10}$ and M-Theory Cosmology,}\\
	{JHEP} {\bf 0610} (2006) 021 [{\tt arXiv:hep-th/0606123}].

\vspace{.3cm}

\item[{\sffamily\bfseries II}]
	M.~Henneaux, M.~Leston, D.~Persson and Ph.~Spindel,\cite{Henneaux:2006bw}\\
	{\it A Special Class of Rank 10 and 11 Coxeter Groups,}\\
	{J. Math. Phys.} {\bf 48} (2007) 053512 [{\tt arXiv:hep-th/0610278}].

\vspace{.3cm}

\item[{\sffamily\bfseries III}]
	M.~Henneaux, D.~Persson and Ph.~Spindel, \cite{LivingReview}\\
	{\it Spacelike Singularities and Hidden Symmetries of Gravity,}\\
	{Living Rev. Rel.} {\bf 1} (2008) 1 [{\tt arXiv:0710.1818}].
	
\vspace{.3cm}

\item[{\sffamily\bfseries IV}]
	L.~Bao, J.~Bielecki, M.~Cederwall, B.E.W.~Nilsson and D.~Persson, \cite{Bao:2007fx}\\
	{\it U-Duality and the Compactified Gauss-Bonnet Term,}\\
	{JHEP} {\bf 0807} (2008) 048 [{\tt arXiv:0710.4907}].

\vspace{.3cm}

\item[{\sffamily\bfseries V}]
	M.~Henneaux, D.~Persson and D.H.~Wesley, \cite{Henneaux:2008me}\\
	{\it Coxeter Group Structure of Cosmological Billiards on Compact Spatial Manifolds,}\\
	{JHEP} {\bf 0809} (2008) 052 [{\tt arXiv:0805.3793}].

\vspace{.3cm}

\item[{\sffamily\bfseries VI}]
	M.~Henneaux, E.~Jamsin, A.~Kleinschmidt and D.~Persson,\cite{Henneaux:2008nr}\\
	{\it On the $E_{10}/$Massive Type IIA Supergravity Correspondence,}\\
	{Phys. Rev.} {\bf D79} (2009) 045008 [{\tt arXiv:0811.4358}].

\vspace{.3cm}

\item[{\sffamily\bfseries VII}]
	B.~Pioline and D.~Persson, \cite{AutomorphicNS5}\\
	{\it The Automorphic NS5-Brane,}\\
	 {Commun. Num. Th. Phys.} {\bf 3}, No. 4 (2009), [{\tt arXiv:0902.3274}].

\vspace{.3cm}

\item[{\sffamily\bfseries VIII}]
	L.~Bao, A~Kleinschmidt, B.E.W.~Nilsson, D.~Persson and B.~Pioline, \cite{Bao:2009fg}\\
	{\it Instanton Corrections to the Universal Hypermultiplet \\ and Automorphic Forms on $SU(2,1)$,}\\
	{Submitted to Commun. Num. Th. Phys.} [{\tt arXiv:0909.4299}].

\end{enumerate}

\newpage

%%%%%%%%%%%%%%%%%%%%%%%%%%%%%%%%%%%%%%%%%
%
% Acknowledgments
%
%%%%%%%%%%%%%%%%%%%%%%%%%%%%%%%%%%%%%%%%%
%\thispagestyle{empty}
%\vspace*{4cm}

\centerline{\ffam\fser\Large Acknowledgments}
\medskip
\smallskip

%\normalsize
{\small
\noindent
I have been given the great privilege of being guided through my PhD with the aid of  two amazing advisors. I am very grateful to Marc Henneaux and Bengt E. W. Nilsson for your unwavering support and encouragement these last four years. Thank you so much Marc for the many enjoyable and rewarding collaborations, helpful discussions and for always making sure that I am busy. Thank you so much Bengt for very close collaboration during the course of six years now, and for our numerous discussions on theoretical physics in general and string theory in particular, during which I have learned so many things. 

In the second half of my PhD, I have also been fortunate enough to work on several projects in close collaboration with Axel Kleinschmidt, who has become much like an extra advisor. I am very grateful for our extensive discussions on everything from $E_{10}$ to number theory, and for many helpful remarks which greatly improved the contents of this thesis. Thank you so much Axel! 

Part II of this thesis had probably not been very coherent if it had not been for the generous help I have received from Boris Pioline during the last two years. I am enormously grateful to Boris for all your explanations and clarifications regarding automorphic forms and their role in string theory, as well as for very enjoyable collaborations, and for numerous useful comments and suggestions on a previous draft of this thesis. 

I want to extend a very special thanks to my two companions Christoffer Petersson and Jakob Palmkvist for your close friendship and support during all these years. To Christoffer, I am particularly grateful for our endless discussions on string theory, and in particular on the topics presented in Chapters 8 and 11 of this thesis. I also want to thank Jakob especially for being the only one who has actually read through this entire manuscript and given numerous helpful suggestions.

I am grateful to Ling Bao, Johan Bielecki, Martin Cederwall, Laurent Houart, Ella Jamsin, Mauricio Leston, Josef Lindman H\"ornlund, Philippe Spindel, Nassiba Tabti and Daniel H. Wesley for very enjoyable and rewarding collaborations. 

A big thanks goes out to the entire theoretical physics groups at Chalmers and ULB, for creating extremely friendly and stimulating research atmospheres. I also wish to thank everyone at the physics department of \'Ecole Normale Sup\'erieure in Paris for kind hospitality during the fall of 2008. 

Finally, I owe thanks to many people with whom I have benefitted from stimulating discussions and correspondence over the years, which has greatly improved my understanding of the contents of this thesis, including Riccardo Argurio, Costas Bachas, Raphael Benichou, Sophie de Buyl, Stephane Detournay, Miranda Cheng, Jarah Evslin, Luca Forte, Matthias Gaberdiel, Gaston Giribet, Ulf Gran, Stefan Hohenegger, Nick Halmagyi, Bernard Julia, Arjan Keurentjes, Chethan Krishnan, Stanislav Kuperstein, Carlo Maccaferri, Yan Soibelman, Dieter Van den Bleeken, Stefan Vandoren, Pierre Vanhove, Alexander Wijns, Niclas Wyllard and Genkai Zhang. 

}

%%\noindent
%%%%%%%%%%%%%%%%%%%%%%%%%%%%%%%%%%%%%%%%%%%%%
%%Ayn Rand
%%
%%%%%%%%%%%%%%%%%%%%%%%%%%%%%%%%%%%%%%%%%%%%
\newpage
\mbox{}%
\\
\\
\\
\\
\\
\\
\\
\\
\\
\\
\large{
\begin{center}
\noindent \textit{``Man cannot survive except by gaining knowledge,}\\
\noindent \textit{and reason is his only means to gain it.}\\
\noindent \textit{Reason is the faculty that perceives, identifies and integrates}\\
\noindent \textit{ the material provided by his senses.}\\%
\noindent \textit{The task of his senses is to give him the evidence of existence,}\\
\noindent \textit{but the task of identifying it belongs to his reason,}\\
\noindent \textit{his senses tell him only that something is, }\\%
\noindent \textit{but what it is must be learned by his mind.''}\\
\end{center}
}%
\normalsize{$\phantom{m}$\hspace{9 cm}-Ayn Rand}

%%%%%%%%%%%%%%%%%%%%%%%%%%%%%%%%%%%%%%%%%%%%%%%
%%
%% Johanna
%%
%%%%%%%%%%%%%%%%%%%%%%%%%%%%%%%%%%%%%%%%%%%%%%

\newpage
\mbox{}%
\\
\\
\\
\\
\\
\\
\\
\\
\\
\\
\begin{flushright}
{\LARGE {\it Till Johanna}}
\end{flushright}

%\newpage
%%%%%%%%%%%%%%%%%%%%%%%%%%%%%%%%%%%%%%%%%%%%%%
%
% Table of contents
%
%%%%%%%%%%%%%%%%%%%%%%%%%%%%%%%%%%%%%%%%%%%%%
%\pagestyle{plain}
%\cleardoublepage
\tableofcontents
%\pagestyle{empty}

%\cleardoublepage
\pagestyle{fancy}
\renewcommand{\chaptermark}[1]{\markboth{Chapter \thechapter\ \ \ #1}{#1}}
\renewcommand{\sectionmark}[1]{\markright{\thesection\ \ #1}}
\lhead[\fancyplain{}{\sffamily\thepage}]%
  {\fancyplain{}{\sffamily\rightmark}}
\rhead[\fancyplain{}{\sffamily\leftmark}]%
  {\fancyplain{}{\sffamily\thepage}}
\cfoot{}
\setlength\headheight{14pt}

\rctr

%%%%%%%%%%%%%%%%%%%%%%%%%%%%%%%%%%%%%%%%%%%%
%
% Part 1
%
%%%%%%%%%%%%%%%%%%%%%%%%%%%%%%%%%%%%%%%%%%%%
%\setcounter{section}{0}
\chapter{Introduction -- A Tour Through the Dualities of String Theory}
\label{Chapter:Introduction}

\section{The Principle of Beauty}
\label{Section:Beauty}
To unravel the fundamental laws of nature is the ultimate goal of theoretical physics. In order to advance in this ambitious undertaking, it is desirable to have a set of underlying principles to guide us through the  mist. A striking example is provided by Einstein's \emph{principle of equivalence}, which states that  the concepts of inertial and gravitational mass are merely two sides of the same coin. With this principle as a guiding light, Einstein was led to the discovery of the general theory of relativity. 

It was argued by Dirac that the most basic principle that should serve as a lamppost for the pursuit of truth in fundamental laws, is the \emph{principle of beauty} \cite{Dirac}. Beauty is of course a subjective notion, but, nonetheless, physicists and mathematicians alike tend to agree on whether a certain theory, or a mathematical equation, is beautiful. This principle of beauty in fundamental laws should \emph{not} be confused with \emph{simplicity}. The mathematical formulation of the laws of nature exhibits a remarkably rich and complicated structure, for which ``simple'' is not a fitting description. Dirac therefore restated the basic principle by saying that we should pursue \emph{mathematical beauty}, since this has often, perhaps surprisingly, proven to go hand in hand with an accurate description of nature. 

In addition to the general theory of relativity, another great achievement of the twentieth century was the discovery of quantum mechanics. This relied heavily on what may in retrospect be referred to as a \emph{principle of symmetry}, stating that the fundamental laws should look the same irrespectively of whether one derives them standing upright or hanging upside down. More accurately, the principle of symmetry demands that the mathematical laws of a theory should remain invariant under a set of \emph{transformations}. In mathematical terminology, such a set of transformations form a \emph{group}, the properties of which are described in terms of \emph{group theory}. These mathematical techniques were vital in the development of the standard model of particle physics, the framework in which the fundamental constituents of nature are described. This theory is to a large extent determined by specifying that its mathematical laws should remain invariant under symmetry transformations described by the group $SU(3)\times SU(2)\times U(1)$. This shows that the principle of symmetry is extremely powerful in constraining the possible descriptions of nature. Of course, to uncover which particular symmetry group is relevant for describing a certain property of nature is in general a difficult problem; but it is a \emph{well defined} problem, and therefore sets the physicist onto a clear path forward.

The underlying theme of this thesis is the quest for finding the fundamental symmetry of \emph{string theory}, which is a leading candidate for a theory unifying the two descriptions of nature mentioned above: general relativity -- describing the large-scale structure of the universe -- and quantum mechanics -- the framework in which we describe interactions between the fundamental building blocks of nature, the \emph{elementary particles}. The basic idea of string theory is that these elementary particles correspond to different vibrational modes of a single fundamental object, a string. Originally developed in the 1970's, string theory has grown into a vast subject of research, by now encompassing large areas of both physics and mathematics. String theory appears to be a consistent physical theory; whether or not it describes our universe is yet to be verified. Despite the frustrating lack of experimental evidence for string theory, it exhibits a truly beautiful and intricate structure which may potentially answer some of the most difficult questions we can ask: How did the universe begin? What is the fundamental structure of spacetime? What are the microscopic constituents of a black hole? 

One of the outstanding problems in string theory is to find a satisfactory fundamental formulation of the theory from first principles, analogous to the way Einstein formulated general relativity based on the principle of equivalence. At present, such a formulation is not known, but rather we have variety of different descriptions which are valid in certain limits and regimes. The remarkable property of the theory, as realized in the works of Hull \& Townsend \cite{HullTownsend} and Witten \cite{Witten}, is that all of these different descriptions are related in various ways to each other, creating an intricate web of \emph{dualities} which ties the whole structure together into a robust and consistent framework. These dualities reveal an elaborate generalization of the principle of symmetry: although the various descriptions of the theory change under \emph{duality transformations}, the complete structure is left invariant. It is the purpose of this thesis to further explore the notion of duality in string theory from several different viewpoints, with the hope that this line of research will eventually lead us to uncover its basic underlying symmetry.  

As a motivational prequel to the remainder of this thesis, we shall now embark on a brief guided tour through the duality web of string theory. The bulk of the thesis is then divided into two main parts -- referred to as {\bf Part~I} and 
{\bf Part~II} -- and the following introduction provides a general overview of the main concepts and ideas which underlie the analysis. It should be noted that the two main parts of the thesis may be read independently. This being said, it is also the purpose of the following introduction to give the reader a notion of how the physics of {\bf Part~I} and {\bf Part~II} may ultimately be related.

 Before we begin, it is in order to extend an apology, and a warning, to the layman reader because, despite the gentle words of motivation above, the following introduction is written at a fairly technical level, and is primarily aimed towards a theoretical physicist who is well versed in the framework of string theory.
\section{The Duality Web of Type II String Theory}

String theory lives in ten spacetime dimensions, with its ``mother'', M-theory, hovering above in eleven dimensions. There are five different ten-dimensional string theories -- known under the esoteric names type I, type IIA and IIB, heterotic $E_8\times E_8$ and heterotic $SO(32)$ -- which are all related by various non-trivial dualities, often via M-theory \cite{Witten}. We will in our current treatment restrict to the type II theories and their relation with M-theory. The other theories are also interesting in their own right, but do not play an immediate role for the analysis in this thesis. We refer the reader to \cite{Polchinski1,Polchinski2,BeckerBeckerSchwarz} for more information. 

\subsection{Duality and Maximal Supersymmetry}

The main feature that enables us to relate various string theories to each other is through \emph{compactification}, namely formulating the theory on a product space $\mbb{R}^{1,D-n}\times X$, where $X$ is an $n$-dimensional compact internal manifold. One of the prime examples is the relation between ten-dimensional type IIA string theory and M-theory on $\mbb{R}^{1,9}\times S^{1}$ \cite{Witten}. It has been known for a long time that the corresponding low-energy limits, eleven-dimensional supergravity and type IIA supergravity, are related by dimensional reduction on a circle \cite{Campbell:1984zc,Giani:1984wc,Huq:1983im}. Subsequently, Witten realized that this correspondence in fact extends to the full quantum theories, with one of the key features being the interpretation of the type IIA string coupling $g^{{}_{(\A)}}_s=e^{\phi_{(\A)}}$ in terms of the radius of the M-theory circle. This implies that at weak coupling, $g^{_{(\A)}}_s\rightarrow 0$, there is a nice description of type IIA string theory in terms of a perturbative expansion in $g^{_{(\A)}}_s$, while in the opposite, strong-coupling limit, $g^{_{(\A)}}_s\rightarrow \infty $, the compact dimension of the circle appears, and the theory becomes eleven-dimensional. This implies that \emph{strongly-coupled} type IIA string theory is accurately described in terms of \emph{weakly-coupled} eleven-dimensional supergravity, which in turn is the low-energy limit of M-theory. 

 By taking yet another direction to be compact, we end up with M-theory compactified on a torus $T^2$, giving rise to a nine-dimensional theory which is dual to type IIA on $ S^{1}$ in the sense described above. Furthermore, in nine dimensions, type IIA string theory is related to type IIB string theory via T-duality on the circle. More precisely, T-duality relates type IIA on a circle $S^1$ of radius $R$ to type IIB on a different circle $\tilde{S}^1$ of radius $1/R$. 
 
We have seen above that the strongly coupled type IIA theory is eleven-dimensional supergravity. What is the corresponding statement for strongly coupled type IIB string theory? The relation between type IIB and M-theory turns out to be slightly more subtle:  via T-duality in $D=9$ one obtains an equivalence between type IIB string theory on $S^1$ and M-theory on $T^2$. Under this map, the complex structure $\Omega=\Omega_1+i\Omega_2$ of the torus is identified with the ``axio-dilaton'' $\tau=C_{(0)}+i e^{-\phi_{(\B)}}$, which contains the type IIB string coupling $g_s^{_{(\B)}}=e^{\phi_{(\B)}}$. The M-theory formulation exhibits a natural $SL(2,\mbb{Z})$-invariance associated with modular transformations of the torus, which maps to an $SL(2,\mbb{Z})$-action on the axio-dilaton $\tau$ of type IIB in nine dimensions. However, since the axio-dilaton exists already in $D=10$, this suggests that ten-dimensional type IIB string theory should in fact be invariant under $SL(2,\mbb{Z})$ \cite{HullTownsend,Witten}. In particular, the action of the modular group on $\tau$ inverts the type IIB string coupling, $e^{\phi_{(\B)}}\rightarrow e^{-\phi_{(\B)}}$, implying that type IIB is \emph{self-dual} under S-duality. In other words, strongly coupled type IIB string theory is \emph{equivalent} to weakly coupled type IIB string theory. 

So far we have restricted our attention to dualities appearing in nine and ten dimensions. However, more interesting structures appear as we take a larger internal manifold. It has been known since the work of Cremmer and Julia \cite{Cremmer:1978ds,Cremmer:1979up} that maximal supergravity on a torus $T^n$ exhibits a chain of hidden symmetries described by Lie groups $G_{D-n}(\mbb{R})$ in their split real form. The moduli space of scalars in lower dimensions is then given by the coset space $\mc{M}_{D-n}=G_{D-n}(\mbb{R})/K$, where $K=K(G_{D-n})$ is the maximal compact subgroup. For example, in eight dimensions one finds $G_8(\mbb{R})=SL(3,\mbb{R})\times SL(2,\mbb{R})$, while in six dimensions the global symmetry group is $SO(5,5)$. These groups are referred to as ``hidden symmetries'' because they are always larger than the naive isometry $GL(n,\mbb{R})$ of the internal torus. The extra structure which is responsible for the enhancement of the symmetry arises from the dualisation of Ramond-Ramond and NS-NS $p$-form fields as we descend in dimension. Remarkably, starting in $D=5$ these global symmetries are given by the exceptional Lie groups $\mc{E}_n(\mbb{R})$ \cite{Cremmer:1979up}. In particular, for dimensional reduction to $D=3$, all vector fields may be dualized into scalars and the theory is globally invariant under the largest exceptional Lie group $\mc{E}_8(\mbb{R})$.

The groups $G_{D-n}(\mbb{R})$ are exact symmetries only of the \emph{classical} effective action in $D-n$ dimensions, while the embedding into string/M-theory breaks the symmetry. This fact can be understood quite generally by the following argument \cite{HullTownsend,Witten}. In addition to the moduli space of scalars $G_{D-n}(\mbb{R})/K$, the reduced theories contain abelian vector fields arising from the reduction of the $D$-dimensional metric and the various $p$-form fields. The charges associated with these vector fields transform in a representation of $G_{D-n}$ (see, e.g. \cite{Obers:1998fb}). However, through the embedding in string theory, these charges arise from the couplings of the vector fields to the basic objects in the theory, i.e. strings and D-branes, and are therefore subject to Dirac-Zwanziger quantization conditions. One then deduces that a continuous symmetry group $G_{D-n}(\mbb{R})$ can never be an exact symmetry of string theory since it does not preserve the lattice of charges \cite{HullTownsend}.

Then what is the fate of the classical symmetry groups $G_{D-n}(\mbb{R})$ in string theory? It has been conjectured by Hull and Townsend that these are broken to discrete subgroups $G_{D-n}(\mbb{Z})\subset G_{D-n}(\mbb{R})$, and that the full string/M-theory remains invariant under $G_{D-n}(\mbb{Z})$ \cite{HullTownsend}. This is known as \emph{U-duality}, and provides one of the cornerstones for the consistency of the string/M-theory duality web \cite{HullTownsend,Witten} (see also \cite{Obers:1998fb}). In the quantum theory, the exact moduli space is then given by 
\beq
\mc{M}_{D-n}^{\text{exact}}=G_{D-n}(\mbb{Z})\bas G_{D-n}(\mbb{R})/K.
\eeq

Let us discuss this in some more detail for the case of compactifications to $D=4$, where the classical symmetry is given by the exceptional group $\mc{E}_7(\mbb{R})$, and the conjectured U-duality group is $\mc{E}_7(\mbb{Z})$. We want to understand the structure of this group from the general arguments above. In $D=4$, all $p$-form fields can be dualized to axionic scalars or to abelian vector fields. For the case at hand, the bosonic sector of the four-dimensional effective action contains, in addition to the metric, 70 scalar fields parametrizing the coset space $\mc{E}_7(\mbb{R})/(SU(8)/\mbb{Z}_2)$, together with 28 Maxwell fields $\mc{A}_{\mu}^{I}$. The electric and magnetic charges $(q^{I}, p_I)$ associated with these vector fields transform in a 56-dimensional representation of the Lie algebra $E_7$. However, as argued above, in the quantum theory these charges are subject to Dirac-Zwanziger charge quantization. More specifically, for two particles with electric-magnetic charges $(q^{I}, p_I)$ and $(\tilde{q}^{I}, \tilde{p}_I)$ the quantization condition enforces the constraint
\beq
q^{I} \tilde{p}_I-\tilde{q}^{I}p_I\in \mbb{Z}.
\eeq
This implies that in the quantum theory the charges $(q^{I},p_I)$ span a 56-dimensional integral lattice which is invariant under the \emph{electric-magnetic duality group} $Sp(56;\mbb{Z})$. Based on these arguments, it was conjectured in \cite{HullTownsend} that the largest possible subgroup of $\mc{E}_7(\mbb{R})$ that can be preserved in the quantum theory is defined by the subset of transformations which leave the electric-magnetic charge lattice invariant. In other words, the U-duality group $\mc{E}_7(\mbb{Z})$ is \emph{defined} by \cite{HullTownsend}
\beq
\mc{E}_7(\mbb{Z}):= \mc{E}_7(\mbb{R})\cap Sp(56;\mbb{Z}).
\eeq
The U-duality conjecture implies that all couplings in the quantum effective action should be functions on   $\mc{E}_7(\mbb{R})/(SU(8)/\mbb{Z}_2)$, which in addition are completely invariant under the discrete group $\mc{E}_7(\mbb{Z})$ \cite{PiolineKiritsis,ObersPioline}. Such functions are known as \emph{automorphic forms} and will play a leading role in {\bf Part II} of this thesis. All quantum corrections to the effective action must respect the U-duality symmetry and this enforces powerful constraints which in principle are sufficient to determine the action exactly (see for instance \cite{GreenGutperle,Green:1997di,PiolineKiritsis,ObersPioline}). Moreover, this statement is expected to hold in any dimension $4\leq D\leq 11$, with the respective duality groups $G_{D-n}(\mbb{Z})$ encoding the relevant quantum corrections in each dimension. 

\subsection{Duality in $\mc{N}=2$ Theories}
 
The fascinating structures that we have seen emerging above can be viewed as a consequence of supersymmetry. In ten dimensions, the type II theories exhibit $\mc{N}=2$ supersymmetry, corresponding to 32 supercharges. Upon dimensional reduction on a torus $T^{n}$, all of these supercharges are preserved. For example, reduction on $T^6$ gives rise to $\mc{N}=8$ supergravity in $D=4$ \cite{Cremmer:1978ds,Cremmer:1979up}. It is this large amount of supersymmetry which constrains the classical moduli spaces of the reduced theories to be given by symmetric spaces $G_{D-n}(\mbb{R})/K$. When taking more complicated internal manifolds $X$, many of the 32 supercharges are broken, giving rise to theories exhibiting a lesser amount of supersymmetry. 

Of particular relevance for the topics treated in {\bf Part II} of this thesis is the case when $X$ is a Calabi-Yau threefold. This compactification breaks a quarter of the original 32 supercharges and thus preserves $\mc{N}=2$ supersymmetry in four dimensions. Originally, Calabi-Yau compactifications were introduced in order to obtain four-dimensional theories containing chiral fermions \cite{Candelas:1985en,Strominger:1985it}. These are also interesting examples for the purpose of probing the intricate duality web of string theory, because they provide an intermediate step between the simpler compactifications preserving $\mc{N}\geq 4$ supersymmetry in $D=4$ -- for which the moduli space is constrained to be a symmetric space $G/K$ -- and the difficult case of (more realistic) $\mc{N}=1$ theories, for which quantum corrections are more severe. 

The classical limit of type II theories compactified on a Calabi-Yau threefold $X$ is described by pure $\mc{N}=2$ supergravity in four dimensions coupled to vector multiplets and hypermultiplets. One of the complications in describing this theory is the fact that the scalar fields no longer parametrize a symmetric space, but is rather given by a product \cite{Bagger:1983tt,deWit:1984px}
\beq
\mc{M}_4= \mc{M}_{{}_\SK}\times \mc{M}_{{}_\QK},
\label{ModuliSpaceProduct}
\eeq
where $\mc{M}_{{}_\SK}$ is a \emph{special K\"ahler manifold}, parametrized by scalars belonging to vector multiplets, and $\mc{M}_{{}_\QK}$ is a \emph{quaternion-K\"ahlermanifold}, parametrized by hypermultiplet scalars. The same structure appears both for the type IIA and type IIB theories, while the particular details of the effective actions are quite different. In type IIA compactifications, the moduli space $\mc{M}^{{}_{(\A)}}_{{}_{\mathrm{SK}}}$ is $2h_{1,1}$-dimensional and encodes the (complexified) K\"ahler structure deformations of $X$, while the quaternionic manifold $\mc{M}^{_{(\A)}}_{{}_\QK}$ is $4(h_{2,1}+1)$-dimensional and encodes the complex structure deformations of $X$. On the type IIB side the situation is reversed: the vector multiplet moduli space $\mc{M}_{{}_\SK}^{_{(\B)}}$ is $2h_{2,1}$-dimensional, while the hypermultiplet moduli space $\mc{M}_{{}_\QK}^{_{(\B)}}$ is $4(h_{1,1}+1)$-dimensional. 

Although $\mc{N}=2$ supersymmetry does constrain the moduli space $\mc{M}_4$ to always split into a special K\"ahler and a quaternion-K\"ahlermanifold, this constraint is still not very strong: there is a zoo of possible quantum corrections to the four-dimensional effective action which deform the metric on the moduli space $\mc{M}_4$. Nevertheless, even when all quantum corrections are turned on, the decoupling between hypermultiplets and vector multiplets is preserved. This fact provides the basis for yet another powerful string duality: \emph{mirror symmetry}. Consider again type IIA string theory on $X$. Then the mirror symmetry conjecture states that there exists a different \emph{mirror} Calabi-Yau manifold $\tilde{X}$, such that the moduli spaces of type IIA and type IIB are related as follows \cite{Candelas:1990rm,MirrorSymmetry}
\beq
\mc{M}_{{}_\SK}^{_{(\A)}}(X)\equiv \mc{M}_{{}_\SK}^{_{(\B)}}(\tilde{X}), \qquad \mc{M}_{{}_\QK}^{_{(\A)}}(X)\equiv \mc{M}_{{}_\QK}^{_{(\B)}}(\tilde{X}),
\label{MirrorSymmetry}
\eeq
which is a consequence of the fact that the complex structures and K\"ahler structures for a mirror pair $(X, \tilde{X})$ are interchanged: $h_{1,1}(X)=h_{2,1}(\tilde{X})$ and $h_{2,1}(X)=h_{1,1}(\tilde{X})$. We should emphasize that this relation between type IIA on $X$ and type IIB on $\tilde{X}$ is not restricted to a relation between the moduli spaces: the statement of (generalized) mirror symmetry is that the complete theories, including quantum corrections, should be equivalent \cite{BeckerBeckerStrominger}. 

Recall now that type IIA and type IB are simply related by T-duality in $D=9$. Could there exist a version of this in lower dimensions, in addition to mirror symmetry? Indeed, there is such a duality which at the level of the moduli space is known as the \emph{c-map} \cite{Cecotti:1988qn}. To reveal this, we start from the type IIA point of view and consider the further reduction on a circle to $D=3$.\footnote{For a nice discussion of the $c$-map, see \cite{Pioline:2006ni}.} Under this reduction, the moduli space $\mc{M}_{{}_\QK}^{_{(\A)}}$ simply goes along unchanged, while the vector multiplet moduli space $\mc{M}_{{}_\SK}^{_{(\A)}}$ gets enhanced due to the additional scalars arising from the reduction process. More precisely, in addition to the $2h_{1,1}$ scalar fields, the vector multiplets in $D=4$ also contain $h_{1,1}$ abelian vector fields $\mc{A}_\mu^{I}, \hs I=1,\dots, h_{1,1}$. Upon reduction on $S^1$ to $D=3$ these vector fields give rise to $h_{1,1}$ new scalars $\zeta^{I}\equiv \mc{A}^{I}_4$, while the three-dimensional vector fields $\mc{A}_\al^{I}$ can in turn be dualized into an additional set of $h_{1,1}$ scalars $\tilde{\zeta}_{I}$, giving rise to a total of $2h_{1,1}$ new scalar fields $(\zeta^{I}, \tilde\zeta_I)$.  These scalars parametrize a $2h_{1,1}$-dimensional torus $T^{2h_{1,1}}$ which is fibered over the four-dimensional moduli space $\mc{M}_{{}_\SK}^{_{(\A)}}$. Had we been considering the reduction of $D=4$, $\mc{N}=2$ supersymmetric \emph{gauge theory}, this would have been the end of the story (see \cite{Seiberg:1996nz}). 

However, in the present analysis, we are analyzing $\mc{N}=2$ supergravity, and therefore we must also take into account the reduction of the \emph{gravity multiplet} in $D=4$. This corresponds simply to the four-dimensional metric $\g_{\mu\nu}$ together with an additional abelian vector $A_{\mu}$, known as the graviphoton. Similarly to the other vector fields, the reduction of $A_\mu$ gives rise to two axionic scalars $\zeta^{0}$ and $\tilde\zeta_0$, while the reduction of the metric contributes with the radius of the circle $e^{U}$ as well as a Kaluza-Klein vector $A_\al\equiv \g_{\al 4}$ which can be dualized into yet another axionic scalar $\psi$. Together with the $2h_{1,1}$ scalars of $\mc{M}_{{}_\SK}^{_{(\A)}}$, we thus conclude that the theory in $D=3$ contains a total of $4(h_{1,1}+1)$ scalar fields. Recall that this is precisely the dimension of the hypermultiplet moduli space $\mc{M}_{{}_\QK}^{_{(\B)}}$ of type IIB string theory compactified on the \emph{same} Calabi-Yau threefold $X$. In fact, the new moduli space in $D=3$ is also a quaternionic manifold which precisely coincides with $\mc{M}_{{}_\QK}^{_{(\B)}}$. Microscopically, this relation is actually not unexpected: it arises from T-duality along the circle $S^1$, in particular mapping the (inverse) radius $e^{U}$ in type IIA on $X\times S^1$ to the string coupling $e^{\phi_{(\B)}}$ in type IIB on $X\times \tilde{S}^1$. In a similar way, one may deduce that the vector multiplet moduli space $\mc{M}_{{}_\SK}^{_{(\B)}}$ is mapped, via the $c$-map, to the hypermultiplet moduli space $\mc{M}_{{}_\QK}^{_{(\A)}}$ of type IIA on $X$. This is therefore in strong contrast to the case of mirror symmetry, which, as we have seen above, interchanges the vector- and hypermultiplet moduli spaces between type IIA and type IIB on \emph{different} Calabi-Yau threefolds.

The analysis above has revealed that there is a fascinating duality structure within the type II sector of string theory also for compactifications on more complicated manifolds than tori. But there appears to be one piece of the puzzle missing: Where is the relation with M-theory? Previously we have learned that boosting the type IIA string coupling $g_s^{_{(\A)}}$ to infinity takes us to eleven-dimensional supergravity. So what happens if we do the same in $D=4$? To answer this question, we must first note an important point. After the compactification on $X$ the dilaton modulus $e^{\phi_{(\A)}}$ belongs to a hypermultiplet, and therefore parametrizes the quaternionic moduli space $\mc{M}_{{}_\QK}^{_{(\A)}}$. This is, however, not quite the whole truth because for the decoupling between the moduli spaces (\ref{ModuliSpaceProduct}) to work out, one must rescale the ten-dimensional dilaton by the volume of the internal manifold, and define a new four-dimensional dilaton by $e^{-2\phi_4}\equiv \text{Vol}(X) e^{-2\phi_{(\A)}}$. The string coupling constant in $D=4$ is then defined with respect to the rescaled dilaton $g_s^{{}_{(4)}}=e^{\phi_4}$, and the strong-coupling limit corresponds to $g_s^{{}_{(4)}}\rightarrow\infty$.

Now consider M-theory compactified on the same Calabi-Yau manifold $X$. This gives rise to $\mc{N}=1$ supergravity in \emph{five} dimensions, with a hypermultiplet moduli space taking the same form as $\mc{M}_{{}_\QK}^{_{(\A)}}$ in (\ref{ModuliSpaceProduct}), although with a different interpretation of the scalar fields. Upon further reduction on $S^1$ to $D=4$ this moduli space goes along unchanged and the type IIA dilaton $e^{\phi_{(\A)}}$ is again interpreted as the radius of the circle. Hence the statement is that the strong-coupling limit $g_s^{{}_{(4)}}\rightarrow \infty$ of type IIA on $X$ should be interpreted as blowing up the M-theory circle $e^{\phi_{(\A)}}\rightarrow \infty$ while keeping the volume of $X$ fixed. To conclude, in the strong-coupling limit, type IIA on $X$ is equivalent to M-theory on $X\times S^1$, in analogy with the $10/11$-dimensional duality. 

\subsection{Instanton Effects and String Dualities}

Because of the lesser amount of supersymmetry preserved, quantum corrections to the four-dimensional effective action in Calabi-Yau compactifications are considerably less constrained compared to the toroidal case. As we have seen, the string coupling $g_s$ belongs to the quaternionic hypermultiplet moduli space $\mc{M}_{{}_\QK}$ both in type IIA and type IIB. This implies that the hypermultiplet moduli space is sensitive to quantum corrections associated with the perturbative power expansion in $g_s$ corresponding to worldsheets of higher genus. Since we have seen that supersymmetry enforces a complete decoupling between the two moduli spaces (\ref{ModuliSpaceProduct}) these quantum corrections will only affect the hypermultiplet moduli space $\mc{M}_{{}_\QK}$. For the vector multiplet moduli space, on the other hand, the situation is different in type IIA and type IIB. In type IIA, the vector multiplet scalars are associated with K\"ahler structure deformations of $X$, and the associated moduli space $\mc{M}_{{}_\SK}^{{}_{(\A)}}$ only receives quantum corrections with respect to the perturbative expansion in the worldsheet coupling $\alpha^{\prime}$, while being completely insensitive to the corrections in the string coupling $g_s$. In type IIB, the K\"ahler structure deformations are encoded in the hypermultiplet moduli space $\mc{M}_{{}_\QK}^{{}_{(\B)}}$, which therefore also receives $\alpha^{\prime}$-corrections in addition to the corrections in $g_s$. Thus, because all quantum corrections on the IIB side affect the quaternionic manifold $\mc{M}_{{}_\QK}^{{}_{(\B)}}$, the vector multiplet moduli space $\mc{M}_{{}_\SK}^{{}_{(\B)}}$ is tree-level exact both in $\alpha^{\prime}$ and in $g_s$ (see \cite{Aspinwall} for a nice review).

In addition, the metric on $\mc{M}_{{}_\QK}^{{}_{(\A /\B)}}$ receives non-perturbative corrections which are exponentially suppressed of order $e^{-1/g_s}$ in the weak-coupling limit. These non-perturbative effects arise from Euclidean D-branes whose worldvolume wraps cycles in the Calabi-Yau manifold \cite{BeckerBeckerStrominger}. On the type IIA side, such effects are attributed to Euclidean D2-branes wrapping special Lagrangian submanifolds in $X$, while in type IIB they arise from all Euclidean D$p$-branes, with $p=-1,1,3,5$, wrapping even cycles in $X$. In addition, both in type IIA and type IIB there are effects from Euclidean NS5-branes wrapping the entire Calabi-Yau threefold. Such NS5-brane instantons behave differently compared to the D-brane instantons since the tension of the NS5-brane scales as $T_{\text{NS5}}\sim g_s^{-2}$ while for a D-brane we have $T_{\text{D}p}\sim g_s^{-1}$. This implies that NS5-brane instanton effects are exponentially suppressed of order $e^{-1/g_s^2}$ and are therefore subdominant in the weak-coupling limit compared to D-brane instantons. 

To compute the complete quantum corrected metric on the hypermultiplet moduli spaces $\mc{M}_{{}_\QK}^{(\A /\B)}$ has long been an outstanding problem. There is generically no analogue in $\mc{N}=2$ theories of the U-duality groups $G(\mbb{Z})$ which proved to be so powerful for toroidal compactifications. Nonetheless, considerable progress has been made recently by assuming that the $SL(2,\mbb{Z})$-invariance of type IIB string theory should be unaffected by the compactification. Enforcing this constraint directly in the four-dimensional effective action made it possible to sum up all instanton corrections to the moduli space $\mc{M}_{{}_\QK}^{_{(\B)}}$ due to D$(-1)$ and D1-instantons \cite{Vandoren1}. Using the mirror map (\ref{MirrorSymmetry}) between the IIA and IIB moduli spaces, it was also possible to sum up the contributions to the moduli space $\mc{M}_{{}_\QK}^{_{(\A)}}$ due to a subset of the D2-brane instantons \cite{Vandoren2}. Additional progress has also been made recently \cite{Alexandrov:2008ds,Alexandrov:2008nk,Alexandrov:2008gh} using twistor techniques, and this will be discussed in more detail in Chapter \ref{Chapter:UniversalHypermultiplet}. In Chapter \ref{Chapter:UniversalHypermultiplet} we will also see that for certain special cases of Calabi-Yau threefolds, known as \emph{rigid}, there is an appealing candidate for a larger discrete group than $SL(2,\mbb{Z})$ that might potentially constrain completely also the NS5-brane instantons. 

%With these words we end our tour through the duality web of type II string theory. The topics discussed above will all play important roles in the analysis of {\bf Part II} of this thesis, where also many issues touched upon here will be explained in more detail. Next, we turn to discuss a different approach to revealing possible underlying symmetries of string theory. Although the physical setting is somewhat different, we will see that there are intriguing relations with the dualities discussed in the present section. The precise relation, however, remains to be uncovered. 

\section{Hyperbolic Billiards and Infinite-Dimensional Dualities}

\subsection{Infinite-Dimensional U-Duality?}
The observant reader may have noticed an apparent lack of continuity in the previous section. If the various compactifications of the type II string theories reveal such interesting structures and dualities, why stop the reduction process in $D=3$? The naive answer is: No reason! However, below $D=3$ things start to become slightly more subtle. As we have seen in the previous section, the dimension $D=3$ is special since it is the first dimension where all $p$-form fields, originating from ten or eleven dimensions, may be dualized into scalars. In the context of Calabi-Yau compactifications, this feature was responsible for the enhancement of the $2n$-dimensional special K\"ahler moduli space $\mc{M}_{{}_\SK}$ to a quaternionic $4(n+1)$-dimensional manifold $\mc{M}_{{}_\QK}$.\footnote{Recall $n=h_{1,1}, h_{2,1}$ for type IIA, IIB, respectively.} 

For the case of toroidal compactifications to $D=4$ we learned that the classical moduli space is given by $\mc{E}_7(\mbb{R})/(SU(8)/\mbb{Z}_2)$. Recall that this four-dimensional theory also contained 28 Maxwell fields. Upon further reduction to $D=3$ all of these 28 abelian vector fields may be dualized and give rise to 56 new axionic scalars. Moreover, as we have seen, the $D=4$ metric yields a dilaton $e^{U}$ and a Kaluza-Klein scalar $\sigma$ in $D=3$. These scalars always appear during the reduction from $D=4$ to $D=3$ and are independent of which four-dimensional theory we started with. Adding up all the scalars we find a total of $2+56+70=128$ scalar fields in $D=3$. These parametrize the coset space $\mc{E}_8(\mbb{R})/(Spin(16)/\mbb{Z}_2)$, with the global symmetry described by the largest of the exceptional Lie groups $\mc{E}_8(\mbb{R})$. 

So what \emph{does} happen if take the compactification below three-dimensions? It was in fact conjectured long ago by Julia \cite{Julia:1980gr,Julia:1982gx} that the chain of hidden symmetries should persist, with the groups $\mc{E}_9(\mbb{R}), \mc{E}_{10}(\mbb{R})$ and $\mc{E}_{11}(\mbb{R})$ appearing upon toroidal reduction to $D=2, D=1$ and $D=0$, respectively. This is not as obvious as it might sound, because these groups are no longer finite-dimensional Lie groups, but rather infinite-dimensional generalisations of these, known as \emph{Kac-Moody groups} (see \cite{Kac}). This makes it very tricky to understand the statement that these should become symmetries in dimensions below $D=3$. How can a theory with a finite number of degrees of freedom suddenly reveal an infinite-dimensional symmetry? This is indeed a crucial point which is not at all understood for the case of $D=1$ and $D=0$, while for reductions to two dimensions it is possible to make sense of the enhanced symmetry. 

In $D=3$ all $p$-form fields can be dualized to scalars, enhancing the symmetry from $\mc{E}_7(\mbb{R})$ to $\mc{E}_8(\mbb{R})$. It turns out that $D=2$ also has a very special property, namely that \emph{scalars are dual to scalars}, in the sense that any axionic scalar $\chi$ can be dualized into another axionic scalar $\tilde{\chi}$ through the relation $d\chi=\star d\tilde{\chi}$. The symmetry group $\mc{E}_8(\mbb{R})$, which acts on the original scalar $\chi$, has an induced non-linear action on the dual scalar $\tilde{\chi}$ through the duality relation $d\chi=\star d\tilde\chi$. Moreover, the duality can be applied an infinite number of times giving rise to an infinite number of dual axions. It is the infinitely many new axionic scalars which are responsible for enhancing the symmetry to $\mc{E}_9(\mbb{R})$, with the total moduli space in $D=2$ being the infinite-dimensional coset space $\mc{E}_9(\mbb{R})/K(\mc{E}_9)$ \cite{E9Geroch}. This phenomenon was originally uncovered in the simpler setting of pure gravity in $D=4$ by Geroch \cite{Geroch:1970nt}, in which case the relevant infinite-dimensional group was later identified as $SL(2,\mbb{R})^{+}$ \cite{Breitenlohner:1986um}, corresponding to the \emph{affine extension} of the three-dimensional duality group $G_3=SL(2,\mbb{R})$. For this reason, the infinite exceptional group $\mc{E}_9$ is sometimes called the \emph{Geroch group}. 

Similarly to $SL(2,\mbb{R})^{+}$, the group $\mc{E}_9$ is an affine Kac-Moody group which can be obtained through an affine extension of $\mc{E}_8$, and is therefore often denoted $\mc{E}_8^{+}$. Concretely, the extension is performed by adjoining an extra node to the Dynkin diagram of $\mc{E}_8$ in a certain prescribed way which will be explained in detail in Chapter \ref{Chapter:KacMoody}. The groups $\mc{E}_{10}$ and $\mc{E}_{11}$, which are conjectured to appear upon further reduction below $D=2$, may in turn be constructed by extending the Dynkin diagram of $\mc{E}_8$ with one or two extra nodes, respectively \cite{Gaberdiel:2002db}. For this reason, they are also denoted by $\mc{E}_{10}=\mc{E}_8^{++}$ and $\mc{E}_{11}=\mc{E}_{8}^{+++}$. 

The affine $\mc{E}_9$-symmetry of type II supergravities in $D=2$ is by now well-established, and may be attributed to the fact that any gravitational theory becomes completely integrable in two dimensions \cite{Breitenlohner:1986um,E9Geroch}. However, it is still not understood \emph{how}, or even \emph{if}, the infinite Kac-Moody groups $\mc{E}_{10}$ and $\mc{E}_{11}$ are supposed to appear upon further reduction. Nevertheless, these groups encode interesting algebraic structures that might turn out to play important roles in the ultimate formulation of string/M-theory. An important difference compared to the finite Lie groups discussed in the previous section is that the invariant metric on the Cartan subgalgebra $\mf{h}$ of the associated Lie algebras $E_{10}$ and $E_{11}$ is of Lorentzian signature. In contrast, all finite-dimensional Lie algebras have Killing forms which are Euclidean when restricted to the Cartan subalgebra. For this reason, the Kac-Moody algebras $E_{10}$ and $E_{11}$ belong to the class of \emph{Lorentzian Kac-Moody algebras} \cite{Gaberdiel:2002db}.

For the purposes of this thesis, it is the Kac-Moody algebra $E_{10}$ which will play a leading role. This belongs to the small subclass of Lorentzian Kac-Moody algebras known as \emph{hyperbolic}. The nomenclature here refers to the fact that the Weyl group $\mc{W}(E_{10})$ leaves invariant the unit hyperboloid inside the Cartan subalgebra $\mf{h}\subset E_{10}$, analogously to the Weyl groups of finite Lie algebras which leave invariant a sphere at infinity, and are therefore known as \emph{spherical}. If the conjecture that $E_{10}$ appears as a symmetry of eleven-dimensional supergravity on $T^{10}$ is true, then it begs the question of what the fate of this symmetry is when the theory is embedded into M-theory. Following the general arguments of Hull and Townsend \cite{HullTownsend}, discussed extensively in the previous section, one would expect that the continuous symmetry $\mc{E}_{10}(\mbb{R})$ be broken to some discrete subgroup $\mc{E}_{10}(\mbb{Z})$, and that M-theory on $T^{10}$ (or type II string theory on $T^{9}$) is invariant under this infinite U-duality group. In fact, an analysis of the moduli space of M-theory on $T^{10}$ was initiated in \cite{BanksMotl}, where it was verified that the moduli space has Lorentzian signature, signaling the underlying structure of $E_{10}$.\footnote{Extensions of these ideas have also been pursued in \cite{Brown:2004jb}.} If these speculations are true, then we would expect that $\mc{E}_{10}(\mbb{Z})$ encodes information about novel non-perturbative effects in string theory, such as Euclidean D8-branes wrapping the internal torus $T^9$. See also \cite{Ganor:1999uy,Ganor:1999ui,PiolineWaldron,Kleinschmidt:2009cv,Kleinschmidt:2009hv} Êfor various points of view on the role of $\mc{E}_{10}(\mbb{Z})$ in string theory.

\subsection{Hyperbolic Weyl Groups and Cosmological Billiards}
The discussion above fits well into the philosophy of Section \ref{Section:Beauty} regarding \emph{the principle of beauty}: the hyperbolic Kac-Moody algebra $E_{10}$ is without a doubt a beautiful and intriguing mathematical object, and its suggestive appearance opens up new and largely unexplored territories which lie on the borderline between mathematics and physics. However, despite these suggestive speculations the facts remain clear: we do not yet understand the underlying role of $\mc{E}_{10}$ within string theory and M-theory.\footnote{Other appearances of $E_{10}$ in string theory are discussed in \cite{Harvey:1995fq,Harvey:1996gc,Barwald:1997gm,Ganor:1999uy,Ganor:1999ui,Brown:2004jb,Ganor}.} Nevertheless, we shall now discuss a very different physical setting in which the hyperbolic Kac-Moody algebra $E_{10}$ also makes a surprising appearance, giving yet another indication of its possible importance. More precisely, it turns out that the Weyl group of $E_{10}$ controls the dynamics of eleven-dimensional supergravity close to a cosmological singularity \cite{ArithmeticalChaos}. To understand this statement, we shall begin by discussing some of the underlying ideas which form the basis of {\bf Part I} of this thesis.

In the 1970's, Belinskii, Khalatnikov and Lifshitz (BKL) analyzed the generic behaviour of four-dimensional pure gravity close to a cosmological (spacelike) singularity \cite{BKL,BKL1,BKL2}. Their conclusion was that in this limit the temporal dependence of the dynamical degrees of freedom dominate over the spatial dependence, creating an effective \emph{decoupling of spatial points}. The approach to the singularity is described by a sequence of \emph{Kasner epochs}, where the dynamics in each epoch is associated to a certain Kasner solution. It was also found that this oscillating behaviour is \emph{chaotic}, implying that there is an infinite sequence of oscillations before one reaches the singularity. Subsequently, Misner showed that the BKL-dynamics at a generic spatial point can be recast in terms of the dynamics of an auxiliary particle moving in a bounded region of hyperbolic space, called \emph{mixmaster behaviour} \cite{Misner}. 

These results were later extended to higher-dimensional gravitational theories, and it was found that pure gravity exhibits chaotic BKL-oscillations only in dimensions $4\leq D\leq 10$, while pure gravity in  eleven dimensions ceases to be chaotic \cite{Demaret}. However, it it turns out that all supergravities arising as low-energy limits of string theory and M-theory do reveal chaotic dynamics in the ``BKL-limit'' \cite{DH1}. The surprising appearance of chaos in eleven-dimensional supergravity is attributed to the additional 3-form field $C_{(3)}$ which is not present in pure gravity. 

 Following the original ideas of Misner, it has been shown that the dynamics in the BKL-limit for any gravitational theory can be recast into geodesic motion in a region of hyperbolic space, bounded by hyperplanes \cite{DHNReview}. The auxiliary particle undergoes geometric reflections against these hyperplanes, thus giving rise to a billiard-type behaviour. From the point of view of the original BKL-analysis, free-flight motion of the particle represents a single Kasner solution, while a geometric reflection flips between two distinct Kasner solutions. The behaviour close to a spacelilke singularity has therefore been dubbed \emph{cosmological billiards} \cite{DHNReview}. Whether a certain theory exhibits chaotic behaviour is then equivalent to determining whether the region in which the billiard dynamics takes place is of finite volume or not. It is well-known that random motion in a finite-volume hyperbolic billiard is chaotic. 

It turns out that all of these discoveries have a very elegant algebraic explanation \cite{ArithmeticalChaos}. The auxiliary hyperbolic space hosting the billiard may be identified with the Cartan subalgebra of a Lorentzian Kac-Moody algebra $\mf{g}$, while the bounded region in which the billiard ball is confined corresponds to the \emph{fundamental Weyl chamber}. In this new interpretation, it becomes clear that the geometric reflections responsible for the sequence of Kasner oscillations in fact generate the Weyl group $\mc{W}$ of $\mf{g}$. Hence, the dynamics in the BKL-limit is completely controlled by the Weyl group of a Lorentzian Kac-Moody algebra. This gives a powerful way of determining whether a given theory exhibits chaotic dynamics or not: if the Lorentzian Kac-Moody algebra $\mf{g}$ is of \emph{hyperbolic type}, then the fundamental Weyl chamber is of finite volume, and the theory is chaotic \cite{HyperbolicKaluzaKlein}. 

Remarkably, when performing this analysis for eleven-dimensional supergravity one finds that the algebra whose Weyl group controls the dynamics in the BKL-limit is precisely the hyperbolic Kac-Moody algebra $E_{10}$, thus explaining the appearance of chaos \cite{ArithmeticalChaos}. Moreover, it turns out that the Weyl group of $E_{10}$ also appears for the type IIA and type IB supergravities, revealing that these three theories display identical dynamical behaviour in the BKL-limit. This is in perfect accordance with the fact that type IIA, type IIB as well as eleven-dimensional supergravity become equivalent upon reduction on a torus $T^{n}$ as discussed above. We should also emphasize that the Weyl group $\mc{W}(E_{10})$ is in fact a subgroup of the conjectured U-duality group $\mc{E}_{10}(\mbb{Z})$, thus suggesting that the BKL limit is an alternative way of ``unveiling'' a possible underlying symmetry of string theory. {\bf Part I} of this thesis is to a large extent devoted to a careful study of the dynamics in the BKL-limit, with particular emphasis on the underlying algebraic structures. 

\subsection{Geodesic Motion on Infinite-Dimensional Coset Spaces}

The particle dynamics in a finite-volume billiard is only sensitive to the hyperplanes which are ``closest'' to the particle geodesic. These are known as the \emph{dominant walls}. From the algebraic point of view, this translates to the statement that the relevant Weyl-reflections governing the BKL-oscillations are only those with respect to the \emph{simple roots} of the Kac-Moody algebra $\mf{g}$, implying that there is a correspondence between dominant billiard walls and simple roots of $\mf{g}$. However, generically all theories give rise to many more walls than just the dominant ones. Depending on the structure of the theory, there can be a number of different \emph{subdominant walls}Ê which play no role in the strict BKL-limit \cite{DHNReview}. However, when going away from the strict BKL-limit it is expected that the dominant walls become less sharp and subdominant walls come into play. Another way to say this is that the complete decoupling of spatial points is no longer valid, and spatial dependence gradually reappears. On the algebraic side the subdominant walls are associated with non-simple positive roots of the Kac-Moody algebra, and hence probes the possible hidden Kac-Moody symmetry $\mf{g}$ beyond its Weyl group $\mc{W}(\mf{g})$.

These ideas were put on a more concrete footing in the context of eleven-dimesional supergravity by Damour, Henneaux and Nicolai \cite{DHN2}. Inspired by the BKL-analysis discussed above, they constructed a manifestly $\mc{E}_{10}$-invariant non-linear sigma model for a particle moving on the infinite-dimensional coset space $\mc{E}_{10}/K(\mc{E}_{10})$.\footnote{It has also been conjectured by West that the Lorentzian Kac-Moody algebra $E_{11}$ should correspond to a non-linearly realized hidden  symmetry of eleven-dimensional supergravity, or possibly of M-theoryÊ\cite{West:2000ga,E11andMtheory}. A lot of work has been done within the framework of this alternative proposal (see for exampleÊ\cite{West:2002jj,West:2003fc,West:2004st,Riccioni:2007au,Riccioni:2009hi}). In the present work, we are mainly motivated by the cosmological billiard picture, and we will therefore not treat this point of view in more detail.} To make sense of the coset space $\mc{E}_{10}/K(\mc{E}_{10})$, the Kac-Moody algebra $E_{10}$ was sliced up into an infinite gradation, called a \emph{level decomposition}, with finite-dimensional subspaces falling into representations of a distinguished $\mf{sl}(10, \mbb{R})$ subalgebra. At low levels this decomposition reveals tensorial representations matching the field content of eleven-dimensional supergravity, and it was shown that there is a dynamical equivalence between the geodesic motion on $\mc{E}_{10}/K(\mc{E}_{10})$ and a certain truncation of eleven-dimensional supergravity. The match works up to \emph{height 30} on the algebraic side, and therefore probes the Kac-Moody algebra $E_{10}$ far beyond its simple roots, which by definition all have height one.\footnote{The height of a root is defined as the sum of all the coefficients  when this root is written as a linear combination of the simple roots. This is explained in more detail in Chapter \ref{Chapter:KacMoody}.}

This correspondence between $\mc{E}_{10}$ and eleven-dimensional supergravity was later extended also to encompass type IIA \cite{Kleinschmidt:2004dy} and type IIB supergravity \cite{Kleinschmidt:2004rg}.\footnote{This type of construction has also been extended by Englert and Houart to $E_{11}=E_8^{+++}$, as well as to other Lorentzian Kac-Moody algebras of triple-extended type \cite{Englert:2003py,BraneDynamicsEnglert,FromE11toE10}. This framework is particularly powerful for analyzing the algebraic structure of BPS solutions in the context of maximal supergravities \cite{Englert:2004it, IntersectingEnglert,Englert:2007qb}. See also \cite{Tabti:2009xn} for a nice recent review. } In the second half of {\bf Part~I} of this thesis we will discuss the original example in great detail, and we will  extend the correspondence to the case of massive type IIA supergravity. 

It has also been understood how to incorporate fermions into the correspondence between supergravity and infinite-dimensional sigma models. The fermionic degrees of freedom are associated with certain spinorial representations of the maximal compact subalgebra $K(E_{10})$ \cite{deBuyl:2005zy,Damour:2005zs,deBuyl:2005mt,Damour:2006xu}. For exampe, in the case of eleven-dimensional supergravity, the gravitino arises as a 320-dimensional vector-spinor representation of $K(E_{10})$. This is an unfaithful representation of the infinite-dimensional algebra $K(E_{10})$, and it is an outstanding problem to understand the construction of non-trivial faithful representations which could match the infinite gradation involving the bosonic degrees of freedom. Because of this mismatch it has not yet been possible to construct a supersymmetric version of the non-linear sigma model on $\mc{E}_{10}/K(\mc{E}_{10})$.     

With these words we end our tour through the duality web of type II string theory and supergravity. The topics discussed in this introduction will all play important roles in the analysis of {\bf Part~I} and {\bf Part II} of this thesis.

\part{Spacelike Singularities and Hyperbolic Structures in (Super-)Gravity}

\chapter{Kac-Moody Algebras}
\label{Chapter:KacMoody}
 In this chapter we present the basic theory of Kac-Moody algebras, with emphasis on the affine and Lorentzian cases.\footnote{This chapter is largely based on lectures given by the author at the \emph{Third Modave Summer School on Mathematical Physics}, held in Modave, Belgium, August 2008.} Our treatment is aimed towards physicists, and to this end we do not give formal definitions, theorems or proofs, but rather we introduce the reader to a ``toolbox'', whose constituents can, in principle, be mastered relatively quickly. We presuppose a working knowledge of the theory of finite simple Lie algebras, and explain in detail how these structures generalize to arbitrary infinite-dimensional Kac-Moody algebras. The techniques presented here play a central role in the remainder of this thesis, in particular in {\bf Part I} but also to some extent in {\bf Part II}. Recommended references for this chapter are \cite{Kac,Humphreys}. 

\section{Preliminary Example: $A_1$ -- The Fundamental Building Block}
\label{Section:mother}

In this section we consider a simple ``warm-up'' example which nevertheless contains many of the important features encountered later on. Even though being the smallest finite-dimensional simple Lie algebra, $A_1$ plays an important role in the general theory of Kac-Moody algebras. In particular, one may view any simple rank $r$ Kac-Moody algebra $\mathfrak{g}$ as a set of $r$ distinct $A_1$-subalgebras which are intertwined in a non-trivial way. This fact is a cornerstone in the representation theory of Kac-Moody algebras \cite{Kac}. For this reason, $A_1$ may be described as the fundamental building block of all Lie algebras, finite- as well as infinite-dimensional. All terminology introduced in this example will be properly defined in subsequent sections.

$A_1\simeq \mathfrak{sl}(2, \mathbb{C})$ is the algebra of $2\times 2$ complex traceless matrices. This algebra is 3-dimensional and we take as a basis the set of generators $\{T_i\ |\ i=1, 2, 3\}$, subject to the commutation relations
\begin{equation}
[T_i, T_j]=-\varepsilon_{ijk} T_k,
\end{equation}
with $\varepsilon_{123}=1$. Writing this out we find the following relations between the generators:
\begin{equation}
[T_1, T_2]=-T_3, \qquad [T_1, T_3]=T_2, \qquad [T_2, T_3]=-T_1.
\label{PERTABeq:sl2relations}
\end{equation} 
In the fundamental representation we have a matrix realization of the form
\begin{equation}
T_1=\frac{1}{2}\left( \begin{array}{cc}
0 & i \\
i & 0Ê\\
\end{array} \right),
\qquad
T_2=\frac{1}{2}\left( \begin{array}{cc}
0 & 1 \\
-1 & 0Ê\\
\end{array} \right),
\qquad
T_3=\frac{1}{2}\left( \begin{array}{cc}
i & 0 \\
0 & -iÊ\\
\end{array} \right),
\end{equation}
which can easily be seen to satisfy (\ref{PERTABeq:sl2relations}). For the purposes of this thesis, it is useful to switch to another basis, where the three basis elements take the form
\begin{eqnarray}
{}Êe \ \equiv \ \phantom{-()}T_2-iT_1 &=& \left( \begin{array}{cc}
0 & \phantom{-}1 \\
0 & \phantom{-}0 \\
\end{array}Ê\right),
\nonumber\\
{}Êf \ \equiv \ -(T_2+iT_1) & =& \left( \begin{array}{cc}
0 & \phantom{-}0 \\
1 & \phantom{-}0 \\
\end{array}Ê\right),
\nonumber\\
\label{PERTABeq:chevbasisl2}
{}Êh\ \equiv\ \phantom{-(T_2} -2iT_3 &=& \left( \begin{array}{cc}
1 & \phantom{-}0 \\
0 & -1 \\
\end{array}Ê\right).
\end{eqnarray}
The new basis $\{e, f, h\}$ is similar to the familiar basis $\{J^{+}, J^{-}, J^{3}\}$  of  $\mathfrak{su}(2)$. The commutation relations now become
\begin{equation}
[e, f] = h,\qquad [h, e] = 2e,\qquad Ê[h, f] = -2f,
\label{PERTABeq:ChevalleyPresentationsl2}
\end{equation}
implying that the generators $e$ and $f$ can be thought of a ``step operators'', taking us, respectively, ``up'' and ``down'' between the lowest and highest weights of the representation. The basis $\{e, f, h\}$ is called the Chevalley basis, and (\ref{PERTABeq:ChevalleyPresentationsl2}) corresponds to a Chevalley presentation of $A_1$. There are two main reasons for working in this basis. Firstly, it is the starting point for the generalization to arbitrary Kac-Moody algebras which we shall consider in the next section. Secondly, in the Chevalley basis, the matrix realization of the generators only involves \emph{real} traceless matrices. This ensures that simply by restricting all linear combinations of generators to the real numbers, we obtain a \emph{real} Lie algebra, namely the split real form $\mathfrak{sl}(2, \mathbb{R})$, consisting of $2\times 2$ real traceless matrices. 

The Chevalley presentation reveals a natural decomposition of $\mathfrak{sl}(2, \mathbb{R})$ as the following direct sum of vector spaces
\begin{equation}
\mathfrak{sl}(2, \mathbb{R})=\mathbb{R} f\oplus \mathbb{R}h\oplus \mathbb{R}e.
\end{equation}
This is the triangular decomposition of the Lie algebra, which in this case simply means that each matrix can be decomposed as a sum of  a lower triangular, a diagonal and an upper triangular matrix. The two subspaces $\mathbb{R}f$ and $\mathbb{R}e$ form nilpotent subalgebras and $\mathbb{R}h$ form the abelian Cartan subalgebra. Of course, these concepts are all somewhat trivial in this example, but it is nevertheless useful to employ this terminology for later use.

An important concept which we shall encounter numerous times is that of a maximal compact subalgebra. It is well known that the maximal compact subalgebra of $\mathfrak{sl}(2, \mathbb{R})$ is $\mathfrak{so}(2)$, the algebra of $2\times 2$ (traceless) antisymmetric matrices. Let us now try to understand this from a more abstract point of view. The algebra $\mathfrak{so}(2)\subset \mathfrak{sl}(2, \mathbb{R})$ is a so-called involutory subalgebra, meaning that there exists an involution $\omega$ of $\mathfrak{sl}(2, \mathbb{R})$ such that $\mathfrak{so}(2)$ coincides with the set $\{x\in \mathfrak{sl}(2, \mathbb{R}) \ |\ \omega(x)=x\}$, which is pointwise fixed by $\omega$. The involution $\omega$ is called the Chevalley involution and is defined on the Chevalley generators as follows
\begin{equation}
\omega(e)=-f, \qquad \omega(f)=-e, \qquad \omega(h)=-h.
\end{equation}
This is obviously an involution, $\omega^2=1$, and it leaves the commutation relations invariant, so $\omega$ is an automorphism of $\mathfrak{sl}(2, \mathbb{R})$. 

Now note that the combination $e-f$ is fixed by $\omega$ and so spans the one-dimensional maximal compact subalgebra of $\mathfrak{sl}(2, \mathbb{R})$. Indeed we have
\begin{equation}
e-f =\left( \begin{array}{cc}
0 & 1\\
-1 & 0\\
\end{array} \right)\ \in \ \mathfrak{so}(2),
\end{equation}
so that we may write $\mathfrak{so}(2)=\mathbb{R} (e-f)$. The Chevalley involution thus induces another decomposition of $\mathfrak{sl}(2, \mathbb{R})$ into pointwise invariant and anti-invariant subsets under the action of $\omega$. Explicitly, this yields the Cartan decomposition
\begin{equation}
\mathfrak{sl}(2, \mathbb{R})=\mathbb{R}(e-f)\oplus \Big(\mathbb{R}h\oplus \mathbb{R}(e+f)\Big).
\end{equation}
It is important to note that this is a direct sum of vector spaces, and, in particular, that the anti-invariant part $\mathfrak{p}\equiv \mathbb{R}h\oplus \mathbb{R}(e+f)$ is \emph{not} a subalgebra. This is in contrast to the triangular decomposition above, for which each subspace is a subalgebra in itself. It is easy to see that $\mathfrak{p}$ does not form a subalgebra by noticing that the commutation relations do not close, 
\begin{equation}
[h, e+f]=2(e-f)\in \mathfrak{so}(2).
\end{equation}
Moreover we have
\begin{equation}
[h, e-f]=2(e+f)\in \mathfrak{p}, \qquad [e-f, e+f]=2h\in \mathfrak{p},
\end{equation}
revealing that the algebraic structure of the decomposition is 
\begin{equation}
[\mathfrak{p}, \mathfrak{p}]\subset \mathfrak{so}(2), \qquad [\mathfrak{so}(2), \mathfrak{p}]\subset \mathfrak{p}, \qquad [\mathfrak{so}(2), \mathfrak{so}(2)]\subset \mathfrak{so}(2),
\end{equation}
indicating that at the ``group level'' $\mathfrak{p}$ corresponds to a symmetric space, which in this case coincides with the coset space $SL(2, \mathbb{R})/SO(2)$. We shall come back to this in Section \ref{Section:rootsystem}. 

Finally, we consider an additional useful decomposition, known as the Iwasawa decomposition, which will play an important role in what follows. It takes the form (direct sum of vector spaces)
\begin{equation}
\mathfrak{sl}(2, \mathbb{R})=\mathfrak{so}(2)\oplus \mathbb{R}h\oplus \mathbb{R}e, 
\end{equation}
where each subspace is now a subalgebra. Note that in this decomposition, even though $\mathfrak{so}(2)$ is defined as before through the Chevalley involution, the second part, $\mathbb{R}h\oplus \mathbb{R}e$ (the Borel subalgebra) is not anti-invariant under $\omega$.

\section{Basic Definitions}
Kac-Moody algebras are infinite-dimensional generalizations of the finite simple Lie algebras, as classified by Cartan and Killing. A standard treatment of an $l$-dimensional Lie algebra is in terms of a set of generators $\{ T_m\ |\ m=1, \dots, l\}$, subject to the commutation relations
\begin{equation}
[T_m, T_n]={f_{mn}}^{p}T_p,
\end{equation}
where the structure constants ${f_{mn}}^p$ contain all the information of the algebra. This construction is not very convenient if we want to generalize it to cases when $l\rightarrow \infty$. It is for this reason that we in the previous section emphasized the importance of the Chevalley-Serre presentation, a construction which is amenable for generalization. We turn now to discuss the basic properties of Kac-Moody algebras, using the Chevalley-Serre presentation.

\subsection{The Chevalley-Serre Presentation} \label{Section:chevserpres}

Let $(e_i, f_i, h_i)$ be a triple of generators, satisfying the commutation relations of $\mathfrak{sl}(2, \mathbb{R})$:
\begin{equation}
[e_i, f_i] = h_i,\qquad [h_i, e_i] = 2e_i,\qquad Ê[h_i, f_i] = -2f_i.
\label{PERTABeq:sl2Triple}
\end{equation}
We can now construct an algebra $\tilde{\mathfrak{g}}$ by letting the index $i$ run from $1$ to $r$ and ``intertwining'' the $r$ copies of $\mathfrak{sl}(2, \mathbb{R})$ through the following relations
\begin{eqnarray}
{}Ê[e_i,f_j]&=&\delta_{ij}h_j,
\nonumber\\
{}Ê[h_i,e_j]&=&A_{ij}e_j,
\nonumber\\
{}Ê[h_i,f_j]&=&-A_{ij}f_j,
\nonumber\\
{} [h_i,h_j]&=&0.
\label{PERTABeq:Chevalleyrelations}
\end{eqnarray} 
It is important to note that here $r$ is always finite, even if the algebra itself might be infinite-dimensional. The structure of the algebra $\tilde{\mathfrak{g}}$ is encoded in the \emph{Cartan matrix} $A$, whose entries, $A_{ij}$,  determine the commutation relations between the generators of the different $\mathfrak{sl}(2, \mathbb{R})$-subalgebras. We shall discuss the Cartan matrix in more detail below, and for now we just impose that it be non-degenerate, $\text{det}\ A\neq 0$. This condition will be lifted later on. We shall often emphasize the dependence of $\tilde{\mathfrak{g}}$ on the Cartan matrix, and write $\tilde{\mathfrak{g}}(A)$. From now on we also fix the base field of $\tilde{\mathfrak{g}}$ to $\mathbb{R}$. 

Further elements of $\tilde{\mathfrak{g}}$ are obtained by taking multiple commutators as follows
\begin{equation}
[e_{i_1}, [e_{i_2}, \dots, [e_{i_{k-1}}, e_{i_k}]\cdots ]], \qquad [f_{i_1}, [f_{i_2}, \dots, [f_{i_{k-1}}, f_{i_k}]\cdots ]].
\label{PERTABeq:multiplecommutators}
\end{equation}
Note that at this point the algebra $\tilde{\mathfrak{g}}$ is infinite-dimensional because there are no relations between the $e_i\ $Êor $f_i$ that restrict these multicommutators. 

Through the use of the Chevalley relations, (\ref{PERTABeq:Chevalleyrelations}), any commutator involving the $h_i$ may be reduced to one of the form of (\ref{PERTABeq:multiplecommutators}). This gives rise to the so-called \emph{triangular decomposition} of $\tilde{\mathfrak{g}}$, which takes the form (direct sum of vector spaces)
\begin{equation}
\tilde{\mathfrak{g}}=\tilde{\mathfrak{n}}_-\oplus \mathfrak{h}Ê\oplus \tilde{\mathfrak{n}}_+. 
\label{PERTABeq:triangulardecomposition}
\end{equation}
Here, $\mathfrak{h}$ is a real vector space spanned by the $h_i$, 
\begin{equation}
\mathfrak{h}=\sum_{i=1}^{r} \mathbb{R}h_i,
\label{PERTABeq:Cartansubalgebra}
\end{equation}
which forms an abelian subalgebra of $\tilde{\mathfrak{g}}$, called the \emph{Cartan subalgebra}. The subspaces $\tilde{\mathfrak{n}}_+$ and $\tilde{\mathfrak{n}}_-$ are freely generated by the $e_i$'s and the $f_i$'s, respectively. The algebra $\tilde{\mathfrak{g}}$ is not simple, but has a maximal ideal $\mathfrak{i}$, which decomposes as a direct sum of ideals,
\begin{equation}
\mathfrak{i}=(\mathfrak{i}\cap \tilde{\mathfrak{n}}_-)\oplus (\mathfrak{i}\cap\tilde{\mathfrak{n}}_+)\equiv \mathfrak{i}_-\oplus \mathfrak{i}_+,
\end{equation}
where the two subspaces $\mathfrak{i}_{+}$ and $\mathfrak{i}_-$ are ideals in $\tilde{\mathfrak{n}}_+$ and $\tilde{\mathfrak{n}}_-$, respectively. It follows that we have
\begin{equation}
\mathfrak{i}_{\pm}\cap \mathfrak{h}=0, \qquad \mathfrak{i}_+\cap\mathfrak{i}_-=0.
\end{equation}
The two ideals $\mathfrak{i}_{\pm}$ are generated by the subsets of elements $S_{\pm}$ given by
\begin{eqnarray}
{}ÊS_+ & = & \big\{ \text{ad}_{e_i}^{1-A_{ij}}(e_j)\ \big| \ i\neq j, \ i, j =1, \dots, r\ \big\},
\nonumber\\
\nonumber\\
{}ÊS_- &=& \big\{\text{ad}_{f_i}^{1-A_{ij}}(f_j)\ \big| \ i\neq j,\ i, j=1, \dots, r\ \big\},
\end{eqnarray}
where ``$\text{ad}$'' denotes the adjoint action, i.e., for $x, y \in \tilde{\mathfrak{g}}$, $\text{ad}_x (y)=[x, y]$. We shall now take the quotient of $\tilde{\mathfrak{g}}(A)$ by the ideal $\mathfrak{i}$. This gives rise to relations among the generators of $\tilde{\mathfrak{n}}_+$ and $\tilde{\mathfrak{n}}_-$. We thereby define the \emph{Kac-Moody algebra} $\mathfrak{g}(A)$, associated with the Cartan matrix $A$, as follows
\begin{equation}
\mathfrak{g}(A)=\tilde{\mathfrak{g}}(A)\ \big/ \ \mathfrak{i}.
\end{equation}
Since we have chosen the base field to be $\mathbb{R}$, this defines the split real form of the corresponding Kac-Moody algebra over $\mathbb{C}$. 

The rank $r$ Kac-Moody algebra $\mathfrak{g}(A)$ is now generated by the $3r$ generators $e_i, f_i, h_i$ subject to the Chevalley relations, (\ref{PERTABeq:Chevalleyrelations}), and the \emph{Serre relations},
\begin{eqnarray}
{}Ê& & \text{ad}_{e_i}^{1-A_{ij}}(e_j)=[e_i, [e_i, \dots, [e_i, e_j]\cdots ]]=0,
\nonumber\\
\nonumber\\
{} & & \text{ad}_{f_i}^{1-A_{ij}}(f_j)=[f_i, [f_i, \dots, [f_i, f_j]\cdots ]] =0,
\label{PERTABeq:Serrerelations}
\end{eqnarray}
with each relation containing $1-A_{ij}$ commutators. The triangular decomposition of $\mathfrak{g}(A)$ then reads
\begin{equation}
\mathfrak{g}(A)=\mathfrak{n}_-\oplus \mathfrak{h}\oplus \mathfrak{n}_+,
\end{equation}
where 
\begin{equation}
\mathfrak{n}_{\pm}=\tilde{\mathfrak{n}}_{\pm}\ \big/ \ \mathfrak{i}_{\pm}.
\end{equation}
The Serre relations thus impose restrictions on $\mathfrak{n}_{\pm}$ which cut the chains of multiple commutators involving $e_i$ and $f_i$. These restrictions might, or might not, render the algebra $\mathfrak{g}(A)$ finite-dimensional. We shall see in the next section how this depends on the properties of the Cartan matrix.

\subsection{The Cartan Matrix and Dynkin Diagrams} 

So far we have discussed how the algebra $\mathfrak{g}(A)$ is constructed by imposing relations between the Chevalley generators $e_i, f_i, h_i$. These relations are completely determined by the entries of the matrix $A$, an important object which we shall now discuss more closely. 

An $r\times r$ matrix $A=(A_{ij})_{i,j=1, \dots, r}$ is called a \emph{generalized Cartan matrix} if it satisfies the following properties:
\begin{eqnarray}
{} & & A_{ii} \hspace{0.1 cm} Ê=\hspace{0.1 cm}  2, \quad i=1, \dots, r,
\nonumber\\
{}Ê&  & A_{ij} \hspace{0.1 cm}  =\hspace{0.1 cm} 0 \hspace{0.1 cm}   \Leftrightarrow  \hspace{0.1 cm}  A_{ji}\hspace{0.1 cm} =\hspace{0.1 cm} 0,
\nonumber\\
{} & & A_{ij}\hspace{0.1 cm}   \inÊ\hspace{0.1 cm}   \mathbb{Z}_-  \quad  (i\neq j).
\end{eqnarray}
For brevity, we shall in the following refer to $A$ simply as a Cartan matrix. The Cartan matrix is called \emph{indecomposable} if the index set $\mathcal{S}=\{1, \dots, r\}$ can not be divided into two non-empty subsets $\mathcal{I}$ and $\mathcal{J}$ such that $A_{ij}=0$ for $i\in\mathcal{I}$ and $j\in\mathcal{J}$. An important statement is then the following: \emph{when the Cartan matrix $A$ is non-degenerate, $\text{det}\ A\neq 0$, and indecomposable, the Kac-Moody algebra $\mathfrak{g}(A)$ is simple} \cite{Kac}. In the following we shall always assume that $A$ is indecomposable.

It is now possible to provide a (partial) classification of the various types of algebras $\mathfrak{g}(A)$ that can be constructed from a Cartan matrix. There exist three main classes:
\begin{itemize}
\item If $A$ is positive definite, the algebra $\mathfrak{g}(A)$ is finite-dimensional and falls under the Cartan-Killing classification, i.e., it is one of the finite simple Lie algebras $A_n, B_n, C_n,$ $D_n, G_2, F_4, E_6, E_7$ or $E_8$.
\item If $A$ is positive-semidefinite, i.e., $\text{det}\ A=0$ with one zero eigenvalue, the algebra is infinite-dimensional and is said to be an \emph{affine} Kac-Moody algebra.\footnote{Strictly speaking, in the case of $\text{det}\ A=0$ the algebra constructed from the Chevalley-Serre relations only corresponds to the \emph{derived} algebra $\mathfrak{g}^{\prime}=[\mathfrak{g}, \mathfrak{g}]$. We shall come back to this issue in Section \ref{Section:Derivation}.} All affine Kac-Moody algebras are classified \cite{Kac}. 
\item If $A$ is not part of the two classes above, the algebra $\mathfrak{g}(A)$ is infinite-dimensional and is generally called an \emph{indefinite} Kac-Moody algebra, by virtue of the fact that $A$ is of indefinite signature.
\end{itemize}
For the third class above, no general classification exists. We shall, however, mainly be interested in a subclass of the indefinite Kac-Moody algebras, corresponding to the case when the matrix $A$ has one negative eigenvalue and $r-1$ positive eigenvalues. The associated Kac-Moody algebras are called \emph{Lorentzian}, because of the signature $(-++\cdots ++)$ of $A$. A special subclass of the Lorentzian algebras, known as \emph{hyperbolic} Kac-Moody algebras, have in fact been classified. We shall define hyperbolic Kac-Moody algebras in Section \ref{Section:hyperbolic}.

Since most of the entries of the Cartan matrix are zero, it is convenient to encode the non-vanishing entries in a diagram, $\Gamma=\Gamma(A)$, called a \emph{Dynkin diagram}. To this end, we associate a node $\circ$ in $\Gamma$ to each Chevalley triple $(e_i, f_i, h_i)$, and if $A_{ij}\neq0$ for $i\neq j$ the nodes $i$ and $j$ are connected by $\text{max}(|A_{ij}|, |A_{ji}|)$ lines. In addition, when $|A_{ij}|> |A_{ji}|$ we draw an arrow from node $j$ to node $i$. Indecomposability of the Cartan matrix $A$ is equivalent to the statement that the Dynkin diagram $\Gamma(A)$ is connected. 

Let us now discuss some simple examples in order to illustrate the relation between the Kac-Moody algebra $\mathfrak{g}(A)$, its Cartan matrix $A$ and the associated Dynkin diagram $\Gamma(A)$. We begin with the simplest possible case, namely the Lie algebra $A_1=\mathfrak{sl}(2, \mathbb{R})$, discussed at length in Section \ref{Section:mother}. Here there is only one Chevalley triple, $(e, f, h)$, and consequently the Cartan matrix is just the number $(2)$, with Dynkin diagram consisting of one node $\circ$. The Lie algebra $A_2=\mathfrak{sl}(3, \mathbb{R})$, in turn, is described by the Cartan matrix $$\left( \begin{array}{cc} 2 & -1 \\ -1 & 2 \\ \end{array} \right),$$ and Dynkin diagram $\circ$---$\circ$. This corresponds to two copies of $\mathfrak{sl}(2, \mathbb{R})$ which are intertwined through the non-vanishing off-diagonal components of the Cartan matrix. In contrast, if $A_{12}=A_{21}=0$ we have a direct sum of Lie algebras $A_1\oplus A_1$ corresponding to the decomposable Cartan matrix $$\left( \begin{array}{cc} 2 & 0 \\ 0 & 2 \\ \end{array} \right),$$ with Dynkin diagram $\circ\phantom{-}\circ$. This algebra is not simple, since the two $A_1$'s constitute two non-trivial ideals. Later on we shall discuss these examples, and more involved ones, in more detail.

\subsection{The Root System and the Root Lattice}
\label{Section:rootsystem}

A very important notion in the theory of Kac-Moody algebras is that of a \emph{root}. In this section we shall develop the basic theory of roots and examine the vector space which they span, hopefully convincing the reader that these issues are extremely useful for a deeper understanding of Kac-Moody algebras. 

Let us begin by noting that, by virtue of the Chevalley relations, (\ref{PERTABeq:Chevalleyrelations}), the adjoint action of $\mathfrak{h}$ on $\mathfrak{n}_{\pm}$ is diagonal,
\begin{equation}
\text{ad}_h(e_i)=[h, e_i]=\alpha_i(h)e_i, \qquad h\in \mathfrak{h},
\label{PERTABeq:adjointaction}
\end{equation}
and similarly for the action on $f_i$. The eigenvalue $\alpha_i(h)$ represents the value of a linear map from $\mathfrak{h}$ to the real numbers,
\begin{equation}
\alpha_i\hspace{0.1 cm} :\hspace{0.1 cm} \mathfrak{h} \hspace{0.1 cm} \ni \hspace{0.1 cm} h\hspace{0.1 cm} \longmapsto \hspace{0.1 cm} \alpha_i(h)\hspace{0.1 cm} \in \hspace{0.1 cm} \mathbb{R}.
\end{equation}
The linear maps $\alpha_i\ $Êare called \emph{simple roots} and belong to the dual space $\mathfrak{h}^{\star}$. We shall sometimes employ the notation $\left<\alpha, h\right>=\alpha(h)$ for the pairing between a form $\alpha\in\mathfrak{h}^{\star}$ and a vector $h\in\mathfrak{h}$. If the eigenvalue $\alpha(h)$ vanishes, then $\alpha$ is not a root. It is also common to refer to the Cartan generators $h_i $Ê as \emph{simple coroots} to emphasize that they belong to the dual of the space of roots. In this case one also writes $\alpha_i^{\vee}\equiv h_i$. The same analysis can be performed for multiple commutators, e.g, 
\begin{eqnarray}
{}Ê[h, [e_{i}, e_{j}]] & =& -[e_i, [e_j, h]] - [e_j, [h, e_i]]
\nonumber\\
{}Ê&=& (\alpha_i+\alpha_j)(h)[e_i, e_j],
\end{eqnarray}
where in the first line we made use of the Jacobi identity. If the generator $e_{\alpha}\equiv [e_i, e_j]$ is non-vanishing, i.e., is not killed by the Serre relations, then $\alpha \equiv \alpha_i+\alpha_j$ is the root associated with $e_{\alpha}$. We denote by $\Pi=\{\alpha_1, \dots, \alpha_r\}$ the basis of simple roots and by $\Pi^{\vee}=\{\alpha_1^{\vee}, \dots, \alpha_r^{\vee}\}$ the basis of simple coroots. Any root can be expressed as an integer linear combination of the simple roots. We denote by $\Phi$ the complete set of roots. This is called the \emph{root system}. In analogy with (\ref{PERTABeq:Cartansubalgebra}) we also have
\begin{equation}
Ê\mathfrak{h}^{\star}=\sum_{i=1}^{r}\mathbb{R}\alpha_i.
\end{equation}
A root is called \emph{positive} (\emph{negative}) if it can be written as a linear combination of the simple roots $\Pi$ with only non-negative (non-positive) coefficients. From the triangular decomposition it follows that all roots are either positive or negative. Thus, the root system, $\Phi$, splits into a disjoint union of positive and negative roots,
\begin{equation}
\Phi=\Phi_+\cup \Phi_-.
\end{equation}
For $\mathfrak{g}\ni x_{\alpha}\neq 0$ the associated root 
\begin{equation}
\alpha=\sum_{i=1}^{r}m_i\alpha_i
\label{PERTABeq:rootexpansion}
\end{equation}
belongs to $\Phi_+$ if all $m_i\in \mathbb{Z}_{\geq 0}$ and to $\Phi_-$ if all $m_i\in\mathbb{Z}_{\leq 0}$. Let us for definiteness take $\alpha$ to be a positive root. Then $-\alpha$ is necessarily a negative root, and we write
\begin{eqnarray} \label{PERTABeq:positivgene}
{}Êx_{\alpha\phantom{-}}&\equiv& e_{\alpha}\hspace{0.1 cm} \in \hspace{0.1 cm} \mathfrak{n}_+,
\\
{} x_{-\alpha}&\equiv & f_{\alpha} \hspace{0.1 cm} \in \hspace{0.1 cm} \mathfrak{n}_-.
\end{eqnarray}
In the Chevalley basis the eigenvalues $\alpha_i(h)$ are always integers, called the \emph{Cartan integers}, revealing that the set of roots $\Phi$ lie on an $r$-dimensional lattice $Q$ spanned by the simple roots,
\begin{equation}
Q=\sum_{i=1}^{r}\mathbb{Z}\alpha_i\ \subset \ \mathfrak{h}^{\star}.
\end{equation}
All elements of the root system thus belong to $Q$ but the converse is not true, hence 
\begin{equation}
\Phi\subset Q.
\end{equation}
In the Cartan subalgebra $\mathfrak{h}$ we similarly have the dual notion of a \emph{coroot lattice}:
\begin{equation}
Q^{\vee}=\sum_{i=1}^{r}\mathbb{Z}\alpha_i^{\vee}\ \subset \ \mathfrak{h}.
\end{equation}

We may now decompose the algebra $\mathfrak{g}$ into disjoint subsets $\mathfrak{g}_{\alpha}\subset\mathfrak{g}$, where each subset is spanned by those generators $x\in\mathfrak{g}$, whose eigenvalue under the action of $h\in\mathfrak{h}$ is given by $\alpha(h)$. This decomposition is called the \emph{root space decomposition} and reads
\begin{equation}
\mathfrak{g}=\bigoplus_{\alpha\in\Phi} \mathfrak{g}_{\alpha},
\end{equation}
where the subspace $\mathfrak{g}_{\alpha}$ is the \emph{root space} associated to the root $\alpha$. Explicitly these are given by
\begin{equation}
\mathfrak{g}_{\alpha}=\big\{ x\in \mathfrak{g}\hspace{0.1 cm} \big| \hspace{0.1 cm} \forall\ h\in\mathfrak{h}\hspace{0.1 cm} :\hspace{0.1 cm} \text{ad}_h(x) =\alpha(h) x \big\}.
\end{equation}
Note that the zeroth subspace $\mathfrak{g}_0$ coincides with the Cartan subalgebra, $\mathfrak{g}_0=\mathfrak{h}$. Because of the disjoint split $\Phi_+\cup\Phi_-$ of the root system we can write the root space decomposition as follows
\begin{equation}
\mathfrak{g}=\mathfrak{h} \oplus \bigoplus_{\alpha\in\Phi_+}\mathfrak{g}_{\alpha} \oplus \bigoplus_{\alpha\in\Phi_+}\mathfrak{g}_{-\alpha}.
\end{equation}
The dimension of each subspace $\mathfrak{g}_{\alpha}$ is called the \emph{multiplicity}, $\text{mult}(\alpha)$, of the root $\alpha$,
\begin{equation}
\text{mult}(\alpha)\equiv \text{dim}\ \mathfrak{g}_{\alpha}.
\end{equation}
Thus, for a given root $\gamma\in\Phi$, with root space
\begin{equation}
\mathfrak{g}_{\gamma}=\mathbb{R}x^{(1)}_{\gamma}\oplus \mathbb{R}x_{\gamma}^{(2)}\oplus \cdots \oplus \mathbb{R}x_{\gamma}^{(k-1)}\oplus \mathbb{R}x_{\gamma}^{(k)},
\end{equation}
we have 
\begin{equation}
\text{mult}(\gamma)=k\hspace{0.1 cm} \in\hspace{0.1 cm} \mathbb{Z}_+ \verb|\| \{0\}.
\end{equation}
The root spaces corresponding to the simple roots are one-dimensional
\begin{equation}
\mathfrak{g}_{\alpha_i}=\mathbb{R}e_i, \qquad \mathfrak{g}_{-\alpha_i}=\mathbb{R} f_i,
\end{equation}
and, consequently, the multiplicities of the simple roots are one,
\begin{equation}
\text{mult}(\alpha_i)=1.
\end{equation}
For finite-dimensional Lie algebras the root multiplicities are always one. This does not carry over to infinite-dimensional Kac-Moody algebras, for which roots can have arbitrarily large multiplicity. We shall come back to the issue of root multiplicities in Section \ref{Section:WeylGroup} when we discuss the Weyl group. We can now write the full root system as follows
\begin{equation}
\Phi = \{\alpha\in \mathfrak{h}^{\star}\ | \ \alpha\neq 0, \ \mathfrak{g}_{\alpha}\neq 0\}.
\end{equation}
A useful notion is that of the \emph{height} of a root. This is a linear integral map 
\begin{equation}
\text{ht} : \alpha \hspace{0.1 cm} \longmapsto\hspace{0.1 cm} \text{ht}(\alpha)\in \mathbb{Z}
\end{equation}
defined as the sum of the coefficients of $\alpha$ in the basis of simple roots (see (\ref{PERTABeq:rootexpansion})),
\begin{equation}
\text{ht}(\alpha)=\sum_{i=1}^{r}m_i.
\end{equation}
It follows that for $\alpha\in\Phi_+$ we have $\text{ht}(\alpha)> 0$, and vice versa for the negative roots. An important object will turn out to be half the sum of all positive roots, 
\beq
\rho := \f{1}{2} \sum_{\al\in \Phi_+} \al.
\label{WeylVector}
\eeq

We now have a better understanding of the appearance of the Cartan matrix in (\ref{PERTABeq:Chevalleyrelations}). It simply corresponds to the values of the simple roots $\alpha_j\in \mathfrak{h}^{\star}$ acting on the simple coroots $\alpha_i^{\vee}\in\mathfrak{h}$, i.e.,
\begin{equation}
A_{ij}=\alpha_j(\alpha_i^{\vee})=\left<\alpha_j, \alpha_i^{\vee}\right>.
\label{PERTABeq:Cartanmatrix1}
\end{equation}

Finally we shall here develop a more geometric description of the root system, which is very useful for our understanding of the Kac-Moody algebra $\mathfrak{g}(A)$. An arbitrary root $\gamma\in \Phi$ may be seen as a ``vector'' in $\mathfrak{h}^{\star}$ with components given by
\begin{equation}
\gamma_i\equiv \gamma(h_i),
\end{equation}
i.e., the components of $\gamma$ correspond to the different values of the root $\gamma\ $Êacting on the simple coroots $h_i=\alpha_i^{\vee}$. We shall sometimes write the \emph{root vector} $\vec{\gamma}$ as
\begin{equation}
\vec{\gamma}=(\gamma_1, \dots, \gamma_r)\hspace{0.1 cm} \in\hspace{0.1 cm}  \mathfrak{h}^{\star}.
\end{equation}
From this point of view the entries $A_{ij}$ of the Cartan matrix correspond to the components of the root vectors $\vec{\alpha}_i$ associated with the simple roots:
\begin{eqnarray}
{}Ê\vec{\alpha}_i &=& (\alpha_{i(1)}, \alpha_{i(2)}, \dots, \alpha_{i(r)})
\nonumber\\
{}Ê& =& (A_{1i}, A_{2i}, \dots, A_{ri}),
\end{eqnarray}
where we have indicated the component index $(i)$ of the simple roots within parenthesis to distinguish it from the index $i$ labeling the different simple roots. 

We conclude this section by defining an involution $\omega$ on the Kac-Moody algebra $\mathfrak{g}(A)$, known as the \emph{Chevalley involution}. This is defined as follows on the Chevalley generators:
\begin{equation}\label{PERTABeq:Chevalleyinv}
\omega(e_i)=-f_i, \qquad \omega(f_i)=-e_i, \qquad \omega(h_i)=-h_i.
\end{equation}
This involution leaves the Chevalley relations, (\ref{PERTABeq:Chevalleyrelations}), invariant and therefore corresponds to an automorphism of $\mathfrak{g}(A)$. The involution $\om$ acts as on multiple commutators in the standard way, e.g, on $e_3 \equiv [e_1, e_2]\in\mathfrak{n}_+\subset \mathfrak{g}(A)$ one has
\begin{equation}
\omega(e_3)=\omega([e_1, e_2])=[\omega(e_1), \omega(e_2)]=[f_1, f_2]=f_3\in\mathfrak{n}_-\subset \mathfrak{g}(A).
\end{equation}
The subset of $\mathfrak{g}(A)$ which is pointwise fixed under $\omega$ defines the \emph{maximal compact subalgebra}
\begin{equation}
K(\mathfrak{g})=\big\{x\in\mathfrak{g}(A)\ \big|\ \omega(x)=x \big\}\ \subset \ \mathfrak{g}.
\end{equation}
The maximal compact subalgebra is generated by the combinations $e_i-f_i, \ i=1, \dots, r,$ of Chevalley generators. We have the induced \emph{Cartan decomposition} of $\mathfrak{g}(A)$ (direct sum of vector spaces):
\begin{equation}
\mathfrak{g}=K(\mathfrak{g})\oplus \mathfrak{p},
\end{equation}
where the complement $\mathfrak{p}$ is the subset of $\mathfrak{g}$ which is pointwise anti-invariant under $\omega$,
\begin{equation}
\mathfrak{p}=\big\{x\in\mathfrak{g}(A)\ \big|\ \omega(x)=-x \big\}.
\end{equation}
This is not a subalgebra of $\mathfrak{g}$, but elements of $\mathfrak{p}$ transforms in some representation of $K(\mathfrak{g})$. The Cartan decomposition yields the following characteristic properties of a \emph{symmetric space}:
\begin{equation}
[\mathfrak{p}, \mathfrak{p}]\subset K(\mathfrak{g}), \qquad [K(\mathfrak{g}), \mathfrak{p}]\subset \mathfrak{p}, \qquad [K(\mathfrak{g}), K(\mathfrak{g})]\subset K(\mathfrak{g}).
\end{equation}

Let us also note here an additional important decomposition of $\mathfrak{g}(A)$. This is the \emph{Iwasawa decomposition} which reads
\begin{equation}
\mathfrak{g}=K(\mathfrak{g})\oplus \mathfrak{h}\oplus \mathfrak{n}_+.
\end{equation}
In the finite-dimensional case this decomposition reduces to the familiar fact that any matrix can be decomposed into an orthogonal part, a diagonal part and an upper triangular part. The subset 
\begin{equation}
\mathfrak{b}=\mathfrak{h}\oplus \mathfrak{n}_+
\end{equation}
is known as the \emph{Borel subalgebra}. There is an alternative Iwasawa decomposition which instead utilizes the negative nilpotent subspace $\mf{n}_-$:
\beq
\mathfrak{g}= \mathfrak{n}_-\oplus \mathfrak{h}\oplus K(\mathfrak{g}),
\eeq
with an associated negative Borel subalgebra
\beq
\mf{b}_{-}=\mf{n}_{-}\oplus \mf{h}.
\eeq
The first of version of the Iwasawa decomposition will be used frequently in {\bf Part I} of this thesis, while in {\bf Part II} we will make use of the second version. This is purely a matter of convention.  

\subsection{The Invariant Bilinear Form}
\label{Section:BilinearForm}

To proceed with the analysis of the roots of a Kac-Moody algebra it is useful to first define a ``metric'' $(\cdot |\cdot )$ on the space $\mathfrak{h}^{\star}$. This will then be extended to an invariant bilinear form on the entire Kac-Moody algebra $\mathfrak{g}(A)$, and will thereby also play an important role in many of the subsequent developments. 

We shall assume, as before, that the Cartan matrix, is non-degenerate, and, in addition, we shall take it to be \emph{symmetrizable}. The first condition will be lifted later on, while the second condition will be kept throughout the remainder of this thesis.  Symmetrizability of $A$ implies that there exists a diagonal matrix $D=\text{diag}(\epsilon_1, \dots, \epsilon_r)$, with all $\epsilon_i > 0$, such that the Cartan matrix decomposes according to
\begin{equation}
A=DS,
\end{equation}
where $S$ is a symmetric $r\times r$ matrix. The matrix $S=(S_{ij})$ now defines a symmetric invertible bilinear form $(\cdot |\cdot )$ on $\mathfrak{h}^{\star}$ as follows
\begin{equation}
S_{ij}\equiv (\alpha_i|\alpha_j), 
\label{PERTABeq:bilinearformdefinition}
\end{equation}
for $\alpha_i, \alpha_j \in\Pi$. Moreover, by imposing the defining relation $A_{ii}=2$ we find
\begin{equation}
\epsilon_i = \frac{2}{(\alpha_i|\alpha_i)}.
\end{equation}

We have now defined a bilinear form on the space $\mathfrak{h}^{\star}$, which in turn induces a bilinear form on the root lattice $Q$. An important consequence of this is the following:
\begin{itemize}
\item For finite-dimensional Lie algebras the bilinear form $(\cdot |\cdot )$ is of Euclidean signature and, consequently, $Q$ is a Euclidean lattice. In this case the bilinear form coincides with the standard Killing form. 
\item For Lorentzian Kac-Moody algebras the bilinear form $(\cdot |\cdot )$ is a flat metric with signature $(-+\cdots +)$ and, consequently, $Q$ is a Lorentzian lattice. 
\end{itemize}

The bilinear form can now be extended to the full Kac-Moody algebra. Since $(\cdot | \cdot)$ is non-degenerate it defines an isomorphism $\mu : \mathfrak{h}^{\star} \rightarrow  \mathfrak{h}$ as follows:
\begin{equation}
\left<\alpha, \mu(\beta)\right> \equiv (\alpha|\beta), \qquad \beta, \alpha\in \mathfrak{h}^{\star}, \hspace{0.1 cm} \mu(\beta)\in\mathfrak{h},
\label{PERTABeq:isomorphism}
\end{equation}
with the inverse map $\mu^{-1} : \mathfrak{h}\rightarrow \mathfrak{h}^{\star}$ then defines a bilinear form $(\cdot |\cdot)$ on the Cartan subalgebra $\mathfrak{h}$ through
\begin{equation}
\left<\mu^{-1}(\alpha^{\vee}), \beta^{\vee}\right>\equiv (\alpha^{\vee}|\beta^{\vee}), \qquad \alpha^{\vee}, \beta^{\vee}\in\mathfrak{h}, \hspace{0.1 cm} \mu(\alpha^{\vee})\in \mathfrak{h}^{\star}.
\end{equation}
Then, from the definition of $(\cdot|\cdot )$ in (\ref{PERTABeq:bilinearformdefinition}), we find 
\begin{equation}
(\alpha_i|\beta)=\frac{1}{\epsilon_i}\left<\beta, \alpha_i^{\vee}\right>.
\label{PERTABeq:usefulrelation1}
\end{equation}
In addition, by virtue of (\ref{PERTABeq:isomorphism}), we have the relation
\begin{equation}
(\alpha_i|\beta)=\left<\beta, \mu(\alpha_i)\right>,
\label{PERTABeq:usefulrelation2}
\end{equation}
and by equating (\ref{PERTABeq:usefulrelation1}) with (\ref{PERTABeq:usefulrelation2}) we arrive at the explicit expressions
\begin{equation}
\mu(\alpha_i)=\frac{1}{\epsilon_i}\alpha_i^{\vee}\qquad \text{or} \qquad \mu^{-1}(\alpha_i^{\vee})=\epsilon_i\alpha_i.
\end{equation}
We can use this result to find a relation between the bilinear forms on $\mathfrak{h}$ and $\mathfrak{h}^{\star}$:
\begin{equation}
(\alpha_i^{\vee}|\alpha_j^{\vee}) = \epsilon_i\epsilon_j(\alpha_i|\alpha_j),
\end{equation}
and in the special case $i=j$ we thus have
\begin{equation}
\frac{(\alpha_i^{\vee}|\alpha_i^{\vee})}{2}=\frac{2}{(\alpha_i|\alpha_i)}.
\end{equation}
Let us further note that (\ref{PERTABeq:usefulrelation1}) ensures that the Cartan matrix can be expressed solely in terms of the bilinear form 
\begin{equation}
A_{ij}=\frac{2(\alpha_i|\alpha_j)}{(\alpha_i|\alpha_i)},
\end{equation} 
an expression which is very useful for practical purposes.

At this point we have a non-degenerate symmetric bilinear form on the Cartan subalgebra $\mathfrak{h}\subset \mathfrak{g}(A)$. To extend this to the entire algebra, one exploits the \emph{invariance} of $(\cdot |\cdot)$, i.e., the property 
\begin{equation}
([x, y], z)=(x, [y, z]), \qquad x, y, z \in \mathfrak{g}(A).
\end{equation}
For example, by computing $(\alpha_i^{\vee}|[\alpha_k^{\vee}, e_j])=A_{kj}(\alpha_i^{\vee}|e_j)$ and using the invariance on the left hand side we find
\begin{equation}
(\alpha_i^{\vee}|e_j)=0, 
\end{equation}
because of the fact that $[\alpha_i^{\vee}, \alpha_k^{\vee}]=0$. A similar argument gives $(\alpha_i^{\vee}|f_j)=0$. Moreover, by computing $([\alpha_k^{\vee}, e_i]|f_j)$ one finds
\begin{equation}
(e_i|f_j)=\epsilon_i\delta_{ij},
\end{equation}
or, more generally, for two arbitrary generators $x_{\alpha}, x_{\beta}\in \mathfrak{g}(A)$ one has
\begin{equation}
(x_{\alpha}|x_{\beta})\sim \delta_{\alpha, -\beta},
\end{equation}
where the proportionality constant depends on the normalization of the Chevalley generators. 

Before we proceed with some examples, let us discuss some additional features of the root system $\Phi$ of a Kac-Moody algebra $\mathfrak{g}(A)$. In the special case when $\mathfrak{g}(A)$ is a rank $r$ finite-dimensional Lie algebra we have seen that the root lattice is an $r$-dimensional Euclidean lattice, thus implying that all roots have positive norm, $\alpha^{2}> 0,\hspace{0.1 cm} \forall \alpha\in\Phi$. In the general case however, the root lattice can have arbitrary signature, and thus roots can in general have positive, zero or negative norm. We shall adopt the standard terminology and call roots of positive norm \emph{real} roots, and those of zero or negative norm, \emph{imaginary} roots. In this way the root system of a Kac-Moody algebra decomposes into two disjoint sets $\Phi_{\Re}$ and $\Phi_{\Im}$ of real and imaginary roots, respectively. The largest norm squared of the real roots is, by analogy with the finite-dimensional case, restricted to 2, while the imaginary roots can come with arbitrarily large negative norm squared. We can thus describe these two types of roots as follows
\begin{eqnarray}
{}Ê\Phi_{\Re}&=& \big\{\alpha\in\Phi\hspace{0.1 cm} \big|\hspace{0.1 cm} 0< (\alpha|\alpha) \leq 2 \big\},
\nonumber\\
\nonumber\\
{} \Phi_{\Im}&=& \big\{\beta\in\Phi\hspace{0.1 cm} \big|\hspace{0.1 cm} (\beta|\beta)\leq 0\big\},
\end{eqnarray}
and we have
\begin{equation}
\Phi=\Phi_{\Im}\cup\Phi_{\Re}.
\end{equation}

The multiplicity of the real roots is always one, $\text{mult}(\alpha)=1,\ \forall\alpha\in\Phi_{\Re}$, while the imaginary roots generally come with a non-trivial multiplicity, $\text{mult}(\beta)>1,\ \forall\beta\in\Phi_{\Im}$. In particular, for the indefinite Kac-Moody algebras the multiplicity of the imaginary roots grows exponentially with increasing height, thus rendering these algebras very difficult to control. We shall come back to the issue of real and imaginary roots in Section \ref{Section:WeylGroup} after we have learned some of the basic properties of the Weyl group. 

\subsection{Example: $A_2$ versus $A_1^{+}$}
\label{Section:ExampleA2A1+}

Let us now try to make all this a bit more concrete, by introducing and comparing two examples in detail. We shall consider the familiar finite-dimensional Lie algebra $A_2=\mathfrak{sl}(3, \mathbb{R})$ and the infinite-dimensional affine Kac-Moody algebra $A_1^{+}$ (the notation will be explained in Section \ref{Section:Extensions}). Our goal is firstly to understand what it is that makes the first one finite and the second one infinite-dimensional. Secondly, we shall investigate and compare the two different root systems. 

\subsubsection{Serre Relations}

\begin{figure}[ht]
\begin{center}
\includegraphics[width=70mm]{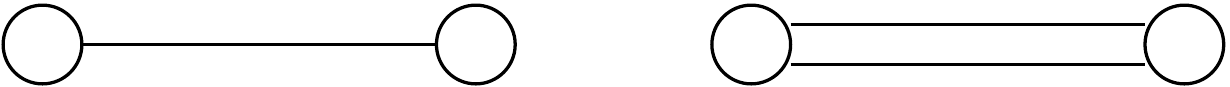}
\caption{On the left the Dynkin diagram of the Lie algebra $A_2=\mathfrak{sl}(3, \mathbb{R})$ and on the right the Dynkin diagram of the affine Kac-Moody algebra $A_1^{+}$.} 
\label{figure:A2&A1+}
\end{center}
\end{figure}

The rank 2 Lie algebras $A_2$ and $A_1^{+}$ are described by the Cartan matrices
\begin{equation}
A[A_2]=\left(\begin{array}{cc}
2 & -1 \\
-1 & 2 \\
\end{array} \right),\qquad A[A_1^{+}]=\left(\begin{array}{cc}
2 & -2 \\
-2 & 2 \\
\end{array}Ê\right),
\end{equation}
with the associated Dynkin diagrams displayed in Figure \ref{figure:A2&A1+}. For simplicity of notation, we shall refer to the two different Cartan matrices simply as $A=A[A_2]$ and $\bar{A}=A[A_1^{+}]$.

The Chevalley generators for $A_2$ are $\{e_1, e_2, f_1, f_2, h_1, h_2\}$ and the ones for $A_1^{+}$ are $\{\bar{e}_1, \bar{e}_2, \bar{f}_1,$ $ \bar{f}_2, \bar{h}_1, \bar{h}_2\}$. The commutation relations follow from the general form of the Chevalley relations in (\ref{PERTABeq:Chevalleyrelations}), with the insertion of the individual Cartan matrix components. For example, we have
\begin{eqnarray}
{} A_2 &: & [h_1, h_2]=0, \quad [h_1, e_2]=-e_2, \quad [h_2, e_1]=-e_1, \quad [e_1, f_1]=h_1,
\nonumber\\
\nonumber\\
{}ÊA_1^{+} & : & [\bar{h}_1, \bar{h}_2]=0, \quad [\bar{h}_1, \bar{e}_2]=-2\bar{e}_2, \quad [\bar{h}_2, \bar{e}_1]=-2\bar{e}_1, \quad [\bar{e}_1, \bar{f}_1]=\bar{h}_1.
\end{eqnarray}
It is clear that the relations in $A_2$ are remarkably similar to those in $A_1^{+}$ with the only difference arising in the relations involving the off-diagonal entries of the Cartan matrices, which, of course, is the only obvious distinction between the algebras at this point. We now want to understand how this seemingly trivial change in the Cartan matrix can render an algebra infinite-dimensional. The answer lies in the Serre relations. 

Let us now proceed to check the Serre relations involving the positive nilpotent generators of ``$e$-type''. The analysis is analogous for the negative ones. For $A_2$ we then have
\begin{equation}
\text{ad}_{e_1}^{1-A_{12}}(e_2)=[e_1, [e_1, e_2]]=0, 
\label{PERTABeq:A2Serrerelation}
\end{equation}
implying that the generator $[e_1, [e_1, e_2]]$ does not exist in the algebra. The commutators $e_3\equiv [e_1, e_2]$, on the other hand, is not killed by (\ref{PERTABeq:A2Serrerelation}) and so corresponds to a new generator. On the negative side, we similarly find the new generator $f_3\equiv -[f_1, f_2]$. No other non-vanishing generators exist in $A_2$ and therefore the algebra is eight-dimensional. We may take as a basis of $A_2$ the eight elements $\{e_1, e_2, e_3, f_1, f_2, f_3, h_1, h_2\}$. This corresponds to the adjoint representation ${\bf 8}$ of $\mathfrak{sl}(3, \mathbb{R})$.

We now turn to $A_1^{+}$. The Serre relation for $\bar{e}_1$ and $\bar{e}_2\ $Êreads
\begin{equation}
\text{ad}_{\bar{e}_1}^{1-\bar{A}_{12}}(\bar{e}_2)=[\bar{e}_1, [\bar{e}_1, [\bar{e}_1, \bar{e}_2]]]=0.
\label{PERTABeq:A1+Serrerelation}
\end{equation}
This condition therefore kills the generator $[\bar{e}_1, [\bar{e}_1, [\bar{e}_1, \bar{e}_2]]]$ in $A_1^{+}$, while there are no restrictions on the following two generators:
\begin{equation}
\bar{e}_3\equiv [\bar{e}_1, \bar{e}_2], \qquad \bar{e}_4=[\bar{e}_1, [\bar{e}_1, \bar{e}_2]].
\end{equation}
In addition, we have the Serre relation for $\bar{e}_2$ acting on $\bar{e}_1$ which yields yet another non-vanishing generator
\begin{equation}
\bar{e}_5\equiv [\bar{e}_2, [\bar{e}_2, \bar{e}_1]].
\end{equation}
It is the existence of $\bar{e}_4$ and $\bar{e}_5$ which renders $A_1^{+}$ infinite-dimensional. For example, consider the following multicommutator, alternating between $\bar{e}_1$ and $\bar{e}_2$,
\begin{equation}
[\bar{e}_1, [\bar{e}_2, [\bar{e}_1, \bar{e}_2]]]=-[\bar{e}_1, [\bar{e}_2, [\bar{e}_2, \bar{e}_1]]\neq 0.
\end{equation}
In $A_2$ this commutator would have been zero because the Serre relations impose $$[e_2, [e_2, e_1]]=0,$$ while in $A_1^{+}$ it corresponds to $-[\bar{e}_1, \bar{e}_5]$ which is unrestricted. It is possible to continue in this way, and any alternating multicommutator is non-vanishing, e.g,
\begin{equation}
[\bar{e}_1, [\bar{e}_2, [\bar{e}_1, [\bar{e}_2, \dots, [\bar{e}_1, \bar{e}_2]\cdots ]]]]\hspace{0.1 cm} \in \hspace{0.1 cm} A_1^{+}.
\end{equation}

\subsubsection{Root Systems}

We shall now proceed to compare the two algebras $A_2$ and $A_1^{+}$ at the level of their respective systems of roots. To $A_2$ we associate the simple roots $\Pi=\{\alpha_1, \alpha_2\}$, and to $A_1^{+}$, $\bar{\Pi}=\{\bar{\alpha}_1, \bar{\alpha}_2\}$. Since the Cartan matrices are symmetric, they give directly the bilinear forms on the space of roots. We have 
\begin{eqnarray}
{} & &Ê (\alpha_1|\alpha_1)=2, \quad (\alpha_2| \alpha_2)=2, \quad (\alpha_1| \alpha_2)=-1,
\nonumber\\
{}Ê& & (\bar{\alpha}_1|\bar{\alpha}_1)=2, \quad (\bar{\alpha}_2|\bar{\alpha}_2)=2, \quad (\bar{\alpha}_1|\bar{\alpha}_2)=-2.
\label{PERTABeq:scalarproductsA1+&A2}
\end{eqnarray}
All roots can be described as integral non-negative or non-positive linear combinations of the simple roots. For $A_2$, we find that $\alpha_1+\alpha_2$ is a root, because the corresponding generator $e_3\equiv [e_1, e_2]$ survives the Serre relations. Thus we define $\alpha_3\equiv \alpha_1+\alpha_2$. Let is now try to add yet another simple root and take, say, $\alpha_3+\alpha_2$. This corresponds to the generator $[e_3, e_2]=[[e_1, e_2], e_2]$ which is zero in $A_2$ because of the Serre relations. In this way we find that the root system $\Phi=\Phi(A)$ of $A_2$ is given by 
\begin{equation}
\Phi=\Phi_+\cup \Phi_-=\big\{ \alpha_1, \alpha_2, \alpha_3\big\}\cup \big\{ -\alpha_1, -\alpha_2, -\alpha_3\},
\end{equation}
revealing that indeed the root system of $A_2$ is finite. Of course, any vector $m \alpha_1+n \alpha_2, \ m, n\in\mathbb{Z}, $ lies on the root lattice $Q$ of $A_2$, even though it is not a root. Because the root $\alpha_3$ is the root with largest height, $\text{ht}(\alpha_3)=2$, of $\Phi$, it is the highest root of the algebra. This is also the highest weight of the adjoint representation. 

\begin{figure}[ht]
\begin{center}
\includegraphics[width=80mm]{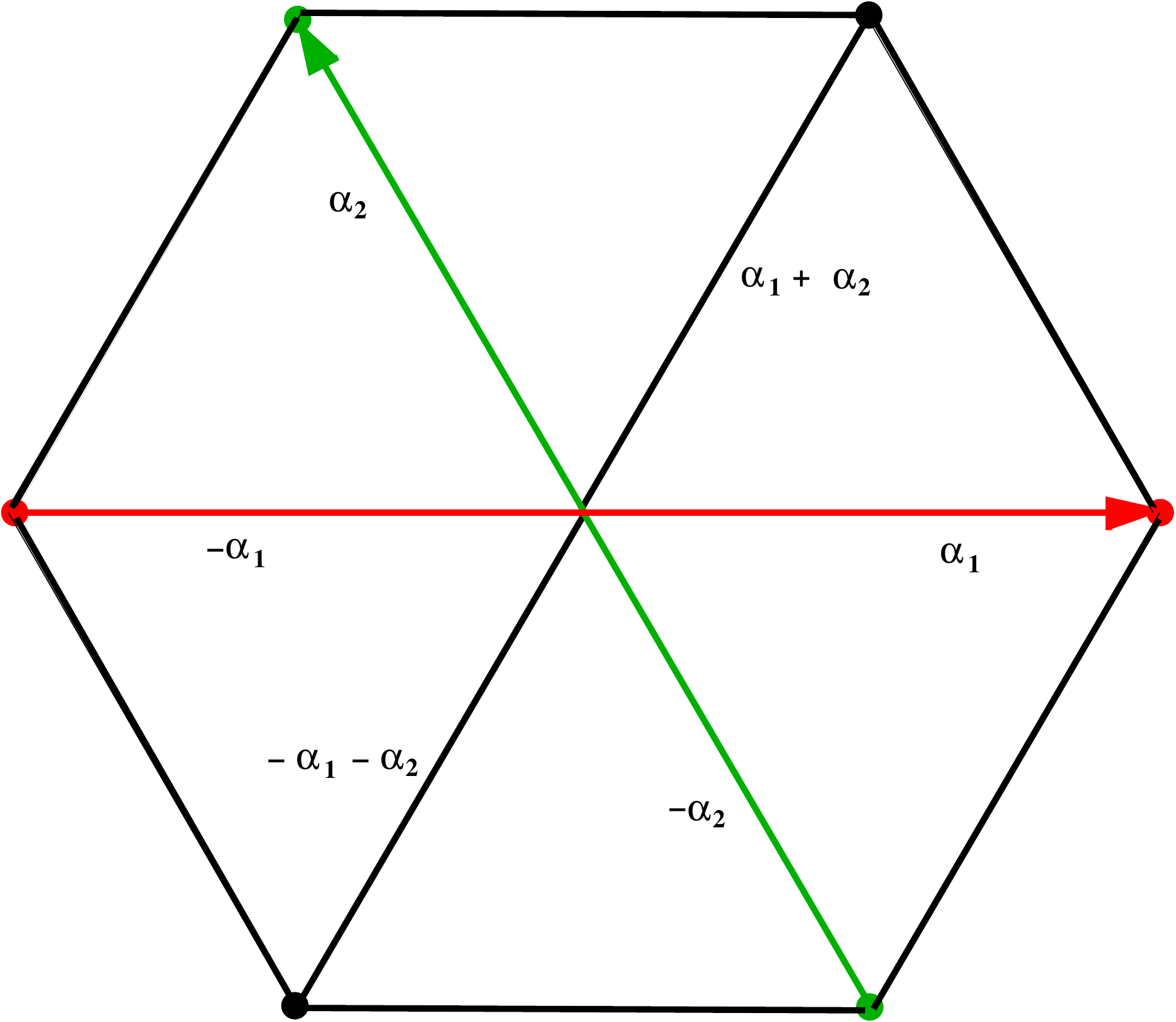}
\caption{The root diagram of $A_2$, representing the adjoint representation ${\bf 8}$. The root $\alpha_1+\alpha_2$ is the highest root corresponding to the highest weight of the representation. } 
\label{figure:RootDiagramA2}
\end{center}
\end{figure}

It is illuminating to draw the root system in a root diagram, which makes it easier to visualize the structure of the algebra. To this end we define, as described in the previous section, the simple root vectors $\vec{\alpha}_1$ and $\vec{\alpha}_2$, with components
\begin{eqnarray}
{}Ê& & \vec{\alpha}_1=(A_{11}, A_{12})=(2, -1),
\nonumber\\
{}Ê& &\vec{\alpha}_2=(A_{21}, A_{22})=(-1, 2).
\end{eqnarray}
The two vectors $\vec{\alpha}_1$ and $\vec{\alpha}_2$ span a two-dimensional Euclidean lattice, with separating angle of $2\pi/3$. We have indicated the root diagram of $A_2\ $Êin Figure \ref{figure:RootDiagramA2}.
Let us now analyze the root system $\bar{\Phi}=\Phi(\bar{A})$ of $A_1^{+}$. We begin by noting that the determinant of the Cartan matrix vanishes
\begin{equation}
\text{det} \left(\begin{array}{cc}
2 & -2 \\
-2 & 2 \\
\end{array}Ê\right)=0,
\label{PERTABeq:determinantA1+}
\end{equation}
as we have seen is the distinguishing feature of affine Kac-Moody algebras. This implies that the bilinear form of the algebra constructed from $A$ is degenerate. We shall discuss how to deal with this feature in Section \ref{Section:Derivation}. For our present purposes, however, we just note that (\ref{PERTABeq:determinantA1+}) implies that there exists a root  $\bar{\delta}\in \bar{\Phi}$, which has zero norm,
\begin{equation}
(\bar{\delta}|\bar{\delta})=0.
\end{equation}
In terms of the simple roots, we have 
\begin{equation}
\bar{\delta}=\bar{\alpha}_1+\bar{\alpha}_2, 
\end{equation}
as follows from (\ref{PERTABeq:scalarproductsA1+&A2}). That $\bar{\delta}$ is indeed a root of $\bar{\Phi}$ can be seen by noting that the associated generator $[\bar{e}_1, \bar{e}_2]$ is non-vanishing. The existence of a null root is indicative of the fact that the algebra, as well as the associated root system, is infinite-dimensional. 

In order to understand the root system it will prove convenient to write $\bar{\alpha}_2=\bar{\delta}-\bar{\alpha}_1$, and treat the root system as a two-dimensional \emph{Lorentzian} space, with basis vectors $\bar{\alpha}_1$ and $\bar{\delta}$. Is $\bar{\delta}-2\bar{\alpha}_1$ a root of the algebra? The answer is no, because the associated generator vanishes,
\begin{equation}
\bar{e}_{\bar{\delta}-2\bar{\alpha}_1}\equiv [\bar{e}_{\bar{\delta}-\bar{\alpha}_1}, \bar{f}_1]=[\bar{e}_2, \bar{f}_1]=0,
\end{equation}
as follows from the Chevalley relations. However, $2\bar{\delta}-\bar{\alpha}_1=\bar{\alpha}_1+2\bar{\alpha}_2$ is a root since it corresponds to the generator $[\bar{e}_2 [\bar{e}_2, \bar{e}_1]] \neq 0$. One can iterate this procedure and find a complete description of the root system. All null roots are multiples of $\bar{\delta}$, while the real roots are combinations of $\pm\bar{\alpha}_1$ with $\bar{\delta}$. As discussed in the previous section, the root system thereby splits into disjoint sets corresponding to the \emph{real} (i.e., spacelike) and the \emph{imaginary} (i.e., lightlike) roots. Explicitly, the root system of $A_1^{+}$ then reads
\begin{equation}
Ê\bar{\Phi} =  \bar{\Phi}_{\Re}\cup\bar{\Phi}_{\Im}= \big\{ \pm\bar{\alpha}_1 +n\bar{\delta}\hspace{0.1 cm} \big|\hspace{0.1 cm} n\in \mathbb{Z}\big\}\cup\big\{ k\bar{\delta}\hspace{0.1 cm} \big|\hspace{0.1 cm} k\in\mathbb{Z}\ \verb|\| \ \{0\} \big\}.
\end{equation}
In addition we have, of course, the usual split of $\bar{\Phi}$ into positive and negative roots.

\subsection{The Weyl Group}
\label{Section:WeylGroup}

A very important concept, which we shall use extensively in subsequent sections, is that of the \emph{Weyl group} $\mathcal{W}=\mathcal{W}(A)$ of the Kac-Moody algebra $\mathfrak{g}(A)$. We begin by constructing the group $\mathcal{W}(A)$ abstractly and then we show how it is related to a Kac-Moody algebra. Fix a set of generators $\mathcal{S}=\{s_1, \dots, s_r\}$ and let $\tilde{\mathcal{W}}$ be the free group generated by $\mathcal{S}$. Let $m=(m_{ij})_{i, j=1, \dots, r}$ be an $r\times r$ matrix satisfying: $(i)$ $m_{ii}=1$, $(ii)$ $m_{ij}\in\mathbb{Z}_{\geq 1}, \ i\neq j, $ and $(iii)$ $m_{ij}=m_{ji}$. The group $\tilde{\mathcal{W}}$ then has a normal subgroup $\mathcal{N}$ generated by the particular combinations \cite{Humphreys}
\begin{equation}
(s_i s_j)^{m_{ij}}.
\end{equation}
The Weyl group $\mathcal{W}$, associated to the set $\mathcal{S}$, is the quotient group
\begin{equation}
\mathcal{W}\equiv \tilde{\mathcal{W}}\ \big/ \ \mathcal{N}.
\end{equation}
This is a particular instance of a \emph{Coxeter group}, and the entries of the matrix $(m_{ij})$ are called \emph{Coxeter exponents}. Our construction implies that the Weyl group, $\mathcal{W}$, is the group generated by the set $\mathcal{S}$ modulo the relations
\begin{equation}
(s_is_j)^{m_{ij}}=1, \quad i, j=1, \dots, r.
\label{PERTABeq:Coxeterrelations}
\end{equation}
The elements $s_i\in\mathcal{S}$ are called the \emph{fundamental reflections}, by virtue of the property
\begin{equation}
s_i^{2}=1,
\end{equation}
as follows from the first condition on $(m_{ij})$ above.

We now show how the group $\mathcal{W}(A)$ enters the story of Kac-Moody algebras. The Weyl group is a group of automorphisms of the root lattice $Q$, 
\begin{equation}
\mathcal{W}\hspace{0.1 cm} : \hspace{0.1 cm} Q \hspace{0.1 cm} \longrightarrow \hspace{0.1 cm} Q,
\end{equation}
with the fundamental reflections $s_i$ being geometrically realized as reflections in the hyperplanes orthogonal to the simple roots $\alpha_i$. More specifically, we associate a fundamental reflection $s_i$ to each simple root $\alpha_i$, such that the action on $\gamma\in Q $ is given by
\begin{equation}
s_i\hspace{0.1 cm} : \hspace{0.1 cm} \gamma \hspace{0.1 cm}\hspace{0.1 cm} \longmapsto \hspace{0.1 cm} s_i(\gamma)= \gamma-\left<\gamma,\alpha_i^{\vee}\right>\alpha_i.
\label{PERTABeq:Weylreflection}
\end{equation}
It is clear from this definition that $s_i$ reverses the sign of $\alpha_i$ and pointwise fixes the hyperplane 
\begin{equation}
T_i=\big\{\beta\in Q\ \big|\ \left<\beta, \alpha_i^{\vee}\right>=0\big\}. 
\end{equation}
Let us also check the condition $s_i^2=1$ explicitly for the geometric realization. Applying (\ref{PERTABeq:Weylreflection}) twice on $\gamma\in Q$ yields
\begin{eqnarray}
{}Ês_i\cdot s_i(\gamma) &=& s_i\big(\gamma-\left<\gamma,\alpha_i^{\vee}\right>\alpha_i\big)
Ê\nonumber\\
{}Ê&=& \gamma-2\left<\gamma, \alpha_i^{\vee}\right>\alpha_i+\left<\gamma, \alpha_i^{\vee}\right>\left<\alpha_i, \alpha_i^{\vee}\right>\alpha_i
\nonumber\\
{}Ê&=& \gamma,
\end{eqnarray}
where, in the last step, we made use of the fact that $\left<\alpha_i, \alpha_i^{\vee}\right>=A_{ii}=2$. 

When realized geometrically in this way, the fundamental reflections are commonly called \emph{Weyl reflections}. When acting on the simple roots themselves, the Weyl reflections become
\begin{equation}\label{PERTABeq:weyreflct}
s_i(\alpha_j)=\alpha_j-\frac{2(\alpha_i|\alpha_j)}{(\alpha_i|\alpha_i)}\alpha_i=\alpha_j-A_{ij}\alpha_i,
\end{equation}
where $(A_{ij})$ is the Cartan matrix. There is a simple relation between the entries of the Cartan matrix and the associated Coxeter exponents. This is displayed inÊ Table \ref{PERTABtable:CoxeterExponents}.
\begin{table}
\begin{center}
\begin{tabular}{ |m{20mm}|m{18mm}|m{18mm}|m{18mm}|m{18mm}|m{18mm}|}
\hline
$A_{ij} A_{ji} $ & $0 $ & $1$ &$ 2$ & $3 $& $\geq 4$ \\
    \hline
$m_{ij}$&$ 2 $& $3 $&$ 4 $&$ 6 $& $\infty $ \\

    \hline
 \end{tabular}
         \caption{The relation between the entries $A_{ij}$ of the Cartan matrix and the Coxeter exponents $m_{ij}$.}
\label{PERTABtable:CoxeterExponents}
\end{center}
\end{table}
When $A_{ij}A_{ji}\geq 4$ the corresponding Coxeter exponents are infinite, implying that there are no relations between the generators $s_i$ and $s_j$ for these particular values of $i$ and $j$.

The bilinear form $(\cdot |\cdot)$ is $\mathcal{W}$-invariant,
\begin{equation}
\big(\omega(\beta)\big|\omega(\beta^{\prime})\big)=\big(\beta\big|\beta^{\prime}\big),
\end{equation}
which follows by direct calculation using (\ref{PERTABeq:Weylreflection}). This implies that the Weyl group is ``orthogonal'' with respect to the bilinear form $(\cdot |\cdot)$ and hence is a discrete subgroup of the isometry group $O(\mathfrak{h}^{\star})$ of $\mathfrak{h}^{\star}$,
\begin{equation}
\mathcal{W}\subset O(\mathfrak{h}^{\star}).
\end{equation}
 
We can now make use of the Weyl group to get a better handle on the structure of the root system. The first important fact is that the root system $\Phi(A)$ of a Kac-Moody algebra $\mathfrak{g}(A)$ is $\mathcal{W}(A)$-invariant, 
\begin{equation}
\mathcal{W}\cdot \Phi=\Phi.
\end{equation}
We can associate a general Weyl reflection $\omega_{\alpha}$ to any root $\alpha\in\Phi$. This will be described by a finite product of the fundamental reflections,
\begin{equation}
\omega_{\alpha}=s_{i_1}s_{i_2}\cdots s_{i_{k-1}}s_{i_k},
\end{equation}
where the minimal number $k$ of fundamental reflections needed to describe $\omega_{\alpha}$ is called the \emph{length} of $\omega_{\alpha}$, and is denoted by $\ell(\alpha)$. By definition, the fundamental reflections have length one, $\ell(s_i)=1$. 

The reflection $\omega_{\alpha}\in\mathcal{W}$ fixes the hyperplane $T_{\alpha}\subset \mathfrak{h}^{\star}$, 
\begin{equation}\label{PERTABeq:planinvinv}
T_{\alpha}=\big\{ \gamma\in\mathfrak{h}^{\star}\hspace{0.1 cm} \big|\hspace{0.1 cm} (\gamma|\alpha)=0, \hspace{0.1 cm} \alpha\in\Phi\big\},
\end{equation}
orthogonal to $\alpha$. By removing all such hyperplanes we may decompose $\mathfrak{h}^{\star}$ into connected subsets, called \emph{chambers}. We choose one such subset $\mathcal{C}\subset \mathfrak{h}^{\star}$ and give it a distinguished status as the \emph{fundamental chamber}. The fundamental chamber is conventionally chosen as the region enclosed by the hyperplanes $T_i$ orthogonal to the simple roots. This implies that $\mathcal{C}$ contains all vectors $\gamma\in\mathfrak{h}^{\star}$ such that $(\gamma|\alpha_i)$ is positive,
\begin{equation}
\mathcal{C}=\big\{ \gamma\in \mathfrak{h}^{\star}\hspace{0.1 cm} |\hspace{0.1 cm} (\gamma|\alpha_i)>0, \forall \ \alpha_i\in\Pi\big\}.
\end{equation}
The chambers in $\mathfrak{h}^{\star}$ correspond to the images $\omega(\mathcal{C})$ for $\omega\in\mathcal{W}$. The union $\mathcal{X}$ of all such images is called the \emph{Tits cone}, 
\begin{equation}
\mathcal{X}=\bigcup_{\omega\in\mathcal{W}}\omega(\mathcal{C}).
\end{equation}
For finite-dimensional Lie algebras, the Tits cone coincides with the space $\mathfrak{h}^{\star}$, while in general, when $\Phi_{\Im}\neq \emptyset$, one has $\mathcal{X}\subset \mathfrak{h}^{\star}$. 

We are now at a stage where we can describe the root system $\Phi$ with more precision. A first important fact is that the sets of real and imaginary roots are separately invariant under the Weyl group,
\begin{eqnarray}
{}Ê\mathcal{W}\cdot \Phi_{\Re}&=& \Phi_{\Re},
\nonumber\\
\nonumber\\
{}Ê\mathcal{W}\cdot \Phi_{\Im}&=& \Phi_{\Im}.
\end{eqnarray}
The sets of real and imaginary roots have a decomposition into disjoint sets of positive and negative roots, and we write
\begin{eqnarray}
{}Ê\Phi_{\Re}&=& \Phi_{\Re +}\cup \Phi_{\Re -},
\nonumber\\
\nonumber\\
{}Ê\Phi_{\Im}&=& \Phi_{\Im +}\cup \Phi_{\Im -},
\end{eqnarray}
with 
\begin{equation}
\Phi_{\Re +}\cap \Phi_{\Im +}=0, \qquad \Phi_{\Re -}\cap \Phi_{\Im -}=0.
\end{equation}
The only root $\alpha\in \Phi_{\Re +}$ for which $s_i(\alpha)\in \Phi_{\Re -}$ is $\alpha=\alpha_i$, implying that 
\begin{equation}
s_i\cdot \Phi_{\Re +}/\{\alpha_i\} =\Phi_{\Re +}/\{\alpha_i\},
\end{equation}
and similarly for the negative real roots. A consequence of this is that, since $\alpha_i\notin\Phi_{\Im}$, the positive and negative imaginary roots are separately invariant under the Weyl group
\begin{equation}
\mathcal{W}\cdot \Phi_{\Im +}=\Phi_{\Im +}, \qquad \mathcal{W}\cdot \Phi_{\Im -}=\Phi_{\Im -}.
\end{equation}

We can now state which elements of the root lattice $Q$ are actually roots. First, all real roots lie in Weyl orbits of the simple roots, and thus we have
\begin{equation}
\Phi_{\Re}=\bigcup_{\omega\in\mathcal{W}}\omega(\Pi).
\end{equation}
We now want to find a similar description for the set of imaginary roots. To this end it is useful to first introduce the notion of the \emph{support}, $\text{supp}(\alpha)$, of an element $\alpha\in Q$. Let 
\begin{equation}
\alpha=\sum_{i=1}^{r}k_i\alpha_i\hspace{0.1 cm} \in \hspace{0.1 cm} Q,
\end{equation}
and introduce the subdiagram $\Xi_{\alpha}(A)\subset \Gamma(A)$, of the Dynkin diagram $\Gamma(A)$, as the diagram consisting only of those vertices $i$ for which $k_i\neq 0$ and of all lines joining these vertices. We then have 
\begin{equation}
\text{supp}(\alpha)\equiv \Xi_{\alpha}(A).
\end{equation}
Next, we define a region $\mathcal{K}\subset \Phi_{\Im +}$ as follows
\begin{equation}
\mathcal{K}=\big\{\alpha\in Q_+ \hspace{0.1 cm} \big|\hspace{0.1 cm} \left<\alpha, \alpha_i^{\vee}\right>\leq 0,\ \forall\ i\hspace{0.1 cm} :\hspace{0.1 cm} \Xi_{\alpha}(A)\ \text{is}\ \text{connected}\big\}.
\end{equation}
The set of positive imaginary roots can now be elegantly described as the union of all images of $\mathcal{K}$ under the Weyl group \cite{Kac},
\begin{equation}
\Phi_{\Im +}=\bigcup_{\omega\in\mathcal{W}}\omega(\mathcal{K}).
\end{equation}
with a similar description of the negative imaginary roots. 

Through the aid of the Weyl group $\mathcal{W}(A)$ we have now obtained a complete description of the root system $\Phi(A)$ of any Kac-Moody algebra $\mathfrak{g}(A)$. In subsequent sections we shall discuss in more detail the root systems for the classes of affine and hyperbolic Kac-Moody algebras, for which some simplifications arise.   

We have defined the Weyl group as the group of reflections with respect to the simple roots. Through a natural generalization of (\ref{PERTABeq:Weylreflection}) one can define reflections $s_{\alpha}\in\mathcal{W}$ through any real root $\alpha\in\Phi_{\Re}$. These reflections act as follows on $\beta\in\mathfrak{h}^{\star}$:
\begin{equation}
s_{\alpha}(\beta)=\beta- \left<\beta, \alpha^{\vee}\right>\alpha=\beta-\frac{2(\beta|\alpha)}{\alpha|\alpha)}\alpha.
\label{PERTABeq:generalizedreflection}
\end{equation}
How are these reflections related to the fundamental reflections $s_i\equiv s_{\alpha_i}$? To answer this question we first note that since $\alpha$ is real we must have $\alpha=\omega(\alpha_i)$ for some $\omega\in\mathcal{W}$ and some $\alpha_i\in \Pi$. Inserting $\alpha=\omega(\alpha_i)$ into (\ref{PERTABeq:generalizedreflection}) then yields
\begin{eqnarray}
{}Ês_{\alpha}(\beta)&=&\beta-\frac{2(\beta|\omega(\alpha_i))}{(\omega(\alpha_i)|\omega(\alpha_i))}\omega(\alpha_i)
\nonumber\\
{}Ê& =& \beta-\frac{2(\omega^{-1}(\beta)|\alpha_i)}{(\alpha_i|\alpha_i)}\omega(\alpha_i)
\nonumber\\
{}Ê&=& \beta-\left<\omega^{-1}(\beta), \alpha_i^{\vee}\right>\omega(\alpha_i).
\label{PERTABeq:generalizedreflection2}
\end{eqnarray}
where we made use of the invariance of $(\cdot |\cdot)$ under the Weyl group. We can rewrite (\ref{PERTABeq:generalizedreflection2}) as 
\begin{eqnarray}
{}Ês_{\alpha}(\beta)&=&\omega\big(\omega^{-1}(\beta)-\left<\omega^{-1}(\beta), \alpha_i^{\vee}\right> \alpha_i\big),
\nonumber\\
{} &=& \omega\cdot  s_i\big( \omega^{-1}(\beta)\big),
\end{eqnarray}
revealing that the generalized reflection $s_{\alpha}$ corresponds to a conjugation of the fundamental reflection $s_i$ by some element $\omega\in\mathcal{W}$:
\begin{equation}
s_{\alpha}=\omega s_i \omega^{-1}\hspace{0.1 cm} \in\hspace{0.1 cm} \mathcal{W}.
\end{equation}

\subsection{Regular Subalgebras of Kac-Moody Algebras}
\label{regular}
In this subsection we discuss some general properties of regular subalgebras of Kac-Moody algebras. The notion of regular subalgebra will be important throughout {\bf Part I} of this thesis. This subsection is based on {\bf Paper I}.

Let $\mf{g}$ be a Kac--Moody algebra, and let $\bar{\mathfrak{g}}$ be a
subalgebra of $\mathfrak{g}$ with triangular decomposition
$\bar{\mathfrak{g}}= \bar{\mf{n}}_- \oplus \bar{\mathfrak{h}} \oplus
\bar{\mf{n}}_+ $. We assume that $\bar{\mathfrak{g}}$ is canonically
embedded in $\mathfrak{g}$, i.e., that the Cartan subalgebra \index{Cartan subalgebra}
$\bar{\mathfrak{h}}$ of $\bar{\mathfrak{g}}$ is a subalgebra of
the Cartan subalgebra $\mathfrak{h}$ of $\mathfrak{g}$,
so that
$\bar{\mathfrak{h}}= \bar{\mathfrak{g}} \cap \mathfrak{h}$. We
say that $\bar{\mathfrak{g}}$ is {\it regularly embedded} in
$\mathfrak{g}$ (and call it a ``regular subalgebra'') if and only if two
conditions are fulfilled: 
\begin{enumerate}
\item the root generators of
$\bar{\mathfrak{g}}$ are root generators of $\mathfrak{g}$,
\item the simple roots of $\bar{\mathfrak{g}}$ are real roots of
$\mathfrak{g}$.
\end{enumerate}
It follows that the Weyl group of
$\bar{\mathfrak{g}}$ is a subgroup of the Weyl group of
$\mathfrak{g}$ and that the root lattice of $\bar{\mathfrak{g}}$
is a sublattice of the root lattice of $\mathfrak{g}$. The second condition is automatic in the finite-dimensional case
where there are only real roots, while it must be separately imposed in
the general case. 

From condition (i) it follows that the
positive root generators of the subalgebra $\bar{\mathfrak{g}}$ might be either positive or negative root generators of $\mathfrak{g}$. In what follows we shall consider only
the case when the positive (respectively, negative) root generators
of $\bar{\mathfrak{g}}$ are also positive (respectively, negative)
root generators of $\mathfrak{g}$, so that $\bar{\mf{n}}_- =
\mf{n}_- \cap \bar{\mathfrak{g}}$ and $\bar{\mf{n}}_+ = \mf{n}_+
\cap \bar{\mathfrak{g}}$. We refer to this as a {\it positive (resp. negative) regular embedding}.

In the finite-dimensional case, there is a useful criterion to
determine regular algebras from subsets of roots. This criterion,
which does not use the highest root, has been generalized to
Kac--Moody algebras in~\cite{Feingold}. It covers also non-maximal
regular subalgebras \index{regular subalgebra} and is captured by the following theorem:

\begin{theorem}
  Let $\Phi_{\Re+}$ be the
  set of positive real roots of a Kac--Moody algebra $\mf{g}$. Let
  $\al_1, \cdots, \al_n \in \Phi_{\Re+}$ be chosen such that none
  of the differences $\al_i - \al_j$ is a root of $\mf{g}$. Assume
  furthermore that the $\al_i$'s are such that the matrix $\bar{A}
  =[\bar{A}_{ij}] = [2 \left( \al_i \vert \al_j \right) /\left(\al_i \vert
    \al_i \right)]$ has non-vanishing determinant. For each $1 \leq i
  \leq n$, choose non-zero root vectors $E_i$ and $F_i$ in the
  one-dimensional root spaces corresponding to the positive real
  roots $\al_i$ and the negative real roots $-\al_i$, respectively,
  and let $H_i = [E_i, F_i]$ be the corresponding element in the
  Cartan subalgebra of $\mf{g}$. Then, the (regular) subalgebra of
  $\mf{g}$ generated by $\{E_i, F_i, H_i\}$, $i= 1, \cdots, n$, is a
  Kac--Moody algebra $\bar{\mf{g}}$ with Cartan matrix \index{Cartan matrix} $\bar{A}$.
  
\end{theorem}

 This theorem is proven in~\cite{Feingold}. Note that
    the Cartan integers $2 \left( \al_i \vert \al_j \right) / 
      \left(\al_i \vert \al_i \right)$ are indeed
    integers (because the $\al_i$'s are positive real roots), which
    are non-positive (because $\al_i - \al_j$ is not a root), so that
    $\bar{A}$ is a Cartan matrix.

\subsubsection*{Comments}

\begin{itemize}
\item When the Cartan matrix is degenerate, the corresponding
  Kac--Moody algebra has nontrivial ideals~\cite{Kac}. Verifying that
  the Chevalley--Serre relations are fulfilled is not sufficient to
  guarantee that one gets the Kac--Moody algebra corresponding to the
  Cartan matrix $[\bar{A}_{ij}]$ since there might be non-trivial
  quotients. Situations in which the algebra generated by the set
  $\{E_i, F_i, H_i\}$ is the quotient of the Kac--Moody algebra with
  Cartan matrix $\bar{A}$ by a non-trivial ideal were discussed
  in~\cite{Henneaux:2006gp}.
\item If the matrix $\bar{A}$ is decomposable, say $\bar{A}=D\oplus E$ with
  $D$ and $E$ indecomposable, then the Kac--Moody algebra
  $\mathbb{KM}(\bar{A})$, constructed from $\bar{A}$, is the direct sum of the Kac--Moody
  algebra $\mathbb{KM}(D)$ and the Kac--Moody algebra
  $\mathbb{KM}(E)$. The subalgebras $\mathbb{KM}(D)$
  and $\mathbb{KM}(E)$ are ideals. If $\bar{A}$ has non-vanishing
  determinant, then both $D$ and $E$ have non-vanishing
  determinant. Accordingly, $\mathbb{KM}(D)$ and $\mathbb{KM}(E)$ are
  simple~\cite{Kac} and hence, either occur faithfully or
  trivially. Because the generators $E_i$ are linearly independent,
  both $\mathbb{KM}(D)$ and $\mathbb{KM}(E)$ occur
  faithfully. Therefore, in the above theorem the only case that
  requires special treatment is when the Cartan matrix $\bar{A}$ has
  vanishing determinant.
\end{itemize}

\section{Affine Kac-Moody Algebras}
\label{Section:affine}
In this section we shall explore the special class of affine Kac-Moody algebras in more detail. This class of algebras corresponds to the first step away from the finite-dimensional Lie algebras, and is the only class of infinite-dimensional Kac-Moody algebras which are well understood. We recall that a Kac-Moody algebra $\mathfrak{g}(A)$ is said to be of affine type if the associated Cartan matrix $A$ is positive semi-definite, $\text{det}\ A=0$, with one zero eigenvalue. Because of the degeneracy of $A$, the bilinear form as constructed in Section \ref{Section:BilinearForm} is degenerate. We shall explain how this problem is circumvented through the inclusion of an additional generator, called the \emph{derivation} $d$, in the Cartan subalgebra. This new generator ensures that the invariant bilinear form on the full algebra is non-degenerate. 

\subsection{The Center of a Kac-Moody Algebra}
\label{Section:Center}

The center $Z$ of a Kac-Moody algebra $\mathfrak{g}(A)$ is defined as follows:
\begin{equation}
Z=\big\{x\in\mathfrak{g}(A)\ \big|\ \forall\ y\in\mathfrak{g}(A)\ :\ [x, y]=0\big\}.
\end{equation}
It is a general result that $Z\neq 0$ if and only if $\text{det}\ A= 0$ \cite{Kac}. This is related to the rank of the matrix $A$. In previous sections we have treated the Cartan matrix as an $r\times r$ matrix of matrix rank equal to the rank $r$ of the associated Kac-Moody algebra $\mathfrak{g}(A)$. Now we shall be more general and let $A=(A_{ij})_{i,j=1, \dots, r}$ be an $r\times r$ matrix of \emph{matrix rank} $n$. In this case, $\text{det}\ A=0$ and the rank $r$ Kac-Moody algebra $\mathfrak{g}(A)$ has a non-trivial center $Z\neq 0$ of dimension
\begin{equation}
\text{dim}\ Z=\text{corank}\ A=r-n.
\end{equation}
For affine Kac-Moody algebras the Cartan matrix has only one zero eigenvalue and hence the corank of $A$ is one, implying that the center is one-dimensional and is spanned by the \emph{central element} $c$,
\begin{equation}
Z=\mathbb{R}c.
\end{equation}
A consequence of this is that affine Kac-Moody algebras are \emph{not} simple, since the center forms a non-trivial ideal in $\mathfrak{g}(A)$. The center $Z$ is always contained in the Cartan subalgebra
\begin{equation}
Z\subset \mathfrak{h},
\end{equation}
implying that the central element $c$ must be expressible as a linear combination $\sum_{i=1}^{r}c_i\alpha_i^{\vee}$ of the simple corrots $\Pi^{\vee}=\{\alpha_1^{\vee}, \dots, \alpha_r^{\vee}\}$. To see this, let $v=(v_1, \dots, v_r)^{T}$ be the non-trivial element of the kernel of the transposed matrix $A^{T}$, i.e.,
\begin{equation}
A^{T}\cdot v=\sum_{j=1}^{r}(A^{T})_{ij} v_j=0.
\end{equation}
We then have $c_i=v_i$, i.e., 
\begin{equation}
c=\sum_{i=1}^{r}v_i\alpha_i^{\vee}\hspace{0.1 cm} \in \hspace{0.1 cm} Z\subset \mathfrak{h}.
\end{equation}
This result follows from the fact that $c$ must commute with all the Chevalley generators $e_i, \hspace{0.1 cm} i=1, \dots, r$, and hence
\begin{equation}
 0 = [c, e_i]=\sum_{j=1}^{r}c_j[\alpha_j^{\vee}, e_i]=\sum_{j=1}^{r}c_j\left<\alpha_i, \alpha_j^{\vee}\right> e_i=\sum_{j=1}^{r}c_jA_{ji}e_i=\big[\sum_{j=1}^{r}(A^{T})_{ij}c_j\big]e_i,
 \end{equation}
which is only satisfied when $\sum_{j=1}^{r}(A^{T})_{ij}c_j=0$, and hence $c_j=v_j$ as announced.

\subsection{The Derived Algebra and the Derivation}
\label{Section:Derivation}

It is now time to define what we mean by an affine Kac-Moody algebra. In fact, the algebra constructed from an ``affine'' Cartan matrix $A$ using the Chevalley-Serre relations is only the \emph{derived} Kac-Moody algebra
\begin{equation}
\mathfrak{g}^{\prime}(A)=[\mathfrak{g}(A), \mathfrak{g}(A)].
\end{equation}
When $\text{det}\ A\neq 0$ the derived algebra $\mathfrak{g}^{\prime}$ coincides with the full algebra $\mathfrak{g}$, i.e., $\mathfrak{g}=[\mathfrak{g}, \mathfrak{g}]$. To understand these statements we must introduce the notion of a \emph{derivation} $d$. The motivation for this is that the complete Kac-Moody algebra must have a well-defined non-degenerate bilinear form, an object which does not exist for the derived algebra $\mathfrak{g}^{\prime}$ because of the degeneracy of the Cartan matrix. For the following discussion it will be convenient to make a slight relabelling of the simple roots and the simple coroots. This is motivated by the fact that for any affine Kac-Moody algebra $\mathfrak{g}$ of rank $r$ one may identify a maximal rank $r-1$ finite-dimensional subalgebra $\bar{\mathfrak{g}}\subset\mathfrak{g}$, and view $\mathfrak{g}$ as an \emph{extension} $\mathfrak{g}\equiv \bar{\mathfrak{g}}^{+}$ of $\bar{\mathfrak{g}}$, where the superscript ``$+$'' indicates that the affine Kac-Moody algebra $\mathfrak{g}$ is obtained by adding a single node to the Dynkin diagram of $\bar{\mathfrak{g}}$ in a prescribed way. We shall discuss the general theory of extensions of Lie algebras in Section \ref{Section:Extensions} but for now this will suffice. To this end we take the $r=k+1$ simple roots of $\mathfrak{g}(A)$ to be $\Pi=\{\alpha_0, \alpha_1, \dots, \alpha_k\}$, with $\bar{\Pi}=\{\alpha_1, \dots, \alpha_k\}$ representing the $k$ simple roots of the finite subalgebra $\bar{\mathfrak{g}}$. The root $\alpha_0$ is called the \emph{affine root}. It is always of the form 
\begin{equation}
\alpha_0=\delta-\theta,
\end{equation}
where $\delta$ is a null root, $(\delta|\delta)=0$, and $\theta$ is the highest root of the finite subalgebra $\bar{\mathfrak{g}}\subset \mathfrak{g}$. 

We now follow Kac \cite{Kac} and add by hand a generator $d\in\mathfrak{h}$ to the Cartan subalgebra, with the property
\begin{equation}
\left<\alpha_i, d\right>=\delta_{i0}, \qquad i=0, 1, \dots, k.
\label{PERTABeq:derivation}
\end{equation}
The basis of the Cartan subalgebra $\mathfrak{h}$ is then taken to be $\Pi^{\vee}=\{\alpha_0^{\vee}, \alpha_1^{\vee}, \dots, \alpha_k^{\vee}, d\}$, on which a non-degenerate bilinear form now exists with the following properties:
\begin{equation}
(c|\alpha_i^{\vee})=0, \quad (c|c)=0, \quad (c|d)=1,
\end{equation}
where the non-degeneracy of $(\cdot|\cdot)$ on $\mathfrak{h}$ follows from the non-vanishing scalar product between the derivation $d$ and the central element $c$. An \emph{affine} Kac-Moody algebra $\mathfrak{g}(A)$ is then defined as the derived algebra $\mathfrak{g}^{\prime}$ augmented with the derivation $d$,
\begin{equation}
\mathfrak{g}=\mathfrak{g}^{\prime} + \mathbb{R}d.
\label{PERTABeq:derivedplusderivation}
\end{equation}
Note that this implies that an affine Kac-Moody algebra of rank $r$ has a Cartan subalgebra of dimension $r+1$. 

Let us now make an effort towards understanding the structure of (\ref{PERTABeq:derivedplusderivation}). Because of its definition, (\ref{PERTABeq:derivation}), the derivation $d\in\mathfrak{h}$ will never appear on the right hand side of any commutator in the algebra. For example, the commutation relations with the positive Chevalley generators are
\begin{eqnarray}
{}Ê[d, e_a] & =& 0, \quad a=1, \dots, k,
\nonumber\\
{}Ê[d, e_0]&=& e_0,
\end{eqnarray}
and similarly for the $f_i$'s. This implies that the derived algebra $\mathfrak{g}^{\prime}$ does not contain the derivation $d$, thus explaining the structure of (\ref{PERTABeq:derivedplusderivation}). In fact, the derivation can be viewed as a ``counting operator'' which counts the number of times the affine generator $e_0$, corresponding to the affine root $\alpha_0$, appears in any commutator. 

\subsection{The Affine Root System}
In the case of affine Kac-Moody algebras, the appearance of imaginary roots does imply a drastic complication. As we have alluded to before, the only ``independent'' imaginary root is the null root $\delta$, of which all other imaginary roots are multiples. The complete root system $\Phi$ is therefore determined by the finite root system $\bar{\Phi}$ of the maximal finite subalgebra $\bar{\mathfrak{g}}\subset\mathfrak{g}$ and the null root $\delta$. As mentioned in the previous section, the Dynkin diagram of $\bar{\mathfrak{g}}$ is obtained by deleting the zeroth node in the Dynkin diagram of the affine algebra $\mathfrak{g}$. 

Let $A$ be the Cartan matrix of an affine Kac-Moody algebra $\mathfrak{g}$, with associated root system $\Phi=\Phi(A)$. We begin by splitting $\Phi$ into its real and imaginary parts,
\begin{equation}
\Phi=\Phi_{\Re}\cup\Phi_{\Im}.
\end{equation}
In the example in Section \ref{Section:ExampleA2A1+} we saw that the real roots of $A_1^{+}$ were given by all roots of the form $\alpha_1+n\delta$, $n\mathbb{Z}$, with $\alpha_1$ being the simple root (and, in fact, the only positive root) of the underlying finite algebra $A_1$. This fact generalizes to any affine Kac-Moody algebra $\mathfrak{g}(A)$ in the following way
\begin{equation}
\Phi_{\Re}=\big\{\alpha+n\delta \ \big|\ \forall \ \alpha\in\bar{\Phi};\ n\in\mathbb{Z}\big\},
\end{equation}
with $\bar{\Phi}$ being, as usual, the root system of the underlying finite-dimensional Lie algebra $\bar{\mathfrak{g}}$. The positive part of the real roots can then be described as follows
\begin{equation}
\Phi_{\Re +}=\big\{\alpha+n\delta\ \big|\ \forall\ \alpha\in\bar{\Phi}; \ n\in\mathbb{Z}_{\geq 0}\big\}\cup \bar{\Phi}_{+}.
\end{equation}
It is well known in Lie theory that if $\alpha$ is a root of a finite-dimensional Lie algebra, then the only multiples of $\alpha$ which are also roots are $\pm \alpha$. This feature carries over to the real part of the root system of general Kac-Moody algebras, while it is no longer true for the imaginary roots. For affine Kac-Moody algebras there exists only one independent imaginary root, and this is the root $\delta$ which appears in the construction of the zeroth simple root $\alpha_0=\delta-\theta$. Any multiple of $\delta$ is also an imaginary root, and hence the imaginary part of the root system of any affine Kac-Moody algebra is very easy to describe:
\begin{equation}
\Phi_{\Im}=\Phi_{\Im+}\cup\Phi_{\Im -}=\big\{n\delta\ \big|\ n\in\mathbb{Z}_{\geq 0}\big\}\cup \{n\delta\ \big|\ n\in\mathbb{Z}_{\leq 0}\big\}.
\end{equation}

\subsection{The Affine Weyl Group}
\label{Section:AffineWeylGroup}

We now want to perform a closer analysis of the Weyl group of an affine Kac-Moody algebra. During our study we will find a natural explanation for where the name \emph{affine} has its origin.

The Weyl group $\mathcal{W}(A)$ associated with an affine Kac-Moody algebra $\mathfrak{g}(A)$ is defined through the geometric action of the fundamental reflections $\mathcal{S}=\{s_0, s_1, \dots, s_k\}$ on the dual space $\mathfrak{h}^{\star}$. Moreover we have seen that one can associate a reflection $s_{\alpha}$ with respect to any real root $\alpha\in\Phi_{\Re}$ as $s_{\alpha}=\omega s_i \omega^{-1}$ for $\omega\in\mathcal{W}$ and $\alpha=\omega(\alpha_i)$. On the other hand, no such construction exists for the imaginary roots since for $\beta\in\Phi_{\Im}$ the pairing $\left<\alpha, \beta^{\vee}\right>$ is not defined. Therefore, although they act \emph{on} $\Phi_{\Im}$, all Weyl reflections are defined \emph{with respect to} real roots only. 

An important new feature owing to the existence of null roots is that, since $(\delta|\alpha)=0$ for all $\alpha\in\Phi_{\Re}$, we have 
\begin{equation}
s_{\alpha}(\delta)=\delta-\frac{2(\delta|\alpha)}{(\alpha|\alpha)}\alpha=\delta.
\end{equation}
This implies that the entire Weyl group acts as the identity on the set of imaginary roots:
\begin{equation}
\omega(\beta)=\beta, \quad \forall\ \beta \ \in\ \Phi_{\Im}.
\end{equation}
Note that this is true only in the affine case for which all roots in $\Phi_{\Im}$ are lightlike, but not in the general case when $\Phi_{\Im}$ also contains timelike roots.

Let $\bar{\mathcal{W}}\subset \mathcal{W}$ be the finite Weyl group of $\bar{\mathfrak{g}}\subset\mathfrak{g}$. A particular feature of affine Weyl groups is the fact that they decompose into a semidirect product of the form
\begin{equation}
\mathcal{W}=\bar{\mathcal{W}}\ltimes \bar{T}^{\vee},
\end{equation}
where $\bar{T}^{\vee}$ denotes the abelian group of translations of the coroot lattice $\bar{Q}^{\vee}$ of $\bar{\mathfrak{g}}$.\footnote{Here we refer to the coroot lattice $Q^{\vee}$ as a lattice in $\mathfrak{h}^{\star}$, in the sense that we can use the non-degenerate bilinerar form $(\cdot |\cdot)$ on $\mathfrak{g}(A)$ to identify $\mathfrak{h}$ with $\mathfrak{h}^{\star}$. Then the coroot lattice is spanned by the simple coroots $\alpha_i^{\vee}\equiv 2\alpha_i/(\alpha_i|\alpha_i)\in\mathfrak{h}^{\star}$ and we have $Q\subset Q^{\vee}$.} We shall explain this phenomenon in detail below in the context of a simple example.

\subsubsection{Example: The Weyl Group of $A_1^{+}$}

Let $\Pi=\{\alpha_0, \alpha_1\}$ be the simple roots of $\mathfrak{g}=A_1^{+}$ and $\bar{\Pi}=\{\alpha_1\}$ the simple root of the underlying finite subalgebra $\bar{\mathfrak{g}}=A_1$. The Weyl group $\bar{\mathcal{W}}$ of $A_1^{+}$ is generated by a single reflection $s_1$ with respect to $\alpha_1$:
\begin{equation}
s_1(\alpha_1)=-\alpha_1,
\end{equation}
implying that $\bar{\mathcal{W}}=\mathbb{Z}_2$. The Cartan matrix of $A_1^{+}$ is $$\left(\begin{array}{cc} 2 & -2 \\ -2 & 2 \\ \end{array} \right)$$ which has a kernel spanned by the column vector $\left( \begin{array}{cc} 1 & 1 \\ \end{array} \right)^{T}.$ Thus, from our discussion of the center of Kac-Moody algebras in Section \ref{Section:Center}, we find that $A_1^{+}$ has a central element given by 
\begin{equation}
c=\alpha_0^{\vee}+\alpha_1^{\vee}.
\end{equation}
It is easy to check that $c$ indeed commutes with all generators of the algebra. For example, we have
\begin{equation}
[c, e_1]=[\alpha_0^{\vee}, e_1]+[\alpha_1^{\vee}, e_1]=(-2+2)e_1=0.
\end{equation}
Recall now from Section \ref{Section:Derivation} that so far we have been dealing only with the derived algebra 
\begin{equation}
\mathfrak{g}^{\prime}=[A_1^{+}, A_1^{+}]
\end{equation}
whose Cartan subalgebra is 
\begin{equation}
\mathfrak{h}^{\prime}=\mathbb{R}\alpha_0^{\vee}+\mathbb{R}\alpha_1^{\vee}.
\end{equation}
However, to understand the Weyl group $\bar{\mathcal{W}}$ it is crucial that we treat the full algebra 
\begin{equation}
A_1^{+}=[A_1^{+}, A_1^{+}]+\mathbb{R}d.
\end{equation}
Thus, we add by hand the derivation $d$ to the algebra, with the properties
\begin{equation}
(c|d)=1, \qquad (d|\alpha_0^{\vee})=0, \qquad (d|d)=0.
\end{equation}
In the following we shall also view the central element $c$ as a basis element of the Cartan subalgebra, instead of the generator $\alpha_0^{\vee}$. This will prove convenient later on. The full Cartan subalgebra now takes the form
\begin{equation}
\mathfrak{h}=\bar{\mathfrak{h}}\oplus\big(\mathbb{R}c+\mathbb{R}d\big),
\end{equation}
where $\bar{\mathfrak{h}}=\mathbb{R}\alpha_1^{\vee}$ is the Cartan subalgebra of $A_1$. 

We shall now proceed to write the simple roots as the root vectors $\vec{\alpha}_0$ and $\vec{\alpha}_1$ with components given by the eigenvalues under the adjoint action of $\mathfrak{h}$ on the Chevalley generators $e_0$ and $e_1$. For $e_0$ we find
\begin{eqnarray}
{}Ê& & [\alpha_1^{\vee}, e_0]=\left<\alpha_0, c\right>e_0=-2e_0,
\nonumber \\
{}Ê& & [c, e_0] = 0,
\nonumber \\
{}Ê& & [d, e_0] = \left<\alpha_0, d\right> e_0 =e_0,
\end{eqnarray}
and for $e_1$ we have
\begin{eqnarray}
{}Ê& & [\alpha_1^{\vee}, e_1]=\left<\alpha_1, \alpha_1^{\vee}\right>e_1=2e_1,
\nonumber \\
{}Ê& & [c, e_1] = 0,
\nonumber \\
{}Ê& & [d, e_1] = \left<\alpha_1, d\right> e_1 =0,
\end{eqnarray}
where we made use of the defining relation, (\ref{PERTABeq:derivation}), for the derivation. Consequently, the component forms of the simple root vectors, in the basis determined by $\alpha_1^{\vee}, c$ and $d$, are
\begin{eqnarray}
{}Ê\vec{\alpha}_0 &=& (-2, 0, 1),
\nonumber \\
{}Ê\vec{\alpha}_1 &=& (2, 0, 0).
\end{eqnarray}
We now have all the ingredients to understand the Weyl group of $A_1^{+}$. The group $\mathcal{W}$ is generated by the two fundamental reflections $s_0$ and $s_1$. We shall compute the action of these generators on an arbitrary vector
\begin{equation}
\vec{\lambda}=(\bar{\lambda}, k, m)\ \in \ \mathfrak{h}^{\star}.
\end{equation}
We begin by computing the action of $s_1$:
\begin{eqnarray}
{}Ês_1(\vec{\lambda})&=& \vec{\lambda}-\frac{2(\vec{\lambda}|\vec{\alpha}_1)}{(\vec{\alpha}_1|\vec{\alpha}_1)}\vec{\alpha}_1
\nonumber \\
{}Ê&=& (\bar{\lambda}, k, m)-(2\bar{\lambda}, 0, 0)
\nonumber \\
{}Ê&=& (-\bar{\lambda}, k, m).
\end{eqnarray}
This result was expected, namely that the action of $s_1$ on the root lattice $\bar{Q}=\mathbb{Z}\alpha_1$ is the same as for the Weyl group of $A_1$, i.e., in component form we have
\begin{equation}
s_1(\bar{\lambda})=-\bar{\lambda}, \qquad s_1(k)=k, \qquad s_1(m)=m.
\end{equation}
Now, let us move on to the more interesting case of the $s_0$-generator. Its action on $\vec{\lambda}$ becomes
\begin{eqnarray}
{}Ês_{0}(\vec{\lambda})&=& \vec{\lambda}-\frac{2(\vec{\lambda}|\vec{\alpha}_0)}{(\vec{\alpha}_0|\vec{\alpha}_0)}\vec{\alpha}_0
\nonumber \\
{}Ê&=& (\bar{\lambda}, k, m)-\big[-\bar{\lambda}+k\big](-2, 0, 1)
\nonumber \\
{}Ê&=& (-\bar{\lambda}+2k, k, m+\bar{\lambda}-k).
\end{eqnarray}
We shall focus on the ``projected'' action of $s_0$ on the root lattice $\bar{Q}$, which reads
\begin{equation}
s_0(\bar{\lambda})=-\bar{\lambda}+2k.
\end{equation}
This corresponds to a reflection of $\bar{\lambda}$, not with respect to the origin, but rather with respect to the point displaced by $k$ away from the origin in $\bar{Q}$. We can make things more clear by noting that the combined reflection $t\equiv s_0\circ s_1$ acts on $\bar{Q}$ as a pure translation:
\begin{equation}
t(\bar{\lambda})=s_0\circ s_1(\bar{\lambda})=\bar{\lambda}+2k.
\end{equation}
Thus the Weyl group of $A_1^{+}$, acting on the Euclidean root lattice $\bar{Q}$ of $\bar{\mathfrak{g}}$, contains translations, i.e., \emph{affine transformations}, thus explaining the origin of the name \emph{affine} Kac-Moody algebras.

The action of $s_0$ can now be written as
\begin{equation}
s_0(\bar{\lambda})=t\circ s_1(\bar{\lambda}),
\end{equation}
which corresponds to the combination of an element $s_1\in\bar{\mathcal{W}}$ and an element $t\in \bar{T}$, with $\bar{T}$ being the abelian group of translations of the root lattice $\bar{Q}$. This group is generated by $t$ with a general element $t^n\in\bar{T}$ acting as follows
\begin{equation}
t^{n}(\bar{\lambda})=\bar{\lambda}+2kn.
\end{equation}
The complete Weyl group of $A_1^{+}$ is therefore
\begin{equation}
\mathcal{W}=\big\{ t^{n}, \ t^n\circ s_1\ \big|\ n\in\mathbb{Z};\ s_1\in\bar{\mathcal{W}}\big\},
\end{equation}
which is equivalent to the semidirect product
\begin{equation}
\mathcal{W}=\bar{\mathcal{W}}\ltimes \bar{T},
\end{equation}
as announced in the beginning of this section. The reason why it is the root lattice $\bar{Q}$ which appears here, and not the coroot lattice $\bar{Q}^{\vee}$, is that for $A_1^{+}$ they coincide, since all roots have the same length. However, in the general case it is the coroot lattice which is relevant. 
\section{Lorentzian Kac-Moody Algebras}
Kac-Moody algebras of indefinite type constitute a vast wasteland of unexplored territory. Not much is known in general about these algebras; most notably there is no indefinite Kac-Moody algebra for which the root multiplicities are known in closed form to arbitrary height. We shall here focus on a particular subclass of indefinite Kac-Moody algebras, namely those for which the invariant bilinear form is of Lorentzian signature. These Lorentzian Kac-Moody algebras similarly constitute an infinite class of unclassified algebras, but they nevertheless has a subclass which is, in a certain sense, under control. We refer here to the Lorentzian algebras which can be obtained through prescribed extensions of finite-dimensional simple Lie algebras \cite{Gaberdiel:2002db}. Extensions of finite Lie algebras can in this way give rise to Lorentzian Kac-Moody algebras, with properties which have proven to be of interest in string and M-theory. The precise extension procedure is described in detail in Section \ref{Section:Extensions}. Then, in Section \ref{Section:hyperbolic}, we analyze a very interesting subclass of the Lorentzian Kac-Moody algebras, known as \emph{hyperbolic}. As was the case with affine Kac-Moody algebras, we shall see that the hyperbolic class also draws its name from properties of the associated Weyl groups. The hyperbolic Kac-Moody algebras are the only examples of indefinite Kac-Moody algebras which have been completely classified. 

\subsection{Extensions of Lie Algebras}
\label{Section:Extensions}

Let $A$ be a symmetrizable generalized Cartan matrix and $S$ its symmetric part. If the bilinear form $(\cdot |\cdot)$ defined by $S$ is of Lorentzian signature then the Kac-Moody algebra $\mathfrak{g}(A)$ is of Lorentzian type. We shall in this section describe how algebras of Lorentzian type can be obtained from finite simple Lie algebras $\bar{\mathfrak{g}}=\mathfrak{g}(\bar{A})$ through double and triple extensions, denoted $\bar{\mathfrak{g}}^{++}$ and $\bar{\mathfrak{g}}^{+++}$, respectively, of the Dynkin diagram $\Gamma(\bar{A})$ associated with $\bar{\mathfrak{g}}$. 

As in Section \ref{Section:affine} we let $\bar{\Pi}=\{\alpha_1, \dots, \alpha_k\}$ be a basis of simple roots for the finite rank $k$ Lie algebra $\bar{\mathfrak{g}}$. The root lattice $\bar{Q}=\sum_{i=1}^{k}\mathbb{Z}\alpha_i$ is Euclidean. Let also $\mathbb{Z}^{1,1}$ be the (unique) even two-dimensional unimodular Lorentzian lattice, spanned by the vectors $u_1$ and $u_2$. We define a non-degenerate bilinear form $(\cdot |\cdot)$ on $\mathbb{Z}^{1,1}$, of signature $(-+)$, by 
\begin{equation}
(u_1|u_2)=1, \qquad (u_1|u_1)=0, \qquad (u_2|u_2)=0.
\end{equation}
This scalar product is induced from the standard Minkowski metric on $\mathbb{R}^{1,1}$ by taking 
\begin{equation}
u_1=(1,0), \qquad u_2=(0, -1),
\end{equation}
and we write
\begin{equation}
\mathbb{R}^{1,1}=\mathbb{R}u_1+\mathbb{R}u_2, \qquad \mathbb{Z}^{1,1}=\mathbb{Z}u_1+\mathbb{Z}u_2.
\end{equation}
We now want to extend the Dynkin diagram $\bar{\Gamma}=\Gamma(\bar{A})$ with one node in such a way that the new diagram $\bar{\Gamma}^{+}$ corresponds to the Dynkin diagram of an affine Kac-Moody algebra $\bar{\mathfrak{g}}^{+}$. This step was actually already discussed in Section \ref{Section:affine}, and we recall that an affine algebra can be obtained from any finite Lie algebra $\bar{\mathfrak{g}}$ by augmenting the set of simple roots with the \emph{affine root}
\begin{equation}
\alpha_0\equiv u_1-\theta,
\end{equation}
where $\theta$ is the highest root of $\bar{\mathfrak{g}}$ and $u_1\in\mathbb{Z}^{1,1}$ corresponds to the null root $\delta$. We now have
\begin{equation}
(\alpha_i|u_1)=0, \quad \text{for}\quad i=1, \dots, k,
\end{equation} 
implying that  the new root lattice 
\begin{equation}
\bar{Q}^{+} =\mathbb{Z}{\alpha}_0+\bar{Q}
\end{equation}
is contained in the direct sum of $\bar{Q}$ and $\mathbb{Z}^{1,1}$,
\begin{equation}
\bar{Q}^+\subset \bar{Q}\oplus \mathbb{Z}^{1,1}.
\end{equation}
We define new indices $A, B=(0, i)$, such that we can write the matrix of scalar products between the new simple roots $\bar{\Pi}^+=\{\alpha_0, \alpha_1, \dots, \alpha_k\}$ as follows
\begin{equation}
\bar{A}_{AB}^+=\frac{2(\alpha_A|\alpha_B)}{(\alpha_A|\alpha_A)},
\end{equation}
which then corresponds to the entries of the Cartan matrix $\bar{A}^+$ of the affine Kac-Moody algebra $\bar{\mathfrak{g}}^+=\mathfrak{g}(\bar{A}^+)$. 

Let us now proceed to include also the second basis vector $u_2\in\mathbb{Z}^{1,1}$. This is done by adding the simple root 
\begin{equation}
\alpha_{-1}\equiv  -u_1-u_2,
\end{equation}
which has non-vanishing scalar product only with $\alpha_0$:
\begin{equation}
(\alpha_{-1}|\alpha_0)=-1, \qquad (\alpha_{-1}|\alpha_i)=0,\quad\forall\ i=1, \dots, k.
\end{equation}
Since we also have $(\alpha_{-1}|\alpha_{-1})=(\alpha_0|\alpha_0)=2$ this implies that the node associated with $\alpha_{-1}$ attaches by a single link to the zeroth node of the Dynkin diagram $\bar{\Gamma}^+$ of $\bar{\mathfrak{g}}^+$. As before, we define new collective indices $I, J=(-1, 0, i)$, and the matrix of scalar products
\begin{equation}
\bar{A}_{IJ}^{++}=\frac{2(\alpha_I|\alpha_J)}{(\alpha_I|\alpha_I)}
\end{equation}
is the Cartan matrix of the \emph{Lorentzian} Kac-Moody algebra $\bar{\mathfrak{g}}^{++}$. The root lattice $\bar{Q}^{++}$ is then of Lorentzian signature and is equivalent to the direct sum of $\bar{Q}$ with $\mathbb{Z}^{1,1}$,
\begin{equation}
\bar{Q}^{++}=\bar{Q}\oplus \mathbb{Z}^{1,1}.
\end{equation}
In order to obtain a triple extension, while still keeping the Lorentzian signature of the root lattice, we introduce yet another two-dimensional Lorentzian lattice $\tilde{\mathbb{Z}}^{1,1}$, spanned by the basis vectors $v_1$ and $v_2$. The scalar product on $\tilde{\mathbb{Z}}^{1,1}$ is of the same form as the one on $\mathbb{Z}^{1,1}$,
\begin{equation}
(v_1|v_2)=1, \qquad (v_1|v_1)=0, \qquad (v_2|v_2)=0.
\end{equation}
We now note that the vector $v_1+v_2\in\tilde{\mathbb{Z}}^{1,1}$ is spacelike, $(v_1+v_2|v_1+v_2)=2$. Thus, by including this vector into the new root lattice we ensure that the Lorentzian signature is preserved, i.e., we do not introduce zero eigenvalues in the bilinear form. We augment the set of simple roots with the ``triple-extended'' root
\begin{equation}
\alpha_{-2}\equiv u_1-(v_1+v_2).
\end{equation}
Again, this root is spacelike, $(\alpha_{-2}|\alpha_{-2})=2$, and the associated node in the Dynkin diagram connects with a single link to the node corresponding to $\alpha_{-1}$,
\begin{equation}
(\alpha_{-2}|\alpha_{-1})=-1, \qquad (\alpha_{-2}|\alpha_A)=0, \quad \forall \ A=0, 1, \dots, k.
\end{equation}
The new Lorentzian root lattice is given by
\begin{equation}
\bar{Q}^{+++}=\mathbb{Z}\alpha_{-2}+\mathbb{Z}\alpha_{-1}+\mathbb{Z}\alpha_0+\bar{Q},
\end{equation}
and we have
\begin{equation}
\bar{Q}^{+++}\ \subset\ \tilde{\mathbb{Z}}^{1,1}\oplus \mathbb{Z}^{1,1}\oplus \bar{Q}.
\end{equation}
Introducing new indices $M, N=(-2, -1, 0, i)$ we can once again organize the scalar products as
\begin{equation}
\bar{A}_{MN}^{+++}=\frac{2(\alpha_M|\alpha_N)}{(\alpha_M|\alpha_M)},
\end{equation}
corresponding to the Cartan matrix $\bar{A}^{+++}$ of a Lorentzian Kac-Moody algebra $\bar{\mathfrak{g}}^{+++}=\mathfrak{g}(\bar{A}^{+++})$. 

Nice examples of extended Lie algebras, which shall be discussed extensively later on, are the Kac-Moody algebras obtained by extending the largest exceptional Lie algebra $E_8$. This gives rise to the following chain of embeddings
\begin{equation}
E_8\ \subset E_9=E_8^{+}\ \subset \ E_{10}=E_{8}^{++}\ \subset \ E_{11}=E_8^{+++},
\end{equation}
of which $E_9$ is affine, $E_{10}$ is hyperbolic and $E_{11}$ is Lorentzian (but not hyperbolic). In the next section we shall focus on the  subclass of hyperbolic Kac-Moody algebras. In particular, the hyperbolic algebra $E_{10}$ will play an important role in {\bf Part I} of this thesis, and is discussed in more detail in Section \ref{section:DecompE10E10} below. See also \cite{Palmkvist:2009bw} for a very nice exposition of the algebras $E_8, E_9$ and $E_{10}$ from a similar point of view.

\subsection{Hyperbolic Kac-Moody Algebras}
\label{Section:hyperbolic}

A hyperbolic Kac-Moody algebra is defined as a Lorentzian algebra which upon removal of any node in the Dynkin diagram yields only finite or (at most one) affine subdiagrams. By this criterion it is easy to understand why $E_{11}$ is not hyperbolic; removal of the node associated with the triple-extended root $\alpha_{-2}$ gives the Dynkin diagram of $E_{10}$ which is neither finite nor affine. In this section we shall see that the class of hyperbolic algebras exhibits some very intriguing features which are unique among all indefinite Kac-Moody algebras. 

In the rest of this section we shall take $\mathfrak{g}(A)$ to be a rank $r$ hyperbolic Kac-Moody algebra, unless otherwise specified. By virtue of its Lorentzian signature the space $\mathfrak{h}^{\star}$ is isomorphic to $r$-dimensional Minkowski space:
\begin{equation}
\mathfrak{h}^{\star}\simeq \mathbb{R}^{1, r-1}.
\end{equation}
An important consequence of this is that there exists a lightcone in $\mathfrak{h}^{\star}$, defined as
\begin{equation}
\mathcal{O}=\big\{x\in\mathfrak{h}^{\star}\ \big|\ (x|x)\leq 0\big\}.
\end{equation}
The lightcone clearly separates real and imaginary roots
\begin{equation}
\Phi_{\Im}=\Phi\cap \mathcal{O}.
\end{equation}
These properties are of course shared among all Lorentzian algebras. However, a unique feature of hyperbolic Kac-Moody algebras is that its root system is in principle known. By this we mean that any element of the root lattice $Q(A)$ is also an element of the root system $\Pi(A)$ if its norm is less than or equal to 2. In this way we find the following description of the root system:
\begin{equation}
\Phi=\big\{\alpha\in Q\ \big|\ (\alpha|\alpha)\leq 2\big\}.
\end{equation}
Furthermore, we recall that the fundamental Weyl chamber $\mathcal{C}$ is defined as the region of $\mathfrak{h}^{\star}$ which is bounded by the hyperplanes $T_i$ Êwhich are orthogonal to the simple roots $\alpha_i$. As a consequence of their definition, hyperbolic Kac-Moody algebras have the special property that all these hyperplanes intersect \emph{inside or on the lightcone}. Thus, for any hyperbolic Kac-Moody algebra the fundamental Weyl chamber is contained inside the lighcone
\begin{equation}
\mathcal{C}\ \subset\ \mathcal{O}.
\end{equation}
Because of the Lorentzian signature, the lightcone decomposes into future and past components, $\mathcal{O}_+$ and $\mathcal{O}_-$, respectively. We shall employ the convention that the simple roots have future temporal directions, implying that the fundamental Weyl chamber, defined as $\{\beta\in\mathfrak{h}^{\star}\ |\ (\beta|\alpha_i)> 0, \ i=1, \dots, r\}$, is actually contained in the \emph{past} lightcone. Since no simple roots are inside the lightcone, the union of all the images of the Weyl group, $\mathcal{W}$, acting on $\mathcal{C}$, will not extend outside of $\mathcal{C}$, and, in fact, we have that the Tits cone, $\mathcal{X}$, coincides with the past lighcone:
\begin{equation}
\mathcal{O}\equiv \mathcal{X}=\bigcup_{\omega\in\mathcal{W}}\omega(\mathcal{C}).
\end{equation}
Because of this, the Weyl chamber $\mathcal{C}$ is not a fundamental domain for the action of $\mathcal{W}$ on all of $\mathfrak{h}^{\star}$, as is the case for finite Lie algebras, but rather is the fundamental domain for the action of $\mathcal{W}$ on the Tits cone, $\mathcal{X}$. 

The Weyl group $\mathcal{W}(A)$ of a hyperbolic Kac-Moody algebra $\mathfrak{g}(A)$ is a discrete subgroup of the isometry group of $\mathfrak{h}^{\star}=\mathbb{R}^{1, r-1}$. Moreover, since all of the hyperplanes $T_i$ are either timelike or lightlike the Weyl group preserves the temporal direction of any $x\in\mathfrak{h}^{\star}$. This implies that the Weyl group of a hyperbolic Kac-Moody algebra is a subgroup of the ortochronous Lorentz group, $O^{\dagger}(1, r-1)$, i.e., the time-preserving part of the isometry group of $\mathbb{R}^{1, r-1}$,
\begin{equation}
\mathcal{W}\ \subset\ O^{\dagger}(1, r-1).
\end{equation}
Because of this fact, the Weyl group preserves spaces of constant negative curvature in $\mathcal{O}$, i.e., the $r$-dimensional hyperbolic space $\mathcal{H}_r$. The hyperplanes $T_i$ project onto hyperplanes in $\mathcal{H}_r$ and because we then have $r+1$ hyperplanes bounding a region in an $r$-dimensional space, the Weyl chamber $\mathcal{C}$ projects onto a simplex of finite volume in $\mathcal{H}_r$. Geometric reflections in the faces of a finite volume simplex in hyperbolic space are elements of a \emph{hyperbolic Coxeter group}. The associated Weyl groups therefore correspond to such hyperbolic Coxeter groups, and it is this fact which is the origin of the name hyperbolic Kac-Moody algebras. 

\subsection{Extended Example: The Weyl Group of $A_1^{++}$}
\label{Section:WeylGroupA1++}
\begin{figure}[ht]
\begin{center}
\includegraphics[width=50mm]{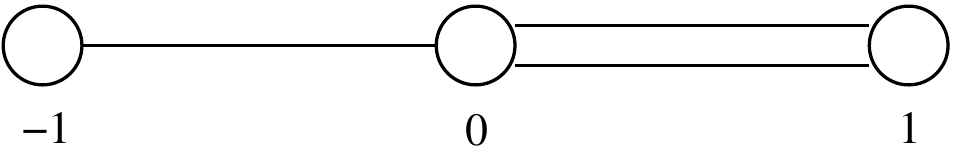}
\caption{The Dynkin diagram of the hyperbolic Kac-Moody algebra $A_1^{++}$.} 
\label{PERTABfig:A1++}
\end{center}
\end{figure}

We have previously discussed the Weyl groups of $\bar{\mathfrak{g}}=A_1$ and $\bar{\mathfrak{g}}^+=A_1^{+}$ in some detail. Recall that we found:
\begin{equation}
\mathcal{W}(\bar{A})=\mathbb{Z}_2, \qquad \mathcal{W}(\bar{A}^+)=\mathbb{Z}_2\ltimes \bar{T},
\end{equation}
where $\bar{T}$ is the abelian group of translations of the Euclidean root lattice $\bar{Q}$. The first group $\mathbb{Z}_2$ is of course of finite order, while the second one, $\mathbb{Z}_2\ltimes \bar{T}$, is of infinite order due to the presence of the translation group. Finite Coxeter groups, such as $\mathbb{Z}_2$, are often called \emph{spherical} because they are reflections about the origin in a Euclidean space and so leaves invariant a sphere at infinity. Similarly the affine Coxeter groups leave the entire Euclidean spaces themselves invariant. Finally, on the other side we have the \emph{hyperbolic} Coxeter groups which leave the hyperbolic space invariant. In this section we shall check this more explicitly by investigating the Weyl group of the hyperbolic Kac-Moody algebra $\mf{g}=A_1^{++}$, i.e., the double extension of $A_1$. The Cartan matrix of this algebra is 
\begin{equation}
A(A_1^{++})=\left( \begin{array}{ccc}
\phantom{-}2 & -1 & \phantom{-}0\\
-1 & \phantom{-}2 & -2 \\
\phantom{-}0 & -2 & \phantom{-}2 \\
\end{array} \right),
\end{equation}
and the associated Dynkin diagram $\Gamma(A_1^{++})$ is displayed in Figure \ref{PERTABfig:A1++}.
We associate a fundamental reflection $s_I$ with each of the simple roots $\alpha_I, \ I=-1, 0, 1$. By making use of  Table \ref{PERTABtable:CoxeterExponents} and (\ref{PERTABeq:Coxeterrelations}) we find that these generators obey
\begin{equation}
(s_1s_0)^{\infty}=1, \qquad (s_0s_{-1})^3=1, \qquad (s_1s_{-1})^2=1.
\end{equation}
The root space of $A_1^{++}$ is three-dimensional and thus the (projected) Weyl chamber $\mathcal{C}$ is a simplex in the two-dimensional hyperbolic plane $\mathcal{H}_2=\{\beta\in\mathfrak{h}^{\star}\ |\ (\beta|\beta)=-1\}$, bounded by the three hyperplanes $T_{-1}, T_0$ and $T_1$. The angle $\vartheta_{IJ}$ between two adjacent faces $I$ and $J$ are related to the Coxeter exponents, $m_{IJ}$, as follows (see, e.g., \cite{Humphreys})
\begin{equation}
\vartheta_{IJ}=\frac{\pi}{m_{IJ}}.
\end{equation}
In our case we thus find that the angle between $T_{1}$ and $T_0$ is zero, the one between $T_1$ and $T_{-1}$ is $\pi/2$ and the one between $T_0$ and $T_{-1}$ is $\pi/3$. This implies that the hyperplanes $T_{1}$ and $T_0$ intersect on the border of the lighcone, while the other walls intersect inside the lightcone. This verifies that $A_1^{++}$ is indeed hyperbolic. 

We now proceed to show that the Weyl group of $A_1^{++}$ is isomorphic to the extended modular group $PGL(2, \mathbb{Z})$ (see, e.g, \cite{Humphreys}). We define the group $PGL(2,\mathbb{Z})$ as the group of $2\times 2$ matrices
\begin{equation}
PGL(2,\mathbb{Z}) \ \ni \ \left( \begin{array}{cc}
a & b \\
c & dÊ\\
\end{array}Ê\right), \qquad a,b,c,d\ \in \ \mathbb{Z},
\label{PERTABeq:SL2Zmatrix}
\end{equation}
with determinant $ad-bc= 1$ and the identification $\{a,b,c,d\}\sim \{-a,-b,-c,-d\}$. [The determinant of any element $X\in GL(2, \mathbb{Z})$ is restricted to $+1$ or $-1$ in order for the inverse $X^{-1}$ to be contained in the group. In the projected group $PGL(2, \mathbb{Z})$ elements with determinant $-1$ are projected out.]

Now, we recall that the hyperbolic plane $\mathcal{H}_2$ has a realization as the complex upper half plane $\mbb{H}$ (see, e.g, \cite{Ratcliffe})
\begin{equation}
\mbb{H}=\{\tau\in\mathbb{C} \ |\ \Im(\tau) > 0\}. 
\end{equation}
The group $PGL(2,\mathbb{Z})$ acts on $\tau\in \mathcal{H}_2$ as 
\begin{equation}
\tau\ \longrightarrow \ \tau^{\prime} \ = \  \frac{au+b}{cu+d},
\label{PERTABeq:actionofPGL2Z}
\end{equation}
where we take 
\begin{equation} 
u = \left\{\begin{array}{ccc}
\tau & ;&ad-bc = 1 \\
\bar{\tau} &; & ad-bc= -1. \\
\end{array} \right.
\label{PERTABeq:parameteru}
\end{equation}
The reason for taking $u=\bar{\tau}$ when $ad-bc=-1$ is to ensure that the upper half plane is preserved, i.e., 
\begin{equation}
\Im(\tau)\ > \ 0 \quad \Longrightarrow \quad \Im(\tau^{\prime})\ >\ 0.
\label{PERTABeq:PreservationofU}
\end{equation}
We can think of the transformation (\ref{PERTABeq:actionofPGL2Z}) as the ordinary action of the modular group $PSL(2,\mathbb{Z})\subset PGL(2,\mathbb{Z})$ together with the action of complex conjugation. Recall that $PSL(2,\mathbb{Z})$ is generated by the translation $\mathcal{T} \ :\ \tau\rightarrow \tau+1$ and the inversion $\mathcal{I}\ :\ \tau\rightarrow -1/\tau$, with the realization
\begin{equation}
\mathcal{T}=\left( \begin{array}{cc}
1 & 1 \\
0 & 1Ê\\
\end{array}Ê\right), \qquad
\mathcal{I}=\left( \begin{array}{cc}
0 & -1 \\
1 & 0Ê\\
\end{array}Ê\right).
\label{PERTABeq:GeneratorsPSL2Z}
\end{equation}
The generators of $PGL(2,\mathbb{Z})$ can now be obtained simply by adding the complex conjugation transformation $\tau\ \longrightarrow\  -\bar{\tau}$, i.e., we obtain the three generators:
\begin{eqnarray}
{}Êr_1 & : & \tau\ \longrightarrow\ -\bar{\tau}
\nonumber \\
{} r_0\equiv r_1\circ \mathcal{T}& : & \tau\ \longrightarrow \ 1-\bar{\tau}
\nonumber \\
{}Êr_{-1}\equiv r_1\circ \mathcal{I} & :& \tau\ \longrightarrow \ \frac{1}{\bar{\tau}}.
\label{PERTABeq:GeneratorsPGL2Z}
\end{eqnarray}
These have the matrix realization:
\begin{equation}
r_1=\left( \begin{array}{cc}
1 & 0 \\
0 & -1Ê\\
\end{array}Ê\right),\qquad
r_0=\left( \begin{array}{cc}
1 & -1 \\
0 & -1Ê\\
\end{array}Ê\right), \qquad 
r_{-1}=\left( \begin{array}{cc}
0 & 1 \\
1 & 0Ê\\
\end{array}Ê\right).
\label{PERTABeq:matrixrealizationGeneratorsPGL2Z}
\end{equation}
Using the explicit action of $PGL(2,\mathbb{Z})$ in (\ref{PERTABeq:actionofPGL2Z}) one may verify, e.g., that
\begin{equation}
(r_0r_{-1})\circ (r_0r_{-1})\circ (r_0r_{-1}) \ :\ \tau\ \longrightarrow\ \tau
\end{equation}
and similarly that $(r_1r_{-1})^2=1$. We also have that no product of $(r_1r_0)$ gives the identity, and so we have $(r_1r_0)^{\infty}$. The group generated by $r_1,r_0$ and $r_{-1}$ therefore coincides with the Weyl group of $A_1^{++}$ and we may conclude that 
\begin{equation}
\mathcal{W}(A_1^{++})\ \simeq \ PGL(2, \mathbb{Z}).
\end{equation} 

Let us finally note that one can see that the groups are the same by comparing the geometric properties of the Weyl chamber with the fundamental domain for the action of $PGL(2, \mathbb{Z})$ on the upper half plane. To this end we write $\tau=x+iy\in\mathfrak{U}$, for $x, y\in\mathbb{R}$, and check that $r_1$ acts as 
\begin{equation}
r_1 \ : \ x+iy  \quad \longrightarrow \quad -x+iy,
\end{equation}
which implies that this is a reflection in the ``hyperplane'' $W_1=\{\tau\in \mathbb{C}\ |\ \Re(\tau)=0\}$, i.e., a reflection in the line $x=0$. By similar arguments one finds that the $s_0$ transformation is a reflection in the line $x=1/2 \ (W_0=\{\tau\in \mathbb{C}\ |\ \Re(\tau)=1/2\})$, and that $s_{-1}$ is a reflection in the unit circle $|\tau|=1\ (W_{-1}=\{\tau\in\mathbb{C}\ |\ |\tau|=1\}$). The angle between $W_1$ and $W_0$ is therefore zero, the angle between $W_1$ and $W_{-1}$ is $\pi/2$ and the angle between $W_0$ and $W_{-1}$ is $\pi/3$. Hence the $PGL(2, \mathbb{Z})$-elements $r_1, r_0$ and $r_{-1}$ generate a Coxeter group with Coxeter exponents  $m_{10}=\infty, m_{1(-1)}=2$ and $m_{0(-1)}=3$, which, again, is the same as for the Weyl group of $A_1^{++}$. 

\subsubsection{Subgroups of $\mc{W}(A_1^{++})$}

There exists a powerful method to obtain regular subalgebras of a Kac-Moody algebra $\mf{g}$ by studying subgroups of its Weyl group $\mc{W}(\mf{g})$. In this section we shall illustrate this procedure for the example of $\mf{g}=A_1^{++}$.

Recall that since $A_1^{++}$ is hyperbolic, all elements of the lightcone $\mc{C}\subset \mf{h}$ are timelike or lightlike. This implies that reflections in these walls preserve the time-orientiation, and hence the Weyl group $\mc{W}(A_1^{++})$  is a subgroup of the ortochronous Lorentz group $O^{+}(2,1)$. As we saw above, the Weyl chamber $\mc{C}$ is contained in the future lightcone $\mc{O}_+\subset{\mf{h}}$, and hence the unit hyperboloid $\mc{H}_2=\{\be \in\mc{O}_+ \hs |Ê\hsÊ(\be |\be )=-1\}$ is invariant under $\mc{W}(A_1^{++})$. We may therefore project the Weyl chamber onto the hyperbolic space $\mc{H}_2$. This can be viewed as a projection from the interior of the future lightcone $\mc{O}_+$ onto the coset space $SL(2,\mbb{R})/SO(2)$. The Weyl group of $A_1^{++}$ is then realized as the group of reflections in $\mc{H}_2$, and the walls of the Weyl chamber correspond to the intersections of the hyperplanes $W_1, W_0$ and $W_{-1}$, defined above, with the unit hyperboloid. We have seen above that this Weyl group is isomorphic to the extended modular group $PGL(2, \mbb{Z})=GL(2, \mbb{Z})/\mbb{Z}_2$.

\begin{figure}[ht]
\begin{center}
\includegraphics[width=60mm]{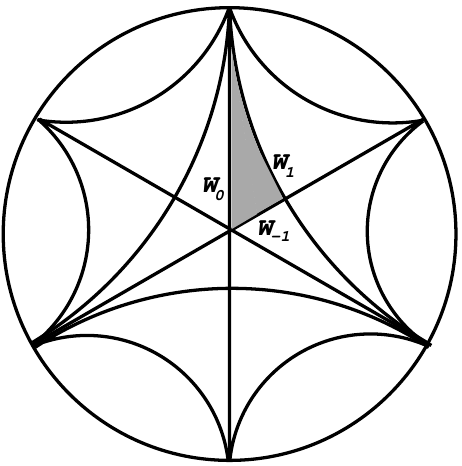}
\caption{Tiling of the hyperbolic plane as defined by the action of the Weyl group $\mc{W}(A_{1,0})$ on the fundamental chamber $\mc{C}=\{\be\in\mf{h}\hs |\hs \al_i(\be)\geq 0, i=1, 0, -1\}$, with $W_1, W_0, W_{-1}$ being the hyperplanes orthogonal to the simple roots $\al_1, \al_0, \al_{-1}$ of $A_1^{++}$.}
\label{figure:A1++TilingFundamentalChamber}
\end{center}
\end{figure}

The action of $PGL(2, \mbb{Z})$ on $\mc{C}$ gives rise to a tiling of the hyperbolic space into an infinite union of subregions, which are all images of the fundamental region $\mc{C}$ under $\mc{W}(A^{++}_{1})$. Each image $\om(\mc{C})\in \mf{h}$, for $\om\in\mc{W}(A_1^{++})$, is a simplex in $\mc{H}_2$, i.e., is bounded by three walls. This is indicated in Figure \ref{figure:A1++TilingFundamentalChamber}.

We shall here use the Weyl group $\mc{W}(A_1^{++})$ to exhibit two explicit embeddings of regular subalgebras of $A_1^{++}$. We consider the rank 3 hyperbolic Kac-Moody algebras $\mf{g}(A_{1,I})$ and $\mf{g}(A_{1,II})$ discussed by Gritsenko and Nikulin in the context of Siegel modular forms \cite{Gritsenko:1996ax}. The second of these, $\mf{g}(A_{1,II})$, has also played an important role recently in understanding the wall-crossing behaviour of BPS states in $D=4$, $\mc{N}=4$ string theory \cite{Dijkgraaf:1996it,Cheng:2008fc,Cheng:2008kt,Cheng:2008gx}.  

%The embedding of $\mf{g}(A_{1,II})$ in $A_1^{++}$ is well known (see, e.g., \cite{FeingoldNicolai,InvarianceUnderCompactification}), while the embedding of  $\mf{g}(A_{1,I})$ appears to be new. 

\begin{figure}[ht]
\begin{center}
\includegraphics[width=30mm]{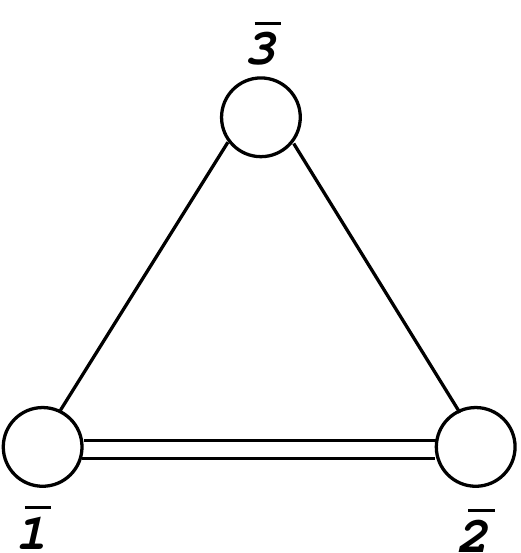}
\caption{The Dynkin diagram of the rank 3 hyperbolic Kac-Moody algebra $\mf{g}(A_{1,I})$. Its embedding into $A_1^{++}$ is defined by $\bar{\al}_1=2\al_1+\al_0, \bar{\al}_2=\al_0, \bar{\al}_3=\al_{-1}$, where $\al_1, \al_0, \al_{-1}$ are the simple roots of $A_1^{++}$.}
\label{figure:A1I}
\end{center}
\end{figure}

\begin{figure}[ht]
\begin{center}
\includegraphics[width=30mm]{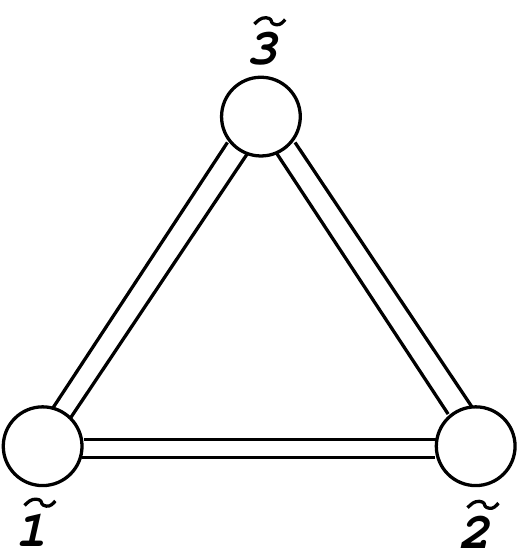}
\caption{The Dynkin diagram of the rank 3 hyperbolic Kac-Moody algebra $\mf{g}(A_{1,II})$. Its embedding into $A_1^{++}$ is defined by $\tilde{\al}_1=\al_1, \tilde{\al}_2=\al_1+2\al_0, \tilde{\al}_3=\al_1+2\al_0+2\al_{-1}$, where $\al_1, \al_0, \al_{-1}$ are the simple roots of $A_1^{++}$.}
\label{figure:A1II}
\end{center}
\end{figure}

The two hyperbolic Kac-Moody algebras $\mf{g}(A_{1,I})$ and $\mf{g}(A_{1,II})$ are constructed from the Cartan matrices
\beq
A_{1,I}=\left(\begin{array}{ccc}
\phantom{-}2 & -2 & -1\\
-2 & \phantom{-}2 & -1\\
-1 & -1 & \phantom{-}2\\
\end{array}\right),\qquad A_{1,II}=\left(\begin{array}{ccc}
\phantom{-}2 & -2 & -2\\
-2 & \phantom{-}2 & -2\\
-2 & -2 & \phantom{-}2\\
\end{array}\right).
\eeq
The associated Dynkin diagrams are displayed in Figures \ref{figure:A1I} and \ref{figure:A1II}. 

\begin{figure}[ht]
\begin{center}
\includegraphics[width=60mm]{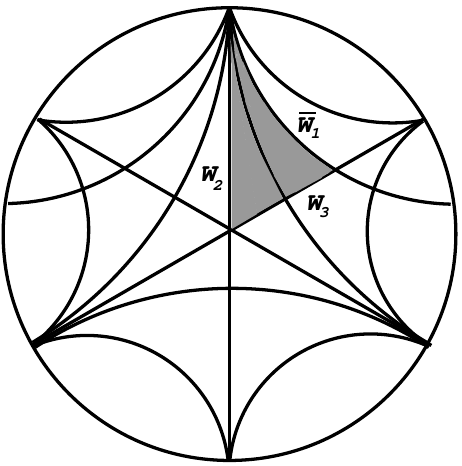}
\caption{The fundamental Weyl chamber $\bar{\mc{C}}$ of $\mc{W}(A_{1,I})$ is shown as the union of $\mc{C}$ and $s_1(\mc{C})$, where $\mc{C}$ is the fundamental Weyl chamber of $\mc{W}(A_{1,0})$.}
\label{figure:A1++TilingZ2Orbifold}
\end{center}
\end{figure}

The Weyl groups $\mc{W}(A_{1,I})$ and $\mc{W}(A_{1,II})$ are hyperbolic Coxeter groups with the following presentations,
\beqa
\mc{W}(A_{1,I})&=&\left< s_i\ |\ s_i^{2}=1,\hs (s_1s_2)^{\infty}=1,\hs (s_1s_3)^{3}=1, \hs (s_2s_3)^{3}=1,\right>,
\nn \\
\mc{W}(A_{1,II})&=&\left< r_i\ |\ r_i^2=1,\hs (r_1r_2)^{\infty}=1,\hs (r_1r_3)^{\infty}=1, \hs (r_2r_3)^{\infty}=1\right>.
\nn \\
\eqa
%\subsubsection{The Embedding $\mf{g}(A_{1,I})\subset A_1^{++}$}
We begin by considering $\mf{g}(A_{1,I})$. Define a new basis of simple roots $\bar{\Pi}\subset \Delta$ as follows
\beq
\bar{\al}_1\equiv 2\al_1+\al_0, \qquad \bar{\al}_2\equiv \al_0, \qquad \bar{\al}_3\equiv \al_{-1}.
\label{embeddingA1I}
\eeq
Clearly, $\bar{\Delta}\subset \Delta$. The scalar products between the new simple roots give rise to the Cartan matrix $A_{1,I}$,
\beq
\f{2(\bar{\al}_i|\bar{\al}_j)}{(\bar{\al}_i|\bar{\al}_i)}=\left(\begin{array}{ccc}
\phantom{-}2 & -2 & -1\\
-2 & \phantom{-}2 & -1\\
-1 & -1 & \phantom{-}2\\
\end{array}\right)=A_{1,I},
\eeq
implying that the algebra $\mf{g}(A_{1,I})$ is a regular subalgebra of $\mf{g}(A_{1,0})=A_1^{++}$, with the embedding defined by (\ref{embeddingA1I}). We further note that the set of simple roots $\bar{\Pi}$ of $\mf{g}(A_{1,I})$ is obtained by replacing $\al_1$ by the image of $\al_0$ under the fundamental reflection $s_1$, 
\beq
s_1(\al_0)=2\al_1+\al_0,
\eeq
implying that the fundamental chamber $\bar{\mc{C}}$ of $\mc{W}(A_{1,I})$ is twice as large as the original fundamental domain $\mc{C}$, and it corresponds to the union
\beq
\bar{\mc{C}}=\mc{C}\cup s_1(\mc{C}),
\eeq
where 
\beq
s_1(\mc{C})=\{h\in\mf{h}\hs |\hs \al_1(h)\leq 0, (2\al_1+\al_0)(h)\geq 0, \al_{-1}(h)\geq 0 \}.
\eeq
Thus, the wall $W_1=\{h\in\mf{h}\hs |\hs \al_1(h)=0\}$ ``slices'' the region $\bar{\mc{C}}$ into two subregions of equal size. This is indicated in Figure \ref{figure:A1++TilingZ2Orbifold}, where the new wall $\bar{W}_1$ is explicitly given by
\beq
\bar{W}_1=\{h\in\mf{h}\ |\ \bar{\al}_1(h)=(2\al_1+\al_0)(h)=0Ê\}.
\eeq
We thus deduce
\beq
\text{Area}\ \bar{\mc{C}} = 2\cdot \text{Area}\ \mc{C}=\f{\pi}{3}.
\eeq
One way to see that the precise numerical factor is $\pi/3$ is to note that the fundamental domain of $\mc{C}$ is half that of the quotient $PSL(2, \mbb{Z})\verb|\|\mc{H}_2$, which is the familiar fundamental domain for the action of $PSL(2, \mbb{Z})$ on the upper half plane $\mc{H}_2$, which is well known to have area $\pi/3$ (see, e.g., \cite{Terras1,Terras2}).

%\subsubsection{The Embedding $\mf{g}(A_{1,II})\subset A_1^{++}$}

We proceed to the case of $\mf{g}(A_{1,II})$. To this end we define yet another set of simple roots $\tilde{\Pi}$ by 
\beq
\tilde{\al}_1=\al_1,\qquad \tilde{\al}_2=\al_1+2\al_0, \qquad \tilde{\al}_3=\al_1+2\al_0+2\al_{-1}.
\label{embeddingA1II}
\eeq
\begin{figure}[ht]
\begin{center}
\includegraphics[width=60mm]{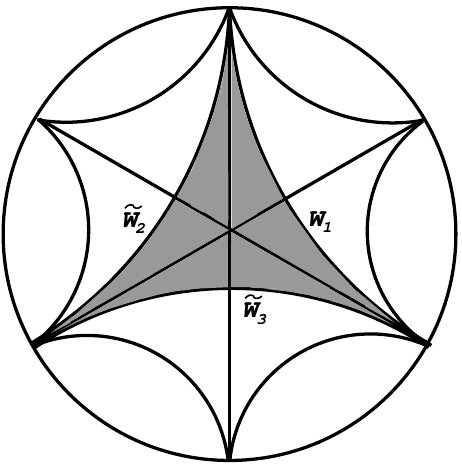}
\caption{The fundamental Weyl chamber $\tilde{\mc{C}}$ of $\mc{W}(A_{1,II})$ is displayed as the ideal triangle inside the Cartan subalgebra of $A_1^{++}$.  }
\label{figure:A1++TilingIdealTriangle}
\end{center}
\end{figure}

The normalized scalar products between these simple roots give rise to the Cartan matrix $A_{1,II}$,
\beq
\f{2(\tilde{\al}_i|\tilde{\al}_j)}{(\tilde{\al}_i|\tilde{\al}_i)}=\left(\begin{array}{ccc}
\phantom{-}2 & -2 & -2\\
-2 & \phantom{-}2 & -2\\
-2 & -2 & \phantom{-}2\\
\end{array}\right)=A_{1,II},
\eeq
and we thereby deduce that $\mf{g}(A_{1,II})$ is a regular subalgebra of $A_1^{++}$ with the embedding defined by (\ref{embeddingA1II}). The structure of the fundamental domain $\tilde{\mc{C}}$ of $\mf{g}(A_{1,II})$ can be analyzed by considering its embedding inside the lightcone of $A_1^{++}$, as above. First note that $\tilde{\mc{C}}$ obviously contains the original fundamental domain $\mc{C}$ of $\mc{W}(A_{1,0})$ as a subregion. In fact, the fundamental Weyl chamber $\tilde{\mc{C}}$ can be embedded into the Cartan subalgebra $\mf{h}\subset A_1^{++}$ as the following union of chambers,
\beq
\tilde{\mc{C}}=\bigcup_{i=1}^{6}\mc{C}_i,
\label{union}
\eeq
where each chamber $\mc{C}_i$ is a copy of the fundamental chamber. Explicitly we have
\beqa
\mc{C}_1&=&\mc{C},
\nn \\
\mc{C}_2&=&s_2(\mc{C}_1),
\nn \\
\mc{C}_3&=& s_3s_2s_3(\mc{C}_2),
\nn \\
\mc{C}_4&=& s_3(\mc{C}_3),
\nn \\
\mc{C}_5&=&s_2(\mc{C}_4),
\nn \\
\mc{C}_6&=&s_3s_2s_3(\mc{C}_5).
\eqa
It is easy to see that the sequence stops at $\mc{C}_6$, since applying $s_3$ on $\mc{C}_6$ takes us back to $\mc{C}$, and any other reflection on $\mc{C}_6$ yields either a chamber outside of $\tilde{\mc{C}}$ or another of the six chambers of $\tilde{\mc{C}}$. We thus find that the size of $\tilde{\mc{C}}$ is
\beq
\text{Area}\hs \tilde{\mc{C}}=6\cdot \text{Area}\hs \mc{C}=\pi.
\eeq
The embedding $\tilde{\mc{C}}\subset\mf{h}$ is displayed in Figure \ref{figure:A1++TilingIdealTriangle}, and it corresponds to the so called ``ideal triangle''. In Chapter \ref{Chapter:Cohomology} we will learn that the union of chambers (\ref{union}) has a nice description in the theory of buildings, and is known as a \emph{gallery}.

\section{Level Decomposition in Terms of Finite Regular Subalgebras}
\label{section:LevelDecomposition}
\setcounter{equation}{0}
%[MODIFY THIS INTRODUCTION]
%We have shown in the previous sections that Weyl groups of
%Lorentzian Kac--Moody algebras naturally emerge when analysing
%gravity in the extreme BKL regime. \index{level decomposition|bb} This has led to the conjecture
%that the corresponding Kac--Moody algebra is in fact a symmetry of
%the theory (most probably enlarged with new
%fields)~\cite{HyperbolicKaluzaKlein}. The idea is that the BKL
%analysis is only the ``revelator'' of that huge symmetry, which would
%exist independently of that limit, without making the BKL
%truncations. Thus, if this conjecture is true, there should be a way
%to rewrite the gravity Lagrangians in such a way that the Kac--Moody
%symmetry is manifest. This conjecture itself was made previously (in
%this form or in similar ones) by other authors on the basis of
%different considerations~\cite{Julia:1980gr,Nicolai:1986jk,E11andMtheory}. To
%explore this conjecture, it is desirable to have a concrete method of
%dealing with the infinite-dimensional structure of a Lorentzian
%Kac--Moody algebra $\mf{g}$. In this section we present such a method.
 
 So far we have discussed general aspects of Kac-Moody algebras, with special emphasis on the Lorentzian subclass. When dealing with these infinite-dimensional structures in a physical context in Chapters \ref{Chapter:Manifest}, \ref{Chapter:GeometricConfigurations} and \ref{Chapter:MassiveIIA} it will prove to be very convenient to slice up the algebras into finite-dimensional subspaces $\mf{g}_{\ell}$. More precisely, following \cite{DHN2}, we will in this section define a so-called
\emph{level decomposition} of the adjoint representation of
$\mf{g}$ such that each level $\ell$ corresponds to a finite
number of representations of a finite regular subalgebra $\mf{r}$
of $\mf{g}$. Generically the decomposition will take the form of
the adjoint representation of $\mf{r}$ plus a (possibly infinite)
number of additional representations of $\mf{r}$. This type of
expansion of $\mf{g}$ will prove to be very useful when
considering sigma models invariant under $\mf{g}$ for which we may
use the level expansion to consistently truncate the theory to any
finite level $\ell$ (see Chapter \ref{Chapter:Manifest}). We begin by illustrating these ideas for the finite-dimensional Lie algebra $\mf{sl}(3,\mbb{R})$ after which we generalize the
procedure to the indefinite case in Sections~\ref{section:FormalC},
\ref{section:DecompAE3AE3} and~\ref{section:DecompE10E10}. This section is based on {\bf Paper III}.

%%%%%%%%%%%%%%%%%%%%%%%%%%%%%%%%%%%%%%%%%%%%%%%%%%%%%%%%%%%%%%%%%%%%%%%%%%%%%%%%%%%
%%%%%%%%%%%%%%%%%%%%%%%%%%%%%%%%%%%%%%%%%%%%%%%%%%%%%%%%%%%%%%%%%%%%%%%%%%%%%%%%%%%

\subsection[A Finite-Dimensional Example: $\mf{sl}(3,\mbb{R})$]
           {A finite-dimensional example:   $\mf{sl}(3,\mbb{R})$}
\label{section:finitedimensionexample}

\begin{figure}[htbp]
  \begin{center}
    \includegraphics[width=35mm]{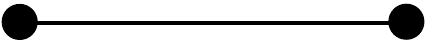}
    \end{center}
    \caption{The Dynkin diagram of $\mf{sl}(3,\mbb{R})$.}
    \label{figure:A2}
    \end{figure}

The rank~2 Lie algebra $\mf{g}=\mf{sl}(3,\mbb{R})$ is
characterized by the Cartan matrix
\begin{equation}
  A[\mf{sl}(3,\mbb{R})]=
  \left(
    \begin{array}{@{}r@{\quad}r@{}}
      2 & -1 \\
      -1 & 2
    \end{array}
  \right),
  \label{CartanMatrixsl(3)}
\end{equation}
whose Dynkin diagram \index{Dynkin diagram} is displayed in Figure~\ref{figure:A2}.

We recall that $\mf{sl}(3, \mbb{R})$ is the split real form of $\mf{sl}(3,\mbb{C})
\equiv A_2$, and is thus defined through the same Chevalley--Serre
presentation as for $\mf{sl}(3,\mbb{C})$, but with all coefficients
restricted to the real numbers.

The Cartan generators $\{ h_1, h_2\} $ will indifferently be denoted
by $ \{ \al_{1}^{\vee}, \al_{2}^{\vee}\}$. As we have seen, they form
a basis of the Cartan subalgebra $\mf{h}$, while the simple roots
$\{\al_{1},\al_{2}\}$, associated with the raising operators $e_1$ and
$e_2$, form a basis of the dual root space $\mf{h}^{\star}$. Any root
$\ga\in \mf{h}^{\star}$ can thus be decomposed in terms of the simple
roots as follows,
\begin{equation}
  \ga=m\al_{1}+\ell\al_{2},
  \label{rootdecompositionsl(3)}
\end{equation}
and the only values of $(m,n)$ are $(1,0)$, $(0,1)$, $(1,1)$ for
the positive roots and minus these values for the negative ones.

\begin{figure}[htbp]
    \centerline{\includegraphics[width=90mm]{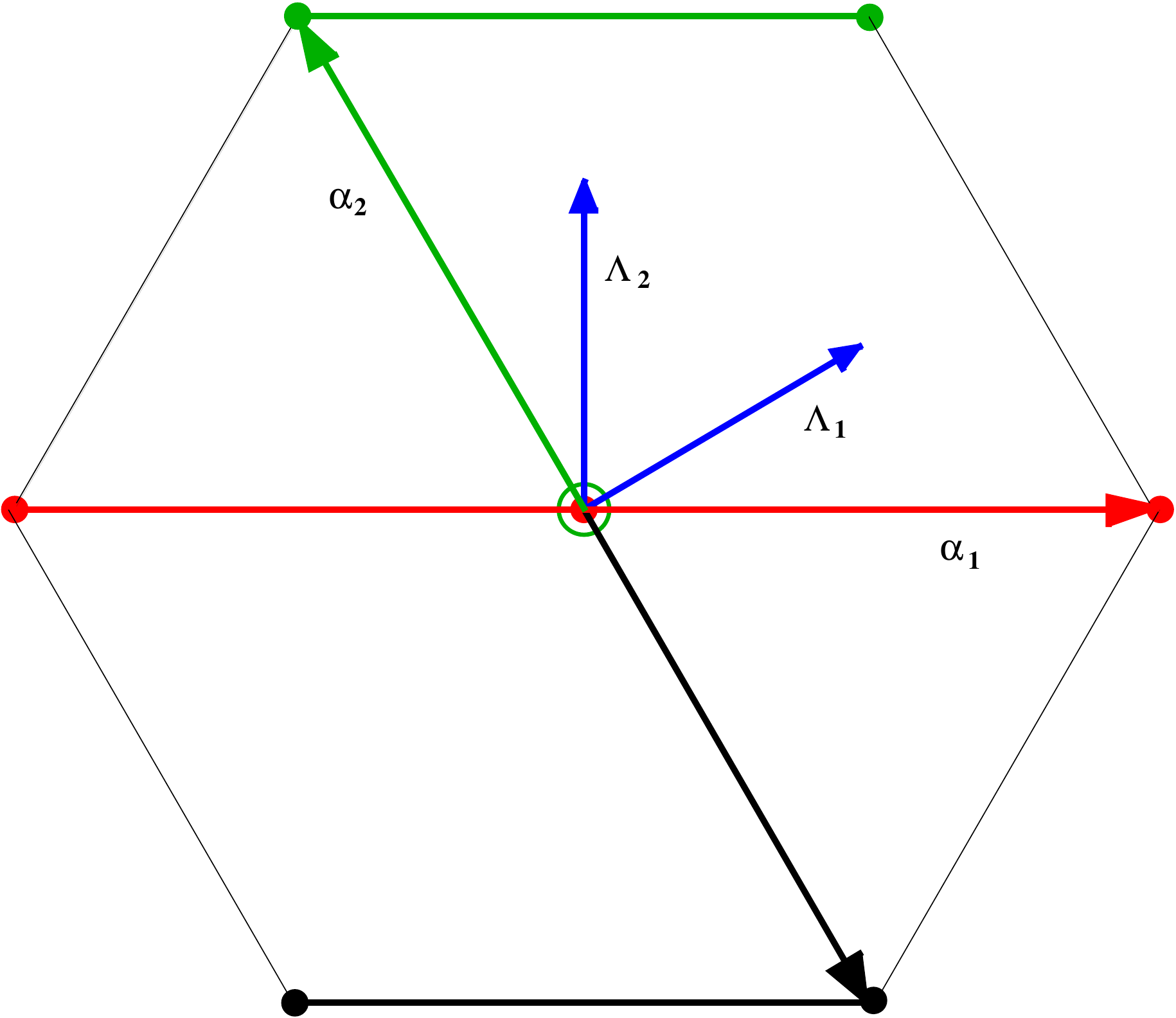}}
    \caption{Level decomposition of the adjoint representation
      $\mc{R}_\mathrm{ad}=\mathbf{8}$ of $\mf{sl}(3,\mbb{R})$ into representations
      of the subalgebra $\mf{sl}(2,\mbb{R})$. The labels $1$ and $2$
      indicate the simple roots $\al_{1}$ and $\al_{2}$. Level zero
      corresponds to the horizontal axis where we find the adjoint
      representation $\mc{R}^{(0)}_\mathrm{ad}=\mathbf{3}_0$ of
      $\mf{sl}(2,\mbb{R})$ (red nodes) and the singlet representation
      $\mc{R}^{(0)}_s=\mathbf{1}_0$ (green circle about the origin). At
      level one we find the two-dimensional representation
      $\mc{R}^{(1)}=\mathbf{2}_1$ (green nodes). The black arrow denotes
      the negative level root $-\al_{2}$ and so gives rise to the
      level $\ell=-1$ representation $\mc{R}^{(-1)}=\mathbf{2}_{(-1)}$. The
      blue arrows represent the fundamental weights $\Lambda_{1}$ and
      $\Lambda_{2}$.}
    \label{figure:AdjointSL3}
  \end{figure}

The algebra $\mf{sl}(3,\mbb{R})$ defines through the adjoint
action a representation of $\mf{sl}(3,\mbb{R})$ itself, called the
adjoint representation, which is eight-dimensional and denoted
$\mathbf{8}$. The weights of the adjoint representation are the
roots, plus the weight $(0,0)$ which is doubly degenerate. The
lowest weight of the adjoint representation is 
\begin{equation}
  \Lambda_{\mf{g}}=-\al_{1}-\al_{2},
  \label{Lowestweightsl(3)}
\end{equation}
corresponding to the generator $[f_1, f_2]$. We display the weights of
the adjoint representation in Figure~\ref{figure:AdjointSL3}.

The idea of the level decomposition \index{level decomposition} is to decompose the adjoint
representation into representations of one of the regular
$\mf{sl}(2,\mbb{R})$-subalgebras associated with one of the two
simple roots $\al_{1}$ or $\al_{2}$, i.e., either
$\{e_1,\al_{1}^{\vee},f_1\}$ or $\{e_2,\al_{2}^{\vee},f_2\}$.
For definiteness we choose the level to count the number $\ell$ of
times the root $\al_{2}$ occurs, as was anticipated by the
notation in Equation~(\ref{rootdecompositionsl(3)}). Consider the subspace
of the adjoint representation spanned by the vectors with a fixed
value of $\ell$. This subspace is invariant under the action of
the subalgebra $\{e_1,\al_{1}^{\vee},f_1\}$, which only changes
the value of $m$. Vectors at a definite level transform
accordingly in a representation of the regular
$\mf{sl}(2,\mbb{R})$-subalgebra 
\begin{equation}
  \mf{r} \equiv
  \mbb{R}e_1\oplus \mbb{R}\al_{1}^{\vee}\oplus \mbb{R}f_1.
\end{equation}

Let us begin by analyzing states at level $\ell=0$, i.e., with
weights of the form $\ga=m\al_{1}$. We see from
Figure~\ref{figure:AdjointSL3} that we are restricted to move along
the horizontal axis in the root diagram. By the defining Lie algebra
relations we know that $\ad_{f_1}(f_1)=0$, implying that
$\Lambda_\mathrm{ad}^{(0)}=-\al_{1}$ is a lowest weight of the
$\mf{sl}(2,\mbb{R})$-representation. Here, the superscript $0$
indicates that this is a level $\ell=0$ representation. The
corresponding complete irreducible module is found by acting on
$f_1$ with $e_1$, yielding 
\begin{equation}
  [e_1,f_1]=\al_{1}^{\vee},
  \qquad
  [e_1,\al_{1}^{\vee}]=-2e_1,
  \qquad [e_1,e_1]=0.
  \label{Representationsl(2)}
\end{equation}
We can then conclude that $\Lambda_\mathrm{ad}^{(0)}=-\al_{1}$ is the lowest
weight of the three-dimensional adjoint representation $\mathbf{3}_0$ of
$\mf{sl}(2,\mbb{R})$ with weights $\{\Lambda_\mathrm{ad}^{(0)}, 0,
-\Lambda_\mathrm{ad}^{(0)}\}$, where the subscript on $\mathbf{3}_0$ again
indicates that this representation is located at level $\ell=0$ in the
decomposition. The module for this representation is
$\mc{L}(\Lambda_\mathrm{ad}^{(0)})=\spn \{f_1,\al_{1}^{\vee},e_1\}$.

This is, however, not the complete content at level zero since we
must also take into account the Cartan generator
$\al_{2}^{\vee}$ which remains at the origin of the root
diagram. We can combine $\al_{2}^{\vee}$ with $\al_{1}^{\vee}$
into the vector 
\begin{equation}
  h=\al_{1}^{\vee}+2\al_{2}^{\vee},
  \label{singletSL2}
\end{equation}
which constitutes the one-dimensional singlet representation
$\mathbf{1}_0$ of $\mf{r}$ since it is left invariant under all
generators of $\mf{r}$, 
\begin{equation}
  [e_1,h]=[f_1,h]=[\al_{1}^{\vee},h]=0,
  \label{singletSL2commutators}
\end{equation}
as follows trivially from the Chevalley relations. Thus level zero
contains the representations $\mathbf{3}_0$ and $\mathbf{1}_0$.

Note that the vectors at level~0 not only transform in a
(reducible) representation of $\mf{sl}(2,\mbb{R})$, but also form
a subalgebra since the level is additive under taking commutators.
The algebra in question is $\mf{gl}(2,\mbb{R}) = \mf{sl}(2,\mbb{R})
\oplus \mathbb{R}$. Accordingly, if the generator $\al_{2}^{\vee}$ is
added to the subalgebra $\mf{r}$, through the combination in
Equation~(\ref{singletSL2}), so as to take the entire $\ell = 0$
subspace, $\mf{r}$ is enlarged from $\mf{sl}(2,\mbb{R})$ to
$\mf{gl}(2,\mbb{R})$, the generator $h$ being the ``trace''
part of $\mf{gl}(2,\mbb{R})$. This fact will prove to be important in
subsequent sections.

Let us now ascend to the next level, $\ell=1$. The weights of
$\mf{r}$ at level 1 take the general form
$\ga=m\al_{1}+\al_{2}$ and the lowest weight is
$\Lambda^{(1)}=\al_{2}$, which follows from the vanishing of the
commutator 
\begin{equation}
  [f_1,e_2]=0.
  \label{vanishingcommutatorlevel1}
\end{equation}
Note that $m \geq 0$ whenever $\ell >0$ since $
m\al_{1}+\ell\al_{2}$ is then a positive root. The complete
representation is found by acting on the lowest weight
$\Lambda^{(1)}$ with $e_1$ and we get that the commutator
$[e_1,e_2]$ is allowed by the Serre relations, \index{Serre relations} while
$[e_1,[e_1,e_2]]$ is killed, i.e.,
\begin{equation}
  \begin{array}{rcl}
    [e_1,e_2] &\neq & 0, \\ [0 em]
    [e_1,[e_1,e_2]]] &=& 0.
  \end{array}
  \label{Serrerelationslevel1} 
\end{equation}
The non-vanishing commutator $e_{\theta}\equiv [e_1,e_2]$ is the
vector associated with the highest root $\theta$ of
$\mf{sl}(3,\mbb{R})$ given by 
\begin{equation}
  \theta=\al_{1}+\al_{2}.
  \label{highestrootsl(3)}
\end{equation}
This is just the negative of the
lowest weight $\Lambda_{\mf{g}}$. The only representation at
level one is thus the two-dimensional representation $\mathbf{2}_1$
of $\mf{r}$ with weights $\{\Lambda^{(1)},\theta\}$. The
decomposition stops at level one for $\mf{sl}(3,\mbb{R})$ because
any commutator with two $e_2$'s vanishes by the Serre relations. \index{Serre relations}
The negative level representations may be found simply by applying
the Chevalley involution and the result is the same as for level
one.

Hence, the total level decomposition \index{level decomposition} of $\mf{sl}(3,\mbb{R})$ in
terms of the subalgebra $\mf{sl}(2,\mbb{R})$ is given by 
\begin{equation}
  \mathbf{8}=\mathbf{3}_0\oplus \mathbf{1}_0 \oplus \mathbf{2}_1 \oplus \mathbf{2}_{(-1)}. 
  \label{leveldecompositionsl(3)}
\end{equation}
Although extremely simple (and familiar), this example illustrates
well the situation encountered with more involved cases below. In the
following analysis we will not mention the negative levels any longer
because these can always be obtained simply through a reflection with
respect to the $\ell =0$ ``hyperplane'', using the Chevalley
involution.

%%%%%%%%%%%%%%%%%%%%%%%%%%%%%%%%%%%%%%%%%%%%%%%%%%%%%%%%%%%%%%%%%%%%%%%%%%%%%%%%%%%
%%%%%%%%%%%%%%%%%%%%%%%%%%%%%%%%%%%%%%%%%%%%%%%%%%%%%%%%%%%%%%%%%%%%%%%%%%%%%%%%%%%

\subsection{Some Formal Considerations}
\label{section:FormalC}

Before we proceed with a more involved example, let us formalize
the procedure outlined above. We mainly follow the excellent
treatment given in~\cite{AxelThesis}, although we restrict ourselves to the
cases where $\mf{r}$ is a \emph{finite} regular subalgebra \index{regular subalgebra} of
$\mf{g}$.

In the previous example, we performed the decomposition of the
roots (and the ensuing decomposition of the algebra) with respect
to one of the simple roots which then defined the level. In
general, one may consider a similar decomposition of the roots of
a rank $r$ Kac--Moody algebra  \index{Kac--Moody algebra} with respect to an arbitrary number
$s<r$ of the simple roots and then the level $\ell$ is generalized
to the ``multilevel'' $\mathbf{\ell}=(\ell_1,\cdots, \ell_s)$.

%%%%%%%%%%%%%%%%%%%%%%%%%%%%%%%%%%%%%%%%%%%%%%%%%%%%%%%%%%%%%%%%%%%%%%%%%%%%%%%%%%%

\subsubsection{Gradation}

We consider a Kac--Moody algebra $\mf{g}$ of rank $r$ and
we let $\mf{r}\subset \mf{g}$ be a finite regular rank $m< r$
subalgebra of $\mf{g}$ whose Dynkin diagram is obtained by
deleting a set of nodes $\mc{N}=\{n_1, \cdots, n_s\} \, (s=r-m)$
from the Dynkin diagram of $\mf{g}$.

Let $\ga$ be a root of $\mf{g}$, 
\begin{equation}
  \ga=\sum_{i \notin \mc{N}}
  m_i \al_{i} + \sum_{a \in \mc{N}}\ell_a \al_{a}.
\end{equation}
To this decomposition of the roots corresponds a decomposition of the
algebra, which is called a \emph{gradation} of $\mf{g}$ and which can
be written formally as 
\begin{equation}
  \mf{g}=\bigoplus_{\mathbf{\ell}\in \mbb{Z}^{s}} \mf{g}_{\mathbf{\ell}},
  \label{gradation}
\end{equation}
where for a given $\ell$, $\mf{g}_{\mathbf{\ell}}$ is the subspace
spanned by all the vectors $e_\gamma$ with that definite value $\ell$
of the multilevel, 
\begin{equation}
  [h,e_\gamma] = \gamma(h) e_\gamma,
  \qquad
  \ell_a(\gamma) = \ell_a.
\end{equation}
Of course, if $\mf{g}$ is finite-dimensional this sum terminates for
some finite level, as in Equation~(\ref{leveldecompositionsl(3)}) for
$\mf{sl}(3,\mbb{R})$. However, in the following we shall mainly be
interested in cases where Equation~(\ref{gradation}) is an infinite
sum.

We note for further reference that the following structure is
inherited from the gradation:
\begin{equation}
  [\mf{g}_{\ell},\mf{g}_{\ell^{\prime}}]\subseteq \mf{g}_{\ell+\ell^{\prime}}. 
  \label{gradation2}
\end{equation}
This implies that for $\ell=0$ we have 
\begin{equation}
  [\mf{g}_0,\mf{g}_{\ell^{\prime}}]\subseteq \mf{g}_{\ell^{\prime}},
\end{equation}
which means that $\mf{g}_{\ell^{\prime}}$ is a representation of
$\mf{g}_0$ under the adjoint action. Furthermore, $\mf{g}_0$ is a
subalgebra. Now, the algebra $\mf{r}$ is a subalgebra of $\mf{g}_0$
and hence we also have 
\begin{equation}
  [\mf{r},\mf{g}_{\ell^{\prime}}]\subseteq \mf{g}_{\ell^{\prime}},
\end{equation}
so that \emph{the subspaces $\mf{g}_{\mathbf{\ell}}$ at definite values of
  the multilevel are invariant subspaces under the adjoint action of
  $\mf{r}$}. In other words, the action of $\mf{r}$ on
$\mf{g}_{\ell}$ does not change the coefficients $\ell_a$.

At level zero, $\mathbf{\ell}=(0,\cdots,0)$, the representation of the
subalgebra $\mf{r}$ in the subspace $\mf{g}_0$ contains the
adjoint representation of $\mf{r}$, just as in the case of
$\mf{sl}(3,\mbb{R})$ discussed in
Section~\ref{section:finitedimensionexample}. All positive and
negative roots of $\mf{r}$ are relevant. Level zero contains in
addition $s$ singlets for each of the Cartan generator associated to
the set $\mc{N}$.

Whenever one of the $\ell_a$'s is positive, all the other ones
must be non-negative for the subspace $\mf{g}_{\mathbf{\ell}}$ to be
nontrivial and only positive roots appear at that value of the
multilevel.

%%%%%%%%%%%%%%%%%%%%%%%%%%%%%%%%%%%%%%%%%%%%%%%%%%%%%%%%%%%%%%%%%%%%%%%%%%%%%%%%%%%

\subsubsection{Weights of $\mf{g}$ and Weights of $\mf{r}$}

Let $V$ be the module of a representation $\mc{R}(\mf{g})$ of $\mf{g}$
and $\Lambda \in \mf{h}^{\star}_{\mf{g}}$ be one of the weights
occurring in the representation. We define the action of $h\in
\mf{h}_{\mf{g}}$ in the representation $\mc{R}(\mf{g})$ on $x\in V$ as 
\begin{equation}
  h \cdot x = \Lambda(h) x 
\end{equation}
(we consider representations of $\mf{g}$
for which one can speak of ``weights''~\cite{Kac}). Any
representation of $\mf{g}$ is also a representation of $\mf{r}$.
When restricted to the Cartan subalgebra $\mf{h}_{\mf{r}}$ of
$\mf{r}$, $\Lambda$ defines a weight $\bar{\Lambda} \in
\mf{h}^{\star}_{\mf{r}}$, which one can realize geometrically as
follows.

The dual space $\mf{h}^{\star}_{\mf{r}}$ may be viewed as the
$m$-dimensional subspace $\Pi$ of $\mf{h}^{\star}_{\mf{g}}$
spanned by the simple roots $\al_{i}$, $i \notin \mc{N}$. The
metric induced on that subspace is positive definite since
$\mf{r}$ is finite-dimensional. This implies, since we assume
that the metric on $\mf{h}^{\star}_{\mf{g}}$ is non-degenerate,
that $\mf{h}^{\star}_{\mf{g}}$ can be decomposed as the direct sum
\begin{equation}
  \mf{h}^{\star}_{\mf{g}} = \mf{h}^{\star}_{\mf{r}} \oplus \Pi^\perp.
\end{equation}
To that decomposition corresponds the decomposition 
\begin{equation}
  \Lambda = \Lambda^\| + \Lambda^\perp
\end{equation}
of any weight, where $\Lambda^\| \in \mf{h}^{\star}_{\mf{r}} \equiv
\Pi$ and $\Lambda^\perp \in \Pi^\perp$. Now, let $h = \sum k_i
\alpha_{i}^\vee \in \mf{h}_{\mf{r}}$ ($i \notin \mc{N}$). One
has $\Lambda(h) = \Lambda^\| (h) + \Lambda^\perp (h) = \Lambda^\|
(h)$ because $\Lambda^\perp (h) = 0$: The component perpendicular
to $\mf{h}^{\star}_{\mf{r}}$ drops out. Indeed, $\Lambda^\perp
(\alpha_{i}^\vee) = \frac{2 (\Lambda^\perp \vert
\alpha_{i})}{(\alpha_{i} \vert \alpha_{i})} = 0$ for $i
\notin \mc{N}$.

It follows that one can identify the weight $\bar{\Lambda} \in
\mf{h}^{\star}_{\mf{r}}$ with the orthogonal projection
$\Lambda^\| \in \mf{h}^{\star}_{\mf{r}}$ of $\Lambda \in
\mf{h}^{\star}_{\mf{g}}$ on $\mf{h}^{\star}_{\mf{r}}$. This is
true, in particular, for the fundamental weights $\Lambda_{i}$.
The fundamental weights $\Lambda_{i}$ project on $0$ for $i \in
\mc{N}$ and project on the fundamental weights
$\bar{\Lambda}_{i}$ of the subalgebra ${\mf{r}}$ for $i \notin
\mc{N}$. These are also denoted $\lambda_{i}$. For a general
weight, one has 
\begin{equation}
  \Lambda=\sum_{i \notin \mc{N}} p_i \Lambda_{i} + \!\!
  \sum_{a \in \mc{N}} k_a \Lambda_{a}
  \label{generalweight}
\end{equation}
and 
\begin{equation}
  \bar{\Lambda} = \Lambda^\| = \sum_{i \notin \mc{N}} p_i \lambda_{i}.
\end{equation}
The coefficients $p_i$ can easily be extracted by taking the scalar
product with the simple roots, 
\begin{equation}
  p_i = \frac{2}{(\alpha_i \vert \alpha_i)}(\alpha_i \vert \Lambda),
\end{equation}
a formula that reduces to 
\begin{equation}
  p_i = (\alpha_i \vert \Lambda)
\end{equation}
in the simply-laced case. Note that $(\Lambda^\| \vert \Lambda^\|)> 0$
even when $\Lambda$ is non-spacelike.

%%%%%%%%%%%%%%%%%%%%%%%%%%%%%%%%%%%%%%%%%%%%%%%%%%%%%%%%%%%%%%%%%%%%%%%%%%%%%%%%%%%

\subsubsection{Outer Multiplicity}
\index{outer multiplicity}

There is an interesting relationship between root
multiplicities in the Kac--Moody algebra $\mf{g}$ and weight
multiplicites of the corresponding $\mf{r}$-weights, which we
will explore here.

For finite Lie algebras, the roots always come with multiplicity
one. This is in fact true also for the real roots of indefinite
Kac--Moody algebras. However, recall from 
Sections~\ref{Section:rootsystem} and \ref{Section:WeylGroup} that imaginary roots can have
arbitrarily large multiplicity. This must therefore be taken into
account in the sum~(\ref{gradation}).

Let $\ga\in\mf{h}^{\star}_{\mf{g}}$ be a root of $\mf{g}$. There
are two important ingredients:
\begin{itemize}
\item The multiplicity $\mult(\ga)$ of each
  $\ga\in\mf{h}^{\star}_{\mf{g}}$ at level $\mathbf{\ell}$ as a
  \emph{root} of $\mf{g}$.
\item The multiplicity $\mult_{\mc{R}^{(\ell)}_{\ga}}(\ga)$ of the
  corresponding weight $\bar{\ga} \in \mf{h}^{\star}_{\mf{r}}$ at
  level $\mathbf{\ell}$ as a \emph{weight} in the representation
  $\mc{R}^{(\ell)}_\ga$ of $\mf{r}$. (Note that two distinct roots at
  the same level project on two distinct ${\mf{r}}$-weights, so that
  given the ${\mf{r}}$-weight and the level, one can reconstruct the
  root.)
\end{itemize}
It follows that the root multiplicity of $\ga$ is given as a sum
over its multiplicities as a weight in the various representations
$\{\mc{R}^{(\ell)}_q \, | \, q=1,\cdots, N_{\ell}\}$ at level $\ell$.
Some representations can appear more than once at each level, and
it is therefore convenient to introduce a new measure of
multiplicity, called the \emph{outer multiplicity} \index{outer multiplicity|bb} 
$\mu(\mc{R}^{(\ell)}_{q})$, which counts the number of times each
representation $\mc{R}^{(\ell)}_q$ appears at level $\ell$. So,
for each representation at level $\mathbf{\ell}$ we must count the
individual weight multiplicities in that representation and also
the number of times this representation occurs. The total
multiplicity of $\ga$ can then be written as 
\begin{equation}
  \mult(\ga)=\sum_{q=1}^{N_{\ell}}\mu(\mc{R}^{(\ell)}_q)
  \mult_{\mc{R}^{(\ell)}_q}(\ga). 
\label{totalmultiplicity}
\end{equation}
This simple formula might provide useful information on which
representations of $\mf{r}$ are allowed within $\mf{g}$ at a given
level. For example, if $\ga$ is a real root of $\mf{g}$, then it
has multiplicity one. This means that in the
formula~(\ref{totalmultiplicity}), only the representations of
$\mf{r}$ for which $\ga$ has weight multiplicity equal to one are
permitted. The others have $\mu(\mc{R}^{(\ell)}_q)=0$.
Furthermore, only one of the permitted representations does
actually occur and it has necessarily outer multiplicity equal to
one, $\mu(\mc{R}^{(\ell)}_q)=1$.

The subspaces $\mf{g}_{\mathbf{\ell}}$ can now be written explicitly
as 
\begin{equation}
  \mf{g}_{\mathbf{\ell}}=\bigoplus_{q=1}^{N_{\mathbf{\ell}}}
  \left[ \bigoplus_{k=1}^{\mu(\mc{R}^{(\ell)}_{q})}
  \mc{L}^{[k]}(\Lambda_q^{(\mathbf{\ell})})\right],
  \label{decompositionsubspaces}
\end{equation}
where $\mc{L}(\Lambda^{(\ell)}_q)$ denotes the module of the
representation $\mc{R}^{(\ell)}_q$ and $N_{\mathbf{\ell}}$ is the
number of inequivalent representations at level $\ell$. It is
understood that if $\mu(\mc{R}^{(\ell)}_{q})=0$ for some $\ell$
and $q$, then $\mc{L}(\Lambda^{(\mathbf{\ell})}_{q})$ is absent from
the sum. Note that the superscript $[k]$ labels multiple modules
associated to the same representation, e.g., if
$\mu(\mc{R}^{(\ell)}_q)=3$ this contributes to the sum with a
term
\begin{equation}
  \mc{L}^{[1]}(\Lambda^{(\ell)}_q)\oplus
  \mc{L}^{[2]}(\Lambda^{(\ell)}_q)\oplus
  \mc{L}^{[3]}(\Lambda^{(\ell)}_q).
\end{equation}

Finally, we mention that the multiplicity $\mult(\al)$ of
a root $\al\in\mf{h}^{\star}$ can be computed recursively using
the \emph{Peterson recursion relation}, defined as~\cite{Kac} 
\begin{equation}
  (\al\vert\al-2\,\rho)c_{\al}= \!\!\!
  \sum_{\scriptsize
    \begin{array}{l}
      \gamma+\gamma'=\al \\
      \gamma,\gamma'\in Q_+
    \end{array}} \!\!\!
  (\gamma\vert\gamma')c_{\gamma}c_{\gamma'}, 
  \label{PetersonRelation} 
\end{equation}
where $Q_+$ denotes the set of all positive integer linear
combinations of the simple roots, i.e., the positive part of the root
lattice, and $\rho$ is the Weyl vector (see Eq. (\ref{WeylVector}). The coefficients $c_{\gamma}$ are
defined as
\begin{equation}
  c_{\ga}=\sum_{k\geq1}\frac 1k\mult\left(\f{\ga}{k}\right),
\end{equation}
and, following~\cite{Bergshoeff:2007qi}, we call this the
\emph{co-multiplicity}. Note that if $\ga/k$ is not a root, this gives
no contribution to the co-multiplicity. Another feature of the
co-multiplicity is that even if the multiplicity of some root $\ga$ is
zero, the associated co-multiplicity $c_{\ga}$ does not necessarily
vanish. Taking advantage of the fact that all real roots have
multiplicity one it is possible, in principle, to compute recursively
the multiplicity of any imaginary root. Since no closed formula exists
for the outer multiplicity $\mu$, one must take a detour via the
Peterson relation and Equation~(\ref{totalmultiplicity}) in order to
find the outer multiplicity of each representation at a given
level.

%%%%%%%%%%%%%%%%%%%%%%%%%%%%%%%%%%%%%%%%%%%%%%%%%%%%%%%%%%%%%%%%%%%%%%%%%%%%%%%%%%%
%%%%%%%%%%%%%%%%%%%%%%%%%%%%%%%%%%%%%%%%%%%%%%%%%%%%%%%%%%%%%%%%%%%%%%%%%%%%%%%%%%%

\subsection{Level Decomposition of   $A_1^{++}$}
\label{section:DecompAE3AE3}

The Kac--Moody algebra  \index{Kac--Moody algebra}$A_1^{++}$ \index{$A_1^{++}$}is one of the simplest
hyperbolic algebras and so provides a nice testing ground for
investigating general properties of hyperbolic Kac--Moody algebras.
From a physical point of view, it is the Weyl group of $A_1^{++}$
which governs the chaotic behavior of pure four-dimensional
gravity close to a spacelike singularity~\cite{HyperbolicKaluzaKlein}. Moreover, as we saw in
Section~\ref{Section:WeylGroupA1++}, the Weyl group of $A_1^{++}$ is isomorphic
to the well-known arithmetic group $PGL(2,\mbb{Z})$.

\begin{figure}[htbp]
    \centerline{\includegraphics[width=50mm]{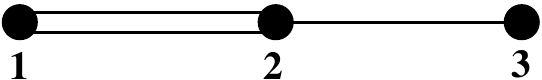}}
    \caption{The Dynkin diagram of the hyperbolic Kac--Moody algebra
      $A_1^{++}\equiv A_1^{++}$. The labels indicate the simple roots
      $\al_{1}, \al_{2}$ and $\al_{3}$. The nodes ``2'' and ``3''
      correspond to the subalgebra $\mf{r}=\mf{sl}(3,\mbb{R})$ with
      respect to which we perform the level decomposition.}
    \label{figure:A1pp2ndversion}
  \end{figure}

The level decomposition \index{level decomposition}of $\mf{g}=A_1^{++}$ follows a similar route
as for $\mf{sl}(3,\mbb{R})$ above, but the result is much more
complicated due to the fact that $A_1^{++}$ is infinite-dimensional.
This decomposition has been treated before in~\cite{DHNReview}.
Recall that the Cartan matrix \index{Cartan matrix}for $A_1^{++}$ is given by 
\begin{equation}
  \left(
    \begin{array}{@{}r@{\quad}r@{\quad}r@{}}
      2 & -2 & 0 \\
      -2 & 2 & -1 \\
      0 & -1 & 2
    \end{array}
  \right),
  \label{CartanMatrixAE3}
\end{equation}
and the associated Dynkin diagram \index{Dynkin diagram}is given in
Figure~\ref{figure:A1pp2ndversion}.

We see that there exist three rank 2 regular subalgebras \index{regular subalgebra}that we
can use for the decomposition: $A_2, A_1\oplus A_1$ or $A_1^{+}$.
We will here focus on the decomposition into representations of
$\mf{r}=A_2=\mf{sl}(3,\mbb{R})$ because this is the one relevant
for pure gravity in four
dimensions~\cite{HyperbolicKaluzaKlein}\footnote{The
  decomposition of $A_1^{++}$ into representations of $A_1^{+}$ was done
  in~\cite{FeingoldFrenkel}.}. The level $\ell$ is then the coefficient
in front of the simple root $\al_{1}$ in an expansion of an arbitrary
root $\ga\in \mf{h}_{\mf{g}}^{\star}$, i.e., 
\begin{equation}
  \ga =\ell \al_{1}+m_2\al_{2}+m_3\al_{3}. 
  \label{rootAE3}
\end{equation}

\begin{figure}
    \centerline{\includegraphics[width=150mm]{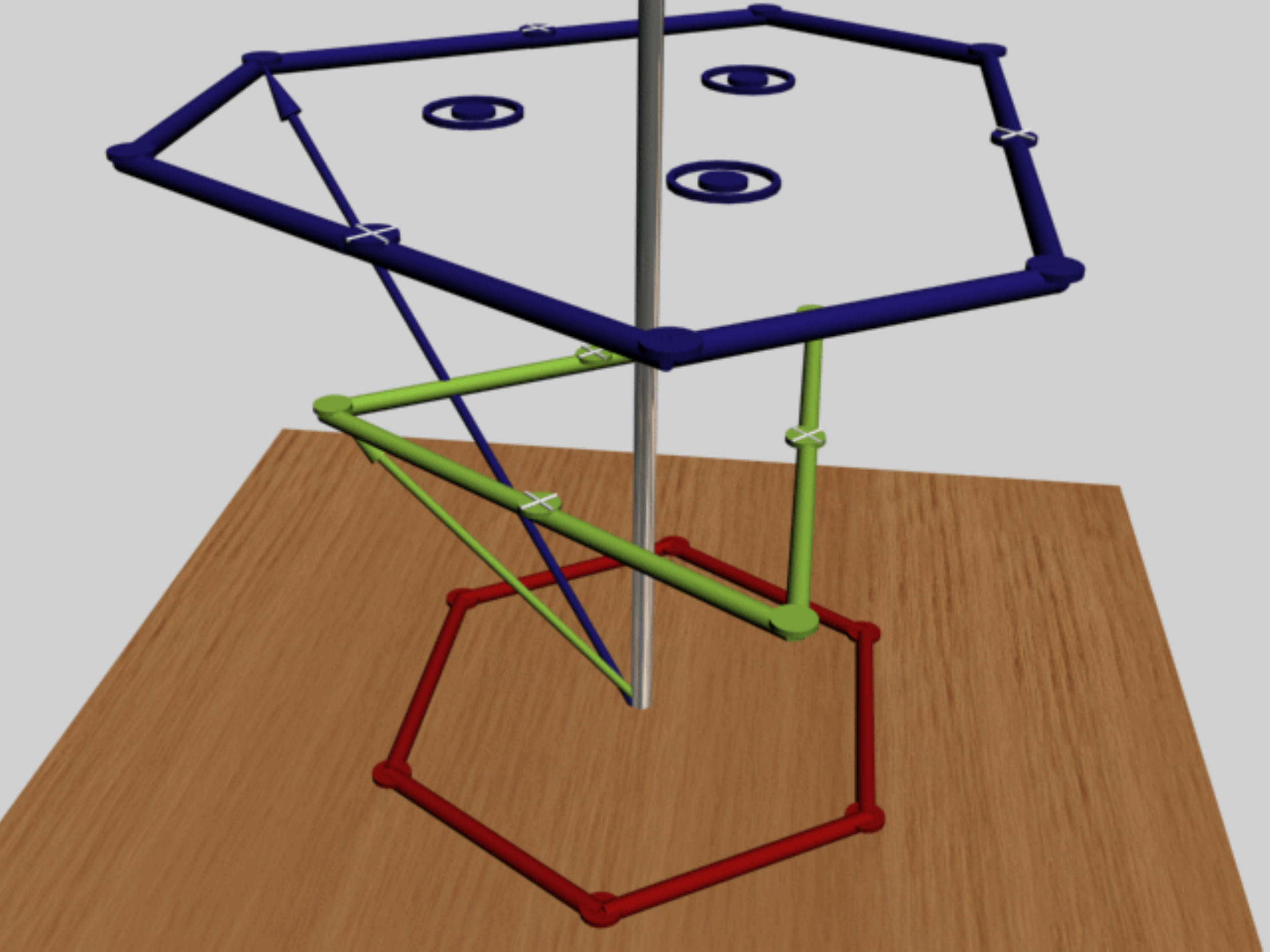}}
    \caption{Level decomposition of the adjoint representation of
      $A_1^{++}$. We have displayed the decomposition up to positive level
      $\ell=2$. At level zero we have the adjoint representation
      $\mc{R}^{(0)}_1=\mathbf{8}_0$ of $\mf{sl}(3,\mbb{R})$ and the singlet
      representation $\mc{R}^{(0)}_2=\mathbf{1}_0$ defined by the simple
      Cartan generator $\al_{1}^{\vee}$. Ascending to level one with
      the root $\al_{1}$ (green vector) gives the lowest weight
      $\Lambda^{(1)}$ of the representation $\mc{R}^{(1)}=\mathbf{6}_1$.
      The weights of $\mc{R}^{(1)}$ labelled by white crosses are on the
      lightcone and so their norm squared is zero. At level two we find
      the lowest weight $\Lambda^{(2)}$ (blue vector) of the
      15-dimensional representation $\mc{R}^{(2)}=\mathbf{15}_2$. Again,
      the white crosses label weights that are on the lightcone. The
      three innermost weights are inside of the lightcone and the rings
      indicate that these all have multiplicity 2 as weights of
      $\mc{R}^{(2)}$. Since these also have multiplicity 2 as
      \emph{roots} of $\mf{h}^{\star}_{\mf{g}}$ we find that the outer
      multiplicity of this representation is one, $\mu(\mc{R}^{(2)})=1$.}
    \label{figure:AE3Dec}
  \end{figure}

We restrict henceforth our analysis to positive levels only,
$\ell\geq 0$. Before we begin, let us develop an intuitive idea of
what to expect. We know that at each level we will have a set of
finite-dimensional representations of the subalgebra $\mf{r}$. The
corresponding weight diagrams will then be represented in a
Euclidean two-dimensional lattice in exactly the same way as in
Figure~\ref{figure:AdjointSL3} above. The level $\ell$ can be
understood as parametrizing a third direction that takes us into
the full three-dimensional root space of $A_1^{++}$. We display the
level decomposition \index{level decomposition}up to positive level two in
Figure~\ref{figure:AE3Dec}.

From previous sections we recall that $A_1^{++}$ is hyperbolic so its
root space is of Lorentzian signature. This implies that there is
a lightcone in $\mf{h}^{\star}_{\mf{g}}$ whose origin lies at the
origin of the root diagram for the adjoint representation of
$\mf{r}$ at level $\ell=0$. The lightcone separates real roots
from imaginary roots and so it is clear that if a representation at
some level $\ell$ intersects the walls of the lightcone, this
means that some weights in the representation will correspond to
imaginary roots of $\mf{h}_{\mf{g}}^{\star}$ but will be real as
weights of $\mf{h}_{\mf{r}}^{\star}$. On the other hand if a
weight lies outside of the lightcone it will be real both as a
root of $\mf{h}_{\mf{g}}^{\star}$ and as a weight of
$\mf{h}_{\mf{r}}^{\star}$.
%\vspace{.2cm}

\subsubsection[Level $\ell=0$]{Level   $\ell=0$}

Consider first the representation content at level zero.
Given our previous analysis we expect to find the adjoint
representation of $\mf{r}$ with the additional singlet
representation from the Cartan generator $\al_{1}^{\vee}$. The
Chevalley generators of $\mf{r}$ are
$\{e_2,f_2,e_3,f_3,\al_{2}^{\vee},\al_{3}^{\vee}\}$ and the
generators associated to the root defining the level are
$\{e_1,f_1,\al_{1}^{\vee}\}$. As discussed previously, the
additional Cartan generator $\al_{1}^{\vee}$ that sits at the
origin of the root space enlarges the subalgebra from
$\mf{sl}(3,\mbb{R})$ to $\mf{gl}(3,\mbb{R})$. A canonical
realisation of $\mf{gl}(3,\mbb{R})$ is obtained by defining the
Chevalley generators in terms of the matrices ${K^{i}}_j \,
(i,j=1,2,3)$ whose commutation relations are 
\begin{equation}
  [{K^{i}}_j,{K^{k}}_l]=\delta^{k}_j {K^{i}}_l-\delta^{i}_l {K^{k}}_j. 
  \label{gl(3)}
\end{equation}
All the defining Lie algebra
relations of $\mf{gl}(3,\mbb{R})$ are then satisfied if we make
the identifications 
\begin{equation}
  \begin{array}{rcl@{\qquad}rcl@{\qquad}rcl}
    && & && & \al_{1}^{\vee}&=& {K^{1}}_1-{K^{2}}_2-{K^{3}}_3, \\
    e_2&=&{K^{2}}_1, &
    f_2&=&{K^{1}}_2, &
    \al_{2}^{\vee}&=&{K^{2}}_2-{K^{1}}_1, \\
    e_3&=&{K^{3}}_2, &
    f_3&=&{K^{2}}_3, &
    \al_{3}^{\vee}&=&{K^{3}}_3-{K^{2}}_2.
  \end{array}
  \label{realisationgl(3)} 
\end{equation}
Note that the trace
${K^{1}}_1+{K^{2}}_2+{K^{3}}_3$ is equal to $-4 \al_{2}^{\vee} - 2
\al_{3}^{\vee} - 3 \al_{1}^{\vee}$. The generators $e_1$ and $f_1$
can of course not be realized in terms of the matrices ${K^{i}}_j$
since they do not belong to level zero. The invariant bilinear
form $(\ |\ )$ at level zero reads 
\begin{equation}
  ({K^{i}}_j|{K^{k}}_l)=
  \delta^{i}_l\delta^{k}_j-\delta^{i}_j\delta^{k}_l,
  \label{Killingformlevelzero}
\end{equation}
where the coefficient in front of the second term on the right hand
side is fixed to $-1$ through the embedding of $\mf{gl}(3, \mbb{R})$
in $A_1^{++}$.

The commutation relations in Equation~(\ref{gl(3)}) characterize the
adjoint representation of $\mf{gl}(3,\mbb{R})$ as was expected at
level zero, which decomposes as the representation
$\mc{R}^{(0)}_\mathrm{ad} \oplus \mc{R}^{(0)}_{s}$ of
$\mf{sl}(3,\mbb{R})$ with $\mc{R}^{(0)}_\mathrm{ad}=\mathbf{8}_0$ and
$\mc{R}^{(0)}_{s}=\mathbf{1}_0$.

\subsubsection{Dynkin Labels}

It turns out that at each positive level $\ell$, the
weight that is easiest to identify is the lowest weight. For
example, at level one, the lowest weight is simply $\alpha_1$ from
which one builds all the other weights by adding appropriate
positive combinations of the roots $\alpha_2$ and $\alpha_3$. It
will therefore turn out to be convenient to characterize the
representations at each level by their (conjugate) Dynkin labels \index{Dynkin labels|bb}
$p_2$ and $p_3$ defined as the coefficients of minus the
(projected) lowest weight $-\bar{\Lambda}^{(\ell)}_\mathrm{lw}$ expanded
in terms of the fundamental weights $\lambda_{2}$ and
$\lambda_{3}$ of $\mf{sl}(3,\mbb{R})$ (blue arrows in
Figure~\ref{figure:15ofSL3}),
\begin{equation}
  -\bar{\Lambda}^{(\ell)}_\mathrm{lw} = p_2 \lambda_{2} + p_3\lambda_{3}. 
  \label{DynkinLabels}
\end{equation}
Note that for any weight $\Lambda$ we have the inequality 
\begin{equation}
  (\Lambda|\Lambda)\leq (\bar{\Lambda}|\bar{\Lambda})
  \label{projectionInequality}
\end{equation}
since $(\Lambda|\Lambda)= (\bar{\Lambda}|\bar{\Lambda}) - \vert
(\Lambda^\perp|\Lambda^\perp) \vert$.

The Dynkin labels \index{Dynkin labels}can be computed using the
scalar product $(\ |\ )$ in $\mf{h}_{\mf{g}}^{\star}$ in the following way:
\begin{equation}
  p_2=-(\al_{2}|\Lambda^{(\ell)}_\mathrm{lw}),
  \qquad
  p_3=-(\al_{3}|\Lambda^{(\ell)}_\mathrm{lw}). 
  \label{DynkinLabels2}
\end{equation}
For the level zero sector we therefore have 
\begin{equation}
  \begin{array}{rcl}
    \mathbf{8}_0 & : & [p_2,p_3]=[1,1],
    \\
    \mathbf{1}_0 & : & [p_2,p_3]=[0,0].
  \end{array}
  \label{DynkinLabelslevelzero}
\end{equation}

The module for the representation $\mathbf{8}_0$ is realized by the
eight traceless generators ${K^{i}}_j$ of $\mf{sl}(3,\mbb{R})$
and the module for the representation $\mathbf{1}_0$ corresponds to
the ``trace'' $\al_{1}^{\vee}$.

Note that the highest weight $\Lambda_\mathrm{hw}$ of a given
representation of $\mf{r}$ is not in general equal to minus the
lowest weight $\Lambda$ of the same representation. In fact,
$-\Lambda_\mathrm{hw}$ is equal to the lowest weight of the
\emph{conjugate} representation. This is the reason our Dynkin
labels are really the conjugate Dynkin labels \index{Dynkin labels}in standard
conventions. It is only if the representation is self-conjugate
that we have $\Lambda_\mathrm{hw}=-\Lambda$. This is the case for example
in the adjoint representation $\mathbf{8}_0$.

It is interesting to note that since the weights of a
representation at level $\ell$ are related by Weyl reflections to
weights of a representation at level $-\ell$, it follows that the
negative of a lowest weight $\Lambda^{(\ell)}$ at level $\ell$ is
actually equal to the \emph{highest} weight
$\Lambda_\mathrm{hw}^{(-\ell)}$ of the conjugate representation at level
$-\ell$. Therefore, the Dynkin labels at level $\ell$ as defined
here are the standard Dynkin labels of the representations at
level $- \ell$.

\subsubsection{Level   $\ell=1$}

We now want to exhibit the representation content at the
next level $\ell=1$. A generic level one commutator is of the form
$[e_1,[\cdots [\cdots ]]]$, where the ellipses denote (positive)
level zero generators. Hence, including the generator $e_1$
implies that we step upwards in root space, i.e., in the direction
of the forward lightcone. The root vector $e_1$ corresponds to a
lowest weight of $\mf{r}$ since it is annihilated by $f_2$ and
$f_3$, 
\begin{equation}
  \begin{array}{rcl}
    \ad_{f_2}(e_1) & = & [f_2,e_1] = 0,
    \\
    \ad_{f_3}(e_1) & = & [f_3,e_1]  = 0,
  \end{array}
  \label{leveloneadjointaction} 
\end{equation}
which follows from the defining relations of $A_1^{++}$.

Explicitly, the root associated to $e_1$ is simply the root
$\al_{1}$ that defines the level expansion. Therefore the lowest
weight of this level one representation is 
\begin{equation}
  \Lambda^{(1)}_\mathrm{lw}=\bar{\al}_{1}, 
  \label{lowestweightlevelone}
\end{equation}
Although $\al_{1}$ is a real \emph{positive} root of
$\mf{h}_{\mf{g}}^{\star}$, its projection $\bar{\al}_{(1)}$ is a
\emph{negative} weight of $\mf{h}_{\mf{r}}^{\star}$. Note that
since the lowest weight $\Lambda^{(1)}_1$ is real, the
representation $\mc{R}^{(1)}$ has outer multiplicity \index{outer multiplicity} one,
$\mu(\mc{R}^{(1)})=1$.

Acting on the lowest weight state with the raising operators of
$\mf{r}$ yields the six-dimensional representation
$\mc{R}^{(1)}=\mathbf{6}_1$ of $\mf{sl}(3,\mbb{R})$. The root
$\al_{1}$ is displayed as the green vector in
Figure~\ref{figure:AE3Dec}, taking us from the origin at level zero to
the lowest weight of $\mc{R}^{(1)}$. The Dynkin labels of this
representation are 
\begin{equation}
  \begin{array}{rcl}
    p_2(\mc{R}^{(1)})&=&-(\al_{2}|\al_{1}) = 2,
    \\
    p_3(\mc{R}^{(1)})&=&-(\al_{3}|\al_{1}) = 0,
  \end{array}
  \label{Dynkinlabelslevel1} 
\end{equation}
which follows directly from the
Cartan matrix of $A_1^{++}$. Three of the weights in $\mc{R}^{(1)}$
correspond to roots that are located on the lightcone in root
space and so are null roots of $\mf{h}^{\star}_{\mf{g}}$. These
are $\alpha_{1}+ \alpha_{2}$, $\alpha_{1}+ \alpha_{2} +
\alpha_{3}$ and $\alpha_{1}+ 2\alpha_{2} + \alpha_{3}$ and
are labelled with white crosses in Figure~\ref{figure:AE3Dec}. The
other roots present in the representation, in addition to 
$\alpha_{1}$, are $\alpha_{1}+ 2\alpha_{2}$ and $\alpha_{1}+
2\alpha_{2} +2 \alpha_{3}$, which are real. This representation
therefore contains no weights inside the lightcone.

The $\mf{gl}(3,\mbb{R})$-generator encoding this representation is
realized as a symmetric 2-index tensor $E^{ij}$ which indeed
carries six independent components. In general we can easily
compute the dimensionality of a representation given its Dynkin
labels using the \emph{Weyl dimension formula} which for
$\mf{sl}(3,\mbb{R})$ takes the form~\cite{Fuchs} 
\begin{equation}
  d_{\Lambda_\mathrm{hw}}\left(\mf{sl}(3,\mbb{R})\right)=
  (p_2+1)(p_3+1)\left(\f{1}{2}(p_2+p_3)+1\right).
  \label{WeylDimensionFomulaA2}
\end{equation}
In particuar, for
$(p_2,p_3)=(2,0)$ this gives indeed $d_{\Lambda^{(1)}_\mathrm{hw,1}}\!\!=6$.

It is convenient to encode the Dynkin labels, \index{Dynkin labels}and, consequently,
the index structure of a given representation module, in a Young
tableau. We follow conventions where the first Dynkin label gives
the number of columns with 1 box and the second Dynkin label gives
the number of columns with 2 boxes\footnote{Since we are, in fact, 
using conjugate Dynkin labels, these conventions are equivalent to
the standard ones if one replaces covariant indices by
contravariant ones, and vice-versa.}. For the representation $\mathbf{6}_1$ the first Dynkin
label is 2 and the second is 0, hence the associated Young
tableau is
\begin{equation}
  \mathbf{6}_1
  \quad \Longleftrightarrow \quad
  {\footnotesize
    \setlength{\tabcolsep}{0.55 em}
    \begin{tabular}{cc}
      \cline{1-2}
      \multicolumn{1}{|c|}{} &
      \multicolumn{1}{c|}{} \\
      \cline{1-2}
    \end{tabular}}\,.
  \label{YoungtableauLevel1}
\end{equation}

At level $\ell=-1$ there is a corresponding negative generator
$F_{ij}$. The generators $E^{ij}$ and $F_{ij}$ transform
contravariantly and covariantly, respectively, under the level
zero generators, i.e., 
\begin{equation}
  \begin{array}{rcl}
    [{K^{i}}_j,E^{kl}]&=&\delta^{k}_{j}E^{il}+\delta^{l}_{j}E^{ki},
    \\ [0 em]
    [{K^{i}}_j,F_{kl}]&=& -\delta^{i}_k F_{jl}-\delta^{i}_l F_{kj}.
  \end{array}
\end{equation}
The internal commutator on level one can be obtained by first
identifying 
\begin{equation}
  e_1\equiv E^{11},
  \qquad
  f_1\equiv F_{11}, 
  \label{levelonegenerators}
\end{equation}
and then by demanding $[e_1,f_1]=\al_{1}^{\vee}$ we find 
\begin{equation}
  [E^{ij},F_{kl}]=2\delta^{(i}_{(k}{K^{j)}}_{l)}-\delta^{(i}_{k}\delta^{k)}_{l}
  ({K^{1}}_1+{K^{2}}_2+{K^{3}}_3),
\end{equation}
which is indeed compatible with the realisation of $\al_{1}^{\vee}$
given in Equation~(\ref{realisationgl(3)}). The Killing form at
level~1 takes the form 
\begin{equation}
  \left( F_{ij}|E^{kl}\right)=\delta_i^{(k}\delta_j^{l)}.
  \label{KillingformLevel1}
\end{equation}

\subsubsection{Constraints on Dynkin Labels}
\index{Dynkin labels}

As we go to higher and higher levels it is useful to
employ a systematic method to investigate the representation
content. It turns out that it is possible to derive a set of
equations whose solutions give the Dynkin labels for the
representations at each level~\cite{DHN2}.

We begin by relating the Dynkin labels to the expansion
coefficients $\ell,m_2$ and $m_3$ of a root $\ga\in
\mf{h}_{\mf{g}}^{\star}$, whose projection $\bar{\ga}$ onto
$\mf{h}_{\mf{r}}^{\star}$ is a lowest weight vector for some
representation of $\mf{r}$ at level $\ell$. We let $a=2,3$ denote
indices in the root space of the subalgebra $\mf{sl}(3,\mbb{R})$
and we let $i=1,2,3$ denote indices in the full root space of
$A_1^{++}$. The formula for the Dynkin labels then gives 
\begin{equation}
  p_{a}=-(\al_{a}|\ga)=-\ell A_{a1}-m_2A_{a2}-m_3A_{a3},
  \label{DynkinlabelsArbitraryLevel}
\end{equation}
where $A_{ij}$ is the
Cartan matrix for $A_1^{++}$, given in Equation~(\ref{CartanMatrixAE3}).
Explicitly, we find the following relations between the
coefficients $m_2, m_3$ and the Dynkin labels: 
\begin{equation}
  \begin{array}{rcl}
    p_2 &=& 2\ell-2m_2+m_3,
    \\
    p_3 &=& m_2-2m_3.
  \end{array}
  \label{Dynkinlabelrelation} 
\end{equation}
These formulae restrict the possible Dynkin labels for each $\ell$ since
the coefficients $m_2$ and $m_3$ must necessarily be non-negative
integers. Therefore, by inverting Equation~(\ref{Dynkinlabelrelation}) we
obtain two Diophantine equations that restrict the possible Dynkin
labels, 
\begin{equation}
  \begin{array}{rcl}
    m_2&=& \displaystyle \f{4}{3}\ell-\f{2}{3}p_2-\f{1}{3}p_3 \geq 0,
    \\
    \\
    m_3&=& \displaystyle \f{2}{3}\ell-\f{1}{3}p_2-\f{2}{3}p_3 \geq 0.
  \end{array}
  \label{DiophantineAE3}
\end{equation}
In addition to these constraints we
can also make use of the fact that we are decomposing the adjoint
representation of $A_1^{++}$. Since the weights of the adjoint
representation are the roots of the algebra we know that the
lowest weight vector $\Lambda$ must satisfy 
\begin{equation}
  (\Lambda | \Lambda)\leq 2. 
  \label{RootConstraint}
\end{equation}
Taking $\Lambda=\ell
\al_{1}+m_2 \al_{2}+m_3 \al_{3}$ then gives the following
constraint on the coefficients $\ell, m_2$ and $m_3$:
\begin{equation}
  (\Lambda|\Lambda)=2\ell^2+2m_2^2+2m_3^2-4\ell m_2-2m_2 m_3 \leq 2. 
  \label{RootConstraint2}
\end{equation}
We are interested in
finding an equation for the Dynkin labels, so we insert
Equation~(\ref{DiophantineAE3}) into Equation~(\ref{RootConstraint2}) to obtain
the constraint
\begin{equation}
  p_2^2+p_3^2+p_2 p_3-\ell^2\leq 3.
  \label{ConstraintDynkinLabels}
\end{equation}
The inequalities in Equation~(\ref{DiophantineAE3}) and
Equation~(\ref{ConstraintDynkinLabels}) are sufficient to determine
the representation content at each level $\ell$. However, this
analysis does not take into account the outer multiplicities, which
must be analyzed separately by comparing with the known root
multiplicities of $A_1^{++}$ as given in
Table $H_3$ on page 215 of \cite{Kac}. We shall return to this issue below.

\subsubsection{Level   $\ell=2$}

Let us now use these results to analyze the case for
which $\ell=2$. The following equations must then be satisfied:
\begin{equation}
  \begin{array}{rcl}
     8-2p_2-p_3&\geq & 0,
    \\
    4-p_2-2p_3&\geq & 0,
    \\
    p_2^2+p_3^2+p_2 p_3& \leq & 7.
  \end{array}
  \label{Level2Constraints} 
\end{equation}
The only admissible solution is $p_2=2$ and $p_3=1$. This
corresponds to a 15-dimensional representation $\mathbf{15}_2$ with
the following Young tableau
\begin{equation}
  \mathbf{15}_2
  \quad \Longleftrightarrow \quad
  {\footnotesize
    \setlength{\tabcolsep}{0.55 em}
    \begin{tabular}{ccc}
      \cline{1-3}
      \multicolumn{1}{|c|}{} &
      \multicolumn{1}{c|}{} &
      \multicolumn{1}{c|}{} \\
      \cline{1-3}
      \multicolumn{1}{|c|}{} \\
      \cline{1-1}
    \end{tabular}}\,.
  \label{YoungtableauLevel2}
\end{equation}
Note that $p_2=p_3=0$ is also a solution to
Equation~(\ref{Level2Constraints}) but this violates the constraint
that $m_2$ and $m_3$ be integers and so is not allowed.

\begin{figure}
    \centerline{\includegraphics[width=90mm]{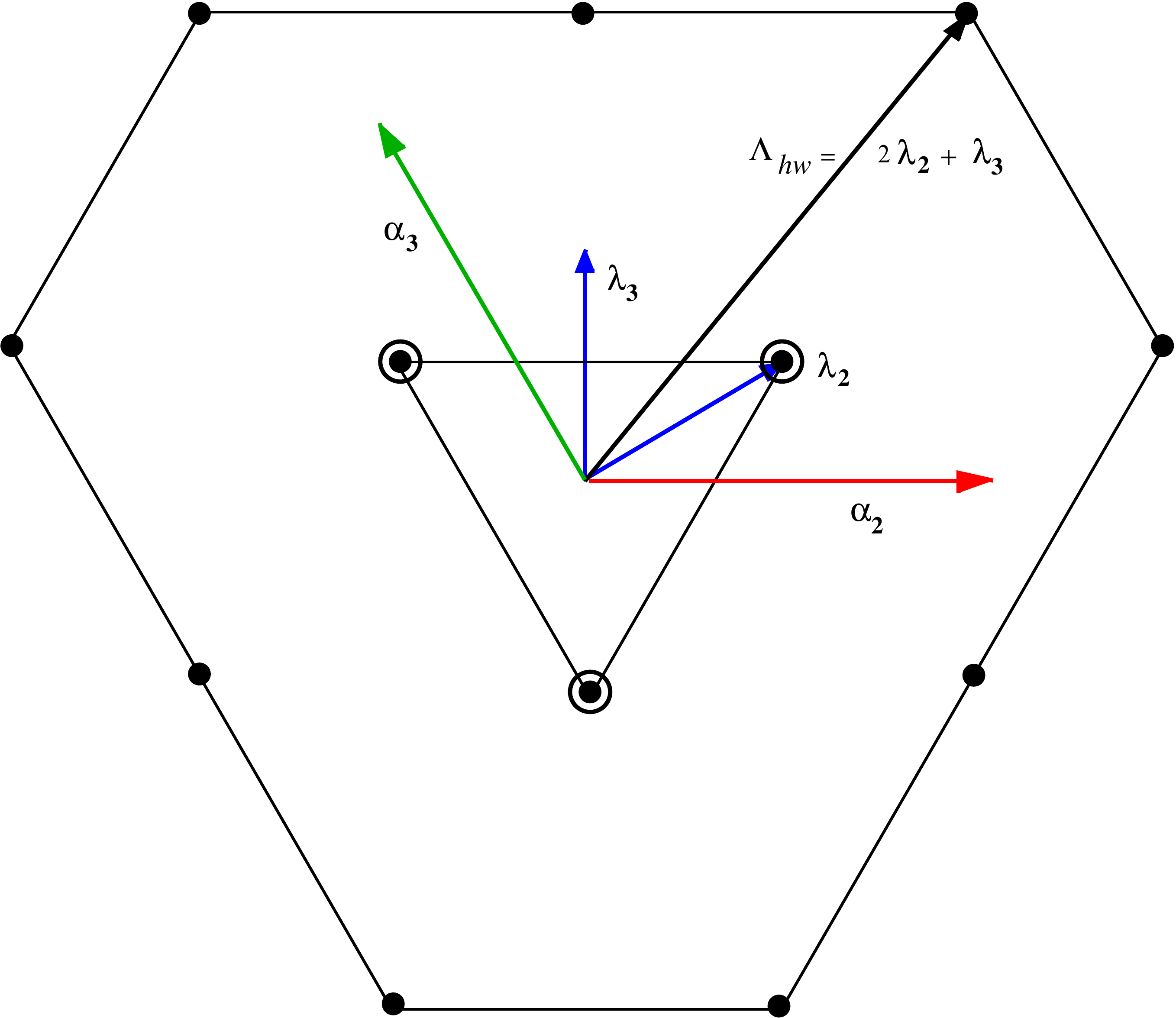}}
    \caption{The representation $\mathbf{15}_2$ of $\mf{sl}(3,\mbb{R})$
      appearing at level two in the decomposition of the adjoint
      representation of $A_1^{++}$ into representations of
      $\mf{sl}(3,\mbb{R})$. The lowest leftmost node is the lowest
      weight of the representation, corresponding to the real root
      $\Lambda^{(2)}=2\al_{1}+\al_{2}$ of $A_1^{++}$. This
      representation has outer multiplicity one. }
    \label{figure:15ofSL3}
  \end{figure}

Moreover, the representation $[p_2,p_3]=[0,2]$ is also a solution
to Equation~(\ref{Level2Constraints}) but has not been taken into account
because it has vanishing outer multiplicity. This can be
understood by examining Figure~\ref{figure:15ofSL3} a little
closer. The representation $[0,2]$ is six-dimensional and has
highest weight $2\lambda_{3}$, corresponding to the middle node
of the top horizontal line in Figure~\ref{figure:15ofSL3}. This
weight lies outside of the lightcone and so is a real root of
$A_1^{++}$. Therefore we know that it has root multiplicity one and
may therefore only occur once in the level decomposition. Since
the weight $2\lambda_{3}$ already appears in the larger
representation $\mathbf{15}_2$ it cannot be a highest weight in
another representation at this level. Hence, the representation
$[0,2]$ is not allowed within $A_1^{++}$. A similar analysis reveals that
also the representation $[p_2, p_3]=[1, 0]$, although allowed by
Equation~(\ref{Level2Constraints}), has vanishing outer multiplicity.

The level two module is realized by the tensor ${E_i}^{jk}$ whose
index structure matches the Young tableau above. Here we have used
the $\mf{sl}(3,\mbb{R})$-invariant antisymmetric tensor
$\epsilon^{abc}$ to lower the two upper antisymmetric indices
leading to a tensor ${E_{i}}^{jk}$ with the properties
\begin{equation}
  {E_i}^{jk} = {E_i}^{(jk)},
  \qquad
  {E_i}^{ik}=0.
\end{equation}
This corresponds to a positive root generator and by the Chevalley
involution we have an associated negative root generator
${F^i}_{jk}$ at level $\ell=-2$. Because the level decomposition
gives a gradation of $A_1^{++}$ we know that all higher level
generators can be obtained through commutators of the level one
generators. More specifically, the level two tensor ${E_i}^{jk}$
corresponds to the commutator
\begin{equation}
  [E^{ij},E^{kl}]=\epsilon^{mk(i}{E_m}^{j)l}+\epsilon^{ml(i}{E_m}^{j)k},
\end{equation}
where $\epsilon^{ijk}$ is the
totally antisymmetric tensor in three dimensions. Inserting the
result $p_2=2$ and $p_3=1$ into Equation~(\ref{DiophantineAE3}) gives
$m_2=1$ and $m_3=0$, thus providing the explicit form of the root
taking us from the origin of the root diagram in
Figure~\ref{figure:AE3Dec} to the lowest weight of $\mathbf{15}_2$ at
level two:
\begin{equation}
  \Lambda^{(2)}=2\al_{1}+\al_{2}.
  \label{LowestweightLevel2}
\end{equation}
This is a real root of $A_1^{++}$,
$(\ga |\ga)=2$, and hence the representation $\mathbf{15}_2$ has
outer multiplicity one. We display the representation $\mathbf{15}_2$
of $\mf{sl}(3,\mbb{R})$ in Figure~\ref{figure:15ofSL3}. The lower
leftmost weight is the lowest weight $\Lambda^{(2)}$. The
expansion of the lowest weight $\Lambda_\mathrm{lw}^{(2)}$ in terms of
the fundamental weights $\lambda_{2}$ and $\lambda_{3}$ is
given by the (conjugate) Dynkin labels
\begin{equation}
  -\Lambda_\mathrm{hw}^{(2)}=
  p_2\lambda_{2} + p_3\lambda_{3}=2\lambda_{2}+\lambda_{3}.
  \label{HighestWeightLevel2}
\end{equation}
The three innermost weights all
have multiplicity 2 as weights of $\mf{sl}(3,\mbb{R})$, as
indicated by the black circles. These lie inside the lightcone of
$\mf{h}^{\star}_{\mf{g}}$ and so are timelike roots of $A_1^{++}$.

\subsubsection[Level $\ell=3$]{Level   $\ell=3$}

We proceed quickly past level three since the analysis
does not involve any new ingredients. Solving
Equation~(\ref{DiophantineAE3}) and
Equation~(\ref{ConstraintDynkinLabels}) for $\ell=3$ yields two
admissible $\mf{sl}(3,\mbb{R})$ representations, $\mathbf{27}_3$ and
$\mathbf{8}_3$, represented by the following Dynkin labels and Young
tableaux:
\begin{equation}
  \begin{array}{rcl}
    \mathbf{27}_3 &:& [p_2,p_3]=[2,2]
    \quad \Longleftrightarrow \quad
    {\footnotesize
      \setlength{\tabcolsep}{0.55 em}
      \begin{tabular}{cccc}
        \cline{1-4}
        \multicolumn{1}{|c|}{} &
        \multicolumn{1}{c|}{} &
        \multicolumn{1}{c|}{} &
        \multicolumn{1}{c|}{} \\
        \cline{1-4}
        \multicolumn{1}{|c|}{} &
        \multicolumn{1}{c|}{} \\
        \cline{1-2}
      \end{tabular}}\,,
    \\ [1.0 em]
    \mathbf{8}_3 &:& [p_2,p_3]=[1,1]
    \quad \Longleftrightarrow \quad
    {\footnotesize
      \setlength{\tabcolsep}{0.55 em}
      \begin{tabular}{cc}
        \cline{1-2}
        \multicolumn{1}{|c|}{} &
        \multicolumn{1}{c|}{} \\
        \cline{1-2}
        \multicolumn{1}{|c|}{} \\
        \cline{1-1}
      \end{tabular}}\,.
  \end{array}
  \label{RepresentationsLevel3} 
\end{equation}
The lowest weight vectors for
these representations are 
\begin{equation}
  \begin{array}{rcl}
    \Lambda^{(3)}_{\mathbf{15}}& =& 3\al_{1}+2\al_{2},
    \\
    \Lambda^{(3)}_{\mathbf{8}} &=& 3\al_{1}+3\al_{2}+\al_{3}.
  \end{array}    
  \label{LowestWeightsLevel3} 
\end{equation}
The lowest weight vector for
$\mathbf{27}_3$ is a real root of $A_1^{++}$, $(\Lambda^{(3)}_{\mathbf{27}}
|\Lambda^{(3)}_{\mathbf{27}} )=2$, while the lowest weight vectors
for $\mathbf{8}_3$ is timelike,
$(\Lambda^{(3)}_{\mathbf{8}}|\Lambda^{(3)}_{\mathbf{8}})=-4$. This
implies that the entire representation $\mathbf{8}_3$ lies inside the
lightcone of $\mf{h}^{\star}_{\mf{g}}$. Both representations have
outer multiplicity one.

Note that $[0,3]$ and $[3,0]$ are also admissible solutions but
have vanishing outer multiplicities by the same arguments as for
the representation $[0,2]$ at level 2.

\subsubsection[Level $\ell=4$]{Level   $\ell=4$}

At this level we encounter for the first time a
representation with non-trivial outer multiplicity. It is a
15-dimensional representation with the following Young tableau
structure:
\begin{equation}
  \mathbf{\bar{15}}_4 : [p_2,p_3]=[1,2]
  \quad \Longleftrightarrow \quad
  {\footnotesize
    \setlength{\tabcolsep}{0.55 em}
    \begin{tabular}{ccc}
      \cline{1-3}
      \multicolumn{1}{|c|}{} &
      \multicolumn{1}{c|}{} &
      \multicolumn{1}{c|}{} \\
      \cline{1-3}
      \multicolumn{1}{|c|}{} &
      \multicolumn{1}{c|}{} \\
      \cline{1-2}
    \end{tabular}}\,.
  \label{RepresentationLevel4}
\end{equation}
The lowest weight vector is
\begin{equation}
  \Lambda^{(4)}_{\mathbf{\bar{15}}}=4\al_{1}+4\al_{2}+\al_{3},
  \label{LowestWeightLevel4}
\end{equation}
which is an imaginary root of $A_1^{++}$,
\begin{equation}
  (\Lambda^{(4)}_{\mathbf{\bar{15}}}|\Lambda^{(4)}_{\mathbf{\bar{15}}})=-6.
  \label{NormLowestWeightLevel4}
\end{equation}
Table $H_3$ of \cite{Kac} we find that this root has
multiplicity 5 as a root of $A_1^{++}$,
\begin{equation}
  \mult(\Lambda^{(4)}_{\mathbf{\bar{15}}})=5.
  \label{MultiplicityLevel4}
\end{equation}
In order for Equation~(\ref{decompositionsubspaces}) to make sense,
this multiplicity must be matched by the total multiplicity of
$\Lambda^{(4)}_{\mathbf{\bar{15}}}$ as a weight of $\mf{sl}(3,\mbb{R})$
representations at level four. The remaining representations at this
level are 
\begin{equation}
  \begin{array}{rcl}
    \mathbf{24}_4 &:& [3,1]
    \quad \Longleftrightarrow \quad
    {\footnotesize
      \setlength{\tabcolsep}{0.55 em}
      \begin{tabular}{cccc}
        \cline{1-4}
        \multicolumn{1}{|c|}{} &
        \multicolumn{1}{c|}{} &
        \multicolumn{1}{c|}{} &
        \multicolumn{1}{c|}{} \\
        \cline{1-4}
        \multicolumn{1}{|c|}{} \\
        \cline{1-1}
      \end{tabular}}\,,
    \\ [1 em]
    \mathbf{\bar{3}}_4 &:& [0,1]
    \quad \Longleftrightarrow \quad
    {\footnotesize
      \setlength{\tabcolsep}{0.55 em}
      \begin{tabular}{c}
        \cline{1-1}
        \multicolumn{1}{|c|}{} \\
        \cline{1-1}
        \multicolumn{1}{|c|}{} \\
        \cline{1-1}
      \end{tabular}}\,,
    \\ [1 em]
    \mathbf{6}_4 &:& [2,0]
    \quad \Longleftrightarrow \quad
    {\footnotesize
      \setlength{\tabcolsep}{0.55 em}
      \begin{tabular}{cc}
        \cline{1-2}
        \multicolumn{1}{|c|}{} &
        \multicolumn{1}{c|}{} \\
        \cline{1-2}
      \end{tabular}}\,,
    \\ [1 em]
    \mathbf{42}_4 &:& [2,3]
    \quad \Longleftrightarrow \quad
    {\footnotesize
      \setlength{\tabcolsep}{0.55 em}
      \begin{tabular}{cccccc}
        \cline{1-6}
        \multicolumn{1}{|c|}{} &
        \multicolumn{1}{c|}{} &
        \multicolumn{1}{c|}{} &
        \multicolumn{1}{c|}{} &
        \multicolumn{1}{c|}{} &
        \multicolumn{1}{c|}{} \\
        \cline{1-6}
        \multicolumn{1}{|c|}{} &
        \multicolumn{1}{c|}{} &
        \multicolumn{1}{c|}{} \\
        \cline{1-3}
      \end{tabular}}\,.
  \end{array}
  \label{MoreRepresentationsLevel4} 
\end{equation}
By drawing these representations explicitly, one sees that the root
$4\al_{1}+4\al_{2}+\al_{3}$, representing the weight
$\Lambda^{(4)}_{\mathbf{\bar{15}}}$, also appears as a weight (but not as
a lowest weight) in the representations $\mathbf{42}_4$ and
$\mathbf{24}_4$. It occurs with weight multiplicity 1 in the $\mathbf{24}_4$
but with weight multiplicity 2 in the $\mathbf{42}_4$. Taking also into
account the representation $\mathbf{\bar{15}}_4$ in which it is the
lowest weight we find a total weight multiplicity of 4. This implies
that, since in $A_1^{++}$
\begin{equation}
  \mult(4\al_{1}+4\al_{2}+\al_{3})=5,
  \label{RootmultiplicityLevel4}
\end{equation}
the outer multiplicity of $\mathbf{\bar{15}}_4$ must be 2, i.e., 
\begin{equation}
  \mu\left(\Lambda^{(4)}_{\mathbf{\bar{15}}}\right)=2.
  \label{OutermultiplicityLevel4}
\end{equation}
When we go to higher and higher levels, the outer multiplicities of
the representations located entirely inside the lightcone in
$\mf{h}_{\mf{g}}$ increase exponentially.

%%%%%%%%%%%%%%%%%%%%%%%%%%%%%%%%%%%%%%%%%%%%%%%%%%%%%%%%%%%%%%%%%%%%%%%%%%%%%%%%%%%
%%%%%%%%%%%%%%%%%%%%%%%%%%%%%%%%%%%%%%%%%%%%%%%%%%%%%%%%%%%%%%%%%%%%%%%%%%%%%%%%%%%

\subsection[Level Decomposition of $E_{10}$]%
           {Level decomposition of   $E_{10}$}
\label{section:DecompE10E10}
In this section we shall describe in detail the level decomposition of the Kac--Moody algebra \index{Kac--Moody algebra} 
$E_{10}$ which is one of the four hyperbolic algebras of maximal rank \cite{Kac}; the
others being $BE_{10}, DE_{10}$ and $CE_{10}$ (see, e.g., {\bf Paper III} for more information). As already mentioned in Section \ref{Section:Extensions}, $E_{10}$ can be constructed
as an overextension \index{overextension} of $E_{8}$ and is therefore
often denoted by $E_{8}^{++}$. Similarly to $E_8$ in the rank 8 case,
$E_{10}$ is the unique indefinite rank 10 algebra with an even
self-dual root lattice, namely the Lorentzian lattice $\Pi_{1,9}$.

Our first encounter with $E_{10}$ in a physical application will be in Chapter \ref{Chapter:Billiards},  where we will show that the Weyl group of $E_{10}$ describes the chaotic dynamics that
emerges when studying eleven-dimensional supergravity close to a
spacelike singularity~\cite{ArithmeticalChaos}.

In Chapter~\ref{Chapter:Manifest}, we will also discuss how to
construct a Lagrangian manifestly invariant under global
$\mc{E}_{10}$-transformations, and compare its dynamics to that of
eleven-dimensional supergravity. The level decomposition 
\index{level decomposition}associated with the removal of the
``exceptional node'' labelled ``10'' in Figure~\ref{figure:E10a} will
be central to the analysis. It turns out that the low-level structure
in this decomposition precisely reproduces the bosonic field content
of eleven-dimensional supergravity~\cite{DHN2}.

Moreover, decomposing $E_{10}$ with respect to different regular
subalgebras reproduces also the bosonic field contents of the Type
IIA and Type IIB supergravities. The fields of the IIA
theory are obtained by decomposition in terms of representations
of the $D_9=\mf{so}(9,9,\mbb{R})$ subalgebra obtained by removing
the first simple root $\al_{1}$~\cite{E10andIIA}. An alternative way of revealing the degrees of freedom of type IIA supergravity is to decompose $E_{10}$ with respect to an $\mf{sl}(9, \mbb{R})$-subalgebra. This viewpoint has a natural interpretation as describing the emergence of type IIA supergavity from a dimensional reduction of eleven-dimensional supergravity. This was analyzed in detail in {\bf Paper VI} and is discussed in Chapter \ref{Chapter:MassiveIIA}.

Similarly the degrees of freedom of type IIB supergavity appear at low levels in a decomposition of $E_{10}$ with respect to the $A_9\oplus
A_{1}=\mf{sl}(9,\mbb{R})\oplus\mf{sl}(2,\mbb{R})$ subalgebra found
upon removal of the second simple root $\al_{2}$~\cite{E10andIIB}. The
extra $A_1$-factor in this decomposition ensures that the global 
$SL(2,\mbb{R})$-symmetry of IIB supergravity is recovered. This $SL(2,\mbb{R})$-symmetry, and its arithmetic subgroup $SL(2,\mbb{Z})\subset SL(2,\mbb{R})$ will also play an important role in {\bf Part II} of this thesis.

%%%%%%%%%%%%%%%%%%%%%%%%%%%%%%%%%%%%%%%%%%%%%%%%%%%%%%%%%%%%%%%%%%%%%%%%%%%%%%%%%%%

\subsubsection{Decomposition with Respect to   $\mf{sl}(10,\mbb{R})$}

\begin{figure}[t]
    \centerline{\includegraphics[width=110mm]{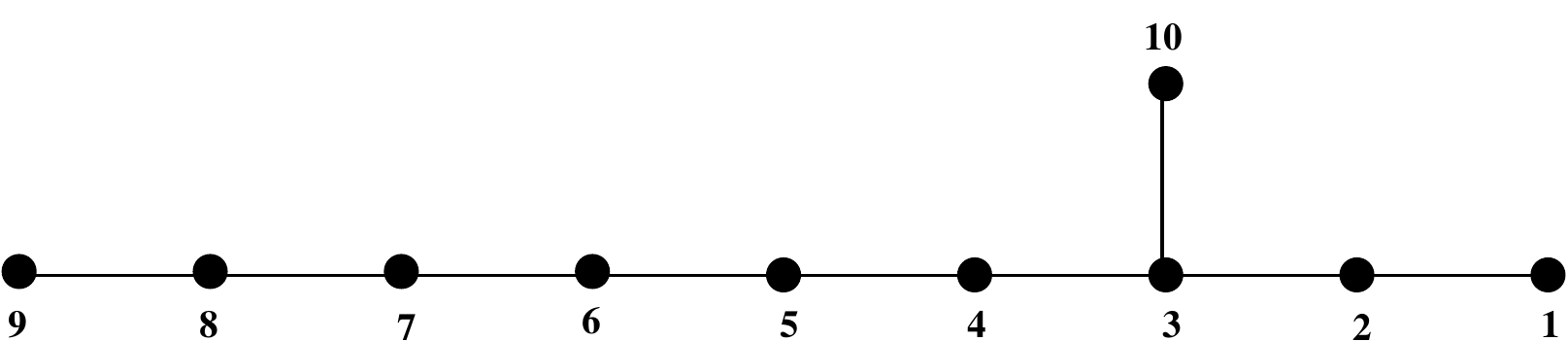}}
    \caption{The Dynkin diagram of $E_{10}$. Labels $i=1,\cdots, 9$
      enumerate the nodes corresponding to simple roots $\al_{i}$ of the
      $\mf{sl}(10,\mathbb{R})$ subalgebra and ``$10$'' labels the
      exceptional node.}
    \label{figure:E10a}
  \end{figure}

Let $\al_{1},\cdots,\al_{10}$ denote the simple roots of
$E_{10}$ \index{$E_{10}$}and $\al_{1}^{\vee}, \cdots, \al_{10}^{\vee}$ the
Cartan generators. These span the root space $\mf{h}^{\star}$ and
the Cartan subalgebra $\mf{h}$, respectively. Since $E_{10}$ is
simply laced the Cartan matrix \index{Cartan matrix}is given by the scalar products
between the simple roots:
\begin{equation}
  A_{ij}[E_{10}]=(\al_{i}|\al_{j})= \left(
    \begin{array}{@{}r@{\quad}r@{\quad}r@{\quad}r@{\quad}r@{\quad}r@{\quad}r@{\quad}r@{\quad}r@{\quad}r@{}}
      2 & -1 & 0 & 0 & 0 & 0 & 0 & 0 & 0 & 0 \\
      -1 & 2 & -1 & 0 & 0 & 0 & 0 & 0 & 0 & 0 \\
      0 & -1 & 2 & -1 & 0 & 0 & 0 & 0 & 0 & -1 \\
      0 & 0 & -1 & 2 & -1 & 0 & 0 & 0 & 0 & 0 \\
      0 & 0 & 0 & -1 & 2 & -1 & 0 & 0 & 0 & 0 \\
      0 & 0 & 0 & 0 & -1 & 2 & -1 & 0 & 0 & 0 \\
      0 & 0 & 0 & 0 & 0 & -1 & 2 & -1 & 0 & 0 \\
      0 & 0 & 0 & 0 & 0 & 0 & -1 & 2 & -1 & 0 \\
      0 & 0 & 0 & 0 & 0 & 0 & 0 & -1 & 2 & 0 \\
      0 & 0 & -1 & 0 & 0 & 0 & 0 & 0 & 0 & 2 \\
    \end{array}
  \right).
\end{equation}
The associated Dynkin diagram \index{Dynkin diagram}is displayed in
Figure~\ref{figure:E10a}. We will perform the decomposition with
respect to the $\mf{sl}(10,\mbb{R})$ subalgebra represented by the
horizontal line in the Dynkin diagram so the level $\ell$ of an
arbitrary root $\al\in\mf{h}^{\star}$ is given by the coefficient in
front of the exceptional simple root, i.e.,
\begin{equation}
  \ga=\sum_{i=1}^{9}m^{i}\al_{i}+\ell\al_{10}.
  \label{rootE10}
\end{equation}

As before, the weight that is easiest to identify for each
representation $\mc{R}(\Lambda^{(\ell)})$ at positive level $\ell$
is the lowest weight $\Lambda^{(\ell)}_\mathrm{lw}$. We denote by
$\bar{\Lambda}_\mathrm{lw}^{(\ell)}$ the projection onto the spacelike
slice of the root lattice defined by the level $\ell$. The
(conjugate) Dynkin labels $p_1,\cdots, p_9$ characterizing the
representation $\mc{R}(\Lambda^{(\ell)})$ are defined as before as
minus the coefficients in the expansion of
$\bar{\Lambda}_\mathrm{lw}^{(\ell)}$ in terms of the fundamental weights
$\lambda^{i}$ of $\mf{sl}(10,\mbb{R})$:
\begin{equation}
  -\bar{\Lambda}_\mathrm{lw}^{(\ell)}=\sum_{i=1}^{9}p_{i}\lambda^{i}.
  \label{Dynkinlabels}
\end{equation}

The Killing form on each such slice is positive definite so the
projected weight $\bar{\Lambda}_\mathrm{hw}^{(\ell)}$ is of course real.
The fundamental weights of $\mf{sl}(10,\mbb{R})$ can be computed
explicitly from their definition as the duals of the simple roots:
\begin{equation}
  \lambda^{i}=\sum_{j=1}^{9}B^{ij}\al_{j},
  \label{fundamentalweightsA9}
\end{equation}
where $B^{ij}$ is the inverse of the Cartan matrix of $A_9$,
\begin{equation} \left(B_{ij}[A_{9}]\right)^{-1}= \f{1}{10}\left(
    \begin{array}{@{}r@{\quad}r@{\quad}r@{\quad}r@{\quad}r@{\quad}r@{\quad}r@{\quad}r@{\quad}r@{}}
      9 & 8 & 7 & 6 & 5 & 4 & 3 & 2 & 1 \\
      8 & 16 & 14 & 12 & 10 & 8 & 6 & 4 & 2 \\
      7 & 14 & 21 & 18 & 15 & 12 & 9 & 6 & 3 \\
      6 & 12 & 18 & 24 & 20 & 16 & 12 & 8 & 4 \\
      5 & 10 & 15 & 20 & 25 & 20 & 15 & 10 & 5 \\
      4 & 8 & 12 & 16 & 20 & 24 & 18 & 12 & 6 \\
      3 & 6 & 9 & 12 & 15 & 18 & 21 & 14 & 7 \\
      2 & 4 & 6 & 8 & 10 & 12 & 14 & 16 & 8 \\
      1 & 2 & 3 & 4 & 5 & 6 & 7 & 8 & 9 \\
    \end{array}
  \right).
  \label{CartanMatrixA9}
\end{equation}
Note that all the entries of $B^{ij}$ are positive which will prove to
be important later on. As we saw for the $A_1^{++}$ case we want to find
the possible allowed values for $(m_1,\cdots, m_9)$, or, equivalently,
the possible Dynkin labels $[p_1,\cdots,p_9]$ for each level $\ell$.

The corresponding diophantine equation, Equation~(\ref{DiophantineAE3}),
for $E_{10}$ was found in~\cite{DHN2} and reads
\begin{equation}
  m^{i}=B^{i3}\ell-\sum_{j=1}^{9}B^{ij}p_j \geq 0.
  \label{DiophantineE10}
\end{equation}
Since the two sets $\{p_i\}$ and $\{m^{i}\}$ both consist of
non-negative integers and all entries of $B^{ij}$ are positive, these
equations put strong constraints on the possible representations that
can occur at each level. Moreover, each lowest weight vector
$\Lambda^{(\ell)}=\ga$ must be a root of $E_{10}$, so we have the
additional requirement
\begin{equation}
  (\Lambda^{(\ell)}|\Lambda^{(\ell)})=
  \sum_{i,j=1}^{9}B^{ij}p_i p_j-\f{1}{10}\ell^2 \leq 2. 
  \label{rootconstraintE10}
\end{equation}

The representation content at each level is represented by
$\mf{sl}(10,\mbb{R})$-tensors whose index structure are encoded in
the Dynkin labels \index{Dynkin labels}$[p_1,\cdots,p_9]$. At level $\ell=0$ we have the
adjoint representation of $\mf{sl}(10,\mbb{R})$ represented by the
generators ${K^{a}}_b$ obeying the same commutation relations as
in Equation~(\ref{gl(3)}) but now with $\mf{sl}(10,\mbb{R})$-indices.

All higher (lower) level representations will then be tensors
transforming contravariantly (covariantly) under the level
$\ell=0$ generators. The resulting representations are displayed
up to level~3 in Table~\ref{table:A9reps}. We see that the level~1
and 2 representations have the index structures of a 3-form and a
6-form respectively. In the $E_{10}$-invariant sigma model, to be
constructed in Chapter \ref{Chapter:Manifest}, these
generators will become associated with the time-dependent physical
``fields'' $A_{abc}(t)$ and $A_{a_1\cdots a_6}(t)$ which are
related to the electric and magnetic component of the 3-form in
eleven-dimensional supergravity. Similarly, the level 3 generator
$E^{a|b_1\cdots b_9}$ with mixed Young symmetry will be associated
to the dual of the spatial part of the eleven-dimensional
vielbein. This field is therefore sometimes referred to as the
``dual graviton''.

\begin{table}
  \caption{The low-level representations in a decomposition of the
    adjoint representation of $E_{10}$ into representations of its
    $A_9$ subalgebra obtained by removing the exceptional node in the
    Dynkin diagram in Figure~\ref{figure:E10a}.}
  \renewcommand{\arraystretch}{1.2}
  \vspace{0.5 em}
  \centering
  \begin{tabular}{|c|c|c|c|c|}
  %  \toprule
  \hline
    $\ell$ & $\Lambda^{(\ell)}=[p_1,\cdots, p_9]$ & 
    $\Lambda^{(\ell)}=(m_1,\cdots,m_{10})$ &
    $A_9$-representation &
    $E_{10}$-generator \\
    \hline 
    \hline
   % \midrule
    1 &
    $[0,0,1,0,0,0,0,0,0]$ &
    $(0,0,0,0,0,0,0,0,0,1)$ &
    $\mathbf{120}_1$ &
    $E^{abc}$ \\
    \hline
    2 &
    $[0,0,0,0,0,1,0,0,0]$ &
    $(1,2,3,2,1,0,0,0,0,2)$ &
    $\mathbf{210}_2$ &
    $E^{a_1\cdots a_6}$ \\
    \hline
    3 &
    $[1,0,0,0,0,0,0,1,0]$ &
    $(1,3,5,4,3,2,1,0,0,3)$ &
    $\mathbf{440}_3$ &
    $E^{a|b_1\cdots b_8}$ \\
    \hline
   % \bottomrule
  \end{tabular}
  \label{table:A9reps}
  \renewcommand{\arraystretch}{1.0}
\end{table}

\subsubsection*{Algebraic Structure at Low Levels}

Let us now describe in a little more detail the
commutation relations between the low-level generators in the level
decomposition of $E_{10}$ (see Table \ref{table:A9reps}). We recover
the Chevalley generators of $A_9$ through the following realisation:
\begin{equation}
  e_{i}={K^{i+1}}_{i},
  \qquad
  f_{i}={K^{i}}_{i+1},
  \qquad
  h_{i}={K^{i+1}}_{i+1}-{K^{i}}_{i}
  \qquad
  (i=1,\cdots , 9),
  \label{A9generators}
\end{equation}
where, as before, the ${K^{i}}_j$'s obey the commutation relations
\begin{equation}
  [{K^{i}}_j,{K^{k}}_l]=\delta^{k}_j {K^{i}}_l-\delta^{i}_l{K^{k}}_j. 
  \label{gl(10)}
\end{equation}
At levels $\pm 1$ we have the positive root generators $E^{abc}$ and
their negative counterparts $F_{abc}=-\tau(E^{abc})$, where $\tau$
denotes the Chevalley involution as defined in
Section~\ref{Section:rootsystem}. Their transformation properties under
the $\mf{sl}(10,\mbb{R})$-generators ${K^{a}}_b$ follow from the
index structure and reads explicitly 
\begin{equation}
  \begin{array}{rcl}
    [{K^{a}}_{b},E^{cde}] & = & \displaystyle
    3\delta^{[c}_{b}E^{de]a},
    \\ [0.5 em]
    [{K^{a}}_{b},F_{cde}] & = & \displaystyle
    -3{\delta^{a}}_{[c}F_{de]b},
    \\ [0.25 em]
    [E^{abc},F_{def}] & = & \displaystyle
    18\delta^{[ab}_{[de}{K^{c]}}_{f]}-
    2\delta^{abc}_{def}\sum_{a=1}^{10}{K^{a}}_{a},
  \end{array}
  \label{level1commutationrelations}
\end{equation}
where we defined
\begin{equation}
  \begin{array}{rcl}
    \delta^{ab}_{cd}&=& \displaystyle
    \f{1}{2}(\de^{a}_{c}\de^{b}_{d}-\de^{b}_{c}\de^{a}_{d})
    \\ [0.75 em]
    \delta^{abc}_{def}&=& \displaystyle
    \f{1}{3!}(\delta^{a}_{d}\de^{b}_{e}\de^{c}_{f}\pm 5
    \mathrm{\ permutations}).
  \end{array}
  \label{deltadefinitions} 
\end{equation}
The ``exceptional'' generators $e_{10}$ and $f_{10}$ are fixed by
Equation~(\ref{A9generators}) to have the following realisation:
\begin{equation}
  e_{10}=E^{123},
  \qquad
  f_{10}=F_{123}.
  \label{realisationexceptionalgenerator}
\end{equation}
The corresponding Cartan generator is obtained by requiring
$[e_{10},f_{10}]=h_{10}$ and upon inspection of the last equation in
(\ref{level1commutationrelations}) we find
\begin{equation}
  h_{10}=-\f{1}{3} \!\! \sum_{i\neq 1,2,3} \!\!
  {K^{a}}_{a}+\f{2}{3}({K^{1}}_{1}+{K^{2}}_{2}+{K^{3}}_{3}),
  \label{ExceptionalCartanGenerator}
\end{equation}
enlarging $\mf{sl}(10,\mbb{R})$ to $\mf{gl}(10,\mbb{R})$.

The bilinear form at level zero is
\begin{equation}
  ({K^{i}}_j|{K^{k}}_l)=\delta^{i}_l\delta^{k}_j-\delta^{i}_j\delta^{k}_l
  \label{KillingformlevelzeroForE10}
\end{equation}
and can be extended level by level to the full algebra by using its
invariance, $([x,y]|z)=(x|[y,z])\,$ for $ x,y,z\in E_{10}$ (see
Section~\ref{Section:BilinearForm}). For level~1 this yields
\begin{equation}
  \left(E^{abc}|F_{def}\right) = 3! \de^{abc}_{def}, 
  \label{BilinearformLevel1}
\end{equation}
where the normalization was chosen such that
\begin{equation}
  \left(e_{10}|f_{10}\right)=\left(E^{123}|F_{123}\right)=1.
  \label{BilinearformExceptionalGenerators}
\end{equation}

Now, by using the graded structure of the level decomposition we
can infer that the level~2 generators can be obtained by commuting
the level~1 generators
\begin{equation}
  [\mf{g}_1,\mf{g}_1]\subseteq \mf{g}_2.
  \label{GradingLevel2}
\end{equation}
Concretely, this means that the level~2 content should be found from
the commutator
\begin{equation}
  [E^{a_1a_2a_3},E^{a_4a_5a_6}].
\end{equation}
We already know that the only representation at this level is $\mathbf{210}_2$, realized by an
antisymmetric 6-form. Since the normalization of this generator is
arbitrary we can choose it to have weight one and hence we find
\begin{equation}
  E^{a_1\cdots a_6}=[E^{a_1a_2a_3},E^{a_4a_5a_6}]. 
\end{equation}
The bilinear form is lifted to level 2 in a similar way as before
with the result
\begin{equation}
  \left(E^{a_1\cdots a_6}|F_{b_1\cdots b_6}\right)=
  6! \delta^{a_1\cdots a_6}_{b_1\cdots b_6}.
  \label{BilinearFormLevel2}
\end{equation}
Continuing these arguments, the level~3-generators can be obtained
from
\begin{equation}
  [[\mf{g}_1,\mf{g}_1],\mf{g}_1]\subseteq \mf{g}_3.
  \label{GradingLevel3}
\end{equation}
From the index structure one would expect to find a 9-form generator
$E^{a_1\cdots a_9}$ corresponding to the Dynkin labels
$[0,0,0,0,0,0,0,0,1]$. However, we see from Table~\ref{table:A9reps}
that only the representation $[1,0,0,0,0,0,0,1,0]$ appears at
level~3. The reason for the disappearance of the representation
$[0,0,0,0,0,0,0,0,1]$ is because the generator $E^{a_1\cdots a_9}$ is
not allowed by the Jacobi identity. A detailed explanation for this
can be found in~\cite{Fischbacher:2005fy}. The right hand side of
Equation~(\ref{GradingLevel3}) therefore only contains the index
structure compatible with the generators $E^{a|b_1\cdots b_8}$,
\begin{equation}
  [[E^{ab_1b_2},E^{b_3b_4b_5}],E^{b_6b_7b_8}]=-E^{[a|b_1b_2]b_3\cdots b_8}, 
  \label{Level3Generator}
\end{equation}
where the minus sign is purely conventional.

For later reference, we list here some additional commutators that
are useful~\cite{DamourNicolaiLowLevels}: 
\begin{equation}
  \begin{array}{rcl}
    [E^{a_1\cdots a_6}, F_{b_1b_2b_3}] &=& \displaystyle
    -5!\delta^{[a_1a_2a_3}_{b_1b_2b_3}E^{a_1a_2a_3]},
    \\ [0.5 em]
    [E^{a_1\cdots a_6}, F_{b_1\cdots b_6}]&=& \displaystyle
    6\cdot 6!\delta^{[a_1\cdots a_5}_{[b_1\cdots b_5}
    {K^{a_6]}}_{b_6]}-\f{2}{3}\cdot 6!\delta^{a_1\cdots
    a_6}_{b_1\cdots b_6} \sum_{a=1}^{10}{K^{a}}_a,
    \\ [1 em]
    [E^{a_1|a_2\cdots a_9}, F_{b_1b_2b_3}]&=& \displaystyle
    -7\cdot 48 \left(\delta^{a_1[a_2a_3}_{b_1b_2b_3}E^{a_4\cdots
    a_9]}-\delta^{[a_2a_3a_4}_{b_1b_2b_3}E^{a_5\cdots a_9]a_1}\right),
    \\ [1 em]
    [E^{a_1|a_2\cdots a_9}, F_{b_1\cdots b_6}] &=& \displaystyle
    -8! \left(\delta^{a_1[a_2\cdots a_6}_{b_1\cdots b_6}E^{a_7a_8a_9]}-
    \delta^{[a_2\cdots a_7}_{b_1\cdots b_6}E^{a_8a_9]a_1}\right).
  \end{array}
\end{equation}

\chapter{Chaotic Cosmology and Hyperbolic Coxeter Groups}
\label{Chapter:Billiards}
In this chapter we will see explicitly how the mathematical structures presented in Chapter \ref{Chapter:KacMoody} appears in a physical context. To this end we begin by analyzing gravity close to a spacelike singularity and explain in detail the reformulation of the dynamics in terms of a hyperbolic billiard. The key steps in this analysis are to work within a Hamiltonian framework, and to perform an Iwasawa decomposition of the spatial metric. This separates diagonal and off-diagonal degrees of freedom and leads in a natural way to the billiard description. Next, we make the connection with Kac-Moody algebras by showing that the billiard dynamics takes place within a bounded region of the Cartan subalgebra of a Lorentzian Kac-Moody algebra. This reveals that the approach to the singularity is controlled by the Weyl group of this Kac-Moody algebra, and provides a natural explanation for the (non-)existence of chaos. This chapter is based on {\bf Paper III}, written in collaboration with Marc Henneaux and Philippe Spindel. 
\section{Cosmological Billiards}
\label{Section:CosmologicalBilliards}
Here we present a lightning review of the billiard interpretation of the dynamics in the BKL limit. We emphasize features which are relevant for understanding the connection between the billiards and the underlying Kac-Moody algebraic structure, a concept which is discussed in more detail in Section \ref{Section:KacMoodyBilliards}. For more details, we refer the reader to {\bf Paper III} or the review \cite{DHNReview}.

Our starting point is the conjecture of Belinskii, Khalatnikov and Lifshitz (BKL) that the dynamics of gravity close to a spacelike (or ``big bang type'') singularity becomes ultralocal due to a complete decoupling of spatial points \cite{BKL,BKL1,BKL2}. We will here make use of Misner's reformulation of the dynamics in terms of a a billiard-type particle motion in an auxiliary hyperbolic space, originally called ``mixmaster behaviour'' \cite{Misner,Misnerb}.

This conjecture was originally framed in the context of pure four-dimensional gravity, but has subsequently been generalised to any dimension and to gravity coupled to any number of $p$-forms. For our purposes it is useful to divide the conjecture into two parts: 
\begin{itemize}

\item The dynamics in the vicinity of the singularity is inherently ``local''. This means that the partial differential equations describing gravity reduce to a collection of ordinary differential equations, at each spatial point, with respect to proper time. This feature is sometimes referred to as a ``decoupling of spatial points''.

\item Each collection of ordinary differential equations can be equivalently described by a billiard motion in a region of hyperbolic space. If this region is of finite volume then the billiard dynamics is chaotic, and if the region is of infinite volume then the dynamics is non-chaotic. 

\end{itemize} 
For what follows we will assume these conjectures to be true.
\subsection{General Considerations}

Our main interest in this thesis will be to analyze the BKL-limit in the context of gravitational theories which arise as low-energy limits of string theory and M-theory. However, the methods applied in this section are completely general and can be applied to any gravitational theory coupled to bosonic matter fields. The only known bosonic matter
fields that consistently couple to gravity are $p$-form fields and ``dilatonic'' scalars, so we begin by considering a general action in $D=d+1$ dimensions of the form 
\begin{equation}
  S\left[ g_{\mu \nu}, \phi, A^{(p)}\right] =
  \int d^D x \, \sqrt{-G}
  \left[R - \partial_\mu \phi \, \partial^\mu \phi -
  \frac{1}{2} \sum_p \frac{e^{\lambda^{(p)} \phi}}{(p+1)!}
  F^{(p)}_{\mu_1 \cdots \mu_{p+1}}
  F^{(p) \, \mu_1 \cdots \mu_{p+1}} \right] , 
  \label{keyaction}
\end{equation}
where we have chosen units such that $16 \pi G = 1$, and we have chosen the Einstein metric
$G_{\mu \nu}$ to have Lorentzian signature $(-, +, \cdots, +)$. We further 
assume that among the scalars, there is only one
dilaton\footnote{This is done mostly for notational convenience.
  If there were other dilatons among the 0-forms, these should be
  separated off from the $p$-forms because they play a distinct
  role. They would appear as additional scale factors and would
  increase the dimensions of the relevant hyperbolic billiard (they
  define additional spacelike directions in the space of scale
  factors).}, denoted $\phi$, whose kinetic term is normalized with weight 1 
with respect to the Ricci scalar. The real parameter
$\lambda^{(p)}$ measures the strength of the coupling to the
dilaton. The other scalar fields, collectively referred to as \emph{axions}, are
denoted $A^{(0)}$ and have dilaton coupling $\lambda^{(0)}
\not=0$. The integer $p \geq 0$ labels the various $p$-forms
$A^{(p)}$ present in the theory, with field strengths $F^{(p)}=dA^{(p)}$.
We assume that the degree $p$ of $A^{(p)}$ is strictly smaller than $D-1$, since a
$(D-1)$-form in $D$ dimensions carries no local degree of freedom.
Furthermore, if $p = D-2$ the $p$-form is dual to a scalar and we
impose also $\lambda^{(D-2)} \not=0$.

The field strength, $F^{(p)}=dA^{(p)}$, of each $p$-form could be modified by
additional coupling terms of Yang--Mills or Chapline--Manton
type~\cite{pvnetal, Chapline:1982ww}\footnote{For example, $F_C = dC^{(2)} - C^{(0)}
dB^{(2)}$ for two 2-forms $C^{(2)}$ and $B^{(2)}$ and a 0-form
$C^{(0)}$, as it occurs in ten-dimensional type IIB supergravity.}, but
we neglect such terms in the action since they are of no consequence in the BKL-limit. For the same reason, we also neglect Chern-Simons terms in what follows.

We shall at this stage consider arbitrary dilaton couplings and
menus of $p$-forms. The following discussion remains
valid no matter what these are; all theories described by the
general action (\ref{keyaction}) lead to the billiard picture.
However, it is only for particular $p$-form menus, spacetime
dimensions and dilaton couplings that the billiard region is
regular and associated with a Kac--Moody algebra. This will be
discussed in Section~\ref{Section:KacMoodyBilliards}. Note that the action,
(\ref{keyaction}), contains as particular cases the bosonic
sectors of all known supergravity theories.

%%%%%%%%%%%%%%%%%%%%%%%%%%%%%%%%%%%%%%%%%%%%%%%%%%%%%%%%%%%%%%%%%%%%%%%%%%%%%%%%%%%
%%%%%%%%%%%%%%%%%%%%%%%%%%%%%%%%%%%%%%%%%%%%%%%%%%%%%%%%%%%%%%%%%%%%%%%%%%%%%%%%%%%

\subsection{Hamiltonian Treatment}
For the purposes of analyzing the action (\ref{keyaction}) in the BKL-limit, it is convenient to first reformulate it in Hamiltonian form. To this end, we assume that there is a spacelike singularity at a finite
distance in proper time. We adopt a spacetime slicing adapted to
the singularity, which ``occurs'' on a slice of constant time. The slicing is then built up starting from the singularity by taking pseudo-Gaussian
coordinates defined by $N = \sqrt{\g}$ and $N^i = 0$, where $N$ is
the lapse and $N^i$ is the shift~\cite{DHNReview}. Here, $\g
\equiv \det (\g_{ij})$, with $\g_{ij}$ being the spatial metric. Thus, in some spacetime patch, our metric ansatz reads\footnote{Note that we have for convenience chosen
  to work with a coordinate coframe $dx^{i}$, with the imposed
  constraint $N= \sqrt{\g}$. In general, one may of course use an
  arbitrary spatial coframe, say $\theta^{i}(x)$, for which the
  associated gauge choice reads $N=w(x)\sqrt{\g}$, with $w(x)$ being a
  density of weight $-1$. This general kind of
  spatial coframe was also used extensively in the recent
  work~\cite{Damour:2007nb}.}
\begin{equation}
  ds^2 = - \g (dx^0)^2 + \g_{ij}(x^0, x^i) \, dx^i \, dx^j ,
\end{equation}
where the local volume $\g$ collapses at each spatial
point as $x^0 \rightarrow +\infty$, in such a way that the proper
time $dT = - \sqrt{\g}\, dx^0$ remains finite (and tends
conventionally to $0^+$). Here we have assumed the singularity to
occur in the past, as in the original BKL analysis, but a similar
discussion holds for future spacelike singularities.

%%%%%%%%%%%%%%%%%%%%%%%%%%%%%%%%%%%%%%%%%%%%%%%%%%%%%%%%%%%%%%%%%%%%%%%%%%%%%%%%%%%

%\subsubsection{Action in canonical form}

In the Hamiltonian description of the dynamics, the canonical
variables are the spatial metric components $\g_{ij}$, the dilaton
$\phi$, the spatial $p$-form components $A^{(p)}_{m_1 \cdots m_p}$
and their respective conjugate momenta $\pi^{ij}$, $\pi_\phi$ and
$\pi_{(p)}^{m_1 \cdots m_p}$. The Hamiltonian action in the
pseudo-Gaussian gauge is given by 
\begin{equation}
  S \left[ \g_{ij}, \pi^{ij}\!, \phi, \pi_\phi,
  A^{(p)}, \pi_{(p)}\right] \!=\!
  \int \! dx^0 \left[\int d^d x \left( \! \pi^{ij} \dot{\g_{ij}} +
  \pi_\phi \dot{\phi} + \sum_p \pi_{(p)}^{m_1 \cdots m_p}
  \dot{A}^{(p)}_{m_1 \cdots m_p} \! \right) - H \right]\!, \quad
  \label{GaussAction}
\end{equation}
where the Hamiltonian is 
\begin{equation}
  \begin{array}{rcl}
    H &=& \displaystyle
    \int d^dx \, \ch =\int d^d x \big(K' + V'\big),
        \\ [0.2 em]
    K' &=& \displaystyle
    \pi^{ij}\pi_{ij} - \frac{1}{d-1} (\pi^i_i)^2 +
    \frac{1}{4} (\pi_\phi)^2 +
    \sum_p \frac{(p!) e^{- \lambda^{(p)} \phi}}{2} \,
    \pi_{(p)}^{m_1 \cdots m_p} \pi_{(p) \, m_1 \cdots m_p},
    \\ [0.5 em]
    V' &=& \displaystyle
    - R \g + \g^{ij} \g \, \partial_i \phi \, \partial_j \phi +
    \sum_p \frac{e^{ \lambda^{(p)} \phi}}{2 \, (p+1)!} \, \g \,
    F^{(p)}_{m_1 \cdots m_{p+1}} F^{(p) \, m_1 \cdots m_{p+1}}. 
  \end{array}
\end{equation}
In addition to
imposing the coordinate conditions $N = \sqrt{\g}$ and $N^i = 0$,
we have also set the temporal components of the $p$-forms equal to
zero (``temporal gauge'').

The dynamical equations of motion are obtained by varying the
above action w.r.t.\ the canonical variables. Moreover, there are
additional constraints on the dynamical variables, which are preserved by the dynamical evolution and need only be imposed at some ``initial'' time, say $x^0=0$. The detailed form of these constraints can be found in \cite{DHNReview,LivingReview}, and here we shall only be concerned with the Hamiltonian constraint, 
\beq
\mc{H}=0,
\eeq
obtained by varying the action with respect to the lapse function $N$, before imposing the gauge condition $N=\sqrt{\g}$.

%%%%%%%%%%%%%%%%%%%%%%%%%%%%%%%%%%%%%%%%%%%%%%%%%%%%%%%%%%%%%%%%%%%%%%%%%%%%%%%%%%%

\subsection{Iwasawa Decomposition of the Metric}

In order to study the dynamical behavior of the fields as $x^0
\rightarrow \infty$ ($\g \rightarrow 0$) and to exhibit the billiard
\index{cosmological billiard}
picture, it is particularly convenient to separate diagonal from off-diagonal components of the spatial metric. Effectively, this corresponds to performing an Iwasawa decomposition of the spatial vielbein. Let $\mathrm{e}(x^{0}, x^{i})$ be the spatial vielbein, such that the spatial metric can be written as $\g(x^{0}, x^{i})=\mathrm{e}(x^{0}, x^{i})^{T} \mathrm{e}(x^{0}, x^{i})$. We then perform the following Iwasawa decomposition of the vielbein\footnote{See Chapter \ref{Chapter:KacMoody} for a definition of the Iwasawa decomposition.}
\beq
\mathrm{e}=\mc{K}\mc{A}\mc{N}, \qquad \mc{K}\in SO(d, \mbb{R}),
\eeq
where $\cn=\cn(x^{0}, x^{i})$ is an upper triangular matrix with $1$'s on the diagonal
($\cn_{ii} =1$, $\cn_{ij} = 0$ for $i>j$) and $\ca=\ca(x^{0}, x^{i})$ is a
diagonal matrix with positive elements, which we parametrize as
\begin{equation}
  \ca = \exp(-\beta),
  \qquad \beta = \diag (\beta^1, \beta^2, \cdots, \beta^d).
\end{equation}
The spatial metric then reads
\begin{equation}
  \g = \mathrm{e}^{T} \mathrm{e}=\cn^{T} \ca^2 \, \cn, 
  \label{IwasawaII}
\end{equation}
or, in components, 
\begin{equation}
  d\sigma^2 = \g_{ij} \, dx^i \, dx^j =
  \sum_{k=1}^d e^{( -2\beta^k)} (\omega^k)^2
\end{equation}
with
\begin{equation}
  \omega^k = \sum_i \cn_{k \, i} \, dx^i.
\end{equation}
The variables $\beta^i$ of the Iwasawa decomposition give the
(logarithmic) scale factors in the new, orthogonal, basis. The
variables $\cn_{ij}$ characterize the change of basis that
diagonalizes the metric and hence they parametrize the off-diagonal
components of the original $\g_{ij}$.

We extend the transformation Equation~(\ref{IwasawaII}) in configuration
space to a canonical transformation in phase space through the
formula 
\begin{equation}
  \pi^{ij} d\g_{ij} = \pi^i d \beta_i + \sum_{i<j} P_{ij} \, d\cn_{ij} .
\end{equation}

Since the scale factors and the off-diagonal variables play very
distinct roles in the asymptotic behavior, we split off the
Hamiltonian as a sum of a kinetic term for the scale factors
(including the dilaton), 
\begin{equation}
  K = \frac{1}{4} \left[ \sum_{i=1}^d \pi_i^2 - \frac{1}{d-1}
  \left( \sum_{i=1}^d \pi_i \right)^2 + \pi_\phi^2 \right],
\end{equation}
plus the rest, denoted by $V$, which will
act as a potential for the scale factors. The Hamiltonian then becomes
\beq 
 \ch = K + V,
 \eeq
 with
\begin{equation}
  \begin{array}{rcl}
    V &=& \displaystyle
    V_S + V_G + \sum_p V_{p} + V_\phi,
    \\ [1.5 em]
    V_S &=& \displaystyle
    \frac{1}{2} \sum_{i<j} e^{-2(\beta^j - \beta^i)}
    \Bigl( \sum_m P_{im}\cn_{jm} \Bigr)^2,
    \\ [1.5 em]
    V_G &=& \displaystyle
    - R \g,
    \\ [0.5 em]
    V_{(p)} &=& \displaystyle
    V_{(p)}^\mathrm{el} + V_{(p)}^\mathrm{magn},
    \\ [1.0 em]
    V_{(p)}^\mathrm{el} &=& \displaystyle
    \frac{p! e^{- \lambda^{(p)} \phi}} {2} \,
    \pi_{(p)}^{m_1 \cdots m_p} \pi_{(p) \, m_1 \cdots m_p},
    \\ [1.0 em]
    V_{(p)}^\mathrm{magn} &=&  \displaystyle
    \frac{e^{\lambda^{(p)} \phi}}{2 \, (p+1)!}
    \, \g \, F^{(p)}_{m_1 \cdots m_{p+1}} F^{(p) \, m_1 \cdots m_{p+1}},
    \\ [1.5 em]
    V_\phi &=& \displaystyle
    \g^{ij} \g \, \partial_i \phi \, \partial_j \phi.
  \end{array}
\end{equation}
The kinetic term $K$ is quadratic in the momenta conjugate to the
scale factors and defines the inverse of a metric in the space of
the scale factors. The space of scale factors will play a crucial role in what follows and we will denote it by $\mc{M}_{\be}$. The metric $\ga_{\mu\nu}$ on $\mc{M}_{\be}$ can be extracted by inverting the metric in the kinetic term $K$ and one finds
\begin{equation}
 ds^2_{\mc{M}_{\be}}=\ga_{\mu\nu}d\be^{\mu} d\be^{\nu}= \sum_i(d\beta^i)^2 - \Bigl( \sum d \beta^i \Big)^2 + (d \phi)^2.
  \label{metricscalefactors}
\end{equation}
Since the metric coefficients do not depend on the scale factors, this metric is always flat and, moreover,  is of Lorentzian signature. A conformal transformation where all scale
factors are scaled by the same number ($\beta^i \rightarrow \beta^i +
\epsilon$) defines a timelike direction. It will be convenient in the
following to collectively denote all the scale factors (the $\beta^i$'s
and the dilaton $\phi$) as $\beta^\mu$, i.e., $(\beta^\mu) = (\beta^i,
\phi)$. We will discuss these important properties of $\mc{M}_{\be}$ later on.

The analysis is further simplified if we take for new $p$-form
variables the components of the $p$-forms in the Iwasawa basis of
the $\omega^k$'s, 
\begin{equation}
  {\cal A}^{(p)}_{i_1 \cdots i_p} = \!\!\!\!
  \sum_{m_1, \cdots,m_{p}} \!\!\!\!
  (\cn^{-1})_{m_1 i_1} \cdots
  (\cn^{-1})_{m_{p} i_{p}} A_{(p) m_1 \cdots m_{p}},
\end{equation}
and again extend this configuration space transformation to a point
canonical transformation in phase space,
\begin{equation}
  \left( \cn_{ij}, P_{ij}, A^{(p)}_{m_1 \cdots m_p},
  \pi_{(p)}^{m_1 \cdots m_p} \right)
  \quad \rightarrow \quad
  \left( \cn_{ij}, P'_{ij}, {\cal A}^{(p)}_{m_1 \cdots m_p},
  \ce_{(p)}^{i_1 \cdots i_p} \right),
\end{equation}
using the formula $\sum p \, dq = \sum p'\, dq'$, which reads 
\begin{equation}
  \sum_{i<j} P_{ij} \dot{\cn}_{ij} + \sum_p \pi_{(p)}^{m_1 \cdots m_p}
  \dot{A}^{(p)}_{m_1 \cdots m_p} =
  \sum_{i<j} P'_{ij} \dot{\cn}_{ij} + \sum_p \ce_{(p)}^{i_1 \cdots i_p}
  \dot{{\cal A}}^{(p)}_{m_1 \cdots m_p}.
\end{equation}
Note that the scale factor variables are unaffected, while the momenta
$P_{ij}$ conjugate to $\cn_{ij}$ get redefined by terms involving
$\ce$, $\cn$ and $\ca$ since the components ${\cal A}^{(p)}_{m_1
 \cdots m_p}$ of the $p$-forms in the Iwasawa basis involve the
$\cn$'s. On the other hand, the new $p$-form momenta, i.e., the
components of the electric field $\pi_{(p)}^{m_1 \cdots m_p}$ in the
basis $\{\omega^k \}$ are simply given by 
\begin{equation}
  \ce^{i_1 \cdots i_p}_{(p)} = \!\!\!\!
  \sum_{m_1,\cdots,m_p} \cn_{i_1 m_1} \ 
  \cn_{i_2 m_2} \cdots \cn_{i_p m_p} \pi_{(p)}^{m_1 \cdots m_p}.
\end{equation}
In terms of the new variables, the electromagnetic potentials
become 
\begin{equation}
  \begin{array}{rcl}
    V_{(p)}^\mathrm{el} &=& \displaystyle
    \frac{p!}{2} \!\!
    \sum_{i_1, i_2,\cdots, i_p} \!\!\!\!\!\!
    e^{-2 e_{i_1 \cdots i_p}(\beta)} (\ce^{i_1 \cdots i_p}_{(p)})^2,
    \\ [1.5 em]
    V_{(p)}^\mathrm{magn} &=& \displaystyle
    \frac{1}{2 \, (p+1)!}
    \sum_{i_1, i_2, \cdots,i_{p+1}} \!\!\!\!\!\!
    e^{-2 m_{i_{1} \cdots i_{p+1}}(\beta)} (\cf_{(p) \, i_1\cdots i_{p+1}})^2.
  \end{array}
\end{equation}
Here, $e_{i_1 \cdots i_p}(\beta)$ are the electric linear forms 
\begin{equation}
  e_{i_1 \cdots i_p}(\beta) =
  \beta^{i_1} + \cdots + \beta^{i_p} + \frac{\lambda^{(p)}}{2} \phi
\end{equation}
(the indices $i_j$ are all distinct because $\ce^{i_1 \cdots i_p}_{(p)}$ is
completely antisymmetric) while $\cf_{(p) \, i_1 \cdots i_{p+1}}$
are the components of the magnetic field $F_{(p) m_1 \cdots
m_{p+1}}$ in the basis $\{\omega^k\}$, 
\begin{equation}
  \cf_{(p) \, i_1 \cdots i_{p+1}} = \!\!\!\!
  \sum_{m_1, \cdots,m_{p+1}} \!\!\!\! (\cn^{-1})_{m_1 i_1} \cdots
  (\cn^{-1})_{m_{p+1} i_{p+1}} F_{(p) m_1 \cdots m_{p+1}},
\end{equation}
and $m_{i_{1} \cdots i_{p+1}}(\beta)$ are the magnetic linear forms
\begin{equation}
  m_{i_{1} \cdots i_{p+1}}(\beta) = \!\!\!\!\!\!
  \sum_{j \notin \{i_1,i_2,\cdots i_{p+1} \}} \!\!\!\!\!\!
  \beta^j -\frac{\lambda^{(p)}}{2} \phi.
\end{equation}
One sometimes rewrites $m_{i_{1} \cdots i_{p+1}}(\beta)$ as
$b_{i_{p+2} \cdots i_d}(\beta)$, where $\{i_{p+2}, i_{p+3},
\cdots, i_d \}$ is the set complementary to $\{i_1,i_2, \cdots
i_{p+1} \}$, e.g., 
\begin{equation}
  b_{1 \, 2 \, \cdots \, d-p-1}(\beta) = \beta^1 + \cdots +
  \beta^{d-p-1} -\frac{\lambda^{(p)}}{2} \phi = m_{d-p \, \cdots \, d}.
\end{equation}
The exterior derivative $\cf$ of
$\ca$ in the non-holonomic frame $\{\omega^k\}$ involves of course
the structure coefficients $C^i{}_{jk}$ in that frame, i.e.,
\begin{equation}
  \cf_{(p) \, i_1 \cdots i_{p+1}} =
  \partial_{[i_1} {\cal A}_{i_2 \cdots i_{p+1}]} +
  \mbox{``}C{\cal A}\mbox{''-terms},
\end{equation}
where 
\begin{equation}
  \partial_{i_1} \equiv
  \sum_{m_1} (\cn^{-1})_{m_1 i_1} (\partial/\partial x^{m_1})
\end{equation}
is here the frame derivative. Similarly, the potential $V_\phi$ reads 
\begin{equation}
  V_\phi = \sum_i e^{-2 \bar{m}_i(\beta)} (\cf_i)^2, 
\end{equation}
where $\cf_i$ is 
\begin{equation}
  \cf_i=(\cn^{-1})_{ji}\partial_j \phi
\end{equation}
and 
\begin{equation}
  \bar{m}_i(\beta) = \sum_{j\not=i} \beta^j. 
\end{equation}

\subsection{Decoupling of Spatial Points}

So far we have only redefined the variables without making any
approximation. We now start the discussion of the BKL-limit, \index{BKL-limit|bb} which
investigates the leading behavior of the degrees of freedom as $x^0 \rightarrow
\infty$ ($\g \rightarrow 0$). As already discussed in the introduction to Section \ref{Section:CosmologicalBilliards}, it is illuminating to separate two aspects
of the BKL conjecture.\footnote{The heuristic
  derivation of~\cite{DHNReview} in the Hamiltonian framework shares many features in common with
  the work of~\cite{Kirillov1993, IKM94, IvKiMe94, KiMe}, extended to
  some higher-dimensional models in~\cite{IvMe, Ivashchuk:1999rm}. The
  central feature of~\cite{DHNReview} is the Iwasawa decomposition \index{Iwasawa decomposition}
  which enables one to clearly see the role of off-diagonal
  variables.}

The first aspect is that the spatial points decouple in the limit $x^0
\rightarrow \infty$, in the sense that one can replace the Hamiltonian
by an effective ``ultralocal'' Hamiltonian $H^\mathrm{UL}$ involving no
spatial gradients, and hence leading at each spatial point to a set of
dynamical equations that are ordinary differential equations with
respect to time. The effective ultralocal Hamiltonian has a form
similar to that of the Hamiltonian governing certain spatially
homogeneous cosmological models, as will be explained below.

The second aspect of the BKL-limit is to take the sharp wall limit of
the ultralocal Hamiltonian. This leads then directly to the billiard \index{cosmological billiard}
description of the dynamics.

%%%%%%%%%%%%%%%%%%%%%%%%%%%%%%%%%%%%%%%%%%%%%%%%%%%%%%%%%%%%%%%%%%%%%%%%%%%%%%%%%%%

\subsubsection{Spatially Homogeneous Models}
\label{section:SpatialHomogeneity}

In spatially homogeneous models, the fields depend only on time in
invariant frames, e.g., for the metric
\begin{equation}
  ds^2 = \g_{ij}(x^0) \psi^i \psi^j,
  \label{coframe}
\end{equation}
where the invariant forms fulfill $$d\psi^i = -\frac{1}{2}
f^i{}_{jk} \psi^j \wedge \psi^k .$$ Here, the $f^i{}_{jk}$ are the
structure constants of the spatial homogeneity group.
Similarly, for a $1$-form and a $2$-form, 
\begin{equation}
  A^{(1)} = A_i(x^0) \psi^i,
  \qquad
  A^{(2)} = \frac{1}{2} A_{ij}(x^0) \psi^i \wedge \psi^j,
  \qquad \mbox{etc.}
\end{equation}
The Hamiltonian constraint yielding the field equations in the spatially
homogeneous context\footnote{This Hamiltonian exists if
  $f^i{}_{ik}= 0$, as we shall assume from now on.} is obtained by
substituting the form of the fields in the general Hamiltonian
constraint and contains, of course, no explicit spatial gradients since the
fields are homogeneous. Note, however, that the structure constants
$f^i{}_{ik}$ contain implicit spatial gradients. The Hamiltonian can
now be decomposed as before and reads
\begin{equation}
  \begin{array}{rcl}
    \ch^\mathrm{UL} &=& \displaystyle
    K + V^\mathrm{UL},
    \\ [0.5 em]
    V^\mathrm{UL} &=& \displaystyle
    V_S + V_G^\mathrm{UL} +
    \sum_p \left( V_{(p)}^\mathrm{el} + V_{(p)}^\mathrm{UL,magn} \right), 
  \end{array}
  \label{calH1}
\end{equation}
where $K$, $V_S$ and
$V_{(p)}^\mathrm{el}$, which do not involve spatial gradients, are
unchanged and where $V_\phi$ disappears since $\pa_i \phi = 0$. The
potential $V_G$ is given by~\cite{Demaret} 
\begin{equation}
  V_G \equiv -\g R =
  \frac{1}{4} \!\! \sum_{i\not= j,i \not= k, j \not= k} \!\!\!\!
  e^{-2G_{ijk}(\beta)} (C^i{}_{jk})^2 +
  \frac{1}{2}\sum_j e^{-2 \bar{m}_j(\beta)}
  \left(C^i{}_{jk} \, C^k{}_{ji} +\cdots \right), 
  \label{formulaforR}
\end{equation}
where the linear forms $G_{ijk}(\beta)$ (with $i, j, k$ distinct)
read 
\begin{equation}
  G_{ijk}(\beta) = 2 \beta^i + \!\!\!\!\!\!
  \sum_{m \, : \, m \not= i, m \not= j, m \not= k} \!\!\!\!\!\! \beta^m, 
\end{equation}
and where the ellipsis represent the terms in the first sum that arise upon
taking $i=j$ or $i=k$. The structure constants in the Iwasawa
frame (with respect to the coframe in Equation~(\ref{coframe})) are
related to the structure constants $f^i{}_{jk}$ through 
\begin{equation}
  C^i{}_{jk} = \sum_{i',j',k'} f^{i'}{}_{j'k'}\cn^{-1}_{ii'} \cn_{jj'} \cn_{kk'}
\end{equation}
 and depend therefore
on the dynamical variables. Similarly, the potential
$V_{(p)}^\mathrm{magn}$ becomes 
\begin{equation}
  V_{(p)}^\mathrm{magn} = \frac{1}{2 \,(p+1)!}
  \!\! \sum_{i_1, i_2, \cdots, i_{p+1}} \!\!\!\!
  e^{-2 m_{i_{1} \cdots i_{p+1}}(\beta)} (\cf^h_{(p) \, i_1 \cdots i_{p+1}})^2,
\end{equation}
 where
the field strengths $\cf^h_{(p) \, i_1 \cdots i_{p+1}}$ reduce to
the ``$A C$'' terms in $dA$ and depend on the potentials and the
off-diagonal Iwasawa variables.

%%%%%%%%%%%%%%%%%%%%%%%%%%%%%%%%%%%%%%%%%%%%%%%%%%%%%%%%%%%%%%%%%%%%%%%%%%%%%%%%%%%

\subsubsection{The Ultralocal Hamiltonian}

Let us now come back to the general, inhomogeneous case and
express the dynamics in the frame $\{dx^0, \psi^i\}$ where the
$\psi^i$'s form a ``generic'' non-holonomic frame in space, 
\begin{equation}
  d\psi^i = -\frac{1}{2} f^i{}_{jk}(x^m) \, \psi^j \wedge \psi^k.
\end{equation}
Here the $f^i{}_{jk}$'s are in general
space-dependent. In the non-holonomic frame, the exact Hamiltonian
takes the form 
\begin{equation}
  \ch = \ch^\mathrm{UL} + \ch^\mathrm{gradient},
\end{equation}
where the ultralocal part $\ch^\mathrm{UL}$ is given by
Equations~(\ref{calH1}) and~(\ref{formulaforR}) with the relevant
$f^i{}_{jk}$'s, and where $\ch^\mathrm{gradient}$ involves the spatial
gradients of $f^i{}_{jk}$, $\beta^m$, $\phi$ and $\cn_{ij}$.

The first part of the BKL conjecture states that one can drop
$\ch^\mathrm{gradient}$ asymptotically; namely, the dynamics of a
generic solution of the Einstein--$p$-form-dilaton equations (not
necessarily spatially homogeneous) is asymptotically determined,
as one goes to the spatial singularity, by the ultralocal
Hamiltonian 
\begin{equation}
  H^\mathrm{UL} = \int d^dx \, \ch^\mathrm{UL},
\end{equation}
provided that the phase space constants
$f^{i}{}_{jk}(x^m) = - f^{i}{}_{kj}(x^m)$ are such that all
exponentials in the above potentials
do appear. In other words, the $f$'s must be chosen such that none
of the coefficients of the exponentials, which involve $f$ and the
fields, identically vanishes -- as would be the case, for example,
if $f^{i}{}_{jk} = 0$ since then the potentials $V_G$ and
$V_{(p)}^\mathrm{magn}$ are equal to zero. This is always possible
because the $f^{i}{}_{jk}$, even though independent of the
dynamical variables, may in fact depend on $x$ and so are not
required to fulfill relations ``$ff = 0$'' analogous to the Bianchi
identity since one has instead ``$\partial f + ff = 0$''.

\subsection{The Sharp Wall Limit}
\label{section:DynamicsBilliardHyp}

The second step in the BKL-limit is to take the
sharp wall limit of the potentials.\footnote{In this thesis we will exclusively
  restrict ourselves to considerations involving the sharp wall
  limit. However, in recent work \cite{Damour:2007nb} it was
  argued that in order to have a rigorous treatment of the dynamics
  close to the singularity also in the chaotic case, it is necessary
  to go beyond the sharp wall limit. This implies that one should
  retain the exponential structure of the dominant walls.} This leads
to the billiard \index{cosmological billiard|bb} picture. It is
crucial here that the coefficients in front of the dominant walls are
all positive. Again, just as for the first step, this limit has not
been rigorously justified. 

To summarize, we first recall that the ultralocal Hamiltonian reads explicitly
\beq
\mc{H}^{\mathrm{UL}}= K+V_{S}+V_G^{\mathrm{UL}}+ \sum_p \left( V_{(p)}^\mathrm{el} + V_{(p)}^\mathrm{UL,magn} \right)
\eeq
where the potentials are completely determined by exponentials containing linear forms $\om(\be)$ in the scale factors $\be$. We can then write the ultralocal Hamiltonian as
\beqa
\mc{H}^{\mathrm{UL}}&=& K+ \sum_{i<j} c_{S}\ e^{-2 s_{ji}(\beta)}+\sum_{i\not= j,i \not= k, j \not= k} \!\!\!\!
 c_{G}\ e^{-2G_{ijk}(\beta)}
 \nn \\
 & & \phantom{K}+\sum_{i_1< i_2< \cdots< i_p} \!\! 
 c_{E}\ e^{-2 e_{i_1 \cdots i_p}(\beta)}+\sum_{i_1< i_2< \cdots< i_{p+1}} \!\!\!\!
 c_{M}\ e^{-2 m_{i_{1} \cdots i_{p+1}}(\beta)},
 \label{UltraLocalHamiltonian}
\eqa
where the coefficients are irrelevant for the dynamics.

The idea is now that as one goes to the singularity, the exponential
potentials get sharper and sharper, and in the strict limit $x^{0}\rightarrow \infty$ they can be replaced by  corresponding $\Theta$-functions, defined by 
\beq
\Theta(x) := \left\{ \begin{array}{cc}
0 & x<0\\ 
\infty & x>0.\\
\end{array} \right.
\eeq
Since $a \Theta(x) = \Theta(x)$ for all $a>0$, the coefficients in (\ref{UltraLocalHamiltonian}) may be neglected and the ``sharp wall'' Hamiltonian becomes 
\begin{eqnarray}
  \ch^\mathrm{sharp} &=&
  K + \sum_{i<j} \Theta \left(-2 s_{ji}(\beta)\right) + \!\!\!\!
  \sum_{i\not= j,i \not= k, j \not= k} \!\!\!\!
  \Theta(-2\alpha_{ijk}(\beta))
  \nonumber
  \\
  && + \!\!\!\! \sum_{i_1< i_2< \cdots< i_p} \!\!\!\!
  \Theta(-2 e_{i_1 \cdots i_p}(\beta)) + \!\!\!\!
  \sum_{i_1< i_2< \cdots< i_{p+1}} \!\!\!\!
  \Theta(-2 m_{i_{1} \cdots i_{p+1}}(\beta)).
  \label{sharp}
\end{eqnarray}
For the readers convenience we list here all of the linear forms appearing in the arguments of the step functions:
\begin{itemize}
\item \emph{Symmetry walls}, originating from off-diagonal components of the spatial metric,
\beq
s_{ji}(\be)= \be^{j}-\be^{i}, \qquad j>i.
\eeq
\item \emph{Gravity walls}, originating from the spatial curvature,
\begin{equation}
  G_{ijk}(\beta) = 2 \beta^i + \!\!\!\!\!\!
  \sum_{m \, : \, m \not= i, m \not= j, m \not= k} \!\!\!\!\!\! \beta^m.
\end{equation}
\item \emph{Electric walls}, originating from the electric components of each $p$-form,
\begin{equation}
  e_{i_1 \cdots i_p}(\beta) =
  \beta^{i_1} + \cdots + \beta^{i_p} + \frac{\lambda^{(p)}}{2} \phi.
\end{equation}
\item \emph{Magnetic walls}, originating from the magnetic components of each $p$-form,
\begin{equation}
  m_{i_{1} \cdots i_{p+1}}(\beta) = \!\!\!\!\!\!
  \sum_{j \notin \{i_1,i_2,\cdots i_{p+1} \}} \!\!\!\!\!\!
  \beta^j -\frac{\lambda^{(p)}}{2} \phi.
  \label{ListWallForms}
\end{equation}
\end{itemize}
We shall refer to these linear forms as \emph{wall forms} and denote them collectively by $\om(\be)=\om_\mu\be^{\mu}$. Each wall form defines a co-dimension one subspace $W$ in the space of scale factors $\mc{M}_{\be}$, which is defined by the locus where the associated wall form vanishes, i.e.
\beq
W[\om] := \big\{ \be\in \mc{M}_{\be} \ \big|\ \om(\be)=0\big\}.
\label{Wall}
\eeq

It is now clear that the complete dynamics of any gravitational theory, described by the action (\ref{keyaction}), reduces in the limit $x^{0}\rightarrow \infty $ to the dynamics of the scale factors $\be(x^{0}, {\bf x})$ at each spatial point ${\bf x}$. Because the
potential walls \index{billiard wall} are infinite (and positive), the motion of the scale factors is constrained to the region of $\mc{M}_{\be}$ where the arguments of all
$\Theta$-functions are negative, i.e., to 
\begin{equation}
  s_{ji}(\beta) \geq 0 \hs (i<j),
  \qquad
  G_{ijk}(\beta) \geq 0,
  \qquad
  e_{i_1 \cdots i_p}(\beta) \geq 0,
  \qquad
  m_{i_{1} \cdots i_{p+1}}(\beta) \geq 0.
\end{equation}
This corresponds to the ``positive side'' of all the walls $W[\om]$. In this region, the dynamics is governed by the kinetic term $K$, implying that the motion is a geodesic for the metric in the space of scale factors $\mc{M}_{\be}$. Since
that metric is flat, this is a straight line. In addition, the
Hamiltonian constraint $\ch=0$ reduces to $K=0$ away from the potential
walls and therefore forces the straight line to be null. In other words, the dynamics of the particle $\be^{\mu}(x^{0})$ is given by \emph{lightlike} motion in $\mc{M}_{\be}$. We shall furthermore assume that the time orientation in the space of the scale factors is such that this lightlike motion is future oriented ($\g \rightarrow 0$ in the future).

\subsection{Dynamics as a Billiard in Hyperbolic Space}
\label{Section:HyperbolicBilliard}

Our analysis so far has led us to the realization that the original gravity-$p$-form action in (\ref{keyaction}) has a very simple description in the limit $x^{0}\rightarrow \infty$ (proper time $T \rightarrow 0^{+}$) in terms of the lightlike linear motion of the scale factors $\be^{\mu}(x^{0}, {\bf x})$ in a region of the space $\mc{M}_{\be}$. This implies that in the BKL-limit we may replace the action (\ref{keyaction}) by an effective geodesic action at each spatial point ${\bf x}$ given by
\beq
S_{\mathrm{BKL}}(\be) := \int dx^{0}\Big[\ga_{\mu\nu} \f{\pa\be^{\mu}}{\pa x^{0}}\f{\pa \be^{\nu}}{\pa x^{0}} - V(\be)\Big],
\label{BKLaction}
\eeq
which is a non-linear sigma model describing maps from the worldline parametrized by $x^{0}$ into the target space $\mc{M}_{\be}$. In defining (\ref{BKLaction}) we have set $N=\sqrt{\g}$ as before. All the remaining non-trivial information about the original action (\ref{keyaction}) is now stored in the potential $V(\be)$, which takes the general form
\beq
V(\be)=\sum_{\om \in \{\mathrm{wall}\ \mathrm{forms}\}} e^{-2\om(\be)},
\eeq
where we sum over all possible wall forms. 

Now let us analyze this action in more detail. Firstly, one may check that all the walls $W[\om]$, defined by Equations (\ref{ListWallForms}) and (\ref{Wall}), are timelike hyperplanes in $\mc{M}_{\be}$. This follows from the fact that the squared norms of all the wall forms $\om$ in (\ref{ListWallForms}) are positive:
\beqa
  (\om\vert \om)& := &\ga^{\mu\nu} \om_{\mu}\om_{\nu}
  \nn \\
&=&   \sum_i \om_i \om_i -
  \frac{1}{d-1} \Big(\sum_i \om_i \Big)\Big(\sum_j \om_j\Big) +
  \om_{\phi}\om_{\phi} \hs >\hs 0,
\eqa
where we have used the inverse metric $\ga^{\mu\nu}$ in $\mc{M}_{\be}$. Explicitly, for the wall forms defined in (\ref{ListWallForms}) one finds 
\begin{equation}
  \begin{array}{rcl}
    (s_{ji} \vert s_{ji} ) &=& 2,
    \\ [0.5 em]
    (G_{ijk} \vert G_{ijk}) &=& 2,
    \\ [0.25 em]
    (e_{i_1 \cdots i_p} \vert e_{i_1 \cdots i_p}) &=& \displaystyle
    \frac{p (d-p-1)}{d-1} + \frac{\left( \lambda^{(p)} \right)^2}{4},
    \\ [0.75 em]
    (m_{i_{1} \cdots i_{p+1}} \vert m_{i_{1} \cdots i_{p+1}}) &=& \displaystyle
    \frac{p (d-p-1)}{d-1} + \frac{\left( \lambda^{(p)} \right)^2}{4}.
  \end{array}
\end{equation}

Because these potential walls are timelike, they have a non-empty intersection with the forward light cone in the space of the scale factors. The motion thus corresponds to lightlike free flight motion in $\mc{M}_{\be}$, occasionally interrupted by collisions with the timelike hyperplanes $W[\om]\subset \mc{M}_{\be}$.  When the null straight line representing the evolution of the scale factors hits one of the walls, it gets reflected
according to the rule~\cite{DH1} 
\begin{equation}
  v^\mu \quad \rightarrow \quad
  v^\mu - 2 \frac{v^\nu \om_{\nu}}{(\om \vert \om)} \om^\mu,
  \label{billiardreflectionrule}
\end{equation}
where $v^{\mu}$ is the velocity vector associated with particle $\be^{\mu}(x^{0})$ (see Section \ref{Section:Kasner}). Since the hyperplanes are timelike, this reflection preserves the time orientation of the motion and therefore belong to the orthochronous Lorentz group $O^\uparrow(d,1)$ which is a subgroup of the isometry group of $\mc{M}_{\be}$.\footnote{If there are $k$ dilatons, this changes to $O^\uparrow(d-1+k,1)$.} 

The motion is thus a succession of future-oriented null straight
line segments interrupted by reflections 
\index{geometric reflection|bb} against the walls, where the motion
undergoes a reflection belonging to $O^\uparrow(k,1)$. We can simplify the description of the dynamics further by noting that not all of the scale-factors are independent. This follows from the Hamiltonian constraint, which enforces 
\beq
\sum_{i} \f{\pa \be^{i}}{\pa x^{0}}\f{\pa \be^{i}}{\pa x^{0}}-\Big(\sum_i \f{\pa \be^{i}}{\pa x^{0}}\Big)\Big(\sum_j \f{\pa \be^{j}}{\pa x^{0}}\Big)+\f{\pa \phi}{\pa x^{0}}\f{\pa \phi}{\pa x^{0}} =0.
\eeq
Since the collisions against the hyperplance $W[\om]$ preserves the time-orientation, it is consistent to use the Hamiltonian constraint to restrict to a physical subset of scale factors. This is conveniently done by projecting the motion onto the unit hyperboloid $\mbb{H}_{d}$, defined by
\beq
\mbb{H}_d := \big\{ \be\in\mc{M}_{\be}\ \big|\ (\be|\be)=-1\big\}.
\eeq
Since we have chosen the time-orientation such that $\sum_i \be^{i} > 0$ the motion is now confined to the positive sheet $\mbb{H}_d^{+}$ of the unit hyperboloid $\mbb{H}_d$. We then recall that the positive sheet of the unit hyperboloid provides a model for \emph{hyperbolic space} (see,
e.g.,~\cite{Ratcliffe}). The straight line motion on $\mc{M}_{\be}$ projects to lightlike geodesic motion on $\mbb{H}_d^{+}$, while the walls $W[\om]$ project onto hyperplanes in hyperbolic space. In this new picture the dynamics corresponds to that of a particle bouncing around in a bounded region of hyperbolic space, very much like a billiard ball bouncing around on a billiard table. This is the origin of the name \emph{cosmological billiards}. 

%Whether the
%collisions eventually stop or continue forever is better visualized by
%projecting the motion radially on the positive sheet of the unit
%hyperboloid, as was done first in the pioneering work of Chitre and
%Misner~\cite{Chitre, Misnerb} for pure gravity in four spacetime
%dimensions. 
%$\sum (\beta^i)^2 - (\sum \beta^i)^2 + \phi^2 = -1$, $\sum \beta^i
%>0$, provides a model of hyperbolic space . \index{hyperbolic space|bb}

The description of the dynamics in terms of a hyperbolic billiard is very useful for determining the nature of the approach to the singularity. In the original work of BKL, it was found that the approach to the singularity may exhibit chaotic behaviour. This can be most easily understood in our present framework. 

As mentioned above, the intersection of a timelike hyperplane $W[\om]$ with the unit hyperboloid $\mbb{H}_d$ defines a hyperplane in hyperbolic space $\mbb{H}_d^{+}$. The region in $\mbb{H}_d^{+}$ on the positive side of all hyperplanes is the allowed dynamical
region and is called the ``billiard table'', denoted $\mc{B}_{\be}\subset \mbb{H}_d^{+}$. It is never compact in
the cases relevant to gravity, but it may or may not have finite
volume. In this picture it is also clear that not all of the walls $W[\om]$ will have dynamical relevance, since some of the walls will be ``behind'' others, and will thus never be seen by the geodesic motion. The billiard table is therefore uniquely defined by a subset $\{W^{\prime}\}\subset\{ W\}$ of \emph{dominant walls}, and can be described as follows
\beq
\mc{B}_{\be}:=\big\{ \be \in \mc{M}_{\be}\ \big|Ê\ \om_{A}(\be)\geq 0, \ A=1, \dots, d+1\big\}, 
\eeq 
where $\om_A$ denotes the $d+1$ dominant walls. The billiard table thus defines a \emph{simplex} in hyperbolic space, i.e. a region in a $d$-dimensional space bounded by $d+1$ walls. If the intersection between two dominant hyperplanes $W$ and $W^{\prime}$ in $\mc{M}_{\be}$ lies outside of the lightcone, then the projected hyperplanes in $\mbb{H}_d^{+}$ will never intersect. As a consequence, the projected region $\mc{B}_{\be}$ will be of infinite volume. In contrast, when all of the dominant walls intersect inside or on the lightcone, then the billiard table $\mc{B}_{\be}$ will be of finite volume. 

When the volume of the billiard table is finite, the collisions with
the potential walls never end (for generic
initial data) and the motion is chaotic. On the other hand, when the
volume is infinite, generic initial data lead to a motion that
ultimately freely runs away to infinity since after a few collisions there will be no wall to interrupt the motion. The dynamics is then non-chaotic. In Section \ref{Section:KacMoodyBilliards} we will describe a powerful mathematical technique which makes it straightforward to determine whether or not the volume of $\mc{B}_{\be}$ is finite, and hence to determine the type of dynamics. 

Let us now for convenience list which of the possible wall forms in (\ref{ListWallForms}) are dominant. These are

\begin{itemize}
\item dominant symmetry walls ($i=1, 2, \cdots , d-1$):
  \begin{equation}
    \beta^{i+1} - \beta^i = 0, 
    \label{symmetryW}
  \end{equation}
 \item dominant gravity walls: 
  \begin{equation}
    2 \beta^1 + \beta^2 + \cdots + \beta^{d-2} = 0,
    \label{curvatureW}
  \end{equation}
\item dominant electric wall: 
  \begin{equation}
    \beta^1 + \cdots + \beta^p + \frac{\lambda^{(p)}}{2} \phi=0,
    \label{electricW}
  \end{equation}
\item  dominant magnetic wall: 
  \begin{equation}
    \beta^1 + \cdots + \beta^{d-p-1} -\frac{\lambda^{(p)}}{2} \phi = 0.
    \label{magneticW}
  \end{equation}
\end{itemize}

%%%%%%%%%%%%%%%%%%%%%%%%%%%%%%%%%%%%%%%%%%%%%%%%%%%%%%%%%%%%%%%%%%%%%%%%%%%%%%%%%%%
%%%%%%%%%%%%%%%%%%%%%%%%%%%%%%%%%%%%%%%%%%%%%%%%%%%%%%%%%%%%%%%%%%%%%%%%%%%%%%%%%%%

\subsection{Recovering the Kasner Solution}
\label{Section:Kasner}

We conclude this section by discussing the relation between the billiard description of the BKL-limit and the more conventional description in terms of a (possibly infinite) sequence of different Kasner solutions. The free motion between two bounces in the billiard is a straight line in the
space of the scale factors. In terms of the original metric
components, this simply corresponds to a Kasner solution with dilaton. To see this, we note that a straight line in $\mc{M}_{\be}$ is described by
\begin{displaymath}
  \beta^\mu = v^\mu x^{0} + \beta^\mu_0,
\end{displaymath}
where the ``velocities'' $v^\mu$ are subject to
\begin{displaymath}
  \sum_i (v^i)^2 - \Big(\sum_i v^i\Big)\Big(\sum_j v^j\Big) + v_\phi^2 = 0,
\end{displaymath}
since the motion is lightlike by the Hamiltonian
constraint. The proper time $dT = -
\sqrt{\g}\, d x^{0}$ is then $T = B \exp(-K x^{0})$, with $K = \sum_i v^i$ for
some constant $B$ (we assume, as before, that the singularity is at $T=0^+$). By the following redefinition
\begin{displaymath}
  p^\mu := \frac{v^\mu}{\sum_i v^i} 
\end{displaymath}
we then obtain the standard Kasner solution 
\begin{eqnarray}
  ds^2 &=& - dT^2 + \sum_i T^{2 p^i} \left(dx^i \right)^2,
  \\
  \phi &=& - p_\phi \ln T + A, 
  \label{KasnerSolution}
\end{eqnarray}
subject to the constraints
\begin{equation}
  \sum_i p^i = 1,
  \qquad
  \sum_i (p^i)^2 + p_\phi^2 = 1,
\end{equation}
where $A$ is a constant of integration and where the coordinates $x^i$
have been suitably rescaled (if necessary). We may therefore conclude that each segment of free flight motion of the billiard ball is in one-to-one correspondence with a certain Kasner solution defined by (\ref{KasnerSolution}). Moreover, each reflection against a wall has the effect of modifying the dynamics to a different Kasner solution. Hence, a finite volume billiard indeed leads to an infinite sequence of Kasner solutions as we approach the singularity, while a finite volume billiard leads to dynamics which eventually settle down in one specific Kasner regime all the way to $T=0^{+}$.

\section{Kac-Moody Billiards}
\label{Section:KacMoodyBilliards}
We shall now reinterpret the results of previous sections within the mathematical framework introduced in Chapter \ref{Chapter:KacMoody}. In particular, we explore the
correspondence between Weyl groups of Lorentzian Kac-Moody algebras and the limiting
behavior of the dynamics of gravitational theories close to a
spacelike singularity.

We have seen in Section~\ref{Section:HyperbolicBilliard} that in the BKL-limit, 
\index{BKL-limit} the dynamics of gravitational theories is equivalent
to a billiard dynamics in a region of hyperbolic space. 
\index{hyperbolic space} In the generic
case, the billiard region has no particular feature. However, as shown in \cite{ArithmeticalChaos}, in the special cases of the supergravities arising as low-energy limits of string theory or M-theory, the billiard exhibits nice regularity properties which are linked to underlying hyperbolic Kac-Moody algebras.

In fact, this feature arises for all
gravitational theories whose toroidal dimensional reduction to three
dimensions exhibits hidden symmetries, in the sense that the reduced
theory can be reformulated as three-dimensional gravity coupled to a
nonlinear sigma-model based on a coset space $K_3\bas G_3$, where
$K_3$ is the maximal compact subgroup 
\index{maximal compact subgroup} of $G_3$. The
``hidden'' symmetry group $G_3$ is also called, by a
generalization of language, ``the U-duality
group''~\cite{Obers:1998fb}. This situation covers the cases of pure
gravity in any spacetime dimension, as well as all known supergravity
models. In all these cases, the billiard region is the fundamental
domain of a Lorentzian Coxeter group \index{Coxeter group} (``Coxeter
billiard''). Furthermore, the Coxeter group in question is
crystallographic and turns out to be the Weyl group of a Lorentzian
Kac--Moody algebra. The billiard table is then the fundamental Weyl
chamber of a Lorentzian Kac--Moody algebra~\cite{ArithmeticalChaos,
  HyperbolicKaluzaKlein} and the billiard is also called a ``Kac--Moody
billiard''. This enables one to reformulate the dynamics as a motion
in the Cartan subalgebra of the Lorentzian Kac--Moody algebra, hinting
at the potential -- and still conjectural at this stage -- existence
of a deeper, infinite-dimensional symmetry of the theory.

%%%%%%%%%%%%%%%%%%%%%%%%%%%%%%%%%%%%%%%%%%%%%%%%%%%%%%%%%%%%%%%%%%%%%%%%%%%%%%%%%%%
%%%%%%%%%%%%%%%%%%%%%%%%%%%%%%%%%%%%%%%%%%%%%%%%%%%%%%%%%%%%%%%%%%%%%%%%%%%%%%%%%%%

%%%%%%%%%%%%%%%%%%%%%%%%%%%%%%%%%%%%%%%%%%%%%%%%%%%%%%%%%%%%%%%%%%%%%%%%%%%%%%%%%%%

\subsection{The Kac-Moody Billiard of Pure Gravity}

\begin{figure}[htbp]
    \centerline{\includegraphics[width=110mm]{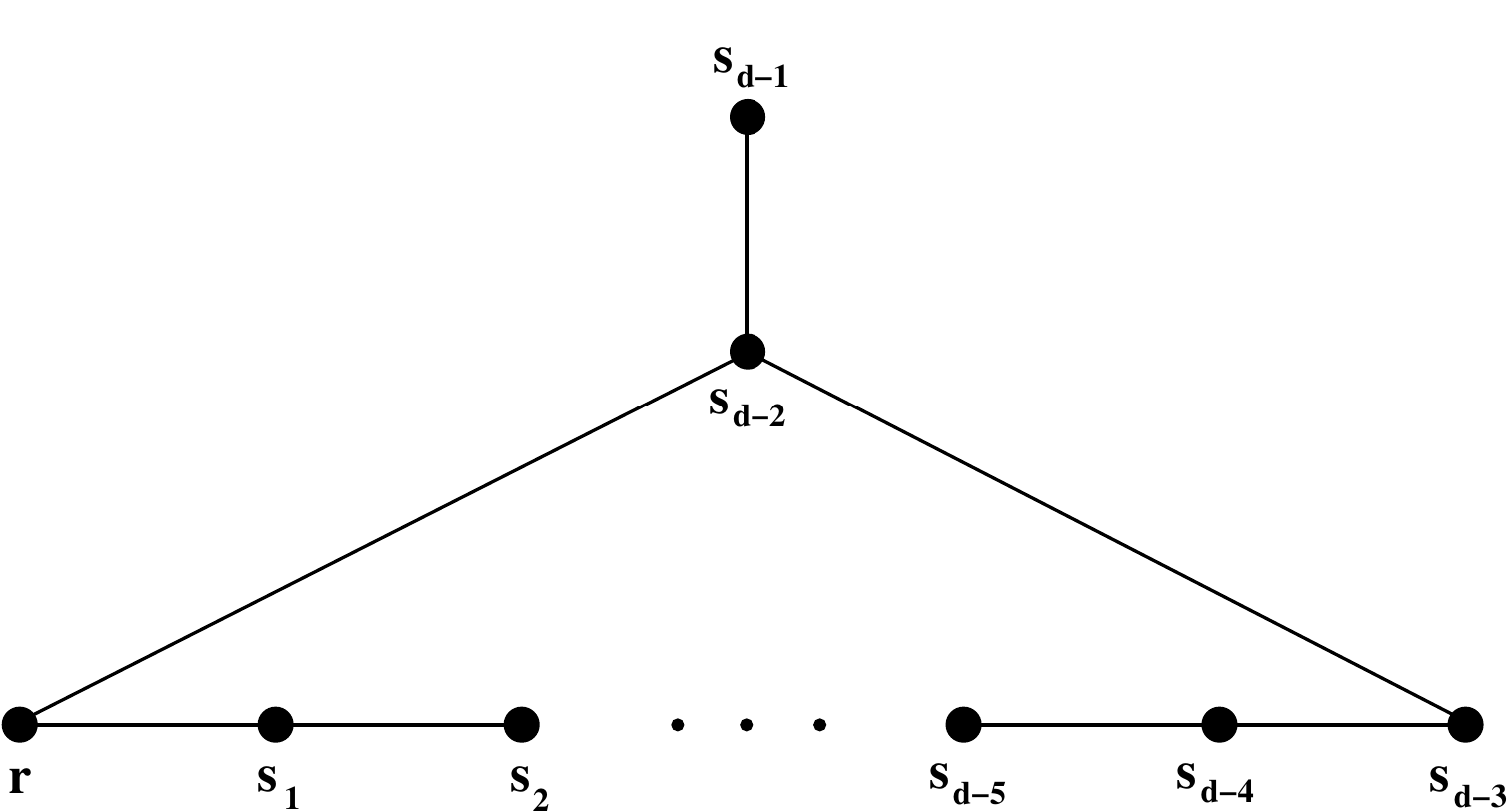}}
    \caption{The Dynkin diagram of the hyperbolic Kac--Moody algebra
      $A_{d-2}^{++}$ which controls the billiard dynamics of pure
      gravity in $D=d+1$ dimensions. The nodes $s_{1}, \cdots, s_{d-1}$
      represent the ``symmetry walls'' arising from the off-diagonal
      components of the spatial metric, and the node $r$ corresponds to
      a ``curvature wall'' coming from the spatial curvature. The
      horizontal line is the Dynkin diagram of the underlying
      $A_{d-2}$-subalgebra and the two topmost nodes, $s_{d-2}$ and
      $s_{d-1}$, give the affine- and overextension, respectively.}
    \label{figure:AnppNumbered}
  \end{figure}

We start by providing some examples of
theories leading to regular billiards, focusing first on pure
gravity in any number of $D$ ($>3$) spacetime dimensions. In this
case, there are $d=D-1$ scale factors $\beta^i$ and the relevant
walls are the symmetry walls, Equation~(\ref{symmetryW}), 
\begin{equation}
  s_i(\beta) \equiv \beta^{i+1} - \beta^i= 0
  \qquad
  (i=1, 2, \cdots, d-1),
\end{equation}
and the curvature wall, Equation~(\ref{curvatureW}),
\begin{equation}
 r(\beta) \equiv 2 \beta^1 + \beta^2 + \cdots + \beta^{d-2}= 0.
\end{equation}
There are thus $d$ relevant walls, \index{billiard wall} which define
a simplex in ($d-1$)-dimensional hyperbolic
space $\mc{H}_{d-1}$. The scalar products of the linear forms defining these
walls are easily computed. One finds as non-vanishing products
\begin{equation}
  \begin{array}{rcl}
    (s_i \vert s_i) &=& \phantom{-}2
    \qquad
    (i = 1, \cdots, d-1),
    \\ [0.25 em]
    (r \vert r) &=& \phantom{-}2,
    \\ [0.25 em]
    (s_{i+1} \vert s_i) &=& -1
    \qquad
    (i = 2, \cdots, d-1)
    \\ [0.25 em]
    (r \vert s_1) &=& -1,
    \\ [0.25 em]
    (r \vert s_{d-2}) &=& -1.
  \end{array}
\end{equation}
The matrix of the scalar products of the wall forms is thus the Cartan
matrix of the (simply-laced) Lorentzian Kac--Moody algebra
$A_{d-2}^{++}$ with Dynkin diagram as in
Figure~\ref{figure:AnppNumbered}. The roots of the underlying
finite-dimensional algebra $A_{d-2}$ are given by $s_i$ ($i = 1,
\cdots, d-3$) and $r$. The affine root is $s_{d-2}$ and the
overextended root is $s_{d-1}$.

Accordingly, in the case of pure gravity in any number of
spacetime dimensions, one finds also that the billiard region is
regular. This provides new examples of Coxeter billiards, with
Coxeter groups $A_{d-2}^{++}$, which are also Kac--Moody billiards
since the Coxeter groups are the Weyl groups of the Kac--Moody
algebras $A_{d-2}^{++}$.

%%%%%%%%%%%%%%%%%%%%%%%%%%%%%%%%%%%%%%%%%%%%%%%%%%%%%%%%%%%%%%%%%%%%%%%%%%%%%%%%%%%

\subsection{The Kac-Moody Billiard for the Coupled Gravity-3-Form System}

Let us review the conditions that must be fulfilled in
order to get a Kac--Moody billiard and let us emphasize how
restrictive these conditions are. The billiard region computed
from any theory coupled to gravity with $n$ dilatons in $D=d+1$
dimensions always defines a convex polyhedron in a $(d+n-1)$-dimensional
hyperbolic space $\mc{H}_{d+n-1}$. In the general case,
the dihedral angles between adjacent faces of $\mc{H}_{d+n-1}$ can
take arbitrary continuous values, which depend on the dilaton
couplings, the spacetime dimensions and the ranks of the $p$-forms
involved. However, only if the dihedral angles are integer
submultiples of $\pi$ (meaning of the form $\pi/k$ for $k\in
\mbb{Z}_{\geq 2}$) do the reflections in the faces of
$\mc{H}_{d+n-1}$ define a Coxeter group. \index{Coxeter group} In this special case the
polyhedron is called a \emph{Coxeter polyhedron}. This Coxeter
group is then a (discrete) subgroup of the isometry group of
$\mc{H}_{d+n-1}$.

In order for the billiard region to be identifiable with the
fundamental Weyl chamber \index{Weyl chamber} of a Kac--Moody algebra, the Coxeter
polyhedron should be a \emph{simplex}, i.e., bounded by $d+n$
walls in a $d+n-1$-dimensional space. In general, the Coxeter
polyhedron need not be a simplex.

There is one additional condition. The angle $\vartheta$ between
two adjacent faces $i$ and $j$ is given, in terms of the Coxeter
exponents, by 
\begin{equation}
  \vartheta= \f{\pi}{m_{ij}}.
  \label{anglebetweenadjacentfaces}
\end{equation}
Coxeter groups that correspond to Weyl groups of Kac--Moody algebras 
are the \emph{crystallographic} Coxeter groups for which $m_{ij}\in
\{2,3,4,6,\infty\}$. So, the requirement for a gravitational
theory to have a Kac--Moody algebraic description is not just that
the billiard region is a Coxeter simplex but also that the angles
between adjacent walls are such that the group of reflections in
these walls is crystallographic.

These conditions are very restrictive and hence gravitational
theories which can be mapped to a Kac--Moody algebra in the
BKL-limit \index{BKL-limit} are rare.

\subsubsection*{The Kac-Moody  Billiard of Eleven-Dimensional Supergravity}

Consider for instance the action~(\ref{keyaction}) for
gravity coupled to a single three-form in $D=d+1$ spacetime
dimensions. We assume $D \geq 6$ since in lower dimensions the
3-form is equivalent to a scalar ($D = 5$) or has no degree of
freedom ($D<5$).

Whenever a $p$-form ($p \geq 1$) is present, the curvature wall is
  subdominant as it can be expressed as a linear combination with
  positive coefficients of the electric and magnetic walls of the
  $p$-forms. (These walls are all listed in
  Section~\ref{Section:HyperbolicBilliard}.) To see this, note the following properties which can be easily checked by direct computation:
  \begin{itemize}
  \item the dominant electric wall is (assuming the
    presence of a dilaton) 
    \begin{equation}
      e_{1\cdots p}(\be)\equiv
      \be^{1}+\be^{2}+\cdots +\be^{p}-\f{\la_p}{2}\phi = 0,
      \label{electricwallcondition}
    \end{equation}
    \item the dominant magnetic wall reads 
    \begin{equation}
      m_{1, p+1, \cdots , d-2} (\be) \equiv
      \be^{1}+\be^{p+1}+\cdots +\be^{d-2}+\f{\la_p}{2}\phi = 0,
    \end{equation}
    \item the dominant curvature wall is given by the sum
    \begin{equation}
    e_{1\cdots p}(\be) + m_{1, p+1, \cdots , d-2} (\be).
    \eeq
\end{itemize}

It follows that in the case of gravity coupled to a
single three-form in $D=d+1$ spacetime dimensions, the relevant walls
are the symmetry walls, Equation~(\ref{symmetryW}), 
\begin{equation}
  s_i(\beta) \equiv \beta^{i+1} - \beta^i= 0,
  \qquad
  i=1, 2, \cdots , d-1
\end{equation}
(as always) and the electric wall 
\begin{equation}
  e_{123}(\beta) \equiv \beta^1 + \beta^2 + \beta^{3} = 0
\end{equation}
($D \geq 8$) or the magnetic wall
\begin{equation}
  m_{1 \cdots D-5}(\beta) \equiv
  \beta^1 + \beta^2 + \cdots \beta^{D-5} = 0
\end{equation}
($D \leq 8$). Indeed, one can express the
magnetic walls as linear combinations with (in general
non-integer) positive coefficients of the electric walls for $D
\geq 8$ and vice versa for $D \leq 8$. Hence the billiard table
is always a simplex (this would not be true had one a dilaton and
various forms with different dilaton couplings).

However, it is only for $D=11$ that the billiard is a Coxeter
billiard. In all the other spacetime dimensions the angle
between the relevant $p$-form wall and the symmetry wall is not an integer submultiple of
$\pi$. More precisely, the angle between

\begin{itemize}
\item the magnetic wall $\beta^1$ and the symmetry wall
  $\beta^2 - \beta^1$ ($D=6$),
\item the magnetic wall $\beta^1 + \beta^2$ and the symmetry wall
  $\beta^3 - \beta^2$ ($D=7$), and
\item the electric wall $\beta^1 + \beta^2 + \beta^3$ and the symmetry
  wall $\beta^4 - \beta^3$ ($D \geq 8$),
\end{itemize}

is easily verified to be an integer submultiple of $\pi$ only for
$D=11$, for which it is equal to $\pi/3$.

From the point of view of the regularity of the billiard, the
spacetime dimension $D=11$ is thus privileged. This is of course
also the dimension privileged by supersymmetry. It is quite
intriguing that considerations \emph{a priori} quite different
(billiard regularity on the one hand, supersymmetry on the other
hand) lead to the same conclusion that the gravity-3-form
system is quite special in $D=11$ spacetime dimensions.

For completeness, we here present the wall system relevant for the
special case of $D=11$. We obtain ten dominant wall forms, which
we rename $\al_1, \cdots, \al_{10}$, 
\begin{equation}
  \begin{array}{rcl}
    \al_{m}(\be)&=& \be^{m+1}-\be^m
    \qquad
    (m=1, \cdots, 10),
    \\ [0.25 em]
    \al_{10}(\be)&=& \be^1+\be^2+\be^3.
  \end{array}
\end{equation}
Then, defining a new
collective index $i=(m, 10)$, we see that the scalar products
between these wall forms can be organized into the matrix 
\begin{equation}
  A_{ij}=2\f{(\al_{i}|\al_{j})}{(\al_{i}|\al_{i})}= \left(
    \begin{array}{@{}c@{\quad}c@{\quad}c@{\quad}c@{\quad}c@{\quad}c@{\quad}c@{\quad}c@{\quad}c@{\quad}c@{}}
      2 & -1 & 0 & 0 & 0 & 0 & 0 & 0 & 0 & 0 \\
      -1 & 2 & -1 & 0 & 0 & 0 & 0 & 0 & 0 & 0 \\
      0 & -1 & 2 & -1 & 0 & 0 & 0 & 0 & 0 & -1 \\
      0 & 0 & -1 & 2 & -1 & 0 & 0 & 0 & 0 & 0 \\
      0 & 0 & 0 & -1 & 2 & -1 & 0 & 0 & 0 & 0 \\
      0 & 0 & 0 & 0 & -1 & 2 & -1 & 0 & 0 & 0 \\
      0 & 0 & 0 & 0 & 0 & -1 & 2 & -1 & 0 & 0 \\
      0 & 0 & 0 & 0 & 0 & 0 & -1 & 2 & -1 & 0 \\
      0 & 0 & 0 & 0 & 0 & 0 & 0 & -1 & 2 & 0 \\
      0 & 0 & -1 & 0 & 0 & 0 & 0 & 0 & 0 & 2 \\
    \end{array}
  \right),
\end{equation}
which can be identified with the Cartan matrix \index{Cartan matrix}
of the hyperbolic Kac--Moody algebra \index{Kac--Moody algebra}
$E_{10}$ that we have encountered in
Section~\ref{section:DecompE10E10}. We again display the
corresponding Dynkin diagram \index{Dynkin diagram} in
Figure~\ref{figure:E10}, where we point out the explicit relation
between the simple roots and the walls of the Einstein--3-form
theory. It is clear that the nine dominant symmetry wall forms
correspond to the simple roots \index{root} $\al_{m}$ of the
subalgebra $\mf{sl}(10,\mbb{R})$. The enlargement to $E_{10}$ is due
to the tenth exceptional root realized here through the dominant
electric wall form $e_{123}$.

  \begin{figure}[htbp]
    \centerline{\includegraphics[width=100mm]{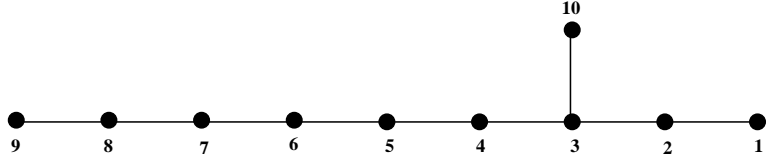}}
    \caption{The Dynkin diagram of $E_{10}$. Labels $m=1,\cdots, 9$
      enumerate the nodes corresponding to simple roots, $\al_{m}$, of the
      $\mf{sl}(10,\mathbb{R})$ subalgebra and the exceptional node,
      labeled ``$10$'', is associated to the electric wall
      $\al_{10}=e_{123}$.}
    \label{figure:E10}
  \end{figure}

%%%%%%%%%%%%%%%%%%%%%%%%%%%%%%%%%%%%%%%%%%%%%%%%%%%%%%%%%%%%%%%%%%%%%%%%%%%%%%%%%%%
%%%%%%%%%%%%%%%%%%%%%%%%%%%%%%%%%%%%%%%%%%%%%%%%%%%%%%%%%%%%%%%%%%%%%%%%%%%%%%%%%%%

\subsection{Dynamics in the Cartan Subalgebra}
\label{section:CSAdynamics}

We have just learned that, in some cases, the group of reflections 
\index{geometric reflection} that describe the (possibly chaotic)
dynamics in the BKL-limit \index{BKL-limit} is a Lorentzian Coxeter
group. In this section we fully exploit this algebraic fact
and show that whenever this Coxeter group is \emph{crystallographic}, the
dynamics takes place in the Cartan subalgebra $\mf{h}$ of the
Lorentzian Kac--Moody algebra  \index{Kac--Moody algebra} $\mf{g}$,
for which the relevant Coxeter group is identified with the Weyl group $\mc{W}(\mf{g})$. Moreover, we show that the
``billiard table'' can be identified with the fundamental Weyl chamber
in $\mf{h}$.

%%%%%%%%%%%%%%%%%%%%%%%%%%%%%%%%%%%%%%%%%%%%%%%%%%%%%%%%%%%%%%%%%%%%%%%%%%%%%%%%%%%

\subsubsection{Scale Factor Space and the Wall System}

Let us first briefly review some of the salient features
encountered so far in the analysis. In the following we denote by
$\mc{M}_{\be}$ the Lorentzian ``scale factor''-space (or
$\be$-space) in which the billiard dynamics takes place. Recall
that the metric in $\mc{M}_\be$, induced by the Einstein--Hilbert
action, is a flat Lorentzian metric, whose explicit form in terms
of the (logarithmic) scale factors reads 
\begin{equation}
  \ga_{\mu\nu} \, d\be^{\mu} \, d\be^{\nu} =
  \sum_{i=1}^{d} d\be^{i} \, d\be^{i} -
  \left(\sum_{i=1}^{d}d\be^{i}\right)
  \left(\sum_{j=1}^{d}d\be^{j}\right) + d\phi \, d\phi, 
\label{metricinbetaspace}
\end{equation}
where $d$ counts the number of physical spatial dimensions (see
Section~\ref{Section:HyperbolicBilliard}). The role of all other ``off-diagonal''
variables in the theory is to interrupt the free-flight motion of the
particle, by adding walls in $\mc{M}_{\be}$ that
confine the motion to a limited region of scale factor space, namely a
convex cone bounded by timelike hyperplanes. When projected onto the
unit hyperboloid, this region defines a simplex in hyperbolic space
which we refer to as the ``billiard table''.

One has, in fact, more than just the walls. The theory provides
these walls with a specific normalization through the Lagrangian,
which is crucial for the connection to Kac--Moody algebras. Let us
therefore discuss in somewhat more detail the geometric properties of
the wall system. The metric, Equation~(\ref{metricinbetaspace}), in
scale factor space can be seen as an extension of a flat Euclidean
metric in Cartesian coordinates, and reflects the Lorentzian nature of
the vector space $\mc{M}_{\be}$. In this space we may identify a pair
of coordinates $(\be^{i}, \phi)$ with the components of a vector
$\be\in \mc{M}_{\be}$, with respect to a basis $\{\bar{u}_{\mu}\}$ of
$\mc{M}_{\be}$, such that
\begin{equation}
  \bar{u}_{\mu}\cdot \bar{u}_{\nu}=\ga_{\mu\nu}.
\end{equation}
The walls themselves are then defined by hyperplanes in this linear
space, i.e., as linear forms $\om=\om_{\mu}\underline{\si}^{\mu}$, for
which $\om=0$, where $\{\underline{\si}^{\mu}\}$ is the basis dual to
$\{\bar{u}^{\mu}\}$. The pairing $\om(\be)$ between a vector $\be\in
\mc{M}_\be$ and a form $\om\in\mc{M}_{\be}^{\star}$ is sometimes also
denoted by $\left< \om, \be \right>$, and for the two dual bases we
have, of course, 
\begin{equation}
  \left< \underline{\si}^{\mu}, \bar{u}_{\nu}\right> =\delta^{\mu}_{\nu}.
\end{equation}
We therefore find that the walls can be written as linear forms in the
scale factors:
\begin{equation}
  \om(\be)=\sum_{\mu, \nu}\om_{\mu}\be^{\nu}
  \left<\underline{\si}^{\mu}, \bar{u}_{\nu}\right>=
  \sum_{\mu}\om_{\mu}\be^{\mu}=
  \sum_{i=1}^{d}\om_{i}\be^{i}+\om_{\phi}\phi.
  \label{wallforms}
\end{equation}
We call $\om(\be)$ \emph{wall forms}. With this interpretation they
belong to the dual space $\mc{M}_{\be}^{\star}$, i.e., 
\begin{equation}
  \begin{array}{rcl}
    \mc{M}_{\be}^{\star} \, \ni\, \om :
    \mc{M}_{\be} &\quad \longrightarrow& \quad \mbb{R},
    \\ [0.25 em]
    \be &\quad \longmapsto& \quad \om(\be).
  \end{array}
  \label{dualspaces}
\end{equation}
From Equation~(\ref{dualspaces}) we may conclude that the walls
bounding the billiard are the hyperplanes $\om=0$ through the origin
in $\mc{M}_{\be}$ which are orthogonal to the vector with components
$\om^{\mu}=\ga^{\mu\nu}\om_{\nu}$.

It is important to note that it is the wall
forms that the theory provides, as arguments of the exponentials in
the potential, and not just the hyperplanes on which these forms $\om$
vanish. The scalar products between the wall forms are computed using
the metric in the dual space $\mc{M}_{\be}^{\star}$, whose explicit
form was given in Section~\ref{Section:HyperbolicBilliard}, 
\begin{equation}
  (\om |\om^{\prime})\equiv \ga^{\mu\nu}\om_{\mu}\om_{\nu}=
  \sum_{i=1}^{d}\om_{i}\om_{i}^{\prime}-\f{1}{d-1}
  \left(\sum_{i=1}^{d}\om_{i}\right)
  \left(\sum_{j=1}^{d}\om_{j}^{\prime}\right)+
  \om_{\phi}\ \om_{\phi}^{\prime},
  \qquad
  \om, \om^{\p}\in\mc{M}_{\be}. 
  \label{inversemetric}
\end{equation}

\subsubsection{Scale Factor Space and the Cartan Subalgebra}

The crucial additional observation is that (for the ``interesting''
theories) the matrix $A$ associated with the relevant walls
$\om_A$, 
\begin{equation}
  A_{AB} = 2 \frac{(\om_A \vert \om_B)}{(\om_A \vert \om_A)}
\end{equation}
is a Cartan matrix, i.e., besides having 2's on its
diagonal, which is rather obvious, it has as off-diagonal entries
non-positive integers (with the property $A_{AB} \not= 0
\Rightarrow A_{BA} \not=0$). This Cartan matrix is of course
symmetrizable since it derives from a scalar product.

For this reason, one can usefully identify the space of the scale
factors with the Cartan subalgebra $\mf{h}$
of the Kac--Moody algebra $\mf{g}(A)$ defined by $A$. In that
identification, the wall forms become the simple roots,
which span the vector space
$\mf{h}^{\star}=\spn\{\al_{1},\cdots,\al_{r}\}$ dual to the Cartan
subalgebra. The rank $r$ of the algebra is equal to the number of
scale factors $\beta^\mu$, including the dilaton(s) if any
($(\beta^\mu) \equiv (\be^{i}, \phi)$). This number is also equal to
the number of walls since we assume the billiard to be a simplex. So,
both $A$ and $\mu$ run from $1$ to $r$. The metric in $\mc{M}_\be$,
Equation~(\ref{metricinbetaspace}), can be identified with the
invariant bilinear form of $\mf{g}$, restricted to the Cartan
subalgebra $\mf{h}\subset \mf{g}$. The scale factors $\beta^\mu$ of
$\mc{M}_\be$ become then coordinates $h^{\mu}$ on the Cartan
subalgebra $\mf{h}\subset\mf{g}(A)$.

The Weyl group of a Kac--Moody algebra has been defined first in
the space $\mf{h}^{\star}$ as the group of reflections in the
walls orthogonal to the simple roots. Since the metric is non-degenerate, one can equivalently define by duality the Weyl group
in the Cartan algebra $\mf{h}$ itself. For each reflection $r_i$ on
$\mf{h}^{\star}$ we associate a dual reflection $r_i^{\vee}$ as
follows,
\begin{equation}
  r_i^{\vee}(\be)=\be-\left< \al_i, \be\right > \al_i^{\vee},
  \qquad
  \be, \al_i^{\vee}\in\mf{h},
  \label{Cartanreflection}
\end{equation}
which is the reflection relative to
the hyperplane $\al_i(\be)=\left< \al_i, \be \right> =0$. This
expression can be rewritten, 
\begin{equation}
  r_i^{\vee}(\be)=
  \be-\f{2(\be|\al_i^{\vee})}{(\al_i^{\vee}|\al_i^{\vee})}\al_i^{\vee},
  \label{Cartanreflection2}
\end{equation}
or, in terms of the scale factor coordinates $\beta^\mu$, 
\begin{equation}
  \be^{\mu} \longrightarrow {\be^{\mu}}^{\p} =
  \be^{\mu}-\f{2(\be\vert \om^{\vee})}{(\om^\vee \vert \om^\vee)}\om^{\vee\mu}. 
  \label{Geometricreflection}
\end{equation}
This is precisely the billiard reflection
Equation~(\ref{billiardreflectionrule}) found in
Section~\ref{section:DynamicsBilliardHyp}.

Thus, we have the following correspondence:
\begin{equation}
  \begin{array}{rcl}
    \mc{M}_{\be} &\equiv& \mf{h},
    \\ [0.25 em]
    \mc{M}_{\be}^{\star} &\equiv& \mf{h}^{\star},
       \\ [0.25 em]
    \om_{A}(\be) &\equiv& \al_{A}(h),
    \\ [0.25 em]
    \mbox{billiard wall reflections} &\equiv&
    \mbox{fundamental Weyl reflections},
  \end{array}
  \label{identifications} 
\end{equation}
As we have also seen, the Kac--Moody algebra $\mf{g}(A)$ is Lorentzian
since the signature of the metric Equation~(\ref{metricinbetaspace})
is Lorentzian. This fact will be crucial in the analysis of subsequent
sections and is due to the presence of gravity, where conformal
rescalings of the metric define timelike directions in scale factor
space.

We thereby arrive at the following important
result~\cite{ArithmeticalChaos, HyperbolicKaluzaKlein, DHNReview}:

\begin{center}
  \begin{tabular}{c}
    \hline \hline \\
    \emph{The dynamics of certain gravity-$p$-form systems can in the BKL-limit be mapped}  \\
    \emph{to a billiard motion in the Cartan subalgebra
      $\mf{h}$ of a Lorentzian Kac--Moody algebra $\mf{g}$.} \\ [1 em]
    \hline \hline
  \end{tabular}
\end{center}

%%%%%%%%%%%%%%%%%%%%%%%%%%%%%%%%%%%%%%%%%%%%%%%%%%%%%%%%%%%%%%%%%%%%%%%%%%%%%%%%%%%

\subsection{Hyperbolicity Implies Chaos}
Let us now point out the main consequence of the identifications in (\ref{identifications}). As in Section \ref{Section:HyperbolicBilliard}, let $\mc{B}_{\be}$ denote the region in scale factor
space to which the billiard motion is confined, i.e. the ``billiard table'', 
\begin{equation}
  \mc{B}_{\be} =\{\be \in \mc{M}_\be\, | \, \om_{A}(\be)\geq 0\},
  \label{billiardtable}
\end{equation}
where the index $A$ runs over all
dominant walls. On the algebraic side, recall from Section \ref{Section:hyperbolic} that the fundamental Weyl chamber $\mc{C}_{\mf{h}}\subset \mf{h}$ is the closed convex (half) cone given by 
\begin{equation}
  \mc{C}_{\mf{h}}=\{h \in \mf{h}\, | \, \al_{A}(h) \geq 0;
  \, A=1,\cdots, \rank \mf{g} \},
  \label{WeylChamber}
\end{equation}
where we have put a subscript on $\mc{C}_{\mf{h}}$ to emphasize that this is the fundamental chamber in the \emph{Cartan subalgebra} $\mf{h}$, in contrast to the fundamental chamber $\mc{C}$ of the dual root space $\mf{h}^{\star}$. We see that the conditions $\al_{A}(h) \geq 0$ defining
$\mc{C}_{\mf{h}}$ are equivalent, upon examination of
Equation~(\ref{identifications}), to the conditions $\om_{A}(\be)\geq
0$ defining the billiard table $\mcBe$. We may therefore make the crucial identification 
\begin{equation}
 \mc{C}_{\mf{h}}\equiv \mc{B}_{\mc{M}_\be},
  \label{WeylChamberandBilliardRegion}
\end{equation}
which means that the particle geodesic is confined to move within the
fundamental Weyl chamber of $\mf{h}$. From the
billiard analysis in Section~\ref{Section:HyperbolicBilliard} we know that the
piecewise motion in scale-factor space is
controlled by geometric reflections with
respect to the walls $\om_{A}(\be)=0$. By comparing with the dominant
wall forms and using the correspondence in
Equation~(\ref{identifications}) we may then conclude that the
geometric reflections of the coordinates $\be^{\mu}(\tau)$ are
controlled by the Weyl group in the Cartan subalgebra of $\mf{g}(A)$.

%%%%%%%%%%%%%%%%%%%%%%%%%%%%%%%%%%%%%%%%%%%%%%%%%%%%%%%%%%%%%%%%%%%%%%%%%%%%%%%%%%%

Now recall further that the BKL dynamics is chaotic if and only if
the billiard table is of finite volume when projected onto the
unit hyperboloid. From our discussion of hyperbolic Coxeter 
groups in Section~\ref{Section:hyperbolic}, and from the identification (\ref{WeylChamberandBilliardRegion}), we deduce 
that this feature is equivalent to hyperbolicity of the
corresponding Kac--Moody algebra. This leads to the crucial
statement~\cite{ArithmeticalChaos,HyperbolicKaluzaKlein,DHNReview}:

\begin{center}
  \begin{tabular}{c}
    \hline \hline \\
    \emph{If the billiard region of a gravity-$p$-form system can be identified with the fundamental } \\
    \emph{Weyl chamber of a hyperbolic Kac--Moody algebra,
      then the dynamics is chaotic.} \\ [1 em]
    \hline \hline
  \end{tabular}
\end{center}

As we have also discussed above, hyperbolicity can be rephrased in
terms of the fundamental weights $\Lambda_i$ defined as 
\begin{equation}
  \left<\Lambda_{j}, \al_i^{\vee}\right> =
  \f{2(\Lambda_j|\al_i)}{(\al_i|\al_i)} \equiv \delta_{ij},
  \qquad
  \al_i^{\vee}\in \mf{h},\, \Lambda_i\in\mf{h}^{\star}.
  \label{fundamentalweightcondition}
\end{equation}
Just as the fundamental
Weyl chamber in $\cH^\star$ can be expressed in terms of the
fundamental weights, the
fundamental Weyl chamber in $\cH$ can be expressed in a similar
fashion in terms of the fundamental coweights: 
\begin{equation}
  \mc{C}_{\mf{h}}=\{\be\in\mf{h}\, |\, \be=
  \sum_i a_i \Lambda_i^{\vee}, \, a_i\in\mbb{R}_{\geq 0}\}.
\end{equation}
Thus, hyperbolicity holds if and
only if none of the fundamental weights are spacelike, 
\begin{equation}
  (\Lambda_{i}|\Lambda_{i})\leq 0,
  \label{timelikecondition}
\end{equation}
for all $ i\in \{1, \cdots, \rank \mf{g}\}$.

\chapter{Compactification, Cohomology and Coxeter Groups}
\label{Chapter:Cohomology}

In the previous chapter we have seen that many gravitational theories of physical interest exhibit chaotic behavour in the BKL-limit, which in turn is related to an underlying hyperbolic Kac-Moody algebraic structure. In this chapter we shall analyze this in more detail, and in particular study the modification of the billiard dynamics when there is a spacetime splitting of the form ${\Sigma}=\mbb{R}\times {X}$, with ${X}$ being a smooth compact internal manifold of arbitrary topology. This question is of particular interest for clarifying the role of BKL-chaos within big bang/big crunch scenarios which typically relies on a smooth collapsing phase in the approach towards the singularity, in apparent contradiction with the chaotic BKL-behaviour exhibited by, for example, string theory related supergravities. We will learn that for certain special internal manifolds, chaos is removed by the compactification, thus providing a possible reconciliation with the aforementioned analysis. Along the way we will also discover fascinating new algebraic structures of the modified billiards which are described using the theory of \emph{buildings}. This chapter is based on {\bf Paper V}, written in collaboration with Marc Henneaux and Daniel H. Wesley. For explicit examples of the results presented in this chapter, as well as a complete classification, the reader is referred to {\bf Paper V}.

\section{Intermediate Asymptotics}

The original BKL analysis is classical and has been pushed all the way to the singularity \cite{BKL,BKL2}.  As such it is valid for any spatial topology. In the approach to a spacelike singularity there is an asymptotic decoupling of spatial points, and the dynamics becomes ``ultralocal".  These results (decoupling of spatial points and chaotic oscillations) are by now well supported by extensive analytical and numerical evidence; a non-exhaustive list of references include \cite{Berger:1998vxa,Berger:1998us,Andersson:2000cv,Ringstrom:2000mk,DHRW02,Garfinkle:2003bb,Uggla,Garfinkle,Damour:2007nb}.

The classical analysis has however obvious limitations and it is not clear what becomes of the BKL results for energy scales above the Planck scale, where quantum gravity effects cannot be ignored. In the standard BKL analysis, which ignores quantum effects, no walls are removed as the big crunch is approached.  But, when some spatial dimensions are compact, there is a wide range of initial conditions for which walls corresponding to massive modes are always subdominant until the universe enters the quantum regime, and these walls are not relevant for the billiard analysis while the universe is described by classical physics \cite{Wesley2005bd,Wesley2006cd}. For this broad set of initial conditions, there is no epoch in which the usual classical BKL analysis, with the full set of walls, applies.

For this reason, it is of interest to consider the regime of \emph{intermediate asymptotics} where the curvature is much smaller than the Planck curvature but where the billiard analysis applies (see \cite{Damour:2005zb}).  In that pre-Planck regime, it is not true that the topology of spacetime is irrelevant \cite{Wesley2005bd}.  A modification might arise in the presence of $p$-forms because the massless spectrum of $p$-forms in the lower-dimensional theory depends on the de Rham cohomology of the internal manifold, and since massless degrees of freedom dominate in the BKL-limit before reaching the Planck scale \cite{Wesley2005bd}, non-trivial compactification eliminates billiard walls corresponding to degrees of freedom which are rendered massive in the compactification.  Depending on which walls are removed by the compactification, a chaotic theory can be rendered non-chaotic. 

%Note that the suppression of massive modes exhibited in \cite{Wesley2005bd} is a classical result, which follows from the virial theorem.  However, one could argue that it is in fact also true quantum mechanically (in the intermediate regime where the geometry can be treated as classical but the matter fields are quantum-mechanical), for standard methods reveal that the expected energy density $\rho_{\rm pp}$ of pair-produced particles is at most  \cite{Bir84}
%\begin{equation}
%\rho_{\rm pp} \sim H^2 e^{-m/H},
%\end{equation}
%where $H$ is the effective Hubble parameter and $m$ the particle mass.  The exponential suppression of massive mode pair production follows the point particle intuition \cite{Gubser:2003vk}, with an even greater suppression expected for string pair production \cite{Tolley:2005ak}.

There are a number of reasons to better understand the interplay between BKL dynamics and compactification.  On the physics side, the BKL-limit with compact internal spaces is relevant for certain types of cyclic or ``pre-big bang" cosmological models built from string or M-theory (see, e.g., \cite{Khoury:2001wf,Steinhardt:2001st,Buonanno:1998bi,Gasperini:2002bn} and references therein).  Cosmological models with a  big crunch/big bang transition rely on a smooth collapsing phase as they approach the big crunch singularity, hence chaotic BKL oscillations close to the singularity are a potential problem for these models.  If chaos can be removed by interpreting our four-dimensional world as an effective description of a more fundamental higher-dimensional theory where the ``troublesome" billiard walls are eliminated through the cohomology of the internal space, the problems with BKL chaos in these models may be circumvented.  On the mathematical side, the billiard regions and the reflection groups that emerge after compactification possess a rich structure which deserves investigation.

\section{The ``Uncompactified Billiard"}

We shall here recall some useful features of the billiards which were discussed in Chapter \ref{Chapter:Billiards}. This corresponds to the case when all the walls are switched on, as it occurs when no internal dimension is compact \cite{Wesley2005bd}.  This is also the case relevant when the BKL analysis is pushed all the way to the singularity.  The billiard region is then the smallest possible one in the sense that the billiard region of all the other cases will contain the billiard region of the uncompactified theory. We shall denote by $\om_{A^{\p}}$ the dominant walls of the uncompactified theory. While the number of dominant walls relevant to the compact cases might not be equal to the dimension $M$ of $\mc{M}_{\be}$, it turns out that for all theories whose dimensional reduction to three dimensions is described by a symmetric space, the number of dominant walls is equal to $M$ \cite{DHNReview}.

In this case the dominant walls confine the billiard motion to the region $\mc{B}_{\be}\subset \mc{M}_{\be}$ defined by (see Section \ref{Section:HyperbolicBilliard})
\beq
\mc{B}_{\be}=\big\{\be\in\mc{M}_{\be}\hs \big|\hs \om_{A^{\p}}(\be)\geq 0, \hs A^{\p}=1, \cdots , M\big\}, \label{originalbilliard}
\eeq
The billiard table is a simplex in $\mbb{H}_m$. We shall call somewhat improperly the region (\ref{originalbilliard}) the ``uncompactified billiard region" and its projection on $\mbb{H}_m$ the ``uncompactified billiard table".

The scalar products $(\om_{A^{\prime}}|\om_{B^{\prime}})$ between the dominant walls can be organized into a matrix,
\beq
A_{A^{\prime}B^{\prime}}=\f{2(\om_{A^{\prime}}|\om_{B^{\prime}})}{(\om_{A^{\prime}}|\om_{A^{\prime}})}.
\eeq
In the noncompact case, the matrix $A$  turns out to possess the properties of a Lorentzian Cartan matrix \cite{ArithmeticalChaos,DHNReview}, thereby identifying the dominant wall forms $\om_{A^{\prime}}$ with the simple roots of the Kac-Moody algebra $\mf{g}(A)$ constructed from  $A$ \cite{Kac}. This Kac-Moody algebra is the ``overextension'' $\mf{g}^{++}$ of the U-duality algebra $\mf{g}$.  The group generated by the reflections in the billiard walls of the uncompactified theory is a Coxeter group, which is the Weyl group of the corresponding Kac-Moody algebra \cite{ArithmeticalChaos,DHNReview}.  We shall denote it by $\mc{W}$.

Recall from Section \ref{Section:hyperbolic}) that the action of the Weyl group $\mc{W}$ on the (future) lightcone $\mc{O}^{+}$ splits up into a disjoint union of chambers, called \emph{Weyl chambers}. One of these chambers is defined by the inequalities $\omega_{A^\prime} \geq 0$ and is called the \emph{fundamental chamber} $\mc{F}$.\footnote{Note that in Chapters \ref{Chapter:KacMoody} and \ref{Chapter:Billiards} the fundamental Weyl chamber was denoted $\mc{C}$. In this chapter we let $\mc{C}$ denote an arbitrary chamber and reserve the notation $\mc{F}$ for the specific choice of fundamental chamber. We trust that the reader will not confuse the two.} Then, all other chambers in $\mc{O}^{+}$ correspond to images of $\mc{F}$ under $\mc{W}$. The action of $\mc{W}$ on the Weyl chambers is simply transitive. When projected onto $\mbb{H}_m$, these chambers become $m$-simplices of finite volume. The fundamental chamber $\mc{F}$ is the uncompactified billiard region in which the chaotic dynamics takes place. The $m+1$ hyperplanes (or dominant walls) which bound the fundamental chamber, correspond to the codimension-one faces of $\mc{F}$ when projected onto $\mbb{H}_n$ (see also Section \ref{section:chambers}).

\section{Compactification and Cohomology}
\label{section:Compactification}
Compactification can modify the billiard, as was shown in \cite{Wesley2005bd,Wesley2006cd}. This occurs because the billiard dynamics in the intermediate regime considered here depends not only on the
$p$--form menu, but also on the topology of the space upon which the theory is
formulated, specifically on the de Rham cohomology $\mc{H}^{p}({X})$ of the compactification manifold ${X}$. The rules for constructing the noncompact billiard system are given above, and here we focus on the ``selection rule" that describes how this billiard is modified after compactification.

\subsection{Selection Rule}

We study situations in which all spatial dimensions are compact, and thus spacetime $\Sigma$ has topology
\beq
\Sigma = \mbb{R}\times{X},
\eeq
where ${X}$ is closed, compact, and orientable. Electric and magnetic walls, $e(\beta)$ and $m(\be)$, arise from the electric and magnetic components, $F_E$ and $F_M$, of a given $p$-form $F$. On a compact manifold ${X}$ the $p$-form fields which remain massless during the compactification correspond to solutions of the equations of motion of the form
\beq
F_E = f_E(t)\, \omega_p \wedge dt, \qquad
F_M = f_B(t)\, \omega_{p+1},
\eeq
where $\omega_p$ and $\omega_{p+1}$ are representatives of the de Rham cohomology classes $\mc{H}^p({X})$ and $\mc{H}^{p+1}({X})$, respectively. When ${X}$ is compact, solutions that yield electric and magnetic billiard walls can therefore only be found when the de Rham cohomology classes are nontrivial.

We may now state the influence of the topology of ${X}$ on the billiard structure simply in terms of a \emph{selection rule}. This rule makes use of the Betti numbers $b_j({X})$, which are the dimensions of the de Rham cohomology classes $\mc{H}^p({X})$. The selection rule reads as follows \cite{Wesley2005bd,Wesley2006cd}:

\begin{itemize}
\item {\bf Selection Rule}: \emph{When $b_s ({X})= 0$ for some $s$, we remove all billiard walls corresponding to electric $s$--forms, or  magnetic $(s-1)$--forms.}
\end{itemize}

\noindent  The selection rule is established using the same assumptions as the noncompact BKL analysis, namely that we are in a regime where classical gravity is valid, and studying a sufficiently ``generic" spacetime solution.

It has been known for some time that the algebraic structure of the billiard is invariant under Kaluza--Klein reduction on tori \cite{InvarianceUnderCompactification}. The selection rule is compatible with these results, since tori have no vanishing Betti numbers. Note also that none of the symmetry walls (or gravity walls) is eliminated by the compactification.  Hence, among the dominant walls of the compactified theory we always have the $(d-1)$ dominant symmetry walls $\beta^2 - \beta^1$, ..., $\beta^d - \beta^{d-1}$.

\section{Gram Matrices and Coxeter Groups}
In order to understand the structure of the reflection group that emerges when some $p$-form walls are switched off as well as the features of the corresponding billiard domain, it is useful to recall a few facts about polyhedra and reflection groups in hyperbolic space. The main reference for this section is \cite{Vinberg}.

\subsection{Convex Polyhedra}

We shall consider convex polyhedra $P$ of hyperbolic space, i.e., regions of the form  \beq P = \bigcap_{s = 1}^N H_s^+,\eeq where  $H_s^+$ is a half-space bounded by the hyperplane $H_s$, and $N$ is the number of such bounding hyperplanes.  In our case, $H_s$ is one of the walls of the relevant dominant wall system,
\beq
H_s=\{\be\in\mc{M}_{\be}\hs |\hs \om_s(\be)=0\},
\eeq
and $H_s^+$ is defined by
\beq
H_s^{+}=\{\be\in\mc{M}_{\be}\hs |\hs \om_s(\be)\geq 0\}.
\eeq
The polyhedra $P$ therefore contain the fundamental domain (\ref{originalbilliard}), defined by the dominant walls of the uncompactified theory. Hence it is clear that $P$ has non-vanishing volume.

\subsection{Relative Positions of Walls in Hyperbolic Space}
It is customary to associate with the convex polyhedron $P$ a matrix $G(P)$, the so-called \emph{Gram matrix}, which differs from the matrix $A$ by normalization. The construction proceeds as follows. To each hyperplane $H_s$ we associate a unit spacelike vector $e_s$ pointing inside $P$, i.e., pointing towards the billiard region (which is thus defined by $(\beta , e_s) \geq 0$).  We then construct the $N \times N$ matrix $G(P)$ of scalar products $(e_s | e_{s'})$. Four cases can occur for the scalar product $(e_s | e_{s'})$ between a given pair of distinct vectors $e_s$ and $e_{s'}$ \cite{Vinberg}\footnote{Note that one cannot have $e_s = - e_{s'}$ since then the region $H_s^+ \cap H_{s'}^+ = H_s$ has vanishing volume, which is excluded as we observed above.}: \begin{enumerate} \item $-1  \leq (e_s| e_{s'}) \leq 0. $ In this case, the hyperplanes $H_s$ and $H_{s'}$ intersect and form an acute angle. The limiting case $(e_s| e_{s'}) = -1$ means that the hyperplanes intersect at infinity, i.e., are parallel.  The other limiting case $(e_s| e_{s'}) = 0$ means that the hyperplanes form a right angle, which is both acute and obtuse.  \item $0  \leq (e_s| e_{s'}) \leq 1. $ In this case, the hyperplanes also intersect, but form an obtuse angle. The limiting case $(e_s| e_{s'}) = 1$ corresponds again to parallel hyperplanes meeting at infinity. \item $(e_s| e_{s'}) < -1$.  \item $(e_s| e_{s'}) >\phantom{-} 1$.   \end{enumerate} In the latter two cases, the hyperplanes diverge, i.e., do not meet even at infinity. The difference between these two cases is that while the condition $H_s^+ \cap H_{s'}^+$ defines a non-empty region when $(e_s| e_{s'}) \leq -1$, one has $H_s^+ \cap H_{s'}^+ = \emptyset$ whenever $(e_s | e_{s'}) \geq 1$.  The fourth case is thus excluded from our analysis, as is the limiting case $(e_s| e_{s'}) = 1$ ($s \not= s'$). We denote also the dihedral angle between the hyperplanes $H_s$ and $H_{s^{\prime}}$ by $H_s^{+}\cap H_{s^{\prime}}^{+}$.  When the hyperplanes intersect, the value of the dihedral angle $H_s^+ \cap H_{s'}^+$ can be found from the relation \beq \cos (H_s^{+} \cap H_{s'}^{+}) = - (e_s| e_{s'}). \eeq

\subsection{Acute-Angled Polyhedra}
If the number of dominant walls is strictly smaller than the dimension $M$ of $\mc{M}_{\be}$, the billiard table in $\mbb{H}_m$ has infinite volume and the motion is non-chaotic.  After a finite number of collisions, the billiard ball escapes to infinity \cite{Demaret,DH2}. We shall therefore assume that the number of dominant walls is greater than or equal to the dimension $M$ of $\mc{M}_{\be}$, a case that needs a more detailed analysis.  The Gram matrix is then of rank $M$ because among the $S$ dominant wall forms, one can find a subset of $M$ of them that defines a basis, namely the $(d-1)$ symmetry walls and one of the other dominant walls if there is no dilaton (or two linearly independent dominant non-symmetry walls if there is a dilaton etc.).  The convex polyhedron defined by the dominant walls is therefore non-degenerate (see \cite{Vinberg}). We shall also assume that the Gram matrix is indecomposable (cannot be written as a direct sum upon reordering of the $e_i$'s), as the decomposable case can be analysed in terms of the indecomposable one.

If the number of dominant walls is exactly equal to $M$, the billiard table is a simplex in hyperbolic space.  Otherwise, one has a non-simplex billiard table, with the number of faces exceeding dim $\mbb{H}_n+1=n+1$.

A crucial notion in the study of reflection groups is that of \emph{acute-angled polyhedra}.  A convex polyhedron is said to be {\em acute-angled} if for any pair of distinct hyperplanes defining it, either the hyperplanes do not intersect, or, if they do, the dihedral angle $H_s^+ \cap H_{s'}^+$ does not exceed $\frac{\pi}{2}$.  The Gram matrix, which has 1's on the diagonal, has then negative entries off the diagonal.  While non-degenerate, indecomposable acute-angled polyhedra on the sphere or on the plane are necessarily simplices, this is not the case on hyperbolic space.

\subsection{Coxeter Polyhedra and the Billiard Group}
\label{section:CoxeterPolyhedra}

We have seen that the dynamics of gravity is described in all cases (uncompactified or compactified) by the motion of a billiard ball in a region of hyperbolic space.  The reflections $s_s$ ($s=1, \dots, N$) with respect to the billiard walls generate a discrete reflection group which we want to characterize.  This group, which we shall call the \emph{billiard group} and denote by $\mf{B}$, is a subgroup of the Coxeter group relevant to the uncompactified case, where the total number of walls is maximum and the billiard region the smallest (and contained in all other billiard regions).  The billiard group is a crystallographic Coxeter group since it is generated by reflections \cite{Vinberg} and since it preserves the root lattice of the Kac-Moody algebra of the uncompactified case.  Its presentation in terms of the billiard walls might, however, be non-standard. The billiard group $\mf{B}$ will be examined more carefully in Sections \ref{section:BilliardGroup} and \ref{section:NonstandardPresentations}.

The billiard table has an important property which it inherits from the complete wall system of the theory. Consider the dihedral angle $H_i^+ \cap H_{j}^+$ between two different walls $H_i$ and $H_j$ that intersect.  If this angle is acute, then it is an integer submultiple of $\pi$, i.e., of the form $\pi/m_{ij}$ where $m_{ij}$ is an integer greater than or equal to $2$.  If this angle is obtuse, then $\pi - H_i^+ \cap H_{j}^+$ is an integer submultiple of $\pi$, i.e., $\pi - H_i^+ \cap H_{j}^+ = \pi/m_{ij}$, where $m_{ij}\in \mbb{Z}_{\geq 2}$. If the walls do not intersect, they are parallel and one has $m_{ij} = \infty$. Thus, given any pair of distinct walls, one can associate to it an integer $m_{ij} = m_{ji} \geq 2$.

In the case when all the angles are acute, and hence integer submultiples of $\pi$, the polyhedron is called a \emph{Coxeter polyhedron}.  Coxeter polyhedra are thus acute-angled.  In hyperbolic space, they may or may not be simplices.

\section{General Results}
\label{section:GeneralResults}

In this section we describe our new results concerning general features of the billiard structures after compactification.  In Section \ref{section:rules}, we use features of the wall system to explain why the billiard table need not be a Coxeter polyhedron after compactification. In Section \ref{section:FundamentalDomains} we describe the billiard region after compactification in terms of galleries, which we explain. Finally, we describe in Section \ref{section:ChaoticProperties} our methods for determining the chaotic properties for all possible compactifications.

\subsection{Rules of the Game}
\label{section:rules}

We show why the billiard region need not be a Coxeter polyhedron after compactification with the aid of two facts about billiard wall systems:

\begin{itemize}

\item \textbf{Fact 1}: \emph{If both an electric and a magnetic wall are present for
a given $p$--form, then the gravitational walls are subdominant.}  This was proven in \cite{DHRW02,InvarianceUnderCompactification}, by noticing that for any $p$ we have
\beq
G_{ijk} = e^{[p]}_{r_1 \cdots r_p} + m_{s_1\cdots s_{d-p-1}}^{[p]},
\eeq
where one of the $r_q$ and one of the $s_q$ are equal to $i$, and neither $j$ nor $k$ appears in either the $r_q$ or $s_q$.

\item \textbf{Fact 2}: \emph{The inner product between a gravitational wall and a $p$--form wall is unity for}
\begin{itemize}
\item \emph{electric walls with} $p \leq D-3$,
\item \emph{magnetic walls with} $p \geq 1$,
\end{itemize}
\emph{and the inner product vanishes for}
\begin{itemize}
\item \emph{electric walls with} $p=D-2$,
\item \emph{magnetic walls with} $p=0$.
\end{itemize}
Typically, $p$ forms with $p > \lfloor D/2 \rfloor - 1$ are dualised so that we may safely assume that $p \leq \lfloor D/2 \rfloor - 1$.  (This is a stronger condition than $p \leq D-3$ for $D \geq 4$).  Then, we have that the inner product between gravitational and $p$--form walls is always positive, except when the $p$--form is a scalar (axion) and the inner product vanishes.  This fact is proven by computing the relevant inner products (which are independent of the dilaton coupling(s) of the $p$--form) using the metric between wall forms.

This fact is significant in combination with the requirement that a system of dominant walls define an acute-angled polyhedron if the inner products between each pair of walls are either zero or negative. Therefore we can conclude:

\emph{If the dominant wall set includes a gravitational wall and any non-scalar $p$--form wall, then the dominant wall system does not define a Coxeter polyhedron.}

\end{itemize}
\noi For compactifications on tori, only {\bf Fact 1} is relevant since all components of the same $p$-forms are preserved in the compactification \cite{InvarianceUnderCompactification}. Hence, electric and magnetic walls are always present in pairs, ensuring that the gravity wall is always subdominant. More general compactifications can eliminate one of the electric or magnetic walls of a given $p$--form, while leaving the other intact.  Unlike the noncompact case, it is possible for both gravitational and $p$--form walls to appear simultaneously in the set of dominant walls. {\bf Fact 2} tells us that when this happens, we no longer have a Coxeter polyhedron.

In our analysis, we never eliminate gravitational walls (although there are some partial results regarding their selection rules that were given in \cite{Wesley2005bd}).  This means that after compactification there is always a gravitational wall in the root system, though it is not necessarily dominant.  Therefore all compactifications we study fall into one of three classes:

\begin{itemize}

\item A pair of electric and magnetic walls from the noncompact theory has not been eliminated by compactification, so the gravitational wall is subdominant.  The resulting billiard table may be a Coxeter polyhedron, depending on the details of the $p$-form menu and couplings in the theory.

\item One (or both) members of each pair of electric/magnetic walls are eliminated, so the gravitational wall is exposed.  If there are any other $p$-form walls left, because of {\bf Fact~2}, the
billiard table cannot be a Coxeter polyhedron. However, it may occur that ``coincidentally" two walls from different $p$-forms can combine and make the gravitational wall subdominant.

\item All of the $p$-form walls are eliminated by compactification, and so only the gravitational wall is left.  In this case one always obtains $A_n^{++}$, the billiard corresponding to pure gravity.  This billiard sits at the bottom of every list when all possible Betti numbers are set to zero.  Occasionally there are also direct summands corresponding to scalar fields that are never eliminated by compactification. 

\end{itemize}

For examples of these three cases we refer the reader to {\bf Paper V}. Only the first of the cases described above arises in the noncompact theory, where gravitational walls are never dominant, except when they are the only non-symmetry walls.

\subsection{Describing the Billiard Group After Compactification}
\label{section:BilliardGroupAfterCompactification}

We shall now begin to assemble the various structures described so far in order to analyze the group-theoretical properties of the billiard dynamics after compactification. This involves understanding the relation between the billiard table and the fundamental domain of the associated reflection group. In this context it is important to distinguish between the formal Coxeter group and the billiard group. We consider these concepts in turn in the following two sections, and explain how they are related in Section \ref{section:NonstandardPresentations}.

\subsubsection{The Formal Coxeter Group}
Recall from Section \ref{section:CoxeterPolyhedra} that the reflections $s_i$ with respect to the dominant walls $H_i$ generate a Coxeter group. To describe this group we must characterize the relations among the reflections $s_i$. Being reflections, they clearly satisfy \beq s_i^2 = 1 \label{Cox1}. \eeq  Consider next the reflections $s_i$ and $s_j$ with respect to two different hyperplanes $H_i$ and $H_j$. Then, the product $s_i s_j$ is a rotation by the angle $\frac{2 \pi}{m_{ij}}$ (where the integers $m_{ij}$ were introduced in Section \ref{section:CoxeterPolyhedra}) and so
\beq (s_i s_j)^{m_{ij}} = 1. \label{Cox2} \eeq

These relations alone define a Coxeter group, which we shall call the \emph{formal Coxeter group} $\mf{C}$ associated with the billiard. One can describe $\mf{C}$ more precisely as follows. Let $\tilde{\mf{C}}$ be the group freely generated by the elements of the set $\mathcal{S}=\{r_1, \dots, r_N\}$, and let $\mf{N}$ be the normal subgroup generated by $(r_i r_j)^{m_{ij}}$, where the \emph{Coxeter exponents} $m_{ij}$ satisfy \cite{Kac,Humphreys} (see also \cite{LivingReview})
\beqa
{}Êm_{ii}&=&1,
\nn \\
 {}Êm_{ij}&\in& \mathbb{Z}_{\geq 2}, \hs  i\neq j,
 \nn \\
{}Êm_{ij}&=& m_{ji}.
\eqa
Then $\mf{C}$ is defined as the quotient group $\tilde{\mf{C}}/\mf{N}$ and has the following standard presentation:
\beq
\mf{C} =  \left< r_1,\dots, r_n\hs |\hs (r_ir_j)^{m_{ij}}=1, \hs i,j=1,\dots, N \right>.
\eeq
One can associate a Coxeter graph with the formal Coxeter group, i.e., with the set of $m_{ij}$'s.  Each $r_i$ defines a node of the Coxeter graph, and two different nodes $i$ and $j$ are connected by a line whenever $m_{ij} >2$, with $m_{ij}$ explicitly written over the line whenever $m_{ij} >3$ \cite{Humphreys}. To a Coxeter graph, one can further associate a symmetric matrix defined by \beq B_{ij} = -\cos \Big(\f{\pi}{m_{ij}}\Big) \eeq with 1's on the diagonal and non-positive elements otherwise.

\subsubsection{The Billiard Group and its Fundamental Domain}
\label{section:BilliardGroup}
In Section \ref{section:CoxeterPolyhedra} we introduced the concept of the \emph{billiard group}Ê $\mf{B}$. Here we shall elaborate on this object and elucidate the structure of its fundamental domain.

The billiard group $\mf{B}$ is defined as the group generated by the reflections $s_i$ with respect to the dominant walls of the billiard table. This group coincides with the formal Coxeter group $\mf{C}$ (with $r_i \equiv s_i$) if and only if there are no additional relations among the $s_i$'s. Two cases must be considered.

\begin{enumerate}
\item {\bf Acute-angled billiard tables.}  There is no additional relations among the $s_i$'s if and only if the billiard table is acute-angled, i.e., is a Coxeter polyhedron \cite{Vinberg}.  In that case, the matrix $B=(B_{ij})$ coincides with the Gram matrix $G(P)$.  Furthermore, the billiard table is a fundamental domain \cite{Vinberg}.
    
    In the hyperbolic case relevant here, the billiard table need not be a simplex.  When it is a simplex, however, there is further structure.  The Gram matrix is non-degenerate and the action of the Coxeter group on the $\beta$-space $\mc{M}_{\be}$ coincides with the standard geometric realization considered in \cite{Humphreys}.  In addition, the matrix \beq
A_{ss^{\prime}}=\f{2(\om_{s}|\om_{s^{\prime}})}{(\om_{s}|\om_{s})},
\eeq has non-positive integers off the diagonal and hence is a non-degenerate Cartan matrix.  It is obvious that the off-diagonal entries are integers since the walls correspond to some roots of the Kac-Moody algebra of the uncompactified case. Moreover, it follows from the fact that the billiard table is acute-angled that they are negative. The billiard group is then the Weyl group of a simple Kac-Moody algebra.\footnote{In the case of a non-simplex, acute-angled billiard table, the matrix $A$ is also a valid Cartan matrix, but it is degenerate and the corresponding Kac-Moody algebra is not simple. Furthermore, the standard geometric realization is defined in a space of dimension larger than $M$, while the billiard realization is defined in the $M$-dimensional $\beta$-space.}  When the matrix $A$ is a Cartan matrix, one can associate to it a Dynkin diagram.

\item {\bf Non acute-angled billiard tables.} If the billiard table is not acute-angled (and hence not a Coxeter polyhedron), there are further relations among the $s_i$'s and the billiard group is therefore the quotient of the formal Coxeter group $\mf{C}$ by these additional relations.  Moreover, the billiard table is not a fundamental domain of the billiard group. One may understand this feature as follows.  Consider a dihedral angle $H_i^+ \cap H_{j}^+$ of the polyhedron that is obtuse.  The rotation $s_i s_j$ by the angle $2 \pi / m_{ij}$ is an element of the group and hence maps reflection hyperplanes to reflection hyperplanes. The image by this rotation of $H_j$ is the hyperplane $s_i H_j$ that intersects the interior of the billiard table.\footnote{When the Coxeter group is crystallographic, the converse is also true: if the angle between $H_i$ and $H_j$ is acute, then the image $s_iH_j$ does \emph{not} intersect the interior of the billiard table.} The reflection $s_i s_j s_i$ with respect to this hyperplane belongs to the group, and hence the billiard table cannot be a fundamental region of the billiard group since the orbit of a point sufficiently close to $s_i H_j$ intersects the billiard table at least twice.  Fundamental regions are obtained by considering all the mirrors (reflection hyperplanes) associated with the group (most of which are not billiard walls), which decompose the space into equivalent chambers that are permuted by the group (homogeneous decomposition). Each of these chambers is a fundamental domain. The billiard group $\mf{B}$ is generated by the reflections in the mirrors of the fundamental domain, which provide a standard presentation of the group, and the billiard table is a union of chambers \cite{Vinberg}. Examples of the occurrence of this phenomenon will be discussed below.

Although the billiard table is not a fundamental domain, it can be naturally described as a gallery defined by the Coxeter group of the uncompactified theory. This is described in Section \ref{section:BilliardGallery}.

\end{enumerate}

\subsubsection{Non-Standard Presentations of Coxeter Groups}
\label{section:NonstandardPresentations}

We have seen how one can associate a formal Coxeter group $\mf{C}$ to the billiard region using the Coxeter presentation $\mf{C}=\tilde{\mf{C}}/\mf{N}$. The billiard group $\mf{B}$ -- which is also a Coxeter group -- differs from $\mf{C}$ when the billiard table possesses obtuse angles because the reflections in the walls of the billiard then fulfill additional relations. This yields a non-standard presentation of the billiard group, which can be formally described as follows.

Let $\mc{B}_{\be}$ be the billiard region after compactification, let the elements of the set $\mc{S}=\{s_i\hs |\hs i=1, \dots, N\}$ be the reflections in the walls $W_i$ bounding $\mc{B}_{\be}$, and let $\tilde{\mf{C}}$ be the formal group freely generated by $\mc{S}$. The dihedral angles between the $W_i$ give rise to a set of Coxeter exponents $m_{ij}$, with associated normal subgroup $\mf{N}\subset \tilde{\mf{C}}$. Suppose now that the region $\mc{B}_{\be}$ is not a fundamental domain of $\mf{B}$, and denote by $\mf{J}$ the normal subgroup of $\tilde{\mf{C}}$ generated by $(s_is_j)^{m_{ij}}$ and any other non-standard relations between the elements of $\mc{S}$. Note that we have $\mf{N}\subset \mf{J}$. The billiard group $\mf{B}$ is then the quotient
\beq
\mf{B}=\tilde{\mf{C}} / \mf{J}.
\eeq
Equivalently, if we denote by $\mf{F}$ the normal subgroup of $\tilde{\mf{C}}$ generated only by the non-standard relations between the elements of $\mc{S}$, we may describe the billiard group as
\beq
\mf{B}=\mf{C}/\mf{F}.
\eeq
Neither of these presentations is a Coxeter presentation.

 In all cases we consider in this analysis, the uncompactified billiard is described by the Weyl group $\mc{W}[\mf{g}^{++}]$ of a Lorentzian Kac-Moody algebra $\mf{g}^{++}$. For general compactifications, the billiard group $\mf{B}$ is therefore a Coxeter subgroup of $\mc{W}[\mf{g}^{++}]$. However, in {\bf Paper V} examples were found of cases when the formal Coxeter group $\mf{C}$ after compactification differs from $\mf{g}^{++}$, while the billiard group $\mf{B}$ is actually isomorphic $\mf{g}^{++}$, with a non-standard presentation. We refer to {\bf Paper V} for more details.

\subsection{Fundamental Domains, Chamber Complexes and Galleries}
\label{section:FundamentalDomains}

We have seen that the billiard table need not be a Coxeter polyhedron upon compactification.  When it is not a Coxeter polyhedron, it no longer corresponds a fundamental domain of the billiard group $\mf{B}$. Moreover, $\mf{B}$ is the quotient by nontrivial extra relations of the formal Coxeter group associated with the billiard table. In this section, we describe how the billiard region relevant to the compactified case can then be built as a union of images of the uncompactified billiard region.  This is achieved using the theory of buildings, in terms of chambers and galleries.

\subsubsection{Chambers and Galleries}
\label{section:chambers}
The analysis in this section makes use of the treatment of Coxeter groups as a theory of \emph{buildings}, a formalism mainly developed by J. Tits. Introductions and references may be found in \cite{Garrett,Brown}.

The basic idea is to study Coxeter groups in geometric language by defining them in terms of the objects on which they act nicely. The \emph{buildings} are the fundamental objects which then defines the associated Coxeter group. For example, finite Coxeter groups act on so-called \emph{spherical buildings}, because these groups preserve the unit sphere. We are interested in Coxeter groups which act on \emph{hyperbolic buildings}, i.e., hyperbolic Coxeter groups, which preserve the hyperbolic space.

An $n$-\emph{simplex} $\mc{X}$ in hyperbolic space is determined by its $n+1$ vertices.  A $1$-simplex is determined by its two endpoints, a $2$-simplex (a triangle) is determined by its three vertices etc.  For this reason, it is convenient to identify an $n$-simplex $\mc{X}$ with the set of its $n+1$ vertices, $\mc{X}=\{\mbox{set of}\ n+1\ \mbox{vertices}\}$.  A \emph{face} $f$ of $\mc{X}$ is a simplex corresponding to a non-empty subset $f\subset \mc{X}$. The \emph{codimension} of $f$ with respect to $\mc{X}$ is given by $\text{dim}\ \mc{X}-\text{dim}\ f$. For example, there are three codimension-one faces in a 2-simplex, which are the three edges of the triangle.

\begin{figure}[ht]
\begin{center}
\includegraphics[width=100mm]{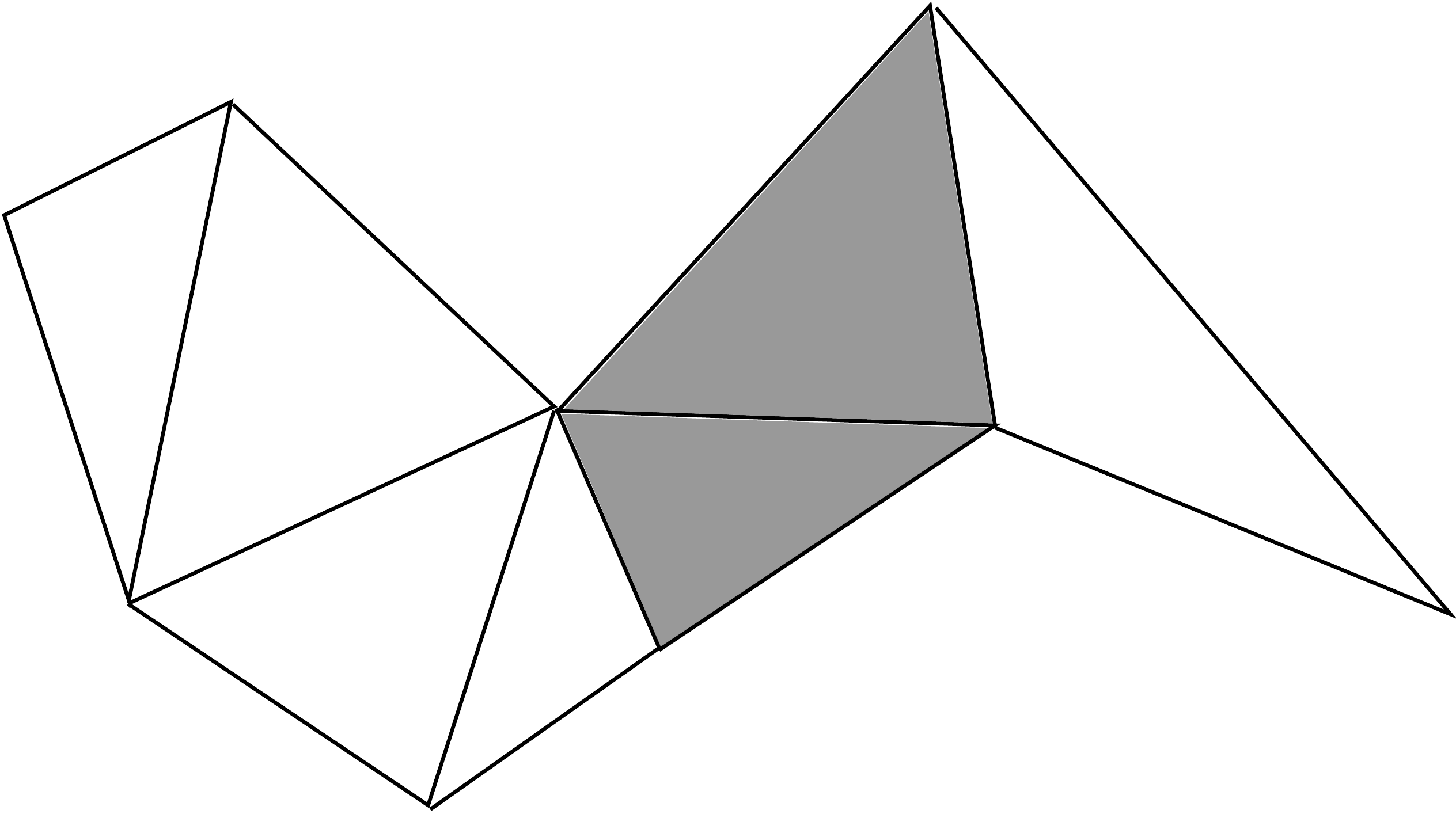}
\caption{A simplicial complex of $2$-simplices (triangles). The two shaded regions represent adjacent (maximal) simplices.}
\label{figure:Simplices}
\end{center}
\end{figure}

Next we define the notion of a simplicial complex. Let $V$ be a set of vertices, and $\mc{K}$ a set of finite subsets $\mc{X}_k\subset V$. We assume that the subsets containing a single vertex of $V$ are all elements of $\mc{K}$. Then $\mc{K}$ is called a \emph{simplicial complex} if it is such that given $\mc{X}\in \mc{K}$ and a face $f$ of $\mc{X}$, then $f\in \mc{K}$. The elements $\mc{X}_k$ of $\mc{K}$ are the simplices in the simplicial complex. Two simplices $\mc{X}_1$ and $\mc{X}_2$ of the same dimension in $\mc{K}$ are called \emph{adjacent} if they share a codimension-one face, i.e., if they are separated by a common wall. Figure \ref{figure:Simplices} illustrates a simplicial complex of $2$-simplices.

\begin{figure}[t]
\begin{center}
\includegraphics[width=140mm]{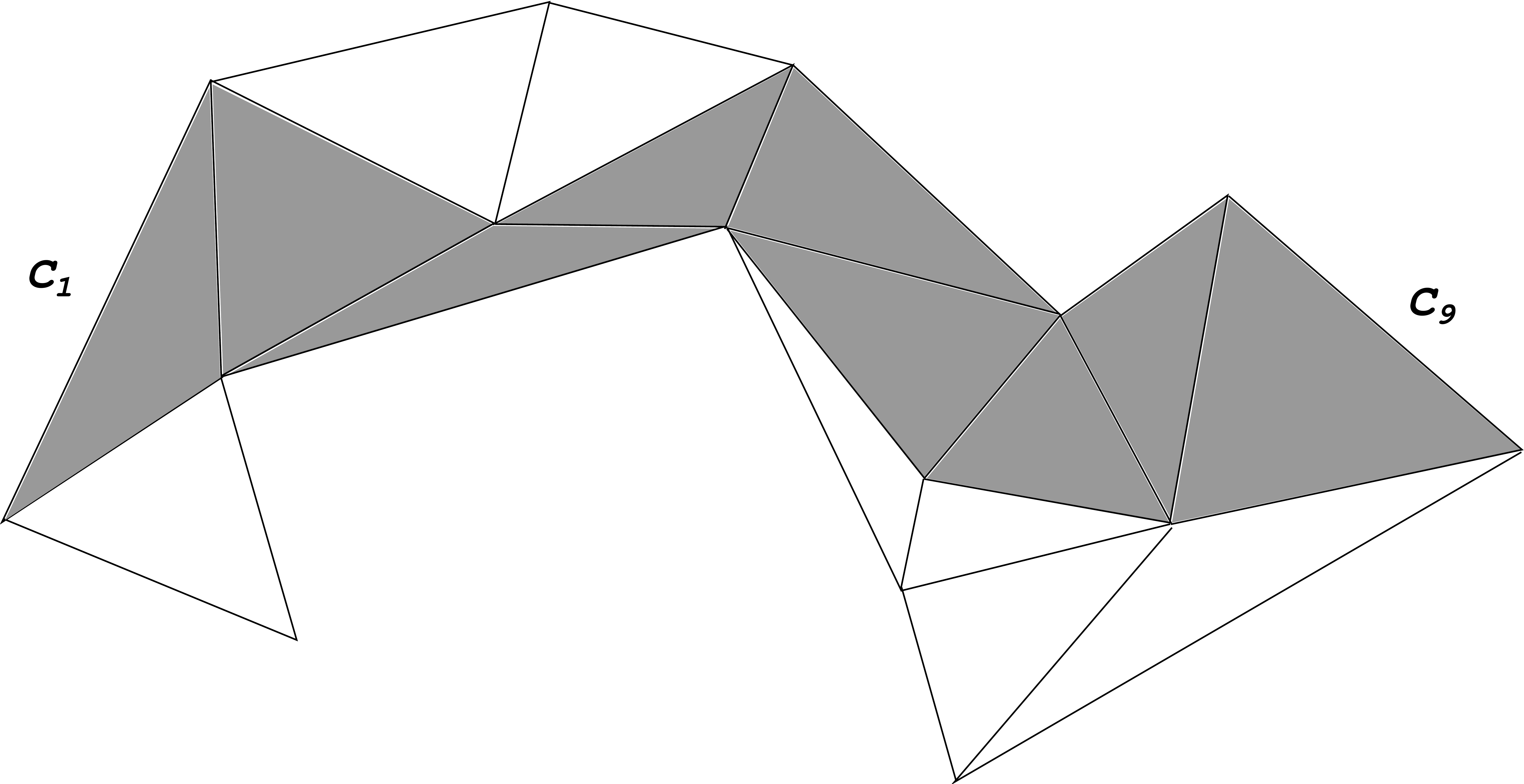}
\caption{A chamber complex with a gallery $\Gamma : \mc{C}_1,\dots, \mc{C}_9$, represented by the shaded region. The length of the gallery is $k=9$, which is also the distance between $\mc{C}_1$ and $\mc{C}_9$ since $\Gamma$ is the shortest gallery connecting $\mc{C}_1$ and $\mc{C}_9$.}
\label{figure:GalleryExample}
\end{center}
\end{figure}

A \emph{maximal simplex} $\mc{C}$ in $\mc{K}$ is such that $\mc{C}$ does not correspond to the face of another simplex in $\mc{K}$. Maximal simplices in a simplicial complex are called \emph{chambers} and they shall be our main objects of study. A sequence of chambers $ \mc{C}_1, \dots, \mc{C}_k$, such that any two consecutive chambers $\mc{C}_i$ and $\mc{C}_{i+1}$ are adjacent is called a \emph{gallery} $\Gamma$. Thus, a gallery corresponds to a connected path between two chambers $\mc{C}_1$ and $\mc{C}_k$ in $\mc{K}$, and we write
\beq
\Gamma \hs :\hs  \mc{C}_1, \mc{C}_2, \dots, \mc{C}_{k-1}, \mc{C}_k.
\eeq
The \emph{length} of $\Gamma$ is $k$, and the \emph{distance} between $\mc{C}_1$ and $\mc{C}_k$ is the length of the shortest gallery connecting them. If any two chambers in $\mc{K}$ are connected by a gallery, then the simplicial complex is called a \emph{chamber complex}. A simple example of a gallery inside a chamber complex is displayed in Figure \ref{figure:GalleryExample}.

\subsubsection{The Billiard Region as Gallery}
\label{section:BilliardGallery}

We describe the billiard regions of compactified gravity-dilaton-$p$-form theories in the language presented above. This is achieved by expressing them in terms of the billiard region of the uncompactified theory (or compactified on a torus) which we recall is the fundamental Weyl chamber $\mc{F}$ of the Weyl group $\mc{W}[\mf{g}^{++}]$ of the Kac-Moody algebra $\mf{g}^{++}$, whose Cartan matrix is defined through the scalar products between the dominant wall forms.

Compactification amounts to a process of removing dominant walls, and so the billiard table is enlarged.  The inequalities $\omega_{A^\prime} \geq 0$ associated with the simple roots of the underlying Kac-Moody algebra are indeed replaced by weaker inequalities.  The bigger region so defined can be written as a union of Weyl chambers of $\mc{W}[\mf{g}^{++}]$.  We shall illustrate this phenomenon on the example of the familiar Lie algebra $A_3$, whose Weyl group is a spherical Coxeter group. The fundamental Weyl chamber $\mc{F}$ is defined by \beq \om_1 (\be) \geq 0,\; \; \om_2 (\be) \geq 0 , \; \; \om_3 (\be) \geq 0, \eeq corresponding to the three simple roots.  The non simple roots are $\om_1 + \om_2$, $\om_2 + \om_3$ and $\om_1 + \om_2 + \om_3$.
Suppose that the single dominant wall $W_{1}$ defined by $\om_1(\be) = 0$ is suppressed. Effectively, this implies that the particle geodesic may cross the wall $W_{1}$. Thus it moves from the region where $\om_1(\be) \geq 0$ into the region where
\beq
\om_{1}(\be)\leq 0.
\eeq
We shall consider two cases. (i) The new billiard region is defined by \beq  \om_1 (\be) + \om_2 (\be) \geq 0,\; \; \om_2 (\be) \geq 0 , \; \; \om_3 (\be) \geq 0, \eeq i.e., the wall $W_1$ is replaced by the wall $\om_1 (\be) + \om_2 (\be) = 0$;  (ii) In the second case, we suppose that also the wall defined by $\om_1(\be)+\om_2(\be)=0$ is suppressed. Then the new billiard region is defined by the inequalities \beq \om_1 (\be) + \om_2 (\be) + \om_3(\be) \geq 0,\; \; \om_2 (\be) \geq 0 , \; \; \om_3 (\be) \geq 0 \eeq i.e., the wall $W_1$ is replaced by the wall $\om_1 (\be) + \om_2 (\be) + \om_3(\be) = 0$.

In the first case, one can write the new billiard region as the union $\mc{F} \cup \mc{A}_2$ where $\mc{A}_2$ is defined by  the inequalities \beq  \om_1 (\be) \leq 0,\; \; \om_1(\be) + \om_2 (\be) \geq 0 , \; \;  \om_3 (\be) \geq 0 \eeq (which imply $\om_2 (\be) \geq 0$). The region $\mc{A}_2$ is the Weyl chamber obtained by reflecting the fundamental Weyl chamber across the wall $W_1$ since
\beqa
{}s_1(\om_1)&=&-\om_1,
\nn \\
{}s_1(\om_2)&=&\om_1+\om_2,
\nn \\
{}s_1(\om_3)&=&\om_3.
\eqa
Hence,
\beq
s_1\cdot \mc{F}=\{\be \in\mf{h}\hs|\hs \om_1(\be)\leq 0, (\om_1+\om_2)(\be)\geq 0, \om_3(\be)\geq 0\} = \mc{A}_2.
\eeq
A reflection of this type which maps a chamber $\mc{C}$ to an adjacent chamber $\mc{C}^{\prime}$ is known as an \emph{adjacency transformation} \cite{Vinberg}.
By removing the single dominant $W_1$, we therefore get in the first case a new region which is precisely twice as large as the original fundamental region.

\begin{figure}[h]
\begin{center}
\includegraphics[width=70mm]{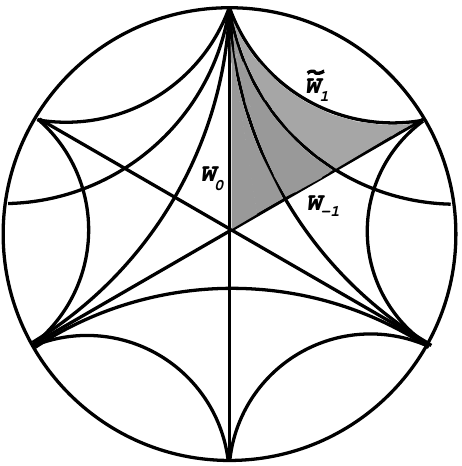}
\caption{Here we illustrate a gallery $\Gamma\hs :\hs \mc{C}_1, \mc{C}_2, \mc{C}_3$ of length 3 for the case of the hyperbolic Kac-Moody algebra $A_1^{++}$. The two walls $W_{0}$ and $W_{-1}$ are associated with the affine and overextended simple roots $\al_0$ and $\al_{-1}$, respectively. The original fundamental Weyl chamber $\mc{C}_1=\{h\in\mf{h}\ |\ \al_1(h)\geq 0, \hs \al_0(h)\geq 0, \hs \al_{-1}(h)\geq 0 \}$ corresponds to the leftmost shaded region. We have removed the wall $W_1=\{h\in\mf{h}\ |\ \al_1(h)=0\}$ as well as the wall $(\al_0+\al_1)(h)=0$. The far end of the billiard region is now bounded by the new wall $\tilde{W}_1=\{h\in\mf{h}\ |\ (2\al_0+3\al_1)(h)=0\}$. Each of the three chambers is clearly a copy of the fundamental chamber, and the total region is of finite volume. See, e.g., \cite{Feingold,LivingReview} for more detailed discussions of the Weyl group of $A_1^{++}$. }
\label{figure:A1++TilingGallery}
\end{center}
\end{figure}

In the second case, although we also remove a single dominant wall of the original billiard, we get a larger region.  This is because we also remove the subdominant wall $\om_1 (\be) + \om_2(\be) = 0$ which is exposed once $W_1$ is removed.  Indeed, the new billiard region can now be written as the union $\mc{F} \cup \mc{A}_2 \cup \mc{A}_3$ where $\mc{A}_3$ is defined by  the inequalities \beq \om_2(\be) \geq 0, \; \; \om_1(\be) + \om_2 (\be) \leq 0 , \; \;   \om_1(\be) + \om_2 (\be)+ \om_3 (\be) \geq 0 \eeq (which imply $\om_1 (\be) \leq 0$ and $\om_3 (\be) \geq 0$).  The region $\mc{A}_3$ is again a Weyl chamber, obtained from $\mc{A}_2$ by acting with the reflection $s$ with respect to $\om_1+\om_2$ since
\beqa
{}s(-\om_1)&=&\om_2
\nn \\
{}s(\om_1 + \om_2)&=&- (\om_1+\om_2),
\nn \\
{}s(\om_3)&=&\om_1 + \om_2 + \om_3.
\eqa
Hence,
\beq
s\cdot \mc{A}_2=\{\be \in\mf{h}\hs|\hs \om_2(\be)\geq 0, (\om_1+\om_2)(\be)\leq 0, \om_1(\be) + \om_2 (\be) + \om_3(\be)\geq 0\} = \mc{A}_3.
\eeq
Note that $\mc{A}_2$ and $\mc{A}_3$ are adjacent.  Thus, while in the second case we also remove a single dominant wall, we now obtain a region three times larger than the fundamental Weyl chamber. This new region can be described as a union of three pairwise adjacent Weyl chambers. In Figure \ref{figure:A1++TilingGallery} we describe pictorially a similar example for the case of the hyperbolic Coxeter group $A_1^{++}$.

By extrapolating this analysis to the general case where we remove an arbitrary number $r\leq n+1$ of dominant walls, we may conclude that the new billiard region corresponds to a union of images of the fundamental Weyl chamber. This naturally has the structure of a simplicial complex, and moreover, by a suitable ordering of the chambers, one sees that it corresponds to a gallery covering the whole complex.  In conclusion, we have found the following: \emph{the billiard region $\mc{B}$ obtained by compactification on a manifold of non-trivial topology is described by a gallery $\Gamma$ inside the Cartan subalgebra $\mf{h}$ of the original hyperbolic Kac-Moody algebra $\mf{g}$.}

The dynamics after compactification is chaotic if the new billiard region is a finite union of images of the fundamental chamber, i.e., if the gallery $\Gamma$ has finite length, while if this union is infinite the particle motion will eventually settle down in a single asymptotic Kasner solution, and chaos is removed.  Since the Coxeter reflections preserve the volume, the volume of $\mc{B}$ is
\beq
\text{vol}\ \mc{B}=k\cdot \text{vol}\ \mc{F},
\eeq
where $k$ is the length of the gallery $\Gamma$ associated with $\mc{B}$.

\subsection{Determining the Chaotic Properties After Compactification}
\label{section:ChaoticProperties}

The selection rules described in Section \ref{section:rules} provide a straightforward means to determine the billiard system after compactification.   Determining whether or not this billiard system is chaotic, i.e., computing the volume of the billiard table, is somewhat more involved because finding explicitly the corresponding galleries might be intricate.   In most cases it is possible to answer this question purely analytically without working out the gallery, although there are several different techniques that work for different types of billiard system. In this section we describe the various methods we employ.

\begin{figure}[ht]
\begin{center}
\includegraphics[width=150mm]{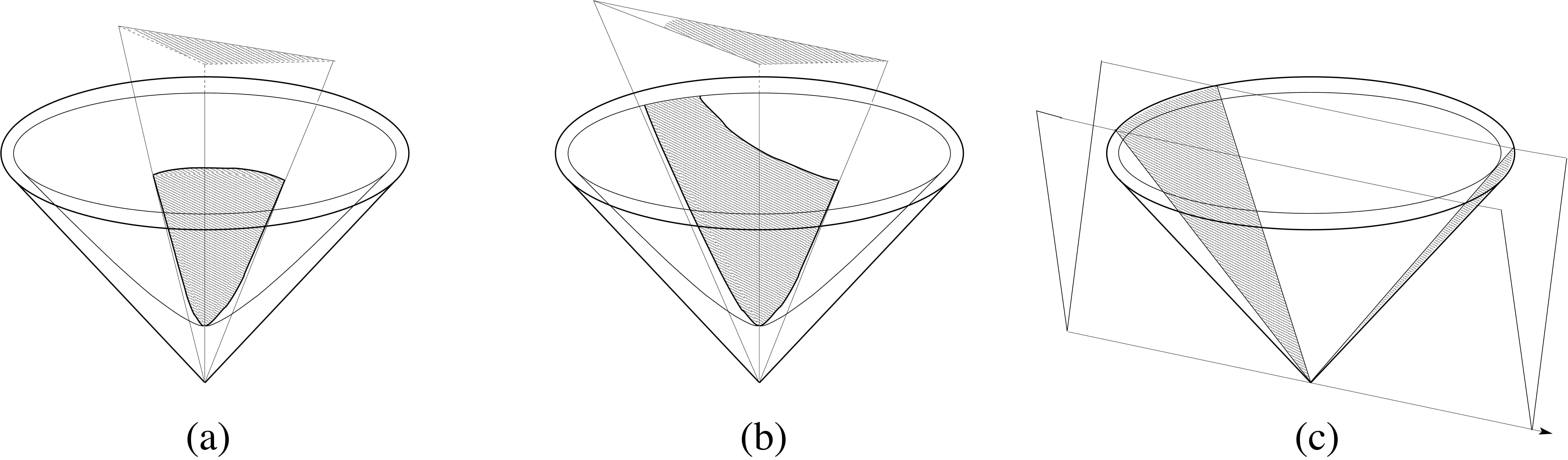}
\caption{Some examples of the wall systems and their chaotic properties: ({\bf a}) the wall system corresponding to a hyperbolic Kac-Moody algebra, ({\bf b}) the wall system of a non-hyperbolic Kac-Moody algebra, or one for which the coweight construction is possible, ({\bf c}) a wall system with fewer walls than the dimension of the $\beta$-space. }
\label{figure:WallCones}
\end{center}
\end{figure}

The simplest case is when the billiard table is a Coxeter simplex.  The matrix $\bar{A}_{ab}$ is then a Cartan matrix. The associated Kac-Moody algebra $\mf{g}(\bar{A})$ is by construction a regular subalgebra
%\footnote{A subalgebra $\bar{\mf{g}}\subset\mf{g}$ is \emph{regularly embedded} in $\mf{g}$ if and only if two conditions are
%fulfilled: (i) the root vectors of $\bar{\mf{g}}$ are root vectors of
%$\mf{g}$; and (ii) the simple roots of $\bar{\mf{g}}$ are real roots of
%$\mf{g}$. Moreover, the embedding is \emph{positive regular} if the positive root vectors of $\bar{\mf{g}}$ are positive root vectors of $\mf{g}$. See, e.g., \cite{Feingold,Henneaux:2006gp} for more detailed discussions on regular subalgebras of Kac-Moody algebras.}
of the Kac-Moody algebra $\mf{g}(A)$ (see Section \ref{regular} for the definition of a regular embedding), whose Weyl group controlled the uncompactified billiard. The dynamics of the compactified billiard is described by the Weyl group $\bar{\mc{W}}$ of $\mf{g}(\bar{A})$, and the billiard region $\bar{\mc{B}}$ coincides with the fundamental domain $\bar{\mc{F}}$ of $\bar{\mc{W}}$. Thus, if $\mf{g}(\bar{A})$ is hyperbolic, then $\bar{\mc{B}}$ is of finite volume, yielding chaotic dynamics. If the Kac-Moody algebra $\mf{g}(\bar{A})$ is Lorentzian but not hyperbolic, then the billiard is non-chaotic. This is illustrated in Figure \ref{figure:WallCones}.

In many cases the billiard table is a simplex, but some dihedral angles are obtuse (positive inner product between two different walls) and the matrix $A_{ab}$ is not a proper Cartan matrix. It is however non-degenerate, so that it is possible to define a set of dominant ``coweights" $\Lambda^{A^{\p}\mu}$ such that
\beq
\omega_{A^{\p}\mu}\Lambda^{B^{\p}\mu} = {\delta^{B^{\p}}}_{A^{\p}},
\eeq
where $\omega$ is a dominant wall labelled by $A^{\p}$.  As in the standard Kac-Moody algebra case, these ``coweights" span the space of rays that lie within the wall cone, provided we only combine ``coweights" using non-negative coefficients.  A non-chaotic solution to the equations of motion, which corresponds to a null ray within the wall cone, exists if and only if there is at least one timelike and one spacelike ``coweight".  This condition is readily checked once the ``coweights" are in hand.  This technique was employed in \cite{Wesley2006cd}.

When the billiard table is not a simplex and the number of walls is less than the dimension of $\mc{M}_{\be}$, then the theory is not chaotic, essentially because there are too few walls to prevent a ray from reaching infinity.

\begin{figure}[ht]
\begin{center}
\includegraphics[width=90mm]{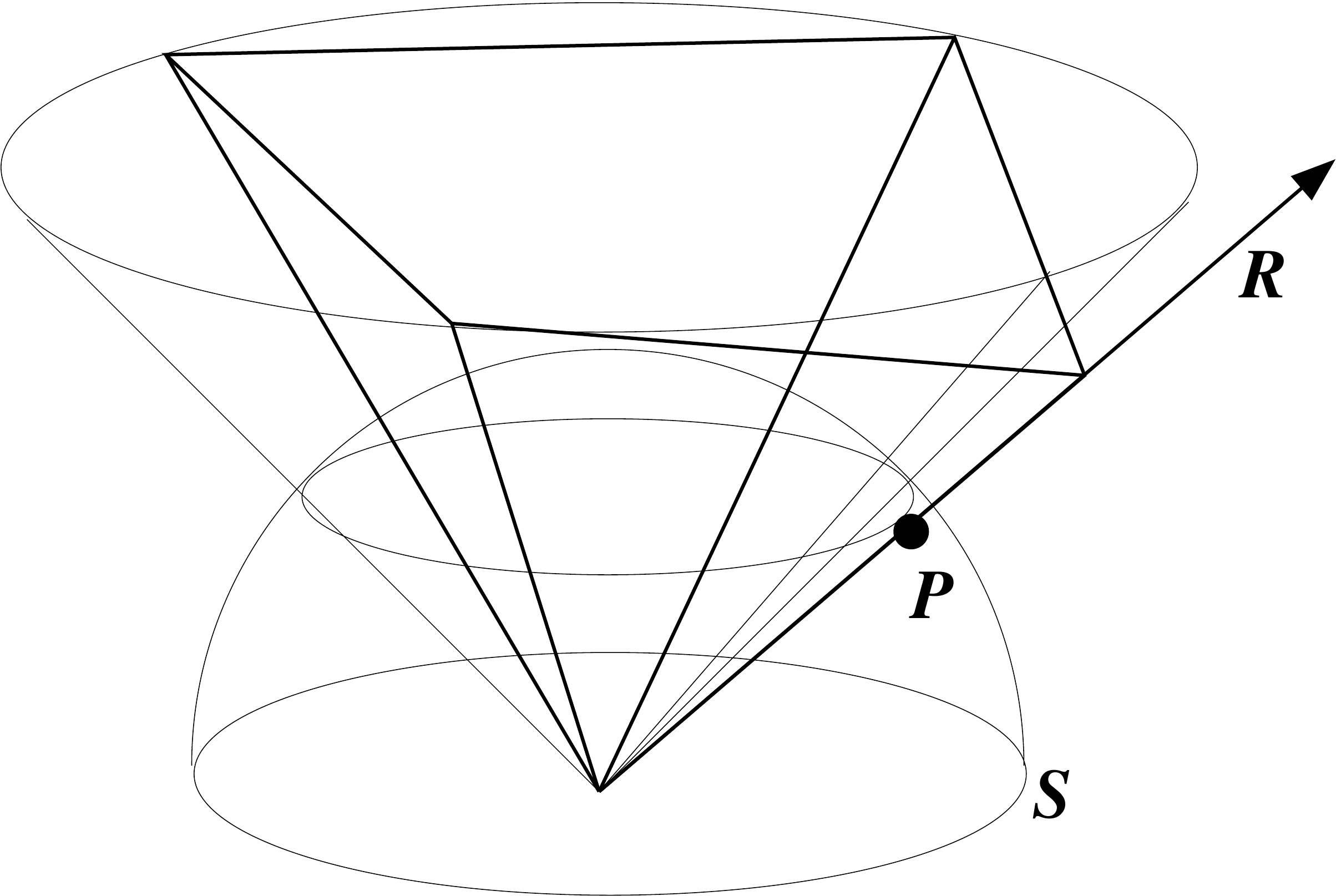}
\caption{The figure illustrates a non-simplex billiard table. To determine whether the theory is chaotic, it is sufficient to locate a spacelike ray. This is equivalent to maximising the Lorentzian norm on the unit sphere surrounding the origin.}
\label{figure:TooManyWalls}
\end{center}
\end{figure}

When the billiard table is not a simplex and the number of walls is greater than the dimension of $\mc{M}_{\be}$, the analysis becomes more complex.
This situation is illustrated in Figure \ref{figure:TooManyWalls}. One method to determine whether chaos is present is to successively remove dominant walls until the billiard region is again a simplex. If there is (at least) one way to do this such that the resulting structure is hyperbolic, then we can conclude that the full region is of finite volume, since reinserting the walls that were removed can never render the volume infinite. In a small number of cases, all wall removals lead to non-chaotic billiards and one cannot conclude immediately whether or not the volume of the billiard table is finite.  Another method is then to search numerically for whether a spacelike direction in the wall cone exists. We do this by maximising the Lorentzian norm of points on the unit sphere that lie within the wall cone.  If the maximal norm is negative, then no spacelike direction exists and the system is chaotic.

Also, as we have described in Section \ref{section:chambers} it is sometimes possible to compute the volume of the billiard region exactly by making use of certain properties of the Weyl group $\mc{W}[\mf{g}]$, associated with the uncompactified theory and construct the associated gallery.

As a byproduct of the ``coweight construction" mentioned above, we can easily prove the following useful fact:

\begin{itemize}
\item {\bf Fact 3:} \emph{Whenever the billiard is described by a direct product $\mf{B}_{\text{fin}}\times \mf{B}_{\text{hyp}}$ of a finite and a hyperbolic Coxeter group, then the dynamics is non-chaotic.}
\end{itemize}

\noi This follows by noting that the metric is a direct sum of the metric of $\mf{B}_{\textit{fin}}$ and that of $\mf{B}_{\textit{hyp}}$ so the coweights associated with each factor define orthogonal subspaces.  Since the coweights associated with the finite factor are spacelike, there will always exist at least one spacelike intersection in the region inside the dominant walls. A different intuitive way to see this is to recall that the volume of the fundamental Weyl chamber of the first factor is finite after projection on the sphere and that of the second factor after projection on the hyperboloid. Two projections are needed to have a finite volume but there is only one here (on a hyperboloid living in the product space).

%%%%%%%%%%%%%%%%%%%%%%%%%%%%%%%%%%%%%%%%%%%%%%%%%%%
%											                  
%\chapter{Beyond the Weyl Group -- Infinite Symmetries Made Manifest}
%
%%%%%%%%%%%%%%%%%%%%%%%%%%%%%%%%%%%%%%%%%%%%%%%%%%%

\chapter{Beyond the Weyl Group -- Infinite Symmetries Made Manifest}
\label{Chapter:Manifest}
%\section{Motivation: Toroidal Reduction and Enhanced Symmetries}
 
We have learned that there is an intimate connection between gravitational theories and hyperbolic Kac-Moody algebras, which is revealed when studying the theory close to a spacelike singularity. It is then natural to speculate whether this indicates the existence of a huge hidden hyperbolic symmetry of gravity. The aim of this chapter is to describe one line of research intended to put these speculations to the test. This approach is
directly inspired by the results obtained through toroidal dimensional
reduction of gravitational theories, where the scalar fields form
coset manifolds exhibiting explicitly larger and larger symmetries as
one goes down in dimensions. In the case of eleven-dimensional
supergravity, reduction on an $n$-torus $T^{n}$ reveals a chain of
exceptional U-duality symmetries
$\mc{E}_{n}$~\cite{Cremmer:1978ds, Cremmer:1979up}, culminating
with $\mc{E}_{8}$ in three dimensions~\cite{Marcus:1983hb}. This
has lead to the conjecture~\cite{Julia:1980gr} that the chain of
enhanced symmetries should in fact remain unbroken and give rise to
the infinite-dimensional duality groups $\mc{E}_{9},
\mc{E}_{10}$ and $\mc{E}_{11}$, as one reduces the theory to
two, one and zero dimensions, respectively.

The connection between the symmetry groups controlling the billiards
in the BKL-limit, and the symmetry groups appearing in toroidal
dimensional reduction to three dimensions has led to the attempt to reformulate eleven-dimensional
supergravity as a one-dimensional nonlinear sigma model with target space given by 
 the infinite-dimensional coset space
$\mc{K}(\mc{E}_{10})\bas \mc{E}_{10}$~\cite{DHN2}. This sigma
model describes the geodesic flow of a particle moving on
$\mc{K}(\mc{E}_{10})\bas \mc{E}_{10}$, whose dynamics can be seen
to match the dynamics of the associated (suitably truncated)
supergravity theory. 

We begin by describing some general aspects of nonlinear sigma models
for finite dimensional coset spaces. We then explain how to generalize
the construction to the infinite-dimensional case. We finally apply
the construction in detail to the case of eleven-dimensional
supergravity where the conjectured symmetry group is
$\mc{E}_{10}$.

This chapter is based on {\bf Paper III}, written in collaboration with Marc Henneaux and Philippe Spindel.

\section{Nonlinear Sigma Models on Finite-Dimensional Coset Spaces}
\label{section:nonlinearsigmamodels}
\index{coset space}

A nonlinear sigma model  \index{nonlinear sigma model|bb} describes
maps $\xi$ from one Riemannian
space $X$, equipped with a metric $\ga$, to another
Riemannian space, the ``target space'' $M$, with metric $g$.
Let $x^{m}\, (m=1, \cdots, p=\dim X)$ be coordinates on $X$ and
$\xi^{\al} \, (\al=1, \cdots, q=\dim M)$ be coordinates on $M$.
Then the standard action for this sigma model is
\begin{equation}
  S=\int_{X} d^{p}x \, \sqrt{\ga}\, \ga^{mn}(x)\,
  \pa_m \xi^{\al}(x)\, \pa_n \xi^{\be}(x)\, g_{\al\be}\left(\xi(x)\right). 
  \label{sigmamodel}
\end{equation}
Solutions to the equations of motion resulting from this action
will describe the maps $\xi^{\al}$ as functions of $x^{m}$.

A familiar example, of direct interest to the analysis below, is
the case where $X$ is one-dimensional, parametrized by the
coordinate $t$. Then the action for the sigma model reduces to
\begin{equation}
  S_{\mathrm{geodesic}}=\int dt \, A \,
  \f{d\xi^{\al}(t)}{dt}\f{d\xi^{\be}(t)}{dt}
  g_{\al\be}\left(\xi(t)\right),
\end{equation}
where $ A$ is $\gamma^{11}\sqrt{\ga}$ and ensures reparametrization
invariance in the variable $t$. Extremization with respect to $A$
enforces the constraint
\begin{equation} 
  \f{d\xi^{\al}(t)}{dt}\f{d\xi^{\be}(t)}{dt} g_{\al\be}\left(\xi(t)\right)= 0,
\end{equation}
ensuring that solutions to this model are null geodesics
on $M$. We have already encountered such a sigma model before,
namely as describing the free lightlike motion of the billiard
ball in the $(\dim M-1)$-dimensional scale-factor space. \index{scale-factor space} In that
case $A$ corresponds to the inverse ``lapse-function'' $N^{-1}$
and the metric $ g_{\al\be}$ is a constant Lorentzian metric.

%%%%%%%%%%%%%%%%%%%%%%%%%%%%%%%%%%%%%%%%%%%%%%%%%%%%%%%%%%%%%%%%%%%%%%%%%%%%%%%%%%%

\subsection{The Cartan Involution and Symmetric Spaces}
\label{section:Involutive}

In what follows, we shall be concerned with sigma models on
symmetric spaces $ K\bas G$ where $ G$ is a Lie
group with semi-simple real Lie algebra $\cG$ in its split real form and $ K=K( G)$
its maximal compact subgroup \index{maximal compact subgroup|bb} with
real Lie algebra $\cK$ (sometimes also denoted $K(\mf{g})$) corresponding to the maximal compact
subalgebra \index{maximal compact subalgebra|bb} of $\cG$. Since
elements of the coset are obtained by factoring out
$ K$, this subgroup is referred to as the ``local
gauge symmetry group'' (see below). Our aim is to provide an
algebraic construction of the metric on the coset space $K\bas G$ and explain how to obtain the
associated sigma model Lagrangian.

We recall from Section \ref{sec:generalities} that for any maximally split real Lie algebra $\mf{g}$, the Chevalley involution \index{Cartan involution}$\omega$ induces a Cartan decomposition of $\cG$
into even and odd eigenspaces:
\begin{equation}
  \mf{g}=\mf{k}\oplus\mf{p}
  \label{CartanDecomposition}
\end{equation}
(direct sum of vector spaces), where 
\begin{equation}
  \begin{array}{rcl}
    \mf{k}&=& \{ x\in\mf{g}\, |\, \omega(x)=x \},
    \\ [0.25 em]
    \mf{p}&=& \{ y\in \mf{g}\, |\, \omega(y)=-y\}
  \end{array}
\end{equation}
play central roles. The decomposition~(\ref{CartanDecomposition}) is
orthogonal, in the sense that
$\mf{p}$ is the orthogonal complement of $\mf{k}$ with respect to the
invariant bilinear form $(\cdot |\cdot ) \equiv B(\cdot ,\cdot )$,
\begin{equation}
  \mf{p}=\{y\in \mf{g}\, |\,\forall x\in \mf{k} : (y|x)=0 \}.
\end{equation}
The commutator relations split in a way characteristic for symmetric
spaces,
\begin{equation}
  [\mf{k}, \mf{k}]\subset\mf{k},
  \qquad
  [\mf{k}, \mf{p}]\subset\mf{p},
  \qquad
  [\mf{p}, \mf{p}]\subset \mf{k}.
\end{equation}
The subspace $\mf{p}$ is not a subalgebra. Elements of $\mf{p}$
transform in some representation of $\mf{k}$, which depends on the Lie
algebra $\mf{g}$. We stress that if the commutator $[\mf{p}, \mf{p}]$
had also contained elements in $\mf{p}$ itself, this would not have
been a symmetric space.

The left coset space $ K\bas G $  \index{coset space|bb} is defined as the set of
equivalence classes $[g]$ of $ G$ defined by the equivalence
relation 
\begin{equation}
  g\sim g^{\p}
  \qquad
  \mbox{iff } g {g^{\p}}^{-1}\in  K \mbox{ and } g, g^{\p}\in  G, 
\end{equation}
i.e.,
\begin{equation}
  [g]=\{k g \, | \, \forall k\in  K\}.
\end{equation}

%\subsection*{Example: The coset space   $SO(n)\bas SL(n,\mbb{R})$}

%As an example to illustrate the Cartan involution we
%consider the coset space  \index{coset space} $SO(n)\bas SL(n,\mbb{R})$. The group $SL(n,
%\mbb{R})$ contains all $n\times n$ real matrices with determinant
%equal to one. The associated Lie algebra $\mf{sl}(n,\mbb{R})$ thus
%consists of real $n\times n$ traceless matrices. In this case the
%Cartan involution is simply (minus) the ordinary matrix transpose
%$(\, )^{T}$ on the Lie algebra elements:
%%
%\begin{equation}
%  \tau : a
%  \quad \longmapsto \quad
%  -a^{T}\qquad a\in\mf{sl}(n,\mbb{R}).
%\end{equation}
%%
%This implies that all antisymmetric traceless $n\times n$ matrices
%belong to $\mf{k}=\mf{so}(n)$. The Cartan involution $\theta$ is the
%differential at the identity of an involution $\Theta$ defined on the
%group itself, such that for real Lie groups (real or complex matrix
%groups), $\theta$ is just the inverse conjugate transpose. Defining 
%%
%\begin{equation}
%    K =\{ g \in G \, |\, \Theta g=g \}
%\end{equation}
%%
%then gives in this example the group $  K =SO(n)$. The
%Cartan decomposition of $\mf{sl}(n, \mbb{R})$ thus splits all elements
%into symmetric and antisymmetric matrices, i.e., for $a\in \mf{sl}(n,
%\mbb{R})$ we have 
%%
%\begin{equation}
%  \begin{array}{rcl}
%    a-a^{T} &\in& \mf{so}(n),
%    \\ [0.25 em]
%    a+a^{T} &\in& \mf{p}.
%  \end{array}
%\end{equation}

%

%%%%%%%%%%%%%%%%%%%%%%%%%%%%%%%%%%%%%%%%%%%%%%%%%%%%%%%%%%%%%%%%%%%%%%%%%%%%%%%%%%%

\subsection{Nonlinear Realisations}
\index{nonlinear realisation|bb}

The group $ G$ naturally acts through right 
multiplication on the quotient space $  K\bas G $ as
\begin{equation}
  [h] \mapsto [h g].
  \label{rigidsymmetry}
\end{equation}
This definition 
makes sense because if $h \sim h'$, i.e., $h' = k h$ for some $k \in
  K $, then $h'g \sim hg$ since $h'g = (kh)g = k(hg)$
(left and right multiplications commute).

In order to describe a dynamical theory on the quotient space
$  K\bas G $, it is convenient to introduce as
dynamical variable the group element $\mathrm{V}(x)\in  G$ and
to construct the action for $\mathrm{V}(x)$ in such a way that the equivalence relation
\begin{equation}
  \forall k(x)\in  K : \mathrm{V}(x) \sim k(x) \mathrm{V}(x)
  \label{gaugesymmetryK}
\end{equation}
corresponds to a gauge symmetry. The physical (gauge invariant)
degrees of freedom are then parametrized by points of the coset
space. We also want to impose Equation~(\ref{rigidsymmetry}) as a rigid
symmetry. Thus, the action should be invariant under
\begin{equation}
  \mathrm{V} (x)
  \quad \longmapsto \quad
  k(x)\mathrm{V}(x) g,
  \qquad k(x)\in   K , \, g\in G. 
\label{LinearRealization}
\end{equation}

One may develop the formalism without fixing the
$  K $-gauge symmetry, or one may instead fix the gauge
symmetry by choosing a specific coset representative $\mathrm{V} (x)\in
 K\bas G$. When $ K$ is a maximal compact subgroup \index{maximal compact subgroup} of $ G$ there are
no topological obstructions, and a standard choice, which is always
available, is to take $\mathrm{V} (x)$ to be of upper triangular form as
allowed by the Iwasawa decomposition. \index{Iwasawa decomposition} This is usually called the
\emph{Borel gauge} and will be discussed in more detail later. In this
case an arbitrary global transformation,
\begin{equation}
  \mathrm{V} (x)
  \quad \longmapsto \quad
  \mathrm{V} (x)^{\p}=\mathrm{V} (x) g, \qquad g\in G,
\end{equation}
will destroy the gauge choice because $\mathrm{V} ^{\p}(x)$
will generically not be of upper triangular form. Then a
compensating local $ K$-transformation is needed that restores
the original gauge choice. The total transformation is thus
\begin{equation}
  \mathrm{V}(x)
  \quad \longmapsto \quad
  \mathrm{V}  (x)^{\p\p}=
  k\left(\mathrm{V}  (x), g \right) \mathrm{V} (x) g,
  \qquad
  k\left(\mathrm{V}(x), g \right)\in  K,\, g\in G,
\end{equation}
where $\mathrm{V}^{\pp}(x)$ is again in the upper triangular
gauge. Because now $k\left(\mathrm{V} (x), g \right)$ depends nonlinearly
on $\mathrm{V} (x)$, this is called a \emph{nonlinear realisation}
\index{nonlinear realisation} of $ G$.

%%%%%%%%%%%%%%%%%%%%%%%%%%%%%%%%%%%%%%%%%%%%%%%%%%%%%%%%%%%%%%%%%%%%%%%%%%%%%%%%%%%

\subsection{Three Ways of Writing the Quadratic  
  $ K\times  G$-Invariant Action}
  \label{Section:GInvariantAction}

Given the field $\mathrm{V}(x)$, we can form the Lie algebra valued
one-form (Maurer--Cartan form) \index{Maurer--Cartan form|bb} 
\begin{equation}
  d\mathrm{V} (x) \, \mathrm{V} (x)^{-1} =
  dx^{\mu} \, \pa_{\mu}\mathrm{V} (x)\mathrm{V} (x)^{-1}. 
\end{equation}
Under the Cartan decomposition, this element
splits according to Equation~(\ref{CartanDecomposition}),
\begin{equation}
  \pa_{\mu}\mathrm{V} (x) \, \mathrm{V} (x)^{-1} =
  \mathrm{Q}_{\mu}(x)+\mathrm{P}_{\mu} (x),
\end{equation}
where $\mathrm{Q}_{\mu}(x)\in \mf{k}$ and
$\mathrm{P}_{\mu}(x)\in\mf{p}$. We can use the Cartan involution $\theta$ to
write these explicitly as projections onto the odd and even
eigenspaces as follows: 
\begin{equation}
  \begin{array}{rcl}
    \mathrm{Q}_{\mu}(x)&=& \displaystyle
    \f{1}{2}\left[\pa_{\mu}\mathrm{V} (x)\mathrm{V} (x)^{-1}+
    \theta\left(\pa_{\mu}\mathrm{V} (x)\mathrm{V} (x)^{-1}\right)\right] \in \mf{k},
    \\ [0.75 em]
    \mathrm{P}_{\mu}(x)&=& \displaystyle
    \f{1}{2}\left[\pa_{\mu}\mathrm{V} (x)\mathrm{V} (x)^{-1}-
    \theta\left(\pa_{\mu}\mathrm{V} (x)\mathrm{V} (x)^{-1}\right)\right] \in \mf{p}. 
  \end{array}
  \label{CartanProjections}
\end{equation}
If we define a \emph{generalized transpose} $\mc{T}$ by
\begin{equation}
  (\, )^{\mc{T}}\equiv -\theta (\, ),
\end{equation}
then $\mathrm{P}_{\mu}(x)$ and $\mathrm{Q}_{\mu}(x)$ correspond to
symmetric and antisymmetric elements, respectively,
\begin{equation}
  \mathrm{P}_{\mu}(x)^{\mc{T}}=\mathrm{P} _{\mu}(x),
  \qquad
  \mathrm{Q}_{\mu}(x)^{\mc{T}}=-\mathrm{Q}_{\mu}(x).
\end{equation}
Of course, in the special case when $\mf{g}=\mf{sl}(n, \mbb{R})$ and
$\mf{k}=\mf{so}(n)$, the generalized transpose $(\, )^{\mc{T}}$
coincides with the ordinary matrix transpose $(\, )^{T}$. The Lie
algebra valued  one-forms with components $\pa_{\mu}\mathrm{V}
(x)\mathrm{V} (x)^{-1}$, $\mathrm{Q}_{\mu}(x)$ and $\mathrm{P} _{\mu}(x)$
are invariant under rigid right multiplication, $\mathrm{V} (x) \mapsto
\mathrm{V} (x) g$.

Being an element of the Lie algebra of the gauge group,
$\mathrm{Q}_{\mu}(x)$ can be interpreted as a gauge connection for
the local symmetry $ K$. Under a local transformation
$k(x)\in K$, $\mathrm{Q}_{\mu}(x)$ transforms as
\begin{equation}
   K : \mathrm{Q}_{\mu}(x)
  \quad \longmapsto \quad
  k(x)\mathrm{Q}_{\mu}(x)k(x)^{-1}+\pa_{\mu}k(x) k(x)^{-1},
\end{equation}
which indeed is the characteristic transformation property of a gauge
connection. On the other hand, $\mathrm{P}_{\mu}(x)$ transforms
covariantly,
\begin{equation}
   K : \mathrm{P}_{\mu}(x)
  \quad \longmapsto \quad
  k(x)\mathrm{P} _{\mu}(x)k(x)^{-1},
\end{equation}
because the element $\pa_{\mu}k(x) k(x)^{-1}$ is projected out due to
the negative sign in the construction of $\mathrm{P}_{\mu}(x)$ in
Equation~(\ref{CartanProjections}).

Using the bilinear form $(\cdot |\cdot)$ we can now form a manifestly
$  K \times  G $-invariant expression by
simply ``squaring'' $\mathrm{P} _{\mu}(x)$, i.e., the $p$-dimensional
action takes the form (cf.~Equation~(\ref{sigmamodel}))
\begin{equation}
  S_{ K\bas G}=\int_{X} d^{p}x \, \sqrt{\ga}\ga^{\mu\nu}
  \left(\mathrm{P}_{\mu}(x) \big|\mathrm{P}_{\nu}(x)\right).
  \label{Firstaction}
\end{equation}

We can rewrite this action in a number of ways. First, we note
that since $\mathrm{Q}_{\mu}(x)$ can be interpreted as a gauge
connection we can form a ``covariant derivative'' $\mathrm{D}_{\mu}$
in a standard way as
\begin{equation}
  \mathrm{D}_{\mu}\mathrm{V} (x)\equiv
  \pa_{\mu}\mathrm{V} (x)-\mathrm{Q}_{\mu}(x)\mathrm{V} (x),
\label{covariantderivate}
\end{equation}
which, by virtue of Equation~(\ref{CartanProjections}), can
alternatively be written as
\begin{equation}
  \mathrm{D}_{\mu}\mathrm{V} (x)=\mathrm{P}_{\mu}(x)\mathrm{V}(x).
\end{equation}
We see now that the action can indeed be interpreted as a gauged
nonlinear sigma model, in the sense that the local invariance is
obtained by minimally coupling the globally $ G$-invariant
expression $(\pa_{\mu}\mathrm{V} (x)\mathrm{V} (x)^{-1}| \pa^{\mu}\mathrm{V}
(x)\mathrm{V} (x)^{-1})$ to the gauge field $\mathrm{Q}_{\mu}(x)$ through
the ``covariantization'' $\pa_{\mu}\rightarrow \mathrm{D}_{\mu}$,
\begin{equation}
  \left(\pa_{\mu}\mathrm{V} (x)\mathrm{V} (x)^{-1}
  \big| \pa^{\mu}\mathrm{V} (x)\mathrm{V} (x)^{-1}\right)
  \quad \longrightarrow \quad
  \left(\mathrm{D}_{\mu}\mathrm{V} (x)\mathrm{V} (x)^{-1}
  \big| \mathrm{D}^{\mu}\mathrm{V} (x)\mathrm{V} (x)^{-1}\right) =
  \left(\mathrm{P}_{\mu}(x) \big|\mathrm{P}_{\nu}(x)\right).
\end{equation}
Thus, the action then takes the form
\begin{equation}
  S_{ K\bas G}=\int_{X} d^{p}x \, \sqrt{\ga}\ga^{\mu\nu}
  \left(\mathrm{D}_{\mu}\mathrm{V} (x)\mathrm{V} (x)^{-1}\big|
  \mathrm{D}_{\nu}\mathrm{V} (x)\mathrm{V} (x)^{-1}\right).
\end{equation}

We can also form a generalized ``metric'' $\mathrm{M}(x)$
\index{generalized metric|bb} that does not
transform at all under the local symmetry, but only transforms under
rigid $ G$-transformations. This is done, using the generalized
transpose (extended from the algebra to the group through the
exponential map~\cite{Helgason}), in the following way,
\begin{equation}
  \mathrm{M} (x)\equiv \mathrm{V} (x)^{\mc{T}}\mathrm{V} (x),
  \label{generalizedmetric}
\end{equation}
which is clearly invariant under local transformations
\begin{equation}
   K : \mathrm{M} (x)
  \quad \longmapsto \quad
  \left(k(x)\mathrm{V} (x)\right)^{\mc{T}}
  \left(k(x)\mathrm{V} (x)\right)=
  \mathrm{V} (x)^{\mc{T}}\left(k(x)^{\mc{T}}k(x)\right)\mathrm{V} (x)=
  \mathrm{M} (x)
\end{equation}
for $k(x)\in K$, and transforms as follows under global
transformations on $\mathrm{V} (x)$ from the right,
\begin{equation}
   G : \mathrm{M} (x)
  \quad \longmapsto \quad
  g^{\mc{T}}\mathrm{M} (x) g,
  \qquad
  g\in  G.
\end{equation}
A short calculation shows that the relation
between $\mathrm{M} (x)\in  G$ and $\mathrm{P} (x)\in \mf{p}$ is
given by 
\begin{eqnarray}
  \f{1}{2}\mathrm{M} (x)^{-1}\pa_{\mu}\mathrm{M} (x) &=&
  \f{1}{2}\left(\mathrm{V} (x)^{\mc{T}}\mathrm{V} (x)\right)^{-1}
  \pa_{\mu}\mathrm{V} (x)^{\mc{T}}\mathrm{V} (x)+
  \left(\mathrm{V} (x)^{\mc{T}}\mathrm{V} (x)\right)^{-1}
  \mathrm{V}(x)^{\mc{T}}\pa_{\mu}\mathrm{V} (x)
  \nonumber
  \\
  &=& \f{1}{2}\mathrm{V} (x)^{-1}\left[\left(\pa_{\mu}\mathrm{V}(x)
  \mathrm{V} (x)^{-1}\right)^{\mc{T}}+\pa_{\mu}\mathrm{V} (x)
  \mathrm{V}(x)^{-1}\right]\mathrm{V} (x)
  \nonumber
  \\
  &=& \mathrm{V} (x)^{-1}\mathrm{P}_{\mu}(x)\mathrm{V} (x).
  \label{MPrelation} 
\end{eqnarray}
Since the factors of $\mathrm{V} (x)$ drop out in the squared
expression,
\begin{equation}
  \left( \mathrm{V} (x)^{-1}\mathrm{P} _{\mu}(x)\mathrm{V} (x)\big|
  \mathrm{V} (x)^{-1}\mathrm{P}^{\mu}(x)\mathrm{V} (x)\right)=
  \left(\mathrm{P}_{\mu}(x)\big|\mathrm{P} ^{\mu}(x)\right),
\end{equation}
Equation~(\ref{MPrelation}) provides a third way to write the
$  K \times  G $-invariant action,
completely in terms of the generalized metric $\mathrm{M} (x)$,
\begin{equation}
  S_{ K\bas G}=\f{1}{4}\int_{X} d^{p}x \,
  \sqrt{\ga}\ga^{\mu\nu}\left(\mathrm{M} (x)^{-1}
  \pa_{\mu}\mathrm{M} (x) \big|\mathrm{M} (x)^{-1}
  \pa_{\nu}\mathrm{M} (x)\right). 
  \label{Maction}
\end{equation}
We call $\mathrm{M}$ a ``generalized metric'' because in the
$SO(n)\bas GL(n, \mbb{R})$-case, it does correspond to the metric, the field
$\mathrm{V}$ being the ``vielbein''; see Section~\ref{GL10/SL10Sigma}.

All three forms of the action are manifestly gauge invariant under
$  K $. If desired, one can fix the gauge, and 
thereby eliminating the redundant degrees of freedom.

%%%%%%%%%%%%%%%%%%%%%%%%%%%%%%%%%%%%%%%%%%%%%%%%%%%%%%%%%%%%%%%%%%%%%%%%%%%%%%%%%%%

\subsection{Equations of Motion and Conserved Currents}

Let us now take a closer look at the equations of motion resulting
from an arbitrary variation $\delta \mathrm{V} (x)$ of the action
in Equation~(\ref{Firstaction}). The Lie algebra element $\delta \mathrm{V}
(x)\mathrm{V} (x)^{-1}\in \mf{g}$ can be decomposed according to the
Cartan decomposition,
\begin{equation}
  \delta \mathrm{V} (x)\mathrm{V} (x)^{-1}=\Sigma(x) +\Lambda(x),
  \qquad
  \Sigma(x)\in \mf{k}, \, \Lambda(x)\in \mf{p}.
\end{equation}
The variation $\Sigma(x)$ along the gauge orbit will be automatically
projected out by gauge invariance of the action. Thus we can set
$\Sigma(x)=0$ for simplicity. Let us then compute $\de \mathrm{P}
_{\mu}(x)$. One easily gets
\begin{equation}
  \de \mathrm{P} _{\mu}(x)= \pa_{\mu}\Lambda(x)-[\mathrm{Q}_{\mu}(x), \Lambda(x)].
\end{equation}
Since $\Lambda(x)$ is a Lie algebra valued scalar we can freely set
$\pa_{\mu}\Lambda(x)\rightarrow \nabla_{\mu}\Lambda(x)$ in the
variation of the action below, where $\nabla^{\mu}$ is a covariant
derivative on $X$ compatible with the Levi--Civita connection. Using
the symmetry and the invariance of the bilinear form one then finds 
\begin{equation}
  \delta S_{ K\bas G}=
  \int_{X}d^{p}x \, \sqrt{\ga}\ga^{\mu\nu}2
  \left[\left(-\nabla_{\nu}\mathrm{P}_{\mu}(x)+
  [\mathrm{Q}_{\nu}(x), \mathrm{P}_{\mu}(x)]\big|\Lambda(x)\right)\right].
\end{equation}
The equations of motion are therefore equivalent to
\begin{equation}
  \mathrm{D}^{\mu}\mathrm{P}_{\mu}(x)=0,
\end{equation}
with
\begin{equation}
 \mathrm{D}_{\mu}\mathrm{P}_{\nu}(x)=
 \nabla_{\mu}\mathrm{P}_{\nu}(x)-[\mathrm{Q}_{\mu}(x), \mathrm{P}_{\nu}(x)],
\end{equation}
and simply express the covariant conservation of $\mathrm{P} _{\mu}(x)$.

It is also interesting to examine the dynamics in terms of the
generalized metric $\mathrm{M}(x)$. The equations of motion for
$\mathrm{M}(x)$ are
\begin{equation}
  \f{1}{2}\nabla^{\mu}\left(\mathrm{M}(x)^{-1}\pa_{\mu}\mathrm{M} (x)\right)=0.
\end{equation}
These equations ensure the conservation of the current
\begin{equation}
  \mc{J}_{\mu}\equiv \f{1}{2} \mathrm{M} (x)^{-1}
  \pa_{\mu}\mathrm{M} (x)=
  \mathrm{V}(x)^{-1}\mathrm{P}_{\mu}(x)\mathrm{V}(x),
  \label{ConservedCurrent}
\end{equation}
i.e.,
\begin{equation}
  \nabla^{\mu}\mc{J}_{\mu}=0.
\end{equation}
This is the conserved Noether current associated with the rigid
$ G$-invariance of the action.

\subsection[Parametrization of $ K\bas G$]%
              {Parametrization of   $ K\bas G$}
\label{Section:ParametrizationG/K}

The Borel gauge choice is always accessible when the group $ K$ is
the maximal compact subgroup of $ G$. In the noncompact case this is
no longer true since one cannot invoke the Iwasawa decomposition (see,
e.g.~\cite{Keurentjes:2005jw} for a discussion of the subtleties
involved when $ K$ is noncompact). This point will, however, not be
of concern to us in this thesis. We shall now proceed to write down the
sigma model action in the Borel gauge for the coset space $ K\bas G$,
with $ K$ being the maximal compact subgroup. Let
$\Pi=\{\al_1^{\vee}, \cdots, \al_n^{\vee}\}$ be a basis of the Cartan
subalgebra $\mf{h}\subset\mf{g}$, and let $\Phi_+\subset
\mf{h}^{\star}$ denote the set of positive roots. The Borel subalgebra 
\index{Borel subalgebra} of $\mf{g}$ can then be written as
\begin{equation}
  \mf{b}=\sum_{i=1}^{n}\mbb{R}\al_i^{\vee}+\!\!
  \sum_{\al\in\Phi_+} \!\! \mbb{R}E_{\al},
\end{equation}
where $E_{\al}$ is the positive root generator spanning the
one-dimensional root space $\mf{g}_{\al}$ associated to the root
$\al$. The coset representative is then chosen to be
\begin{equation}
  \mathrm{V} (x)=\mathrm{V}_1(x)\mathrm{V}_2(x)=
  \Exp \left[\sum_{i=1}^{n}\phi_{i}(x)\al_{i}^{\vee}\right]\,
  \Exp \left[\sum_{\al\in\Phi_+}\chi_{\al}(x)E_{\al}\right] \in  K\bas G. 
  \label{GeneralCosetRepresentative}
\end{equation}
Because $\mf{g}$ is a finite Lie algebra, the sum over positive roots
is finite and so this is a well-defined construction.

From Equation~(\ref{GeneralCosetRepresentative}) we may compute the Lie
algebra valued one-form $\mfVV$ explicitly. Let us do this in some
detail. First, we write the general expression in terms of
$\tV_1(x)$ and $\tV_2(x)$,
\begin{equation}
  \mfVV=\pa_{\mu}\tV_1(x) \tV_1(x)^{-1}+
  \tV_1(x)\left(\pa_{\mu}\tV_2(x) \tV_2(x)^{-1}\right)\tV_1(x)^{-1}. 
  \label{Oneformsplit}
\end{equation}
To compute the individual terms in this expression we need to make use
of the Baker--Hausdorff formulas: 
\begin{equation}
  \begin{array}{rcl}
    \pa_{\mu}e^{A}e^{-A}&=& \displaystyle
    \pa_{\mu}A+\f{1}{2!}[A, \pa_{\mu}A]+\f{1}{3!}[A, [A, \pa_{\mu}A]]+\cdots,
    \\ [0.5 em]
    e^{A}B e^{-A}&=& \displaystyle
    B+[A, B]+\f{1}{2!}[A, [A, B]]+\cdots.
  \end{array}
  \label{BakerHausdorff} 
\end{equation}
The first term in Equation~(\ref{Oneformsplit}) is easy to compute
since all generators in the exponential commute. We find
\begin{equation}
  \pa_{\mu}\tV_1(x) \tV_1(x)^{-1}=
  \sum_{i=1}^{n}\pa_{\mu}\phi_i(x)\al_{i}^{\vee} \in \mf{h}.
\end{equation}
Secondly, we compute the corresponding expression for $\tV_2(x)$. Here
we must take into account all commutators between the positive root
generators $E_{\al}\in\mf{n}_+$. Using the first of the
Baker--Hausdorff formulas above, the first terms in the series become
\begin{eqnarray}
  \pa_{\mu}\tV_2(x) \tV_2(x)^{-1}&=&
  \pa_{\mu}\Exp \left[\sum_{\al\in\Phi_+}\chi_{\al}(x)E_{\al}\right]\,
  \Exp \left[-\!\!\!\sum_{\al^{\p}\in\Phi_+}\!\!\!
  \chi_{\al^{\p}}(x)E_{\al^{\p}}\right]
  \nonumber
  \\
  &=& \!\! \sum_{\al\in\Phi_+} \!\! \pa_{\mu}\chi_{\al}(x)E_{\al}+
  \f{1}{2!} \!\! \sum_{\al, \al^{\p}\in\Phi_+} \!\!\!\!
  \chi_{\al}(x)\pa_{\mu}\chi_{\al^{\p}}(x)[E_{\al}, E_{\al^{\p}}]
  \nonumber
  \\
  & & + \f{1}{3!} \!\! \sum_{\al, \al^{\p}, \al^{\pp}\in \Phi_+}
  \!\!\!\! \chi_{\al}(x) \chi_{\al^{\p}}(x)\pa_{\mu}
  \chi_{\al^{\p\p}}(x)[E_{\al}, [E_{\al^{\p}}, E_{\al^{\p\p}}]]+\cdots.
  \nn \\
  \label{longsum}
\end{eqnarray}
Each multi-commutator $[E_{\al}, [E_{\al^{\p}}, \cdots]\cdots,
  E_{\al^{\p\p\p}}]$ corresponds to some new positive root generator,
say $E_{\ga}\in \mf{n}_+$. However, since each term in the
expansion~(\ref{longsum}) is a sum over all positive roots, the
specific generator $E_{\ga}$ will get a contribution from all
terms. We can therefore write the sum in ``closed form'' with the
coefficient in front of an arbitrary generator $E_{\ga}$ taking the
form
\begin{equation}
  \mc{R}_{\ga, \mu}(x)\equiv
  \pa_{\mu}\chi_{\ga}(x)+\f{1}{2!}
  \underbrace{\chi_{\zeta}(x)\pa_{\mu}\chi_{\zeta^{\p}}(x)}_{\zeta+\zeta^{\p}=\ga}+
  \cdots+\f{1}{k_{\ga}!}
  \underbrace{\chi_{\eta}(x)\chi_{\eta^{\p}}(x)\cdots
  \chi_{\eta^{\p\p}}(x)\pa_{\mu}\chi_{\eta^{\p\p\p}}(x)}_{\eta+\eta{\p}+
  \cdots+\eta^{\pp}+\eta^{\pp\p}=\ga},
\end{equation}
where $k_{\ga}$ denotes the number corresponding to the last term in
the Baker--Hausdorff expansion in which the generator $E_{\ga}$
appears. The explicit form of $\mc{R}_{\ga, \mu}(x)$ must be computed
individually for each root $\ga\in\Phi_+$.

The sum in Equation~(\ref{longsum}) can now be written as
\begin{equation}
  \pa_{\mu}\tV_2(x) \tV_2(x)^{-1}=
  \!\! \sum_{\al\in\Phi_+}\!\! \mc{R}_{\al,\mu}(x)E_{\al}. 
  \label{sumclosedform}
\end{equation}
To proceed, we must conjugate this expression with $\tV_1(x)$ in order
to compute the full form of Equation~(\ref{Oneformsplit}). This
involves the use of the second Baker--Hausdorff formula in
Equation~(\ref{BakerHausdorff}) for each term in the sum,
Equation~(\ref{sumclosedform}). Let $h$ denote an arbitrary element of
the Cartan subalgebra,
\begin{equation}
  h=\sum_{i=1}^{n}\phi_{i}(x)\al_{i}^{\vee} \in \mf{h}.
\end{equation}
Then the commutators we need are of the form
\begin{equation}
  [h, E_{\al}]=\al(h)E_{\al},
\end{equation}
where $\al(h)$ denotes the value of the root $\al\in\mf{h}^{\star}$
acting on the Cartan element $h\in\mf{h}$,
\begin{equation}
  \al(h)=\sum_{i=1}^{n}\phi_i(x)\al(\al_i^{\vee})=
  \sum_{i=1}^{n}\phi_i(x)\left<\al,\al_i^{\vee}\right>\equiv
  \sum_{i=1}^{n}\phi_i(x)\al_i.
\end{equation}
So, for each term in the sum in Equation~(\ref{sumclosedform}) we
obtain 
\begin{eqnarray}
  \tV_1(x)E_{\al}\tV_1(x)^{-1}&=&
  E_{\al}+\sum_{i}\phi_i(x)\al_i E_{\al} +
  \f{1}{2}\sum_{i,j} \phi_i(x)\phi_j(x)\al_i\al_j E_{\al}+\cdots
  \nonumber
  \\
  &=& \Exp \left[\sum_i \phi_i(x)\al_i\right] E_{\al}
  \nonumber
  \\
  &=& {e}^{\al(h)} E_{\al}. 
\end{eqnarray}
We can now write down the complete expression for the element $\mfVV$,
\begin{equation}
  \mfVV=\sum_{i=1}^{n}\pa_{\mu}\phi_i(x)\al_{i}^{\vee}+
  \!\! \sum_{\al\in\Phi_+} \!\! {e}^{\al(h)} \mc{R}_{\al, \mu}(x)E_{\al}.
\end{equation}
Projection onto the coset $\mf{p}$ gives (see
Equation~(\ref{CartanProjections}))
\begin{equation}
  \mathrm{P}_{\mu}(x)=
  \sum_{i=1}^{n}\pa_{\mu}\phi_i(x)\al_i^{\vee}+
  \f{1}{2}\!\sum_{\al\in\Phi_+} \!\! e^{\al(h)}\mc{R}_{\al, \mu}(x)
  \left(E_{\al}+E_{-\al}\right),
\end{equation}
where we have used that $E_{\al}^{\mc{T}}=E_{-\al}$ and
$(\al_i^{\vee})^{\mc{T}}=\al_i^{\vee}$.

Next we want to compute the explicit form of the action in
Equation~(\ref{Firstaction}). Choosing the following normalization for the
root generators,
\begin{equation}
  (E_{\al}|E_{\al^{\p}})=
  \delta_{\al,-\al^{\p}},
  \qquad
  (\al_i^{\vee}|\al_j^{\vee})= \delta_{ij},
  \label{NormalizationRootGenerators}
\end{equation}
which implies
\begin{equation}
  (E_{\al}| E_{\al^{\p}}^{\mc{T}})=
  (E_{\al}|E_{-\al^{\p}})=\delta_{\al,\al^{\p}}
\end{equation}
one finds the form of the $  K  \times
  G $-invariant action in the parametrization of
Equation~(\ref{GeneralCosetRepresentative}),
\begin{equation}
  S_{ K\bas G}=\int_{X} d^{p}x \, \sqrt{\ga}\ga^{\mu\nu}
  \left[\sum_{i=1}^{n}\pa_{\mu}\phi_i(x)\pa_{\nu}\phi_i(x)+
  \f{1}{2} \!\sum_{\al\in\Delta_{+}} \!\!
  e^{2\al(h)}\mc{R}_{\al, \mu}(x)\mc{R}_{\al, \nu}(x)\right]. 
  \label{FiniteSigmaModel}
\end{equation}

\section{Geodesic Sigma Models on Infinite-Dimensional Coset Spaces}
\label{section:InfiniteSigmaModels}

In the following we shall both ``generalize and specialize'' the
construction from Section~\ref{section:nonlinearsigmamodels}. The
generalization amounts to relaxing the restriction that the algebra
$\mf{g}$ be finite-dimensional. Although in principle we could
consider $\mf{g}$ to be any indefinite Kac--Moody algebra, we shall be
focusing on the case where it is of Lorentzian type. The analysis
will also be a specialization, in the sense that we consider only
\emph{geodesic} sigma models, meaning that the Riemannian space
$X$ is the one-dimensional worldline of a particle, parametrized
by one variable $t$. This restriction is of course motivated by
the billiard description of gravity close to a spacelike
singularity, where the dynamics at each spatial point is
effectively described by a particle geodesic in the fundamental
Weyl chamber of a Lorentzian Kac--Moody algebra.

The motivation is that the construction of a geodesic sigma model
that exhibits this Kac--Moody symmetry in a manifest way, would
provide a link to understanding the role of the full algebra
$\mf{g}$ beyond the BKL-limit.

%%%%%%%%%%%%%%%%%%%%%%%%%%%%%%%%%%%%%%%%%%%%%%%%%%%%%%%%%%%%%%%%%%%%%%%%%%%%%%%%%%%

\subsection{Formal Construction}
\label{subsection:parametrization}
For definiteness, we consider only the case when the Lorentzian
algebra $\mf{g}$ is a split real form, although this is not really
necessary as the Iwasawa decomposition \index{Iwasawa decomposition} 
holds also in the non-split case (see, e.g., \cite{Riccioni:2008jz,Houart:2009ed}).

A very important difference from the finite-dimensional case is
that we now have nontrivial \emph{multiplicities} of the
imaginary roots (see Section~\ref{Section:rootsystem}).
Recall that if a root $\al\in\Delta$ has multiplicity $m_{\al}$,
then the associated root space $\mf{g}_{\al}$ is
$m_{\al}$-dimensional. Thus, it is spanned by $m_{\al}$ generators
$E_{\al}^{[s]}\, (s=1, \cdots, m_{\al})$,
\begin{equation}
  \mf{g}_{\al}=\mbb{R}E_{\al}^{[1]}+ \cdots + \mbb{R} E_{\al}^{[m_{\al}]}.
\end{equation}
The root multiplicities are not known in
closed form for any indefinite Kac--Moody algebra, but must be
computed recursively as described in
Section~\ref{section:LevelDecomposition}.

Our main object of study is the coset representative $\mc{V}(t)\in
 K\bas G$, which must now be seen as ``formal'' exponentiation of the
infinite number of generators in $\mf{p}$. We can then proceed as
before and choose $\mc{V}(t)$ to be in the Borel gauge, i.e., of
the form
\begin{equation}
  \mc{V}(t)=
  \Exp \left[\sum_{\mu=1}^{\dim \mf{h}} \be^{\mu}(t)\al_{\mu}^{\vee}\right]\,
  \Exp \left[\sum_{\al\in\Phi_+}\sum_{s=1}^{m_{\al}}
  \xi^{[s]}_{\al}(t)E_{\al}^{[s]}\right] \in  K\bas G. 
\label{InfiniteCosetRepresentative}
\end{equation}
Here, the index $\mu$ does not correspond to ``spacetime'' but instead
is an index in the Cartan subalgebra \index{Cartan subalgebra}$\mf{h}$, or, equivalently,
in ``scale-factor space'' \index{scale-factor space} (see Section~\ref{Section:HyperbolicBilliard}).
In the following we shall dispose of writing the sum over $\mu$
explicitly. The second exponent in
Equation~(\ref{InfiniteCosetRepresentative}) contains a formal infinite
sum over all positive roots $\Phi_+$. We will describe in detail
in subsequent sections how it can be suitably truncated. The coset
representative $\mc{V}(t)$ corresponds to a nonlinear realisation
\index{nonlinear realisation} of $ G$ and transforms as
\begin{equation}
   G : \mc{V}(t)
  \quad \longmapsto \quad
  k\left(\mc{V}(t), g\right)\mc{V}(t) g,
  \qquad
  k\left(\mc{V}(t), g\right)\in K,\, g\in G.
\end{equation}

A $\mf{g}$-valued ``one-form'' can be constructed analogously to
the finite-dimensional case,
\begin{equation}
  \pa \mc{V}(t)\mc{V}(t)^{-1}=\mc{Q}(t)+\mc{P}(t),
\end{equation}
where $\pa\equiv \pa_t$. The first term on the right hand side
represents a $\mf{k}$-connection that is fixed under the Chevalley
involution,
\begin{equation}
  \tau(\mc{Q})=\mc{Q},
\end{equation}
while $\mc{P}(t)$ lies in the orthogonal complement $\mf{p}$ and so is
anti-invariant,
\begin{equation}
  \tau(\mc{P})=-\mc{P}
\end{equation}
(for the split form, the Cartan involution coincides with the
Chevalley involution). \index{Chevalley involution} Using the
projections onto the coset $\mf{p}$
and the compact subalgebra $\mf{k}$, as defined in
Equation~(\ref{CartanProjections}), we can compute the forms of
$\mc{P}(t)$ and $\mc{Q}(t)$ in the Borel gauge, and we find 
\begin{equation}
  \begin{array}{rcl}
    \mc{P}(t)&=& \displaystyle
    \pa \be^{\mu}(t)\al_{\mu}^{\vee}+\f{1}{2}\sum_{\al\in\Phi_+}
    \sum_{s=1}^{m_{\al}}e^{\al(\be)}\mf{R}^{[s]}_{\al}(t)
    \left(E^{[s]}_{\al}+E^{[s]}_{-\al}\right),
    \\ [0.75 em]
    \mc{Q}(t)&=& \displaystyle
    \f{1}{2}\sum_{\al\in\Phi_+}
    \sum_{s=1}^{m_{\al}}e^{\al(\be)}\mf{R}^{[s]}_{\al}(t)
    \left(E^{[s]}_{\al}-E^{[s]}_{-\al}\right),
  \end{array}
  \label{InfiniteQandP} 
\end{equation}
where $\mf{R}^{[s]}_{\al}(t)$ is the analogue of $\mc{R}_{\al}(x)$ in
the finite-dimensional case, i.e., it takes the form
\begin{equation}
  \mf{R}^{[s]}_{\al}(t)=\pa \xi^{[s]}_{\al}(t)+
  \f{1}{2}\underbrace{\xi_{\zeta}^{[s]}(t)
  \pa \xi^{[s]}_{\zeta^{\p}}(t)}_{\zeta+\zeta^{\p}=\al}+\cdots,
  \label{InfiniteCosetCoefficients}
\end{equation}
which still contains a finite number of terms for each positive root
$\al$. The value of the root $\al\in\mf{h}^{\star}$ acting on
$\be=\be^{\mu}(t)\al_{\mu}^{\vee}\in\mf{h}$ is
\begin{equation}
  \al(\be)=\al_{\mu}\be^{\mu}.
\end{equation}
Note that here the notation $\al_{\mu}$ does not correspond to a
simple root, but denotes the components of an arbitrary root vector
$\al\in\mf{h}^{\star}$.

The action for a particle moving on the infinite-dimensional coset
space $K\bas G$ can now be constructed using the invariant bilinear
form $(\cdot |\cdot )$ on $\mf{g}$,
\begin{equation}
  S_{ K\bas G}=\int dt\,n(t)^{-1}\left(\mc{P}(t)|\mc{P}(t)\right),
\end{equation}
where $n(t)$ ensures invariance under reparametrizations of $t$. The
variation of the action with respect to $n(t)$ constrains the motion
to be a \emph{null geodesic} on $ K\bas G$,
\begin{equation}
  \left(\mc{P}(t)|\mc{P}(t)\right)=0.
\end{equation}
Defining, as before, a covariant derivative $\mf{D}$ with respect to
the local symmetry $ K$ as
\begin{equation}
  \mf{D}\mc{P}(t)\equiv \pa \mc{P}(t)-\left[\mc{Q}(t), \mc{P}(t)\right],
\end{equation}
the equations of motion read simply
\begin{equation}
  \mf{D}\left(n(t)^{-1} \mc{P}(t)\right)=0.
  \label{InfiniteEOM}
\end{equation}
The explicit form of the action in the parametrization of
Equation~(\ref{InfiniteCosetRepresentative}) becomes
\begin{equation}
  S_{ K\bas G}=\int dt \, n(t)^{-1}
  \left[G_{\mu\nu}\,\pa\be^{\mu}(t)\,\pa\be^{\nu}(t)+
  \f{1}{2}\sum_{\al\in\Phi_+}\sum_{s=1}^{m_{\al}}e^{2\al(\be)}\,
  \mf{R}^{[s]}_{\al}(t)\,\mf{R}^{[s]}_{\al}(t)\right],
  \label{InfiniteSigmaModel}
\end{equation}
where $G_{\mu\nu}$ is the flat Lorentzian metric, defined by the
restriction of the bilinear form $(\cdot |\cdot )$ to the Cartan
subalgebra $\mf{h}\subset\mf{g}$. The metric $G_{\mu\nu}$ is exactly
the same as the metric in scale-factor space \index{scale-factor space} (see
Section~\ref{Section:HyperbolicBilliard}), and the kinetic term for the Cartan
parameters $\be^{\mu}(t)$ reads explicitly
\begin{equation}
  G_{\mu\nu}\,\pa\be^{\mu}(t)\,\pa\be^{\nu}(t)=
  \sum_{i=1}^{\dim \mf{h}-1}\pa \be^{i}(t)\,\pa\be^{i}(t)-
  \left(\sum_{i=1}^{\dim \mf{h}-1} \!\! \pa\be^{i}(t)\right)
  \left(\sum_{j=1}^{\dim \mf{h}-1} \!\! \pa\be^{j}(t)\right)+
  \pa\phi(t)\,\pa\phi(t).
  \label{MetricCartanSubalgebra}
\end{equation}

Although $\mf{g}$ is infinite-dimensional we still have the notion
of ``formal integrability'', owing to the existence of an infinite
number of conserved charges, defined by the equations of motion in
Equation~(\ref{InfiniteEOM}). We can define the generalized metric
\index{generalized metric} for any $\mf{g}$ as
\begin{equation}
  \mc{M}(t)\equiv \mc{V}(t)^{\mc{T}}\mc{V}(t),
\end{equation}
where the transpose $(\, )^{\mc{T}}$ is defined as before in
terms of the Chevalley involution,
\begin{equation}
  (\,)^{\mc{T}}=-\tau(\,).
\end{equation}
Then the equations of motion $\mf{D}\mc{P}(t)=0$ are equivalent to the
conservation $\pa \mf{J}=0$ of the current
\begin{equation}
  \mf{J}\equiv \f{1}{2}\mc{M}(t)^{-1}\pa \mc{M}(t).
\end{equation}
This can be formally solved in closed form
\begin{equation}
  \mc{M}(t)=e^{t\mf{J}^{\mc{T}}}\mc{M}(0)e^{t\mf{J}},
\end{equation}
and so an arbitrary group element $g\in G$ evolves according to
\begin{equation}
  g(t)=k(t)g(0)e^{t\mf{J}},
  \qquad
  k(t)\in K.
  \label{GeneralSolution}
\end{equation}

Although the explicit form of $\mc{P}(t)$ contains infinitely many
terms, we have seen that each coefficient $\mf{R}_{\al}^{[s]}(t)$
can, in principle, be computed exactly for each root $\al$. This,
however, is not the case for the current $\mf{J}$. To find the
form of $\mf{J}$ one must conjugate $\mc{P}(t)$ with the coset
representative $\mc{V}(t)$ and this requires an infinite number of
commutators to get the correct coefficient in front of any
generator in $\mf{J}$.

%%%%%%%%%%%%%%%%%%%%%%%%%%%%%%%%%%%%%%%%%%%%%%%%%%%%%%%%%%%%%%%%%%%%%%%%%%%%%%%%%%%

\subsection{Consistent Truncations}
\label{section:ConsistentTruncations}

One method for dealing with infinite expressions like
Equation~(\ref{InfiniteQandP}) consists in considering successive
finite expansions allowing more and more terms, while still respecting
the dynamics of the sigma model. 

This leads us to the concept of a \emph{consistent truncation} of the
sigma model for $ K\bas G$. We take as definition of such a truncation
any sub-model $S^{\p}$ of $S_{ K\bas G}$ whose solutions are also
solutions of the original model.

There are two main approaches to finding suitable truncations that
fulfill this latter criterion. These are the so-called
\emph{subgroup truncations} and the \emph{level truncations},
which will both prove to be useful for our purposes, and we
consider them in turn below.

\subsubsection*{Subgroup Truncation}

The first consistent truncation we shall treat is the
case when the dynamics of a sigma model for some global group
$ G$ is restricted to that of an appropriately chosen subgroup
$\bar{ G}\subset G$. We consider here only subgroups
$\bar{ G}$ which are obtained by exponentiation of regular
subalgebras $\bar{\mf{g}}$ of $\mf{g}$. The concept of regular
embeddings of Lorentzian Kac--Moody algebras 
\index{Kac--Moody algebra} is discussed in detail in
{\bf Paper I}.

To restrict the dynamics to that of a sigma model based on the
coset space $K(\bar{ G})\bas \bar{ G}$, we first assume that
the initial conditions $g(t)\big|_{t=0}=g(0)$ and $\pa g(t)\big|_{t=0}$
are such that the following two conditions are satisfied:

\begin{enumerate}
\item The group element $g(0)$ belongs to $\bar{ G}$.
\item The conserved current $\mf{J}$ belongs to $\bar{\mf{g}}$.
\end{enumerate}

When these conditions hold, then $g(0)e^{t\mf{J}}$ belongs to $\bar{ G}$
for all $t$. Moreover, there always exists $\bar{k}(t)\in
 K(\bar{ G})$ such that
\begin{equation}
  \bar{g}(t)\equiv \bar{k}(t) g(0) e^{t\mf{J}} \in K(\bar{G})\bas \bar{G},
\end{equation}
i.e, $\bar{g}(t)$ belongs to the Borel subgroup of $\bar{ G}$.
Because the embedding is regular, $\bar{k}(t)$ belongs to $ K$
and we thus have that $\bar{g}(t)$ also belongs to the Borel subgroup
of the full group $ G$. 

Now recall that from Equation~(\ref{GeneralSolution}), we know that
$\bar{g}(t)=\bar{k}(t)g(0)e^{t\mf{J}}$ is a solution to the
equations of motion for the sigma model on
$K(\bar{G})\bas \bar{G}$. But since we have found that
$\bar{g}(t)$ preserves the Borel gauge for $ K\bas G$, it follows
that $\bar{k}(t)g(0)e^{t\mf{J}}$ is a solution to the equations of
motion for the full sigma model. Thus, the dynamical evolution of the
subsystem $S^{\p}=S_{K(\bar{G})\bas \bar{G}}$ preserves the
Borel gauge of $ G$. These arguments show that initial conditions in $\bar{ G}$
remain in $\bar{ G}$, and hence the dynamics of a sigma model on
$ K\bas G$ can be consistently truncated to a sigma model on
$K(\bar{G})\bas \bar{G}$.

Finally, we recall that because the embedding
$\bar{\mf{g}}\subset\mf{g}$ is regular, the restriction of the
bilinear form on $\mf{g}$ coincides with the bilinear form on
$\bar{\mf{g}}$. This implies that the Hamiltonian constraints
\index{Hamiltonian constraint} for
the two models, arising from time reparametrization invariance of
the action, also coincide. 

We shall make use of subgroup truncations in
Chapter~\ref{Chapter:GeometricConfigurations}.

\subsubsection{Level Truncation and Height Truncation}

Alternative ways of consistently truncating the
infinite-dimensional sigma model rest on the use of
\emph{gradations} of $\mf{g}$,
\begin{equation}
  \mf{g}=\cdots + \mf{g}_{-2}+\mf{g}_{-1}+\mf{g}_0+\mf{g}_1+\mf{g}_2+\cdots,
\end{equation}
where the sum is infinite but each subspace is finite-dimensional.
One also has
\begin{equation}
  [\mf{g}_{\ell^{\p}}, \mf{g}_{\ell^{\p\p}}]\subseteq \mf{g}_{\ell^{\p}+\ell^{\p\p}}.
\end{equation}
Such a gradation was for instance constructed in
Section~\ref{section:LevelDecomposition} and was based on a so-called
\emph{level decomposition} of the adjoint representation of $\mf{g}$
into representations of a finite regular subalgebra \index{regular subalgebra}
$\mf{r}\subset \mf{g}$. We will now use this construction to truncate
the sigma model based of $ K\bas G$, by ``terminating'' the gradation of
$\mf{g}$ at some finite level $\bar{\ell}$. More specifically, the
truncation will involve setting to zero all coefficients
$\mf{R}_{\al}^{[s]}(t)$, in the expansion of $\mc{P}(t)$,
corresponding to roots $\al$ whose generators $E_{\al}^{[s]}$ belong
to subspaces $\mf{g}_{\ell}$ with $\ell>\bar{\ell}$. Part of this
section draws inspiration from the treatment in~\cite{DHN2, DHNReview,
  AxelThesis}.

The level $\ell$ might be the height, or it might count the number
of times a specified single simple root appears. In that latter
case, the actual form of the level decomposition must of course be
worked out separately for each choice of algebra $\mf{g}$ and each
choice of decomposition. We will do this in detail in
Section~\ref{section:E10SigmaModel} for a specific level decomposition
of the hyperbolic algebra $E_{10}$. Here, we shall display the general
construction in the case of the \emph{height truncation}, which exists
for any algebra.

Let $\al$ be a positive root, $\al\in\Phi_+$. It has the
following expansion in terms of the simple roots
\begin{equation}
  \al=\sum_{i}m_i\al_i
  \qquad
  (m_i\geq 0).
\end{equation}
Then the \emph{height} of $\al$ is defined as (see
Section~\ref{Section:rootsystem})
\begin{equation}
  \htx (\al)=\sum_{i}m_i.
\end{equation}
The height can thus be seen as a linear integral map $\htx : \Phi
\rightarrow \mbb{Z}$, and we shall sometimes use the notation
$\htx (\al)=h_{\al}$ to denote the value of the map
$\htx $ acting on a root $\al\in \Phi$.

To achieve the height truncation, we assume that the sum over all
roots in the expansion of $\mc{P}(t)$, Equation~(\ref{InfiniteQandP}), is
ordered in terms of increasing height. Then we can consistently
set to zero all coefficients $\mf{R}_{\al}^{[s]}(t)$ corresponding
to roots with greater height than some, suitably chosen, finite
height $\bar{h}$. We thus find that the finitely truncated coset
element $\mc{P}_0(t)$ is
\begin{equation}
  \mc{P}_0(t)\equiv \mc{P}(t)\big|_{\htx \leq \bar{h}}=
  \pa \be^{\mu}(t)\al_{\mu}^{\vee}+
  \f{1}{2}\sum_{\substack{\al \in \Delta_{+}\\ \htx (\al) \leq \bar{h}}}
  \sum_{s=1}^{m_{\al}}e^{\al(\be)}\mf{R}^{[s]}_{\al}(t)
  \left(E^{[s]}_{\al}+E^{[s]}_{-\al}\right),
\end{equation}
which is equivalent to the statement
\begin{equation}
  \mf{R}_{\ga}^{[s]}(t)=0
  \qquad
  \forall \ga\in\Phi_+ ,\, \htx (\ga)>\bar{h}.
\end{equation}

For further use, we note here some properties of the coefficients
$\mf{R}_{\al}^{[s]}(t)$. By examining the structure of
Equation~(\ref{InfiniteCosetCoefficients}), we see that
$\mf{R}_{\al}^{[s]}(t)$ takes the form of a temporal derivative
acting on $\xi_{\al}^{[s]}(t)$, followed by a sequence of terms
whose individual components, for example $\xi_{\zeta}^{[s]}(t)$,
are all associated with roots of \emph{lower} height than $\al$,
$\htx (\zeta)< \htx (\al)$. It will prove useful to think
of $\mf{R}_{\al}^{[s]}(t)$ as representing a kind of
``generalized'' derivative operator acting on the field
$\xi_{\al}^{[s]}$. Thus we define the operator $\mc{D}$ by
\begin{equation}
  \mc{D}\xi_{\al}^{[s]}(t)\equiv
  \pa \xi^{[s]}_{\al}(t)+\mc{F}_{\al}^{[s]}
  \left(\xi\pa\xi, \xi^2\pa\xi,\cdots \right),
\end{equation}
where $\mc{F}_{\al}^{[s]}(t)$ is a polynomial function of the
coordinates $\xi(t)$, whose explicit structure follows from
Equation~(\ref{InfiniteCosetCoefficients}). It is common in the
literature to refer to the level truncation as ``setting all higher
level covariant derivatives to zero'', by which one simply means that
all $\mc{D}\xi_{\ga}^{[s]}(t)$ corresponding to roots $\ga$ above a
given finite level $\bar{\ell}$ should vanish. Following~\cite{DHN2}
we shall call the operators $\mc{D}$ ``covariant derivatives''.

It is clear from the equations of motion $\mf{D}\mc{P}(t)=0$, that if
all covariant derivatives $\mc{D}\xi_{\ga}^{[s]}(t)$ above a given
height are set to zero, this choice is preserved by the dynamical
evolution. Hence, the height (and any level) truncation is indeed a
consistent truncation. Let us here emphasize that it is \emph{not}
consistent by itself to merely put all fields $\xi_{\ga}^{[s]}(t)$
above a certain level to zero, but one must take into account the fact
that combinations of lower level fields may parametrize a higher level
generator in the expansion of $\mc{P}(t)$, and therefore it is crucial
to define the truncation using the derivative operator
$\mc{D}\xi_{\ga}^{[s]}(t)$.

\section{Eleven-Dimensional Supergravity and $K(\mc{E}_{10})\bas \mc{E}_{10}$}
\label{section:E10SigmaModel}

We shall now illustrate the results of the previous sections by
explicitly constructing an action for the coset space
$K(\mc{E}_{10})\bas \mc{E}_{10}$. \index{$E_{10}$} We employ the level
decomposition of $E_{10}=\mathrm{Lie}\, \mc{E}_{10}$ in terms of its
regular $\mf{sl}(10, \mbb{R})$-subalgebra (see
Section~\ref{section:LevelDecomposition}), to write the coordinates
on the coset space  \index{coset space} as (time-dependent) $\mf{sl}(10, \mbb{R})$-tensors. It is
then shown that for a truncation of the sigma model  \index{nonlinear sigma model} at level
$\ell=3$, these fields can be interpreted as the physical fields
of eleven-dimensional supergravity. This ``dictionary'' ensures
that the equations of motion arising from the sigma model on
$K(\mc{E}_{10})\bas \mc{E}_{10}$ are equivalent to the (suitably
truncated) equations of motion of eleven-dimensional
supergravity~\cite{DHN2}.

%%%%%%%%%%%%%%%%%%%%%%%%%%%%%%%%%%%%%%%%%%%%%%%%%%%%%%%%%%%%%%%%%%%%%%%%%%%%%%%%%%%

\subsection{Low-Level Fields}
\label{section:lowlevelfields}

We perform the level decomposition \index{level decomposition}of $E_{10}$ with respect to the
$\mf{sl}(10, \mbb{R})$-subalgebra obtained by removing the
exceptional node in the Dynkin diagram in
Figure~\ref{figure:E10a}. This procedure was described in
Section~\ref{section:LevelDecomposition}. When using this
decomposition, a sum over (positive) roots becomes a sum over all $\mf{sl}(10,
\mbb{R})$-indices in each (positive) representation appearing in
the decomposition.

We recall that up to level three the following representations
appear 
\begin{equation}
  \begin{array}{l}
    \ell=0:
    \qquad
    {K^{a}}_b,
    \\
    \ell=1:
    \qquad
    E^{abc}=E^{[abc]},
    \\
    \ell=2:
    \qquad
    E^{a_1\cdots a_6}=E^{[a_1\cdots a_6]},
    \\
    \ell=3:
    \qquad
    E^{a|b_1\cdots b_8}=E^{a|[b_1\cdots b_8]},
  \end{array}
  \label{LowLevelRepsE10} 
\end{equation}
where all indices are $\mf{sl}(10,
\mbb{R})$-indices and so run from 1 to 10. The level zero
generators ${K^{a}}_b$ correspond to the adjoint representation of
$\mf{sl}(10, \mbb{R})$ and the higher level generators correspond
to an infinite tower of raising operators of $E_{10}$. As
indicated by the square brackets, the level one and two
representations are completely antisymmetric in all indices, while
the level three representation has a mixed Young tableau symmetry:
It is antisymmetric in the eight indices $b_1 \cdots b_8$ and is
also subject to the constraint
\begin{equation}
  E^{[a|b_1\cdots b_8]}=0.
  \label{LevelThreeConstraint}
\end{equation}
In the scale factor space ($\be$-basis), the roots of $E_{10}$
corresponding to the above generators act as follows on $\be\in
\mf{h}$: 
\begin{equation}
  \begin{array}{rcl}
    {K^{a}}_{b} & \quad \Longleftrightarrow & \quad
    \al_{ab}(\be)=\be^{a}-\be^{b}
    \qquad (a>b),
    \\
    E^{abc} & \quad \Longleftrightarrow & \quad
    \al_{abc}(\be) =\be^{a}+\be^{b}+\be^c,
    \\
    E^{a_1\cdots a_6} & \quad \Longleftrightarrow & \quad
    \al_{a_1\cdots a_6}(\be)=\be^{a_1}+\cdots +\be^{a_6},
    \\
    E^{a|a b_1\cdots b_7 } & \quad \Longleftrightarrow & \quad
    \al_{ab_1\cdots b_7}(\be) =2\be^{a}+\be^{b_1}+\cdots +\be^{b_7},
    \\
    E^{a_1|a_2\cdots a_9} & \quad \Longleftrightarrow & \quad
    \al_{a_1\cdots a_9}(\be) =\be^{a_1}+\cdots + \be^{a_9}.
  \end{array}
\end{equation}
We can use the scalar product in root space,
$\mf{h}^{\star}$, to compute the norms of these roots. The metric on
$\mf{h}^{\star}$ is the inverse of the metric in
Equation~(\ref{MetricCartanSubalgebra}), and for $E_{10}$ it takes
the form 
\begin{equation}
  (\om|\om)=G^{ij}\om_{i}\om_{j}=
  \sum_{i=1}^{10}\om_i \om_i -
  \f{1}{9}\left(\sum_{i=1}^{10}\om_i\right)\left(\sum_{j=1}^{10}\om_j\right),
  \qquad
  \om\in\mf{h}^{\star}.
\end{equation}
The level zero, one and two generators correspond to real roots of
$E_{10}$,
\begin{equation}
  (\al_{ab}|\al_{cd})=2,
  \qquad
  (\al_{abc}|\al_{def})=2,
  \qquad
  (\al_{a_1\cdots a_6}|\al_{b_1\cdots b_6})=2.
\end{equation}
We have split the roots corresponding to the level three generators
into two parts, depending on whether or not the special index $a$
takes the same value as one of the other indices. The resulting two
types of roots correspond to real and null roots, respectively,
\begin{equation}
  (\al_{ab_1\cdots b_7}|\al_{cd_1\cdots d_7})=2,
  \quad
  (\al_{a_1\cdots a_9}|\al_{b_1\cdots b_9})=0.
\end{equation}
Thus, the first time that generators corresponding to imaginary roots
appear in the level decomposition is at level three. This will prove
to be important later on in our analysis.

%%%%%%%%%%%%%%%%%%%%%%%%%%%%%%%%%%%%%%%%%%%%%%%%%%%%%%%%%%%%%%%%%%%%%%%%%%%%%%%%%%%

\subsection{The   $SO(10)\bas GL(10, \mbb{R})  $-Sigma Model}
\label{GL10/SL10Sigma}
\index{nonlinear sigma model}

Because of the importance and geometric significance of level zero, we
shall first develop the formalism for the $SO(10)\bas GL(10, \mbb{R})  $-sigma model. A general group element $H$ in the
subgroup $GL(10, \mbb{R})$ reads 
\begin{equation}
  H = \Exp \left[{h_{a}}^b {K^{a}}_b \right]
\end{equation}
where ${h_{a}}^b$ is a $10\times 10$ matrix (with
$a$ being the row index and $b$ the column index). Although the
${K^{a}}_b $'s are generators of $E_{10}$ and can, within this framework, at best be
viewed as infinite matrices, it will prove convenient -- for streamlining
the calculations -- to view them in the present section also as
$10\times 10$ matrices, since we confine our attention to the
finite-dimensional subgroup $GL(10, \mbb{R})$. Namely, ${K^{a}}_b $ is
treated as a 
$10\times 10$ matrix with 0's everywhere except 1 in position
$(a,b)$. The final formulation in terms of the variables
${h_{a}}^b(t)$ -- which are $10\times 10$ matrices irrespectively as to 
whether one deals with $GL(10, \mbb{R})$ \emph{per se} or as a subgroup of
$E_{10}$ -- does not depend on this interpretation.

It is also useful to describe $GL(10, \mbb{R})$ as the set of linear
combinations ${m_i}^j K{^i}_{j}$ where the $10\times 10$ matrix
${m_i}^j$ is invertible. The product of the $K{^i}_{j}$'s is
given by
\begin{equation}
K{^i}_{j}\, K{^k}_{m} = {\delta^k}_j \, {K^i}_m.
\end{equation}
One easily verifies that if $M = {m_i}^j K{^i}_{j}$ and $N
= {n_i}^j K{^i}_{j}$ belong to $GL(10, \mbb{R})$, then $MN =
{(mn)_i}^j K{^i}_{j}$ 
where $mn$ is the standard product of the $10\times 10$ matrices $m$
and $n$. Furthermore, $\Exp \left({h_i}^j {K^i}_j \right) =
{\left(e^h\right)_i}^j {K^i}_j$ where $e^h$ is the standard matrix
exponential.

Under a general transformation, the representative $H(t)$ is
multiplied from the left by a time-dependent $SO(10)$ group
element $R$ and from the right by a constant linear $GL(10, \mbb{R})$-group element $L$.
Explicitly, the transformation takes the form (suppressing the
time-dependence for notational convenience)
\begin{equation} H
  \rightarrow H' = R H L. 
  \label{transformationvielbein}
\end{equation}
In terms of components, with $H = {e_a}^b {K^a}_b$, ${e_a}^b =
{(e^h)_a}^b$, $R = {R_a}^b {K^a}_b$ and $L = {L_a}^b {K^a}_b$, one
finds
\begin{equation}
  {e'_a}^b = {R_a}^c {e_c}^d {L_d}^b,
\end{equation}
where we have
set $H' = {e'_a}^b {K^a}_b$. The indices on the coset
representative have different covariance properties. To emphasize
this fact, we shall write a bar over the first index, ${e_a}^b
\rightarrow {e_{\bar{a}}}^b$. Thus, barred indices transform under
the local $SO(10)$ gauge group and are called ``local'', or also
``flat'', indices, while unbarred indices transform under the
global $GL(10, \mbb{R})$ and are called ``world'', or also ``curved'',
indices. The gauge invariant matrix product $M= H^T H$ is equal to
\begin{equation}
  M = g^{ab} K_{ab},
\end{equation}
with $K_{ab} \equiv {K^c}_b \delta_{ac}$ and
\begin{equation}
  g^{ab} = \sum_{\bar{c}} {e_{\bar{c}}}^a {e_{\bar{c}}}^b. 
  \label{sigmamodelMetric}
\end{equation}
The $g^{ab}$ do not transform under local $SO(10)$-transformations and
transform as a (symmetric) contravariant tensor under rigid $GL(10,
\mbb{R})$-transformations,
\begin{equation}
  {g'}^{ab}= g^{cd} {L_c}^a {L_d}^b.
\end{equation}
They are components of a nondegenerate symmetric matrix that can be
identified with an inverse Euclidean metric.

Indeed, the coset space $SO(10)\bas GL(10, \mbb{R})  $ can be
identified with the space of symmetric tensors of Euclidean
signature, i.e., the space of metrics. This is because two
symmetric tensors of Euclidean signature are equivalent under a
change of frame, and the isotropy subgroup, say at the identity,
is evidently $SO(10)$. From this point of view, the coset representative
${e_a}^b$ is the spatial vielbein.

The action for the coset space $SO(10)\bas GL(10, \mbb{R})  $ with the
metric of Equation~(\ref{KillingformlevelzeroForE10}) is easily found to
be
\begin{equation}
  \mc{L}_0=\f{1}{4}\left(g^{ac}(t)g^{bd}(t)-g^{ab}(t)g^{cd}(t)\right)
  \pa {g}_{ab}(t) \, \pa {g}_{cd}(t).
 \label{LagrangianforGL10SO10}
\end{equation}
Note that the quadratic form multiplying the time derivatives is just
the ``De Witt supermetric''~\cite{DeWitt:1967yk}. Note also for future
reference that the invariant form $\pa H \, H^{-1}$ reads explicitly
\begin{equation}
  \pa {H}\, H^{-1} = {\pa {e}_{\bar{a}}}^{\ph{a} b} \,
  {e_b}^{\bar{c}} \, {K^a}_c,
\end{equation}
where ${e_b}^{\bar{n}}$ is the inverse vielbein.

%%%%%%%%%%%%%%%%%%%%%%%%%%%%%%%%%%%%%%%%%%%%%%%%%%%%%%%%%%%%%%%%%%%%%%%%%%%%%%%%%%%

\subsection{Sigma Model Fields and  
  $SO(10)_{\mathrm{local}}\times GL(10,\mbb{R})_{\mathrm{rigid}}$-Covariance}

We now turn to the full nonlinear sigma model  \index{nonlinear sigma model} for 
$K(\mc{E}_{10})\bas \mc{E}_{10}$. Rather than exponentiating the
Cartan subalgebra separately as in
Equation~(\ref{InfiniteCosetRepresentative}), it will here prove
convenient to instead single out the level zero subspace
$\mf{g}_{0}=\mf{gl}(10, \mbb{R})$. This permits one to control
easily $SO(10)_{\mathrm{local}}\times GL(10, \mbb{R})_{\mathrm{rigid}}$-covariance. 
To make this level zero 
covariance manifest, we shall furthermore assume that the Borel
gauge has been fixed only for the non-zero levels, and we keep all
level zero fields present. The residual gauge freedom is then just
multiplication by an $SO(10)$ rotation from the left.

Thus, we take a coset representative of the form
\begin{equation}
  \mc{V}(t)=H(t)\,
  \Exp \left[\f{1}{3!}\mc{A}_{abc}(t)E^{abc}+
  \f{1}{6!}\mc{A}_{a_1\cdots a_6}(t)E^{a_1\cdots a_6}+
  \f{1}{9!}\mc{A}_{a|b_1\cdots b_8}(t)E^{a|b_1\cdots b_8}+\cdots \right],
  \label{E10cosetrepresentative}
\end{equation}
where the sum in the first exponent would be restricted to all $a\geq
b$ if we had taken a full Borel gauge also at level zero. The
parameters $\mc{A}_{abc}(t), \mc{A}_{a_1\cdots a_6}(t)$ and
$\mc{A}_{a|b_1\cdots b_8}(t)$ are coordinates on the coset space
$K(\mc{E}_{10})\bas \mc{E}_{10}$ and will eventually be interpreted
as physical time-dependent fields of eleven-dimensional supergravity.

How do the fields transform under $SO(10)_{\mathrm{local}}\times
GL(10,\mbb{R})_{\mathrm{rigid}}$? Let $R \in SO(10)$, $L \in GL(10,
\mbb{R})$ and decompose $\mc{V}$ according to
Equation~(\ref{E10cosetrepresentative}) as the product 
\begin{equation}
  \mc{V} = H T, 
\end{equation}
with 
\begin{equation}
  \begin{array}{rcl}
    H &= & \displaystyle
    \Exp \left[{h_{a}}^b(t) {K^{a}}_b \right] \in GL(10, \mbb{R}),
    \\ [1 em]
    T & = & \displaystyle
    \Exp \left[\f{1}{3!}\mc{A}_{abc}(t)E^{abc}+
    \f{1}{6!} \mc{A}_{a_1\cdots a_6}(t)E^{a_1\cdots a_6}+
    \f{1}{9!}\mc{A}_{a|b_1\cdots b_8}(t)E^{a|b_1\cdots b_8}+\cdots \right].
  \end{array}
\end{equation}
One has
\begin{equation}
  \mc{V} \rightarrow \mc{V}' = R (H T) L = (R H L) (L^{-1} T L).
\end{equation}
Now, the first matrix $H' = RHL$ clearly belongs to $GL(10, \mbb{R})$,
since it is the product of a rotation matrix by two $GL(10,
\mbb{R})$-matrices. It has exactly the same transformation as in
Equation~(\ref{transformationvielbein}) above in the context of the
nonlinear sigma model for $SO(10)\bas GL(10, \mbb{R})  $. Hence, the
geometric interpretation of ${e_{\bar{a}}}^b = {(e^h)_{\bar{a}}}^b$ as
the vielbein remains.

Similarly, the matrix $T' \equiv L^{-1} T L$ has exactly the same
form as $T$, 
\begin{eqnarray}
  T' &=&\Exp \left( L^{-1} \left[\f{1}{3!}\mc{A}_{abc}(t)E^{abc}+
  \f{1}{6!}\mc{A}_{a_1\cdots a_6}(t)E^{a_1\cdots a_6} +
  \f{1}{9!}\mc{A}_{a|b_1\cdots b_8}(t)E^{a|b_1\cdots b_8}+\cdots \right] L \right)
  \nonumber
  \\
  &=&\Exp \left[\f{1}{3!}\mc{A}'_{abc}(t)E^{abc}+
  \f{1}{6!}\mc{A}'_{a_1\cdots a_6}(t)E^{a_1\cdots a_6} +
  \f{1}{9!}\mc{A}'_{a|b_1\cdots b_8}(t)E^{a|b_1\cdots b_8}+\cdots \right], 
\end{eqnarray}
where the variables
$\mc{A}'_{abc}$, $\mc{A}'_{a_1\cdots a_6}$, ..., are obtained from
the variables $\mc{A}_{abc}$, $\mc{A}_{a_1\cdots a_6}$, ..., by
computing $L^{-1} E^{abc} L$, $L^{-1} E^{a_1 \cdots a_6} L$, ...,
using the commutation relations with ${K^a}_b$. Explicitly, one
gets
\begin{equation}
  \mc{A}'_{abc} = {(L^{-1})_a}^e {(L^{-1})_b}^f
  {(L^{-1})_c}^g \mc{A}_{efg},
  \qquad
  \mc{A}'_{a_1\cdots a_6} =
  {(L^{-1})_{a_1}}^{b_1} \cdots {(L^{-1})_{a_6}}^{b_6}
  \mc{A}_{b_1\cdots b_6},
  \qquad
  \mbox{ etc.}
\end{equation}
Hence, the fields
$\mc{A}_{abc}$, $\mc{A}_{a_1\cdots a_6}$, ... do not transform
under local $SO(10)$ transformations. However, they do transform under
rigid $GL(10, \mbb{R})$-transformations as tensors of the type
indicated by their indices. Their indices are world indices and
not flat indices.

%%%%%%%%%%%%%%%%%%%%%%%%%%%%%%%%%%%%%%%%%%%%%%%%%%%%%%%%%%%%%%%%%%%%%%%%%%%%%%%%%%%

\subsection{``Covariant Derivatives''}

The invariant form $\pa {\mc{V}} \, \mc{V}^{-1}$ reads
\begin{equation}
  \pa {\mc{V}} \, \mc{V}^{-1} =
  \pa {H} \, H^{-1} + H (\pa {T} \, T^{-1}) H^{-1}.
  \label{invariantform10.15}
\end{equation}
The first term is the
invariant form encountered above in the discussion of the level
zero nonlinear sigma model for $SO(10)\bas GL(10, \mbb{R})$. So let us focus on
the second term. It is clear that $\pa {T} \, T^{-1}$ will contain
only positive generators at level $\geq 1$. So we set, in a
manner similar to Equation~(\ref{sumclosedform}),
\begin{equation}
  \pa {T} \, T^{-1} =
  \sum_{\alpha \in \Phi_+} \sum_{s_\alpha=1}^{\text{mult}\ \alpha}\mc{D} \mc{A}_{\alpha, s_{\alpha}} \, E_{\alpha,s_{\alpha}},
\end{equation}
 where the sum is over positive roots at levels
one and higher and takes into account the root multiplicities. The expressions $\mc{D} \mc{A}_{\alpha, s_{\alpha}}$ are
linear in the time derivatives $\pa {\mc{A}}$. As before, we call them
``covariant derivatives''. They are computed by making use of the
Baker--Hausdorff formula, as in
Section~\ref{subsection:parametrization}. Explicitly, up to level 3,
one finds
\begin{equation}
  \partial T \, T^{-1} =
  \f{1}{3!}\mc{D}\mc{A}_{abc}(t)E^{abc}+
  \f{1}{6!}\mc{D}\mc{A}_{a_1\cdots a_6}(t)E^{a_1\cdots a_6}+
  \f{1}{9!}\mc{D}\mc{A}_{a|b_1\cdots b_8}(t)E^{a|b_1\cdots b_8}+\cdots,
\end{equation}
with 
\begin{equation}
  \begin{array}{rcl}
    \mc{D}\mc{A}_{abc}(t) &=& \displaystyle
    \pa \mc{A}_{abc}(t),
    \\ [0.25 em]
    \mc{D}\mc{A}_{a_1\cdots a_6}(t)&=& \displaystyle
    \pa \mc{A}_{a_1\cdots a_6}(t) +10\mc{A}_{[a_1a_2a_3}(t)
    \pa\mc{A}_{a_4a_5a_6]}(t),
    \\ [0.25 em]
    \mc{D}\mc{A}_{a|b_1\cdots b_8}(t) &=& \displaystyle
    \pa \mc{A}_{a|b_1\cdots b_8}(t)+42\mc{A}_{\left< ab_1b_2\right.}(t)
    \pa \mc{A}_{\left. b_3\cdots b_8\right>}(t)-42
    \pa\mc{A}_{\left< ab_1b_2\right.}(t)\mc{A}_{\left. b_3\cdots b_8\right>}(t),
    \\ [0.25 em]
    & & \displaystyle
    + 280 \mc{A}_{\left< ab_1b_2\right.}(t)
    \mc{A}_{b_3b_4b_5}(t)\pa\mc{A}_{\left. b_6b_7b_8\right> }(t),
  \end{array}
  \label{CovariantDerivatives}
\end{equation}
as computed in~\cite{DHN2}. The notation $\left< a_1 \cdots
a_k\right>$ denotes projection onto the Young tableaux symmetry
carried by the field upon which the covariant derivative
acts\footnote{As an example, consider the projection
$P_{\al\be\ga}\equiv T_{\left<\al\be\ga\right>}$ of a three index
tensor $T_{\al\be\ga}$ onto the Young tableaux
\begin{displaymath}
  {\footnotesize
    \setlength{\tabcolsep}{0.55 em}
    \begin{tabular}{cc}
      \cline{1-2}
      \multicolumn{1}{|c|}{} &
      \multicolumn{1}{c|}{} \\
      \cline{1-2}
      \multicolumn{1}{|c|}{} \\
      \cline{1-1}
    \end{tabular}}\,.
\end{displaymath}
This projection is given by
\begin{displaymath}
  P_{\al\be\ga}=\f{1}{3}(T_{\al\be\ga}+T_{\be \al\ga}-
  T_{\ga\be\al}-T_{\be\ga\al}),
\end{displaymath}
which clearly satisfies
\begin{displaymath}
  P_{\al\be\ga}=-P_{\ga\be \al},
  \qquad
  P_{[\al\be\ga]}=0.
\end{displaymath}
Note also that $P_{\al\be\ga}\neq P_{\be\al\ga}$.}. It should be
stressed that the covariant derivatives $\mc{D}\mc{A}$ have the same
transformation properties as $\mc{A}$ under $SO(10)$ (under which they are inert)
and $GL(10, \mbb{R})$ since the $GL(10, \mbb{R})$
transformations do not depend on time.

%%%%%%%%%%%%%%%%%%%%%%%%%%%%%%%%%%%%%%%%%%%%%%%%%%%%%%%%%%%%%%%%%%%%%%%%%%%%%%%%%%%

\subsection{The   $K(\mc{E}_{10})\times \mc{E}_{10}$-Invariant
  Action at Low Levels}

The action can now be computed using the bilinear form $(\cdot
|\cdot )$ on $E_{10}$,
\begin{equation}
  S_{K(\mc{E}_{10})\bas \mc{E}_{10}}= \int dt\, n(t)^{-1}
  \left(\mc{P}(t)\big|\mc{P}(t)\right),
\end{equation}
where $\mc{P}$ is
obtained by projecting orthogonally onto the maximal compact subalgebra
$\mf{k}\subset E_{10}$ by using the generalized transpose,
\begin{equation}
  ({K^{a}}_b)^{\mc{T}}={K^{b}}_a,
  \qquad
  (E^{abc})^{\mc{T}}=F_{abc}, \cdots \mbox{etc.},
\end{equation}
where
as above $(\ )^{\mc{T}}=-\om(\ )$ (with $\om$ being the Chevalley
involution). \index{Chevalley involution}
We shall compute the action up to, and including, level~3,
\begin{equation}
  S_{K(\mc{E}_{10})\bas \mc{E}_{10}}= \int dt\, n(t)^{-1}
  \left(\mc{L}_0+\mc{L}_1+\mc{L}_2+\mc{L}_3+ \cdots\right).
\end{equation}

From Equation~(\ref{invariantform10.15}) and the fact that generators at
level zero are orthogonal to generators at levels $\not=0$, we see
that $\mc{L}_0$ will be constructed from the level zero part
$\partial H \, H^{-1}$ and will coincide with the Lagrangian
(\ref{LagrangianforGL10SO10}) for the nonlinear sigma model
$SO(10)\bas GL(10, \mbb{R})  $,
\begin{equation}
  \mc{L}_0=\f{1}{4}\left(g^{ac}(t)g^{bd}(t)-g^{ab}(t)g^{cd}(t)\right)
  \pa g_{ab}(t) \, \pa g_{cd}(t).
\end{equation}

To compute the other terms, we use the following trick. The
Lagrangian must be a $GL(10, \mbb{R})$ scalar. One can easily compute it
in the frame where $H= 1$, i.e., where the metric $g_{ab}$ is
equal to $\delta_{ab}$. One can then covariantize the resulting
expression by replacing everywhere $\delta_{ab}$ by $g_{ab}$. To
illustrate the procedure consider the level $1$ term. One has, for
$H = 1$ and at level $1$, $\pa {\mc{V}} \, \mc{V}^{-1} =
\f{1}{3!}\mc{D}\mc{A}_{abc}(t)E^{abc}$ and thus, with the same gauge 
conditions, $\mc{P}(t) = \f{1}{2 \cdot
3!}\mc{D}\mc{A}_{abc}(t)\left(E^{abc} + F^{abc} \right)$ (where we
have raised the indices of $F_{abc}$ with $\delta^{ab}$, $F_{123}
\equiv F^{123}$ etc). Using $(E^{a_1 a_2 a_3}|F^{b_1 b_2 b_3})=
\delta^{a_1 b_1} \delta^{a_2 b_2} \delta^{a_3 b_3} \pm $
permutations that make the expression antisymmetric (3!
terms; see Section~\ref{section:DecompE10E10}), one then gets $\mc{L}_1 = \f{1}{2 \cdot
3!}\mc{D}\mc{A}_{abc}(t) \, \mc{D}\mc{A}_{def}(t) \, \delta^{ad}
\delta^{be} \delta^{cf}$ in the frame where $g_{ab} =
\delta_{ab}$. This yields the level 1 Lagrangian in a general
frame, 
\begin{eqnarray}
  \mc{L}_1 &=& \f{1}{2\cdot 3!} g^{a_1c_1}g^{a_2c_2}g^{a_3c_3}
  \mc{D}\mc{A}_{a_1a_2a_3}(t)\,\mc{D}\mc{A}_{c_1c_2c_3}(t)
  \nonumber
  \\
  &=& \f{1}{2\cdot 3!}\mc{D}\mc{A}_{a_1a_2a_3}(t)\,
  \mc{D}\mc{A}^{a_1a_2a_3}(t). 
\end{eqnarray}
By a similar analysis, the level~2 and 3 contributions are 
\begin{equation}
  \begin{array}{rcl}
    \mc{L}_2 &=& \displaystyle
    \f{1}{2\cdot 6!}\,\mc{D}\mc{A}_{a_1\cdots a_6}(t)\,
    \mc{D}\mc{A}^{a_1\cdots a_6}(t),
    \\ [0.75 em]
    \mc{L}_3 &=& \displaystyle
    \f{1}{2\cdot 9!}\,\mc{D}\mc{A}_{a|b_1\cdots b_8}(t)\,
    \mc{D}\mc{A}^{a|b_1\cdots b_8}(t).
  \end{array}
\end{equation}
Collecting all terms, the final form of the action for
$K(\mc{E}_{10})\bas \mc{E}_{10}$ up to and including level
$\ell=3$ is 
\begin{eqnarray}
  S_{K(\mc{E}_{10})\bas \mc{E}_{10}}&=&
  \int dt\, n(t)^{-1}\left[ \f{1}{4}
  \left(g^{ac}(t)g^{bd}(t)-g^{ab}(t)g^{cd}(t)\right)
  \pa g_{ab}(t)\,\pa g_{cd}(t) \right.
  \nonumber
  \\
  & & \qquad \qquad \quad~~
  + \f{1}{2\cdot 3!}\mc{D}\mc{A}_{a_1a_2a_3}(t)\,
  \mc{D}\mc{A}^{a_1a_2a_3}(t)+ \f{1}{2\cdot 6!}
  \mc{D}\mc{A}_{a_1\cdots a_6}(t)\,\mc{D}\mc{A}^{a_1\cdots a_6}(t)
  \nonumber
  \\
  & & \qquad \qquad \quad~~
  \left. +\f{1}{2\cdot 9!}\mc{D}\mc{A}_{a|b_1\cdots b_8}(t)\,
  \mc{D}\mc{A}^{a|b_1\cdots b_8}(t)+\cdots \right],
  \label{E10action}
\end{eqnarray}
which agrees with the action found in~\cite{DHN2}.

%%%%%%%%%%%%%%%%%%%%%%%%%%%%%%%%%%%%%%%%%%%%%%%%%%%%%%%%%%%%%%%%%%%%%%%%%%%%%%%%%%%

\subsection{The Correspondence}
\label{correspondence}

We shall now relate the equations of motion for the
$K(\mc{E}_{10})\bas \mc{E}_{10}$ sigma model to the equations of
motion of eleven-dimensional supergravity. As the precise
correspondence is not yet known, we shall here only
sketch the main ideas. These work remarkably well at low levels
but need unknown ingredients at higher levels.

We have seen that the sigma model for 
$K(\mc{E}_{10})\bas \mc{E}_{10}$ can be consistently truncated
level by level. More precisely, one can consistently set equal to zero all
covariant derivatives of the fields above a given level and get a
reduced system whose solutions are solutions of the full system.
We shall show here that the consistent truncations of
$K(\mc{E}_{10})\bas \mc{E}_{10}$ at levels 0, 1 and 2
yields equations of motion that coincide with the equations of
motion of appropriate consistent truncations of eleven-dimensional
supergravity, using a prescribed dictionary presented below. We will
also show that the correspondence extends up to parts of level~3.

We recall that in the gauge $N^i=0$ (vanishing shift) and
$A_{0bc}=0$ (temporal gauge), the bosonic fields of eleven-dimensional
supergravity are the spatial metric $\g_{ab}(x^{0},x^{i})$, the
lapse $N(x^{0},x^{i})$ and the spatial components
$A_{abc}(x^0, x^{i})$ of the vector potential 3-form. The
physical field is $F = dA$ and its electric and magnetic
components are, respectively, denoted $F_{0abc}$ and $F_{abcd}$.
The electric field involves only time derivatives of
$A_{abc}(x^0, x^{i})$, while the magnetic field involves spatial
gradients.

\subsubsection*{Levels 0 and 1}

If one keeps only levels zero and one, the sigma model
action~(\ref{E10action}) reduces to
\begin{eqnarray}
  S[g_{ab}(t), \mc{A}_{abc}(t), n(t)] &=& \int dt\,n(t)^{-1}
  \left[\f{1}{4}\left(g^{ac}(t)\,g^{bd}(t)-g^{ab}(t)\,g^{cd}(t)\right)
  \pa {g}_{ab}(t)\,\pa {g}_{cd}(t) \right.
  \nonumber
  \\
  && \qquad \qquad \quad~~
  \left. + \f{1}{2\cdot 3!}\pa {\mc{A}}_{a_1a_2a_3}(t)\,
  \pa {\mc{A}}^{a_1a_2a_3}(t)\right].
  \label{levels0and1}
\end{eqnarray}

Consider now the consistent homogeneous truncation of
eleven-dimensional supergravity in which the spatial metric, the
lapse and the vector potential depend only on time (no spatial
gradient). Then the reduced action for this truncation is
precisely Equation~(\ref{levels0and1}) provided one makes the
identification $t=x^{0}$ and 
\begin{eqnarray}
  g_{ab}(t)&=& \g_{ab}(t),
  \\
  \mc{A}_{abc}(t) &=& A_{abc}(t),
  \\
  n(t) &=& \frac{N(t)}{\sqrt{\g(t)}}
\end{eqnarray}
(see, for instance,~\cite{Demaret}). Also the Hamiltonian
constraints (the only one left) coincide. Thus, there is a perfect
match between the sigma model truncated at level one and
supergravity ``reduced on a 10-torus''. If one were to drop level
one, one would find perfect agreement with pure gravity. In the
following, we shall make the gauge choice $N = \sqrt{\g}$, equivalent
to $n=1$.

\subsection*{Level 2}

At levels 0 and 1, the supergravity fields $\g_{ab}$ and $A_{abc}$
depend only on time. When going beyond this truncation, one needs
to introduce some spatial gradients. Level~2 introduces spatial
gradients of a very special type, namely allows for a homogeneous
magnetic field. This means that $A_{abc}$ acquires a space
dependence, more precisely, a linear one (so that its gradient
does not depend on $x$). However, because there is no room for
$x$-dependence on the sigma model side, where the only independent
variable is $t$, we shall use the trick to describe the magnetic
field in terms of a dual potential $A_{a_1 \cdots a_6}$. Thus,
there is a close interplay between duality, the sigma model
formulation, and the introduction of spatial gradients.

There is no tractable, fully satisfactory variational formulation
of eleven-dimensional supergravity where both the 3-form
potential and its dual appear as independent variables in the
action, with a quadratic dependence on the time derivatives (this
would be double-counting, unless an appropriate self-duality
condition is imposed~\cite{Dualisation1,Dualisation2}). This means that from now on, we shall not compare
the actions of the sigma model and of supergravity but, rather, only
their respective equations of motion. As these involve the
electromagnetic field and not the potential, we rewrite the
correspondence found above at levels 0 and 1 in terms of the metric
and the electromagnetic field as 
\begin{equation}
  \begin{array}{rcl}
    g_{ab}(t)&=& \g_{ab}(t),
    \\ [0.25 em]
    \mc{D}\mc{A}_{abc}(t) &=& F_{0abc}(t).
  \end{array}
\end{equation}
The equations of motion for the nonlinear sigma model, obtained from
the variation of the Lagrangian Equation~(\ref{E10action}), truncated
at level two, read explicitly 
\begin{equation}
  \begin{array}{rcl}
    \displaystyle
    \f{1}{2}\pa \left(n^{-1}g^{ac}\,\pa g_{cb}\right) &=&
    \displaystyle
    \f{n^{-1}}{4}\left(\mc{D}\mc{A}^{ac_1c_2}\,
    \mc{D}\mc{A}_{bc_1c_2}-\f{1}{9}
    {\delta^{a}}_b \mc{D}\mc{A}^{c_1c_2c_3}\,
    \mc{D}\mc{A}_{c_1c_2c_3} \right)
    \\ [1 em]
    & & \displaystyle
    +\f{n^{-1}}{2\cdot 5!}
    \left(\mc{D}\mc{A}^{a c_1\cdots c_5}\,
    \mc{D}\mc{A}_{b c_1\cdots c_5}-\f{1}{9} {\delta^{a}}_b
    \mc{D}\mc{A}^{c_1\cdots c_6}\,
    \mc{D}\mc{A}_{c_1\cdots c_6}\right),
    \\ [1 em]
    \pa\left(n^{-1}\mc{D}\mc{A}^{a_1a_2a_3}\right)&=&
    \displaystyle
    - \f{1}{3!}\, n^{-1} \mc{D}\mc{A}^{a_1\cdots a_6}\,
    \mc{D}\mc{A}_{a_4a_5a_6},
    \\ [1 em]
    \pa\left(n^{-1}\mc{D}\mc{A}^{a_1\cdots a_6}\right)&=& 0.
  \end{array}
  \label{EOMLevel2}
\end{equation}

In addition, we have the constraint obtained by varying $n$, 
\begin{eqnarray}
  \left(\mc{P}|\mc{P}\right)& =& 
  \f{1}{4}\left(g^{ac}\,g^{bd}-g^{ab}\,g^{cd}\right)
  \pa g_{ab}\,\pa g_{cd}
  \nonumber
  \\
  & & + \f{1}{2\cdot 3!}\mc{D}\mc{A}^{a_1a_2a_3}\,
  \mc{D}\mc{A}_{a_1a_2a_3}+ \f{1}{2\cdot 6!}
  \mc{D}\mc{A}^{a_1\cdots a_6}\,\mc{D}\mc{A}_{a_1\cdots a_6} 
  \nonumber
  \\
  &=&0.
  \label{HamiltonianConstraintLevel2}
\end{eqnarray}

On the supergravity side, we truncate the
equations to metrics $\g_{ab}(t)$ and electromagnetic fields
$F_{0abc}(t)$, $F_{abcd}(t)$ that depend only on time. We take, as in
Section~\ref{Section:HyperbolicBilliard}, the spacetime metric to be of the form
\begin{equation}
  ds^2=-N^2(t)\,dt^2+\g_{ab}(t)\,dx^{a}\,dx^{b},
\end{equation}
but now with $x^0\equiv t$. In the following we use Greek letters $\lambda,
\sigma, \rho, \cdots $ to denote eleven-dimensional spacetime indices,
and Latin letters $a, b, c, \cdots $ to denote ten-dimensional spatial
indices.

The equations of motion and the Hamiltonian constraint for
eleven-dimensional supergravity have been explicitly written
in~\cite{Demaret}, so they can be expediently compared with the
equations of motion of the sigma model. The dynamical equations for
the metric read 
\begin{eqnarray}
  \f{1}{2}\pa\left(\sqrt{\g}N^{-1}\g^{ac}\pa \g_{cb}\right) &=&
  \frac{1}{12} N \sqrt{\g} F^{a \rho \sigma \tau}F_{b \rho \sigma \tau} -
  \frac{1}{144}N \sqrt{\g} \, {\delta^a}_b\, F^{\lambda \rho \sigma \tau}
  F_{\lambda \rho \sigma \tau}
  \nonumber
  \\
  &=& \f{1}{4}N^{-1}\sqrt{\g}F^{0ac_1c_2}F_{0bc_1c_2}-
  \f{1}{36}N^{-1}\sqrt{\g}\, {\delta^{a}}_b\, F^{0c_1c_2c_3}F_{0c_1c_2c_3}
  \nonumber
  \\
  & & +\f{1}{12}N\sqrt{\g}F^{ac_1c_2c_3}F_{bc_1c_2c_3}-
  \f{1}{144}N\sqrt{\g}\, {\delta^{a}}_b\, F^{c_1c_2c_3c_4}F_{c_1c_2c_3c_4},
  \nn \\
  \label{Einstein00}
\end{eqnarray}
and for the electric and magnetic fields we have, respectively, the
equations of motion and the Bianchi identity,
\begin{equation}
  \begin{array}{rcl}
    \displaystyle
    \pa\left(F^{0abc} N \sqrt{\g} \right) &=&
    \displaystyle
    \frac{1}{144} \varepsilon^{0 a b c d_1 d_2 d_3 e_1 e_2 e_3 e_4}
    F_{0 d_1 d_2 d_3} F_{e_1 e_2 e_3 e_4},
    \\ [0.75 em]
    \displaystyle
    \pa F_{a_1 a_2 a_3 a_4} &=& 0.
  \end{array}
  \label{ChernSimons} 
\end{equation}
Furthermore we have the Hamiltonian constraint \index{Hamiltonian constraint} 
\begin{equation}
  \f{1}{4}\left(\g^{ac}\g^{bd}-\g^{ab}\g^{cd}\right)
  \pa \g_{ab}\,\pa \g_{cd} + \frac{1}{12} F^{0abc}F_{0abc} +
  \frac{1}{48}N^2 F^{abcd}F_{abcd} = 0.
  \label{HamiltonianC0}
\end{equation}
We will not discuss any of the other constraints in this thesis. See \cite{Damour:2007dt,Damour:2009ww} for interesting recent results regarding the role of constraints in the geodesic $\mc{E}_{10}$-invariant sigma model.

One finds again perfect agreement between the sigma model equations, (\ref{EOMLevel2}) and~(\ref{HamiltonianConstraintLevel2}),
and the equations of eleven-dimensional supergravity, (\ref{Einstein00}) and (\ref{HamiltonianC0}), provided one
extends the above dictionary through~\cite{DHN2}
\begin{equation}
  \mc{D}\mc{A}^{a_1 \cdots a_6}(t) = - \frac{1}{4!}
  \varepsilon^{a_1 \cdots a_6 b_1 b_2 b_3 b_4}F_{b_1 b_2 b_3 b_4}(t).
\end{equation}
This result appears to be quite remarkable, because the Chern--Simons
term in (\ref{ChernSimons}) is in particular reproduced with
the correct coefficient, which in eleven-dimensional supergravity is
fixed by invoking supersymmetry.

\subsection*{Level 3}

Level~3 should correspond to the introduction of further
controlled spatial gradients, this time for the metric. Because
there is no room for spatial derivatives as such on the sigma
model side, the trick is again to introduce a dual graviton field.
When this dual graviton field is non-zero, the metric does depend
on the spatial coordinates.

Satisfactory dual formulations of non-linearized gravity do not exist. At
the linearized level, however, the problem is well understood
since the pioneering work by Curtright~\cite{Curtright}. In
eleven spacetime dimensions, the dual graviton field is {\em
described precisely by a tensor $\mc{A}_{a \vert b_1 \cdots b_8}$
with the mixed symmetry of the Young tableau $[1, 0, 0, 0, 0, 0,
0, 1,0]$ appearing at level~3 in the sigma model description}.
Exciting this field, i.e., assuming $\mc{D}\mc{A}_{a \vert b_1
\cdots b_8} \not=0$ amounts to introducing spatial gradients for the
metric -- and, for that matter, for the other fields as well -- as
follows. Instead of considering fields that are homogeneous on a
torus, one considers fields that are homogeneous on non-abelian
group manifolds. This introduces spatial gradients (in coordinate
frames) in a well controlled manner.

Let $\theta^{a}$ be the group invariant one-forms, with structure
equations
\begin{equation}
  d \theta^a = \frac{1}{2} {C^a}_{bc} d \theta^b \wedge d \theta^c.
\end{equation}
We shall assume that ${C^a}_{ac} = 0$ (``Bianchi class
A''). Truncation at level~3 assumes that the metric and the electric
and magnetic fields depend only on time in this frame and that the
${C^a}_{bc}$ are constant (corresponding to a group). The
supergravity equations have been written in that case
in~\cite{Demaret} and can be compared with the sigma model
equations. There is almost a complete match between both sets of
equations provided one extends the dictionary at level~3 through
\begin{equation}
  \mc{D}\mc{A}^{a| b_1\cdots b_8}(t)=
  \f{3}{2}\varepsilon^{b_1\cdots b_8 cd}{{C}^{a}}_{cd}
\end{equation}
(with the equations of motion of the sigma model implying that
$\mc{D}\mc{A}^{a| b_1\cdots b_8}$ does not depend on time). Note that
to define an invertible mapping between the level three fields and the
${C^{a}}_{bc}$, it is important that ${C^{a}}_{bc}$ be traceless;
there is no ``room'' on level three on the sigma model side to
incorporate the trace of ${C^{a}}_{bc}$.

With this correspondence, the match works perfectly for real roots up
to, and including, level three. However, it fails for fields
associated with imaginary roots (level~3 is the first time imaginary
roots appear, at height 30)~\cite{DHN2}. In fact, the terms that match
correspond to ``$SL(10, \mbb{R})$-covariantized $E_8$'', i.e., to
fields associated with roots of $E_8$ and their images under the Weyl
group of $SL(10, \mbb{R})$.

Since the match between the sigma model equations and supergravity
fails at level~3 under the present line of investigation, we shall
not provide the details but refer instead to~\cite{DHN2} for more
information. The correspondence up to level~3 was also checked
in~\cite{DamourNicolaiLowLevels} through a slightly different
approach, making use of a formulation with local frames, i.e., using
local flat indices rather than global indices as in the present
treatment.

Let us note here that higher level fields of $E_{10}$, corresponding
to imaginary roots, have been considered from a different point of
view in~\cite{Brown:2004jb}, where they were associated with certain
brane configurations (see also~\cite{Brown:2004ar,Carlevaro}).

\subsection*{The Dictionary}

One may view the above failure at level~3 as a serious flaw
to the sigma model approach to exhibiting the $E_{10}$
symmetry\footnote{This does not exclude that other approaches
  would be successful. That $E_{10}$, or perhaps $E_{11}$, does encode
  a lot of information about M-theory is a fact, but that this should
  be translated into a sigma model reformulation of the theory appears
  to be questionable.}. Let us, however, be optimistic for a moment
and assume that these problems will somehow get resolved, perhaps by
changing the dictionary or by including higher order terms. So, let us
proceed.

What would be the meaning of the higher level fields? As discussed in
Section~\ref{section:higherlevels}, there are indications that fields
at higher levels contain higher order spatial gradients and therefore
enable us to reconstruct completely, through something similar to a
Taylor expansion, the most general field configuration from the fields
at a given spatial point.

From this point of view, the relation between the supergravity degrees
of freedom $\g_{ij}(t,x)$ and $F_{(4)}(t,x)=dA_{(3)}(t, x)$ would be
given, at a specific spatial point $x=\mathbf{x}_0$ and in a suitable
spatial frame $\theta^{a}(x)$ (that would also depend on $x$), by the
following ``dictionary'':
\begin{equation}
  \begin{array}{rcl}
    g_{ab}(t) &=& \g_{ab}(t, \mathbf{x}_0),
    \\ [0.75 em]
    \mc{D}\mc{A}_{abc}(t)& =& F_{tabc}(t, \mathbf{x}_0),
    \\ [0.75 em]
    \mc{D}\mc{A}^{a_1\cdots a_6}(t) &=& \displaystyle
    -\f{1}{4!}\varepsilon^{a_1\cdots a_6 bcde}F_{bcde}(t, \mathbf{x}_0),
    \\ [1 em]
    \mc{D}\mc{A}^{a| b_1\cdots b_8}(t)&=& \displaystyle
    \f{3}{2}\varepsilon^{b_1\cdots b_8 cd}{{C}^{a}}_{cd}(\mathbf{x}_0),
  \end{array}
\end{equation}
which reproduces in the homogeneous case what we have seen up to
level~3.

This correspondence goes far beyond that of the algebraic
description of the BKL-limit \index{BKL-limit}in terms of Weyl reflections in the
simple roots of a Kac--Moody algebra. \index{Kac--Moody algebra}
Indeed, the dynamics of the
billiard is controlled entirely by the walls associated with {\em
simple} roots and thus does not transcend height one. Here, we go
to a much higher height and successfully extend (unfortunately
incompletely) the intriguing connection between eleven-dimensional
supergravity and $E_{10}$.

%%%%%%%%%%%%%%%%%%%%%%%%%%%%%%%%%%%%%%%%%%%%%%%%%%%%%%%%%%%%%%%%%%%%%%%%%%%%%%%%%%%

\subsection{Higher Levels and Spatial Gradients}
\label{section:higherlevels}

We have seen that the correspondence between the
$\mc{E}_{10}$-invariant sigma model and eleven-dimensio\-nal
supergravity fails when we include spatial gradients beyond first
order. It is nevertheless believed that the information about
spatial gradients is somehow encoded within the algebraic
description: One idea is that space is ``smeared out'' among
the infinite number of fields contained in $\mc{E}_{10}$ and it is
for this reason that a direct dictionary for the inclusion of
spatial gradients is difficult to find. If true, this would imply
that we can view the level expansion on the algebraic side as
reflecting a kind of ``Taylor expansion'' in spatial gradients on
the supergravity side. Below we discuss some speculative ideas
about how such a correspondence could be realized in practice.

\subsection*{The ``Gradient Conjecture''}

One intriguing suggestion put forward in~\cite{DHN2}
was that fields associated to certain ``affine representations''
of $E_{10}$ \index{$E_{10}$}could be interpreted as spatial derivatives acting on
the level one, two and three fields, thus providing a direct
conjecture for how space ``emerges'' through the level
decomposition of $E_{10}$. The representations in question
are those for which the Dynkin label associated with the
overextended root of $E_{10}$ vanishes, and hence these
representations are realized also in a level decomposition \index{level decomposition}of the
regular $E_9$-subalgebra obtained by removing the overextended
node in the Dynkin diagram of $E_{10}$.

The affine representations were discussed in
Section~\ref{section:LevelDecomposition} and we recall that they are
given in terms of three infinite towers of generators, with the
following $\mf{sl}(10, \mbb{R})$-tensor structures,
\begin{equation}
  {E^{a_1a_2a_3}}_{b_1\cdots b_k},
  \qquad
  {E^{a_1\cdots a_6}}_{b_1\cdots b_k},
  \qquad
  {E^{a_1|a_2\cdots a_9}}_{b_1\cdots b_k},
\end{equation}
where the upper indices have the same Young tableau symmetries as the
$\ell=1, 2$ and $3$ representations, while the lower indices are all
completely symmetric. In the sigma model these generators of
$\mc{E}_{10}$ are parametrized by fields exhibiting the same index
structure, i.e., ${\mc{A}_{a_1a_2a_2}}^{b_1\cdots b_k}(t)$,
${\mc{A}_{a_1\cdots a_6}}^{b_1\cdots b_k}(t)$ and
${\mc{A}_{a_1|a_2\cdots a_9}}^{b_1\cdots b_k}(t)$.

The idea is now that the three towers of fields have precisely the
right index structure to be interpreted as spatial gradients of
the low level fields 
\begin{equation}
  \begin{array}{rcl}
    {\mc{A}_{a_1a_2a_2}}^{b_1\cdots b_k}(t)&=&
    \pa^{b_1}\cdots \pa^{b_k}\mc{A}_{a_1a_2a_3}(t),
    \\ [0.25 em]
    {\mc{A}_{a_1\cdots a_6}}^{b_1\cdots b_k}(t)&=&
    \pa^{b_1}\cdots \pa^{b_{k}}\mc{A}_{a_1\cdots a_6}(t),
    \\ [0.25 em]
    {\mc{A}_{a_1|a_2\cdots a_9}}^{b_1\cdots b_k}(t)&=&
    \pa^{b_1}\cdots \pa^{b_{k}}\mc{A}_{a_1|a_2\cdots a_9}(t).
  \end{array}
\end{equation}

Although appealing and intuitive as it is, this conjecture is
difficult to prove or to check explicitly, and not much progress
in this direction has been made since the original proposal.
Let us point out, however, that recently~\cite{Englert:2007qb} this problem was attacked
from a rather different point of view with some very interesting
results, indicating that the gradient conjecture may need modification.

%%%%%%%%%%%%%%%%%%%%%%%%%%%%%%%%%%%%%%%%%%%%%%%%%%%
%
%\chapter{Geometric Configurations and Cosmological Solutions from $E_{10}$}
%
%%%%%%%%%%%%%%%%%%%%%%%%%%%%%%%%%%%%%%%%%%%%%%%%%%%

\chapter{Geometric Configurations and Cosmological Solutions from $\mc{E}_{10}$}
\label{Chapter:GeometricConfigurations}

In this chapter we will show that the low level
equivalence between the $K(\mc{E}_{10})\bas \mc{E}_{10}$ sigma
model and eleven-dimensional supergravity can be put to practical
use for finding exact solutions. We will restrict our analysis to a cosmological sector where it is
assumed that all spatial gradients can be neglected so that all
fields depend only on time. Moreover, we impose diagonality of the
spatial metric. These conditions must of course be compatible with
the equations of motion; if the conditions are imposed initially,
they should be preserved by the time evolution.

A large class of solutions to eleven-dimensional supergravity
preserving these conditions were found in~\cite{Demaret}. These
solutions have zero magnetic field but have a restricted number of
electric field components turned on. Surprisingly, it was found
that such solutions have an elegant interpretation in terms of so
called \emph{geometric configurations}, 
\index{geometric configuration} denoted $(n_m, g_3)$, of
$n$ points and $g$ lines (with $n \leq 10$) drawn on a plane with
certain pre-determined rules. That is, for each geometric
configuration (whose definition is recalled below) one can
associate a diagonal solution with some non-zero electric field
components $F_{tijk}$, determined by the configuration. In this
section we re-examine this result from the point of view of the
sigma model based on $K(\mc{E}_{10})\bas \mc{E}_{10}$.

We show, following~\cite{Henneaux:2006gp}, that each configuration
$(n_m, g_3)$ encodes information about a (regular) subalgebra
$\mgb$ of $E_{10}$, \index{$E_{10}$} and the supergravity solution
associated to the configuration $(n_m g_3)$ can be obtained by
restricting the $\mc{E}_{10}$-sigma model to the subgroup $\bar{G}$
whose Lie algebra is $\mgb$. Therefore, we will here make use of both
the level truncation and the subgroup truncation simultaneously; first
by truncating to a certain level and then by restricting to the
relevant $\mgb$-algebra generated by a subset of the representations
at this level. 

This chapter is based on {\bf Papers I} and {\bf II}, written in collaboration with Marc Henneaux, Mauricio Leston and Philippe Spindel, as well as on {\bf Paper III}, written in collaboration with Marc Henneaux and Philippe Spindel.

%%%%%%%%%%%%%%%%%%%%%%%%%%%%%%%%%%%%%%%%%%%%%%%%%%%%%%%%%%%%%%%%%%%%%%%%%%%%%%%%%%%
%%%%%%%%%%%%%%%%%%%%%%%%%%%%%%%%%%%%%%%%%%%%%%%%%%%%%%%%%%%%%%%%%%%%%%%%%%%%%%%%%%%

\section{Bianchi~I Models and Eleven-Dimensional Supergravity}

On the supergravity side, we will restrict the metric and the
electromagnetic field to depend on time only, 
\begin{equation}
  \begin{array}{rcl}
    ds^2 & = & - N^2(t) \, dt^2 + \g_{ab}(t)\,dx^a\,dx^b,
    \\ [0.25 em]
    F_{\lambda \rho \sigma \tau} & =&
    F_{\lambda \rho \sigma \tau}(t).
  \end{array}
\end{equation}
Recall from Section~\ref{section:E10SigmaModel} that with these
ans\"atze the dynamical equations of motion of eleven-dimensional
supergravity reduce to~\cite{Demaret}
\begin{eqnarray}
  \f{1}{2}\pa\left(\sqrt{\g}N^{-1}\g^{ac}\pa \g_{cb}\right) &=&
  \frac{1}{12} N \sqrt{\g} F^{a \rho \sigma \tau} F_{b \rho \sigma \tau} -
  \frac{1}{144}N \sqrt{\g} \, {\delta^a}_b\, F^{\lambda \rho \sigma \tau}
  F_{\lambda \rho \sigma \tau},
  \label{Einstein0}
  \\
  \pa\left(F^{tabc} N \sqrt{\g} \right) &=&
  \frac{1}{144} \varepsilon^{t a b c d_1 d_2 d_3 e_1 e_2 e_3 e_4}
  F_{t d_1 d_2 d_3} F_{e_1 e_2 e_3 e_4},
  \\
  \pa F_{a_1 a_2 a_3 a_4} &=& 0. 
\end{eqnarray}
This corresponds to the truncation of the sigma model at
level~2 which, as we have seen, completely matches the
supergravity side. We also defined $\pa\equiv \pa_t$ as in
Section~\ref{section:E10SigmaModel}. Furthermore we have the following
constraints,
\begin{eqnarray}
  \f{1}{4}\left(\g^{ac}\g^{bd}-\g^{ab}\g^{cd}\right)
  \pa\g_{ab}\,\pa\g_{cd} + \frac{1}{12} F^{tabc}F_{tabc} +
  \frac{1}{48}N^2 F^{abcd}F_{abcd} &=& 0,
  \label{HamiltonianC}
  \\
  \frac{1}{6} N F^{tbcd}F_{abcd} &=& 0,
  \label{momentum}
  \\
  \varepsilon^{tabc_1 c_2 c_3 c_4 d_1 d_2 d_3 d_4}
  F_{c_1 c_2 c_3 c_4} F_{d_1 d_2 d_3 d_4} &=& 0,
\end{eqnarray}
which are, respectively, the Hamiltonian constraint, momentum
constraint and Gauss' law. Note that Greek indices $\al, \be, \ga,
\cdots$ correspond to the full eleven-dimensional spacetime, while
Latin indices $a, b, c, \cdots$ correspond to the ten-dimensional
spatial part.

We will further take the metric to be purely time-dependent and
diagonal,
\begin{equation}
  ds^2 = - N^{2}(t) \, dt^2 + \sum_{i = 1}^{10} a_i^2(t) (dx^i)^2.
  \label{diagonalmetric}
\end{equation}
This form of the metric has manifest invariance under the ten distinct
spatial reflections 
\begin{equation}
  \begin{array}{rcl}
    x^j &\rightarrow & -x^j,
    \\ [0.25 em]
    x^{i\neq j} & \rightarrow & x^{i\neq j},
  \end{array}
\end{equation}
and in order to ensure compatibility with the Einstein equations, the
energy-momentum tensor of the 4-form field strength must also be
diagonal.

%%%%%%%%%%%%%%%%%%%%%%%%%%%%%%%%%%%%%%%%%%%%%%%%%%%%%%%%%%%%%%%%%%%%%%%%%%%%%%%%%%%

\subsection{Diagonal Metrics and Geometric Configurations}

Assuming zero magnetic field (this restriction will be lifted below),
one way to achieve diagonality of the energy-momentum tensor is to
assume that the non-vanishing components of the electric field
$F^{\perp a b c}=N^{-1}F_{tabc}$ are determined by \emph{geometric
  configurations} $(n_m,g_3)$ with $n \leq 10$~\cite{Demaret}.

A geometric configuration \index{geometric configuration|bb} $(n_m,g_3)$ is a set of $n$ points and $g$
lines with the following incidence rules~\cite{Kantor, Hilbert, Page}:

\begin{enumerate}
\item Each line contains three points. \label{rule_1}
\item Each point is on $m$ lines. \label{rule_2}
\item Two points determine at most one line. \label{rule_3}
\end{enumerate}

It follows that two lines have at most one point
in common. It is an easy exercise to verify that $m n = 3 g $. An
interesting question is whether the lines can actually be realized
as straight lines in the (real) plane, but, for our purposes, it
is not necessary that it should be so; the lines can be bent.

Let $(n_m,g_3)$ be a geometric configuration with $n \leq 10$
points. We number the points of the configuration $1, \cdots, n$.
We associate to this geometric configuration a pattern of
electric field components $F^{\perp a b c}$ with the following
property: $F^{\perp a b c}$ can be non-zero only if the triple
$(a,b,c)$ is a line of the geometric configuration. If it is
not, we take $F^{\perp a b c} = 0$. It is clear that this
property is preserved in time by the equations of motion (in the
absence of magnetic field). Furthermore, because of Rule~\ref{rule_3}
above, the products $F^{\perp a b c} F^{\perp a' b' c'} g_{bb'}
g_{cc'}$ vanish when $a \not= a'$ so that the energy-momentum tensor
is diagonal.

%%%%%%%%%%%%%%%%%%%%%%%%%%%%%%%%%%%%%%%%%%%%%%%%%%%%%%%%%%%%%%%%%%%%%%%%%%%%%%%%%%%
%%%%%%%%%%%%%%%%%%%%%%%%%%%%%%%%%%%%%%%%%%%%%%%%%%%%%%%%%%%%%%%%%%%%%%%%%%%%%%%%%%%

\section{Geometric Configurations and Regular Subalgebras of   $E_{10}$}
\label{geometric}

We prove here that the conditions on the electric field embodied
in the geometric configurations $(n_m, g_3)$ have a direct
Kac--Moody algebraic \index{Kac--Moody algebra}interpretation. They simply correspond to
a consistent truncation of the $E_{10}$ nonlinear sigma model to a
$\bar{\mf{g}}$ nonlinear sigma model, where $\bar{\mf{g}}$ is a
rank $g$ Kac--Moody subalgebra of $E_{10}$ (or a quotient of such a
Kac--Moody subalgebra by an appropriate ideal when the relevant
Cartan matrix has vanishing determinant), with three crucial
properties: (i) It is regularly embedded in $E_{10}$ (see Section \ref{regular} for the definition of regular embedding), (ii) it is generated by electric roots only, and (iii)
every node $P$ in its Dynkin diagram \index{Dynkin diagram}$\mathbb{D}_{\mgb}$ is linked to
a number $k$ of nodes that is independent of $P$ (but depend on the
algebra). We find that the Dynkin diagram $\mathbb{D}_{\mgb}$ of
$\mgb$ is the \emph{line incidence diagram} \index{line incidence diagram|bb} of the geometric
configuration $(n_m, g_3)$, in the sense that (i) each line of $(n_m,
g_3)$ defines a node of $\mathbb{D}_{\mgb}$, and (ii) two nodes of
$\mathbb{D}_{\mgb}$ are connected by a single bond iff the
corresponding lines of $(n_m, g_3)$ have no point in common. This
defines a geometric duality between a configuration $(n_m, g_3)$ and
its associated Dynkin diagram $\mathbb{D}_{\mgb}$. In the following we
shall therefore refer to configurations and Dynkin diagrams
\index{Dynkin diagram}related in this way as \emph{dual}.

None of the algebras $\mgb$ relevant to the truncated models turn out
to be hyperbolic: They can be finite, affine, or Lorentzian with
infinite-volume Weyl chamber. \index{Weyl chamber} Because of this,
the solutions are non-chaotic. After a finite number of collisions,
they settle asymptotically into a definite Kasner regime (both in the
future and in the past).

%%%%%%%%%%%%%%%%%%%%%%%%%%%%%%%%%%%%%%%%%%%%%%%%%%%%%%%%%%%%%%%%%%%%%%%%%%%%%%%%%%%

\subsection{General Considerations}

In order to match diagonal Bianchi~I cosmologies with the sigma
model, one must truncate the
$\mathcal{K}(\mathcal{E}_{10})\bas \mathcal{E}_{10}$ action in such a
way that the sigma model metric $g_{ab}$ is diagonal. This will be
the case if the subalgebra $\mgb$ to which one truncates has no
generator $K^i{}_{j}$ with $i \not=j$. The off-diagonal
components of the metric are precisely the exponentials of the
associated sigma model fields. The set of simple roots of $\mgb$
should therefore not contain any root at level zero.

Consider ``electric'' regular subalgebras of $E_{10}$, for which
the simple roots are all at level one, where the 3-form electric
field variables live. These roots can be parametrized by three
indices corresponding to the indices of the electric field, with
$i_1<i_2<i_3$. We denote them $\alpha_{i_1i_2i_3}$. For
instance, $\alpha_{123} \equiv \alpha_{10}$. In terms of the
$\beta$-parametrization of~\cite{ArithmeticalChaos, DHNReview}, one
has $\alpha_{i_1i_2i_3} = \beta^{i_1} + \beta^{i_2} +
\beta^{i_3}$.

Now, for $\mgb$ to be a regular subalgebra, \index{regular subalgebra}it must fulfill the condition that the difference between any two of
its simple roots is not a root of $E_{10}$: $\alpha_{i_1i_2i_3} -
\alpha_{i'_1i'_2i'_3} \notin \Phi_{E_{10}}$ for any pair
$\alpha_{i_1i_2i_3}$ and $\alpha_{i'_1i'_2i'_3}$ of simple roots
of $\mgb$. But one sees by inspection of the commutator of
$E^{i_1i_2i_3}$ with $F_{i'_1i'_2i'_3}$ in
Equation~(\ref{level1commutationrelations}) that $\alpha_{i_1i_2i_3} -
\alpha_{i'_1i'_2i'_3}$ is a root of $E_{10}$ if and only if the sets
$\{i_1, i_2, i_3\}$ and $\{i'_1, i'_2, i'_3\}$ have exactly two points
in common. For instance, if $i_1= i'_1$, $i_2 = i'_2$ and $i_3 \not=
i'_3$, the commutator of $E^{i_1i_2i_3}$ with $F_{i'_1i'_2i'_3}$
produces the off-diagonal generator $K^{i_3}{}_{i'_3}$
corresponding to a level zero root of $E_{10}$. In order to fulfill
the required condition, one must avoid this case, i.e., one must
choose the set of simple roots of the electric regular subalgebra
$\mgb$ in such a way that given a pair of indices $(i_1, i_2)$, there
is at most one $i_3$ such that the root $\alpha_{i j k}$ is a simple
root of $\mgb$, with $(i,j,k)$ being the re-ordering of $(i_1, i_2, i_3 )$
such that $i<j<k$.

To each of the simple roots $\alpha_{i_1i_2i_3}$ of $\mgb$, one can
associate the line $(i_1,i_2,i_3)$ connecting the three points
$i_1$, $i_2$ and $i_3$. If one does this, one sees that the above
condition is equivalent to the following statement: \emph{The set of
points and lines associated with the simple roots of $\mgb$ must
fulfill the third rule defining a geometric configuration,
namely, that two points determine at most one line}. Thus, this
geometric condition has a nice algebraic interpretation in terms
of regular subalgebras of $E_{10}$.

The first rule, which states that each line contains 3 points, is
a consequence of the fact that the $E_{10}$-generators at level
one are the components of a 3-index antisymmetric tensor. The
second rule, that each point is on $m$ lines, is less fundamental
from the algebraic point of view since it is not required to hold
for $\mgb$ to be a regular subalgebra. It was imposed
in~\cite{Demaret} in order to allow for solutions isotropic in the
directions that support the electric field. We keep it here as it
yields interesting structure.

%%%%%%%%%%%%%%%%%%%%%%%%%%%%%%%%%%%%%%%%%%%%%%%%%%%%%%%%%%%%%%%%%%%%%%%%%%%%%%%%%%%

\subsection{Incidence Diagrams and Dynkin Diagrams}
\label{section:incidenceanddynkin}

We have just shown that each geometric configuration \index{geometric configuration} $(n_m,g_3)$
with $n \leq 10$ defines a regular subalgebra $\mgb$ of $E_{10}$. In
order to determine what this subalgebra $\mgb$ is, one needs, according to the theorem recalled in Section \ref{regular}, to
compute the Cartan matrix \index{Cartan matrix}
\begin{equation}
  C = [C_{i_1 i_2 i_3, i'_1 i'_2 i'_3}] =
  \left[\left(\alpha_{i_1 i_2 i_3} \vert \alpha_{i'_1 i'_2 i'_3}\right)\right]
\end{equation}
(the real roots of $E_{10}$ have length squared equal to 2).
According to that same theorem, the algebra $\mgb$ is then just the
rank $g$ Kac--Moody algebra with Cartan matrix $C$, unless $C$ has zero
determinant, in which case $\mgb$ might be the quotient of that
algebra by a nontrivial ideal.

Using for instance the root parametrization
of~\cite{ArithmeticalChaos, DHNReview} and the expression of the
scalar product in terms of this parametrization, one
easily verifies that the scalar product is equal to
\begin{equation}
  \left(\alpha_{i_1 i_2 i_3} \vert \alpha_{i'_1 i'_2 i'_3}\right) =
  \left\{
    \begin{array}{l@{\qquad}l}
      2 & \mbox{if all three indices coincide},
      \\
      1 & \mbox{if two and only two indices coincide},
      \\
      0 & \mbox{if one and only one index coincides},
      \\
      -1 & \mbox{if no indices coincide}.
    \end{array}
  \right.
\end{equation}
The second possibility does not arise in our case since we deal with
geometric configurations. For completeness, we also list the scalar
products of the electric roots $\alpha_{ijk}$ ($i<j<k$) with the
symmetry roots $\alpha_{\ell m}$ ($\ell < m$) associated with the
raising operators $K^m {}_{\ell}$: 
\begin{equation}
  \left(\alpha_{ijk} \vert \alpha_{\ell m}\right) =
  \left\{
    \begin{array}{l@{\qquad}l}
      -1 & \mbox{if }\ell\in \{i,j,k\}\mbox{ and }m \notin \{i,j,k\},
      \\
      0 & \mbox{if }\{\ell, m\} \subset \{i,j,k\}\mbox{ or }\{\ell, m\}
      \cap \{i,j, k\} = \emptyset,
      \\
      1 & \mbox{if }\ell\notin \{i,j,k\}\mbox{ and }m \in \{i,j,k\},
    \end{array}
  \right.
\end{equation}
as well as the scalar products of the symmetry
roots among themselves, 
\begin{equation}
  \left(\alpha_{ij} \vert \alpha_{\ell m}\right) =
 \left\{
    \begin{array}{l@{\qquad}l}
      - 1 & \mbox{if }j = \ell\mbox{ or }i = m,
      \\
      0 & \mbox{if }\{\ell, m\} \cap \{i,j\} = \emptyset,
      \\
      1 & \mbox{if }i = \ell\mbox{ or }j \not= m,
      \\
      2 & \mbox{if }\{\ell, m\} = \{i,j\}.
    \end{array}
  \right.
\end{equation} 
Given a geometric configuration \index{geometric configuration} $(n_m,g_3)$, one can
associate with it a ``line incidence diagram'' \index{line incidence diagram} that encodes the
incidence relations between its lines. To each line of
$(n_m,g_3)$ corresponds a node in the incidence diagram. Two nodes
are connected by a single bond if and only if they correspond to
lines with no common point (``parallel lines''). Otherwise, they
are not connected\footnote{One may also consider a point incidence
diagram defined as follows: The nodes of the point incidence
diagram are the points of the geometric configuration. Two nodes
are joined by a single bond if and only if there is no straight
line connecting the corresponding points. The point incidence
diagrams of the configurations $(9_3,9_3)$ are given
in~\cite{Hilbert}. For these configurations, projective duality
between lines and points lead to identical line and point incidence
diagrams. Unless otherwise stated, the expression ``incidence
diagram'' will mean ``line incidence diagram''. \index{line incidence diagram}}. By inspection of the
above scalar products, we come to the important conclusion that {\em
  the Dynkin diagram of the regular, rank $g$, Kac--Moody subalgebra
  $\mgb$ associated with the geometric configuration $(n_m,g_3)$ is
  just its line incidence diagram.} We shall call the Kac--Moody
algebra $\mgb$ the algebra ``dual'' to the geometric configuration
$(n_m,g_3)$.

Because the geometric configurations have the property that the
number of lines through any point is equal to a constant $m$, the
number of lines parallel to any given line is equal to a number
$k$ that depends only on the configuration and not on the line.
This is in fact true in general and not only for $n\leq 10$ as can
be seen from the following argument. For a configuration with $n$
points, $g$ lines and $m$ lines through each point, any given line
$\Delta$ admits $3(m-1)$ true secants, namely, $(m-1)$ through
each of its points\footnote{A true secant is here defined as a
line, say $\Delta^{\prime}$, distinct from $\Delta$ and with a
non-empty intersection with $\Delta$.}. By definition, these
secants are all distinct since none of the lines that $\Delta$
intersects at one of its points, say $P$, can coincide with a
line that it intersects at another of its points, say
$P^{\prime}$, since the only line joining $P$ to $P^{\prime}$ is
$\Delta$ itself. It follows that the total number of lines that
$\Delta$ intersects is the number of true secants plus $\Delta$
itself, i.e., $3(m-1)+1$. As a consequence, each line in the
configuration admits $k=g-[3(m-1)+1]$ parallel lines, which is
then reflected by the fact that each node in the associated Dynkin
diagram has the same number $k$ of adjacent nodes.

%%%%%%%%%%%%%%%%%%%%%%%%%%%%%%%%%%%%%%%%%%%%%%%%%%%%%%%%%%%%%%%%%%%%%%%%%%%%%%%%%%%
%%%%%%%%%%%%%%%%%%%%%%%%%%%%%%%%%%%%%%%%%%%%%%%%%%%%%%%%%%%%%%%%%%%%%%%%%%%%%%%%%%%

\section{Cosmological Solutions With Electric Flux}

\begin{figure}[htbp]
    \centerline{\includegraphics[width=50mm]{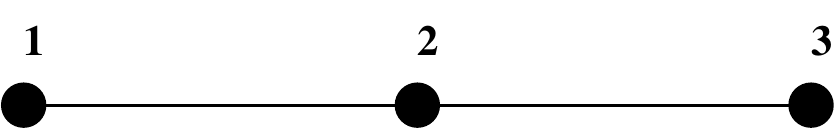}}
    \caption{$(3_{1},1_{3})$: The only allowed configuration for $n=3$.}
    \label{figure:G31}
  \end{figure}

Let us now make use of these considerations to construct some
explicit supergravity solutions. We begin by analyzing the
simplest configuration $(3_{1},1_{3})$, of three points and one
line. It is displayed in Figure~\ref{figure:G31}. This case
is the only possible configuration for $n=3$.

This example also exhibits some subtleties associated with the
Hamiltonian constraint and the ensuing need to extend $\mgb$ when the
algebra dual to the geometric configuration is
finite-dimensional. We will come back to this issue below.

%%%%%%%%%%%%%%%%%%%%%%%%%%%%%%%%%%%%%%%%%%%%%%%%%%%%%%%%%%%%%%%%%%%%%%%%%%%%%%%%%%%

\subsection{General Discussion}

In light of our discussion, considering the geometric
configuration $(3_{1},1_{3})$ is equivalent to turning on only the
component $\mc{A}_{123}(t)$ of the 3-form that parametrizes the
generator $E^{123}$ in the coset representative $\mc{V}(t)\in
K(\mc{E}_{10})\bas \mc{E}_{10}$. Moreover, in order to have the
full coset description, we must also turn on the diagonal metric
components corresponding to the Cartan generator $h = [E^{123},
F_{123}]$. The algebra has thus basis $\{e,f,h\}$ with
\begin{equation}
  e\equiv E^{123},
  \qquad
  f\equiv F_{123},
  \qquad
  h=[e,f] = - \! \f{1}{3}\sum_{a\neq 1,2,3} \!\!\!
  {K^{a}}_{a}+\f{2}{3}({K^{1}}_{1}+{K^{2}}_{2}+{K^{3}}_{3}),
  \label{A1generators}
\end{equation}
where the form of $h$ followed directly from the general commutator
between $E^{abc}$ and $F_{def}$ in
Section~\ref{section:LevelDecomposition}. The Cartan matrix is just
$(2)$ and is nondegenerate. It defines an $A_1=\mf{sl}(2, \mbb{R})$
regular subalgebra. The Chevalley--Serre relations, which are
guaranteed to hold according to the general argument, are easily
verified. The configuration $(3_{1},1_{3})$ is thus dual to $A_{1}$,
\begin{equation}
  \mf{g}_{(3_1,3_1)}=A_1.
\end{equation}
This $A_1$ algebra is simply the $\mf{sl}(2, \mbb{R})$-algebra
associated with the simple root $\alpha_{10}$. Because the
Killing form of $A_1$ restricted to the Cartan subalgebra
$\mf{h}_{A_1}=\mbb{R}h$ is positive definite, one cannot find a
solution of the Hamiltonian constraint if one turns on only the
fields corresponding to $A_1$. One needs to enlarge $A_1$ (at
least) by a one-dimensional subalgebra $\mathbb{R} l$ of
$\mh_{E_{10}}$ that is timelike. As will be discussed further
below, we take for $l$ the Cartan element ${K^4}_4 + {K^5}_5 +
{K^6}_6 + {K^7}_7 + {K^8}_8 + {K^9}_9 + {K^{10}}_{10}$, which
ensures isotropy in the directions not supporting the electric
field. Thus, the appropriate regular subalgebra of $E_{10}$ in
this case is $A_1 \oplus \mathbb{R}l$.

The need to enlarge the algebra $A_1$ was discussed in the
paper~\cite{CosmologyNicolai} where a group theoretical interpretation
of some cosmological solutions of eleven-dimensional supergravity was
given. In that paper, it was also observed that $\mathbb{R}l$ can be
viewed as the Cartan subalgebra \index{Cartan subalgebra}of the (non-regularly embedded)
subalgebra $A_1$ associated with an imaginary root at level~21, but
since the corresponding field is not excited, the relevant subalgebra
is really $\mathbb{R}l$.

%%%%%%%%%%%%%%%%%%%%%%%%%%%%%%%%%%%%%%%%%%%%%%%%%%%%%%%%%%%%%%%%%%%%%%%%%%%%%%%%%%%

\subsection{The Solution}
\label{section:solution}

In order to make the above discussion a little less abstract we
now show how to obtain the relevant supergravity solution by
solving the $\mc{E}_{10}$-sigma model equations of motion and then
translating these, using the dictionary from
Section~\ref{correspondence}, to supergravity solutions. For this
particular example the analysis was done in~\cite{CosmologyNicolai}.\footnote{See also \cite{Damour:2009zc} for some recent related work on ``fermionic Kac-Moody billiards''.}

In order to better understand the role of the timelike generator
$l\in \mf{h}$ we begin the analysis by omitting it. The
truncation then amounts to considering the coset representative
\begin{equation}
  \mc{V}(t)=e^{\phi(t)h}\, e^{\mc{A}_{123}(t)E^{123}} \in
  K(\mc{E}_{10})\bas \mc{E}_{10}.
\end{equation}
The projection $\mc{P}(t)$ onto the coset becomes
\begin{eqnarray}
  \mc{P}(t)&=& \f{1}{2}\left[\pa \mc{V}(t) \, \mc{V}(t)^{-1}+
  \left(\pa\mc{V}(t) \, \mc{V}(t)^{-1}\right)^{\mc{T}}\right]
  \nonumber
  \\
  &=& \pa\phi(t) h +\f{1}{2}e^{2\phi(t)}\pa\mc{A}_{123}(t)
  \left(E^{123}+F_{123}\right), 
\end{eqnarray}
where the exponent is the linear form $\al(\phi)=2\phi$ representing
the exceptional simple root $\al_{123}$ of $E_{10}$. More precisely,
it is the linear form $\al$ acting on the Cartan generator $\phi(t)
h$, as follows:
\begin{equation}
  \al(\phi h)=\phi \left< \al, h\right>=
  \phi \left<\al, \al^{\vee}\right>=\al^2 \phi=2\phi.
\end{equation}
The Lagrangian becomes 
\begin{eqnarray}
  \mc{L}&=& \f{1}{2}\left(\mc{P}(t)|\mc{P}(t)\right)
  \nonumber
  \\
  &=& \pa\phi(t)\pa\phi(t)+\f{1}{4}e^{4\phi(t)}\,
  \pa \mc{A}_{123}(t)\,\pa\mc{A}_{123}(t). 
  \label{A1truncation} 
\end{eqnarray}
For convenience we have chosen the gauge $n=1$ of the free parameter
in the $K(\mc{E}_{10})\bas \mc{E}_{10}$-Lagrangian. Recall that for the level one
fields we have $\mc{D}\mc{A}_{abc}(t)=\pa \mc{A}_{abc}(t)$, which is
why only the partial derivative of $\mc{A}_{123}(t)$ appears in the
Lagrangian.

The reason why this simple looking model contains information
about eleven-dimensional supergravity is that the $A_{1}$
subalgebra represented by $(e, f, h)$ is embedded in $E_{10}$
through the level~1-generator $E^{123}$, and hence this
Lagrangian corresponds to a consistent subgroup truncation of the
$\mc{E}_{10}$- sigma model.

Let us now study the dynamics of the Lagrangian in
Equation~(\ref{A1truncation}). The equations of motion for
$\mc{A}_{123}(t)$ are
\begin{equation}
  \pa \left(\f{1}{2} e^{4\phi(t)}\pa \mc{A}_{123}(t)\right)=0
  \quad \Longrightarrow \quad
  \f{1}{2} e^{4\phi(t)}\pa \mc{A}_{123}(t)=a,
  \label{A123Equation}
\end{equation}
where $a$ is a constant. The equations for the $\ell=0$ field
$\phi$ may then be written as
\begin{equation}
  \pa^{2}\phi(t)=2a^2 e^{-4\phi(t)}.
\end{equation}
Integrating once yields
\begin{equation}
  \pa\phi(t) \, \pa\phi(t) + a^2 e^{-4\phi(t)}=E,
\end{equation}
where $E$ plays the role of the energy for the dynamics of
$\phi(t)$. This equation can be solved exactly with the
result~\cite{CosmologyNicolai}
\begin{equation}
  \phi(t)=\f{1}{2}\ln \left[\f{2a}{\sqrt{E}}\cosh \sqrt{E} t\right] \equiv
  \f{1}{2}\ln H(t).
  \label{phisolution}
\end{equation}
We must also take into account the Hamiltonian constraint
\begin{equation}
  \mc{H}=\left(\mc{P}|\mc{P} \right) = 0,
\end{equation}
arising from the variation of $n(t)$ in the $\mc{E}_{10}$-sigma
model. The Hamiltonian becomes 
\begin{eqnarray}
  \mc{H} & =& 2\pa \phi(t) \, \pa \phi(t) +
  \f{1}{2}e^{4\phi(t)}\pa \mc{A}_{123}(t) \, \pa \mc{A}_{123}(t)
  \nonumber
  \\
  &=& 2\left(\pa \phi(t)\pa \phi(t)+a^2 e^{-4\phi(t)}\right)
  \nonumber
  \\
  &=& 2E. 
\end{eqnarray}
It is therefore impossible to satisfy the Hamiltonian constraint
unless $E=0$. This is the problem which was discussed above, and the
reason why we need to enlarge the choice of coset representative to
include the timelike generator $l\in \mf{h}$. We choose $l$
such that it commutes with $h$ and $E^{123}$,
\begin{equation}
  [l, h]=[l, E^{123}]=0,
\end{equation}
and such that isotropy in the directions not supported by the electric
field is ensured. Most importantly, in order to solve the problem of
the Hamiltonian constraint, $l$ must be timelike, 
\begin{equation}
  l^2= \left( l | l \right) < 0, 
  \label{timelikegenerator}
\end{equation}
where $(\cdot |\cdot )$ is the scalar product in the Cartan subalgebra
of $E_{10}$. The subalgebra to which we truncate the sigma model is
thus given by 
\begin{equation}
  \mgb=A_{1}\oplus \mathbb{R} l \subset E_{10},
  \label{A1plustimelike}
\end{equation}
and the corresponding coset representative is
\begin{equation}
  \tilde{\mc{V}}(t) =
  e^{\phi(t)h + \tilde{\phi}(t)l}e^{\mc{A}_{123}(t)E^{123}}.
  \label{timelikecoset}
\end{equation}
The Lagrangian now splits into two disconnected parts, corresponding
to the direct product $SO(2)\bas SL(2, \mbb{R})\times \mbb{R}$,
\begin{equation}
  \tilde{\mc{L}}=
  \left(\pa\phi(t)\,\pa\phi(t)+\f{1}{4}e^{4\phi(t)}
  \pa\mc{A}_{123}(t)\,\pa \mc{A}_{123}(t)\right) +
  \f{l^2}{2}\pa\tilde{\phi}(t)\,\pa\tilde{\phi}(t).
  \label{newLagrangian}
\end{equation}
The solution for $\tilde{\phi}$ is therefore simply linear in time,
\begin{equation}
  \tilde{\phi}=|l^2|\sqrt{\tilde{E}}\, t. 
  \label{phitildesolution}
\end{equation}
The new Hamiltonian now gets a contribution also from the Cartan
generator $l$,
\begin{equation}
  \tilde{\mc{H}}=2E-|l^2|\tilde{E}.
  \label{newHamiltonian}
\end{equation}
This contribution depends on the norm of $l$ and since $l^2<0$, it is
possible to satisfy the Hamiltonian constraint, provided that we set 
\begin{equation}
  \tilde{E}=\f{2}{|l^2|}E.
\end{equation}

We have now found a consistent truncation of the
$K(\mc{E}_{10})\times \mc{E}_{10}$-invariant sigma model which
exhibits $SL(2, \mbb{R})\times SO(2)\times \mbb{R}$-invariance. We
want to translate the solution to this model,
Equation~(\ref{phisolution}), to a solution of eleven-dimensional
supergravity. The embedding of $\mf{sl}(2, \mbb{R})\subset E_{10}$
in Equation~(\ref{A1generators}) induces a natural ``Freund--Rubin'' type
($1+3+7$) split of the coordinates in the physical metric, where
the 3-form is supported in the three spatial directions
$x^{1},x^{2},x^{3}$. We must also choose an embedding of the
timelike generator $l$. In order to ensure isotropy in the
directions $x^{4}, \cdots, x^{10}$, where the electric field has
no support, it is natural to let $l$ be extended only in the
``transverse'' directions and we take~\cite{CosmologyNicolai}
\begin{equation}
  l ={K^{4}}_{4}+\cdots +{K^{10}}_{10},
  \label{timelikegeneratorembedding}
\end{equation}
which has norm
\begin{equation}
  (l | l )=\left( {K^{4}}_{4}+\cdots +{K^{10}}_{10} |
  {K^{4}}_{4}+\cdots +{K^{10}}_{10}\right ) =-42. 
  \label{timelikenorm}
\end{equation}
To find the metric solution corresponding to our sigma model, we first
analyze the coset representative at $\ell=0$,
\begin{equation}
  \tilde{\mc{V}}(t)\big|_{\ell=0}=
  \Exp \left[\phi(t) h +\tilde{\phi}(t)l\right].
\end{equation}
In order to make use of the dictionary from
Section~\ref{correspondence} it is necessary to rewrite this in a way
more suitable for comparison, i.e., to express the Cartan generators
$h$ and $l$ in terms of the $\mf{gl}(10, \mbb{R})$-generators
${K^{a}}_b$. We thus introduce parameters ${\xi^{a}}_b(t)$ and
$\tilde{\xi}^{a}{}_{b}(t)$ representing, respectively, $\phi$ and
$\tilde{\phi}$ in the $\mf{gl}(10, \mbb{R})$-basis. The level zero
coset representative may then be written as 
\begin{eqnarray}
  \tilde{\mc{V}}(t)\big|_{\ell=0} &=&\Exp \left[\sum_{a=1}^{10}
  \left({\xi^{a}}_{a}(t)+\tilde{\xi}^{a}{}_{a}(t)\right){K^{a}}_{a}\right]
  \nonumber
  \\
  &=&\Exp \left[\sum_{a=4}^{10}\left({\xi^{a}}_{a}(t)+
  \tilde{\xi}^{a}{}_{a}(t)\right){K^{a}}_{a}+
  \left({\xi^{1}}_{1}(t){K^{1}}_{1}+{\xi^{2}}_{2}(t){K^{2}}_{2}+
  {\xi^{3}}_{3}(t){K^{3}}_{3}\right)\right],
  \nn \\
  \label{levelzero} 
\end{eqnarray}
where in the second line we have split the sum in order to highlight
the underlying spacetime structure, i.e., to emphasize that
$\ti{\xi}^{a}{}_{b}$ has no non-vanishing components in the
directions $x^{1}, x^{2}, x^{3}$. Comparing this to
Equation~(\ref{A1generators}) and
Equation~(\ref{timelikegeneratorembedding}) gives the diagonal
components of ${\xi^{a}}_{b}$ and $\tilde{\xi}^{a}{}_{b}$,
\begin{equation}
  {\xi^{1}}_{1}={\xi^{2}}_{2}={\xi^{3}}_{3}=2\phi/3,
  \qquad
  {\xi^{4}}_{4}= \cdots={\xi^{10}}_{10}=-\phi/3,
  \qquad
  {\tilde{\xi}^{4}}{}_{4}=\cdots = {\tilde{\xi}^{10}}{}_{10}=\ti{\phi}.
  \label{chiandxicomponents}
\end{equation}
Now, the dictionary from Section \ref{correspondence} identifies
the physical spatial metric as follows:
\begin{equation}
  \g_{ab}(t)=
  {e_{a}}^{\bar{a}}(t){e_{b}}^{\bar{b}}(t) \delta_{\bar{a}\bar{b}}=
  {(e^{\xi(t)+\ti{\xi}(t)})_{a}}^{\bar{a}}
  {(e^{\xi(t)+\ti{\xi}(t)})_{b}}^{\bar{b}}\delta_{\bar{a}\bar{b}}
\end{equation}
By observation of Equation~(\ref{chiandxicomponents}) we find the
components of the metric to be 
\begin{equation}
  \begin{array}{rcl}
    \g_{11}&=&\g_{22}=\g_{33}=e^{4\phi/3},
    \\ [0.25 em]
    \g_{44}&=&\cdots=\g_{(10)(10)}=e^{-2\phi/3+2\ti{\phi}}.
  \end{array}
  \label{metriccomponents} 
\end{equation}
This result shows clearly how the embedding of $h$ and $l$ into
$E_{10}$ is reflected in the coordinate split of the metric. The gauge
fixing $N=\sqrt{\g}$ (or $n=1$) gives the $\g_{tt}$-component of the
metric,
\begin{equation}
  \g_{tt}=N^2=e^{14\ti{\phi}-2\phi/3}. 
  \label{ttcomponent}
\end{equation}
Next we consider the generator $E^{123}$. The dictionary tells us that
the field strength of the 3-form in eleven-dimensional supergravity
at some fixed spatial point $\mathbf{x}_0$ should be identified as
\begin{equation}
  F_{t123}(t,\mathbf{x}_{0})=\mc{D}\mc{A}_{123}(t)=\pa \mc{A}_{123}(t). 
\end{equation}
It is possible to eliminate the $\mc{A}_{123}(t)$ in favor of the
Cartan field $\phi(t)$ using the first integral of its equations
of motion, Equation~(\ref{A123Equation}),
\begin{equation}
  \f{1}{2}e^{-4\phi(t)}\pa\mc{A}_{123}(t)= a. 
  \label{eomA}
\end{equation}
In this way we may write the field strength in terms of $a$ and the
solution for $\phi$,
\begin{equation}
  F_{t123}(t,\mathbf{x}_{0})=2 a e^{4\phi(t)}=2 a H^{-2}(t).
  \label{Fsolution}
\end{equation}
Finally, we write down the solution for the spacetime metric
explicitly:
\begin{eqnarray}
  ds^{2}&=&-e^{14\ti{\phi}+2\phi/3}\,dt^{2}+
  e^{4\phi/3}\left[(dx^1)^{2}+(dx^2)^2+(dx^3)^2\right]+
  e^{2\ti{\phi}-2\phi/3}\sum_{\bar{a}=4}^{10}(dx^{\bar{a}})^{2}
  \nonumber
  \\
  &=& - H^{1/3}(t)e^{\f{1}{3}\sqrt{\ti{E}t}}\,dt^{2}+
  H^{-2/3}(t)\left[(dx^1)^{2}+(dx^2)^2+(dx^3)^2\right]+
  H^{1/3}(t)e^{\f{\sqrt{\ti{E}}}{21}t}\sum_{\bar{a}=4}^{10}(dx^{\bar{a}})^{2},
  \nonumber
  \\
  \label{SM2Brane} 
\end{eqnarray}
where
\begin{equation}
  H(t)=\f{2a}{\sqrt{E}}\cosh \sqrt{E} t.
\end{equation}
This solution coincides with the cosmological solution first found
in~\cite{Demaret} for the geometric configuration $(3_1, 3_1)$, and it
is intriguing that it can be exactly reproduced from a manifestly
$\mc{E}_{10}\times K(\mc{E}_{10})$-invariant action, a priori
unrelated to any physical model.

Note that in modern terminology, this solution is an SM2-brane
solution (see, e.g.,~\cite{AcceleratingOhta} for a review) since
it can be interpreted as a spacelike (i.e., time-dependent)
version of the M2-brane solution. From this point of view the
world volume of the SM2-brane is extended in the directions
$x^1, x^2$ and $x^3$, and so is Euclidean.

In the BKL-limit \index{BKL-limit}this solution describes two asymptotic Kasner
regimes, at $t\rightarrow \infty$ and at $t\rightarrow -\infty$.
These are separated by a collision against an electric wall,
corresponding to the blow-up of the electric field
$F_{t123}(t)\sim H^{-2}(t)$ at $t=0$. In the billiard
\index{cosmological billiard}picture the
dynamics in the BKL-limit is thus given by free-flight motion
interrupted by one geometric reflection against the electric wall,
\begin{equation}
  e_{123}(\be)=\be^1+\be^2+\be^3,
\end{equation}
which is the exceptional simple root of $E_{10}$. This indicates that
in the strict BKL-limit, electric walls and SM2-branes are actually
equivalent.

%%%%%%%%%%%%%%%%%%%%%%%%%%%%%%%%%%%%%%%%%%%%%%%%%%%%%%%%%%%%%%%%%%%%%%%%%%%%%%%%%%%

\subsection{Intersecting Spacelike Branes From Geometric Configurations}

\begin{figure}[htbp]
    \centerline{\includegraphics[width=50mm]{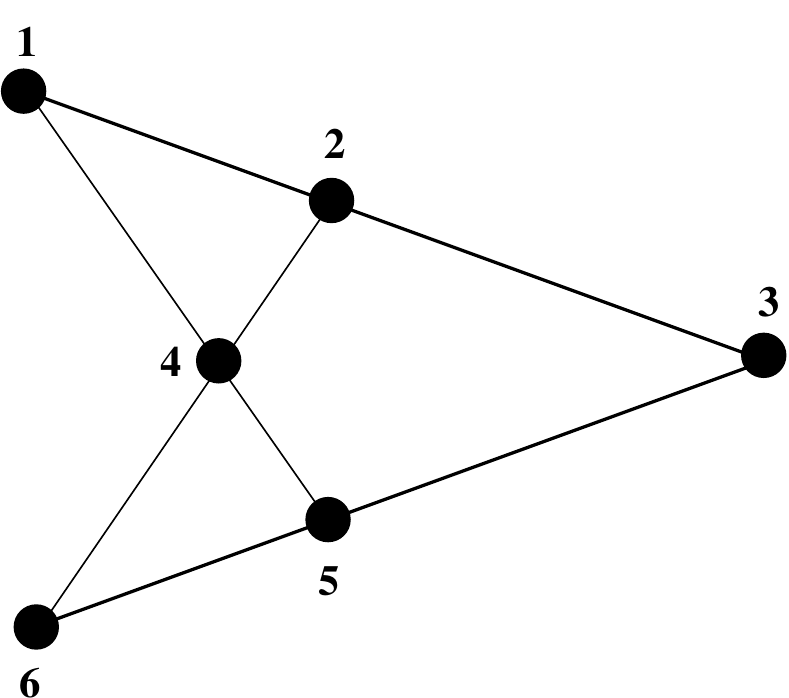}}
    \caption{The configuration $(6_{2},4_{3})$, dual to the Lie
      algebra $A_1\oplus A_1\oplus A_1\oplus A_1$.}
    \label{figure:G62}
  \end{figure}

Let us now examine a slightly more complicated example. We consider
the configuration $(6_{2},4_{3})$, shown in
Figure~\ref{figure:G62}. This configuration has four lines and six
points. As such the associated supergravity model describes a
cosmological solution with four components of the electric field
turned on, or, equivalently, it describes a set of four intersecting
SM2-branes~\cite{Henneaux:2006gp}.

From the configuration we read off the Chevalley--Serre
generators associated to the simple roots of the dual algebra:
\begin{equation}
  e_{1}=E^{123},
  \qquad
  e_{2}=E^{145},
  \qquad
  e_{3}=E^{246},
  \qquad e_{4}=E^{356}.
  \label{G62generators}
\end{equation}
The first thing to note is that all generators have one index in
common since in the graph any two lines share one node. This implies
that the four lines in $(6_{2},4_{3})$ define four \emph{commuting}
$A_1$ subalgebras,
\begin{equation}
  (6_{2},4_{3})
  \quad \Longleftrightarrow \quad
  \mf{g}_{(6_{2},4_{3})}=A_{1}\oplus A_{1} \oplus A_{1}\oplus A_{1}.
  \label{A1^4}
\end{equation}
One can make sure that the Chevalley--Serre relations are indeed
fulfilled for this embedding. For instance, the Cartan element $h =
[E^{b_1 b_2 b_3}, F_{b_1 b_2 b_3}]$ (no summation on the fixed,
distinct indices $b_{1},b_{2},b_{3}$) reads
\begin{equation}
  h=-\f{1}{3}\!\sum_{a\neq b_{1},b_{2},b_{3}} \!\!\!
  {K^{a}}_{a}+\f{2}{3}({K^{b_{1}}}_{b_{1}}+
  {K^{b_{2}}}_{b_{2}}+{K^{b_{3}}}_{b_{3}}).
  \label{GeneralCartanElement}
\end{equation}
Hence, the commutator $[h,E^{b_{i} cd}]$ vanishes whenever $E^{b_{i}
  cd}$ has only one $b$-index, 
\begin{eqnarray}
  [h,E^{b_{i} cd}]&=&-\f{1}{3}[({K^{c}}_{c}+{K^{d}}_{d}),E^{b_{i}cd}]+
  \f{2}{3}[({K^{b_{1}}}_{b_{1}}+{K^{b_{2}}}_{b_{2}}+{K^{b_{3}}}_{b_{3}}),E^{b_{i}cd}]
  \nonumber
  \\
  &=& \left(-\f{1}{3}-\f{1}{3}+\f{2}{3}\right)E^{b_{i}cd}=0
  \qquad
  (i=1,2,3). 
  \label{vanishingcommutator} 
\end{eqnarray}
Furthermore, multiple commutators of the step operators are
immediately killed at level $2$ whenever they have one index or more
in common, e.g.,
\begin{equation}
  [E^{123},E^{145}]=E^{123145}=0.
  \label{vanishingcommutator2}
\end{equation}
To fulfill the Hamiltonian constraint, one must extend the algebra
by taking a direct sum with $\mathbb{R}l$, $l={K^7}_7 + {K^8}_8 +
{K^9}_9 + {K^{10}}_{10}$. Accordingly, the final algebra is
$A_{1}\oplus A_{1} \oplus A_{1}\oplus A_{1} \oplus \mathbb{R}l$.
Because there is no magnetic field, the momentum constraint and
Gauss' law are identically satisfied.

By investigating the sigma model solution corresponding to the
algebra $\mf{g}_{(6_2, 4_3)}$, augmented with the timelike
generator $l$,
\begin{equation}
  \mgb=A_1\oplus A_1\oplus A_1 \oplus A_1\oplus \mbb{R}l,
\end{equation}
we find a supergravity solution which generalizes the one found
in~\cite{Demaret}. The solution describes a set of four intersecting
SM2-branes, with a five-dimensional transverse spacetime in the
directions $t, x^{7},x^{8},x^{9},x^{10}$.

Let us write down also this solution explicitly. The full set of
generators for $\mf{g}_{(6_2, 4_3)}$ is 
\begin{equation}
  \begin{array}{rcl@{\qquad}rcl@{\qquad}rcl@{\qquad}rcl}
    e_{1}&=&E^{123}, &
    e_{2}&=&E^{145}, &
    e_{3}&=&E^{246}, &
    e_{4}=E^{356}
    \\ [0.25 em]
    f_{1}&=&F_{123}, &
    f_{2}&=&F_{145}, &
    f_{3}&=&F_{246}, &
    f_{4}=F_{356}
    \\ [0.5 em]
    h_{1}&=&\multicolumn{10}{l}{\displaystyle
    -\f{1}{3}\!\sum_{a\neq 1,2,3}\!\!\!
    {K^{a}}_{a}+\f{2}{3}({K^{1}}_{1}+{K^{2}}_{2}+{K^{3}}_{3}),}
    \\ [0.5 em]
    h_{2}&=&\multicolumn{10}{l}{-\displaystyle
    \f{1}{3}\!\sum_{a\neq 1,4,5}\!\!\!
    {K^{a}}_{a}+\f{2}{3}({K^{1}}_{1}+{K^{4}}_{4}+{K^{5}}_{5}),}
    \\ [0.5 em]
    h_{3}&=&\multicolumn{10}{l}{\displaystyle
    -\f{1}{3}\!\sum_{a\neq 2,4,6}\!\!\!
    {K^{a}}_{a}+\f{2}{3}({K^{2}}_{2}+{K^{4}}_{4}+{K^{6}}_{6}),}
    \\ [0.5 em]
    h_{4}&=&\multicolumn{10}{l}{\displaystyle
    -\f{1}{3}\!\sum_{a\neq3,5,6}\!\!\!
    {K^{a}}_{a}+\f{2}{3}({K^{3}}_{3}+{K^{5}}_{5}+{K^{6}}_{6}).}
  \end{array}
  \label{(6_2,4_3)generators} 
\end{equation}
The coset element for this configuration then becomes
\begin{equation}
  \mc{V}(t)=e^{\phi_{1}(t)h_{1}+\phi_{2}(t)h_{2}+\phi_{3}(t)h_{3}+
  \phi_{4}(t)h_{4}+\ti{\phi}(t)l}\,
  e^{\mc{A}_{123}(t)E^{123}+\mc{A}_{145}(t)E^{145}+\mc{A}_{246}(t)E^{246}+
  \mc{A}_{356}(t)E^{356}}.
  \label{fourA1coset}
\end{equation}
We must further choose the timelike Cartan generator, $l\in
\mf{h}$, appropriately. Examination of
Equation~(\ref{(6_2,4_3)generators}) reveals that the four electric
fields are supported only in the spatial directions
$x^{1},\cdots,x^{6}$ so, again, in order to ensure isotropy in the
directions transverse to the $S$-branes, we choose the timelike Cartan
generator as follows:
\begin{equation}
  l={K^{7}}_{7}+{K^{8}}_{8}+{K^{9}}_{9}+{K^{10}}_{10},
  \label{timelikeCartan}
\end{equation}
which implies
\begin{equation}
  l^2=\left( l | l \right)=
  \left({K^{7}}_{7}+{K^{8}}_{8}+{K^{9}}_{9}+{K^{10}}_{10}|
  {K^{7}}_{7}+{K^{8}}_{8}+{K^{9}}_{9}+{K^{10}}_{10}\right) = -12. 
  \label{qminus}
\end{equation}
The Lagrangian for this system becomes
\begin{equation}
  \mc{L}_{(6_2,4_3)}=\mc{L}_{1}+\mc{L}_{2}+\mc{L}_{3}+\mc{L}_{4}+
  \f{l^2}{2}\pa \ti{\phi}(t)\,\pa \ti{\phi}(t), 
  \label{Lagrangian4A1}
\end{equation}
where $\mc{L}_1, \mc{L}_2, \mc{L}_3$ and $\mc{L}_4$ represent the
$SL(2, \mbb{R})\times SO(2)$-invariant Lagrangians corresponding to
the four $A_1$-algebras. The solutions for $\phi_1(t), \cdots,
\phi_4(t)$ and $\ti{\phi}(t)$ are separately identical to the ones for
$\phi(t)$ and $\ti{\phi}(t)$, respectively, in
Section~\ref{section:solution}. From the embedding into $E_{10}$,
provided in Equation~(\ref{(6_2,4_3)generators}), we may read off the
solution for the spacetime metric,
\begin{eqnarray}
  ds^{2}_{(6_2,4_3)}&=&
  -(H_{1}H_{2}H_{3}H_{4})^{1/3}e^{\f{2}{3}\sqrt{E}t}\,dt^{2}+
  (H_{1}H_{4})^{-2/3}(H_{2}H_{3})^{1/3}(dx^{1})^{2}
  \nonumber
  \\
  & &+(H_{1}H_{3})^{-2/3}(H_{2}H_{4})^{1/3}(dx^{2})^{2}+
  (H_{1}H_{2})^{-2/3}(H_{3}H_{4})^{1/3}(dx^{3})^{2}
  \nonumber
  \\
  & &+(H_{3}H_{4})^{-2/3}(H_{1}H_{2})^{1/3}(dx^{4})^{2}+
  (H_{2}H_{4})^{-2/3}(H_{1}H_{3})^{1/3}(dx^{5})^{2}
  \nonumber
  \\
  & &+(H_{2}H_{3})^{-2/3}(H_{1}H_{4})^{1/3}(dx^{6})^{2}+
  (H_{1}H_{2}H_{3}H_{4})^{1/3}e^{\f{1}{6}\sqrt{E}t}
  \sum_{\bar{a}=7}^{10}(dx^{\bar{a}})^2.
  \nn \\
  \label{fourSM2branes} 
\end{eqnarray}
As announced, this describes four intersecting SM2-branes with a
$1+4$-dimensional transverse spacetime. For example the brane that
couples to the field associated with the first Cartan generator is extended
in the directions $x^{1},x^{2},x^{3}$. By restricting to the case
$\phi_{1}=\phi_2=\phi_3=\phi_4\equiv \phi$ the metric simplifies to 
\begin{eqnarray}
  ds^{2}_{(6_2,4_3)}&=&
  -\left(\f{2a}{\sqrt{E}}\right)^{4/3}\cosh^{4/3}\sqrt{E}t\,
  e^{\f{2}{3}\sqrt{\ti{E}}t}dt^{2}+
  \left(\f{2a}{\sqrt{E}}\right)^{-2/3}\cosh^{-2/3}\sqrt{E}t
  \sum_{a^{\prime}=1}^{6}(dx^{a^{\prime}})^{2}
  \nonumber
  \\
  & &+\left(\f{2a}{\sqrt{E}}\right)^{4/3} \cosh^{4/3}\sqrt{E} t \,
  e^{\f{1}{6}\sqrt{\ti{E}}t}\sum_{\bar{a}=7}^{10}(dx^{\bar{a}})^{2},
  \label{restrictedsolution} 
\end{eqnarray}
which coincides with the cosmological solution found in~\cite{Demaret}
for the configuration $(6_{2},4_3)$. We can therefore conclude that
the algebraic interpretation of the geometric configurations found in
this chapter generalizes the solutions given in the aforementioned
reference.

In a more general setting where we excite more roots of $E_{10}$,
the solutions of course become more complex. However, as long as
we consider \emph{commuting} subalgebras there will naturally be
no coupling in the Lagrangian between fields parametrizing
different subalgebras. This implies that if we excite a direct sum
of $m$ $A_{1}$-algebras the total Lagrangian will split according
to
\begin{equation}
  \mc{L}=\sum_{k=1}^{m} \mc{L}_{k} + \ti{\mc{L}},
  \label{LagrangianSplitting}
\end{equation}
where $\mc{L}_{k}$ is of the same form as
Equation~(\ref{A1truncation}), and $\ti{\mc{L}}$ is the Lagrangian for
the timelike Cartan element, needed in order to satisfy the
Hamiltonian constraint. It follows that the associated solutions are 
\begin{equation}
  \begin{array}{rcl}
    \phi_{k}(t)&=&\displaystyle
    \f{1}{2}\ln\left[\f{a_{k}}{E_{k}}\cosh \sqrt{E_{k}} t\right]
    \qquad
    (k=1,\cdots,m),
    \\ [1.0 em]
    \ti{\phi}(t)&=& |l^2|\sqrt{\ti{E}}t.
  \end{array}
  \label{intersectingsolutions} 
\end{equation}
Furthermore, the resulting structure of the metric depends on the
embedding of the $A_{1}$-algebras into $E_{10}$, i.e., which
level~1-generators we choose to realize the step-operators and hence
which Cartan elements that are associated to the $\phi_{k}$'s. Each
excited $A_{1}$-subalgebra will turn on an electric 3-form that
couples to an SM2-brane and hence the solution for the metric will
describe a set of $m$ intersecting SM2-branes.

\begin{figure}[htbp]
    \centerline{\includegraphics[width=60mm]{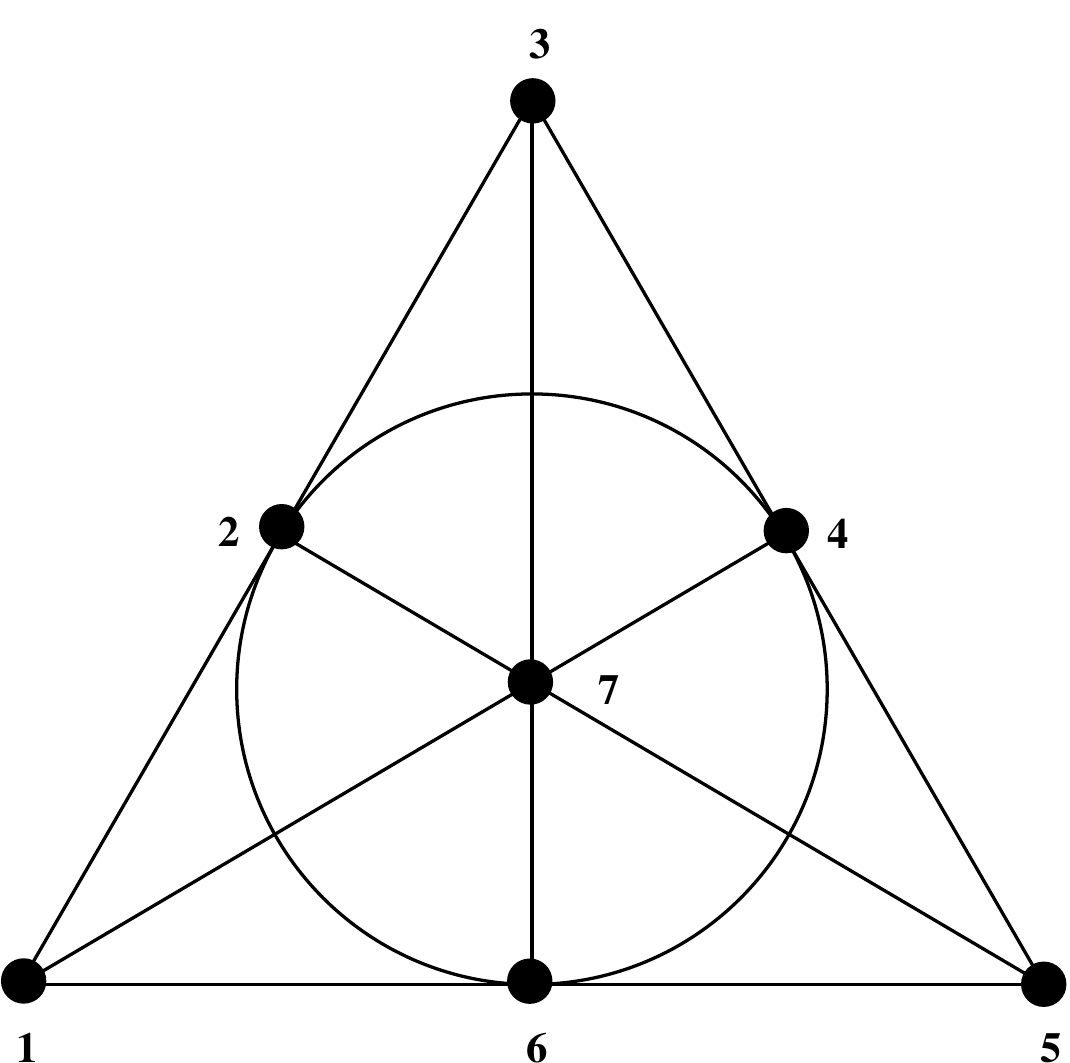}}
    \caption{The Fano Plane, $(7_3, 7_3)$, dual
      to the Lie algebra $A_1\oplus A_1\oplus A_1\oplus A_1\oplus A_1\oplus A_1\oplus A_1$.}
    \label{figure:Fano}
  \end{figure}

As an additional nice example, we mention here the configuration
$(7_3, 7_3)$, also known as the \emph{Fano plane}, which consists of
7 lines and 7 points (see Figure~\ref{figure:Fano}). This
configuration is well known for its relation to the octonionic
multiplication table~\cite{Baez}. For our purposes, it is interesting
because none of the lines in the configuration are parallell. Thus,
the algebra dual to the Fano plane is a direct sum of seven
$A_1$-algebras and the supergravity solution derived from the sigma
model describes a set of seven intersecting SM2-branes.

%%%%%%%%%%%%%%%%%%%%%%%%%%%%%%%%%%%%%%%%%%%%%%%%%%%%%%%%%%%%%%%%%%%%%%%%%%%%%%%%%%%

\subsection{Intersection Rules for Spacelike Branes}

For multiple brane solutions, there are rules for how these branes may
intersect in order to describe allowed
BPS-solutions~\cite{IntersectingArgurio}. These intersection rules \index{intersection rules|bb} also
apply to spacelike branes~\cite{IntersectingOhta} and hence they apply
to the solutions considered here. In this section we will show that the
intersection rules for multiple $S$-brane solutions are encoded in the
associated geometric configurations~\cite{Henneaux:2006gp}.

For two spacelike $q$-branes, $A$ and $B$, in $M$-theory the rules
are
\begin{equation}
  SMq_{A}\cap SMq_{B}=\f{(q_{A}+1)(q_{B}+1)}{9}-1.
  \label{intersectionrules}
\end{equation}
So, for example, if we have two SM2-branes the result is
\begin{equation}
  SM2\cap SM2=0,
  \label{intersectingSM2}
\end{equation}
which means that they are allowed to intersect on a 0-brane. Note
that since we are dealing with spacelike branes, a zero-brane is
extended in one spatial direction, so the two SM2-branes may
therefore intersect in one spatial direction only. We see from
Equation~(\ref{fourSM2branes}) that these rules are indeed fulfilled
for the configuration $(6_2,4_3)$.

In~\cite{IntersectingEnglert} it was found in the context of
$\mf{g}^{+++}$-algebras that the intersection rules \index{intersection rules} for extremal
branes are encoded in orthogonality conditions between the various
roots from which the branes arise. This is equivalent to saying
that the subalgebras that we excite are commuting, and hence the
same result applies to $\mf{g}^{++}$-algebras in the cosmological
context\footnote{This was also pointed out
  in~\cite{CosmologyNicolai}.}. From this point of view, the
intersection rules can also be read off from the geometric
configurations in the sense that the configurations encode information
about whether or not the algebras commute.

The next case of interest is the Fano plane, $(7_3,7_3)$. As
mentioned above, this configuration corresponds to the direct sum
of 7 commuting $A_{1}$ algebras and so the gravitational
solution describes a set of 7 intersecting SM2-branes. The
intersection rules \index{intersection rules} are guaranteed to be
satisfied for the same reason as before.

%%%%%%%%%%%%%%%%%%%%%%%%%%%%%%%%%%%%%%%%%%%%%%%%%%%%%%%%%%%%%%%%%%%%%%%%%%%%%%%%%%%
%%%%%%%%%%%%%%%%%%%%%%%%%%%%%%%%%%%%%%%%%%%%%%%%%%%%%%%%%%%%%%%%%%%%%%%%%%%%%%%%%%%

\section{Cosmological Solutions With Magnetic Flux}

\begin{figure}[htbp]
    \centerline{\includegraphics[width=100mm]{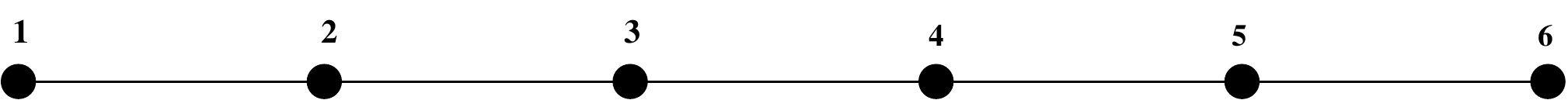}}
    \caption{The simplest ``magnetic configuration'' $(6_1, 1_6)$,
      dual to the algebra $A_1$. The associated supergravity solution
      describes an SM5-brane, whose world volume is extended in the
      directions $x^{1}, \cdots, x^{6}$.}
    \label{figure:MagneticConfiguration}
  \end{figure}

We will now briefly sketch how one can also obtain the SM5-brane
solutions from geometric configurations and regular subalgebras of
$E_{10}$. In order to do this we consider ``magnetic'' subalgebras
of $E_{10}$, constructed only from simple root generators at level
two in the level decomposition of $E_{10}$. To the best of our
knowldege, there is no theory of geometric configurations
developed for the case of having 6 points on each line, which
would be needed here. However, we may nevertheless continue to
investigate the simplest example of such a configuration, namely
$(6_1, 1_6)$, displayed in Figure~\ref{figure:MagneticConfiguration}.

The algebra dual to this configuration is an
$A_1$-subalgebra of $E_{10}$ with the following generators:
\begin{equation}
  \begin{array}{rcl}
    e&=&E^{123456}=F_{123456},
    \\ [0.25 em]
    h&\equiv&\displaystyle
    [E^{123456},F_{123456}]=-\f{1}{6}\sum_{a\neq 1,\cdots,6}
    {K^{a}}_{a}+ \f{1}{3}({K^{1}}_{1}+\cdots+{K^{6}}_{6}).
  \end{array}
  \label{SM5generators} 
\end{equation}
Although the embedding of this algebra
is different from the electric cases considered previously, the
sigma model solution is still associated to an $SL(2,
\mbb{R})/SO(2)$ coset space and therefore the solutions for
$\phi(t)$ and $\ti{\phi}(t)$ are the same as before. Because of
the embedding, however, the sigma model translates to a different
type of supergravity solution, namely a spacelike five-brane whose
world volume is extended in the directions $x^{1}, \cdots, x^{6}$.
The metric is given by
\begin{equation}
  ds^{2}=- H^{-4/3}(t)\,e^{\f{2}{3}\sqrt{E_{-}} t}\,
  dt^{2}+H^{-1/3}(t)\sum_{a^{\prime}=1}^{6}(dx^{a^{\prime}})^2+
  H^{1/6}(t)e^{\f{1}{6}\sqrt{E_{-}}t}\sum_{\bar{a}=7}^{10}(dx^{\bar{a}})^2.
  \label{SM5brane}
\end{equation}
This solution coincides with the SM5-brane found by Strominger and
Gutperle in~\cite{Strominger}\footnote{In~\cite{Strominger}
  they were dealing with a hyperbolic internal space so there was an
  additional $\sinh$-function in the transverse spacetime.}. Note that
the correct power of $H(t)$ for the five-brane arises here entirely
due to the embedding of $h$ into $E_{10}$ through
Equation~(\ref{SM5generators}).

Because of the existence of electric-magnetic duality on the
supergravity side, it is suggestive to expect the existence of a
duality between the two types of configurations $(n_m, g_3)$ and
$(n_m, g_6)$, of which we have here seen the simplest realisation
for the configurations $(3_1, 1_3)$ and $(6_1, 1_6)$.

%%%%%%%%%%%%%%%%%%%%%%%%%%%%%%%%%%%%%%%%%%%%%%%%%%%%%%%%%%%%%%%%%%%%%%%%%%%%%%%%%%%
%%%%%%%%%%%%%%%%%%%%%%%%%%%%%%%%%%%%%%%%%%%%%%%%%%%%%%%%%%%%%%%%%%%%%%%%%%%%%%%%%%%

\section{The Petersen Algebra and the Desargues Configuration}

We want to end this section by considering an example which is
more complicated, but very interesting from the algebraic point of
view. There exist ten geometric configurations of the form
$(10_3, 10_3)$, i.e., with exactly ten points and ten
lines. In~\cite{Demaret}, these were associated to supergravity
solutions with ten components of the electric field turned on. This
result was re-analyzed by some of the present authors
in~\cite{Henneaux:2006gp} where it was found that many of these
configurations have a dual description in terms of Dynkin diagrams of
rank~10 Lorentzian Kac--Moody subalgebras of $E_{10}$. One would
therefore expect that solutions of the sigma models for these algebras
should correspond to new solutions of eleven-dimensional
supergravity. However, since these algebras are infinite-dimensional,
the corresponding sigma models are difficult to solve without further
truncation. Nevertheless, one may argue that explicit solutions should
exist, since the algebras in question are all non-hyperbolic, so we know
that the supergravity dynamics is non-chaotic.

\begin{figure}[htbp]
\centerline{\includegraphics[width=70mm]{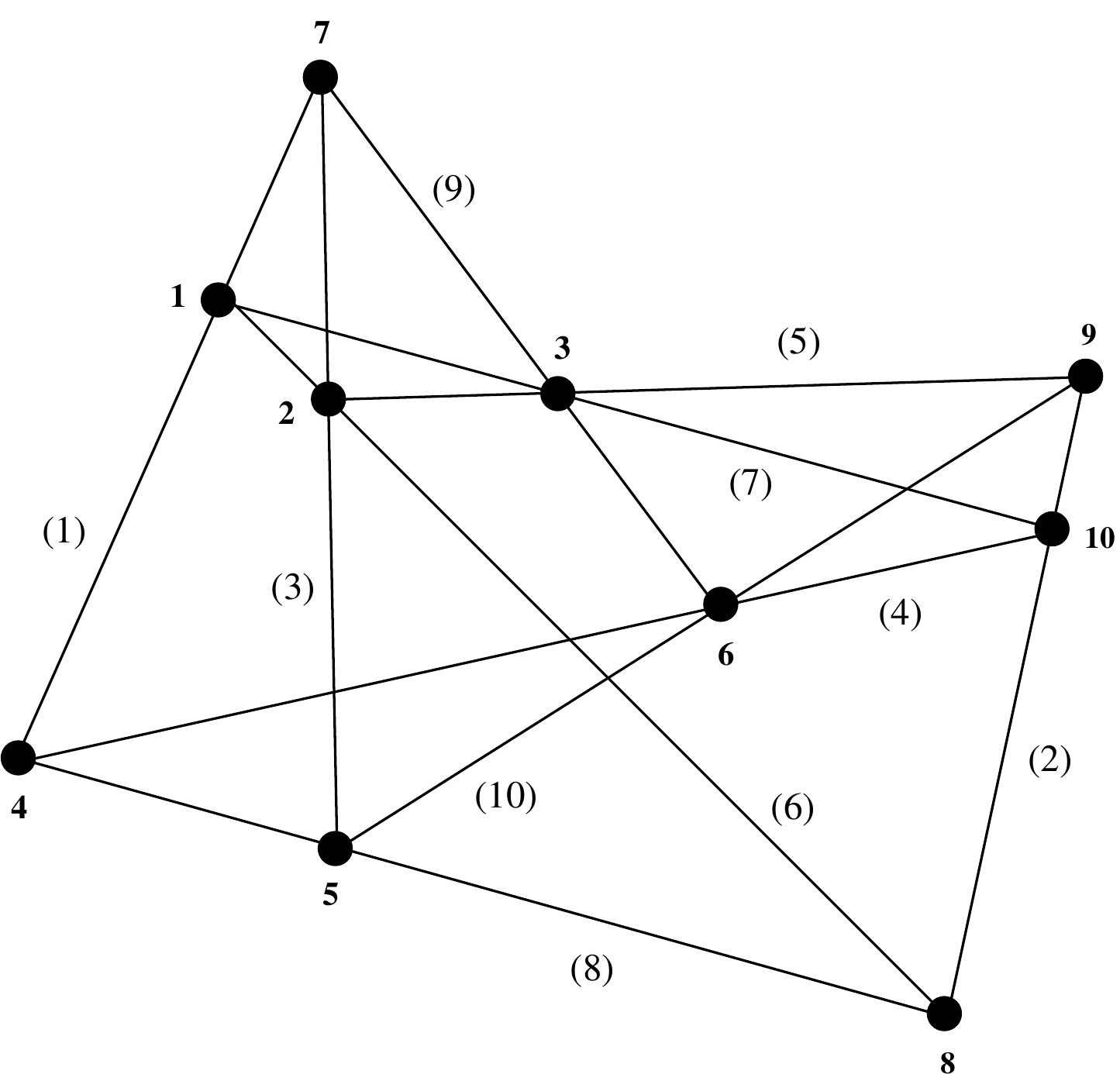}}
\caption{$(10_{3},10_{3})_{3}$: The Desargues configuration, dual
to the Petersen graph.} \label{figure:GDesargues}
\end{figure}

We shall here consider one of the $(10_3, 10_3)$-configurations in
some detail, referring the reader to~\cite{Henneaux:2006gp} for a
discussion of the other cases. The configuration we will treat is
the well known \emph{Desargues configuration}, displayed in
Figure~\ref{figure:GDesargues}. The Desargues configuration is
associated with the 17th century French mathematician
\emph{G\'{e}rard Desargues} to illustrate the following ``Desargues
theorem'' (adapted from~\cite{Page}):

\begin{quote}
  \emph{Let the three lines defined by $\{4,1\},\{5,2\}$ and
    $\{6,3\}$ be concurrent, i.e., be intersecting at one point, say
    $\{7\}$. Then the three intersection points $8\equiv
    \{1,2\}\cap\{4,5\}, 9\equiv \{2,3\}\cap\{5,6\}$ and $10\equiv
    \{1,3\}\cap\{4,6\}$ are colinear. }
\end{quote}

Another way to say this is that the two triangles
$\{1,2,3\}$ and $\{4,5,6\}$ in Figure~\ref{figure:GDesargues} are
in perspective from the point $\{7\}$ and in perspective from the
line $\{8,10,9\}$.

\begin{figure}[htbp]
    \centerline{\includegraphics[width=60mm]{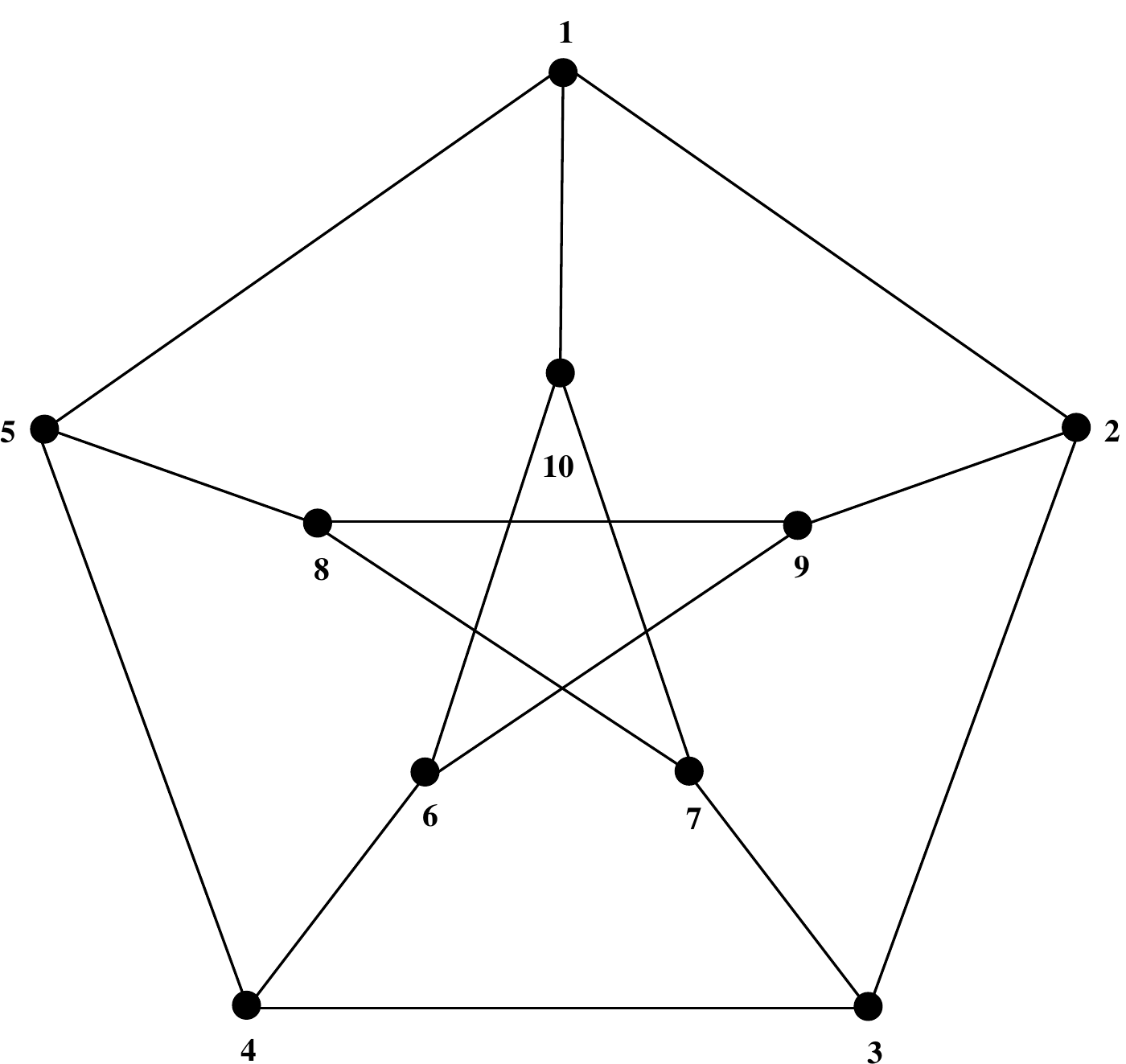}}
    \caption{This is the so-called Petersen graph. It is the
      Dynkin diagram dual to the Desargues configuration, and is in fact a
      geometric configuration itself, denoted $(10_{3},15_{2})$.}
    \label{figure:D2(Petersen)}
  \end{figure}

As we will see, a new fascinating feature emerges for this case, namely
that the Dynkin diagram dual to this configuration \emph{also}
corresponds in itself to a geometric configuration. In fact, the
Dynkin diagram dual to the Desargues configuration turns out to be
the famous \emph{Petersen graph}, denoted $(10_{3},15_{2})$, which
is displayed in Figure~\ref{figure:D2(Petersen)}.

To construct the Dynkin diagram we first observe that each line in
the configuration is disconnected from three other lines, e.g., 
$\{4,1,7\}$ have no nodes in common with the lines $\{2,3,9\}$,
$\{5,6,9\}$, $\{8,10,9\}$. This implies that all nodes in the
Dynkin diagram will be connected to three other nodes. Proceeding
as in Section~\ref{section:incidenceanddynkin} leads to the Dynkin
diagram in Figure~\ref{figure:D2(Petersen)}, which we identify as the
Petersen graph. The corresponding Cartan matrix \index{Cartan matrix} is
\begin{equation}
  A(\mf{g}_\mathrm{Petersen})=\left(
    \begin{array}{@{}r@{\quad}r@{\quad}r@{\quad}r@{\quad}r@{\quad}r@{\quad}r@{\quad}r@{\quad}r@{\quad}r@{}}
      2 & -1 & 0 & 0 & 0 & 0 & 0 & 0 & -1 & -1 \\
      -1 & 2 & -1 & 0 & 0 & -1 & 0 & 0 & 0 & 0 \\
      0 & -1 & 2 & -1 & 0 & 0 & 0 & -1 & 0 & 0 \\
      0 & 0 & -1 & 2 & -1 & 0 & 0 & 0 & 0 & -1 \\
      0 & 0 & 0 & -1 & 2 & -1 & 0 & 0 & -1 & 0 \\
      0 & -1 & 0 & 0 & -1 & 2 & -1 & 0 & 0 & 0 \\
      0 & 0 & 0 & 0 & 0 & -1 & 2 & -1 & 0 & -1 \\
      0 & 0 & -1 & 0 & 0 & 0 & -1 & 2 & -1 & 0 \\
      -1 & 0 & 0 & 0 & -1 & 0 & 0 & -1 & 2 & 0 \\
      -1 & 0 & 0 & -1 & 0 & 0 & -1 & 0 & 0 & 2 \\
    \end{array}
  \right),
  \label{PetersenCartanMatrix}
\end{equation}
which is of Lorentzian signature with
\begin{equation}
  \det A(\mf{g}_\mathrm{Petersen})=-256.
  \label{PetersenDeterminant}
\end{equation}
The Petersen graph was invented
by the Danish mathematician \emph{Julius Petersen} in the end of
the 19th century. It has several embeddings on the
plane, but perhaps the most famous one is as a star inside a
pentagon as depicted in Figure~\ref{figure:D2(Petersen)}. One of its
distinguishing features from the point of view of graph theory is
that it contains a \emph{Hamiltonian path} but no
\emph{Hamiltonian cycle}\footnote{We recall that a
Hamiltonian path is defined as a path in an undirected graph which
intersects each node once and only once. A Hamiltonian cycle is
then a Hamiltonian path which also returns to its initial node.}.

  \begin{figure}[htbp]
    \centerline{\includegraphics[width=45mm]{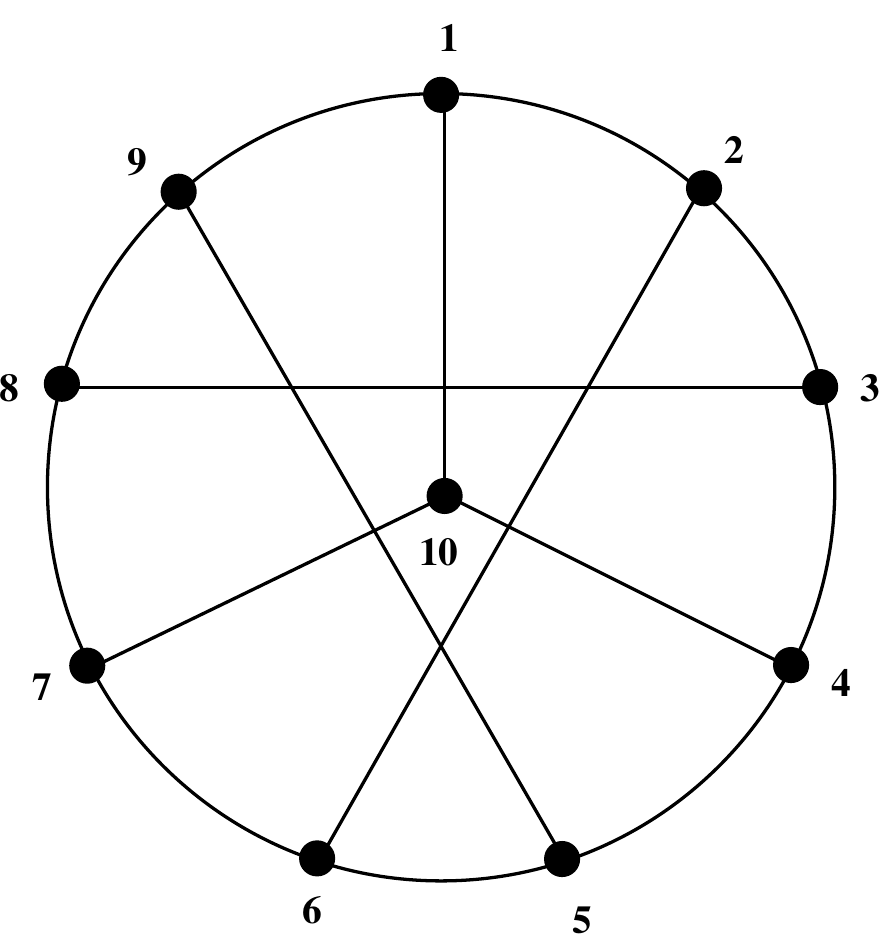}}
    \caption{An alternative drawing of the Petersen graph in the
      plane. This embedding reveals an $S_3$ permutation symmetry about
      the central point.}
    \label{figure:D2(Petersen)2}
  \end{figure}

Because the algebra is Lorentzian (with a metric that coincides
with the metric induced from the embedding in $E_{10}$), it does
not need to be enlarged by any further generator to be compatible
with the Hamiltonian constraint. Recently, cosmological solutions associated with the Petersen algebra were further analyzed in \cite{Ivashchuk:2008ns}. 

Some mathematical aspects of the present analysis was extended in {\bf Paper II}, were all rank 10 and 11 Coxeter groups whose Coxeter graphs have incidence index 3 and 4, respectively, were classified. For the rank 10 class, it was found that 11 of the 21 Coxeter groups with $\mc{I}=3$ can be embedded in to $E_{10}$, while the remaining correspond to indefinite Kac-Moody algebras which are not part of $E_{10}$. In the case of the rank 11 Coxeter groups, it was found that 28 of the 266 Coxeter groups with $\mc{I}=4$ could be embedded into $E_{11}$.

%%%%%%%%%%%%%%%%%%%%%%%%%%%%%%%%%%%%%%%%%%%%%%%%%%%
%
%\chapter{$E_{10}$ and Massive Type IIA Supergravity}
%
%%%%%%%%%%%%%%%%%%%%%%%%%%%%%%%%%%%%%%%%%%%%%%%%%%%

\chapter{$\mc{E}_{10}$ and Massive Type IIA Supergravity}
\label{Chapter:MassiveIIA}

In the previous chapter, we discussed in detail the example of eleven-dimensional supergravity, and how its associated dynamics in a cosmological regime is reproduced from a geodesic sigma model with target space $K(\mc{E}_{10})\bas \mc{E}_{10}$. This correspondence relied heavily on the level deomposition of the Lie algebra $E_{10}$ with respect to the finite-dimensional subalgebra $\mf{sl}(10, \mbb{R})$, which at low levels reveals representations matching the tensor structure of the $p$-form fields of eleven-dimensional supergravity. As discussed in some detail in Section \ref{section:DecompE10E10}, slicing up $E_{10}$ with respect to different finite subalgebras gives rise to the field contents also of the type IIA and IIB supergravities in ten dimensions. In this chapter we shall investigate this correspondence in the context of massive type IIA supergravity, originally discovered by Romans \cite{Romans:1985tz}, whose field content arises from a decomposition of $E_{10}$ with respect to a specific $\mf{sl}(9, \mbb{R})$-subalgebra. Here it is particularly interesting to note that from the $E_{10}$ point of view, the mass deformation arises from a representation at $\mf{sl}(10,\mbb{R})$-level 4 in the terminology of Sections \ref{section:DecompE10E10} and \ref{section:E10SigmaModel}. Although this level 4 representation has no interpretation in the eleven-dimensional theory, we shall find that it matches precisely with the mass-parameter in the type IIA context, thus revealing an algebraic and dynamic unification of massive IIA supergravity and eleven-dimensional supergravity with the framework of the $K(\mc{E}_{10})\bas \mc{E}_{10}$-sigma model at low levels. In this chapter we will spell out this relation in detail, including both the bosonic and fermionic sectors of massive IIA supergravity. To include fermions in the correspondence, we must extend the $K(\mc{E}_{10})\bas \mc{E}_{10}$-sigma model of Section \ref{section:E10SigmaModel} to incorporate spinorial representations of $K(\mc{E}_{10})$. This is developed in Section \ref{sec:E10} following earlier work \cite{deBuyl:2005zy,Damour:2005zs,deBuyl:2005mt,Damour:2006xu}. To improve readability, many of the details of the calculations are relegated to Appendix \ref{Appendix:MassiveIIA}. This chapter is based on {\bf Paper VI}, written in collaboration with Marc Henneaux, Ella Jamsin and Axel Kleinschmidt. 
\section{Massive IIA supergravity}
\label{sec:massiveIIA}

Massive type IIA supergravity was first constructed by Romans \cite{Romans:1985tz} by deforming the standard type IIA supergravity through a St\"uckelberg mechanism, giving a mass to the two-form potential through the replacement $F_{\mu\nu}\rightarrow F_{\mu\nu}+mA_{\mu\nu}$, where $F_{\mu\nu}=2\partial_{[\mu}A_{\nu]}$. The potential $A_{\mu}$ can then be gauged away by a gauge transformation of $A_{\mu\nu}$. This process unfortunately obscures the massless limit to recover the standard IIA theory~\cite{Giani:1984wc,Campbell:1984zc,Huq:1983im} as some of the supersymmetry variations involve a coefficient $m^{-1}$. This is remedied by a field redefinition presented in \cite{Bergshoeff:1996ui,Lavrinenko:1999xi}, that we will make use of in this chapter. 

\begin{figure}
\begin{center}
\includegraphics{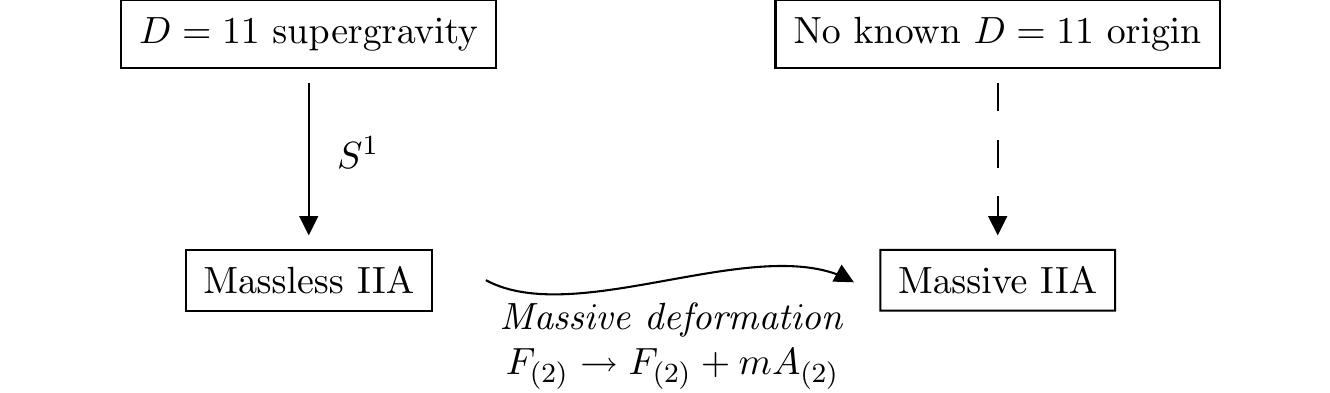}
\end{center}
\caption{\label{mIIA} \sl Massive IIA supergravity from $D=11$
    supergravity: Massive type IIA supergravity is obtained as a deformation of the standard type IIA supergravity, but unlike the latter, it does not possess any known eleven-dimensional origin. See Figure \ref{E10mIIA} for a pictorial description of how $D=11$ supergravity and massive IIA supergravity are unified inside $\mc{E}_{10}$.}
\end{figure}

Moreover, a more democratic version of massive type IIA is given in \cite{Lavrinenko:1999xi,Bergshoeff:2001pv}, in which every form field comes with its dual. In particular, it involves a nine-form as a (potential) dual to the mass (seen as a field strength). This democratic formulation makes explicit the striking feature of massive IIA supergravity that it allows the existence of a D$8$-brane, that is known to exist in type IIA string theory. Indeed, this feature requires a nine-form potential that couples to the D$8$-brane, and accordingly does not appear in massless IIA supergravity. Although in our analysis we will not use directly the democratic formulation, we will employ the field redefinitions used in \cite{Bergshoeff:1996ui} in order to clarify the massless limit.

\begin{figure}[t!]
\begin{center}
\includegraphics{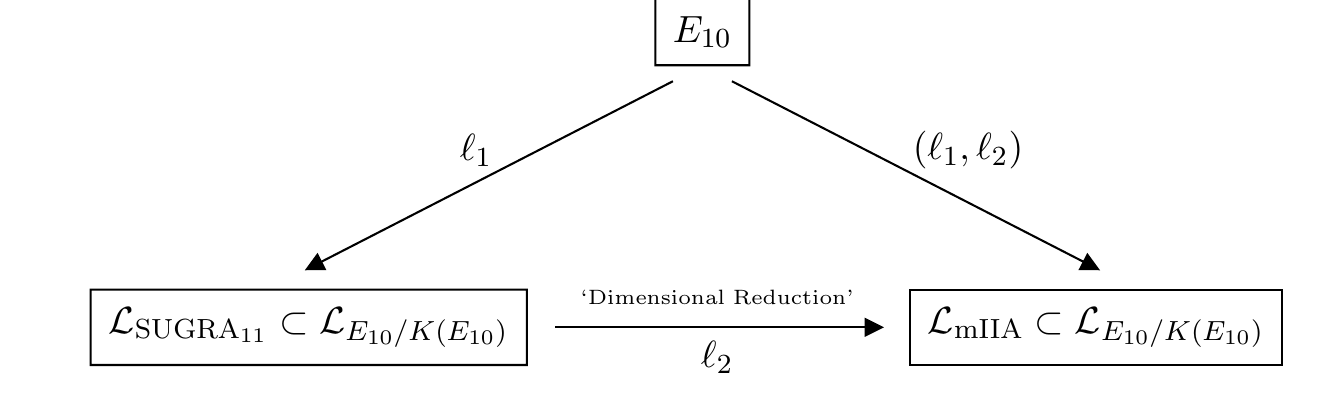}
\end{center}
\caption{\label{E10mIIA} \sl This picture describes the common $\mc{E}_{10}$ origin
  of eleven-dimensional supergravity and massive type IIA supergravity. First,
  if one considers a level $\ell_1$ decomposition of $\mc{E}_{10}$ with respect to
  $A_9$ (cf. Figure \ref{figure:E10}), one sees that the first levels
  ($\ell_1=0$ to $\ell_1=3$) of an $ K(\mc{E}_{10})\bas \mc{E}_{10} $ sigma-model correspond
  to a truncated version of eleven-dimensional supergravity, with Lagrangian
  $\mc{L}_{\mathrm{SUGRA}_{11}}$ \cite{Damour:2002cu}. Taking this as a
  starting point, we can perform an additional level $\ell_2$ decomposition on
  the sigma model. On the lower $\ell_1$ levels ($\ell_1=0$ to $\ell_1=3$)
  this is equivalent to a dimensional reduction of eleven-dimensional
  supergravity, which gives massless IIA supergravity (cf. Figure
  \ref{mIIA}). However, if one includes one of the generators appearing at
  $\ell_1 = 4$, this leads to a theory that coincides with a truncated version
  of massive IIA supergravity, with Lagrangian
  $\mathcal{L}_{\mathrm{mIIA}}$. This procedure is equivalent to a multi-level
  $(\ell_1,\ell_2)$ decomposition of $\mc{E}_{10}$ with respect to $A_8$. More
  details on this are given in Section \ref{sectionlevelE10}. By similar
  arguments, one could add IIB supergravity to this picture, in the sense that
  it has the same $\mc{E}_{10}$ origin  in a level decomposition with respect to
  $A_8\times A_1$, i.e. with respect to node 8 in the Dynkin diagram in Figure
  \ref{figure:E10} \cite{Kleinschmidt:2004rg}.}  
\end{figure}

In addition, as a motivation for the present work, one also finds a nine-form in a certain decomposition of $\mc{E}_{10}$, as will be developed in the next section, and this nine-form appears in the $ K(\mc{E}_{10})\bas \mc{E}_{10} $ Lagrangian in the same way as the mass term in the massive IIA Lagrangian.
Massive IIA supergravity has in common with many other deformed maximal
supergravities that it does not possess any known higher dimensional origin,
as illustrated in Figure \ref{mIIA}. A consequence of the present work is to
show that, although they are not related by dimensional reduction, eleven
dimensional supergravity and massive IIA supergravity have the same $\mc{E}_{10}$
origin as displayed in Figure \ref{E10mIIA}, see
also~\cite{InvarianceUnderCompactification,Kleinschmidt:2003mf,West:2004st}.\footnote{We note
  that the search for an eleven-dimensional origin of the Romans mass
  parameter, and hence the D$8$-brane, has led to studies of an M-theory
  M$9$-brane which is meant to exist in the presence of one Killing direction
  and then reduces to the IIA D$8$-brane, see
  e.g.~\cite{Bergshoeff:1997ak,Bergshoeff:1999bx}.} 

\subsection{Romans' Theory}
In this section, and throughout the chapter, the curved ten-dimensional space-time indices are denoted by $\mu$ and decompose into time $t$ and nine spatial indices $m$. The flat indices are similarly denoted by $\alpha=(0,a)$. Our space-time signature is $(-+\ldots +)$. More details on the conventions can be found in Appendix~\ref{Appendix:mIIAdetails}.

The complete supersymmetric Lagrangian (up to second order in fermions) in
Einstein frame is given by the sum of a bosonic and a fermionic part,
$\mathcal L=\mathcal L^{[B]}+\mathcal L^{[F]}$. The bosonic sector contains a
metric $G_{\mu\nu}$, a dilaton $\phi$, a one-form $A_{(1)}$ with field
strength $F_{(2)}$, a two-form $A_{(2)}$ with field strength $F_{(3)}$, a
three-form $A_{(3)}$ with field strength $F_{(4)}$, and a real mass parameter
$m$. The bosonic part of the Lagrangian reads {\allowdisplaybreaks
\begin{subequations}
\label{lag}
\beqa
\mathcal L^{[B]} &=& \sqrt {-G} \Bigg[
  R-\f12|\partial\phi|^2-\f14e^{3\phi/2}|F_{(2)}|^2-\f1{12}e^{-\phi}|F_{(3)}|^2
    -\f1{48}e^{\phi/2}|F_{(4)}|^2-\f12m^2e^{5\phi/2}\Bigg]\nn\\  
&+&\varepsilon^{\mu_1\ldots\mu_{10}}\Bigg[\f1{144}
 \partial_{\mu_1}A_{\mu_2\mu_3\mu_4}\partial_{\mu_5}A_{\mu_6\mu_7\mu_8}A_{\mu_9\mu_{10}}  
 +\f{m}{288} \partial_{\mu_1}A_{\mu_2\mu_3\mu_4} A_{\mu_5\mu_6} A_{\mu_7\mu_8}
    A_{\mu_9\mu_{10}}  \nn\\
&& \ \ \ \ \ \ \ \ \ \ \ \ + \f{m^2}{1280}
A_{\mu_1\mu_2}A_{\mu_3\mu_4}A_{\mu_5\mu_6}A_{\mu_7\mu_8}A_{\mu_9\mu_{10}}\Bigg]. 
\eqa
Here, we already point out that the massive deformation induces a positive definite potential on the scalar sector. This is in marked contrast to what happens for gauged supergravity in lower dimensions where the scalar potential is generically indefinite~\cite{Nicolai:2000sc,deWit:2008ta}, causing also problems in the connection to $\mc{E}_{10}$~\cite{Bergshoeff:2008}.

On the fermionic side, we have two gravitini, combined in a single $10\times32$ component
vector-spinor $\psi_\mu$, and two dilatini, combined in a single $32$
component Dirac-spinor $\lambda$, which decompose into two
fields of opposite chirality under $SO(1,9)$. For this sector the Lagrangian
takes the form 
\beqa
\mathcal L^{[F]} &=& i \sqrt {-G}\ \Bigg[ - 2\bpt_{\mu_1}\Gamma^{\mu_1\ldots\mu_3} D_{\mu_2}\pt_{\mu_{3}}
-\blt\G^{\mu} D_{\mu}\lt + \partial_{\mu_1}\phi\blt\G^{\mu_2}\G^{\mu_1}\pt_{\mu_2} \nn\\
&+&\frac1{4}e^{3\phi/4}F_{\mu_1\mu_2}\Big(\bpt_{[\nu_1}\G^{\nu_1}\G^{\mu_1\mu_2}\G^{\nu_2}\G_{10}\pt_{\nu_2]} -\frac3{2}\blt\G^\nu\G^{\mu_1\mu_2}\G_{10}\pt_\nu+\frac58\blt\G^{\mu_1\mu_2}\G_{10}\lt\Big)\nn\\
&+&\frac1{12}e^{-\phi/2}F_{\mu_1\ldots\mu_3}\Big(\bpt_{[\nu_1}\G^{\nu_1}\G^{\mu_1\ldots\mu_3}\G^{\nu_2}\G_{10}\pt_{\nu_2]}-\blt\G^\nu\G^{\mu_1\ldots\mu_3}\G_{10}\pt_\nu\Big)\nn\\
&-&\frac1{48}e^{\phi/4}F_{\mu_1\ldots\mu_4}\Big(\bpt_{[\nu_1}\G^{\nu_1}\G^{\mu_1\ldots\mu_{4}}\G^{\nu_2}\pt_{\nu_2]}-\frac12\blt\G^\nu\G^{\mu_1\ldots\mu_{4}}\pt_\nu+\frac38\blt\G^{\mu_1\ldots\mu_{4}}\lt\Big) \nn\\
&+&\frac{21}{8}me^{5\phi/4}\blt\lt-\frac{1}{2}me^{5\phi/4}\bpt_{\mu_1}\G^{\mu_1\mu_2}\pt_{\mu_2}+\frac{5}{4}me^{5\phi/4}\blt\G^{\mu}\pt_{\mu}\Bigg].
\eqa
\end{subequations}
The last line contains the explicit mass terms for the fermions. There are
also implicit mass deformations in the definitions of the field strength in
terms of the gauge potentials

\beqa
\label{curvs}
F_{\mu_1\mu_2} &=& 2\partial_{[\mu_1} A_{\mu_2]} + m A_{\mu_1\mu_2}\,,\nn\\
F_{\mu_1\mu_2\mu_3} &=& 3\partial_{[\mu_1} A_{\mu_2\mu_3]}\,,\nn\\
F_{\mu_1\mu_2\mu_3\mu_4} &=& 4\partial_{[\mu_1} A_{\mu_2\mu_3\mu_4]} 
   + 4 A_{[\mu_1}F_{\mu_2\mu_3\mu_4]} 
   + 3 m A_{[\mu_1\mu_2} A_{\mu_3\mu_4]}\,.
\eqa
In the Lagrangian, they are contracted without additional factors, for example
\beq |F_{(2)}|^2=F_{\mu_1\mu_2}F^{\mu_1\mu_2}.
\eeq 
In (\ref{curvs}) we see the
characteristic feature of a deformed theory that the usual tensor hierarchy of 
gauge fields is broken: there are forms coupling to potentials of higher
degree but only through terms proportional to the deformation parameter.
The tangent space field strengths are defined as usual by
conversion with the (inverse) vielbein $e_\al{}^\mu$, for example
$F_{\al_1\al_2}=e_{\al_1}{}^{\mu_1} e_{\al_2}{}^{\mu_2}F_{\mu_1\mu_2}$. Flat
indices are raised and lowered with the Minkowski metric and we will often
write contracted flat spatial indices on the same level. 
\subsection{Supersymmetry Variations}

The supersymmetry variations leaving (\ref{lag}) invariant, up to total
derivatives and higher order fermion terms, are listed in this section. For the fermions, they read

\beqa
\delta_{\vet}\pt_\mu&=& D_\mu\vet-\frac1{32}me^{5\phi/4}\Gamma_\mu\vet-\frac1{64}e^{3\phi/4}F_{\nu\rho}({\Gamma_{\mu}}^{\nu\rho}-14\delta_{\mu}^{[\nu}\Gamma^{\rho]}\big)\G_{10}\vet\nn\\
&&+\frac1{96}e^{-\phi/2}F_{\nu\rho\sigma}\big({\Gamma_{\mu}}^{\nu\rho\sigma}-9\delta_{\mu}^{[\nu}\Gamma^{\rho\sigma]}\big)\Gamma_{10}\vet\nn\\
\label{deltapsi}
&&+\frac1{256}e^{\phi/4}F_{\nu\rho\sigma\gamma}\big({\Gamma_{\mu}}^{\nu\rho\sigma\gamma}-\f{20}{3}\delta_{\mu}^{[\nu}\Gamma^{\rho\sigma\gamma]}\big)\vet,
\eqa
and 
\beqa
\delta_{\vet}\lt&=&\frac1{2}\partial_\mu\phi\G^\mu\vet+\frac{5}{8}e^{5\phi/4}m\,\vet-\frac{3}{16}e^{3\phi/4}F_{\mu\nu}\G^{\mu\nu}\G_{10}\vet\nn\\
&&-\frac1{24}e^{-\phi/2}F_{\mu\nu\rho}\G^{\mu\nu\rho}\G_{10}\vet+\frac1{192}e^{\phi/4}F_{\mu\nu\rho\sigma}\G^{\mu\nu\rho\sigma}\vet,
\eqa
while for the bosons we have
\beq
\delta_{\vet}{e_\mu}^\alpha= i \bvet\G^\alpha\pt_\mu,\ \ \ \ \ \delta_{\vet}\phi= i \blt\vet,
\eeq
and
\beq\label{susybos}
\delta_{\vet} A_{\mu}=\theta_{\mu},\ \ \ \ \ \delta_{\vet} A_{\mu\nu}=\theta_{\mu\nu},\ \ \ \ \ \delta_{\vet} A_{\mu\nu\rho}=\theta_{\mu\nu\rho}+6 A_{[\mu}\theta_{\nu\rho]},
\eeq
where
\beqa
\theta_\mu&:=&i
e^{-3\phi/4}\left(-\bpt_\mu-\f3{4}\blt\G_\mu\,\right)\G_{10}\vet,\nn\\ 
\theta_{\mu\nu}&:=&i e^{\phi/2}\left(\, 2\
  \bpt_{[\mu}\G_{\nu]}-\f1{2}\blt\G_{\mu\nu}\,\right)\G_{10}\vet,\nn \\ 
\theta_{\mu\nu\rho}&:=&i e^{-\phi/4}\left(\, 3\ \bpt_{[\mu}\G_{\nu\rho]}+\f1{4}\blt\G_{\mu\nu\rho}\,\right)\vet.
\eqa
Note that the mass only enters in the supersymmetry variations of the fermions.
\section{On $\mc{E}_{10}$ and the Geodesic Sigma Model for $ K(\mc{E}_{10})\bas \mc{E}_{10} $}
\label{sec:E10}

In this section we give some basic properties of the Kac--Moody
algebra $E_{10}$ and explain how to construct a non-linear sigma model
for geodesic motion on the infinite-dimensional coset space
$ K(\mc{E}_{10})\bas \mc{E}_{10} $. To this end we shall slice up the adjoint representation
of $E_{10}$ in a multi-level decomposition, suitable to reveal the field
content of massive IIA supergravity~\cite{Schnakenburg:2002xx,Kleinschmidt:2003mf}. In order to incorporate also the
fermionic sector in the sigma model, we will analyse the relevant (unfaithful) 
Dirac-spinor and vector-spinor representations of $K(E_{10})$ up to
the desired level.\footnote{The vector-spinor representation $\Psi$ that we construct (following \cite{deBuyl:2005zy,Damour:2005zs,deBuyl:2005mt,Damour:2006xu}) is unfaithful in the sense that it is a finite-dimensional representation of the infinite-dimensional algebra $K(E_{10})$. However, we stress that in the spirit of the original `gradient conjecture' of \cite{Damour:2002cu} it would be more natural to regard $\Psi$ as the first component in a faithful (infinite-dimensional) representation $\hat{\Psi}:=(\Psi, \pa_a\Psi, \dots )$ of $K(E_{10})$, where the remaining components encode spatial gradients of the gravitino $\Psi$ \cite{Damour:2005zs,deBuyl:2005mt}. As the employed ${\bf 320}$ representation is a fully consistent representation of $K(\mc{E}_{10})$, its transformations (that will be shown to be in good agreement with supergravity below) will never leave this $320$-dimensional (invariant) representation space. A reconciliation of this with the gradient conjecture is beyond the scope of the present analysis.} For a more detailed discussion of the general $E_{10}$ methods employed here we
refer to the review papers \cite{DHNReview,LivingReview}, for more
information about $K(E_{10})$ see
\cite{deBuyl:2005zy,Damour:2005zs,deBuyl:2005mt,Damour:2006xu},
and for a mathematical introduction to Kac-Moody algebras the canonical reference
is \cite{Kac}.  

\subsection{Generalities of the Kac-Moody Algebra $E_{10}$}
\label{sec:generalities}

Here we discuss the salient features of the hyperbolic Kac--Moody algebra $E_{10}$, the group of which we shall denote by $\mc{E}_{10}$. We will furthermore only be concerned with the split real form $E_{10}:=\mf{e}_{10(10)}(\mbb{R})$ of the complex Lie algebra $E_{10}(\mbb{C})$. The split real form is generated by ten triples $(e_i, f_i, h_i)$, $i=1,\dots ,10$, of Chevalley generators, each triple making up a distinguished subalgebra, 
\beq
\mf{sl}_i(2, \mbb{R})=\mbb{R}f_i \oplus \mbb{R}h_i \oplus \mbb{R}e_i \subset E_{10}.
\eeq 
These subalgebras are intertwined inside $E_{10}$ according to the stucture of the Dynkin diagram in Figure \ref{figure:E10}. The full structure of the algebra follows from multiple commutators of the form $[e_{i_1}, [e_{i_2}, \cdots [e_{i_{k-1}}, e_{i_k}]\cdots ]]$ (and similarly for the $f_i$'s) modulo the so-called Serre relations. We have the standard triangular decomposition
\beq
E_{10}=\mf{n}_-\oplus \mf{h} \oplus \mf{n}_+,
\eeq
where $\mf{h}=\sum_{i}\mbb{R}h_i$ is the Cartan subalgebra and the nilpotent parts $\mf{n}_{\pm}$ are generated by the $e_i$'s and $f_i$'s, respectively. In other words, the subspace $\mf{n}_+$ contains the positive step operators, while $\mf{n}_-$ contains the negative step operators. 
\begin{figure}[t]
\centering
\includegraphics{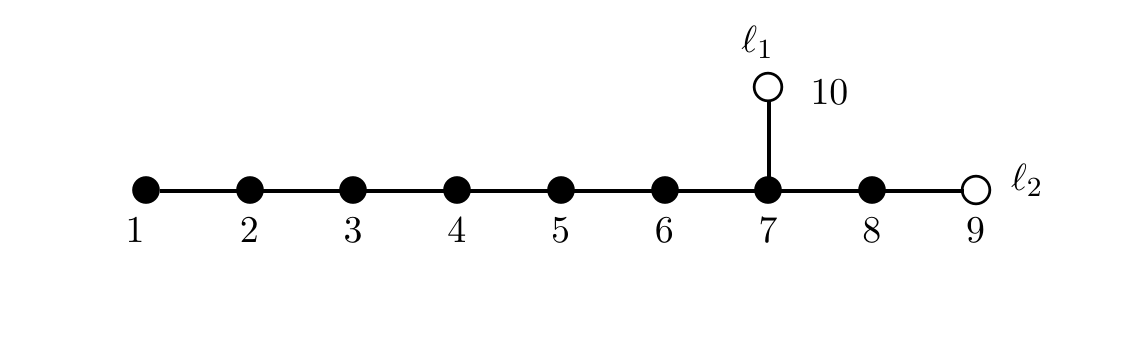}
\caption{\label{e10dynk}\sl The Dynkin diagram of $E_{10}$ with the nodes associated with the level decomposition indicated in white.}
\label{figure:E10}
\end{figure}

The maximal compact subalgebra $K(E_{10})\subset E_{10}$ is defined as the subalgebra which is pointwise fixed by the Chevalley involution $\om$,
\beq\label{ke10}
K(E_{10}):=\{ x\in E_{10} \ |\  \om(x)=x\},
\eeq
where $\om$ is defined through its action on each triple $(e_i, f_i, h_i)$:
\beq
\om(e_i)=-f_i, \qquad \om(f_i)=-e_i, \qquad \om(h_i)=-h_i.
\eeq
By virtue of the existence of $K(E_{10})$ we have the standard Iwasawa  and Cartan decompositions (direct sums of vector spaces),
\beqa
E_{10} &=& K(E_{10})\oplus \mf{h} \oplus \mf{n}_+,\quad \hs \text{(Iwasawa)}
\nn \\
E_{10} &=& K(E_{10})\oplus \mf{p}, \qquad \qquad \text{(Cartan)},
\label{CartanIwasawa}
\eqa
which will both be of importance in subsequent sections. The generators belonging to $\mf{p}$ are those which are anti-invariant under $\om$. Note that the subspace $\mf{p}$ does not close under the Lie bracket, but rather transforms in a representation of $\mf{k}$. On the other hand, the subspace $\mf{b}:=\mf{h}\oplus \mf{n}_+$ is a subalgebra of $E_{10}$, known as a \emph{Borel subalgebra}. Using the Chevalley involution we project an arbitrary generator $x\in E_{10}$ onto the different subspaces of  the Cartan decomposition according to
\beqa
\mc{Q}&:=& \f{1}{2}\big[x+\om(x)\big] \hs \in \hs   K(E_{10}), 
\nn \\
\mc{P}&:=& \f{1}{2}\big[x-\om(x)\big] \hs \in \hs \mf{p}.
\label{CartanProjection}
\eqa

The Cartan matrix $A$ of $E_{10}$, deduced from the Dynkin diagram in Figure \ref{figure:E10}, induces an indefinite and non-degenerate bilinear form $\left( \cdot |\cdot \right)  $ on $\mf{h}$ as follows
\beq
\left( h_i|h_j\right)  :=A_{ij}.
\eeq
By invariance, this bilinear form can be extended to all of $E_{10}$, and in particular for the Chevalley generators we have
\beq
\left( e_i|f_j\right)  =\delta_{ij}, \qquad \left( h_i|e_j\right)  =0, \qquad \left( h_i|f_j\right)  =0.
\eeq
This bilinear form will be used in subsequent sections to construct a manifestly $E_{10}$-invariant Lagrangian.

\subsection{The IIA Level Decomposition of $E_{10}$ and $K(E_{10})$}
\label{sectionlevelE10}
To elucidate the relation between the infinite-dimensional algebra $E_{10}$ and the field content of massive type IIA supergravity, we shall perform a decomposition of the adjoint representation of $E_{10}$ into representations of the finite-dimensional subalgebra $A_8\cong \mf{sl}(9, \mbb{R})$, defined by nodes $1, \dots, 8$ in the Dynkin diagram in Figure \ref{figure:E10} (see also~\cite{Kleinschmidt:2003mf}). Each generator of $E_{10}$ is then represented as an $\mf{sl}(9, \mbb{R})$-tensor, say $X_{a_1\cdots a_k}\in E_{10}$, where the indices are interpreted as the flat spatial indices of the ten-dimensional supergravity theory and transform in a given irreducible representation described by a set of Dynkin indices, or equivalently by a Young tableau. Since $E_{10}$ is of rank 10 and $A_8$ is of rank 8, this decomposition is a multilevel $\ell:=(\ell_1, \ell_2)$-decomposition, with $\ell_1$ being associated with the exceptional node in the Dynkin diagram, while $\ell_2$ corresponds to the rightmost node. 

A useful alternative point of view on this decomposition is to first perform only the $\ell_1$ decomposition with respect to the horizontal $A_{9}\cong \mf{sl}(10, \mbb{R})$ subalgebra consisting of nodes $1$ through to $9$ in Figure~\ref{e10dynk}. As is well-known, at low $\ell_1$ levels, this decomposition gives rise to the field content of eleven-dimensional supergravity \cite{Damour:2002cu,DamourNicolaiLowLevels}. We can then view the additional decomposition with respect to $\ell_2$ as a `dimensional reduction' from $D=11$ to $D=10$. It is this point of view which enables us to relate the mass term in massive type IIA supergravity to a generator of $E_{10}$ at level $\ell_1=4$ in the standard `M-theory decomposition' of \cite{Damour:2002cu,DamourNicolaiLowLevels,West:2004st}. This is intriguing since in $D=11$ the matching between supergravity and $E_{10}$ has only been successful up to $\ell_1=3$. Thus, the mass term in $D=10$ provides a non-trivial check of $E_{10}$ beyond its `$\mf{sl}(10, \mbb{R})$-covariantized $\mf{e}_8$' subset, i.e. the generators of $\mf{e}_8$ and their images under (the Weyl group of) $\mf{sl}(10, \mbb{R})$.

The relation between $E_{10}$ and $\mf{e}_{11}$ and the mass deformation parameter of type IIA supergravity has been pointed out in various earlier references. In \cite{Schnakenburg:2002xx} the massive type IIA theory was reformulated as a non-linear realization where the mass term was associated with a certain nine-form generator of $E_{11}$. Shortly after, in \cite{InvarianceUnderCompactification}, it was observed that the mass term in Romans' theory corresponds to a positive real root of $E_{10}$. This was further elaborated upon in \cite{Kleinschmidt:2004dy} where a truncated version of massive type IIA supergravity in an $SO(9,9)$-covariant formulation was shown to be equivalent to the sigma model for $E_{10}$ in a level decomposition with respect to a $D_9\cong \mf{so}(9,9)$-subalgebra. This analysis mainly focused on the bosonic sector, but a preliminary analysis of the fermionic sector was initiated by restricting to the zeroth level of the $E_{10}$-decomposition. It was also pointed out in \cite{Kleinschmidt:2003mf,West:2004st,LivingReview} that the mass term has a natural interpretation as a generator at level four in a decomposition of $E_{10}$ with respect to $A_8\cong \mf{sl}(9, \mbb{R})$. From this point of view, the kinetic term in the sigma model associated with this generator naturally gives rise to the mass term of type IIA upon dimensional reduction. This is the viewpoint which we extend in the present work.

\subsubsection{Generators of $E_{10}$}
The level decomposition of $E_{10}$ under the $A_8\cong \mf{sl}(9,\mbb{R})$ subalgebra up to level $(\ell_1, \ell_2)=(4,1)$ is shown in Table~\ref{iiadec}. At level $(\ell_1, \ell_2)=(0,0)$ there is a copy of $\mf{gl}(9, \mbb{R})=\mf{sl}(9, \mbb{R})\oplus \mbb{R}$, as well as a scalar generator associated with the dilaton. The commutation relations at this level are ($a,b=1,\ldots, 9$)
\beqa
\lb K^a{}_b , K^c{}_d \rb &=& \delta^c_b K^a{}_d - \delta^a_d K^c{}_b,
\nn \\
\lb T, K^a{}_b \rb &=& 0,
\eqa
and the bilinear form reads
\beq
\left( K^a{}_b| K^c{}_d\right)   = \delta^a_d\delta^c_b- \delta^a_b\delta^c_d, \qquad \left( T|T\right)  =\f{1}{2}, \qquad \left( T|{K^{a}}_b\right)  =0.
\eeq
\begin{table}
\centering
\begin{tabular}{|c|c|c|}
\hline
$(\ell_1,\ell_2)$ & $\mf{sl}(9, \mbb{R})$ Dynkin labels & Generator of $E_{10}$ \\
\hline
\hline
$(0,0)$ & $[1,0,0,0,0,0,0,1]\oplus[0,0,0,0,0,0,0,0]$ & $K^a{}_b$ \\
$(0,0)$ & $[0,0,0,0,0,0,0,0]$ & $T$ \\
$(0,1)$ & $[0,0,0,0,0,0,0,1]$ & $E^{a}$\\
$(1,0)$ & $[0,0,0,0,0,0,1,0]$ & $E^{a_1a_2}$\\
$(1,1)$ & $[0,0,0,0,0,1,0,0]$ & $E^{a_1a_2a_3}$\\
$(2,1)$ & $[0,0,0,1,0,0,0,0]$ & $E^{a_1\ldots a_5}$\\
$(2,2)$ & $[0,0,1,0,0,0,0,0]$ & $E^{a_1\ldots a_6}$\\
$(3,1)$ & $[0,1,0,0,0,0,0,0]$ & $E^{a_1\ldots a_7}$\\
$(3,2)$ & $[1,0,0,0,0,0,0,0]$ & $E^{a_1\ldots a_8}$\\
$(3,2)$ & $[0,1,0,0,0,0,0,1]$ & $E^{a_0|a_1\ldots a_7}$\\
$(4,1)$ & $[0,0,0,0,0,0,0,0]$ & $E^{a_1\ldots a_9}$\\
\hline
\end{tabular}
\caption{\label{iiadec}\sl IIA Level decomposition of $E_{10}$ under $\mf{sl}(9, \mbb{R})$.}
\end{table}

All objects transform as $\mf{gl}(9, \mbb{R})$ tensors in the obvious way. The positive level generators are obtained through multiple commutators between the (fundamental) generators $E^{a}$ and $E^{ab}$ on levels $(0,1)$ and $(1,0)$, respectively. For example, the generator on level $(1,1)$ is obtained simply as the commutator
\beq
[E^{ab}, E^{c}]=E^{abc}.
\label{commutatorexample}
\eeq
All the remaining relevant commutators up to level $(4,1)$ are given in Appendix \ref{app:commutators}. No generators of $E_{10}$ appear on mixed positive and negative levels, meaning that for any $X_{\ell}\in E_{10}$ the levels $\ell_1$ and $\ell_2$ are either both non-positive or both non-negative. In terms of the multilevel $\ell=(\ell_1, \ell_2)$ we shall denote this by $\ell\leq 0$ or $\ell \geq 0$, respectively. This implies that the level decomposition induces a grading of $E_{10}$ into an infinite set of finite-dimensional subspaces $\mf{g}_{\ell}$ with respect to the multilevel $\ell$. Let $E_{\ell}$ and $E_{\ell^{\prime}}$ be arbitrary root vectors in the subspaces $\mf{g}_{\ell}$ and $\mf{g}_{\ell^{\prime}}$ of $E_{10}$. Then a generic commutator, generalizing (\ref{commutatorexample}), takes the form
\beq 
[E_{\ell}, E_{\ell^{\prime}}]=E_{\ell+\ell^{\prime}} \hs \in \hs \mf{g}_{\ell+\ell^{\prime}}.
\eeq
The negative level generators are simply obtained using the Chevalley involution, and for an arbitrary positive level generator $E_{\ell}\in \mf{g}_{\ell}$ we define the associated negative generator as
\beq
F_{\ell} :=-\om(E_{\ell}) \in \mf{g}_{-\ell}, \qquad   \ell\in \mbb{Z}^2_{\ge 0}.
\eeq
Because of the graded structure, commutators between generators in $\mf{g}_{\ell}$ and $\mf{g}_{-\ell}$ belong to the zeroth subspace $\mf{g}_0$. For example, for the fundamental generator $E^{ab}\in \mf{g}_{(0,1)}$ we have 
\beq
[E^{a_1a_2},F_{b_1b_2}]=-\frac12 \delta^{a_1a_2}_{b_1b_2} K 
    +4 \delta^{[a_1}_{[b_1} K^{a_2]}_{\,\,\,\,\, b_2]}   
    -2 \delta^{a_1a_2}_{b_1b_2} T.
\eeq
The explicit form of the remaining commutators can be found in Appendix
\ref{app:commutators}. 

\subsubsection{The ``Mass Generator''}

Let us now discuss in more detail how the generator associated with the mass term appears in the level decomposition of $E_{10}$. First consider the decomposition with respect to $\ell_1$. At level $\ell_1=4$ one finds a generator corresponding to an $\mf{sl}(10, \mbb{R})$-tensor with 12 indices of the form $E^{\dot{a}|\dot{b}|\dot{c}_1\cdots \dot{c}_{10}}$ (with $\dot{a}=(10, a),$ etc.), which has mixed Young symmetry, i.e. it is antisymmetric in the block of indices $\dot{c}_1, \dots, \dot{c}_{10}$ and symmetric in $\dot a, \dot b$. This generator has no physical interpretation in $D=11$ supergravity. However, consider now the further level decomposition with respect to $\ell_2$. This can be realized by doing a dimensional reduction on $E^{\dot{a}|\dot{b}|\dot{c}_1\cdots \dot{c}_{10}}$ along the direction `10'. We are interested in the generator obtained in this way by fixing the first three indices to $\dot{a}=\dot{b}=\dot{c}_1=10$, which yields \cite{West:2004st,LivingReview}
\beq\label{massgen}
E^{a_1\cdots a_{9}}:= \f18E^{10|10|10 a_1\cdots a_{9}}.
\eeq
The resulting tensor corresponds to the $(\ell_1, \ell_2)=(4,1)$-generator $E^{a_1\cdots a_9}$ in Table \ref{iiadec}. This $\mf{sl}(9, \mbb{R})$-tensor is a nine-form, and is therefore associated with a supergravity potential $A_{a_1\cdots a_9}$ whose field-strength is a top-form in $D=10$. This is the generator associated with the mass term. From this analysis it is clear that the mass deformation of type IIA supergravity probes $E_{10}$ beyond the realm of what has been successfully verified previously in the context of $D=11$ supergravity. This is especially interesting due to the fact that the $\ell_1=4$-generator $E^{\dot{a}|\dot{b}|\dot{c}_1\cdots \dot{c}_{10}}$ is a `genuine' $E_{10}$ element, with no components contained in $E_8$ nor in $E_9$.

\subsubsection{Generators of $K(E_{10})$}

The level decomposition of $E_{10}$ induces a decomposition of its maximal compact subalgebra $K(E_{10})$ which was defined in (\ref{ke10}). Using the Cartan decomposition, (\ref{CartanIwasawa}) and (\ref{CartanProjection}), we define compact and noncompact generators, $J_{\ell}\in K(E_{10})$ and $S_{\ell}\in \mf{p}$, as follows
\beq
J_{\ell}:= E_{\ell}-F_{\ell}, \qquad S_{\ell}:= E_{\ell}+F_{\ell}, \qquad \ell \in \mbb{Z}_{\ge 0}^2\backslash \{(0,0)\}.
\label{CompactAndNoncompactGenerators}
\eeq
Furthermore, at level $(0,0)$ we have the $\mf{so}(9)$ Lorentz generators
\beq\label{00gens}
M^{ab}:={K^a}_b-{K^b}_a.
\eeq
This decomposition of $K(E_{10})$ is not a gradation, but rather corresponds to a \emph{filtration} \cite{Damour:2006xu}, in the sense that an arbitrary commutator between two `positive level' generators exhibits a graded structure modulo lower-level generators only,
\beq
[J_{\ell}, J_{\ell^{\prime}}]=J_{|\ell-\ell^{\prime}|}+J_{\ell+\ell^{\prime}},
\eeq 
where it is understood that $J_{|\ell-\ell^{\prime}|}\neq 0$ if and only if $(\ell-\ell^{\prime})\geq 0$ or $(\ell-\ell^{\prime})\leq 0$. In other words, $J_{|\ell-\ell^{\prime}|}$ is non-zero only when the difference $(\ell-\ell^{\prime})$ involves no mixing between negative and positive levels. 

Using (\ref{CompactAndNoncompactGenerators}), together with the $E_{10}$ commutators in Appendix \ref{app:commutators}, we may deduce the abstract $K(E_{10})$-relations at each `level'. Let us consider a few examples to illustrate the procedure. We begin by defining the $K(E_{10})$-generators associated with level $(0,0)$ and the fundamental generators at level $(0,1)$ and $(1,0)$:
\beq
 J^{a}:= E^{a}-F_a, \qquad J^{ab}:=E^{ab}-F_{ab}.
\eeq 
Since there are no generators of mixed level $(1,-1)$ the commutator $[E^{a}, F_b]$ vanishes, and the commutator between $J^{ab}$ and $J^{c}$ simply gives
\beq
[J^{ab}, J^{c}]=J^{abc},
\eeq
with $J^{abc}:=E^{abc}-F_{abc}$. Proceeding in the same way for $J^{a_1 a_2}$ and $J^{a_1a_2a_3}$, we obtain $J^{a_1\cdots a_5}$ modulo lower level terms, 
\beq
\lb J^{a_1a_2}, J^{a_3a_4a_5}\rb =J^{a_1\cdots a_5}-6 \delta_{a_1a_2}^{[a_3a_4}J^{a_5]}.
\eeq
Note that one may project onto $J^{a_1\cdots a_5}$ as follows,
\beq
J^{a_1\cdots a_5}= \lb J^{[a_1a_2}, J^{a_3a_4a_5]}\rb\,.
\eeq
Here, all indices are $\mf{so}(9)$ vector indices and are raised and lowered with the Euclidean $\delta_{ab}$ so that the position does not matter.
The other relevant $K(E_{10})$-commutators are listed in Appendix \ref{app:SpinorReps}. 

\subsection{Spinorial Representations}

The fermionic degrees of freedom in the sigma model for $ K(\mc{E}_{10})\bas \mc{E}_{10} $ transform in spinorial representations of $K(E_{10})$. The two relevant $K(E_{10})$ representations are finite-dimensional (unfaithful) and of dimensions $32$ and $320$, respectively.
In the decomposition of $K(E_{10})$ associated with eleven-dimensional supergravity these representations transform as a $32$-dimensional Dirac spinor representation $\epsilon$ of $\mf{so}(10)\subset K(E_{10})$ and a $320$-dimensional vector-spinor representation  $\Psi_{\dot{a}}, \hs \dot{a}=(10, a)$, of $\mf{so}(10)\subset K(E_{10})$, identified with the gravitino \cite{deBuyl:2005zy,Damour:2005zs,deBuyl:2005mt,Damour:2006xu}. Upon reduction to the IIA theory (through the additional level decomposition with respect to $\ell_2$) the gravitino decomposes into a 32-dimensional spinor $\Psi_{10}$ and a 288-dimensional vector spinor $\Psi_a$ of $\mf{so}(9)$. However, because they both descend from $\Psi_{\dot{a}}$, these two representations will mix under $K(E_{10})$~\cite{Kleinschmidt:2006tm}. No such complication arises for the supersymmetry parameter $\epsilon$, for which we keep the same notation in the IIA picture. The spinor $\Psi_{10}$ will be associated with the ten-dimensional dilatino, while the vector-spinor $\Psi_a$ is related to the gravitino. 

For the first three levels, the transformation properties of the spinor $\epsilon$ are
\beqa\label{vards}
M^{ab}\cdot \epsilon &=& \f{1}{2}\Gamma^{ab}\epsilon,
\nn \\
J^{a}\cdot \epsilon &=& \f{1}{2}\Gamma_{10}\Gamma^{a}\epsilon,
\nn \\
J^{ab}\cdot \epsilon &=& \f{1}{2}\Gamma_{10}\Gamma^{ab}\epsilon,
\eqa
where $\Gamma^{a}$ and $\Gamma^{ab}$ are $\mf{so}(9)$ $\Gamma$-matrices (our $\Gamma$-matrix conventions are given in Appendix \ref{app:conventions}). The higher level transformations are now defined through the abstract $K(E_{10})$-relations; for example:
\beqa
J^{a_1a_2a_3}\cdot \epsilon &:=& \Big[J^{a_1a_2}, J^{a_3}\Big]\cdot \epsilon \qquad =\f{1}{2}\Gamma^{a_1a_2a_3}\epsilon,
\nn \\
J^{a_1\cdots a_5}\cdot \epsilon &:=& \Big[J^{[a_1a_2}, J^{a_3a_4a_5]}\Big] \cdot \epsilon = \f{1}{2}\Gamma_{10}\Gamma^{a_1\cdots a_5}\epsilon.
\eqa
Similarly, we have for the spinor component $\Psi_{10}$ the following low-level transformations:
\beqa\label{varpsi10}
M^{a_1a_2}\cdot  \Psi_{10} &=&\f12\G^{a_1a_2}\Psi_{10}\nn\\
J^{a}\cdot \Psi_{10} &=& \f12 \G_{10}\G^a\Psi_{10}+\Psi^a,\nn\\
J^{a_1a_2}\cdot \Psi_{10} &=& \f16 \G_{10}\G^{a_1a_2}\Psi_{10}+\f43 \G^{[a_1}\Psi^{a_2]},
\eqa
showing explicitly the mixing between $\Psi_{10}$ and $\Psi_{a}$ under $K(E_{10})$. Finally, we also have for $\Psi_a$:
\beqa\label{varpsia}
M^{a_1a_2}\cdot  \Psi_{b} &=& \f12\G^{a_1a_2}\Psi_{b}+2\delta_b^{[a_1}\Psi^{a_2]},\nn\\
J^{a}\cdot \Psi_{b}&=&\f12 \G_{10}\G^a\Psi_{b}-\delta^a_b\Psi_{10},\nn\\
J^{a_1a_2}\cdot \Psi_{b} &=& \f12\G_{10}\G^{a_1a_2}\Psi_b -\f43\G_{10}\delta^{[a_1}_b\Psi^{a_2]}+\f23\G_{10}\G_b^{\ [a_1}\Psi^{a_2]}\nn\\
&&+\f43\delta_b^{[a_1}\G^{a_2]}\Psi_{10}-\f13\G_b^{\ a_1a_2}\Psi_{10}.
\eqa
More details on these $K(E_{10})$-representations can be found in Appendix \ref{app:SpinorReps}. We note that we could also have redefined the $SO(9)$ spinor $\Psi_a$ by a shift with $\G_a\G^{10}\Psi_{10}$ as one does in Kaluza--Klein reduction (cf. (\ref{kkferm})) but refrain from doing so here.

\subsection{The Non-Linear Sigma Model for $ K(\mc{E}_{10})\bas \mc{E}_{10} $}
\label{sec:sigmamod}

A non-linear sigma model with rigid $\mc{E}_{10}$-invariance and local $K(\mc{E}_{10})$-invariance may now be constructed using the properties of $E_{10}$ described in previous sections. By virtue of the Iwasawa decomposition we can always choose a coset representative in the partial `Borel gauge' by taking 
\beq
\cV (t) :=\mc{V}= \mc{V}_0  e^{\phi T} e^{A\star E}  \hs \in \hs  K(\mc{E}_{10})\bas \mc{E}_{10} ,
\eeq
where $\mc{V}_0=\text{exp}\ {{h^{a}}_b{K^{b}}_a}$ represents the $GL(9,
\mbb{R})$ inverse vielbein ${e_a}^m$, while $e^{A\star E}$ contains the
positive step operators of $E_{10}$. We call this a partial Borel gauge
since $\mc{V}_0$ is not constrained to contain only positive step operators of
$\mf{gl}(9,\mbb{R})$. In other words, $\mc{V}_0$ is an arbitrary $GL(9,
\mbb{R})$-matrix and not a representative of the coset $SO(9)\bas GL(9, \mbb{R})$. With some abuse of terminology, we shall sometimes refer to
the part of $\mc{E}_{10}$ parametrized by $\mc{V}$ as the `Borel subgroup', and
denote it by $\mc{E}_{10}^{+}$ \footnote{This differs from the `true' Borel subgroup $B:=
\text{exp}\ \mf{b}\subset \mc{E}_{10}$ through the negative root generators in
$\mc{V}_0$ (see Section \ref{sec:generalities} for the definition of
$\mf{b}$). $\mc{E}_{10}^{+}$ is only a parabolic subgroup of $\mc{E}_{10}$.}. The positive
level part $ e^{A \star E}$ of $\mc{V}$ is defined as  
\beq \label{posexp}
 e^{A \star E}:=\text{exp}\big[A_{m}(t)E^m]\text{exp}\big[\frac{1}{2} A_{m_1m_2}(t) E^{m_1m_2}]\text{exp}\big[\frac{1}{3!} A_{m_1m_2m_3}(t) E^{m_1m_2m_3}\big]\cdots 
\eeq
with similar exponentials occurring for the higher levels. Note that in this expression the indices $m_1, m_2, \dots$ are $\mf{sl}(9, \mbb{R})$-indices, and hence correspond to curved spatial indices from a supergravity  point of view. 

The coset representative $\mc{V}$ transforms under global $g\in \mc{E}_{10}$-transformations from the right and local $k\in K(\mc{E}_{10})$-transformations from the left:
\beq
\mc{V} \longmapsto k\mc{V}g.
\label{transformationcoset}
\eeq
From $\mc{V}$ we construct the Lie algebra-valued Maurer--Cartan form 
\beq\label{mcform}
\partial_t \cV \cV^{-1} = \cP +\cQ,
\eeq
where we also employed the Cartan decomposition, according to (\ref{CartanProjection}). The transformation property of $\mc{V}$ in (\ref{transformationcoset}) implies that the coset part $\mc{P}\in \mf{p}$ of the Maurer-Cartan form is globally $\mc{E}_{10}$-invariant while transforms covariantly under $K(\mc{E}_{10})$:
\beq
K(\mc{E}_{10}) \hs : \hs \mc{P}\hs  \longmapsto\hs  k\mc{P}k^{-1}.
\eeq
On the other hand, $\mc{Q}\in K(E_{10})$ properly transforms as a connection,
\beq
K(\mc{E}_{10}) \hs : \hs \mc{Q}\hs \longmapsto\hs  k\mc{Q}k^{-1} +\pa_t k k^{-1}.
\eeq

Using the bilinear form on $E_{10}$, we can now construct a manifestly $\mc{E}_{10}\times K(\mc{E}_{10})_{\text{local}}$-invariant Lagrangian as follows \cite{Damour:2002cu,DamourNicolaiLowLevels}
\beq\label{e10boslag}
\mc{L}^{[B]}_{ K(\mc{E}_{10})\bas \mc{E}_{10} }=\f{1}{4}n(t) ^{-1}\left( \mc{P}\big|\mc{P}\right)  ,
\eeq
where the lapse function $n(t)$ ensures invariance under reparametrizations of the geodesic parameter $t$. We have also included a superscript $[B]$ to emphasize that this Lagrangian is only the bosonic part of a sigma model which also includes the fermionic degrees of freedom in the ${\bf 320}$ of $K(\mc{E}_{10})$ to be introduced shortly. The equations of motion for $n(t)$ enforces a lightlike (Hamiltonian) constraint on the dynamics,
\beq 
\left( \mc{P}\big|\mc{P}\right)  =0,
\eeq
while the equations of motion for $\mc{P}$ (in the gauge $n=1$) read
\beq
\mc{D}\mc{P}:= \pa_t \mc{P}-[\mc{Q}, \mc{P}]=0,
\label{bosoniceom}
\eeq
where we defined the $K(\mc{E}_{10})$-covariant derivative $\mc{D}$. Equation (\ref{bosoniceom}) encodes the dynamics of the bosonic sector of the sigma model and is written out in detail in Appendix~\ref{Appendix:EOMBosonic}.

The fermionic degrees of freedom are included in the Lagrangian through the spinor representation $\Psi$ as follows \cite{Damour:2005zs,deBuyl:2005mt,Damour:2006xu}
\beq\label{e10fermlag}
\mc{L}^{[F]}_{ K(\mc{E}_{10})\bas \mc{E}_{10} }=-\f{i}{2}\left( \Psi \big| \mc{D} \Psi\right)  ,
\eeq
where the bracket now denotes an invariant inner product on the representation space. The associated `Dirac equation' reads
\beq
\mc{D}\Psi:= \pa_t\Psi-\mc{Q}\cdot \Psi=0,
\label{fermioniceom}
\eeq
where it is understood that the connection $\mc{Q}$ acts on $\Psi$ in the vector-spinor representation constructed in the previous section. Equation (\ref{fermioniceom}) is written in terms of the full $320$-dimensional spinor $\Psi$, which encodes both the `gravitino' $\Psi_a$ and the `dilatino' $\Psi_{10}$. The separate equations of motion for these fields can be written out as follows
\beqa
\mc{D}\Psi_a&:=& \pa_t \Psi_{a\phantom{0}}-\mc{Q}\cdot \Psi_{a\phantom{0}}=0,
\nn \\
\mc{D}\Psi_{10}&:=& \pa_t\Psi_{10}-\mc{Q}\cdot \Psi_{10}=0.
\label{GravitinoDilatinoEOM}
\eqa
Recall that the $K(E_{10})$-action on $\Psi_a$ contains $\Psi_{10}$-terms, and vice versa. This is important since the same mixing between the gravitino and the dilatino occurs in the corresponding supergravity equations of motion (see Appendix~\ref{app:sugraf}). The fermionic equations of motion in (\ref{GravitinoDilatinoEOM}) are written out in more detail in Appendix \ref{app:FermionicSigmaModelEOM}.

The bosonic equations of motion (\ref{bosoniceom}) were written for the gauge choice $n=1$. The lapse function $n$ has a superpartner $\Psi_t$, which is a Dirac spinor under $K(E_{10})$, and the associated supersymmetry transformations are
\beqa \label{SusyTransfSigmaModel}
\delta_{\epsilon} n &=& i\epsilon^T\Psi_t,\nn \\
\delta_{\epsilon} \Psi_t &=& \mc{D}\epsilon.
\eqa
The fermionic equations of motion are then valid in the `supersymmetric gauge' $\Psi_t=0$. The associated constraint (analogous to the Hamiltonian constraint) that should be imposed is the supersymmetry constraint which states that the spin $\Psi$ is orthogonal to the velocity $\mc{P}$. The full form of this constraint is unknown due to the fact that it is not known how to supersymmetrize the $\mc{E}_{10}$ model correctly. Nevertheless, low level expressions have been obtained in~\cite{Damour:2005zs,Damour:2006xu}, which the reader should also consult for further discussions on this point.

Let us now analyse these properties of the geodesic sigma model in more detail by utilizing the level decomposition of $E_{10}$ with respect to $\mf{sl}(9, \mbb{R})$. Expanding the coset element $\mc{P}$ up to level $(3,2)$ we obtain 
\beq
\cP = \partial_t\phi T + \frac12 p_{ab} (K^a{}_b+K^b{}_a) +  \sum_{\ell> 0}P_{\ell}\star S_{\ell}   \label{Pexpanded}
\eeq
and the associated expansion for $\mc{Q}$ is given by
\beq
\mc{Q}=\frac12 q_{ab}M^{ab}+\sum_{\ell > 0}Q_{\ell}\star J_{\ell},
\eeq
where $\star$ schematically denotes the coupling between the dynamical fields $P_{\ell}$ and the associated generators $S_{\ell}$. The explicit form of this expansion will be given below. In  Borel gauge, the fields at non-zero levels are the same in $\mc{Q}$ and $\mc{P}$:
\beq
Q_{\ell}\equiv P_{\ell}, \qquad \ell \in \mathbb{Z}^2_{\ge 0}\backslash \{(0,0)\}\,.
\eeq
The explicit form of the higher level terms in the expansion of $\mc{P}$ reads
\beqa\label{cmform}
\sum_{\ell> 0}P_{\ell}\star S_{\ell}&:=&e^{3\phi/4} P_{a_1} S^{a_1} 
   + \frac12 e^{-\phi/2} P_{a_1a_2} S^{a_1a_2}
   + \frac1{3!} e^{\phi/4} P_{a_1a_2a_3} S^{a_1a_2a_3}\nn\\
&&   + \frac1{5!} e^{-\phi/4} P_{a_1\ldots a_5} S^{a_1\ldots a_5}
   + \frac1{6!} e^{\phi/2} P_{a_1\ldots a_6} S^{a_1\ldots a_6}
   + \frac1{7!} e^{-3\phi/4} P_{a_1\ldots a_7} S^{a_1\ldots a_7}\nn\\
&& + \frac1{8!} P_{a_1\ldots a_8}S^{a_1\ldots a_8} 
   +\frac1{8!}P_{a_0|a_1\ldots a_7}S^{a_0|a_1\ldots a_7} 
   + \frac1{9!} e^{-5\phi/4} P_{a_1\ldots a_9} S^{a_1\ldots a_9}+\ldots
   \nn \\
   \eqa
with a similar expression for $\mc{Q}$ with the $S_{\ell}$'s replaced by the corresponding $J_{\ell}$'s.   
   
{}From a given parametrization of $\mc{V}$ as in (\ref{posexp}) one can work out explicit expressions for $\cP$ in terms of the `potentials' $A_{\ell}$ (which {\em a fortiori} carries flat indices as it transforms under $K(E_{10})$) which appear in the construction of $\mc{V}$. For example, one finds
\beq 
P_a = \frac12 e_a{}^m\partial_t A_m =\frac12 e_a{}^m DA_m
\eeq
in terms of the `covariant derivatives' of \cite{Damour:2002cu}. Here, we have written $\mc{V}_0=e_a{}^m$ as an inverse $GL(9, \mbb{R})$ vielbein. We do not require the exact expressions for the higher level components of $\mc{P}$ for establishing the correspondence and their expression will  be more complicated due to the appearance of additional terms, arising when expanding the Maurer-Cartan form $\pa_t\mc{V}\mc{V}^{-1}$ using the Baker--Campbell--Hausdorff formula $d e^X e^{-X} = dX + \frac12 [X,dX] +\ldots$. 

The geodesic equations $\partial_t\cP =\lb\cQ,\cP\rb$ can be written conveniently by treating the $\cV_0$ contribution separately in a partially covariant derivative $\cD$, see~\cite{DamourNicolaiLowLevels}. For example, for the fundamental generators at level $(0,1)$ and $(1,0)$ the equations of motion become (for $n=1$)
\beqa
\cD (e^{3\phi/2} P_a) &=&
    -e^{\phi/2} P_{ac_1c_2} P_{c_1c_2}   + \frac2{5!} e^\phi P_{ac_1\ldots c_5}P_{c_1\ldots c_5}     +\frac{12}{8!} P_{ac_1\ldots c_7}P_{c_1\ldots c_7} \nn\\
 & &      +\frac{1}{4\cdot 7!}P_{a|c_1\ldots c_7}P_{c_1\ldots c_7}
\label{eompa} \\
\cD (e^{-\phi} P_{a_1a_2}) &=& 
    2 e^{\phi/2}  P_{a_1a_2c}P_c +\frac13 e^{-\phi/2} P_{a_1a_2c_1c_2c_3}P_{c_1c_2c_3}\nn\\
 & &  +\frac2{5!}e^{-3\phi/2} P_{a_1a_2c_1\ldots c_5}P_{c_1\ldots c_5} 
     +\frac2{7!} e^{-5\phi/2}P_{a_1a_2c_1\ldots c_7}P_{c_1\ldots c_7} \nn\\
 & & +\frac1{6!}P_{a_1a_2c_1\ldots c_6}P_{c_1\ldots c_6} + \frac{1}{4\cdot 5!} P_{c_1|c_2\ldots c_6a_1a_2}P_{c_1\ldots c_6}.
\eqa
In Section \ref{sec:correspondence} we will show that these equations are equivalent to the equations of motion for the electric fields $F_{ta}$ and $F_{tab}$, respectively. The equations for all the remaining levels are given in Appendix \ref{Appendix:EOMBosonic}. The  fermionic equations of motion can be found in~\ref{app:FermionicSigmaModelEOM} where the connection term $\cQ$ is evaluated in the vector-spinor representation. The supersymmetry variation (\ref{SusyTransfSigmaModel}) requires the evaluation of $\cQ$ in the Dirac-spinor representation, the resulting expression can be found in~\ref{app:susyvarcoset}.

\section{The Correspondence}
\label{sec:correspondence}

In this section we make the $\mc{E}_{10}/$massive IIA correspondence explicit by giving a dictionary between the dynamical variables of the $ K(\mc{E}_{10})\bas \mc{E}_{10} $-sigma model and the fields of massive IIA supergravity. The comparison cannot be done at the level of the respective Lagrangians since the $\mc{E}_{10}$ sigma model naturally incorporates kinetic terms for all fields as well as their duals, which is not the case for the IIA Lagrangian. The matching is rather done at the level of the equations of motion, where we will see that bosonic equations of motion and Bianchi identities on the supergravity side all become associated with geodesic equations of motion on the $\mc{E}_{10}$ side. Similarly, the fermionic supergravity equations will be associated with the Dirac equation of the spinning coset particle.

\subsection{Bosonic Equations of Motion and Truncation}
\label{EOMandTruncations}
To be able to compare the equations of motion on the supergravity side
(\ref{eoms}) and (\ref{lev0}) to the ones on the sigma model side (\ref{bosoniceom}), spelt out in Appendix \ref{Appendix:EOMBosonic},
we need to rewrite the former. As is customary in the correspondence between $\mc{E}_{10}$ and supergravity we split the
indices $\al=(0,a)$ into temporal and spatial indices and also adopt
a pseudo-Gaussian gauge for the ten-dimensional vielbein:
\beq
\label{10v}
{e_{\mu}}^{\al}=\left(\begin{array}{cc}
N & 0 \\
0 & {e_m}^{a} \\
\end{array}\right)\,.
\eeq
In addition we demand that the spatial trace of the spin connection vanishes (see Appendix~\ref{app:conventions} for our conventions) 
\beq\label{spintrace}
\omega_{a\,ab}=0 \quad\Rightarrow\quad \Omega_{ba\,a}=0
\eeq 
and write the tracefree spin connection as $\tilde{\omega}_{a\,bc}$ as a reminder. We also choose temporal gauges for all supergravity gauge
potentials,
\beq\label{tempgauge}
A_{t} = 0\,,\quad A_{tm} = 0\,,\quad A_{tm_1m_2} =0\,.
\eeq

Moreover, we can only expect that a truncated version of the supergravity
equation corresponds to the coset model equations. 
This truncation was originally devised in the context of eleven-dimensional
supergravity, where it was strongly motivated by the billiard analysis of the
theory close to a spacelike singularity (the `BKL-limit')
\cite{Damour:2002cu,DamourNicolaiLowLevels}. Recall from Chapter \ref{Chapter:Billiards} that in this limit spatial points decouple and
the dynamics becomes effectively time-dependent, ensuring that the truncation
is a valid one in this regime. In this chapter, we analyse the same question in
the context of massive IIA supergravity, and an identical procedure requires
the truncation of a set of spatial gradients. These can be obtained from a
BKL-type analysis of massive IIA and their full list is presented in
Appendix~\ref{truncation}. Notice that except for the expression involving the
mass, all the spatial gradients to be truncated away can be obtained by
dimensional reduction of the eleven-dimensional truncation. 

Let us illustrate the implications of this truncation on the supergravity
equations in some detail for an explicit example. The following truncation of
massive IIA supergravity can be deduced from the billiard analysis: 
\beq
\label{trunc}
\partial_a\left(N e^{3\phi/4} F_{b_1b_2}\right)=0.
\eeq
The effect of this truncation appears in the equation of motion for the
two-form field strength $F_{\alpha\beta}$ (see (\ref{eom1}) in Appendix
\ref{SugraBeom}). After splitting time and space indices, the space component
of this equation reads\footnote{$D^{(0)}$ here contains the contributions from
the time derivatives of the spatial vielbein $e_m{}^a$ and will be identified
with the corresponding $\mc{E}_{10}$ coset derivative operator below.} 
\beqa
D^{(0)}(e^{3\phi/2}F_{tb})&=&-\frac12e^{\phi/2}F_{ta_1a_2b}F_{ta_1a_2}
+\frac1{3!}e^{\phi/2}N^{2}F_{a_1\ldots a_3b}F_{a_1\ldots a_3}\nn\\
&&+\frac34e^{3\phi/2}N^2\partial_a\phi F_{ab}
-\frac12e^{3\phi/2}N^2\Omega_{a_1a_2b}F_{a_1a_2}\nn\\
&&+Ne^{3\phi/4} \partial_a\left(N e^{3\phi/4} F_{ab}\right)\,.
\label{electricequation}
\eqa
Using the truncation (\ref{trunc}), the last term on the right hand side
vanishes, and consequently the equation to compare with the $\mc{E}_{10}$ equation
of motion (\ref{eompa}) is (\ref{electricequation}) without the
bottom line. Let us emphasize that the truncation (\ref{trunc}) also follows
from dimensional reduction of the associated truncation in the equation of
motion for the electric field $F_{0\dot{a}_1\dot{a}_2\dot{a}_3}$ in eleven
dimensions \cite{Damour:2002cu,DamourNicolaiLowLevels}. Moreover, this example makes
clear the important comment that the truncation we impose is \emph{not}
equivalent to discarding all spatial gradients on the supergravity side, since
it is clear that spatial derivatives of the one-form potential $A_a$ are
implicitly contained in $F_{a_1a_2}$. 

From comparing in detail the supergravity equations (Appendix
\ref{Appendix:mIIAdetails}) with the truncations applied with the bosonic
equations of the coset model (Appendix \ref{Appendix:EOMBosonic}) one can
derive a correspondence between the components of the coset velocity $\cP$ and
the fields of supergravity. The dictionary is given in Table \ref{dicoeom}
where, in the last line,  $e=\text{det}\ {e_m}^{a}=\sqrt{\text{det}\
  g_{mn}}=\sqrt{g}$. It is perfectly consistent with identifying the form
equations of motion (\ref{eoms}) with the sigma model expressions
(\ref{boseom1}) to (\ref{boseom3}), the Bianchi identities (\ref{Bianchis})
with the equations (\ref{Bianchi1}) to (\ref{Bianchi3}) and the dilaton
equation of motion (\ref{dilsugra}) with (\ref{dilatoneom}). 

\begin{table}[t]
\centering
\begin{tabular}{|c|c|c|}
\hline
$(\ell_1,\ell_2)$&$\mc{E}_{10}$ fields& Bosonic fields of supergravity\\
\hline
\hline
$(0,0)$&$p_{ab}$ & $-N\om_{a\,b0}$\\[.2cm]
$(0,0)$&$q_{a_1a_2}$ & $-N\om_{0\,a_1a_2}$\\[.2cm]
$(0,1)$&$P_a$ & $\f12 N F_{0a}$\\[.2cm]
$(1,0)$&$P_{a_1a_2}$ & $\f{1}{2}N F_{0a_1a_2}$\\[.2cm]
$(1,1)$&$P_{a_1a_2a_3}$ & $\f{1}{2}N F_{0a_1a_2a_3}$ \\[.2cm]
$(2,1)$&$P_{a_1\cdots a_5}$ & $\f{1}{2\cdot 4!}Ne^{\phi/2}{\epsilon_{a_1\cdots
    a_5}}^{b_1b_2b_3b_4}F_{b_1b_2b_3b_4}$ \\[.2cm]
$(2,2)$&$P_{a_1\cdots a_6}$ & $\f{1}{2\cdot 3!}N e^{-\phi}{\epsilon_{a_1\cdots
    a_6}}^{b_1b_2b_3}F_{b_1b_2b_3}$ \\[.2cm]
$(3,1)$&$P_{a_1\cdots a_7}$ & $-\f{1}{2\cdot 2!}Ne^{3\phi/2}{\epsilon_{a_1\cdots
    a_7}}^{b_1b_2}F_{b_1b_2}$ \\[.2cm]
$(3,2)$&$P_{a_1\cdots a_8}$ & $-\frac12N{\epsilon_{a_1\cdots a_8}}^{b}\partial_{b}\phi$\\[.2cm]
$(3,2)$&$P_{a_0|a_1\cdots a_7}$ & $2N{\epsilon_{a_1\cdots a_7}}^{b_1b_2}\tilde{\Omega}_{b_1b_2\,a_0}$ \\[.2cm]
$(4,1)$&$P_{a_1\cdots a_9}$ & $\frac12 N e^{5\phi/2}\eps_{a_1\ldots a_9}m$\\[.2cm]
\hline
\hline
-&$n$&$Ne^{-1}$\\
\hline
\end{tabular}
\caption{\label{dicoeom}\sl Bosonic dictionary: This table shows the correspondence between $\mc{E}_{10}$ fields and  bosonic fields of massive IIA supergravity that one obtains by considering the equations of motion and the Bianchi equations. }
\end{table}

The remaining equation, the Einstein equation (\ref{Einsteinsugra}), does not fit perfectly in this picture. More precisely, two terms do not match completely with (\ref{Einstein}). One is similar to the mismatch in
the $A_9$ decomposition relevant to $D=11$ supergravity and is a contribution
to the Ricci tensor $R_{ab}$ of the form $\Omega_{a\,cd}\Omega_{b\,dc}$ 
\cite{DamourNicolaiLowLevels}. The other term is a mismatch of the coefficient in the
energy momentum tensor of the dilaton $T_{ab}\sim \partial_a\phi \partial_b\phi$; the
coefficient is off by a factor two. This can be traced back to $D=11$ where
both mismatches were part of the $D=11$ Ricci tensor. In this sense this is
not a new discrepancy but a known one. It is to be noted that all the terms
involved in the mismatch are related to contributions to the Lagrangian which
would give rise to walls corresponding to imaginary roots in the cosmological
billiards picture \cite{Damour:2002cu}. There is no mismatch in the equation
of motion of the dilaton $\phi$ since in the reduction the missing term in
$D=11$ does not contribute to this equation. 

Let us study the effect of the mass term more closely. In the bosonic
equations, the mass appears in five places: the dilaton equation
(\ref{dilsugra}), the Einstein equation (\ref{Einsteinsugra}), the equation of
motion for $F_{ta_1a_2}$ (\ref{eom2}), the Bianchi equation for $F_{a_1a_2}$
(\ref{SugraBianchi1}) and of course in its own equation of motion
$\partial_tm=0$. It is remarkable that the contribution of the real root
corresponding to the mass deformation enters all equivalent sigma-model
equations correctly, even though it is beyond the realm of $E_8$ generators
and above height $30$. In particular, the $\mc{E}_{10}$-invariant sigma model
produces the right potential for the scalar $\phi$. This is possible since the
supergravity potential is positive definite in agreement with positive
definiteness of the $\mc{E}_{10}$-invariant Lagrangian (\ref{e10boslag}) away from
the Cartan subalgebra. By contrast, for gauged deformations in lower
dimensions where the supergravity potential is
indefinite and not reproduced fully by $\mc{E}_{10}$~\cite{Bergshoeff:2008}. From
this point of view the massive Romans theory seems to be special since the $\mc{E}_{10}$ model reproduces correctly all
the effects of the deformation. 

\subsection{The Truncation Revisited} 

The truncation we applied to supergravity, using billiard arguments, also
proves useful for ensuring the consistency of the dictionary. Indeed,
notice that all the sigma model variables depend on (sigma model) `time' only,
and hence we must demand that their spatial gradients vanish: 
\beq
 \pa_a\mc{P}(t)=0. 
 \label{GeneralTruncation}
 \eeq  
Applied to the Maurer--Cartan expansion, this equation translates to
constraints on the supergravity variables upon using the dictionary in
Table~\ref{dicoeom}. Of
course, (\ref{GeneralTruncation}) does not really make sense on the sigma
model side, where spatial gradients have no meaning, but must rather be
understood as a convenient way of encoding the relevant truncations on the
supergravity side. Nevertheless, the truncations encoded in
(\ref{GeneralTruncation}) correspond precisely to the truncations we have
just imposed.  Let us illustrate this for the example of the truncation on 
the magnetic field in (\ref{electricequation}).   
The level $(3,1)$ part of the expansion of the Maurer-Cartan form
contains the following term (see (\ref{cmform}))  
\beq 
\mc{P}(t)\big|_{(3,1)}=\f{1}{7!}e^{-3\phi/4}P_{a_1\cdots a_7}(t)S^{a_1\cdots a_7}.
\eeq 
The coefficient of the generator $S^{a_1\cdots a_7}\in E_{10}$ is a dynamical quantity which is purely time-dependent. This implies that on the supergravity side we must make sure that the corresponding quantity is also purely time-dependent, i.e. we must demand 
\beq 
\pa_a(e^{-3\phi/4}P_{a_1\cdots a_7})=0,
\eeq  
which, after using the dictionary in Table \ref{dicoeom}, precisely yields the truncation in (\ref{trunc}). Similarly, one sees that requiring (\ref{GeneralTruncation}) for each term of the Maurer-Cartan expansion corresponds to the truncations of Appendix \ref{truncation}.

\subsection{Fermionic Equations of Motion} 
\label{feom} 
To make contact between the fermionic equations of Romans' theory, (equations of motion (\ref{eomlam}) and (\ref{eompsi}) and supersymmetry variation (\ref{deltapsi})), and the Kac-Moody side of the story, (equations of motion (\ref{eompsi10}) and (\ref{eomke10}) and supersymmetry variation (\ref{susyexpli})), we must make some further field redefinitions. We redefine the gravitino components, the dilatino and the supersymmetry parameter as
\beqa 
{} \psinew_0&\equiv&  g^{1/4}\Big(\pt_0-\Gamma_0\Gamma^{a}\pt_a\Big),
\nn \\ 
{} \psinew_a&\equiv&  g^{1/4}\Big(\pt_a-\f1{12} \Gamma_{a}\lt\Big),
\nn\\ 
{}\lambdanew&\equiv & \f{2}3 g^{1/4}\lt,
\nn \\ 
{}  \venew &\equiv& g^{-1/4}\vet.
\eqa 
 
With these redefinitions, using the bosonic dictionary obtained in the
previous section and in the gauge 
\beq \label{susygauge}
\psinew_0=0, 
\eeq 
we can show that the
equations of motion of the fermions of massive IIA supergravity (\ref{eomlam})
and (\ref{eompsi}) are equivalent to the Kac-Moody fermionic spinor equations
(\ref{eompsi10}) and (\ref{eomke10}) if we assume a correspondence between the unfaithful
representation of $K(E_{10})$ and the redefined fermionic fields
displayed in the first half of Table \ref{dicofer}.  

\begin{table}[t]
\centering
\begin{tabular}{|c|c|}
\hline
$K(\mc{E}_{10})$ representations&Fermionic fields of supergravity\\
\hline
\hline
$\Psi_a$ & $\psinew_a=g^{1/4}\Big(\pt_a-\f1{12} \Gamma_{a}\lt\Big)$\\[.2cm]
$\Gamma^{10}\Psi_{10}$ & $\lambdanew=\f{2}3 g^{1/4}\lt$\\[.2cm]
\hline
$\epsilon$ & $\venew=g^{-1/4}\vet$\\[.2cm]
$\Psi_t$ & $\psinew_t=n g^{1/4}\Big(\pt_0-\Gamma_0\Gamma^{a}\pt_a\Big)$\\[.2cm]
\hline
\end{tabular}
\caption{\label{dicofer}\sl Fermionic dictionary: This table presents the relation between the spinor and vector spinor unfaithful representations of $K(E_{10})$ and the fermionic fields of massive type IIA supergravity. The first half is obtained by requiring the equations of motion for the fermions to match with the  $K(E_{10})$ equation and the second half comes from the supersymmetry variation of $\psi_t$.}
\end{table}

To illustrate how the correspondence is proved, let us consider the mass terms of the equation of motion for the gravitino. On the supergravity side, one looks at the spatial components of the equation of motion (\ref{eompsi}). We only keep the terms involving the time derivative of $\pt_a$ and the mass:
\beq
\label{ex1}
\partial_0\pt_a+\f1{16}e^{5\phi/4}m\G^0\G_{ab}\pt^b+\f5{16}e^{5\phi/4}m\G^0\pt_a+\f5{64}e^{5\phi/4}m\G^0\G_a\lt+\cdots=0.
\eeq
One the sigma model side, we must consider the terms at level $(4,1)$ of the explicit version of the equation of motion (\ref{eomke10}), that give
\beq
\label{ex2}
\pa_t \Psi_a-\f{1}{2\cdot9!}e^{-5\phi/4}P_{b_1\cdots b_9}\Gamma_{10}\Gamma^{b_1\cdots b_9}\Psi_a
 +\f{12}{9!}e^{-5\phi/4}P_{b_1\cdots b_9}\Gamma_{10}{\Gamma_{a}}^{b_1\cdots b_8}\Psi^{b_9}+\cdots=0.
\eeq
Using the bosonic and fermionic dictionaries given in Tables \ref{dicoeom} and \ref{dicofer}, it is now a purely algebraic exercise to see that (\ref{ex1}) and (\ref{ex2}) are equivalent.

\subsection{Supersymmetry Variations of Fermions}
Considering the supersymmetry variation of the gravitino provides us with a consistency check of the previously obtained bosonic and fermionic dictionaries. To this end it is natural to define 
\beq
\psinew_t \equiv n\psinew_0,
\eeq
where $t$ is considered as a `vector index' along the world line with respect to the einbein $n$.

Using the previous redefinitions yields the following supersymmetry transformation on the redefined gravitino 
\beqa
{} \delta_{\varepsilon}\psinew_t &=& \partial_t \varepsilon +\f{1}{4}g^{-1}\partial_t g \varepsilon+\f{1}{4}N \om_{0ab}\Gamma^{ab}\varepsilon+\f14 Ne^{5\phi/4}m\Gamma_0\varepsilon 
\nn \\
{}& & + \f{1}{4\cdot 4!}e^{\phi/4}N\Gamma^{abcd}\Gamma_0F_{abcd}\varepsilon-\f{1}{4\cdot 3!}e^{\phi/4}N\Gamma^{abc}F_{0abc}\varepsilon
\nn \\
{}& & -\f{1}{8}e^{3\phi/4}N \Gamma^{ab}\Gamma_0\Gamma_{10}F_{ab}\varepsilon+\f{1}{4}e^{3\phi/4}N\Gamma^{a}\Gamma_{10}F_{0a}\varepsilon
\nn \\
{}& & -\f{1}{24}e^{-\phi/2}N\Gamma^{abc}\Gamma_0\Gamma_{10}F_{abc}\varepsilon-\f{1}{8}e^{-\phi/2}N\Gamma^{ab}\Gamma_{10}F_{0ab}\varepsilon
\nn \\
{}& & +\f{1}{4}N\om_{abc}\Gamma^{abc}\Gamma_0\varepsilon -N\Gamma_0\Gamma^{a}\Big[\partial_a \varepsilon+\f{1}{4}g^{-1}\partial_a g \varepsilon\Big].
\label{redefinedgravitinovariation}
\eqa

We can now identify the right hand side of (\ref{redefinedgravitinovariation}) with the $K(\mc{E}_{10})$-covariant derivative acting on the $32$-dimensional spinor representation $\epsilon$ (\ref{SusyTransfSigmaModel}), given explicitly in (\ref{susyexpli}), using the bosonic and fermionic dictionaries already computed (Tables \ref{dicoeom} and \ref{dicofer}). We then obtain the second half of Table \ref{dicofer}, that is, the relations for the supersymmetry parameter and the time component of the gravitino.

In conclusion, we see that all the fermionic equations of motion as well as the supersymmetry variation of $\psi_t$ match with the $ K(\mc{E}_{10})\bas \mc{E}_{10} $ fermionic theory. In particular, we notice that the mass enters these equations correctly everywhere.

%%%%%%%%%%%%%%%%%%%%%%%%%%%%%%%%%%%%%%%%%%%%%%%%

%%%%%%%%%%%%%%%%%%%%%%%%%%%%%%%%%%%%%%%%%%%%
%
% Part 2
%
%%%%%%%%%%%%%%%%%%%%%%%%%%%%%%%%%%%%%%%%%%%%

\part{U-Duality and Arithmetic Structures in String Theory}
\chapter{Aspects of String Duality and Quantum Corrections}
\label{Chapter:Aspects}
This chapter is intended to serve as a motivation for the topics treated in {\bf Part II} of this thesis. The main philosophy that we shall advocate is that quantum corrections in string theory are strongly constrained by invariance under discrete (U-)duality groups $G(\mbb{Z})$, and in many instances may be completely summed up in terms of certain $G(\mbb{Z})$-invariant automorphic forms. The purpose of {\bf Part II} is to understand the construction of such automorphic forms, and how to extract the physical information which they encode. In this first chapter, we shall discuss general aspects of these techniques, as well as consider some specific examples in more detail. The main example which will be used as a source of inspiration throughout the remainder of this thesis is the seminal analysis of \cite{GreenGutperle}, which revealed that an infinite series of instanton corrections to type IIB string theory can be completely summed up in terms of certain $SL(2,\mbb{Z})$-invariant Eisenstein series. This example is discussed in Section \ref{Section:D(-1)Example}, while a more general treatment of Eisenstein series is given in Chapters \ref{Chapter:ConstructingAutomorphicForms} and \ref{Chapter:FourierExpansion}.

\section{Perturbative and Non-Perturbative Effects in String Theory}
We begin here with a general discussion of some important aspects of string perturbation theory, emphasizing in particular the interplay between the worldsheet and spacetime viewpoints. We explain how this analysis naturally leads us to also consider non-perturbative effects, and we discuss the origin of such effects in terms of D-brane instantons. 

\subsection{General Structure of String Perturbation Theory}
\label{Section:StringPerturbation}
String theory is defined perturbatively in terms of an asymptotic expansion in two coupling constants: the worldsheet coupling $\al^{\prime}\equiv \ell_s^{2}$, with $\ell_s$ being the fundamental string length, and the string coupling $g_s\equiv e^{\left<\phi\right>}$, where $e^{\left<\phi\right>}$ denotes the expectation value of the dilaton. Any (closed string) scattering amplitude may then be written schematically as follows
\beq
\mc{A}=\sum_{n=0}^{\infty} \sum_{g=0}^{\infty}  \left(\al^{\prime}\right)^{n-4}g_s^{2(g-1)} \mc{A}_{(n, g)},
\label{schematicamplitude}
\eeq
where $g$ denotes the genus of the string worldsheet and $\mc{A}_{(n,g)}$ is the genus $g$ amplitude of order $n-4$ in $\alpha^{\prime}$. We emphasize that $\alpha^{\prime}$ is dimensionful, a property of the worldsheet expansion which will play an important role in what follows.

The theory is under good control only in either of the two limits $\alpha^{\prime}\rightarrow 0$ or $g_s\rightarrow 0$. In the limit of small string coupling, $g_s\rightarrow 0$, string theory is well described (on-shell) in terms of the two-dimensional worldsheet conformal field theory. On the other hand, when the worldsheet coupling is small, $\alpha^{\prime}\rightarrow 0$, the strings themselves are well approximated by point particles, and the theory is effectively described by spacetime (super-)gravity. Even though morally true, this picture is somewhat oversimplified. In reality there is an interesting interplay between worldsheet effects, governed by $\alpha^{\prime}$, and effects associated with the genus expansion in $g_s$.

 To illustrate the discussion, let us consider in more detail the spacetime point of view of string perturbation theory. To leading order in $\alpha^{\prime}$ and $g_s$, we have an effective description in terms of ten-dimensional Einstein-Hilbert gravity coupled to $p$-forms and fermionic matter fields, with an action of the schematic form
 \beq
 S_{(0,0)}=(\alpha^{\prime})^{-4}\int d^{10}x \sqrt{G^{(s)}}e^{-2\phi}\Big[R-\star d\phi\wedge d\phi-\sum_{p}e^{\lambda_p \phi}\star F_{(p+1)}\wedge F_{(p+1)}+\cdots \Big],
 \label{generalaction}
 \eeq
where $F_{(p+1)}$ denotes the field strengths of whichever $p$-form potentials $A_{(p)}$ the theory contains, and $\lambda_p$ represents the associated dilaton couplings.\footnote{In the Ramond-Ramond sector all of the dilaton couplings are zero, $\lambda_p=0$.} The ellipsis indicate possible Chern-Simons couplings that are not important for the present discussion, and for the same reason we have also suppressed fermionic terms. The action (\ref{generalaction}) is written in \emph{string frame} as is indicative of the dilaton prefactor in front of the bracket. By comparison to the schematic amplitude in Equation (\ref{schematicamplitude}), we note that $e^{-2\phi}$ is the correct power of the string coupling corresponding to the genus zero, $g=0$, part of the sum. This is the leading order effect and corresponds to a tree-level contribution in $g_s$. Further note the overall factor of $(\alpha^{\prime})^{-4}$, which has the correct dimension to compensate for the dimension of the measure.\footnote{This is related to Newton's constant in $D=10$ through $16\pi G_N^{(10)}=(2\pi)^7 (\alpha^{\prime})^{4} g_s^2$.} This is then what is referred to as tree-level in $\alpha^{\prime}$, and hence the action (\ref{generalaction}) is a tree-level contribution with respect to the worldsheet coupling as well as the string coupling. This is indicated by the subscript on $S_{(0,0)}$ in (\ref{generalaction}). 

In general, the spacetime effective action receives perturbative quantum corrections both in $\alpha^{\prime}$ and in $g_s$. Since $\alpha^{\prime}$ has dimension length squared, amplitudes which are higher order in $\alpha^{\prime}$ naturally come with higher powers of momentum insertions. From the point of view of the effective action, these higher powers of momenta are associated with higher derivatives acting on the fundamental fields. Hence, the worldsheet $\alpha^{\prime}$-expansion is tantamount to a \emph{derivative expansion} in spacetime. 

Let us now consider an explicit example, namely type IIB string theory in $D=10$. The tree-level effective action then corresponds to the action of type IIB supergravity. The next to leading order correction in $\alpha^{\prime}$ arises from the $n=1$ contribution in (\ref{schematicamplitude}). This is of order $(\alpha^{\prime})^{-3}$ and is therefore associated with a term in the effective action which is quartic in derivatives. One possible such term contains four derivatives of the metric $G^{(s)}_{\mu\nu}$, and therefore corresponds to a quadratic curvature correction in the effective action. However, it turns out that because of the large amount of supersymmetry $(\mc{N}=2)$ in the theory, the corresponding amplitude vanishes, $\mc{A}_{(1,g)}=0$. In fact, the amplitude $\mc{A}_{(2,g)}$ also vanishes and the first non-trivial $\alpha^{\prime}$-correction arises from the $n=3$ term in (\ref{schematicamplitude}) and is of order $(\alpha^{\prime})^{-1}$. This induces a term in the action with eight derivatives on the metric $G^{(s)}_{\mu\nu}$. This correction enters the effective action in the following way
\beq
S_{(3,0)}= (\alpha^{\prime})^{-4}\int d^{10} x\sqrt{G^{(s)}}e^{-2\phi}\Big[R +(\alpha^{\prime})^{3} \mc{R}^4+\cdots \Big],
\label{R4term1}
\eeq
where $\mc{R}^4$ denotes a specific combination of curvature scalars constructed out of the Ricci scalar $R$, the Ricci tensor $R_{\mu\nu}$ and the Riemann tensor $R_{\mu\nu\rho\sigma}$. The precise kinematical structure will not concern us here, and we refer the interested reader to \cite{Green:1980zg,Green:1981xx,Green:1981ya} for more details. There will also be additional terms in (\ref{R4term1}) of the same order in $\alpha^{\prime}$ corresponding to combinations of curvature terms and $p$-forms (e.g. $R^2F^{4}$), as well as pure $p$-form terms (e.g. $F^8$). These terms also play an important role in the analysis but will for simplicity of argument be neglected in the present discussion. 

The effective action (\ref{R4term1}) now displays the first non-trivial $\alpha^{\prime}$-correction, but still only contains tree-level contributions in the string coupling $g_s$. In principle, there might be an additional infinite series of perturbative corrections in $g_s$ at each order in $\alpha^{\prime}$. However, it turns out that the lowest order terms in $\alpha^{\prime}$ (two-derivative terms) do not receive any corrections in $g_s$. We will see in Section \ref{Section:D(-1)Example} that this is also compatible with S-duality of type IIB string theory. On the other hand, the $\mc{R}^4$-term does receive $g_s$-corrections, and the action (\ref{R4term1}) must therefore be completed to include these additional contributions:
\beq
S_{(3,\infty)}= (\alpha^{\prime})^{-4}\int d^{10} x\sqrt{G^{(s)}}\Big[e^{-2\phi} R +(\alpha^{\prime})^{3}\sum_{g=0}^{\infty} c_g\ e^{2(g-1)\phi} \mc{R}^4+\cdots \Big],
\label{R4term}
\eeq
where $c_g$ are some coefficients which in general depend on the moduli of the theory. To calculate the precise values of these coefficients is a difficult problem, in particular since there might in principle be an infinite series of terms with non-vanishing coefficients $c_g$. In reality, however, it turns out that there are powerful duality arguments which constrain the coefficients such that the moduli-dependent  coefficient in front of the $\mc{R}^{4}$-term may be determined exactly \cite{GreenGutperle}. This will be discussed in detail in Section \ref{Section:D(-1)Example}.

\subsection{Including Non-Perturbative Effects}
\label{Section:NonPerturbativeEffects}

So far we have discussed only perturbative aspects of string theory. In addition, there exist \emph{non-perturbative} effects which are crucial for the consistency of the theory. As argued long ago by Shenker  \cite{Shenker}, to understand this it is important to note that the genus-expansion in (\ref{schematicamplitude}) does not converge, but rather should be interpreted as an \emph{asymptotic series}.\footnote{See, e.g., \cite{Beneke} for a discussion of asymptotic series in perturbation theory.} This follows from the fact that for any fixed order $k$ in $\alpha^{\prime}$ the large genus, $g\rightarrow \infty$, behaviour of $\mc{A}_k$ is \cite{Shenker}
\beq
\lim_{g\rightarrow \infty} \mc{A}_k\equiv \lim_{g\rightarrow \infty} \sum_{g=0}^{\infty} e^{2(g-1)\left< \phi\right>}\mc{A}_{(k, g)}\sim \sum_{g=0}^{\infty} e^{2(g-1)\left< \phi\right>}\ a^{-2g} (2g)!
\eeq
where $a$ is some constant. The factorial growth of $\mc{A}_{(k, g)}\sim (2g)!$ for large genus is due to the growth of the volume of the moduli space of Riemann surfaces $\mc{M}_g$ for large $g$ \cite{HarerZagier,Penner,Shenker}. In \cite{Shenker} it was suggested that in order to cure this divergence, one must include additional effects in (\ref{schematicamplitude}) which are suppressed by a factor $e^{-1/g_s}$ in the limit $g_s\rightarrow 0$. These would therefore correspond to non-perturbative effects from the point of view of string perturbation theory. This prediction of instanton effects in string theory was later verified with the discovery of D-branes \cite{Polchinski}, which are solitonic objects whose tension scales as $T\sim g_s^{-1}$, ensuring that they indeed contribute by exponentially suppressed terms of order $e^{-1/g_s}$, in accordance with the arguments of \cite{Shenker}. More specifically, consider a spacetime splitting of the form $\mbb{R}^{1, 9-r}\times X$, where $X$ is some $r$-dimensional compact internal manifold. Then any Euclidean D$p$-brane wrapping a $p+1$-dimensional submanifold $\mc{C}\subset X$ is completely localized in the external spacetime $\mbb{R}^{1, 10-r}$, and may therefore be interpreted as a \emph{D-brane instanton}. In general, such instanton effects are suppressed by factors of $e^{-\mathrm{Vol}(\mc{C})/g_s}$, where $\mathrm{Vol}(\mc{C})$ denotes the volume of the submanifold $\mc{C}$. In addition, there exist non-perturbative effects from the worldsheet point of view. These are exponentially suppressed by factors $e^{-\mathrm{Vol}(\mc{B})/\alpha^{\prime}}$ in the limit $\alpha^{\prime}\rightarrow 0$, and arises from Euclidean fundamental strings wrapping holomorphic two-cycles $\mc{B}$ in $X$ \cite{Dine:1986zy,Dine:1987bq}. 

These results indicate that the action (\ref{R4term}) is not yet the full story, but rather should be further completed to include possible non-perturbative effects at each order in $\alpha^{\prime}$. More precisely, we should add to the action an infinite series of such contributions, schematically indicated as follows
\beq
S_{(3,\infty)}^{\mathrm{np}}=(\alpha^{\prime})^{-4}\int d^{10} x\sqrt{G^{(s)}}\left[g_s^{-2} R +(\alpha^{\prime})^{3}\left(\sum_{g=0}^{\infty} c_g\ g_s^{2(g-1)}+\sum_{N\neq 0} b_N e^{-|N|/g_s}\right) \mc{R}^4+\cdots \right],
\label{R4termNonPert}
\eeq
where $N$ denotes the ``instanton charge'' and $b_N$ is some moduli-dependent coefficient.\footnote{In this expression we have been deliberately sloppy and replaced the dynamical dilaton field $e^{\phi}$ directly with the string coupling $g_s$ to illustrate the point. It is of course understood that $g_s$ really corresponds to expectation value of the dilaton, as in (\ref{schematicamplitude}).} Since we are in $D=10$ there is no volume dependence in the suppression factor $e^{-|N|/g_s}$. These kinds of effects indeed do appear in type IIB string theory, and may be attributed to D$(-1)$-instantons, which are 0-dimensional objects and hence localized in ten-dimensional spacetime. In the next section, we will see that the coefficients $b_N$ are highly non-trivial and in fact contain an infinite series of additional perturbative excitations for each instanton number $N$.

Let us finally note that recently \cite{PiolineVandoren} the arguments of \cite{Shenker}, regarding the large-order behaviour of string perturbation theory, were generalized to the large-order behaviour of certain infinite series of D-instanton corrections in type II string theory on a Calabi-Yau threefold $X$. From this analysis it was concluded that the D-instanton series should itself be treated asymptotically and, in order for the series to converge, additional non-perturbative effects of order $e^{-1/g_s^2}$ must be included. Such effects are naturally attributed to Euclidean NS5-branes wrapping the entire internal manifold $X$ and it was conjectured that in order to ensure a convergent instanton series, the contributions from NS5-brane instantons must be included. We will explicitly see the effects of NS5-brane instantons arising in the analysis of Chapter \ref{Chapter:UniversalHypermultiplet}, which is based on results from {\bf Paper VIII}.

\section{Summing Up Instantons With Automorphic Forms}
\label{Section:D(-1)Example}
The discussion in the previous section made it clear that there is an intricate interplay between worldsheet effects governed by $\alpha^{\prime}$ and effects from the genus expansion governed by the string coupling $g_s$. While we have seen that perturbative $\alpha^{\prime}$-corrections are to a certain extent constrained by spacetime supersymmetry, no such constraint was put on the $g_s$-corrections. In this section we will see that there exist extra constraints, loosely referred to as ``string dualities'', which severly constrain the form of perturbative and non-perturbative corrections in the string coupling at each order in $\alpha^{\prime}$. These dualities are generally known as S-, T, or U-duality depending on the context. Typically, they are described by discrete groups $G(\mbb{Z})$ which should leave the full quantum effective action invariant. Enforcing this invariance on the lowest order terms in $g_s$ may then potentially fix completely the infinite series of perturbative and non-perturbative corrections in (\ref{R4termNonPert}). In this section we will describe in detail the simplest example where this can be done, namely in type IIB string theory which exhibits invariance under the S-duality group $G(\mbb{Z})=SL(2,\mbb{Z})$. Towards the end of the section we will also discuss extensions of these techniques to lower dimensions where larger duality groups appear.

\subsection{The Exact $\mc{R}^4$-Correction in Type IIB String Theory}

In Section \ref{Section:StringPerturbation} we discussed the existence of an $\alpha^{\prime}$-correction   corresponding to a quartic curvature correction to Einstein gravity in $D=10$. This occurs as the first non-trivial $\alpha^{\prime}$-correction to type IIB supergravity. We will now take a closer look at the possible perturbative and non-perturbative $g_s$-corrections to the $\mc{R}^4$-term. 

Let us begin by reminding the reader about the bosonic field content of the theory. In addition to the metric, there is a dilaton modulus $e^{\phi}$, an NS-NS 2-form $B_{(2)}$, and a set of Ramond-Ramond $p$-form fields $C_{(p)}$ with $p=0, 2, 4, 6, 8$. The Ramond-Ramond forms couple to the D-branes of the theory, namely D$(-1)$, D1, D3, D5, and D7-branes, respectively. In the NS-NS sector, $B_{(2)}$ couples to the fundamental string F1, while its dual $\tilde{B}_{(6)}$ couples to the NS5-brane. These are the basic objects in the theory, and upon compactification they might all give rise to various non-perturbative effects in the lower-dimensional theory. 

For our present purposes, however, we shall focus on the D$(-1)$-branes. These exhibit the special property of having no extension in any of the ten spacetime directions, and therefore simply correspond to points. The D$(-1)$-branes thus have an interpretation as spacetime instantons in $D=10$. This is in marked contrast to the other objects in the theory which have no such interpretation in ten dimensions. The D$(-1)$-instantons are sourced by the zero form $C_{(0)}$, which we shall refer to as the \emph{axion} and denote by $C_{(0)}\equiv \chi$. The axion and the dilaton further combine into a complex scalar 
\beq
\tau=\tau_1+i\tau_2 := \chi+ ie^{-\phi},
\eeq
 which parametrizes the moduli space $\mc{M}_{10}=SL(2,\mbb{R})/SO(2)$. The group $SL(2,\mbb{R})$ is a global symmetry of the classical type IIB action. This invariance is however not manifest when the action is written in string frame as in (\ref{generalaction}). This is apparent by noting that the Einstein-Hilbert term is multiplied by an overall factor $e^{-2\phi}$ which transforms non-trivially under $SL(2, \mbb{R})$. To exhibit the $SL(2,\mbb{R})$-invariance, we perform a Weyl-rescaling of the string frame metric, $G^{(s)}_{\mu\nu}\equiv e^{\phi/2}G_{\mu\nu}$, which then reduces the action to the standard Einstein-Hilbert form
\beq
S_{(0,0)}= (\alpha^{\prime})^{-4} \int d^{10} x \sqrt{G} \Big[R+\cdots  \Big].
\eeq
This is referred to as \emph{Einstein frame} and the rescaled metric $G_{\mu\nu}$ does not transform under $SL(2,\mbb{R})$, ensuring that the Einstein-Hilbert term is now manifestly invariant under $SL(2,\mbb{R})$. The other terms in the action also organize themselves in such a way that the full action in Einstein frame is $SL(2,\mbb{R})$-invariant. It is generally expected in string theory that such classical continuous global symmetries $G(\mbb{R})$ should be broken by quantum effects. However, one expects that there will in fact be some remnant of this classical symmetry preserved in the full quantum theory in the form of a discrete subgroup $G(\mbb{Z})\subset G(\mbb{R})$ \cite{HullTownsend}. 

Let us then investigate this question within the present framework of type IIB supergravity. To this end we  turn on the quartic correction term $\mc{R}^4$ in the effective action, as discussed in Section \ref{Section:StringPerturbation}. We have seen in (\ref{R4term}) that in string frame this term enters with the same power $e^{-2\phi}$ of the dilaton as the Einstein-Hilbert term, signifying a tree-level effect. To make contact with the discussion of $SL(2,\mbb{R})$-invariance above, we convert also the $\mc{R}^{4}$-term to Einstein frame, with the result
\beq
S_{(3,0)}=(\alpha^{\prime})^{-4} \int d^{10} x \sqrt{G} \Big[R+ (\alpha^{\prime})^{3}c_0 \tau_2^{3/2}\mc{R}^4+\cdots  \Big],
\eeq
where we recall that $\tau_2=e^{-\phi}$, and we have also reinstalled the genus zero coefficient from (\ref{R4term}). By explicit string theory calculations \cite{Grisaru:1986dk,GrossWitten,Grisaru:1986kw}, it has been verified that $c_0$ is non-vanishing and takes the precise value $c_0=2\zeta(3)$, where $\zeta(3)$ denotes the value of the Riemann zeta function $\zeta(z)$ for $z=3$. 

In addition, in \cite{Green:1981ya} the next to leading order term, corresponding to genus $g=1$ in (\ref{schematicamplitude}), was calculated and it was found that also this coefficient is non-vanishing, with $c_1=4\zeta(2)$. Hence, the $\mc{R}^4$-term receives at least a 1-loop correction, and up to genus one we then have
\beq
S_{(3,1)}=(\alpha^{\prime})^{-4} \int d^{10} x \sqrt{G} \Big[R+ (\alpha^{\prime})^{3}\Big(2\zeta(3) \tau_2^{3/2}+4\zeta(2) \tau_2^{-1/2}\Big)\mc{R}^4+\cdots  \Big].
\label{R4EinsteinFrameOneLoop}
\eeq
%Let us emphasize that this result is written in Einstein frame, and this is the reason why the dilaton powers do not match the structure in (\ref{schematicamplitude}). However, as we have seen, a Weyl-rescaling to string frame results in an overall factor of $\sqrt{\tau_2}$, revealing the correct powers of the string coupling
%\beq
% S_{(3,1)}=(\alpha^{\prime})^{-4} \int d^{10} x \sqrt{G^{(s)}} \Big[e^{-2\phi} R+ (\alpha^{\prime})^{3}\Big(2\zeta(3)e^{-2\phi} +4\zeta(2) \Big)\mc{R}^4+\cdots  \Big].
%\eeq
From this expression it is clear that the perturbative contributions to the $\mc{R}^4$-correction explicitly break the $SL(2, \mbb{R})$-invariance of the effective action. This follows since in Einstein frame the $\mc{R}^4$-term is itself invariant, while the dilaton-dependent coefficients transform non-trivially. From the generic form of (\ref{R4term}) we may in fact deduce that each term in  the entire infinite series of possible perturbative corrections will separately break the invariance under $SL(2, \mbb{R})$. However, there is by now a large body of evidence \cite{HullTownsend,Witten} that type IIB string theory is self-dual under the S-duality group $SL(2,\mbb{Z})\subset SL(2,\mbb{R})$, i.e. the modular group of integer-valued $2\times 2$ matrices with unit determinant. It is then expected that after all perturbative and non-perturbative quantum corrections have been taken into account, the coefficient in front of $\mc{R}^4$ should in fact be invariant under $SL(2,\mbb{Z})$.

This philosophy was first utilized in the seminal work of Green and Gutperle \cite{GreenGutperle}, where they conjectured that the action (\ref{R4EinsteinFrameOneLoop}) should be completed to an exact expression of the form:
\beq
S_{(3,\mathrm{exact})}=(\alpha^{\prime})^{-4} \int d^{10} x \sqrt{G} \Big[R+ (\alpha^{\prime})^{3}f(\tau) \mc{R}^4+\cdots  \Big],
\label{R4EinsteinFrameExact}
\eeq
where $f(\tau)$ is some $SL(2,\mbb{Z})$-invariant function of the modulus $\tau \in SL(2,\mbb{R})/SO(2)$, i.e. an \emph{automorphic form}.\footnote{We shall generally refer to functions invariant under some discrete group $G(\mbb{Z})$ as \emph{automorphic forms}, while in the special case of $G(\mbb{Z})=SL(2,\mbb{Z})$ these are sometimes called \emph{modular forms}. Let us also note that in the literature on holomorphic modular forms for $SL(2,\mbb{Z})$, it is common to distinguish between modular \emph{forms} and \emph{functions}, the latter being completely $SL(2,\mbb{Z})$-invariant while the former transform with some overall (modular) weight. In this thesis we shall not make this distinction.  } This function is further constrained so that in a weak-coupling expansion, $\tau_2=g_s^{-1}\rightarrow \infty$, it should reproduce the tree-level and one-loop terms in (\ref{R4EinsteinFrameOneLoop}), 
\beq
f(\tau) \sim  2\zeta(3) \tau_2^{3/2}+4\zeta(2) \tau_2^{-1/2}+\cdots ,
\eeq
where the ellipsis indicate possible additional perturbative and non-perturbative contributions to the $\mc{R}^4$-term. It turns out that there is a unique\footnote{A priori, it is possible that so called \emph{cusp forms} (roughly, automorphic forms for which the perturbative part vanishes) could also contribute, as they would not spoil the perfect agreement with the tree-level and one-loop terms. However, this possibility has been ruled out by constraints from supersymmetry \cite{Green:1997di,Berkovits:1997pj,Pioline:1998mn,Green:1998by}.} candidate for $f(\tau)$ given by the non-holomorphic Eisenstein series \cite{GreenGutperle}
\beq
\mc{E}^{SL(2,\mbb{Z})}(\tau; s)=\sum_{(m,n)\neq (0,0)} \f{\tau_2^s}{|m+n\tau|^{2s}},
\label{SL(2)Eisenstein}
\eeq
where the parameter $s$ is known as the \emph{order} of the Eisenstein series. For the special value $s=3/2$ this Eisenstein series has a Fourier expansion of the form
\beq
\mc{E}^{SL(2,\mbb{Z})}(\tau; 3/2)=2\zeta(3) \tau_2^{3/2}+4\zeta(2) \tau_2^{-1/2}+4\pi \sqrt{\tau_2}\sum_{N\neq 0} \mu_{3/2}(N)N K_1\big(2\pi |N| \tau_2\big) e^{-2\pi iN\tau_1},
\label{GreenGutperleExpansion}
\eeq 
where $K_1(x)$ is the modified Bessel function, and $\mu_{3/2}(N)$ is a combinatorial coefficient which plays an important role, and will be discussed in more detail at the end of this section. Remarkably, the expansion (\ref{GreenGutperleExpansion}) reproduces correctly the two perturbative terms in (\ref{R4EinsteinFrameOneLoop}), including the exact genus zero and genus one coefficients $c_0=2\zeta(3)$ and $c_1=4\zeta(2)$. 

The next important thing to note about the expansion (\ref{GreenGutperleExpansion}) is the absence of perturbative contributions beyond one-loop, implying that $SL(2,\mbb{Z})$-invariance enforces $c_{g}=0$ for all coefficients in (\ref{R4term}) with $g\geq 2$. This is a strong prediction of S-duality in type IIB string theory, and lies at the heart of powerful non-renormalisation theorems beyond one-loop for the $\mc{R}^4$-term also in dimensions $D\leq 10$ \cite{GreenGutperle,Green:1997as,Berkovits:1997pj,Green:1999pu,Green:1999pv,Green:2005ba,Green:2006gt,Green:2006yu,Green:2008uj,Green:2008bf}. See also \cite{Antoniadis:1997zt} for an independent verification of (\ref{R4EinsteinFrameExact}) using heterotic/type II duality. 

It remains to discuss the last part of (\ref{GreenGutperleExpansion}), apparently encoding an infinite series of additional non-perturbative corrections. The non-perturbative effects are best revealed by noting that for large argument the modified Bessel function $K_1(x)$ in (\ref{GreenGutperleExpansion}) has an expansion of the form
\beq
 K_1(x)=\sqrt{\f{\pi}{2x}} e^{-x}\Big[1 + \mc{O}(1/x) + \cdots \Big].
\eeq
This expansion is justified in the weak-coupling limit $\tau_2\rightarrow \infty$ and yields
\beq
\mc{E}^{SL(2,\mbb{Z})}(\tau; 3/2)=2\zeta(3) \tau_2^{3/2}+4\zeta(2) \tau_2^{-1/2}+2\pi \sum_{N\neq 0} \mu_{3/2}(N)\sqrt{N} e^{-S_{\mathrm{inst}}(\tau)} \Big[1+\mc{O}(\tau_2)+\cdots \Big].
\label{GreenGutperleExpansion2}
\eeq 
This shows that for each $N$ we have infinite series of perturbative excitations which are exponentially suppressed by a factor $e^{-S_{\mathrm{inst}}(\tau)}$, where we have defined
\beq
S_{\mathrm{inst}}(\tau):= 2\pi |N| \tau_2 + 2\pi i N \tau_1=2\pi |N| e^{-\phi} + 2\pi i N \chi.
\eeq 
This is precisely the Euclidean action for D$(-1)$-instantons \cite{Gibbons:1995vg,GreenGutperle}. The imaginary part of $S_{\mathrm{inst}}(\tau)$ corresponds to the analogue of the ``theta-angle'' in Yang-Mills theory, and encodes the coupling to the Ramond-Ramond axion $\chi$. The infinite series of corrections in (\ref{GreenGutperleExpansion2}), induced by the expansion of the Bessel function, may then be attributed to perturbative excitations around the instanton background \cite{GreenGutperle}. The existence of these perturbative corrections for each $N$ is another prediction of type IIB $SL(2,\mbb{Z})$-duality, albeit one which is very difficult to verify since explicit methods for instanton calculus are not well developed in string theory. We note that the expansion (\ref{GreenGutperleExpansion2}) is indeed of the expected form indicated in (\ref{R4termNonPert}).

The success of this analysis led to the conjecture that the exact $\mc{R}^4$-term be given by \cite{GreenGutperle}
\beq
S_{(3,\mathrm{exact})}=(\alpha^{\prime})^{-4} \int d^{10} x \sqrt{G} \Big[R+ (\alpha^{\prime})^{3}\mc{E}^{SL(2,\mbb{Z})}(\tau; 3/2) \mc{R}^4+\cdots  \Big].
\label{R4EinsteinFrameExact2}
\eeq
This result may be seen as a direct realization of the arguments of Shenker \cite{Shenker}, discussed in Section \ref{Section:NonPerturbativeEffects}, in the sense that the perturbative series in (\ref{R4term}) was ``completed'' by non-perturbative effects into the form (\ref{R4EinsteinFrameExact2}). In fact, the Eisenstein series $\mc{E}^{SL(2,\mbb{Z})}(\tau; s)$ is absolutely convergent for $\Re(s)> 1$, and hence the series encoded in the coefficient in (\ref{R4EinsteinFrameExact2}) indeed converges. 

We conclude this section with a discussion of the important coefficient $\mu_{3/2}(N)$, appearing in the expansion (\ref{GreenGutperleExpansion}). This is a number theoretic quantity, known in the physics literature as the \emph{instanton measure}, which is given explicitly by a sum over divisors of the instanton number $N$ \cite{GreenGutperle,Yi,SethiStern,GreenGutperleInstantonMeasure,Kostov:1998pg,Moore:1998et}
\beq
\mu_{3/2}(N):= \sum_{n|N} n^{-2}.
\label{D(-1)instantonmeasure}
\eeq
Roughly speaking, this counts the degeneracies of charge $N$ D$(-1)$-instantons. More accurately, the non-perturbative effects in (\ref{GreenGutperleExpansion}) are not to be interpreted as multi-instanton contributions, but rather as contributions from a single instanton with arbitrary charge $N$. This is due to the fact that processes involving multiple instantons only contribute to terms which are higher order in $\alpha^{\prime}$, because of the saturation of zero modes required for a non-vanishing contribution \cite{GreenGutperle}. Hence, the proper statement is that $\mu_{3/2}(N)$ should correspond to the measure on the moduli space of a single D$(-1)$-instanton of charge $N$ \cite{GreenGutperleInstantonMeasure}. 

It is illuminating to consider the further reduction of type IIB string theory on a circle $S^{1}(R)$ to $D=9$. By T-duality, the D$(-1)$-instantons are then mapped to Euclidean D0-particles wrapping the circle in type IIA string theory on $S^{1}(R^{-1})$. More precisely, each D$(-1)$-instanton with (real) action  $\Re(S_{\mathrm{inst} })=2\pi |N|e^{-\phi}$ is mapped to a D0-particle with mass $2\pi R |N|e^{-\phi}=2\pi R |\tilde{m}n|  e^{-\phi}$, which has $n$ units of momentum along the circle and winding number $\tm$. Under this duality, there is therefore a simple way to understand the sum over divisors of $N$ in (\ref{D(-1)instantonmeasure}): it counts the number of partitions of $N$ into two integers $\tm$ and $ n$, or, in other words, the number of ways in which a single charge $N$ D$(-1)$-instanton may be constructed from a bound state of D0-particles with momentum $n$ and winding $\tm$ along the Euclidean circle $S^1$ \cite{GreenGutperle}.

\subsection{U-Duality and the Exact $\mc{R}^4$-Correction in $D<10$}
\label{Section:UDualityR4Corrections}
The analysis so far has been restricted to type IIA and type IIB string theory in $D=10$, with the exception of a short detour to $D=9$. We have seen that type IIB already in maximal dimension exhibits non-perturbative effects due to D$(-1)$-instantons, while in type IIA similar effects only appear in nine dimensions after reduction on $S^1$. In fact, as briefly mentioned in Section \ref{Section:NonPerturbativeEffects}, this is a general feature of non-perturbative effects in string theory; as we compactify the theory to lower dimensions new instanton effects appear. These arise because when spacetime splits into an external part $\mbb{R}^{1,10-r}$ and an $r$-dimensional compact internal part $X$, then Euclidean D-branes, or NS5-branes, may wrap submanifolds of $X$, hence becoming completely localized with respect to the external spacetime $\mbb{R}^{1,10-r}$. This also implies that in lower dimensions it becomes increasingly difficult to sum up all instanton corrections to the effective action.

One generic problem that arises when studying the theory on $\mbb{R}^{1,10-r}\times X$ is that the lower-dimensional theory exhibits a complicated moduli space $\mc{M}$ parametrized by the scalar fields of the theory. The geometry of this moduli space is in turn related to the topology of the compact manifold $X$. This often means that the same techniques used in the previous sections for D$(-1)$-instantons may no longer work since it is not known in general what the analogue of the S-duality group $SL(2,\mbb{Z})$ is in lower dimensions. In other words, the power of automorphy is lost. 

The situation improves considerably if we restrict to compact manifolds which preserve a large amount of supersymmetry. In particular, the simplest case is to take $X=T^{r}$ which preserves all of the 32 supercharges of type II string theory. For such toroidal compactifications it turns out that the classical moduli space $\mc{M}$ is always a symmetric space of the form $G/K$, where $G$ is a global symmetry of the classical action and $K$ is the ``$R$-symmetry''. For example, consider type IIB string theory reduced on $T^2$. Under this reduction, the classical moduli space is enhanced from $\mc{M}_{10}=SL(2,\mbb{R})/SO(2)$ to 
\beq
\mc{M}_8=\big[SL(2, \mbb{R})/SO(2)\big]\times \big[SL(3,\mbb{R})/ SO(3)\big]
\eeq  
in $D=8$. It is generally expected that the global symmetry group will again be broken by quantum effects to a discrete subgroup 
\beq
G(\mbb{Z})=SL(2,\mbb{Z})\times SL(3,\mbb{Z}),
\eeq
known as the \emph{U-duality group} \cite{HullTownsend} (see \cite{Obers:1998fb} for a review). Hence, in this case we expect that the power of automorphy is retained, since the exact effective action in $D=8$ should exhibit manifest invariance under $G(\mbb{Z})$. This was indeed shown to be true in \cite{PiolineKiritsis} (see also preceeding work \cite{Green:1997di}), where a proposal for the exact coefficient of the $\mc{R}^4$-term was given in terms of an $SL(3,\mbb{Z})$-invariant Eisenstein series of the form
\beq
\mc{E}^{SL(3,\mbb{Z})}(\mc{K};s)= \sum_{{m}\in\mbb{Z}^{3}\bas \{(0,0,0)\}} \Big[{m}^{T} \cdot \mc{K} \cdot {m}\Big]^{-s},
\label{SL(3)Eisenstein}
\eeq
where the relevant order is again $s=3/2$. The matrix $\mc{K}$ parametrizes the coset space $SL(3,\mbb{R})/SO(3)$. This Eisenstein series is a direct generalization of (\ref{SL(2)Eisenstein}), and some of its properties will be discussed in detail in Sections \ref{Section:SL(3)Eisenstein} and \ref{Section:FourierExpansionSL(3)}. 

In accordance with our general discussion above, new instanton effects appear in $D=8$ from Euclidean fundamental strings and D1-branes wrapping the torus $T^2$. It was shown in \cite{PiolineKiritsis} that these effects are indeed correctly reproduced in the Fourier expansion of $\mc{E}^{SL(3,\mbb{Z})}$, which also encodes the D$(-1)$-instanton sum in (\ref{SL(2)Eisenstein}). Another interesting feature of this example is that the Eisenstein series (\ref{SL(3)Eisenstein}) is only absolutely convergent for $\Re(s)> 3/2$ and in fact exhibits a pole for $s=3/2$. However, after regularization, this pole is replaced by a logarithmic divergence which precisely reproduces the physical infrared logarithmic divergence of the one-loop correction to the $\mc{R}^4$-term in $D=8$ \cite{PiolineKiritsis}.

As also discussed in Chapter \ref{Chapter:Introduction}, when taking a larger and larger internal torus, the duality groups $G(\mbb{R})$ also become larger, culminating in $D=4$ with the group $G(\mbb{R})=\mc{E}_{7}(\mbb{R})$ \cite{CremmerJulia}, and with classical moduli space given by
\beq
\mc{M}_4=\mc{E}_{7}(\mbb{R})/(SU(8)/\mbb{Z}_2).
\eeq 
In addition to the metric and the scalar fields parametrizing the coset space $\mc{M}_4$, the bosonic sector in $D=4$ also contains 28 abelian vector fields, whose electric and magnetic charges together span a symplectic lattice invariant under $Sp(56;\mbb{Z})$. Based on these arguments, it was conjectured in \cite{HullTownsend} that the U-duality group $G(\mbb{Z})$ in $D=4$ should be
\beq
\mc{E}_{7}(\mbb{Z}):= \mc{E}_{7}(\mbb{R})\cap Sp(56;\mbb{Z}).
\eeq
In $D=4$ yet more non-perturbative effects are expected to arise from Euclidean $Dp$-branes ($p=-1,1,3,5$) as well as NS5-branes wrapping even cycles in $T^{6}$. Similarly, from the type IIA perspective these effects arise from Euclidean D$p$-branes ($p=0, 2, 4$) wrapping odd cycles in $T^6$, as well as NS5-branes wrapping the entire six-torus. It was speculated in \cite{ObersPioline} that such effects could possibly be summed up with a certain $\mc{E}_{7}(\mbb{Z})$-invariant Eisenstein series. However, it is not clear at present whether or not it is possible to make this proposal explicit; most notably to extract the Fourier coefficients of such an Eisenstein series appears to be a very hard challenge which has not yet been achieved (see, however, \cite{Green:2010wi} for recent progress in this direction). 

Another problem is related to the perturbative contributions in the Fourier expansion. As mentioned above, there is no reason to believe that the one-loop non-renormalization theorems that are valid for the $\mc{R}^4$-correction in $D=10$ should be violated in lower dimensions. In other words, it is expected that the $\mc{R}^4$-term in $D=4$ should also be perturbatively finite beyond one-loop. However, in general the number of perturbative terms in the Fourier expansion of a $G(\mbb{Z})$-invariant automorphic form is equal to the order of the Weyl group $\mc{W}(\mf{g})$ of the Lie algebra $\mf{g}=\text{Lie}\ G$ \cite{Langlands}.\footnote{These issues will be explained in more detail in Chapter \ref{Chapter:FourierExpansion}.} For example, in the case of $SL(2,\mbb{R})$ the Weyl group is of order 2, $\mc{W}(\mf{sl}(2,\mbb{R}))=\mbb{Z}_2$, and this is the reason for the existence of only two perturbative terms in the expansion (\ref{GreenGutperleExpansion}).  In the case of $\mc{E}_{7}(\mbb{Z})$-invariant Eisenstein series this presents a problem, since the order of the Weyl group of $\mc{E}_{7}$ is very large. A possible way out is that the number of constant terms decreases for ``residues'' of the general Eisenstein series, i.e. corresponding to a specific choice of representation of the duality group $G$. It is therefore plausible that the correct coefficient in front of the $\mc{R}^{4}$-term in $D=4$ should be given by some high-order residue of the general $\mc{E}_{7}(\mbb{Z})$-invariant Eisenstein series proposed in \cite{ObersPioline}.\footnote{I am grateful to Boris Pioline for emphasizing this to me.} In Chapter \ref{Chapter:FourierExpansion}, we will discuss some examples of this phenomenon in the simpler context of $SL(3,\mbb{Z})$, revealing that the Eisenstein series (\ref{SL(3)Eisenstein}) in fact arises from a more general Eisenstein series $\mc{E}^{SL(3,\mbb{Z})}(\mc{K};s_1,s_2)$ in the limit $s_2\rightarrow 0$.

To conclude the discussion, let us provide some words of motivation for subsequent chapters. We have learned that (U-)duality $G(\mbb{Z})$ in any dimension $D\leq 10$ provides powerful constraints on the structure of perturbative and non-perturbative effects in string theory. In particular, $G(\mbb{Z})$-invariant automorphic forms may potentially be used to sum up all quantum corrections to the effective action of string theory on $\mbb{R}^{1,10-r}\times X$, at least in cases where $X$ preserves a sufficient amount of supersymmetry. In addition to the toroidal case discussed in this section, another example is provided by type IIA string theory on $X=K3\times T^2$ which is known to be dual to heterotic string theory on $T^6$ \cite{Duff:1986cx,Aspinwall:1996mn}. In this case the U-duality group is given by $SL(2, \mbb{Z})\times SO(6,22;\mbb{Z})$ and automorphic techniques have been extensively used to constrain the effective action \cite{Gregori:1997hi,Kiritsis:2000zi,Obers:2001sw}. All this motivates us to study the general construction of automorphic forms on coset spaces $G/K$, and in particular to understand how to compute their Fourier expansions. This is the purpose of the remaining chapters of {\bf Part II} of this thesis. We will also see in Chapter \ref{Chapter:UniversalHypermultiplet} that in certain cases automorphic techniques may be used also for compact manifolds $X$ preserving a smaller amount of supersymmetry, e.g. for Calabi-Yau threefolds.  

\chapter{Constructing Automorphic Forms}
\label{Chapter:ConstructingAutomorphicForms}
In the previous chapter, we learned that discrete subgroups $G(\mbb{Z})$ of certain Lie groups $G$ play an important role as underlying symmetry groups in string theory. We have seen how invariance under $G(\mbb{Z})$Ê enforces strong constraints on perturbative and non-perturbative quantum corrections. These contributions to the low-energy effective action are encoded in $G(\mbb{Z})$-invariant functions of the moduli fields, i.e \emph{automorphic forms}. We briefly touched upon one of the simplest examples, namely the well-known non-holomorphic Eisenstein series $\mc{E}^{SL(2,\mbb{Z})}(\tau; s)$, which encodes D$(-1)$-instanton corrections in type IIB string theory. In this section we shall formalize these techniques and give a more general treatment of automorphic forms on arithmetic quotients $G(\mbb{Z})\bas G/K$. We will also consider in detail the Fourier expansion of these automorphic forms, which is the framework in which physical effects are revealed. In this context, it is important to distinguish between two different cases: \emph{abelian} and \emph{non-abelian} Fourier expansions. This refers to whether the nilpotent subgroup $N\subset G$ is abelian or non-abelian. Representative examples are given by Eisenstein series on $SL(2, \mbb{R})$ and $SL(3, \mbb{R})$, which are treated in detail in Sections \ref{Section:FourierExpansionSL(2)} and \ref{Section:FourierExpansionSL(3)}, respectively.

Before we begin, let us give a brief guide to some of the relevant literature on automorphic forms. A canonical mathematical reference for Eisenstein series are the lecture notes by Langlands \cite{Langlands}. Perhaps more accessible are the books by Terras \cite{Terras1,Terras2}, or in the special case of $G=SL(2,\mbb{R})$ a highly recommended reference is the book by Borel \cite{Borel}, while for $G=SL(3,\mbb{R})$ we recommend the thesis by Bump \cite{Bump}, and for the general case of $GL(n,\mbb{R})$ the book by Goldfeld \cite{Goldfeld} is highly recommended. For nice physicist's accounts, upon which the present treatment is mainly based, we recommend the paper \cite{ObersPioline} by Obers and Pioline, the review \cite{Pioline:2003bk} by Pioline and Waldron. See also the recent tour de force analysis of Eisenstein series on $SL(d,\mbb{R})$ and $SO(d,d)$ by Green, Russo and Vanhove \cite{Green:2010wi}.

\section{Constructing Automorphic Forms}
We shall here describe several different methods for constructing automorphic forms on coset spaces $G/K$. We first give a general treatment of these methods, comparing and contrasting the mathematician's viewpoint with more physics-inspired constructions. For the applications considered in this thesis we will mainly be interested in a special class of automorphic forms known as \emph{Eisenstein series}. The general construction of Eisenstein series is therefore treated in detail in Sections \ref{Section:EisensteinPoincareSeries} and \ref{Section:EisensteinLatticeConstruction}. Then in Section \ref{Section:SphericalVector} we discuss a more general framework which in principle allows for constructing any automorphic form on a coset space $G/K$, of which interesting examples include Eisenstein series and theta series. 

\subsection{Eisenstein Series - Mathematicians' Approach}
\label{Section:EisensteinPoincareSeries}

Let $G(\mbb{R})$ be a Lie group in its split real form\footnote{As we shall see in Chapter \ref{Chapter:UniversalHypermultiplet} the construction can be generalized to the case of non-split real forms, but for simplicity we here restrict to the split case.}, and let $\mf{g}$ be the associated Lie algebra of rank $r$. We know from Section \ref{Section:rootsystem}Ê that $G$ exhibits an Iwasawa decomposition of the form
\beq
G=NAK, 
\label{IwasawaRightCoset}
\eeq
where $N$ denotes the nilpotent subgroup, $A$ is the Cartan torus and $K$ is the maximal compact subgroup. Let furthermore $h_i$ ($i=1, \dots, r$) be the generators of the Cartan subalgebra $\mf{h}=\text{Lie}\ A$ and choose a parametrization of the Cartan torus of the following form
\beq
a(\phi)= \exp\Big[- \sum_{i=1}^{r} \phi_{i} h_{i}\Big]:= \prod_{i=1}^{r} \Phi_i^{h_i} \in A,
\eeq
where we defined the variables
\beq
\Phi_i := \exp\big[- \phi_i\big].
\eeq
Now note that by virtue of the Iwasawa decomposition (\ref{IwasawaRightCoset}) the left action of any element $x\in N$ on $g=na(\phi) k\in G$ leaves invariant the variables $\Phi_i$ parametrizing the Cartan torus. A function $E^{G(\mbb{Z})}(g;\lambda_{\mc{R}})$ which is invariant under a discrete subgroup $G(\mbb{Z})\subset G(\mbb{R})$ may now be constructed as follows\footnote{More generally \cite{Langlands}, one may construct $E(g;\lambda_{\mc{R}}) = \sum_{\ga\in P(\mbb{Z}\bas G(\mbb{Z})} e^{(\lambda_{\mc{R}}+\rho)\big(H(\ga\cdot g)\big)}$, where $g\in G(\mbb{R})$, $P(\mbb{Z})$ is a parabolic subgroup of $G(\mbb{Z})$, $H(g)$ is an element of the Cartan subalgebra of $\mf{g}=\text{Lie}\ G$ and $\rho$ is the Weyl vector (half the sum of the positive roots, see (\ref{WeylVector})).}
\beq
{E}^{G(\mbb{Z})}(g;\lambda_{\mc{R}}) := \sum_{\ga\in N(\mbb{Z})\bas G(\mbb{Z})} \prod_{i=1}^{r} \Big(\ga \cdot \Phi_i\Big)^{s_i},
\label{GeneralPoincareSeries}
\eeq
where $\lambda_{\mc{R}}=(s_1,\dots ,s_r)$ is a highest weight vector\footnote{In general, it is not necessary for $\lambda_{\mc{R}}$ to be a highest weight vector, but only some arbitrary vector in weight space. However, for our purposes, it is sufficient to restrict to the case when $\lambda_{\mc{R}}$ is a highest weight, in which case the present construction relates nicely to the one in the following section. I thank Boris Pioline for pointing this out to me.} associated with the representation $\mc{R}$ of $G$. Since the product of all $\Phi_i$ is invariant under $N$, we must in order to ensure convergence sum only over the orbit $N(\mbb{Z})\bas G(\mbb{Z}):= \big(G(\mbb{Z})\cap N\big)\bas G(\mbb{Z})$. The function ${E}^{G(\mbb{Z})}(g;\lambda_{\mc{R}})$ is known as an \emph{Eisenstein series}, and a choice of the parameters $s_i$ is called the \emph{order} of the Eisenstein series. The general method of constructing $G(\mbb{Z})$-invariant functions by summing over images of a coset $\Gamma\bas G(\mbb{Z})$, for $\Gamma$ some discrete subgroup of $G$, is known as \emph{Poincar\'e series}, and may also be used to construct other types of automorphic forms. 

An important remark is that the Eisenstein series ${E}^{G(\mbb{Z})}(g;\lambda_{\mc{R}})$ is an  eigenfunction of all $G$-invariant differential operators on $G/K$. In particular, it is an eigenfunction of the Laplacian on $G/K$, with eigenvalue proportional to the value of the quadratic Casimir in the representation $\mc{R}$:
\beq
\Delta_{G/K}  {E}^{G(\mbb{Z})}(g;\lambda_{\mc{R}})= c \big(\lambda_{\mc{R}}, \lambda_{\mc{R}}+2\rho\big){E}^{G(\mbb{Z})}(g;\lambda_{\mc{R}}),
\label{GeneralLaplacian}
\eeq
where $\rho$ is the Weyl vector (see Eq. (\ref{WeylVector})), and the proportionality constant $c\neq 0$ depends on the conventions. We shall see explicit examples of Eq. (\ref{GeneralLaplacian}) in Sections \ref{Section:SL(2)Eisenstein} and \ref{Section:SL(3)Eisenstein}, as well as in Chapter \ref{Chapter:UniversalHypermultiplet}. 

%The Eisenstein series ${E}^{G(\mbb{Z})}(g;\lambda_{\mc{R}})$ has poles for those values of $\lambda_{\mc{R}}$ where the constant terms diverge \cite{Langlands}.\footnote{I am grateful to Pierre Vanhove for stressing this point to me.}

\subsection{Eisenstein Series - Physicists' Approach}
\label{Section:EisensteinLatticeConstruction}
In this section we shall describe a more ``physics-oriented'' method for constructing non-holomorphic Eisenstein series on arithmetic quotients,
\beq
G(\mbb{Z})\bas G(\mbb{R})/K(G), 
\eeq
where $G(\mbb{Z})$ denotes a discrete subgroup of the continuous Lie group $G(\mbb{R})$. In other words, we are interested in $G(\mbb{Z})$-invariant functions whose argument takes values in the coset $G/K$.  The method we will discuss was developed in \cite{ObersPioline} by Obers and Pioline.

We will be able to use some of the technology discussed in {\bf Part I} of this thesis. First we choose a coset representative $\mc{V}\in G/K$ in the Borel gauge (see Section \ref{Section:ParametrizationG/K}). Note that this is a representative for the right coset $G/K$, rather than the left coset $K\bas G$ which was used extensively in {\bf Part I}. This choice is more convenient for our present purposes. The difference lies in the choice of Iwasawa decomposition, which here reads $G=NAK$, as in (\ref{IwasawaRightCoset}), in contrast to $G=KAN$ used in {\bf Part I}. The main building block of the Eisenstein series will then be the ``generalized metric'' $\mc{K}$, constructed from $\mc{V}$ as follows:
\beq
\mc{K}:=\mc{V}\mc{V}^{\mc{T}}, 
\eeq
where $(\ )^{\mc{T}}$ denotes the ``generalized transpose'' defined in Section \ref{Section:GInvariantAction}, which for the special case of $G=SL(n, \mbb{R})$ reduces to the ordinary matrix transpose. The coset representative $\mc{V}$ transforms by $k^{-1}\in K$ from the right and by $g\in G$ from the left, 
\beq
\mc{V} \quad \longmapsto \quad g\mc{V} k^{-1},
\eeq
which in turn implies that $\mc{K}$ is manifestly invariant under $K$, while transforms covariantly under $G$:
\beq
\mc{K} \quad \longmapsto \quad g \mc{K} g^{\mc{T}} .
\eeq
Any function of $\mc{K}$ will then by construction live on the coset $G/K$. To proceed, we must specify what is meant by the discrete subgroup $G(\mbb{Z})\subset G(\mbb{R})$. For the purposes of this thesis  the group $G(\mbb{Z})$ simply corresponds to the subset of $G(\mbb{R})$ that leaves invariant some discrete lattice $\Lambda_{\mbb{Z}}$. In physical applications, this lattice can for example correspond to the lattice of electric and magnetic charges (see, eg., \cite{HullTownsend}). 

The desired Eisenstein series may then be defined as follows
\beq
\mc{E}^{G(\mbb{Z})}(\mc{K};s):=\sum_{{m}\in \Lambda_{\mbb{Z}}\atop m\wedge {m}=0}^{\quad \hs \prime} \Big[{m}^{\mc{T}}\cdot \mc{K}\cdot {m}\Big]^{-s},
\label{GeneralEisensteinSeries}
\eeq
where the prime on the sum indicates that the term with $m=(m_1,\dots, m_n)=(0,\dots, 0)$ should be omitted. This is indeed a function on $G/K$ because it is constructed from $\mc{K}$, and it is also invariant under $G(\mbb{Z})$ since the action of $\ga\in G(\mbb{Z})$ on $\mc{K}$ will hit the lattice vectors ${m}\in \Lambda_{\mbb{Z}}$ and simply results in a change in summation variables, i.e.
\beq
\mc{E}^{G(\mbb{Z})}(\ga \cdot \mc{K};s)=\sum_{{m}\in \Lambda_{\mbb{Z}}\atop m\wedge {m}=0} \Big[{m}^{\mc{T}}\cdot \ga\cdot \mc{K}\cdot \ga^{\mc{T}} \cdot {m}\Big]^{-s}=\sum_{{m}^{\prime}\in \Lambda_{\mbb{Z}}\atop {m}^{\prime}\wedge {m}^{\prime}=0}\Big[{m}^{\prime \mc{T}}\cdot \mc{K}\cdot {m}^{\prime}\Big]^{-s},
\eeq
where ${m}^{\prime}:= \ga^{\mc{T}}Ê\cdot {m} \in \Lambda_{\mbb{Z}}$. Note that the sum over $m\in \Lambda_{\mbb{Z}}$ is restricted to $m\wedge m=0$Ê which is a certain quadratic constraint which is required in general to make sure that $\mc{E}^{G(\mbb{Z})}(\mc{K};s)$ is an eigenfunction of the Laplacian on $G/K$. For the case of $G=SL(n, \mbb{R})$ this constraint is not needed, but it will appear explicitly in Chapter \ref{Chapter:UniversalHypermultiplet} when we consider the example of $G=SU(2,1)$. For a more detailed discussion of the constraint see \cite{ObersPioline}. We note that for an $n$-dimensional integral square lattice $\Lambda_{\mbb{Z}}=\mbb{Z}^n$, the Eisenstein series (\ref{GeneralEisensteinSeries}) converges absolutely for $\Re(s)> n/2$ \cite{Terras2}.

Let us briefly conclude with a comparison to the construction of the previous section. The Eisenstein series $\mc{E}^{G(\mbb{Z})}(\mc{K};s)$ may be considered as a special case of the Poincar\'e series ${E}^{G(\mbb{Z})}(g;\lambda_{\mc{R}})$ in (\ref{GeneralPoincareSeries}). More precisely, $\mc{E}^{G(\mbb{Z})}(\mc{K};s)$ arises as a certain limiting choice of the parameters $s_i$. This will become more clear in Sections \ref{Section:SL(2)Eisenstein} and \ref{Section:SL(3)Eisenstein} where we consider explicit examples.

\subsection{Constructing Automorphic Forms Using Spherical Vectors}
\label{Section:SphericalVectorsConstruction}
There exists a general method for constructing automorphic forms on $G(\mbb{Z})\bas G/K$ as advocated in \cite{Kazhdan:2001nx,Kazhdan,Pioline:2003bk}. The fundamental object in this construction is the so called \emph{spherical vector} $f_K $ which belongs to a Hilbert space of square-integrable functions. The spherical vector is defined by its invariance under $K(G)$, acting on $\mc{H}$ through a linear representation $\rho$ of $G(\mbb{R})$. To understand this construction we begin by recalling a few facts about the notion of induced representations. 

%See for example \cite{} for a more complete discussion.

\subsubsection{A Note on Parabolic Induction}

We want to describe representations $\rho$ of a Lie group $G$ acting on functions belonging to a Hilbert space $\mc{H}$, which in our case will be identified with the space $L^2(P\bas G)$ of square integrable, real-valued functions on the coset space $P\bas G$, where $P$ is the Borel subgroup of $G$. We will see how $P$ is defined in explicit examples later on.  Elements of $L^2(P\bas G)$ are then functions of the form
\beq
f \hs : \hs  n \hs \longmapsto \hs f(n) \in \mbb{R}, \qquad n\in P\bas G.
\eeq
The group $G$ acts on $n\in P\bas G$ from the right and a compensating transformation of $p\in P$ from the left is required to obtain a new representative $n^{\prime}$ for the coset space $P\bas G$, 
\beq
n\hs \longmapsto \hs p n g = n^{\prime}\in P\bas G.
\eeq
We will then consider representations $\rho$ of $G$ acting on functions $f\in L^2(P\bas G)$ with the property that
\beq
\rho(g)\cdot f(n) = f(ng)= f(p n^{\prime})=\chi(p) f(n^{\prime}),
\eeq
where 
\beq
\chi \hs :\hs p \hs \longmapsto \hs \chi(p)\in \mbb{R}
\eeq
is the so called \emph{infinitesimal character} defining the representation $\rho$. Representations of $G$ obtained in this way by induction from the Heisenberg parabolic $P$ belong to the \emph{principal continuous series} of $G$-representations.

\subsubsection{Automorphic Forms and Spherical Vectors}
\label{Section:SphericalVector}

We shall now apply these techniques to construct an automorphic form $\Psi$ on $G/K$, invariant under $G(\mbb{Z})$. As mentioned above, to this end we need to find a \emph{spherical vector}, i.e. a $K$-invariant function $f_K\in \mc{H}=L^{2}(P\bas G)$ belonging to the representation $\rho$ discussed above. In addition we need two more ingredients, namely a $G(\mbb{Z})$-invariant \emph{distribution} $f_{\mbb{Z}}$ which belongs to the dual space $\mc{H}^{\star}$, as well as an inner product $\left<\hs ,\hs \right>$, provided by the canonical pairing between $\mc{H}$ and $\mc{H}^{\star}$. Putting things together, we may now quite generally define $\Psi$ as \cite{Kazhdan:2001nx,Kazhdan,Pioline:2003bk}
\beq
\Psi(g):=\left< f_{\mbb{Z}}, \rho(g)\cdot f_K\right>.
\label{generalautomorphicform}
\eeq 
The calculation of $\Psi$ can be simplified by invoking the Iwasawa decomposition of $G$ (see (\ref{IwasawaRightCoset})).We thus have the corresponding decomposition of $g\in G$,
\beq
g=nak, \qquad n\in N, \hs a\in A, \hs k\in K.
\eeq
Now we may use the property that $f_K$ is invariant under $K$,
\beq
\rho(k)\cdot f_K =f_K,
\eeq
implying that 
\beq
\rho(g)\cdot f_K=\rho(nak)\cdot f_K=\rho(na)\cdot f_K.
\eeq
Defining the standard coset representative in the Borel gauge as follows
\beq
\mc{V}=na \hs \in \hs G/K,
\eeq
we may write the automorphic form as 
\beq
\Psi(\mc{V})=\left< f_{\mbb{Z}}, \rho(\mc{V})\cdot f_K\right>.
\label{generalautomorphicform}
\eeq
This minor simplification will turn out to be very useful in the construction of $\Psi$. The coset representative $\mc{V}\in G/K$ transforms by $k^{-1}\in K$ on the right and $\ga\in  G(\mbb{Z})$ on the left,
\beq
\mc{V} \longmapsto \ga\mc{V}k^{-1}.
\eeq
On $\Psi$ the left action by $k$ becomes a left action on $f_{K}$, which is invariant, and the right action of $\ga$ becomes a left action on $f_{\mbb{Z}}$, which is also invariant. Hence, $\Psi(\mc{V})$ is by construction a function on the double quotient $G(\mbb{Z})\bas G / K$.

%It is possible to generalize this construction to the case when the automorphic form is not a singlet under the maximal compact subgroup $K$ but transforms in some representation $\mc{R}$. This is done by replacing the spherical vector $f_K$ by a new function $f_{\mc{R}(K)}$, which is no longer invariant under $K(G)$ but transforms in the representation $\mc{R}$. The transforming automorphic form then becomes
%\beq
%\Psi_{\mc{R}(K)}(g)=\left< f_{\mbb{Z}}, \rho(g)\cdot f_{\mc{R}(K)}\right>,
%\eeq
%where we included the full $SL(2,\mbb{R})$ element $g=kan$ since $f_{\mc{R}(K)}$ is no longer invariant under $\rho(k)$. We will discuss such transforming automorphic forms in more detail in Section \ref{Section:TransformingAutomorphicForms}.
\subsubsection{{  $p$}-Adic Automorphic Forms}
Although very appealing, the method described above is complicated by the fact that the distribution $f_{\mbb{Z}}$ is in general difficult to obtain. There is however a powerful mathematical technique, developed in \cite{Kazhdan:2001nx,Kazhdan,Pioline:2003bk}, to compute $f_{\mbb{Z}}$ using $p$-adic number theory. In this approach, the distribution $f_{\mbb{Z}}$ is reinterpreted in terms of a \emph{$p$-adic spherical vector} $f_p$ that can be straightforwardly constructed from its real counterpart $f_{K}$ in a way that will be explained shortly. The automorphic form $\Psi$ can then be written in the following way 
\beq
\Psi(\mc{V})=\sum_{\vec{x}\in \mbb{Q}^n} \Big[\prod_{p<\infty} f_p(\vec{x})\Big]\rho(\mc{V})\cdot f_{K}(\vec{x}),
\label{AdelicAutomorphicForm}
\eeq
where $\vec{x}$ is a vector of rational numbers in $\mbb{Q}^n$, and the product is over all prime numbers $p$. This construction is based on the fact that for the field of $p$-adic numbers the analogue of the maximal compact subgroup $K$ is $G(\mbb{Z}_p)$, where $\mbb{Z}_p$ denotes the ring of $p$-adic integers which is a compact subset of the $p$-adic numbers $\mbb{Q}_p$ (see Eq. \ref{padicinteger}). Hence, in the $p$-adic framework, the $G(\mbb{Z})$-invariant vector $f_{\mbb{Z}}$ simply corresponds to the $G(\mbb{Z}_p)$-invariant ``spherical'' vector $f_p$. Moreover, it was found in \cite{Kazhdan:2001nx,Kazhdan,Pioline:2003bk} that once the $K$-invariant spherical vector $f_K$ is constructed, there is a simple prescription to obtain the $p$-adic spherical vector $f_p$. This prescription will be explained in explicit examples below. We will then also see how to evaluate the somewhat intimidating infinite product over primes in (\ref{AdelicAutomorphicForm}), as well as how to recover the lattice summation appearing in the Eisenstein series in (\ref{GeneralEisensteinSeries}) from the sum over rational numbers in $\Psi(\mc{V})$. This will reveal that for a certain choice of spherical vector $f_K$ the construction in (\ref{AdelicAutomorphicForm}) is equivalent to the construction of $\mc{E}^{G(\mbb{Z})}(\mc{K};s)$ in (\ref{GeneralEisensteinSeries}). In the next section will introduce some of the basic terminology, and in Sections \ref{Section:SL(2)Eisenstein} and \ref{Section:SL(3)Eisenstein} will discuss some detailed examples of how this works.

%The underlying principle behind this construction is known as the \emph{strong approximation theorem} and states that any function on $G(\mbb{Z})\bas G/K$ is equivalent to a function on $G(\mbb{Q})\bas G(\mbb{A})/ K(\mbb{A})$, where $\mbb{A}$ is the field of \emph{adeles}, to be defined in the next section. 

%
%This analysis will also reveal that the constraint $\delta(\vec{m}\wedge \vec{m})$ has a natural interpretation in $\Psi(\mc{V})$ which is related to the particular representation of $G(\mbb{R})$ to which $f_K$ belongs. 

\subsubsection{A Lightning Review of $p$-Adic Numbers}
In order to use the formula (\ref{AdelicAutomorphicForm}) later on, we need to introduce a few basic facts about $p$-adic number theory. This is a vast subject in mathematics, and the following treatment is merely intended to provide the minimum set of tools required for subsequent sections. For a very nice introduction to $p$-adic numbers for physicists we recommend \cite{BrekkeFreund}. A relatively accessible mathematical treatment is also given in \cite{GelbartMiller}.

Perhaps the most intuitive definition of a $p$-adic number $x_p$ is to consider it as a Laurent series in the prime number $p$:
\beq
x_p= c_kp^{k}+c_{k+1}p^{k+1}+\cdots , \qquad k\in\mbb{Z},
\eeq
with the coefficients $c_k$ being strictly positive integral, and bounded by $p$. The field of $p$-adic numbers is denoted by $\mbb{Q}_p$, and may be considered as a ``completion'' of the real numbers $\mbb{R}$ under the $p$-adic norm $|\cdot |_p$. This norm is defined as 
\beq
|x|_p:= p^{-k}, \qquad x\in \mbb{Q},
\eeq
where $k$ is the largest integer such that $x/p^{k}\in \mbb{Z}$. 

We also have the subset of $p$-adic integers $\mbb{Z}_p$ which similary may be considered as the completion of the standard integers $\mbb{Z}$ under the $p$-adic norm. They have the following definition:
\beq
\mbb{Z}_p=\{x\in\mbb{Q}_p\ |\ |x|_p\leq 1\}\subset \mbb{Q}_p.
\label{padicinteger}
\eeq

We can also extend the $p$-adic norm to vectors of rational numbers. Recall that the Euclidean norm $|\vec{v}|_{\infty}=||\vec{v}||$ of a vector $\vec{v}=(v_1, \dots, v_n)\in \mbb{Q}^n$ is defined as
\beq
||\vec{v}||:= \sqrt{v_1^2+\cdots v_n^2}.
\eeq
The $p$-adic counterpart of this is then given by
\beq
|\vec{v}|_p:=\mathrm{max}(|v_1|_p, \dots, |v_n|_p).
\eeq
Note that, in contrast to the Euclidean norm, there is no square root in the definition of $|\vec{v}|_p$.

\section{Eisenstein series on $SL(2,\mbb{R})/SO(2)$}
\label{Section:SL(2)Eisenstein}
In this section we shall illustrate the various constructions discussed above in the context of the standard non-holomorphic $SL(2, \mbb{Z})$-Eisenstein series on $SL(2, \mbb{R})/SO(2)$. In Section \ref{Section:D(-1)Example} we saw that this Eisenstein series takes the form
\beq
\mc{E}^{SL(2,\mbb{Z})}(\tau; s)=\sum_{(m,n)\neq (0,0)}\f{\tau_2^s}{|m+n\tau|^{2s}},
\label{EisensteinSeriesSL(2,Z)2}
\eeq
where the complex parameter $\tau$ parametrizes the upper half plane 
\beq
\mbb{H}=\{\tau \in \mbb{C}\ |\ \tau_2 \geq 0\}.
\eeq 
We will now see how Eq. (\ref{EisensteinSeriesSL(2,Z)2}) may be reproduced using either of the three methods described in Sections \ref{Section:EisensteinPoincareSeries}, \ref{Section:EisensteinLatticeConstruction} and \ref{Section:SphericalVectorsConstruction}. 
\subsection{Poincar\'e Series}
Recall from Section \ref{Section:mother} that the Lie algebra $\mf{sl}(2, \mbb{R})$ is generated by the Chevalley triple $(f,h,e)$, subject to the relations
\beq
[h, e]=2e, \qquad [h, f]=-2f, \qquad [e, f]=h.
\eeq
Following the logic of Section \ref{Section:EisensteinPoincareSeries} we then perform an Iwasawa decomposition of $g\in SL(2,\mbb{R})$:
\beq
g=e^{\chi e}e^{-\tilde{\phi} h} k=\left(\begin{array}{cc}
 1 & \chi \\
  & 1 \\
  \end{array}\right)\left(\begin{array}{cc}
e^{-\tilde\phi} & \\
 & e^{\tilde\phi} \\
 \end{array} \right)k, \qquad k\in SO(2),
  \label{SL(2)Iwasawa}
  \eeq
where
\beq
a(\tilde\phi)=\left(\begin{array}{cc}
e^{-\tilde\phi} & \\
 & e^{\tilde\phi} \\
 \end{array} \right):= \Phi^{H}.
 \eeq
Implementing this in the general formula (\ref{GeneralPoincareSeries}) yields the Eisenstein series
\beq
E^{SL(2,\mbb{Z})}(g;\tilde{s})=\sum_{\ga\in N(\mbb{Z})\bas SL(2,\mbb{Z})} \big(\ga\cdot \Phi\big)^{\tilde{s}}.
\label{PoincareSL(2)}
\eeq 
To make contact with Eq. \ref{EisensteinSeriesSL(2,Z)2} we note that the isomorphism between the upper half plane $\mbb{H}$ and the coset space $SL(2,\mbb{R})$ is given by 
\beq
\tau=\tau_1+i\tau_2:=\chi +i e^{-\phi},
\eeq
where $\phi \equiv 2\tilde\phi$, and hence $\Phi \equiv \tau_2^{2}$. Moreover, the action of $SL(2,\mbb{Z})$ on the upper half plane is given by the standard fractional transformation
\beq
\ga \cdot \tau = \f{a\tau+b}{c\tau+d}, \qquad a, b, c, d\in\mbb{Z}, \quad ad-bc=1,
\eeq
which in particular implies 
\beq
\ga\cdot \Phi  =\left(\f{\tau_2}{|c\tau+d |^2}\right)^{2}.
\label{gammaactionPhi}
\eeq
To understand the sum in Eq. (\ref{PoincareSL(2)}), we must also analyze the structure of the coset $N(\mbb{Z})\bas SL(2,\mbb{Z})$. This corresponds to the equivalence relation
\beq
SL(2,\mbb{Z}) \ni \left(\begin{array}{cc}
a & b \\
c & d \\ 
 \end{array} \right) \sim
\left(\begin{array}{cc}
1 & \ell \\
 & 1 \\ 
 \end{array} \right) \left(\begin{array}{cc}
a & b \\
c & d \\ 
 \end{array} \right)=\left(\begin{array}{cc}
a+\ell c & b+\ell d \\
c & d\\ 
 \end{array} \right), \qquad \ell\in \mbb{Z}.
 \label{equivalence}
 \eeq
Thus, given two coprime integers $(c,d)$, there exists $a,b\in\mbb{Z}$, which are unique up the equivalence defined by (\ref{equivalence}), such that the relation $ad-bc=1$ holds. Hence, the coset $N(\mbb{Z})\bas SL(2,\mbb{Z})$ is completely characterized by the two coprime integers $c$ and $d$. Finally, redefining $\tilde{s}\equiv s/2$ and using Eq. \ref{gammaactionPhi} we find that the Eisenstein series (\ref{PoincareSL(2)}) becomes
 \beq
 E^{SL(2,\mbb{Z})}(\tau; s)=\sum_{(c,d)=1}^{\quad\hs \prime} \f{\tau_2^{s}}{|c\tau+d|^{2s}},
 \eeq
where the prime on the sum indicates that the term with $c=d=0$ should be removed. The additional coprime condition compared to Eq. (\ref{EisensteinSeriesSL(2,Z)2}) can be understood as follows. Consider the sum over $m, n\in \mbb{Z}$ in (\ref{EisensteinSeriesSL(2,Z)2}) and make the change of variables $m\equiv k d$ and $n\equiv k c$, with $k=\text{gcd}(m,n)$ and $(c,d)=1$. Implementing this in (\ref{EisensteinSeriesSL(2,Z)2}) yields
\beq
\sum_{(m,n)\neq  (0,0) }\f{\tau_2^s}{|m+n\tau|^{2s}}=\sum_{(c,d)=1 }^{\quad \hs \prime}\Big[\sum_{k\neq 0} k^{-2s}\Big]\f{\tau_2^s}{|c\tau+d|^{2s}}=2\zeta(2s) \sum_{(c,d)=1 }^{\quad \hs \prime}\f{\tau_2^s}{|c\tau+d|^{2s}},
\eeq
where we recall that the Riemann zeta function is given by
\beq \zeta(2s)=\sum_{k=1}^{\infty} k^{-2s}.
\eeq
We may therefore conclude that the two different Eisenstein series are related in the following simple way
\beq
\mc{E}^{SL(2,\mbb{Z})}(\tau; s)=2\zeta(2s)E^{SL(2,\mbb{Z})}(\tau; s).
\label{SL(2)Relation}
\eeq
\subsection{Lattice Construction}
Let us now see how we can construct $\mc{E}^{SL(2,\mbb{Z})}(\tau; s)$ using the method discussed in Section \ref{Section:EisensteinLatticeConstruction}. Again, invoking the Iwasawa decomposition, we may choose a coset representative $\mc{V}$ in the Borel gauge as follows
\beq
\mc{V}= e^{\chi e}e^{-\f{\phi}{2} h} =\left(\begin{array}{cc}
e^{-\phi/2} & \\
 & e^{\phi/2} \\
 \end{array} \right)\left(\begin{array}{cc}
 1 & \chi \\
  & 1 \\
  \end{array}\right)\in SL(2, \mbb{R})/SO(2).
  \label{SL(2)CosetRepresentative}
  \eeq
 The generalized metric $\mc{K}$ then takes the form
 \beq
 \mc{K}:=\mc{V}\mc{V}^{T}=\left(\begin{array}{cc}
 e^{-\phi}+e^{\phi}\chi^2 & e^{\phi} \chi \\
 e^{\phi}\chi & e^{\phi}	 \\
 \end{array}Ê\right).
 \eeq 
According to the general prescription described in Section \ref{Section:EisensteinLatticeConstruction} the Eisenstein series is obtained by taking a lattice vector $\vec{m}=( n, m)\in \mbb{Z}^2$, corresponding to the fundamental representation of $SL(2, \mbb{Z})$, and then summing over the integers $m, n$ as follows
\beqa
 \mc{E}^{SL(2,\mbb{Z})}(\tau; s)&=&\sum_{\vec{m}\in \mbb{Z}^2}^{\quad \hs \prime} \Big[\vec{m}^T\cdot \mc{K}Ê\cdot \vec{m}\Big]^{-s}
 \nn \\
 &=& \sum_{(m, n)\neq  (0,0)} \Big[\f{e^{\phi}}{n^2+e^{2\phi}(m+n\chi)^2}\Big]^s.
 \eqa
 This result agrees with Eq. (\ref{EisensteinSeriesSL(2,Z)2}) if we identify 
 \beq
 \tau:= \chi + i e^{-\phi},
 \eeq
 which is the standard map between the upper half plane $\mbb{H}$ and the coset space $SL(2, \mbb{R})/SO(2)$.

% For later use it will also prove useful to write the equivalent form of the Eisenstein series in the antifundamental representation of $SL(2, \mbb{Z})$. This is obtained simply by replacing the matrix $\mc{M}$ by its inverse, and we find
% \beqa
% \bar{\mc{E}}_{s}^{SL(2,\mbb{Z})}(\tau, \bar{\tau})&=&\sum_{\vec{n}\in \mbb{Z}\bas (0,0)} \Big[\vec{n}^T\cdot \mc{M}^{-1}\cdot \vec{n}\Big]^s
% \nn \\
% &=& \sum_{(n_1, n_2)\in \mbb{Z}^2\bas (0,0)} \Big[\f{e^{\phi}}{n_2^2+e^{2\phi}(n_1-n_2\chi)^2}\Big]^s,
% \eqa
% where the vector $\vec{n}$ is now to be considered as belonging to the antifundamental representation of $SL(2, \mbb{Z})$. We shall see later that the minus sign appearing in the denominator will play a role when we construct $\bar{\mc{E}}_{s}^{SL(2,\mbb{Z})}(\tau, \bar{\tau})$ in terms of the spherical vector, as discussed in Section \ref{Section:SphericalVectorsConstruction}. Note that , in this simple example, the two Eisenstein series $ \mc{E}^{SL(2,\mbb{Z})}(\tau; s)$ and $\bar{\mc{E}}_{s}^{SL(2,\mbb{Z})}(\tau, \bar{\tau})$ are equivalent under the identifications $m_1\equiv n_1$ and $m_2\equiv -n_2$. 

%%%%%%%%%%%%%%%%%%%%%%%%%%%%%%%%%%%%%%%%%%%%%

\subsection{Spherical Vector}
Finally, we will show how to construct the Eisenstein series ${\mc{E}}_{s}^{SL(2,\mbb{Z})}(\tau, \bar{\tau})$ using the more abstract framework of Section \ref{Section:SphericalVectorsConstruction}. This will be done using induction from the parabolic subgroup 
\beq
P=\left\{\left(\begin{array}{cc}
t_1 & \\
\eta & t_2 \\ 
\end{array}Ê\right) \hs \Big| \hs \eta, t_1, t_2 \in \mbb{R};\hs t_1 t_2=1 \right\}.
\eeq
Note that $P$ coincides with the Borel subgroup $B_{-}\subset SL(2, \mbb{R})$ generated by $\{f, h\}$. A representative for the parabolic coset $P\bas SL(2, \mbb{R})$ can now be chosen as
\beq
n=\left( \begin{array}{cc}
1 & x \\
 & 1Ê\\
 \end{array} \right)\in P\bas SL(2, \mbb{R}).
 \label{ParabolicCosetRepresentative}
 \eeq
 Functions in $L^2\big(P\bas SL(2, \mbb{R})\big)$ corresponding to the principal continuous series then obey
 \beq
 \rho(g) \cdot f(n)=f(ng)=f(pn^{\prime})=\chi_s(p) f(n^{\prime}),
 \label{SL(2)PrincipalContinuous}
 \eeq
 where the character $\chi_s(p)$ is defined as
 \beq
 \chi_s(p) := t_1^{-2s}, \qquad p=\left(\begin{array}{cc}
t_1 & \\
\eta & t_2 \\ 
\end{array}Ê\right)\in P.
 \eeq
  We want to find a suitable spherical vector, i.e. an $SO(2)$-invariant function $f_K$ belonging to $L^2\big(P\bas SL(2, \mbb{R})\big)$. The maximal compact subgroup $SO(2)$ acts on $n\in P\bas SL(2, \mbb{R})$ from the right, and a compensating transformation by $p\in P$ from the left is required to restore the upper-triangular form of (\ref{ParabolicCosetRepresentative}). Since $k\in SO(2)$ acts on $n$ as a rotation, it will by definition leave invariant the Euclidean norms $||\vec{r}_i||$ of the two rows $\vec{r}_1$ and $\vec{r}_2$ in $n$. Note furthermore that the compensating transformation of $p\in P$ acts simply by multiplication on the top row:
 \beq
 p n = \left(\begin{array}{cc}
 t_1 & t_1 x \\
 \eta & t_2+\eta x \\
 \end{array} \right).
 \eeq
 We may therefore take the spherical vector $f_K$ to be defined as the Euclidean norm of the top row of $n$:
 \beq 
 f_K(x):= ||\vec{r}_1||^{-2s}=(1+x^2)^{-s}.
 \eeq
Although the compensating action of $p$ from the left modifies $f_K$ by an overall factor $t_1^{2s}$, we must also include the character $\chi_s(p)=t_1^{-2s}$ since $f_K$ belongs to the principal continuous series. Hence, we find that $f_K(x)$ is indeed invariant, and corresponds to the desired spherical vector. 

To construct the automorphic form $\Psi(\mc{V})$, we must also compute the action of $\rho(\mc{V})$ on $f_K(x)$, where $\mc{V}\in SL(2, \mbb{R})/SO(2)$ is the standard coset representative, Eq. (\ref{SL(2)CosetRepresentative}),  in the Borel gauge. This is straightforward, following the prescription in (\ref{SL(2)PrincipalContinuous}): the action of $\rho(\mc{V})$ on any function $f(n)$ is given by 
\beq
\rho(\mc{V})\cdot f(n)=f(n\mc{V})=f(p_0 n^{\prime})=\chi_s(p_0)f(n^{\prime})=e^{s\phi} f(n^{\prime}),
\eeq
where 
\beq
p_0=\left(\begin{array}{cc}
e^{-\phi/2} & \\
 & e^{\phi/2} \\
 \end{array}Ê\right), \qquad n^{\prime}=\left(\begin{array}{cc}
1 & e^{\phi}(x+\chi)\\
 & 1 \\
 \end{array}Ê\right).
 \eeq
 The spherical vector thus transforms as follows
 \beq
 \rho(\mc{V})\cdot f_K(x)= e^{s\phi} \big( 1+e^{2\phi}(x+\chi)^2\big)^{-s}.
 \label{SL(2)actionSphericalVector}
 \eeq
The last missing piece in the construction of $\Psi(\mc{V})$ is the $p$-adic spherical vector $f_p$. It was found in \cite{Kazhdan:2001nx,Kazhdan,Pioline:2003bk} that $f_p$ can be constructed according to the following simple prescription: replace the Euclidean norm $||\cdot ||$ in $f_K$ by the $p$-adic version $|\cdot |_p$.\footnote{This procedure works because the action of the group is (projective) linear in this representation. I am grateful to Boris Pioline for pointing this out to me.} Hence, we take the following definition of the $p$-adic spherical vector
\beq
f_p(x):= |\vec{r}_1|_p^{-2s}=\mathrm{max}\big(1, |x|_p\big)^{-2s},
\label{SL(2)pAdicSphericalVector}
\eeq
where the $p$-adic norm $|\cdot |_p$ was defined in Section \ref{Section:SphericalVectorsConstruction}. Inserting Eq. (\ref{SL(2)actionSphericalVector}) and Eq. (\ref{SL(2)pAdicSphericalVector}) into Eq. (\ref{AdelicAutomorphicForm}), we find that the automorphic form $\Psi(\mc{V})$ takes the form
\beq
\Psi(\mc{V})=\sum_{x\in \mbb{Q}} \Big[\prod_{p<\infty} \mathrm{max}\big(1, |x|_p\big)^{-2s}\Big] e^{s\phi} \big( 1+e^{2\phi}(x+\chi)^2\big)^{-s}.
\eeq
In order to make sense of this expression, we must evaluate the infinite sum over prime numbers. To this end we first make a change of variables in the summation, and write the rational number $x\in \mbb{Q}$ as
\beq
x=\f{m}{n}, \qquad m, n \in \mbb{Z}, \hs n\neq 0, \hs (m,n)=1.
\label{ChangeofVariable}
\eeq
This yields
\beq
\Psi(\mc{V})=\sum_{m\in\mbb{Z}}\sum_{n\neq 0} \Big[\prod_{p<\infty} \mathrm{max}\Big(1, \Big|\f{m}{n}\Big|_p\Big)^{-2s}\Big] e^{s\phi} \Big( 1+e^{2\phi}\big(\f{m}{n}+\chi\big)^2\Big)^{-s}.
\eeq
Now we can evaluate the product over primes. Suppose first that $m=n$. Then we know by definition of the $p$-adic norm that $|m/n |_p=1$ and hence the product over primes gives unity. On the other hand, when $m\neq n$ the maximum value of $f_p$ for each $p$ will be when $p=n$, for which we have
\beq
\prod_{p< \infty} \mathrm{max}\Big[1, \Big|\f{m}{n}\Big|_p\Big]=n.
\eeq
All other cases, i.e. when $p\neq n$, will again give unity because of the ``max''-condition. Hence, the result is
\beqa
\Psi(\mc{V})&=&\sum_{(m,n )=1\atop  n\neq 0} n^{-2s} e^{s\phi} \left[ 1+e^{2\phi}\Big(\f{m}{n}+\chi\Big)^2\right]^{-s}
\nn \\
&=&\sum_{(m,n )=1\atop n\neq 0} \f{\tau_2^s}{|m+n\tau|^{2s}},
\eqa
where $\tau=\chi+i e^{-\phi}$ as usual. We conclude that indeed for $f_K$ in the principal continuous series, this final form of $\Psi(\mc{V})$ coincides with the non-holomorphic Eisenstein series $E^{SL(2,\mbb{Z})}(\tau;s)$ in Eq. (\ref{EisensteinSeriesSL(2,Z)2}), modulo the missing term corresponding to $n=0$. 

\section{Eisenstein Series on $SL(3,\mbb{R})/SO(3)$}
\label{Section:SL(3)Eisenstein}

We will now consider in detail also the example of $SL(3, \mbb{Z})$-invariant Eisenstein series in the principal continuous series of $SL(3,\mbb{R})$-representations.
\subsection{Poincar\'e Series}

The rank 2 Lie algebra $\mf{sl}(3,\mbb{R})$ is generated by the two Chevalley triples $(e_1, f_1, h_1)$ and $(e_2,f_2,h_2)$ represented in the fundamental representation by the following matrices:
\beq
{}Ê e_1=\left(\begin{array}{ccc}
0 & 1 & 0\\
0 & 0 & 0\\
0 & 0 & 0\\
\end{array}\right),\quad
e_2=\left(\begin{array}{ccc}
0 & 0 & 0\\
0 & 0 & 1\\
0 & 0 & 0\\
\end{array}\right),\quad 
 h_1=\left(\begin{array}{ccc}
1 & \phantom{-}0 & \phantom{-}0\\
0 & -1 & \phantom{-}0\\
0 & \phantom{-}0 & \phantom{-}0\\
\end{array}\right),\quad
h_2=\left(\begin{array}{ccc}
0 & \phantom{-}0 & \phantom{-}0\\
0 &  \phantom{-}1 & \phantom{-}0\\
0 & \phantom{-}0 & -1 \\
\end{array}\right),
\eeq
together with $f_i=(e_i)^{T}, \hs i=1,2$. An arbitrary element $g\in SL(3, \mbb{R})$ may then be represented in Iwasawa form as
\beqa
g&=&\exp\left[v e_1+\f{1}{2}\mc{A}_2 e_2+\Big(\mc{A}_1-\f{1}{2} v \mc{A}_1\Big) e_3\right] \exp\left[ -\tilde{\phi}_1 h_1-\tilde\phi_2 h_2 \right] k 
\nn \\
&=&\left(\begin{array}{ccc}
1 & v & \mc{A}_1 \\
 & 1 & \mc{A}_2 \\
 & & 1\\
 \end{array}\right) \left(\begin{array}{ccc}
 e^{-\tilde\phi_1} & & \\
 & e^{\tilde\phi_1-\tilde\phi_2} & \\
 & & e^{\tilde\phi_2} \\
 \end{array} \right)k,
 \label{SL(3)ElementVTconventions}
 \eqa
where $e_3=[e_1,e_2]$. The Cartan torus is thus parametrized by
\beq
a(\tilde\phi)=\left(\begin{array}{ccc}
 e^{-\tilde\phi_1} & & \\
 & e^{\tilde\phi_1-\tilde\phi_2} & \\
 & & e^{\tilde\phi_2} \\
 \end{array} \right):= \big(\Phi_1\big)^{h_1} \big(\Phi_2\big)^{h_2}.
 \eeq
 From Eq. (\ref{GeneralPoincareSeries}) we then find that an $SL(3,\mbb{Z})$-invariant Eisenstein series in the principal series of $SL(3,\mbb{R})$ is constructed as follows
 \beq
  E^{SL(3,\mbb{Z})}(g; \tilde{s}_1,\tilde{s}_2)=\sum_{\ga \in N(\mbb{Z})\bas SL(3,\mbb{Z})} \Big(\ga\cdot \Phi_1\Big)^{\tilde{s}_1} \Big(\ga\cdot \Phi_2\Big)^{\tilde{s}_2}.
  \label{SL(3)PoincareSeries}
  \eeq
For the purpose of simplifying the analysis in Section \ref{Section:FourierExpansionSL(3)}, it will be useful to make a slight change of variables which is in better accordance with the conventions of \cite{VT}:
\beq
\Phi_1 = y^{-1/6} u^{-1/2}, \qquad \Phi_2= y^{-1/3},
\eeq
and we also redefine the parameters $\tilde{s}_1$ and $\tilde{s}_2$ as follows
\beq
\tilde{s}_1= -2 s_2, \qquad \tilde{s}_2= -2 s_1.
\eeq
Implementing these changes in (\ref{SL(3)PoincareSeries}) yields
\beq
E^{SL(3,\mbb{Z})}(g;s_1,s_2)=\sum_{\ga \in N(\mbb{Z})\bas SL(3,\mbb{Z})} \Big(\ga\cdot y\Big)^{\f{2s_1+s_2}{3}} \Big(\ga\cdot u\Big)^{s_2}.
  \label{SL(3)PoincareSeriesVTvariables}
  \eeq
To write this out explicitly, we consider an arbitrary $SL(3, \mbb{Z})$-element of the form
\beq
\ga=\left(\begin{array}{ccc}
a_1 & a_2 & a_3 \\
b_1 & b_2 & b_3 \\
c_1 & c_2 & c_3 \\
\end{array} \right)\in SL(3,\mbb{Z}),
\eeq
and the Eisenstein series may then be written as
 \beqa
 E^{SL(3,\mbb{Z})}(g; s_1, s_2)&=& y^{\f{s_2-s_1}{3}} \sum_{(c, b)\in \mbb{Z}^6}^{\quad \prime}\bigg\{ \Big[c_1^2 u+\f{1}{u}\big(c_2+c_1 v\big)^2+\f{1}{y}\big(c_1\mc{A}_1+c_2\mc{A}_2+c_3\big)^2\Big]^{-s_1}
 \nn \\
 & & \times \Big[yD_3^2+u\big(D_2-\mc{A}_2 D_3\big)^2+\f{1}{u}\big(D_1-vD_2-(\mc{A}_1-v\mc{A}_2)D_3\big)^2\Big]^{-s_2}\bigg\},
 \nn \\
 \label{SL(3)PoincareSeriesVTvariablesExplicit}
 \eqa
 where we defined 
 \beq
 D_1:= b_2c_3-b_3c_2, \qquad D_2:=b_3c_1-b_1c_3, \qquad D_3:= b_1c_2-b_2c_1.
 \eeq
The summation variables in (\ref{SL(3)PoincareSeriesVTvariablesExplicit}) are further constrained by the structure of the coset $N(\mbb{Z})\bas SL(3,\mbb{Z})$, and it was shown in \cite{VT} that this implies the following relations:
\beqa
(c_1,c_2,c_3)=(D_1,D_2,D_3)&=&1,
\nn \\
c_1 D_1+c_2 D_2+c_3D_3&=&0,
\nn \\
c_1, D_1 &\geq & 0,
\eqa
where we used the notation $(x,y,z)= \text{gcd}(x,y,z)$. 

We will see in Section \ref{Section:FourierExpansionSL(3)} that these variables are natural when computing the Fourier expansion. In particular, it will be important that the complex parameter $z=v+iu$ transforms in the standard fractional way under a certain $SL(2,\mbb{Z})$ subgroup of $SL(3,\mbb{Z})$. 

\subsection{Lattice Construction}

Let us now proceed to the method of Section \ref{Section:EisensteinLatticeConstruction}. To this end we begin by choosing the coset representative
\beq
\mc{V}=
  \left(\begin{array}{ccc}
1 & \mc{A} & \mc{C} \\
 & 1 & \mc{B} \\
  & & 1 \\
  \end{array}Ê\right) \left(\begin{array}{ccc}
1/\eta_1 & & \\
 & \eta_1/\eta_2 & \\
  & & \eta_2\\
  \end{array}Ê\right)\hs \in\hs SL(3,\mbb{R})/SO(3).
  \eeq
We are here using slightly different variables compared to the previous subsection in order to make the comparison with the analysis in Section \ref{Section:SphericalVectorSL(3)} straightforward. Below we will also explain the relation with the variables used in Eq. (\ref{SL(3)PoincareSeriesVTvariablesExplicit}). 

We construct the $SO(3)$-invariant generalized metric
\beq
\mc{K}=\mc{V}\mc{V}^T=\left(\begin{array}{ccc}
\f{1}{\eta_1^2}+\f{\eta_1^2}{\eta_2^2}\mc{A}+\eta_2^2\mc{C}^2 & \f{\eta_1^2}{\eta_2^2}\mc{A}+\eta_2^2\mc{B}\mc{C} & \eta_2^2\mc{C} \\  
  \f{\eta_1^2}{\eta_2^2}\mc{A}+\eta_2^2\mc{B}\mc{C} & \f{\eta_1^2}{\eta_2^2}+\eta_2^2\mc{B}^2 & \eta_2^2\mc{B}\\
  \eta_2^2\mc{C} & \eta_2^2\mc{B} & \eta_2^2 \\
  \end{array}\right),
  \eeq
and we choose a lattice vector $m=(m_3, m_2, m_1)\in \mbb{Z}^3$ in the fundamental representation of $SL(3, \mbb{Z})$. Using Eq. (\ref{GeneralEisensteinSeries}) we then obtain the following Eisenstein series:
\beqa
\mc{E}^{SL(3, \mbb{Z})}(\mc{K};s)&=& \sum_{{m}\in \mbb{Z}^3}^{\quad \hs \prime} \Big[{m}^T\cdot \mc{M}\cdot {m}\Big]^{-s}
\nn \\
&=& \sum_{(m_1, m_2, m_3)\neq (0,0,0)} \Big[\f{m_3^2}{\eta_1^2}+\f{\eta_1^{2}}{\eta_2^2}\big(m_2+m_3\mc{A}\big)^2+\eta_2^2\big(m_1+m_2\mc{B}+m_3\mc{C}\big)^2\Big]^{-s}.
\nn \\
\label{SL(3,Z)MinimalEisensteinSeries1}
\eqa
To compare this result with Eq. (\ref{SL(3)PoincareSeriesVTvariablesExplicit}) we change variables to
\beq
\mc{A}=v, \quad \mc{B}=\mc{A}_2, \quad \mc{C}=\mc{A}_1, \quad \eta_1=y^{-1/6} u^{-1/2}, \quad \eta_2=y^{-1/3},
\eeq
after which (\ref{SL(3,Z)MinimalEisensteinSeries1}) becomes
\beq
\mc{E}^{SL(3,\mbb{Z})}(\mc{K}; s)=y^{-s/3}\sum_{m\in\mbb{Z}^3}^{\quad \hs \prime}  \Big[m_1^2 u+\f{1}{u}\big(m_2+m_1 v\big)^2+\f{1}{y}\big(m_1\mc{A}_1+m_2\mc{A}_2+m_3\big)^2\Big]^{-s}.
\label{SL(3,Z)MinimalEisensteinSeries2}
\eeq
Here we recognize the first summand of Eq. (\ref{SL(3)PoincareSeriesVTvariablesExplicit}), while the range of summation is different because of the extra coprime condition $(c_1,c_2,c_3)=1$ in (\ref{SL(3)PoincareSeriesVTvariablesExplicit}). This can be understood in the same way as in the $SL(2,\mbb{R})$-example in Section \ref{Section:SL(2)Eisenstein}, namely redefine $m_i\equiv k c_i, \hs i=1,2,3,$ with $(c_1,c_2,c_3)=1$ and $k=(m_1,m_2,m_3)$. After this change of summation variables, Eq. (\ref{SL(3,Z)MinimalEisensteinSeries2}) becomes
\beq
\mc{E}^{SL(3,\mbb{Z})}(\mc{K}; s)=2\zeta(2s) y^{-s/3}\sum_{(c_1,c_2,c_3)=1}^{\quad \hs \prime}  \Big[c_1^2 u+\f{1}{u}\big(c_2+c_1 v\big)^2+\f{1}{y}\big(c_1\mc{A}_1+c_2\mc{A}_2+c_3\big)^2\Big]^{-s}.
\label{SL(3,Z)MinimalEisensteinSeries3}
\eeq
The minimal Eisenstein series is an eigenfunction of the Laplacian on $SL(3,\mbb{R})/SO(3)$, with eigenvalue given by (minus) the quadratic Casimir:
\beq
\Delta_{SL(3,\mbb{R})/SO(3)}\mc{E}^{SL(3,\mbb{Z})}(\mc{K}; s)=-\mc{C}_2(s) \mc{E}^{SL(3,\mbb{Z})}(\mc{K}; s),
\eeq
where 
\beq
\mc{C}_2(s)= \frac23 s(2s-3).
\eeq
The Eisenstein series $\mc{E}^{SL(3,\mbb{Z})}(\mc{K}; s)$ is also an eigenfunction under the cubic Casimir operator with eigenvalue
\beq
\mc{C}_3(s)= s(2s-3)(4s-3).
\eeq
A characteristic feature of the minimal representation is further that the generators obey the equation 
\beq
\label{c3c2rel}
4\, \mc{C}_3^2 - 108\, \mc{C}_2^3 - 81\, \mc{C}_2^2 =0 .
\eeq

It is in fact possible to write the full principal series in terms of the method of Section \ref{Section:EisensteinLatticeConstruction}. In this case, one introduces two lattice vectors $m\in \mbb{Z}^3$ and $n\in\mbb{Z}^{3}$ and consider the following series \cite{AutomorphicMembrane}
\beq
\mc{E}^{SL(3,\mbb{Z})}(\mc{K}; s_1,s_2)=\sum_{(m,n)\in\mbb{Z}^6\atop m^T\cdot n=0}^{\quad \hs \prime} \left[m^{T}\cdot \mc{K}\cdot m\right]^{-s_1}\left[n^T \cdot \mc{K}^{-1} \cdot n\right]^{-s_2},
\label{principalEisenstein}
\eeq
where the summation excludes $(m_1,m_2,m_3)=(0,0,0)$ as usual. The relation with the principal series in Eq. (\ref{SL(3)PoincareSeriesVTvariables}) again becomes apparent after eliminating the coprime condition, which then yields a generalized version of Eq. (\ref{SL(2)Relation}):
\beq
\mc{E}^{SL(3,\mbb{Z})}(\mc{K}; s_1,s_2)=4\zeta(2s_1)\zeta(2s_2) E^{SL(3,\mbb{Z})}(g; s_1,s_2).
\eeq
The principal Eisenstein series (\ref{principalEisenstein}) converges absolutely for \cite{Bump}­
\beq
\Re(s_1)> 0, \qquad \Re(s_2)>0,
\eeq
and is an eigenfunction of the Laplacian on $SL(3,\mbb{R})/SO(3)$ with eigenvalue given by (minus) the value of the quadratic Casimir in the principal series of $SL(3,\mbb{R})$:
\beq
\Delta_{SO(3,\mbb{R})/SO(3)}\mc{E}^{SL(3,\mbb{Z})}(\mc{K}; s_1,s_2)=-\mc{C}_2\mc{E}^{SL(3,\mbb{Z})}(\mc{K}; s_1,s_2),
\eeq
with 
\beq
\mc{C}_2(s_1,s_2)=\frac43(s_1^2+s_2^2+s_1 s_2)-2(s_1+s_2).
\eeq
The principal Eisenstein series is of course also an eigenfunction of the cubic Casimir with eigenvalue
\beq
\mc{C}_3(s_1,s_2)=(s_1-s_2)(2s_2+4 s_1-3)(2s_1 + 4 s_2-3).
\eeq

By comparing the explicit structure of the minimal Eisenstein series (\ref{SL(3,Z)MinimalEisensteinSeries3}) with the principal series (\ref{SL(3)PoincareSeriesVTvariablesExplicit}), we deduce that the minimal Eisenstein series arises as the limiting case $s_2\rightarrow 0$, i.e.
\beq
\mc{E}^{SL(3,\mbb{Z})}(\mc{K}; s_1)=\lim_{s_2\rightarrow 0} \left[ 2\zeta(s_1) E^{SL(3,\mbb{Z})}(g; s_1, s_2)\right].
\eeq
It turns out that this is not the only limit in which the principal representation reduces to the minimal representation. The set of values of $(s_1, s_2)$ for which this happens is known as the ``polar divisor'' and are given explicitly by:
\beq
\label{linmin}
s_1=0\ ,\qquad s_1=1 \ ,\qquad s_2=0\ ,\qquad s_2=1 \ ,\qquad 
s_1+s_2=\frac12\ ,\qquad s_1+s_2=\frac32\ .
\eeq
This corresponds to the six lines in the $(s_1, s_2)$-plane where the relation (\ref{c3c2rel}) between the quadratic and cubic Casimirs is obeyed.

\subsection{Spherical Vector} 
\label{Section:SphericalVectorSL(3)}

In this section we will finally reproduce the minimal Eisenstein series in Eq. (\ref{SL(3,Z)MinimalEisensteinSeries1}) using the general approach of Section \ref{Section:SphericalVectorsConstruction}, based on the spherical vector $f_K$. As in the case of $SL(2, \mbb{R})$ above, we shall consider representations $\rho$ induced from the parabolic subgroup consisting of lower-triangular matrices 
  \beq
  P=\Bigg\{ \left(\begin{array}{ccc}
t_1 &  &  \\
\ast & t_2 &  \\
\ast   &\ast  & t_3 \\
  \end{array}Ê\right)\hs \big| \hs t_1t_2t_3=1\Bigg\}\subset SL(3,\mbb{R}).
  \label{SL(3)Parabolic}
  \eeq
A natural representative for $P\bas SL(3,\mbb{R})$ is given by
\beq
n=\left( \begin{array}{ccc}
1 & x & z \\
 & 1 & y \\
 & & 1Ê\\
 \end{array} \right)\hs \inÊ\hs  P\bas SL(3,\mbb{R}).
 \label{SL(3)/Prepresentative}
 \eeq  
We consider the representation $\rho$ of $g\in SL(3,\mbb{R})$ induced from $P$ which acts on functions $f(n)$ on $P\bas SL(3,\mbb{R})$ by
 \beq
 \rho(g)\cdot f(n)=f(ng)=f(pn^{\prime})=\chi_s(p)f(n^{\prime}), 
 \label{ParabolicInduction}
 \eeq
 where $\chi_s(p)$ is the character
 \beq
 \chi_s(p):=t_1^{-2s}.
 \label{character}
 \eeq
 We have here chosen the simplest possible character for the purpose of reproducing the minimal Eisenstein series $\mc{E}^{SL(3,\mbb{Z})}(\mc{K};s)$. The general principal series $\mc{E}^{SL(3,\mbb{Z})}(\mc{K};s_1,s_2)$ may also be constructed by induction from the parabolic subgroup $P$ in (\ref{SL(3)Parabolic}). However, carrying out the full program for this case is much more complicated and we will therefore refrain from doing so here. See however \cite{Pioline:2003bk,AutomorphicMembrane} for a discussion.
 
 The element $ng\in SL(3, \mbb{R})$ is generically not in the form of (\ref{SL(3)/Prepresentative}), but by a compensating transformation of $p\in P$ from the left we obtained $ng=p n^{\prime}$, with $n^{\prime}\in P\bas SL(3, \mbb{R})$. The specific choice of character in (\ref{character}) corresponds to a certain limit of the principal continuous series of $SL(3, \mbb{R})$ representation (see \cite{AutomorphicMembrane}). In fact, it restricts from the reducible representation acting on functions $f(x,y,z)$ of three variables to an irreducible representation acting on functions of two variables, $f(x,z):=f(x,1,z)$, i.e. to the minimal representation. The spherical vector $f_K(x,z)$ is defined by the condition
 \beq
 \rho(k)\cdot f_K(x,z)=f_K(x,z), \qquad k\in SO(3).
 \eeq 
 A simple way of determining $f_K(x,z)$ is to consider the action of $SO(3)$ on $n\in P\bas SL(3,\mbb{R})$. $SO(3)$ acts from the right on $n$ and a compensating $P$-transformation from the left is needed to restore the upper-triangular form. The action of $p\in P$ on $n$ is 
 \beq
 pn=\left(\begin{array}{ccc}
 t_1 & & \\
 p_{21} & t_2 &  \\
 p_{31} & p_{32}Ê& t_3 \\
 \end{array}Ê\right)\left(\begin{array}{ccc}
 1 & x & z\\
  & 1 & y \\
  & Ê& 1 \\
 \end{array}Ê\right)=\left(\begin{array}{ccc}
 t_1 & t_1x & t_1z\\
 p_{21} & t_2+p_{21}x & t_2y+ p_{21}z \\
 p_{31} & p_{32}+p_{31}x Ê& t_3+p_{32}y+p_{31}z \\
 \end{array}Ê\right).
 \label{actionofP}
 \eeq
Since the action of $K=SO(3)$ on $n$ is just a rotation, it preserves the Euclidean norm of each row $\vec{r}_i$ of $n$. If we act by $k\in K$ on $n$ from the right we will however destroy the upper triangular form of $n$, and a compensating transformation by $p\in P$ from the left is required. Moreover, we see that the action of $P$ on the top row $\vec{r}_1=(1,x,z)$ is very simple: $p\cdot \vec{r}_1=t_1\vec{r}_1$. The spherical vector $f_K(x, z)$, in the minimal representation, can therefore be taken as the Euclidean norm of the vector $\vec{r}_1$ raised to the power $-2s$:
\beq
f_K(x,z):=||\vec{r}_1||^{-2s}=(1+x^2+z^2)^{-s}.
\eeq
The action of $k \in SO(3)$ will transform $f_K$ by an overall factor $t_1^{2s}$ because of the compensating $P$-transformation in \ref{actionofP}, but since $f_K$ belongs to the principal continuous series we must also include the character $\chi_s(p)=t_1^{-2s}$ from (\ref{character}), implying that $f_K(x,z)$ is indeed invariant.  

To find the action of $\mc{V}\in SL(3,\mbb{R})/SO(3)$ on the spherical vector in this representation, we simply apply the formula (\ref{ParabolicInduction}) on $f_K$, which yields
\beq 
\rho(\mc{V})\cdot  f_K(n)=f_K(n\mc{V})=f_K(p_0n^{\prime})=\chi_s(p_0)f_K(n^{\prime})=\eta_1^{2s}||\vec{r}_1^{\hs \prime}||^{-2s},
\eeq
where 
\beqa 
n^{\prime}&=&\left( \begin{array}{ccc}
1 & \f{\eta_1^2}{\eta_2}\big(x+\mc{A}\big) & \eta_1\eta_2\big(z+x\mc{B}+\mc{C}\big) \\
 & 1 & \f{\eta_2^2}{\eta_1}\big(y+\mc{B}\big) \\
 & & 1Ê\\
 \end{array} \right), 
 \nn \\
   p& =& \left(\begin{array}{ccc}
 1/\eta_1 & & \\
& \eta_1/\eta_2 &  \\
&Ê& \eta_2 \\
 \end{array}Ê\right).
 \label{nprime}
 \eqa
The transformed spherical vector thus corresponds to the norm squared $||{\vec{r}_1}^{\hs \prime}||^{-2s}$ of the top row $\vec{r}_1^{\hs \prime}$ of the transformed matrix $n^{\prime}\in P\bas SL(3, \mbb{R})$. In addition, we must include the overall character $\eta_1^{2s}$ and the final result is
\beq
\rho(\mc{V})\cdot f_K(x,z)=\eta_1^{2s}\Big[1+\f{\eta_1^{4}}{\eta_2^2}\big(x+\mc{A}\big)^2+\eta_1^2\eta_2^2\big(z+x\mc{B}+\mc{C}\big)^2\Big]^{-s}.
\eeq
The $p$-adic spherical vector is again obtained by replacing the Euclidean norm $||\cdot ||$ in $f_K(x, z)$ by its $p$-adic version, which yields
\beq
f_p(x,z):=|1,x,z|_p^{-2s}=\text{max}\big(1,|x|_p, |z|_p\big)^{-2s}.
\eeq
At this stage, the automorphic form $\Psi(\mc{V})$ in (\ref{AdelicAutomorphicForm}) for the minimal representation of $SL(3, \mbb{R})$ thus reads
\beqa
\Psi(\mc{V})= \sum_{(x,z)\in\mbb{Q}^{2}}\Big[\prod_{p < \infty}\text{max}\big(1,|x|_p, |z|_p\big)^{-2s} \Big]\Big[\f{1}{\eta_1^2}+\f{\eta_1^{2}}{\eta_2^2}\big(x+\mc{A}\big)^2+\eta_2^2(z+x\mc{B}+\mc{C}\big)^2\Big]^{-s}.
\nn \\
\label{SL(3)pAdicAutomorphicForm}
\eqa 
 To proceed, we must again evaluate the infinite product over primes. This can be done by the same method as for the $SL(2, \mbb{R})$-example in the previous subsection. We begin by writing the rational summation variables $x$ and $z$ as follows:
 \beq
 x=\f{m_2}{m_3}, \qquad z=\f{m_1}{m_3},
 \eeq
 where $m_i\in \mbb{Z}$ ($m_3\neq 0$), and $(m_1, m_2, m_3)=1$. By the same reasoning as before the infinite product over the $p$-adic spherical vector becomes
\beq
\prod_{p<\infty}\text{max}\Big(1, \Big|\f{m_1}{m_3}\Big|_p, \Big|\f{m_2}{m_3}\Big|_p\Big)^{-2s}=m_{3}^{-2s}.
\eeq
Inserting this result into Eq. (\ref{SL(3)pAdicAutomorphicForm}) then yields
 \beq
\Psi(\mc{V})=  \sum_{(m_1, m_2, m_3)=1\atop m_3\neq 0} \Big[\f{m_3^2}{\eta_1^2}+\f{\eta_1^{2}}{\eta_2^2}\big(m_2+m_3\mc{A}\big)^2+\eta_2^2\big(m_1+m_2\mc{B}+m_3\mc{C}\big)^2\Big]^{-s}.
 \eeq
This result agrees with the minimal  limit $(s_1,s_2)\rightarrow (s,0)$ of the principal Eisenstein series in (\ref{SL(3)PoincareSeriesVTvariablesExplicit}), up to the missing term in the sum with $m_3=0$. 

\chapter{Fourier Expansion of Automorphic Forms}
\label{Chapter:FourierExpansion}
We have now learned how to construct $G(\mbb{Z})$-invariant automorphic forms on a symmetric space  $G/K$ in a variety of different ways. The remaining task is then to extract the physical information that they contain. For the simplest case of $G(\mbb{Z})=SL(2,\mbb{Z})$ we have seen in Chapter \ref{Chapter:Aspects} that this is done through Fourier expansion, which recasts the automorphic form into an infinite sum, revealing perturbative and non-perturbative quantum effects. In this chapter we shall discuss in more detail how to compute such Fourier expansions, again with particular emphasis on the special examples of $SL(2,\mbb{Z})$ and $SL(3,\mbb{Z})$. We begin in Section \ref{Section:GeneralFourierExpansion} by explaining the main philosophy, stressing certain key points, while in Section \ref{Section:FourierExpansionSL(2)} and \ref{Section:FourierExpansionSL(3)} we consider explicit examples. Section \ref{Section:FourierExpansionSL(3)} is based on {\bf Paper VII}, written in collaboration with Boris Pioline. 
\section{General Considerations}
\label{Section:GeneralFourierExpansion}
In this section we shall discuss some general features of the Fourier expansion of Eisenstein series $E^{G(\mbb{Z})}(g; \lambda_{\mc{R}})$, constructed as in Section \ref{Section:PoincareSeries}. We will be deliberately schematic in order to emphasize the key issues. Details are relegated to the examples in the following sections. We begin by noting the important fact that by virtue of the Iwasawa decomposition $g=nak\in G$, the function $E^{G(\mbb{Z})}(g; \lambda_{\mc{R}})$ is by construction (left) invariant under the nilpotent subgroup $N(\mbb{Z})\subset G(\mbb{Z})$, 
\beq
 E^{G(\mbb{Z})}(x g; \lambda_{\mc{R}})=E^{G(\mbb{Z})}(g; \lambda_{\mc{R}}), \qquad x\in N(\mbb{Z}).
 \label{NilpotentInvariance}
 \eeq
 To understand the implications of this fact, we parametrize the element $n\in N$ in the standard way (see Chapter \ref{Chapter:Manifest} where this was called the ``Borel gauge'')
 \beq
 n = \exp\bigg[ \sum_{\al\in \Phi_+} \chi_{\al} E_{\al}\bigg] \hs \in\hs  N,
 \label{NilpotentParametrization}
 \eeq
 where $\Phi_+$ denotes the positive root system of the Lie algebra $\mf{g}=\text{Lie}\ G$ (see Section \ref{Section:rootsystem}), and $E_{\al}\in \mf{g}_\al$ are the positive root generators. We shall refer to the parameters $\chi_{\al}$ as  ``axions'', by analogy with the axion $\chi$ in the type IIB example of Chapter \ref{Chapter:Aspects}. In this parametrization, it is clear that the left action of $x \in N(\mbb{Z})$ on $g=nak$ will leave the ``dilatonic'' parameters $\phi$ of $a\in A$ invariant, while acting as integer-valued shifts on the axions:
 \beq
 N(\mbb{Z})\hs \ni \hs x \hs : \hs \chi_{\al} \longmapsto \hs \chi_{\al} + \ell_{\al},
 \label{GeneralAxionShift}
 \eeq
 where $\ell_{\al}\in \mbb{Z}$ are the corresponding parameters of $x\in N(\mbb{Z})$. This expression is somewhat schematic but nevertheless illustrates the main point. By comparing (\ref{GeneralAxionShift}) with (\ref{NilpotentParametrization}), we then deduce that $E^{G(\mbb{Z})}(g; \lambda_{\mc{R}})$ exhibits a \emph{periodicity} in the set of axionic moduli $\{ \chi_{\al}\}$, schematically indicated as follows
 \beq
 E^{G(\mbb{Z})}\Big(\phi, \{\chi_{\al}\}; \lambda_{\mc{R}}\Big)=E^{G(\mbb{Z})}\Big(\phi, \{\chi_{\al}+\ell_{\al}\}; \lambda_{\mc{R}}\Big),
 \eeq
 where we collectively denoted the Cartan parameters by $\phi$. This periodicity ensures that $E^{G(\mbb{Z})}(x g; \lambda_{\mc{R}})$ has a Fourier expansion of the general form
 \beq
 E^{G(\mbb{Z})}( g; \lambda_{\mc{R}})=\sum_{\ell_{\al}\in \mbb{Z}^{D}} \mf{C}_{\ell_{\al}}\big(\phi\big) \exp\Big[ 2\pi i \sum_{\al\in \Phi_+} \ell_{\al} \chi_{\al}\Big],
 \label{MoralFourierExpansion}
 \eeq
 where $D=\mathrm{dim}\ \mathrm{Lie}\ N$. From this expression it is clear that the Fourier expansion essentially corresponds to a diagonalization of the action of the nilpotent subgroup $N(\mbb{Z})\subset G(\mbb{Z})$. 
 
% Equivalently, it may be viewed as a decomposition of the Hilbert space $L^2(B_-\bas G)$, where $B_-=N_- A$ is the negative Borel subgroup, into irreducible representations of the nilpotent group $N$.
 
 Although morally true, in the discussion above we have actually suppressed the main subtlety: the nilpotent group $N$ is generically not abelian and therefore can not be diagonalized according to the general form indicated in (\ref{MoralFourierExpansion}). This implies that in the general case when $N$ is non-abelian, Eq. (\ref{GeneralAxionShift}) is actually \emph{wrong}. To remedy this, one must decompose the Fourier expansion into an \emph{abelian} part and a \emph{non-abelian} part:
 \beq
 E^{G(\mbb{Z})}( g; \lambda_{\mc{R}})=E_{A}(g)+E_{NA}(g),
 \label{NonabelianSplit}
 \eeq
 where the abelian term corresponds to the Fourier expansion with respect to the ``abelianized'' nilpotent group $\tilde{N}\equiv N/Z$. Here $Z$ denotes the center of $N$. The non-abelian part $E_{NA}(g)$ then represents the remaining decomposition with respect to the center $Z$. We will see in Section \ref{Section:FourierExpansionSL(3)} that for $G=SL(3,\mbb{R})$, $N$ is isomorphic to a three-dimensional Heisenberg group, and in this case the center $Z$ is one-dimensional, corresponding to the commutator subgroup $[N,N]$. 
 
Let us proceed to write the correct form of (\ref{GeneralAxionShift}) when $N$ is non-abelian. To this end, let $\tilde{\chi}_i,\hs i=1, \dots, \tilde{D}=\text{dim}\ \text{Lie}\ \tilde{N}$ denote the parameters of the abelianized group $\tilde{N}$, and let $z_a, \hs a=1, \dots, d=\text{dim}\ \text{Lie}\ Z$ denote the parameters of the center $Z$. The abelianized discrete group $\tilde{N}(\mbb{Z})$ then acts on the scalars $\tilde{\chi}_i$ in the same way as in (\ref{GeneralAxionShift}),
 \beq
 \tilde{N}(\mbb{Z}) \hs : \hs \tilde{\chi}_{i}\hs \longmapsto \hs \tilde{\chi}_{i} + \tilde{\ell}_{i},
 \label{abelianAxionShift}
 \eeq
while the center $Z$ leaves all $\tilde{\chi}_i$ invariant. On the other hand, the full group $N(\mbb{Z})=\tilde{N}(\mbb{Z})\ltimes Z(\mbb{Z})$ acts non-trivially on the scalars $z_a$:
\beq
\tilde{N}(\mbb{Z})\ltimes Z(\mbb{Z}) \hs :\hs z_a \hs \longmapsto \hs z_a+k_a +\sum_{i,j=1}^{\tilde{D}}c_{ij} \tilde{\ell}_i \tilde{\chi}_j, \qquad k_a\in \mbb{Z},
\label{CenterAxionicShift}
\eeq
 for some numerical coefficients $c_{ij}$ which depend on the structure of the group $N$. As a consequence of (\ref{CenterAxionicShift}), the Fourier expansion will no longer display a complete separation of variables between the Cartan parameters $\phi$ and the axions $\chi_{\al}$, in contrast to the abelian case (\ref{MoralFourierExpansion}). The correct form of the non-abelian Fourier expansion then reads
 \beqa
 E^{G(\mbb{Z})}( g; \lambda_{\mc{R}})&=&\phantom{+}\sum_{\tilde{\ell}\in \mbb{Z}^{\tilde{D}}} \mf{C}_{\tilde{\ell}}\big(\phi\big) \exp\left[ 2\pi i \sum_{i=1}^{\tilde{D}} \tilde{\ell}_i \tilde{\chi}_i\right]
 \nn \\
 & & +\sum_{k \in \mbb{Z}^{d}} \mf{C}_\ell\big(\phi, \tilde{\chi}_1, \dots , \tilde{\chi}_{\tilde{D}}\big) \exp\left[2\pi i \sum_{a=1}^{d} k_a z_a\right].
 \label{NonabelianFourierExpansionGeneralForm}
 \eqa
We emphasize that in the second (non-abelian) term the coefficients $\mf{C}_{\ell}\big(\phi, \tilde{\chi}_i\big)$ are invariant under the center $Z$ of $N$, while transforms with a non-trivial phase factor under the abelianized group $\tilde{N}$, which compensates for the transformation of the $z_a$'s.
 
Let us conclude this section with a brief discussion of the ``perturbative'' part of the Fourier expansion corresponding to the zeroth Fourier coefficients $\mf{C}_{0}$ in (\ref{NonabelianFourierExpansionGeneralForm}). In mathematical terminology, these are known as the \emph{constant terms} and effectively correspond to integrating out all of the axionic scalars:
\beqa
\mf{C}_{0}\big(\phi\big)& =& \int_{N(\mbb{Z})\bas N} E^{G(\mbb{Z})}( g; \lambda_{\mc{R}}) dn
\nn \\
&=& \int_{0}^{1} d\tilde{\chi}_1\cdots \int_0^{1}\tilde{\chi}_{\tilde{D}} \int_0^{1} d z_1\cdots \int_0^{1} d z_d\ E^{G(\mbb{Z})}\big(\phi, \tilde{\chi}_i, z_a ; \lambda_{\mc{R}}\big).
\nn \\
\eqa
Being completely independent of the axionic scalars, these are precisely the terms which encode the perturbative contributions to the Fourier expansion, analogously to the tree-level and one-loop term in Eq. (\ref{GreenGutperleExpansion}). See \cite{Green:2010wi} for many explicit constant term calculations related to string theory. 

We turn now to explicit examples of Fourier expansions, starting with simplest case of $G(\mbb{Z})=SL(2,\mbb{Z})$, in which case $N$ is abelian and the complications discussed above are absent. However, in the example of $SL(3,\mbb{Z})$ considered in Section \ref{Section:FourierExpansionSL(2)} we will explicitly see the structure of (\ref{NonabelianFourierExpansionGeneralForm}) emerging. In Chapter \ref{Chapter:UniversalHypermultiplet} we will also analyze in detail the non-abelian Fourier expansion of Eisenstein series on $SU(2,1)$. 
 
 %\subsection{Abelian Fourier Expansion}
\section{Fourier Expansion of $\mc{E}^{SL(2,\mbb{Z})}(\tau; s)$}
\label{Section:FourierExpansionSL(2)}
As our first example, we shall compute in detail the Fourier expansion of the non-holomorphic Eisenstein series $\mc{E}^{SL(2,\mbb{Z})}(\tau; s)$ discussed in Section \ref{Section:D(-1)Example}. We begin by analyzing the general structure of the expansion by utilizing the fact that $\mc{E}^{SL(2,\mbb{Z})}(\tau; s)$ is an eigenfunction of the Laplacian on $SL(2,\mbb{R})$. In Section \ref{Section:SL(2)PoissonResummation} we then use Poisson resummation to explicitly evaluate the Fourier coefficients.
\subsection{General Structure}

Recall that the Eisenstein series $\mc{E}^{SL(2,\mbb{Z})}(\tau; s)$ takes the form
\beq
\mc{E}^{SL(2, \mbb{Z})}(\tau;s)=\sum_{(m, n)\neq (0,0)} \f{\tau_2^s}{|m+n\tau|^{2s}},
\label{SL(2)Eisenstein}
\eeq
where $\tau=\tau_1+i\tau_2$ parametrizes the coset space space $SL(2,\mbb{R})/SO(2)$ via the standard map discussed in Section \ref{Section:SL(2)Eisenstein}. We have also seen in Section \ref{Section:SL(2)Eisenstein} that $\mc{E}^{SL(2,\mbb{Z})}(\tau; s)$ is by construction invariant under the discrete subgroup $SL(2,\mbb{Z})$ which acts in the following way on the coordinate $\tau$:
\beq
\tau \longmapsto \f{a\tau+b}{c\tau+d}, \qquad a, b, c, d\in \mbb{Z}, \quad ad-bc=1.
\eeq
This group is generated by two transformations $S$ and $T$ (see, e.g. \cite{Serre}, and Section \ref{Section:WeylGroupA1++}), acting on $\tau$ as
\beq
S\hs  : \hs \tau\hs  \longmapsto\hs  -\f{1}{\tau}, \qquad  T\hs :\hs \tau\hs \longmapsto\hs \tau+1.
\eeq
Let us for a moment focus on the shift transformation $T$. Notice that this leaves invariant the imaginary part of $\tau$, while shifts the real part according to $\tau_1\longmapsto \tau_1+1$. Since the Eisenstein series is invariant under the entire group $SL(2, \mbb{Z})$ it will in particular satisfy
\beq
\mc{E}^{SL(2, \mbb{Z})}(\tau_1, \tau_2; s)=\mc{E}^{SL(2, \mbb{Z})}(\tau_1+1, \tau_2; s),
\eeq
implying that it is periodic in the variable $\tau_1$ with period $1$, in accordance with the general discussion of the previous section. This ensures that $\mc{E}^{SL(2, \mbb{Z})}(\tau; s)$ has a Fourier expansion of the form
\beq
\mc{E}^{SL(2,\mbb{Z})}(\tau; s)=\sum_{N\in \mbb{Z}}\mf{C}_N(\tau_2) e^{2\pi i N \tau_1},
\eeq
where the Fourier coefficients $\mf{C}_N(\tau_2)$ do not depend on the real part of $\tau$. It is useful to separate out the zeroth Fourier coefficient $\mf{C}_0$ and write 
\beq
\mc{E}^{SL(2,\mbb{Z})}(\tau; s)=\mf{C}_0(\tau_2)+\sum_{N\neq 0} \mf{C}_N(\tau_2)e^{2\pi i N\tau_1}.
\label{GeneralFourierExpansionSL(2)}
\eeq
The zeroth coefficient corresponds to the \emph{constant term} 
\beq
\mf{C}_N(\tau_2)=\int_{N(\mbb{Z})\bas N} \mc{E}^{SL(2,\mbb{Z})}(\tau; s) dn=\int_0^{1} \mc{E}^{SL(2,\mbb{Z})}(\tau; s) d\tau_1,
\eeq
as discussed in general in Section \ref{Section:GeneralFourierExpansion}. Now recall further that the Eisenstein series is an eigenfunction of the Laplacian on $SL(2, \mbb{R})/SO(2)$:
\beq
\Delta \mc{E}^{SL(2,\mbb{Z})}(\tau;s)=s(s-1)  \mc{E}^{SL(2,\mbb{Z})}(\tau;s),
\label{LaplacianConditionSL(2)}
\eeq
with 
\beq
\Delta=\tau_2^2 \Big(\f{\pa^2}{\pa \tau_1^2}+\f{\pa^2}{\pa \tau_2^2}\Big).
\eeq
This Laplacian condition on $\mc{E}^{SL(2,\mbb{Z})}(\tau;s)$ induces differential equations for the coefficients $\mf{C}_N(\tau_2)$ which fixes their dependence on $\tau_2$. Let us consider the constant term first. Since this term is independent of $\tau_1$, it must obey the equation
\beq
\Delta_{\mc{R}} \mf{C}_0(\tau_2)= s(s-1) \mf{C}_0(\tau_2),
\label{radialequationSL(2)}
\eeq
where we defined the ``radial part'' part of the Laplacian 
\beq
\Delta_{\mc{R}}:= \tau_2^2\f{\pa^2}{\pa \tau_2^2}.
\eeq
The terminology refers to the fact that this is the Laplacian only along the directions of the Cartan subalgebra of $SL(2, \mbb{R})$, corresponding roughly to the ``radius'' of the hyperbolic space $SL(2,\mbb{R})/SO(2)$. From Equation (\ref{radialequationSL(2)}) we deduce that the constant term is of the form
\beq
\mf{C}_0(\tau_2)=A(s) \tau_2^{s}+B(s)\tau_2^{1-s},
\eeq
where $A(s)$ and $B(s)$ are unknown functions that remains to be determined. These are fixed by $SL(2,\mbb{Z})$-invariance of the Fourier expansion, as we shall see explicitly below. Let us now proceed to analyze the remaining Fourier coefficients $\mf{C}_N(\tau_2)$ for $N\neq 0$. Inserting the bulk part of the general Fourier expansion (\ref{GeneralFourierExpansionSL(2)}) into the Laplacian equation (\ref{LaplacianConditionSL(2)}) yields the following equation for the coefficients:
\beq
\left[\tau_2^2\f{\pa^2}{\pa \tau_2^2}-4\pi^2 N^2 \tau_2^2-s(s-1)\right] \mf{C}_N(\tau_2)=0.
\eeq
To solve this equation, it is convenient to redefine $\mf{C}_N(\tau_2)\equiv \sqrt{\tau_2}\ \tilde{\mf{C}}_N(\tau_2)$, after which the equation for $\tilde{\mf{C}}_N$ is identified with a Bessel equation, with solution given by the modified Bessel function:
% which yields
%\beq
%\left[\tau_2^2\f{\pa^2}{\pa \tau_2^2}+\f{1}{2}\tau_2^{3/2}\f{\pa}{\pa\tau_2}-4\pi^2 N^2\tau_2^{5/2}-\Big(s(s-1)+\f{1}{4}\Big)\sqrt{\tau_2}\right]\tilde{\mf{C}}_N(\tau_2)=0.
%\eeq 
%This may now be recognized as a Bessel equation, with a solution given in terms of the modified Bessel function as follows
\beq
\tilde{\mf{C}}_N(\tau_2)=a(N) K_{s-1/2}\big(2 \pi |N| \tau_2\big),
\eeq
 where $a(N)$ is some $\tau_2$-independent prefactor. Recalling that $ \mf{C}_N(\tau_2)=\sqrt{\tau_2}\tilde{\mf{C}}_N(\tau_2)$, we then find that the Fourier expansion (\ref{GeneralFourierExpansionSL(2)}) becomes
 \beq
 \mc{E}^{SL(2, \mbb{Z})}(\tau;s)=A(s) \tau_2^{s}+B(s)\tau_2^{1-s}+\sqrt{\tau_2}\sum_{N\neq 0} a(N) K_{s-1/2}\big(2 \pi |N| \tau_2\big) e^{2\pi i N \tau_1}.
 \label{FourierAlmost}
 \eeq
The remaining coefficients $A(s),  B(s)$ and $a(N)$ contain the missing information required to make this expression $SL(2,\mbb{Z})$-invariant. To determine these unknowns, we shall now consider the manifestly $SL(2, \mbb{Z})$-invariant expression (\ref{SL(2)Eisenstein}) and explicitly compute its Fourier expansion through Poisson resummation techniques. This will cast (\ref{SL(2)Eisenstein}) into the form (\ref{FourierAlmost}), allowing us to deduce the missing pieces. 

\subsection{Explicit Fourier Coefficients from Poisson Resummation}
\label{Section:SL(2)PoissonResummation}

We follow the general method described in \cite{ObersPioline}. Starting from (\ref{SL(2)Eisenstein}), the first step is to extract the term with $n=0$ from the sum, which yields
\beq
\mc{E}^{SL(2, \mbb{Z})}(\tau; s)=2\zeta(2s)\tau_2^{s} +\sum_{m\in\mbb{Z}}\sum_{n\neq 0}\f{\tau_2^{s}}{|m+n\tau|^{2s}},
\label{Eisenstein2}
\eeq
where the Riemann zeta function is defined as follows
\beq
\zeta(t):=\sum_{m=1}^{\infty}\f{1}{m^{t}}.
\label{zeta}
\eeq
By comparing this with (\ref{FourierAlmost}) we may already deduce that $A(s)=2\zeta(2s)$. 
Note further that since $n\neq 0$ in (\ref{Eisenstein2}) the sum over $m$ is unrestricted. To proceed, it is convenient to employ the following integral representation of the summand 
\beq
\f{1}{|m+n\tau|^{2s}}=\f{\pi^s}{\Gamma(s)}\int_0^{\infty}\f{dt}{t^{s+1}}\ e^{-\f{\pi}{t}|m+n\tau|^2},
\label{Integralrepresentation}
\eeq
and we expand the argument of the exponential as follows
\beq
|m+n\tau|^2=(m+n\tau)(m+n\bar{\tau})=(m+n\tau_1)^2+n^2\tau_2^2.
\eeq
Inserting this into (\ref{Eisenstein2}) then yields
\beq
\mc{E}^{SL(2,\mbb{Z})}(\tau; s)=2\zeta(2s)\tau_2^s+\f{\pi^s \tau_2^s}{\Gamma(s)}\sum_m\sum_{n\neq 0}\int_0^{\infty}\f{dt}{t^{s+1}}e^{-\f{\pi}{t}\big[(m+n\tau_1)^2+n^2\tau_2^2\big]}.
\label{Eisenstein3}
\eeq
Since the sum over $m$ is unrestricted we may perform the following Poisson resummation:
\beq
\sum_m e^{-\f{\pi }{t}(m+n\tau_1)^2}=\sqrt{t}\sum_{\tilde{m}}e^{-\pi t \tm^2-2\pi i\tm n \tau_1}.
\eeq
Inserting this into (\ref{Eisenstein3}) ensures that the term involving $\tau_1$ is independent of the integration variable $t$, and we obtain
\beq
{}Ê\mc{E}^{SL(2,\mbb{Z})}(\tau; s)=2\zeta(2s)\tau_2^s+\f{\pi^s \tau_2^s}{\Gamma(s)}\sum_{\tm}\sum_{n\neq 0}e^{-2\pi i \tm n \tau_1}\int_0^{\infty}\f{dt}{t^{s+1/2}}e^{-\pi t \tm^2-\f{\pi}{t}n^2\tau_2^2}.
\eeq
We now extract the $\tm=0$ part of the sum, which allows us to determine the coefficient $B(s)$ in (\ref{FourierAlmost}),
\beqa
{}Ê\mc{E}^{SL(2,\mbb{Z})}(\tau; s)&=&2\zeta(2s)\tau_2^s+2\sqrt{\pi}\f{\Gamma(s-\f{1}{2})}{\Gamma(s)}\zeta(2s-1)\tau_2^{1-s}
\nn \\
{}Ê&+& \f{\pi^s\tau_2^s}{\Gamma(s)}\sum_{\tm\neq 0}\sum_{n\neq 0}e^{-2\pi i \tm n\tau_1}\int_0^{\infty}\f{dt}{t^{s+1/2}}e^{-\pi t \tm^2-\f{\pi}{t}n^2\tau_2^2},
\eqa
where we again made use of (\ref{zeta}) and (\ref{Integralrepresentation}). The remaining integral is of Bessel type, and may be evaluated as follows 
\beq \int_0^{\infty}\f{dt}{t^{s+1/2}}e^{-\pi t \tm^2-\f{\pi}{t}n^2\tau_2^2}=2\left|\f{\tm}{n}\right|^{s-1/2}\tau_2^{1/2-s}K_{s-1/2}(2\pi|\tm n| \tau_2),
\eeq
which yields
\beqa
{}Ê\mc{E}^{SL(2,\mbb{Z})}(\tau; s)&=&2\zeta(2s)\tau_2^s+2\sqrt{\pi}\f{\Gamma(s-\f{1}{2})}{\Gamma(s)}\zeta(2s-1)\tau_2^{1-s}
\nn \\
{}Ê& & +\f{2\pi^s\sqrt{\tau_2}}{\Gamma(s)}\sum_{\tm\neq 0}\sum_{n \neq 0} \left|\f{\tm}{n}\right|^{s-1/2}K_{s-1/2}(2\pi |\tm n|\tau_2)e^{-2\pi i \tm n\tau_1}.
\nnÊ\\
\eqa
To make contact with (\ref{FourierAlmost}) we further change variables in the sum to $N\equiv -\tm n$. Replacing the sum over $\tm$ by a sum over $N\in\mbb{Z}\bas \{0\}$ puts a restriction on the sum over $n$ which enforces the constraint $\tm=N/n\in\mbb{Z}\bas \{0\}$. Effectively this implies that the sum over $n$ becomes a sum over divisors of $N$, written as $\sum_{n|N}$. We thus have
\beq
\sum_{\tm\neq 0}\sum_{n\neq 0}\left|\f{\tm}{n}\right|^{s-1/2}=\sum_{N\neq 0}\sum_{n|N}N^{s-1/2} n^{1-2s}\equiv \sum_{N\neq 0}N^{s-1/2} \mu_{1-2s}(N),
\eeq 
where we defined the quantity 
\beq
\mu_{1-2s}(N):= \sum_{n|N}n^{1-2s},
\label{divisorsum}
\eeq
which is known as the \emph{instanton measure} \cite{GreenGutperle,Yi,SethiStern,GreenGutperleInstantonMeasure} (see Section \ref{Section:D(-1)Example}). The final form of the Fourier expansion is therefore
\beqa
{}Ê\mc{E}^{SL(2,\mbb{Z})}(\tau; s)&=&2\zeta(2s)\tau_2^s+2\sqrt{\pi}\f{\Gamma(s-\f{1}{2})}{\Gamma(s)}\zeta(2s-1)\tau_2^{1-s}
\nn \\
{}Ê& & +\f{2\pi^s\sqrt{\tau_2}}{\Gamma(s)}\sum_{N\neq 0}\mu_{1-2s}(N) N^{s-1/2}K_{s-1/2}(2\pi |N|  \tau_2)e^{2\pi i N\tau_1}.
\nnÊ\\
\eqa
By comparing this expression with (\ref{FourierAlmost}) we finally deduce that the remaining Fourier coefficient $a(N)$ is given by
\beq
a(N)=\f{2\pi^s}{\Gamma(s)}N^{s-1/2} \mu_{1-2s}(N).
\eeq
Let us finally note that we may rewrite this in a slightly more compact way in terms of the so called \emph{completed zeta function}:
\beq
\xi(s):=\pi^{-s/2} \Gamma(s/2) \zeta(s),
\label{CompletedZetaFunction}
\eeq
which obeys the nice functional relation
\beq \xi(s)=\xi(1-s).
\label{CompletedZetaFunctionalRelation}
\eeq
Extracting an overall factor of $2\zeta(2s)$, we may then write the Fourier expansion of the Poincar\'e series discussed in Section (\ref{Section:SL(2)Eisenstein}) in the following way:
\beqa
{}ÊE^{SL(2,\mbb{Z})}(\tau;s)&=& \sum_{\ga \in N(\mbb{Z})\bas SL(2, \mbb{Z})}\big(\ga\cdot \tau_2\big)^{s}\nn \\
&=&\tau_2^s+\f{\xi(2s-1)}{\xi(2s)}\tau_2^{1-s}
 +\f{\sqrt{\tau_2}}{\xi(2s)}\sum_{N\neq 0}\mu_{1-2s}(N) N^{s-1/2}K_{s-1/2}(2\pi |N|  \tau_2)e^{2\pi i N\tau_1}.
\nn \\
\label{FinalFourierSL(2)}
\eqa
Let us finally mention that the functional relation (\ref{CompletedZetaFunctionalRelation}) of the completed zeta function $\xi(s)$ can be extended to the following nice functional relation between Eisenstein series of different order \cite{Langlands}:
\beq
\xi(s)E^{SL(2,\mbb{Z})}(\tau; (1-s)/2)=\xi(-s)E^{SL(2,\mbb{Z})}(\tau; (1+s)/2) .
\eeq

\subsection{Spherical Vector Construction of the Fourier Expansion}
\label{Section:SphericalVectorFourierSL(2)}

From the general formula (\ref{AdelicAutomorphicForm}) it is clear that the two important building blocks of any automorphic form are the spherical vector $f_K$ and the $p$-adic counterpart $f_p$. For the case of $SL(2,\mbb{Z})$, we have seen in Section \ref{Section:SL(2)Eisenstein} that the infinite product over prime numbers in (\ref{AdelicAutomorphicForm}) can be evaluated and gives rise to the correct form of the Eisenstein series $E^{SL(2,\mbb{Z})}(\tau;s)$. It turns out that also the bulk part of the Fourier expansion of $E^{SL(2,\mbb{Z})}(\tau;s)$ may be recast into this form, by identifying a new spherical vector corresponding to the Fourier transform of $f_K(x)=(1+x^2)^{-s}$. Unfortunately the constant terms do not follow from this construction, but may in principle be obtained by enforcing invariance under the Weyl group of $SL(2,\mbb{R})$ \cite{Kazhdan:2001nx,Pioline:2003bk}. 

We begin by noting that, as a consequence of the $p$-adic norm, the sum over primes in (\ref{AdelicAutomorphicForm}) has only support on the integers. Thus, Eq. (\ref{AdelicAutomorphicForm}) can be rewritten as follows
\beq
E^{SL(2,\mbb{Z})}(\tau;s)=\sum_{x\in \mbb{Z}}^{\quad \hs \prime}f_{\mbb{Z}}(x) \rho(\mc{V})\cdot f_K(x),
\label{SL(2)pAdic}
\eeq
where $f_{\mbb{Z}}$ is the $SL(2,\mbb{Z})$-invariant distribution discussed in Section \ref{Section:SphericalVector}, which arises after evaluating the product over primes. To identify the individual building blocks in (\ref{SL(2)pAdic}) from the Fourier expansion of $E^{SL(2,\mbb{Z})}(\tau, s)$ in (\ref{FinalFourierSL(2)}), we begin by restricting to the origin of moduli space, $\tau=i$, in which case (\ref{SL(2)pAdic}) may be written as
\beq
E^{SL(2,\mbb{Z})}(i;s)=\sum_{x\in \mbb{Z}}^{\quad \hs \prime} f_{\mbb{Z}}(x) f_K(x).
\label{SL(2)pAdicOrigin}
\eeq
Similarly, at the origin of the moduli space, the bulk piece of the Fourier expansion (\ref{FinalFourierSL(2)}}) becomes
\beq
E^{SL(2,\mbb{Z})}(i;s)=\f{1}{\xi(2s)}\sum_{x\in \mbb{Z}}^{\quad \hs \prime}\mu_{1-2s}(x) x^{s-1/2}K_{s-1/2}(2\pi |x|).
\label{FinalFourierSL(2)Origin}
\eeq
Comparing (\ref{FinalFourierSL(2)Origin}) to (\ref{SL(2)pAdicOrigin}) we may extract the new spherical vector:
\beq
f_K(x):=x^{s-1/2}K_{s-1/2}(2\pi |x|),
\label{NewSphericalVector}
\eeq
together with the arithmetic coefficient
\beq
f_{\mbb{Z}}(x):= \f{1}{\xi(2s)}\mu_{1-2s}(x),
\eeq
where $\mu_{1-2s}(x)$ is the sum over divisors defined in \ref{divisorsum}. We finally note that (\ref{NewSphericalVector}) is indeed the Fourier transform of the old spherical vector $f_K(x)=(1+x^{2})^{-s}$.

\section{Fourier Expansion of ${E}^{SL(3, \mbb{Z})}(g;s_1,s_2)$}
\label{Section:FourierExpansionSL(3)}
In this section we shall give the complete Fourier expansion of the general Eisenstein series ${E}^{SL(3,\mbb{Z})}(g; s_1,s_2)$ in the principal series. This was computed in detail by Vinogradov and Takhtadzhyan in \cite{VT}, and we will here reproduce their results in a slightly adapted form (see also the thesis by Bump \cite{Bump}). The following section is based on {\bf Paper VII}, which is written in collaboration with Boris Pioline. The Eisenstein series $E(g;s_1,s_2)$ is conjectured to capture D$(-1)$, D5 and NS5-brane instanton corrections to the hypermultiplet moduli space in type IIB string theory compactified on a Calabi-Yau threefold $X$. For the precise physical interpretation of the Fourier expansion see {\bf Paper VII}. 

\subsection{General Structure}

For the readers convenience we begin by recalling the explicit form (\ref{SL(3)PoincareSeriesVTvariablesExplicit}) of the principal Eisenstein series:
 \beqa
 E^{SL(3,\mbb{Z})}(g; s_1, s_2)&=& y^{\f{s_2-s_1}{3}} \sum_{(c, b)\in \mbb{Z}^6}^{\quad \prime}\bigg\{ \Big[c_1^2 u+\f{1}{u}\big(c_2+c_1 v\big)^2+\f{1}{y}\big(c_1\mc{A}_1+c_2\mc{A}_2+c_3\big)^2\Big]^{-s_1}
 \nn \\
 & & \times \Big[yD_3^2+u\big(D_2-\mc{A}_2 D_3\big)^2+\f{1}{u}\big(D_1-vD_2-(\mc{A}_1-v\mc{A}_2)D_3\big)^2\Big]^{-s_2}\bigg\},
 \nn \\
 \label{SL(3)PoincareSeriesVTvariablesExplicit2}
 \eqa
 where 
 \beq
 D_1:= b_2c_3-b_3c_2, \qquad D_2:=b_3c_1-b_1c_3, \qquad D_3:= b_1c_2-b_2c_1.
 \eeq
In order to make contact with the conventions of {\bf Paper VII} we make the following minor change of moduli variables:
\beq
y = \nu^{-1}\ ,\qquad u=\tau_2\ ,\qquad \mc{A}_1=-\psi\ ,\qquad \mc{A}_2 = -c_0 \ ,\qquad v = - \tau_1\ .
\label{ChangeVariables}
\eeq
The nilpotent subgroup $N\subset SL(3,\mbb{R})$ is isomorphic to a three-dimensional Heiseberg group, and therefore, according to the general analysis of Section \ref{Section:GeneralFourierExpansion}, the general structure of the Fourier expansion splits into an abelian and a non-abelian part
\beq
E^{SL(3,\mbb{Z})}(g; s_1, s_2)=E_A(g;s_1, s_2)+E_{NA}(g;s_1, s_2), 
\eeq
where
\beqa
E_A(g;s_1,s_2)&=& \sum_{(p,q)\in \mbb{Z}^2} \mf{C}_{p,q}(\nu, \tau_2) e^{-2\pi i (q\tau_1-p c_0)},
\nn \\
E_{NA}(g;s_1, s_2) &=& \sum_{k\neq 0} \mf{C}_k(\nu, \tau,c_0) e^{2\pi i k\psi}.
\label{(Non-)abelian}
\eqa
Recall that the abelian term represents the Fourier expansion with respect to the abelianized group $\tilde{N}=N/Z$, where $Z$ is the center of $N$, corresponding to the commutator subgroup:
\beq
Z=[N,N]=\left\{\left(\begin{array}{ccc}
1 & & *\\
& 1 & \\
& & 1\\
\end{array}Ê\right) \right\},
\eeq
while non-abelian term is the Fourier expansion with respect to the center $Z$, coordinatized by the axionic scalar $\psi$. It was shown in \cite{VT} that one may further diagonalize the action of one of the shift symmetries $(\tau_1,c_0)\mapsto (\tau_1+a, c_0+b)$, $a,b\in \mbb{Z}$, and restrict the coefficients $\mf{C}_k$ in (\ref{(Non-)abelian}) as follows:
\beq
\mf{C}_k( \nu, \tau_2,\tau_1,c_0)=\sum_{(p,q)\in \mbb{Z}^2} \Psi_{q, d}(\nu, [\tau_2]_{-k, p}) 
\, e^{-2\pi i \big(q [\tau_1]_{-k, p}- p c_0\big)},
\label{nonabelianexplicit}
\eeq
where $d=\gcd (p,k)$. The variables $[\tau_1]_{-k, p}$ and $[\tau_2]_{-k, p}$
denote the real and imaginary parts of the image of $\tau=\tau_1+i \tau_2$ under an $SL(2,\mbb{Z})$-transformation of the form
\beq
\delta=\left(\begin{array}{cc}
\al & \beta \\
-k^{\prime} & p^{\prime} \\
\end{array}\right)\ ,\qquad
\delta\cdot \tau=
\f{\al \tau+\beta}{-k^{\prime} \tau+p^{\prime}}\equiv [\tau_1]_{-k,p}
+i [\tau_2]_{-k,p},
\label{deft12}
\eeq
where $k^{\prime}=k/d$ and $p^{\prime}=p/d$  and 
$\alpha,\beta$ are two integers such that $\alpha p'+\beta k'=1$. Since $k\neq 0$, 
this is usefully rewritten as
\beq
[\tau_2]_{-k, p}=\f{d^2 \tau_2}{|p-k\tau|^2}, \qquad [\tau_1]_{-k, p}=-\frac{d \alpha}{k}+
\f{d^2 (p-k\tau_1)}{k|p-k\tau|^2}\ ,
\label{vtransformed}
\eeq
where we used $\beta=(1-p^{\prime}\al)/k^{\prime}$ to derive the second relation. In terms of the group element $g\in SL(3,\mbb{R})$ (see Eq. (\ref{SL(3)ElementVTconventions})) this $SL(2,\mbb{Z})$ corresponds to the left action on $g$ by the $SL(3,\mbb{Z})$ element
\beq
\ga=\left( \begin{array}{ccc}
\al & \be & 0\\
-k^{\prime} & p^{\prime} & 0 \\
0 & 0 & 1\\
\end{array} \right)\in SL(3,\mbb{Z}),
\eeq
under which $y$ is invariant while $(\mc{A}_1, \mc{A}_2)$ transform linearly. The appearance of the somewhat complicated expressions for $[\tau_1]_{1,2}$ and $[\tau_2]_{1,2}$ in (\ref{nonabelianexplicit}) is a consequence of the non-abelian structure of the Heisenberg group $N$. To summarize, we have found that the general structure of the Fourier expansion reads
\beqa
E^{SL(3,\mbb{Z})}(g; s_1, s_2)&=& \phantom{+} \sum_{(p,q)\in \mbb{Z}^2} \Psi_{p,q}(\nu, \tau_2) e^{-2\pi i (q\tau_1-p c_0)}
\nn \\
& & + \sum_{k\neq 0} \sum_{(p,q)\in \mbb{Z}^2} \Psi_{q, d}(\nu, [\tau_2]_{-k, p}) 
\, e^{-2\pi i \big(q [\tau_1]_{-k, p}- p c_0-k\psi\big)},
\nn \\
\label{SL(3)ExpansionComplete}
\eqa
where we relabeled the abelian coefficients as $\mf{C}_{p,q}\equiv \Psi_{p,q}$ in accordance with the notation of {\bf Paper VII}. Physically, one may understand the $SL(2,\mbb{Z})$ action on $\tau$ in (\ref{SL(3)ExpansionComplete}) as follows. In type IIB string theory, the (D5,NS5)-brane bound states belong to a doublet of $SL(2,\mbb{R})$. This implies that an arbitrary bound state of D5-NS5 branes may be labelled by two integers $(p,k)$ and is often referred to as a $(p,q)$ 5-brane. Thus, physically, once we know how to describe the effects from, say, $p$ D5-branes, the NS5-brane contributions follow simply from enforcing S-duality, i.e. $SL(2,\mbb{Z})$-invariance. This is the physical origin of the somewhat complicated phase factor in (\ref{SL(3)ExpansionComplete}). See {\bf Paper VII} for more information. We shall now study the Fourier coefficients $\Psi_{p,q}$ and $\Psi_{q,d}$ in more detail.

\subsection{Constant Terms}
We begin to consider the zeroth Fourier coefficient $\Psi_{0,0}$. This is given by the constant term 
\beq
 \Psi_{0,0}(\nu, \tau_2; s_1, s_2) = \int_{0}^{1} d\tau_1 \int_{0}^{1} d c_0 \int_{0}^{1} d\psi \ E(\nu, \tau_2, \tau_2, c_0, \psi; s_1, s_2),
 \eeq
 which was evaluated in \cite{VT} with the result
 \beqa
 \Psi_{0,0}(\nu, \tau_2; s_1, s_2)& = &\nu^{-\frac{2s_1+s_2}{3}} \tau_2^{s_2}
 + c(s_1) ,\nu^{-\frac12+\frac{s_1-s_2}{3}} \tau_2^{s_3}
 + c(s_2)  \nu^{-\frac{2s_1+s_2}{3}} \tau_2^{1-s_2}
 \nn \\
& & + c(s_1)\,c(s_3)\,  \nu^{-\frac12+\frac{s_1-s_2}{3}} \tau_2^{1-s_3}
+c(s_2)\,c(s_3)\,  \nu^{\frac{2s_1+s_2}{3}-1} \tau_2^{s_1} 
\nn \\
& & +c(s_1)\,c(s_2)\,c(s_3)\,  \nu^{\frac{2s_1+s_2}{3}-1} \tau_2^{1-s_1}\ ,
\nn \\
 \eqa
 where $s_3=s_1+s_2-1/2$ and we defined 
\beq
c(s):= \f{\xi(2s-1)}{\xi(2s)}.
\eeq
Recall also that $\xi(s)$ is the completed Riemann zeta function (\ref{CompletedZetaFunction}). 
\subsection{Abelian Fourier Coefficients}
Let us now proceed to analyze the abelian Fourier coefficients $\Psi_{p,q}$ in (\ref{SL(3)ExpansionComplete}). Here it is convenient to begin by studying the simplest cases $\Psi_{0,q}$ and $\Psi_{p,0}$. As we shall see, the general abelian coefficient is considerably more involved. From \cite{VT} we deduce
\beq
\begin{split}
\Psi_{0,q}(\nu, \tau_2)&=\phantom{+} \f{2(2\pi)^{1/2-s_1} c(s_2)c(s_3)}{\xi(2s_1)}\nu^{\f{s_1+2s_2}{3}-1} \tau_2^{1-s_1} \,\mu_{1-2s_1}(q) \, \, \mc{K}_{1/2-s_1}\big(2\pi |q| \tau_2\big)\\
&\quad+ \f{2(2\pi)^{1/2-s_2}}{\xi(2s_2)} \nu^{-\f{2s_1+s_2}{3}} \tau_2^{1-s_2}\,
\mu_{1-2s_2}(q) \,  \, \mc{K}_{1/2-s_2}\big(2\pi |q| \tau_2\big) \\
&\quad+ \f{2(2\pi)^{1/2-s_3}c(s_1)}{\xi(2s_3)} \nu^{\f{s_1-s_2-1}{2}} \tau_2^{1-s_3}\,
\mu_{1-2s_3}(q) \,\, \mc{K}_{1/2-s_3}\big(2\pi |q| \tau_2\big)\ ,
\end{split}
\label{abelian1}
\eeq
and 
\beq
\begin{split}
\Psi_{p, 0}(\nu, \tau_2)&=\phantom{+} \f{2(2\pi)^{1/2-s_1}}{\xi(2s_1)} \nu^{-\f{2s_1+s_2}{3}}
\tau_2^{s_2}\mu_{1-2s_1}(p) \,  \, 
\mc{K}_{1/2-s_1}\big(2\pi |p|/\sqrt{ \nu \tau_2} \big)\\
&\quad+ \f{2(2\pi)^{1/2-s_2}c(s_1)c(s_3)}{\xi(2s_2)} 
\nu^{\f{s_1-s_2}{3}-\f{1}{2}}\tau_2^{\f{3}{2}-s_1-s_2}\, 
\mu_{1-2s_2}(p) \,  \, \mc{K}_{1/2-s_2}\big(2\pi |p| /\sqrt{ \nu \tau_2}\big)\\
&\quad+ \f{2(2\pi)^{1/2-s_3}c(s_2)}{\xi(2s_3)} \nu^{-\f{s_2+2s_1}{3}}
\tau_2^{1-s_2} \, \mu_{1-2s_3}(p) \, \, \mc{K}_{1/2-s_3}\big(2\pi |p| /\sqrt{ \nu \tau_2}\big)\ .
\end{split}
\label{abelian2}
\eeq
where we have defined the rescaled modified Bessel function
\beq
\mc{K}_s(x):=x^{-s} K_s(x),
\eeq
and we defined the sum over divisors,
\beq
\mu_s(r):=\sum_{n|r} n^{s}.
\eeq

We now proceed to analyze the coefficients $\Psi_{p,q}$ for $pq\neq0$, which we recall were absent for the minimal Eisenstein series. These coefficients may be written as
\beq
\label{genab}
\begin{split}
\Psi_{p,q}(\nu, \tau_2)&=\nu^{\f{s_2-s_1}{6}-\f{1}{2}} \tau_2^{\f{s_2-s_1}{2}+\f{1}{2}}
\mu(p,q) (pq)^{s_3-\frac12} \,  \mc{I}_{s_1,s_2}\left( R_{p,q}, \x_{p,q} \right)\ ,
\end{split}
\eeq
where
\beq
R_{p,q}\equiv \left(\f{p^2 |q|}{\nu}\right)^{1/3}, \qquad \x_{p,q}\equiv \tau_2^{-1} \left(\frac{p^2}{\nu q^2}\right)^{1/3} = t^2  \left(\frac{|p|}{|q|}\right)^{2/3}\ .
\label{pqvariables}
\eeq
Here, we have also defined the ``double divisor sum''
\beq
\sigma_{\alpha,\beta}(n,m) \equiv \sum_{m=d_1 d_2 d_3,\atop
d_1,d_2,d_3>0, \gcd(d_3,n)=1} d_2^{\alpha} d_3^{\beta}\ ,
\eeq
and the integral
\beq
\label{integral}
 \mc{I}_{s_1,s_2}(R,\x)
\equiv  \int_0^{\infty} K_{s_3-1/2}\big(2\pi R\,  \x^{-1}  \sqrt{1+x}\big) \,K_{s_3-1/2}\Big(2\pi R \, \x^{1/2} \sqrt{1+1/x}\Big) \, x^{\f{s_2-s_1}{2}} \f{dx}{x}\ .
 \eeq
 The summation measure $\mu(p,q)$ may also be extracted with the result\footnote{The normalization of $\mu(p,q)$ differs slightly from the one in {\bf Paper VII}, which defined the summation measure after performing a saddle point approximation of the integral in (\ref{genab}).}
\beq
\mu(p,q):= 
\f{4}{\xi(2s_1)\xi(2s_2)\xi(2s_3)} 
\sum_{d_1|p} \sum_{d_2|\frac{p}{d_1}}  d^{2-s_1-s_2} d_1^{1-2s_3} d_2^{1-2s_2}\sigma_{1-2s_1,1-2s_3}\Big(\f{p}{d_1 d_2}, |q|\Big).
\eeq

\subsection{Non-Abelian Fourier Coefficients}
We now proceed to the non-abelian terms in (\ref{SL(3)ExpansionComplete}) corresponding to $k\neq 0$. These may be written as
\beq
 \label{nonabeliancoefficient}
\Psi_{q,d}(\nu, \tau) =
\nu^{\f{s_2-s_1}{6}-\f{1}{2}}
\sum_{q\in \mbb{Z}}{\mu}(p,q,k)  [\tau_2]_{-k,p}^{\f{s_2-s_1}{2}+\f{1}{2}} \, (d\,q)^{s_3-\frac12}\,
\mc{I}\left( R_{p,q,k},\x_{p,q,k} \right)\  e^{-2\pi i q [\tau_1]_{-k, p}}  ,
\eeq
where 
\beq
R_{p,q,k} \equiv  \left( \frac{ d^2 |q|  }{\nu}\right)^{1/3}, \qquad \x_{p,q,k}\equiv  [\tau_2]_{-k,p}^{-1}\, \left(\f{d^2}{\nu q^2}\right)^{1/3},
\eeq
and $d\equiv\gcd(p,k)>0$ and the integral $\mc{I}(R, \x)$ is the same as in (\ref{integral}). The non-abelian summation measure reads\footnote{Similarly to the abelian measure, the normalization of $\mu(p,q,k)$ differs from the one in {\bf Paper VII}, which defined the measure after extracting the saddle point of the integral in (\ref{nonabeliancoefficient}). In particular, there is an additional phase factor $\exp\{2\pi i d q  \al /k\}$ in the non-abelian measure of {\bf Paper VII}, which arises, via (\ref{vtransformed}), from the phase factor in (\ref{nonabeliancoefficient}). We will also see this explicitly in the next subsection.} 
\beq
{\mu}(p,q,k)\equiv 
\f{4 }{\xi(2s_1)\xi(2s_2)\xi(2s_3)} 
\sum_{d_1|d} \sum_{d_2|\frac{d}{d_1}}  d^{2-s_1-s_2} d_1^{1-2s_3} d_2^{1-2s_2}\sigma_{1-2s_1,1-2s_3}\Big(\f{d}{d_1 d_2}, |q|\Big)\ .
\label{NonabelianMeasure}
\eeq
For more details on this Fourier expansion, and in particular its physical interpretation, we refer the reader to {\bf Paper VII}. 

\subsection{The Exact Spherical Vector in the Principal Series}

As was done in Section \ref{Section:SphericalVectorFourierSL(2)} for the Fourier expansion of $E^{SL(2,\mbb{Z})}(\tau;s)$, we can now cast the non-abelian Fourier expansion of $E^{SL(3,\mbb{Z})}(g;s_1,s_2)$ into the general framework of Section \ref{Section:SphericalVector}. This will allow us to extract the exact spherical vector $f_K$ for the full principal series of $SL(3,\mbb{R})$, whose semi-classical limit was obtained previously in \cite{AutomorphicMembrane}. As above, we follow {\bf Paper VII} closely. 

To make the link with the analysis in \cite{AutomorphicMembrane}, we begin by making the following change of variables \beq
 y=-k\ ,\qquad x_0=p\ ,\qquad
 x_1=(d^2 q)^{1/3}\ , 
 \label{ChangeOfVariables}
 \eeq
and, as in Section \ref{Section:SphericalVectorFourierSL(2)}, we work at the origin of moduli space
where $\tau_1=c_0=\psi=0$ and $\tau_2=\nu=1$, such that
\beq
[\tau_2]^{(0)}_{1,2}=\f{d^2}{y^2+x_0^2}\ , \qquad R=x_1\ ,\qquad  \x=\f{y^2+x_0^2}{x_1^2}\ ,
\eeq
where we have used (\ref{vtransformed}) and (\ref{ChangeVariables}). Moreover, at the origin of moduli space, the phase factor in the non-abelian term of \eqref{nonabeliancoefficient} becomes
\beq
\exp\left[-2\pi i q [\tau_1]^{(0)}_{1,2}\right]=\exp\left[\f{2\pi iq d \al}{k}\right]\ \exp\left[\f{2\pi i\ pq}{k(k^2+p^2)}\right]
= \exp\left[\f{2\pi i x_1^3 \al }{d k}\right] \, \exp\left[ -\f{2\pi i x_0 x_1^3}{y(y^2+x_0^2)}\right].
\label{decomposition}
\eeq
Following the reasoning of Section \ref{Section:SphericalVectorFourierSL(2)}, we may further write the non-abelian part of the expansion at the origin of moduli space as
\beq
E_{NA}(1; s_1, s_2)=\sum_{(y,x_0,x_1)\in \mbb{Z}}^{\quad \prime}
 f_{\mbb{Z}}(y, x_0, x_1) \, f_K(y, x_0, x_1).
\label{NonabelianTermSPApprox}
\eeq
Comparing this expression with (\ref{nonabeliancoefficient}), we deduce that the exact spherical vector is given by
\beq
\label{exactsl3}
\begin{split}
f_K(y,x_0,x_1) &= \, 
(y^2+x_0^2)^{\frac12(s_1-s_2-1)} \, x_1^{3(s_1+s_2-1)} \, e^{-\f{2\pi i x_0 x_1^3}{y(y^2+x_0^2)}}\\
&\times \int_0^{\infty} K_{s_3-\frac12}\Big(\frac{2\pi \,x_1^3}{y^2+x_0^2}\sqrt{1+x}\Big) 
K_{s_3-\frac12}\Big(2\pi \sqrt{(y^2+x_0^2)(1+1/x)}\Big) x^{\f{s_2-s_1}{2}} \f{dx}{x} \ ,
 \end{split}
 \eeq
while the summation measure is
\beq
f_{\mbb{Z}}(y, x_0, x_1)\equiv 
\f{4 \, e^{\f{2\pi i x_1^3 \al }{d k}}}{\xi(2s_1)\xi(2s_2)\xi(2s_3)} 
\sum_{d_1|d} \sum_{d_2|\frac{d}{d_1}}  d^{2-s_1-s_2} d_1^{1-2s_3} d_2^{1-2s_2}\sigma_{1-2s_1,1-2s_3}\Big(\f{d}{d_1 d_2}, \f{x_1^3}{d^{2}}\Big)\ ,
\eeq
where we recall that $d=\gcd(y, x_0)$. 

The spherical vector simplifies considerably in the limit where $y,x_0,x_1$ are scaled to infinity
with fixed ratio: in this case the saddle point approximation of the integral in (\ref{nonabeliancoefficient}) becomes (see {\bf Paper VII} for the derivation)
\beq
\mc{I}(y, x_0, x_1) \sim  \f{(y^2+x_0^2)^{\f{s_2-s_1+1}{2}} x_1^{s_1-s_2-2}}{(y^2+x_0^2+x_1^2)^{1/4}}\exp\left[- \f{2\pi (y^2+x_0^2+x_1^2)^{3/2}}{y^2+x_0^2}\right]\ ,
\eeq
and  the spherical vector simplifies to 
\beq
f_K(y, x_0, x_1)\sim  
 \f{x_1^{4s_1+2s_2-5}}{(y^2+x_0^2+x_1^2)^{1/4}} \exp\left[- \f{2\pi(y^2+x_0^2+x_1^2)^{3/2}}{y^2+x_0^2}
-\f{2\pi i x_0 x_1^3}{y(y^2+x_0^2)}\right]\ .
\label{nonabeliancoefficientSP}
\eeq
As a consistency check, we note that in the special case $(s_1, s_2)=(0,0)$\footnote{In the conventions of \cite{AutomorphicMembrane} this corresponds to$(\lambda_{23}, \lambda_{21})=(1,1)$.}, this result agrees with the semi-classical spherical vector of the principal series representation of $SL(3,\mbb{R})$ obtained in \cite{AutomorphicMembrane}. 

\chapter{Instanton Corrections to the Universal Hypermultiplet }
\label{Chapter:UniversalHypermultiplet}
The analysis in {\bf Part II} of this thesis has so far been restricted to string compactifications on tori, in which case maximal supersymmetry is preserved, giving rise to moduli spaces described by symmetric spaces $\mc{M}=G/K$, where $G$ is a global continuous symmetry of the classical effective action. These moduli spaces are known as \emph{rigid}, since they allow no global deformations that preserve the holonomy $K$ (i.e. is compatible with supersymmetry) except for an overall additional quotient by a discrete group $G(\mbb{Z})\subset G(\mbb{R})$ \cite{Aspinwall}. Hence, for these cases the exact quantum moduli space must be
\beq
\mc{M}_{\mathrm{exact}}=G(\mbb{Z})\bas G(\mbb{R})/K,
\eeq
where $G(\mbb{Z})$ is the U-duality group \cite{HullTownsend}. This implies that all quantum corrections to the effective action are encoded in the duality group $G(\mbb{Z})$, leading to powerful techniques for summing up perturbative and non-perturbative corrections in terms of $G(\mbb{Z})$-invariant automorphic forms, as discussed in Chapter \ref{Chapter:Aspects}. The purpose of the present chapter is to analyze cases where the internal manifold is more complicated than a torus, for which the moduli space is no longer given by a symmetric space. In particular, we shall focus on the special case of type IIA compactifications on \emph{rigid} Calabi-Yau threefolds in which case we propose that a natural candidate for the underlying duality group $G(\mbb{Z})$ is the \emph{Picard modular group} $SU(2,1;\mbb{Z}[i])$. We further construct an $SU(2,1;\mbb{Z}[i])$-invariant Eisenstein series, and show that its Fourier expansion reproduces the expected contributions from D2- and NS5-brane instantons. This chapter is based on {\bf Paper VIII}, which is written in collaboration with Ling Bao, Axel Kleinschmidt, Bengt E. W. Nilsson and Boris Pioline.

Let us also mention that related results were recently obtained in {\bf Paper VII}, which was written in collaboration with Boris Pioline. There it was proposed that for generic (non-rigid) Calabi-Yau compactifications, a candidate for the underlying duality group in $D=4$ is $G(\mbb{Z})=SL(3,\mbb{Z})$. This suggestion is complementary to the discussion in the present chapter since for generic Calabi-Yau manifolds, the Picard group is not expected to play any role, while on the other hand for rigid compactifications $SL(3,\mbb{Z})$ is not relevant. Some mathematical aspects of the analysis of {\bf Paper VII} was presented in Section \ref{Section:FourierExpansionSL(3)}. 

In the following section we will give a general overview of the relevant techniques, after which we begin a more  detailed analysis in Section \ref{Section:UniversalSector}.

\section{Instanton Corrections to Quaternionic Moduli Spaces}

As was briefly mentioned above, and discussed in Chapter \ref{Chapter:Introduction}, for compactifications on more complicated internal manifolds than tori, the moduli space $\mc{M}$ is typically not described in terms of a simple symmetric space, rendering it difficult to find the duality group $G(\mbb{Z})$, if it exists. Particularly interesting examples are type II Calabi-Yau threefold compactifications, which give rise to moduli spaces corresponding to a product of a special K\"ahler manifold $\mc{M}_{{}_{\mathrm{SK}}}$ and a quaternion-K\"ahlermanifold $\mc{M}_{{}_{\mathrm{QK}}}$. Due to the small amount of supersymmetry preserved, these moduli space are subject to quantum corrections which deform the geometry. Recall from the discussion in Chapter \ref{Chapter:Introduction} that the special K\"ahler manifold $\mc{M}_{{}_\SK}$ receives perturbative quantum corrections associated with the worldsheet coupling $\alpha^{\prime}$, as well as non-perturbative corrections from worldsheet instantons, which are exponentially suppressed of order $e^{-1/\alpha^{\prime}}$ in the weak-coupling limit $\alpha^{\prime}\rightarrow \infty$ \cite{Dine:1986zy,Dine:1987bq}. Recall further that the four-dimensional dilaton $e^{\phi}$ belongs to the hypermultiplet moduli space $\mc{M}_{{}_\QK}$, implying that this moduli space is sensitive to perturbative corrections from the genus expansion in $g_s=e^{\phi}$. In addition, $\mc{M}_{{}_\QK}$ receives non-perturbative corrections from D-brane instantons \cite{BeckerBeckerStrominger}. In the type IIA picture, these instantons aries from Euclidean D2-branes wrapping special Lagrangian 3-cycles in the Calabi-Yau manifold $X$, while in type IIB the contributions are due to Euclidean D$p$-branes, $p=-1,1,3,5$, wrapping even cycles in $X$. In both type IIA and type IIB there are also instanton effects due to Euclidean NS5-branes wrapping all of $X$.

Special K\"ahler manifolds are described by a holomorphic prepotential $F$, and it is therefore possible to additively encode quantum corrections to the metric on $\mc{M}_{{}_{\mathrm{SK}}}$ in terms of $F$ \cite{Candelas:1990rm,MirrorSymmetry}. For the quaternion-K\"ahlermetric on $\mc{M}_{{}_\QK}$, however, no such natural description exists and one must resort to more sophisticated techniques to encode deformations of $\mc{M}_{{}_{\mathrm{QK}}}$. It has long been an outstanding problem to compute the instanton-corrected metric on $\mc{M}_{{}_\QK}$, in particular the contributions from NS5-brane instantons. Recently, however, considerable progress has been made in this endeavour by utilizing techniques from twistor theory. These developments will play an important role in this chapter and we shall here give a brief introductory overview. More details are given in Section \ref{Section:InstantonCorrections}.

The basic idea, developed in a series of papers  \cite{VandorenAlexandrov,Vandoren1,Vandoren2,Alexandrov:2008ds,Alexandrov:2008nk,Alexandrov:2008gh}, is that deformations of the quaternionic moduli space $\mc{M}_{{}_{\mathrm{QK}}}$ can be uplifted to deformations of the associated \emph{twistor space} $\mc{Z}$, a $\mbb{C}P^1$-bundle over $\mc{M}_{{}_{\mathrm{QK}}}$. Quantum corrections are then encoded in local complex (Darboux) coordinates on $\mc{Z}$ together with their associated transition functions gluing together local patches in $\mc{Z}$. These complex coordinates are called \emph{twistor lines}, and play a similar role for $\mc{M}_{{}_{\mathrm{QK}}}$ as the holomorphic prepotential for special K\"ahler manifolds. A crucial object in this story is the \emph{contact potential} $e^{\Phi}$ which determines the K\"ahler potential on the twistor space $\mc{Z}$. As shown in \cite{Vandoren1,Vandoren2,Alexandrov:2008gh}, perturbative and non-perturbative quantum corrections can then be elegantly encoded in an $SL(2,\mbb{Z})$-invariant completion of the contact potential $e^{\Phi}$, in a very similar way as the corrections to the $\mc{R}^4$-term discussed in Chapter \ref{Chapter:Aspects}. 

Despite these new techniques, an unsolved problem has been to incorporate the corrections from NS5-brane instantons which start to contribute in $D=4$, in which case $SL(2,\mbb{Z})$-invariance has proven insufficient. It would therefore be desirable to have access to a larger duality group $G(\mbb{Z})$, such that the $G(\mbb{Z})$-invariant completion of $e^{\Phi}$ also sums up all NS5-brane instanton corrections. For generic Calabi-Yau threefolds it is not known what the group $G(\mbb{Z})$ might  be, but in this chapter we shall see that for the simpler case of compactifications of \emph{rigid} Calabi-Yau threefolds there is a very appealing candidate for the underlying duality group. As mentioned above, we argue that for rigid Calabi-Yau compactifications in type IIA string theory, the underlying duality group is given by the Picard modular group $SU(2,1;\mbb{Z}[i])$, which is discrete subgroup of $SU(2,1)$ defined over the Gaussian integers $\mbb{Z}[i]$ (see, e.g., \cite{FrancsicsLax}).

\section{The Universal Sector of Type IIA on Calabi-Yau Threefolds}
\label{Section:UniversalSector}
Let us now present the physical setting in which our analysis
applies. We consider the low-energy effective theory arising from the
compactification of type IIA string theory on a rigid Calabi-Yau threefold
$\mc{X}$. The bosonic sector of this
compactification correponds to $D=4$ Maxwell-Einstein gravity coupled to the
so called \emph{universal hypermultiplet}, with moduli space
$\mc{M}_{{}_{\mathrm{UH}}}$. In the analysis we will disregard the vector multiplets which decouple and do not play any role in what follows.

%We describe the further compactification of this
%theory on $S^1$, in which case the gravity multiplet gives rise to an additional
%moduli space $\mc{M}_{{}_{\mathrm{GM}}}$, which turns out to be identical to
%$\mc{M}_{{}_{\mathrm{UH}}}$. In this context we discuss the $c$-map which relates
%the degrees of freedom of the hypermultiplet sector to the degrees of freedom
%in the gravity sector.

\subsection{Type IIA on Calabi-Yau Threefolds}
Type IIA compactifications on a generic Calabi-Yau threefold $X$ give rise to
$ \mc{N} =2$ supergravity in four dimensions. This theory
splits into three separate pieces: $(1)$ the \emph{gravity multiplet},
consisting of the $D=4$ metric $g_{\mu\nu}$ and an abelian vector
$\mc{A}_{\mu}$ known as the graviphoton; $(2)$ the \emph{vector multiplets},
consisting of $n_V=h_{1,1}(X)$ abelian vectors $A^{I}_{\mu}$ and
$n_V=h_{1,1}(X)$ complex scalars $Z^{I}$; $(3)$ and, finally, the
\emph{hypermultiplets}, consisting of $4n_H=4(h_{2,1}(X)+1)$ real scalars
$\varphi^{i}$. The fermionic degrees of freedom of these multiplets will not be needed in the present work.

The vector multiplet moduli $Z^{I}$ parametrize a complex $2n_V$-dimensional
\emph{special K\"ahler manifold} $\mc{M}_{{}_{\mathrm{V}}} $, while the hypermultiplet moduli
$\varphi^i$ parametrize a real $4n_H$-dimensional \emph{quaternion}-\emph{K\"ahler
  manifold} $\mc{M}_{{}_{\mathrm{H}}} $. The total moduli space $\mc{M}(X)$ splits locally into
a direct product
\beq
\mc{M}(X)=\mc{M}_{{}_{\mathrm{V}}} \times \mc{M}_{{}_{\mathrm{H}}} .
\eeq
The scalars $\phi^{i}$ parametrizing the hypermultiplet moduli space further splits into $h_{2,1}$ complex scalars $z^{a}$ which parametrize the complex structure moduli space $\mc{M}_C(\mc{X})$, together with $2h_{2,1}$ real Ramond-Ramond scalars $(\zeta^{I}, \tilde{\zeta}_I)$, $I=0,1,\dots, h_{2,1}$, arising from the periods of the Ramond-Ramond 3-form $C_{(3)}$, and the NS-scalar $\psi$ which is the dual of the NS 2-form $B_{(2)}$ (see Section \ref{cmap} below for a more detailed account of this in the rigid case). In addition, the four-dimensional dilaton $e^{\phi}$ belongs to the hypermultiplet moduli
space, implying that $\mc{M}_{{}_{\mathrm{H}}} $ is sensitive to quantum corrections
originating from the string genus expansion. At the same time, the vector
multiplet moduli space $\mc{M}_{{}_{\mathrm{V}}} $ receives corrections which are higher order
in $\al^{\prime}$, as well as worldsheet instanton
corrections which scale as $e^{-1/\alpha^{\prime}}$. The quantum corrections
to $\mc{M}_{{}_{\mathrm{V}}} $ can be conveniently encoded in corrections to the prepotential
$F(Z)$, whose exact form is known \cite{Candelas:1990rm}.

The corrections to the hypermultiplet moduli space has, however, proven to be much more complicated to understand. This is partly due to the fact that there is no
analogue of the prepotential for describing the geometry of
quaternion-K\"ahler manifolds. Moreover, since the geometry of $\mc{M}_{{}_{\mathrm{H}}} $ is
sensitive to corrections associated with the string coupling $g_s=e^{-\phi}$,
the moduli space of the hypermultiplets is expected to also receive
contributions which are non-perturbative as $g_s\rightarrow 0$. More
precisely, $\mc{M}_{{}_{\mathrm{H}}} $ receives instanton corrections from Euclidean D2-branes
wrapping special Lagrangian submanifolds in $X$, as well as Euclidean
NS5-branes wrapping all of $X$ \cite{BeckerBeckerStrominger}. The former
scales as $e^{-1/g_s}$, as is characteristic for D-brane instantons, while the
latter contributions scale as $e^{-1/g_s^2}$, in accordance with the tension
of NS5-branes. These quantum effects will be further discussed in Section
\ref{Section:InstantonCorrections}.

As was already mentioned in Chapter \ref{Chapter:Introduction}, topologically the hypermultiplet moduli space $\mc{M}_{{}_{\mathrm{H}}}$ has the structure of a torus bundle over the moduli space $\mc{M}_C(\mc{X})$ of complex structure deformations of  $\mc{X}$. The torus fiber $T^{2h_{2,1}+2}$ is parametrized by the Ramond-Ramond scalars $(\zeta^{I}, \tilde{\zeta}_I)$ and may be identified with the intermediate Jacobian $J(\mc{X})=H^3(\mc{X},\mbb{C})/H^3(\mc{X},\mbb{Z})$ (see the following section for more information on $J(\mc{X})$ in the rigid case). However, this is not the complete story, since the moduli space also contains the axionic scalar $\psi$, which parametrizes an additional circle bundle over $J(\mc{X})$. Equivalently, the axion  $\psi$Ê can be combined with the dilaton into the complex scalar $\psi + i e^{-2\phi}$ which parametrizes the complex line $\mbb{C}^{*}$. Thus we can view the total space as a $\mbb{C}^{*}$-line bundle $\mc{L}$ over the Jacobian $J(\mc{X})$. The complete moduli space therefore splits locally into the product:
\beq
\mc{M}_{{}_{\mathrm{H}}} = \mc{M}_C(\mc{X}) \times {T}^{2h_{2,1}+2} \times \mbb{C}^{*}.
\label{hyperbundle}
\eeq
The line bundle $\mc{L}$ is non-trivial and has first Chern class $c_1(\mc{L})= d\zeta^{I} \wedge d\tilde{\zeta}_I$ (see, e.g., \cite{Gunther:1998sc,RoblesLlana:2006ez,Alexandrov:2008nk,AutomorphicNS5}), which may be identified with the K\"ahler form of $J(\mc{X})$ in a ``principal polarization''. This is in broad accordance with Witten's general analysis of the fivebrane partition function which precisely was shown to be a section of the same line bundle $\mc{L}$ over the intermediate Jacobian \cite{WittenFivebrane}. 

We further note that the first two factors in (\ref{hyperbundle}) correspond to the so called Seiberg-Witten torus fibration which appears in the hypermultiplet moduli space of $D=4,\ \mc{N}=2$ supersymmetric gauge theory reduced to three dimensions \cite{Seiberg:1996nz} (see also the recent related work \cite{GMN}). In the same spirit, an alternative way to view to think of the axions $(\psi, \zeta^{I}, \tilde{\zeta}_I)$ is as coordinates on a ``twisted torus'' $\tilde{T}^{2h_{2,1}+3}$, where the twist arises from the fact that the Peccei-Quinn shift symmetries of the axions generate a non-abelian group, which in turn is isomorphic to a $(2h_{2,1}+3)$-dimensional Heisenberg group with center parametrized by $\psi$. Thus, shifts of $(\zeta^{I}, \tilde{\zeta}_I)$ yield an induced shift on $\psi$, giving rise to a twisted torus. We will discuss this Heisenberg group in more detail in Section \ref{cmap} after we restrict to a rigid Calabi-Yau threefold. 

To summarize, we conclude that the effect of switching on gravity is to turn the Seiberg-Witten torus fibration over the complex structure moduli space into a \emph{twisted} torus fibration over $\mc{M}_C(\mc{X})$. This twist is also responsible for changing the hypermultiplet moduli space from a hyperk\"ahler manifold into a quaternion-K\"ahler manifold.  
\subsection{Restricting to the Rigid Case}

We shall now restrict to the case of interest for our analysis, namely when
$X$ is a rigid Calabi-Yau threefold, i.e. $h_{2,1}(X)=0$. In the following, we
denote by $\mc{X}$ a rigid Calabi-Yau threefold, while reserving
the notation $X$ for generic Calabi-Yau threefolds. In this case the
cohomology group $H^3(\mc{X})$ simplifies considerably:
\beq
H^3(\mc{X})=H^{3,0}(\mc{X})+H^{0,3}(\mc{X}).
\eeq
Since $H^{3,0}(\mc{X})$ and $H^{0,3}(\mc{X})$ are both one-dimensional, we have only two ``universal'' 3-cycles, $\mc{A}$ and $\mc{B}$,
corresponding to the Hodge numbers $h_{3,0}(\mc{X})=h_{0,3}(\mc{X})=1$. No
restriction is imposed on the K\"ahler structure $h_{1,1}(\mc{X})$. Such
Calabi-Yau manifolds have been discussed both in the mathematical (see, e.g.,
\cite{Noriko}) and the physics literature (see, e.g.,
\cite{Candelas:1985en,Strominger:1985it,Candelas:1993nd,Bershadsky:1993cx}).

A peculiarity of compactification on a rigid Calabi-Yau manifold $\mc{X}$ is
that there exists no mirror manifold $\mc{Y}$, such that type IIA on $\mc{X}$
would be dual to type IIB on $\mc{Y}$. This is due to the fact that $\mc{Y}$
would necessarily have $h_{1,1}(\mc{Y})=h_{2,1}(\mc{X})=0$, which is not
possible for a K\"ahler manifold. Nevertheless, there exist dual
non-geometric type IIB compactifications, which are purely described in terms
of the conformal field theory (certain ``Landau-Ginzburg'' models) of the internal manifold
\cite{Candelas:1993nd}. Moreover, for generic Calabi-Yau manifolds $X$ which
admit K3-fibrations, the type IIA theory on $X$ is dual to heterotic
compactifications on $K3\times T^2$ \cite{Aspinwall:1995vk} (see also
\cite{Aspinwall:1996mn}). A curious feature of rigid Calabi-Yau
compactifications is that no such dual heterotic description exists, since
rigid Calabi-Yau threefolds do not admit K3-fibrations. For these reasons we shall in the present analysis solely restrict to the type IIA
picture, where our analysis has a clear geometric interpretation in terms of
the underlying rigid Calabi-Yau manifold $\mc{X}$.

The moduli space of type IIA on $\mc{X}$ is now given by
\beq
\mc{M}(\mc{X})=\mc{M}_{{}_{\mathrm{V}}}  \times \mc{M}_{{}_{\mathrm{UH}}},
\eeq
where the vector multiplet moduli space $\mc{M}_{{}_{\mathrm{V}}} $ is unchanged, while the
total hypermultiplet moduli space $\mc{M}_{{}_{\mathrm{UH}}}$ is a real
4-dimensional quaternion-K\"ahlermanifold corresponding to the
\emph{universal hypermultiplet} \cite{Cecotti:1988qn}. $\mc{M}_{{}_{\mathrm{UH}}}$
is universal in the sense that it does not depend on the details of the
compactification manifold $\mc{X}$. A first important note is that
$\mc{M}_{{}_{\mathrm{UH}}}$ contains the four-dimensional dilaton $e^{\phi}$, and
so is still sensitive to stringy quantum effects. In addition, the
ten-dimensional Ramond-Ramond 3-form $C_{(3)}$ gives rise to a universal
complex scalar $C=\zeta^{0}+i\tilde{\zeta}_0\equiv \chi+i\tilde{\chi}$ in $D=4$, associated with the
reduction of $C_{(3)}$ along the universal holomorphic and anti-holomorphic
3-cycles $\mc{A}$ and $\mc{B}$, respectively (see the next section for more details). Finally, the reduction of the
NS-NS 2-form $B_{(2)}$ gives rise to a 2-form $B_{\mu\nu}$ in $D=4$ which can
further be dualized to a third axionic scalar $\psi$.

It was shown in \cite{Cecotti:1988qn} that the four real scalars $\{e^{\phi},
\chi, \tilde{\chi}, \psi\}$ parametrize the coset space
\beq
\mc{M}_{{}_{\mathrm{UH}}}=SU(2,1)/(SU(2)\times U(1)),
\label{UHmodulispace}
\eeq
which is a quaternionic manifold with holonomy given by $SU(2)\times
U(1)$. Moreover, this coset space has the unusual feature of being
quaternion-K\"ahleras well as K\"ahler, a property not shared by the
hypermultiplet moduli space $\mc{M}_{{}_{\mathrm{H}}} $ appearing for type IIA
compactifications on generic Calabi-Yau threefolds $X$. While the quaternion-K\"ahler condition is preserved by supersymmetry, the
K\"ahler property of $\mc{M}_{{}_{\mathrm{UH}}}$ will generically be broken by
quantum effects \cite{AntoniadisMinasian2}. Hence, for this reason quantum corrections to
$\mc{M}_{{}_{\mathrm{UH}}}$ cannot be encoded directly in terms of corrections to
the classical K\"ahler potential.

At this stage it should be emphasized that even though the scalar fields just mentioned
occur universally in any Calabi-Yau manifold, it is not true that the ``universal hypermultiplet
manifold" \eqref{UHmodulispace} is a universal subsector of the hypermultiplet moduli space 
$\mc{M}_{\mathrm{H}}(X)$ when $X$ is non-rigid \cite{Aspinwall} (which is in contrast to the 
generalized universal hypermultiplet sector introduced in {\bf Paper IIV}).
However, in cases where  $\mc{M}$ is a symmetric space, it can often be written 
as a fiber bundle over $\mc{M}_{\mathrm{UH}}$. One example is type II string theory
compactified on $T^7$, where the moduli space can be written as the fiber bundle
\cite{Cecotti:1988qn,Aspinwall}
\beq
E_{7(7)}/(SU(8)/\mbb{Z}_2) \hs \to \hs
 \big[SU(2,1)/ (SU(2)\times U(1))\big] \times \big[SL(2, \mbb{R})/SO(2)\big]  .
\eeq
A similar decomposition occurs for very special $\mc{N}=2$ supergravity theories,
where the second factor on the r.h.s. is replaced by a non-compact version of the
5-dimensional U-duality group \cite{Gunaydin:2001bt,Gunaydin:2004md}. Thus, it is possible that the construction in this
paper generalizes beyond the case of rigid Calabi Yau threefolds.

\subsection{The Universal Hypermultiplet Metric and the $c$-Map}
\label{cmap}

As explained in Chapter \ref{Chapter:Introduction}, the hypermultiplet moduli space in type IIA on a Calabi-Yau manifold $\mc{X}$ is related to the vector multiplet moduli space in type IIB on $\mc{X}$ via the $c$-map \cite{Cecotti:1988qn}. Effectively this corresponds to performing a reduction on $S^1$ in type IIB and then T-dualizing the circle to the type IIA picture. This implies that classically the metric on the type IIA hypermultiplet moduli space is determined by the prepotential of the special-K\"ahler vector multiplet moduli space in type IIB. Let us now work this out for the case of interest, namely when $\mc{X}$ is rigid. Then, since $h_{2,1}(\mc{X})=0$, there is no complex structure moduli space and the vector multiplet moduli space in type IIB is therefore trivial. Nevertheless, as we shall see, the image of the $c$-map is non-trivial.

Since $\mc{X}$ has no complex structure deformations, its prepotential $F(X)$ is determined
from the special geometry relations
\beq
X=\int_{\mc{A}} \Omega , \qquad \qquad \f{\partial F}{\partial X} = \int_{\mc{B}} \Omega\ 
\label{Periods}
\eeq
to be quadratic, namely 
\beq
F(X)=\tau X^2 /2\ ,\qquad \tau := \f{\int_{\mc{B}} \Omega}{\int_{\mc{A}}\Omega}
\label{tau}
\eeq
where $\Omega\in H^{3,0}(\mc{X})$ is the holomorphic 3-form,  
$(\mc{A}, \mc{B})$ is an integral symplectic basis of  $H_{3}(\mc{X},\mbb{Z})$,
and $\tau$ is a fixed complex number, the period matrix. Fixing a dual cohomology basis $(\al, \be)$ of $H^3(\mc{X}, \mbb{Z})$ we may thus expand $\Omega=X(\al-\tau \be)$. 

The $c$-map then leads to the metric\footnote{Concretely, this metric may be obtained from the general $c$-map metric given in Eq. 4.31 of \cite{Alexandrov:2008nk} by setting $h_{2,1}=0$, $F(X)=\tau X^2/2$ and implementing the change of variables: $r=e^{-2\phi}$, $\zeta^{0}=-2\sqrt{2} \chi$, $\tilde\zeta_0 = 2\sqrt{2} \tilde\chi$, $\sigma =8\psi$.}
\beq ds^2_{\mc{M}_{\mathrm{UH}}}(\tau)= d\phi^2+e^{2\phi} \f{|d\tilde\chi+\tau d\chi|^2}{\Im \tau} + e^{4\phi} \big(d\psi +\chi
d\tilde{\chi}-\tilde{\chi}d\chi\big)^2.
\label{dsuh}
\eeq
In type IIA string theory compactified on $\mc{X}$, $e^{\phi}$ is the four-dimensional string coupling, $\psi$ is the NS-NS axion, dual to the 2-form $B_{(2)}$ in $D=4$,
and $(\chi, \tilde\chi)$ are the periods of the Ramond-Ramond 3-form $C_{(3)}$:
\beq
\chi = \int_{\mc{A}} C_{(3)}, \qquad \qquad \tilde\chi = \int_{\mc{B}} C_{(3)}.
\label{RRperiods}
\eeq
In the dual type IIB string theory on $\mc{X} \times S_1$, $e^{\phi}$ is instead the inverse radius of the circle in 4D Planck units, while $\chi,\tilde{\chi}$ are the components of the ten-dimensional Ramond-Ramond 4-form $C_{(4)}$ on $H^{3}(\mc{X},\mbb{R})\times S^1$. 
 
Classically, the family of metrics \eqref{dsuh}, parametrized by $\tau$, are all locally isometric
to the symmetric space $SU(2,1)/(SU(2)\times U(1))=\mbb{CH}^2$ 
(see Section \ref{Section:SU(2,1)} for details). A standard choice is  to take $\tau=i$, leading to the familiar form of the left-invariant metric on $\mbb{CH}^2$. However, at the quantum level the choice of $\tau$ is not inoccuous. Indeed, as already mentioned above, 
the Ramond-Ramond scalars $(\chi, \tilde\chi)$ are known \cite{Morrison:1995yi} to parametrize 
the intermediate Jacobian
\beq
J(\mc{X})=\f{H^{3}(\mc{X}, \mbb{C})}{H^{3}(\mc{X}, \mbb{Z})} .
\eeq
Equivalently they are subject to discrete identifications 
\beq
\label{chiid}
(\chi, \tilde\chi)\rightarrow (\chi+a, \tilde\chi+b), \hs a,b\in \mbb{Z}\ .
\eeq
If $\mc{X}$ is a rigid Calabi-Yau threefold, its intermediate Jacobian is an elliptic curve 
\beq
J(\mc{X}) = \mbb{C}/ ( \mbb{Z} + \tau \mbb{Z}) 
\eeq
where $\tau$ is the period matrix defined in (\ref{tau}). Thus, different choices of $\tau$
lead to locally isometric but globally inequivalent metrics. In the present work we shall restrict to the particular choice $\tau=i$,
 corresponding to rigid Calabi-Yau threefolds for which the intermediate Jacobian is a square torus $\mbb{C}/\mbb{Z}[i]$, where $\mbb{Z}[i]$  denotes the {Gaussian integers} $\{z\in\mbb{C}\ |\ \Re(z), \Im(z)\in \mbb{Z}\}$, corresponding to the ring of integers in the imaginary quadratic number field $\mbb{Q}(i)$. Mathematically, this choice implies  in particular  that $J(\mc{X})$
admits ``complex multiplication'', a notion that originates from the study of elliptic curves $\mbb{C}/(\mbb{Z}+\tau\mbb{Z})$, which are said to admit complex multiplication (or are of ``CM-type'') if and only if the modular parameter $\tau$ takes values in an imaginary quadratic extension of $\mbb{Q}$. %In this case, the associated $j$-invariant is always an algebraic integer. 
Many but not all rigid Calabi-Yau threefolds admit complex multiplication; a necessary and sufficient criterion is that the intermediate Jacobian of the Calabi-Yau threefold is of CM-type (see, e.g., \cite{Noriko} for a review). An example of a rigid Calabi-Yau threefold that does not admit complex multiplication is provided by  the hypersurface constructed in \cite{Schoen}.\footnote{We also note that complex multiplication has appeared previously in the physics literature in \cite{Moore:1998pn,Gukov:2002nw}.} For examples of rigid Calabi-Yau threefolds that admit complex multiplication by $\mbb{\mbb{Z}}[i]$, as is relevant in the present work, see \cite{Noriko}.

\section{On the Coset Space $SU(2,1)/(SU(2)\times U(1))$}
\label{Section:SU(2,1)}
 This section introduces our conventions for the group $SU(2,1)$ and a convenient parametrization of the coset space $SU(2,1)/(SU(2)\times U(1))$. We will also discuss the isomorphism between $SU(2,1)/(SU(2)\times U(1))$ and complex hyperbolic space $\mbb{CH}^2$. Finally, we introduce the Picard group $SU(2,1;\Zn[i])$ which
acts as a modular group on $\mbb{CH}^2$.

\subsection{The Group $SU(2,1)$ and its Lie Algebra $\mf{su}(2,1)$}
\label{sec:su21sec}

The Lie group $SU(2,1)$ is defined as a subgroup of the group $GL(3,\cx)$ of
invertible $(3\times 3)$ complex matrices via
\beq\label{SU21def}
SU(2,1) = \left\{ g \in GL(3,\cx)\,:\, g^\dagger \eta g = \eta\,\,
  \,\text{and}\, \det(g)=1\right\}\,.
\eeq
Here, the defining metric $\eta$ is given by
\beq\label{invariantmetric}
\eta = \left(\begin{array}{ccc}
0 & 0 &-i\\
0 & 1 & 0 \\
i & 0 & 0
\end{array}\right)
\eeq
and has signature $(++-)$. We note that the condition $g^\dagger \eta g=\eta$
already implies $|\det(g)|=1$ and so we can also think of $SU(2,1)$ as the set
of unitary matrices $U(2,1)$ modulo a pure phase, $SU(2,1)\cong PU(2,1)$, with the
projectivization $P$ referring to the equivalence relation $g\sim g
e^{i\alpha}$ for $\alpha \in [0,2\pi)$. The diagonal matrices
$e^{i\alpha}\text{diag}(1,1,1)$ form the center of the group $U(2,1)$.

The Lie group $SU(2,1)$ as defined in (\ref{SU21def}) has as Lie algebra
\beq\label{su21def}
\mf{su}(2,1) =\left\{ X \in \mf{gl}(3,\cx) \,:\, X^\dagger \eta + \eta X =0
  \,\,\text{and}\,\, \tr(X)=0\right\}\,.
\eeq
It is a Lie algebra of real dimension $8$ and a particular real form of
$\mf{sl}(3,\cx)$. It consists of four compact and four non-compact generators,
the maximal real torus is one-dimensional. Since we will have ample
opportunity to refer to specific generators we define the non-compact and
compact Cartan generators
\beq\label{cartans}
H = \left(\begin{array}{ccc}1&0&0\\0&0&0\\0&0&-1\end{array}\right)\,,\quad
J = \left(\begin{array}{ccc}i&0&0\\0&-2i&0\\0&0&i\end{array}\right)\,,
\eeq
the positive step operators
\beq\label{posstep}
X_1  = \left(\begin{array}{ccc}0&-1+i&0\\0&0&1-i\\0&0&0\end{array}\right)\,,\quad
\tilde{X}_1
=\left(\begin{array}{ccc}0&1+i&0\\0&0&1+i\\0&0&0\end{array}\right)\,,\quad
X_2 =  \left(\begin{array}{ccc}0&0&1\\0&0&0\\0&0&0\end{array}\right)\,,
\eeq\label{negstep}
and the negative step operators
\beq
Y_{-1}
=\left(\begin{array}{ccc}0&0&0\\1+i&0&0\\0&-1-i&0\end{array}\right)\,,\quad
\tilde{Y}_{-1}
=\left(\begin{array}{ccc}0&0&0\\-1+i&0&0\\0&-1+i&0\end{array}\right)\,,\quad
Y_{-2} = \left(\begin{array}{ccc}0&0&0\\0&0&0\\-1&0&0\end{array}\right)\,.
\eeq
The subscript refers to the eigenvalue under the adjoint action of the
non-compact Cartan generator $H$, e.g. $[H,X_1]=X_1$ --- the adjoint action of
the compact Cartan generator $J$ is not diagonalisable over the real
numbers. Furthermore, the generators satisfy
\beq\label{HeisenbergAlgebra}
X_2=-\frac14\left[X_1,\tilde{X}_1\right]\,,
\eeq
such that the positive step operators form a Heisenberg algebra.
Furthermore, the negative step operators $Y$ are minus the Hermitian conjugate
of the positive step operator $X$.

The Lie algebra $\mf{su}(2,1)$ has a natural five grading by the generator $H$
as a direct sum of vector spaces
\beq\label{5grading}
\mf{su}(2,1)=\mf{g}_{-2}\oplus \mf{g}_{-1} \oplus \mf{g}_0 \oplus \mf{g}_1
\oplus \mf{g}_2\,,
\eeq
with
\beqa
\mf{g}_{-2} = \mbb{R}Y_{-2}\,,\,
\mf{g}_{-1} = \mbb{R}Y_{-1}\oplus \mbb{R}\tilde{Y}_{-1}\,,\,
\mf{g}_0 = \mbb{R}H\oplus \mbb{R} J \,,\,
\mf{g}_1  = \mbb{R}X_{1}\oplus \mbb{R}\tilde{X}_{1}\,,\,
\mf{g}_{2} = \mbb{R}X_{2}\,.
\eqa
One sees that the $H$-eigenspaces with eigenvalue $\pm 1$ are degenerate. This
is a characteristic feature of the reduced root system $BC_1$ underlying the
real form $\mf{su}(2,1)$ of $\mf{sl}(3,\cx)$. There is a single root $\alpha$
since the real rank of $\mf{su}(2,1)$ is one, and there are non-trivial root
spaces $\mf{g}_1$ and $\mf{g}_2$ corresponding to $\alpha$ and $2\alpha$,
respectively.\footnote{A discussion of the restricted root
  system can for example be found  in \cite{LivingReview}.}
The $\mf{sl}(2,\mbb{R})$ subalgebra associated with the
$2\alpha$ root space is  canonically normalised and can be given a standard
basis for example with $H$, $E=-X_2$ and $F=Y_{-2}$, so that
$[E,F]=H$. The corresponding $SL(2,\mbb{R})$ subgroup of $SU(2,1)$ is given by
matrices of the form
\beq\label{sl2sub}
\left\{ \left(\begin{array}{ccc}
a & 0 & b\\
0 & 1 & 0\\
c & 0 & d\end{array}\right) \,:\,
a,b,c,d\in \mbb{R}\,\,\text{and}\,\, ad-bc=1\right\}
\subset SU(2,1)\,.
\eeq
Under this embedding, the fundamental representation of $SU(2,1)$ decomposes as $3=2\oplus 1$.
There exists a second, non-regular embedding of $SL(2,\mbb{R})$ inside $SU(2,1)$,  consisting
of matrices of the form
\beq\label{sl2sub2}
SL(2,\mbb{R}) = \left\{ \left(\begin{array}{ccc}
a^2 & (-1+i)a b & i b^2\\
(-1-i)a c & ad+bc & (1-i)b d\\
-i c^2 & (1+i)c d & d^2\end{array}\right) \,:\,
a,b,c,d\in \mbb{R}\,\,\text{and}\,\, ad-bc=1\right\}\,.
\eeq
Under this embedding, the fundamental representation of $SU(2,1)$ remains irreducible.
The two subgroups (\ref{sl2sub}) and (\ref{sl2sub2}) together generate the whole of $SU(2,1)$.

The Iwasawa decomposition of the Lie algebra $\mf{su}(2,1)$ reads
\beq \label{su21iwa}
\mf{su}(2,1)=\mf{n}_+ \oplus \mf{a} \oplus \mf{k} ,
\eeq
where the non-compact (abelian) Cartan subalgebra $\mf{a}=\mbb{R}H$
while the nilpotent subspace
$\mf{n}_+=\mbb{R}X_1\oplus\mbb{R}\tilde{X}_1\oplus X_2$ is spanned by the
positive step operators. The compact subalgebra of
$\mf{su}(2,1)$ is $\mf{k}=\mf{su}(2)\oplus\mf{u}(1)$ as a direct sum of Lie
algebras.\footnote{By contrast, the Iwasawa decomposition (\ref{su21iwa}) is
  only a direct sum of vector spaces and not of Lie algebras.}
The generators of $\mf{su}(2)$ and $\mf{u}(1)$ are given explicitly by the anti-Hermitian matrices
\beqa
\hat{K}_1 &=& \frac14\left(X_{1} + Y_{-1}\right)\,,\quad
\hat{K}_2 = \frac14\left(\tilde{X}_{1} + \tilde{Y}_{-1}\right) \,,\quad
\hat{K}_3 =  \frac14\left(X_{2} + Y_{-2}  + J\right) \,,\nn \\
\hat{J} &=& \frac{3}{4}\left( X_{2} + Y_{-2}\right)-\frac14 J\,.
\eqa
These satisfy $\left[\hat{J} ,\hat{K}_i\right] =0$ and
$\left[\hat{K}_i,\hat{K}_j\right] = -\epsilon_{ijk} \hat{K}_k$.

\subsection{Complex Hyperbolic Space}

The group $SU(2,1)$ has a natural action on complex hyperbolic two-space which
we model here using the unbounded hyperquadric model \cite{FrancsicsLax} of
complex dimension two
\beq\label{chs}
\cxh^2 = \left\{ \cZ=(z_1,z_2)\in \cx^2 \,:\,\cF(\cZ)>0\right\}\,,
\eeq
where the function $\cF:\cx^2\to \mbb{R}$ is defined by
\beq\label{KaehlerArgument}
\cF(\cZ) := \Im(z_1) - \frac12 |z_2|^2 >0\,,
\eeq
and is sometimes referred to as the ``height
function''. The condition $\cF(\cZ)>0$ is the generalization of the usual
condition $\Im(\tau)>0$ for a complex number $\tau$ to lie on the upper half
plane of complex dimension one, and we see that for $z_2=0$ the space $\cxh^2$
contains the usual upper half plane. We will refer to the space $\cxh^2$
defined in (\ref{chs}) as the {\em complex hyperbolic space}, or the {\em complex upper half plane}.
The link to the upper half plane will be made more explicitly below in
Section~\ref{sec:su21coset}.

The action of $SU(2,1)$ on $\cZ\in\cxh^2$ can now be written in the compact
fractional linear form
\beq\label{su21act}
g\cdot \cZ = \frac{A\cZ+B}{C\cZ+D}\quad \quad \text{for}\quad
g=\left(\begin{array}{cc}A&B\\C&D\end{array}\right)\,,
\eeq
where the blocks $A$, $B$, $C$ and $D$ have the sizes $(2\times 2)$, $(2\times 1)$, $(1\times
2)$ and $(1\times 1)$, respectively, so that the denominator is
a complex number. In order to verify that (\ref{su21act}) defines an action of
$SU(2,1)$ on complex hyperbolic space one needs to check that it preserves the
defining condition in (\ref{chs}). This can be seen from computing
\beq\label{trmheight}
\cF(g\cdot \cZ) = \frac{\cF(\cZ)}{|C\cZ+D|^2}\,,
\eeq
which again is a generalization of the usual $SL(2, \mbb{Z})$-transformation of $\Im(\tau)$ on the (real)
upper half plane. Furthermore, one needs to check the group property of the
action which is straight-forward. In fact, when verifying (\ref{trmheight})
one only requires the condition $g^\dagger\eta g =\eta$ so that
(\ref{su21act}) defines an action of all of $U(2,1)$ on complex hyperbolic
two-space. 

The K\"ahler metric on $\cxh^2$ can be written in terms of the following
K\"ahler potential
\beq\label{kahlerpot}
K(\cZ) = -\log \cF(\cZ)\,.
\eeq
Written out in terms of the two complex coordinates $\cZ=(z_1,z_2)$ this gives
the Euclidean metric
\beq\label{metricZ}
ds^2 = \frac14 \cF^{-2}\left[dz_1d\bar{z}_1 + iz_2dz_1d\bar{z}_2
-i \bar{z}_2 dz_2d\bar{z}_1 +2\Im(z_1)dz_2d\bar{z}_2\right]\,.
\eeq
The group $SU(2,1)$ acts isometrically on $\cxh^2$.

\subsection{Relation to the Scalar Coset Manifold $SU(2,1)/(SU(2)\times U(1))$}
\label{sec:su21coset}

The complex hyperbolic upper half plane is also isomorphic to the Hermitian symmetric space
\beq\label{su21cos}
\cxh^2 \cong SU(2,1)/(SU(2)\times U(1))\,,
\eeq
where the right hand side should properly be restricted to the connected
component of the identity. The Hermitian symmetric space is of real dimension
four and can be parametrized by four real variables $\{\phi, \chi, \tilde\chi, \psi\}$ in triangular gauge, using
the Iwasawa decomposition (\ref{su21iwa}), as
\beqa\label{cosetel}
\cV =  e^{ \chi X_1 + \tilde\chi\tilde{X}_1 +2\psi X_2} e^{-\phi  H}
= \left(\begin{array}{ccc}
e^{-\phi} & \tilde\chi-\chi +i (\chi+\tilde\chi) &
e^{\phi}\left(2\psi+i(\chi^2+\tilde\chi^2)\right)\\
0 & 1 & e^\phi \left(\chi+\tilde\chi+i(\tilde\chi-\chi)\right)\\
0 & 0 & e^{\phi}
\end{array}\right)\,.
\eqa
The symmetric space is a right coset in our conventions and the element $\cV$
transforms as
$\cV\to g\cV k^{-1}$ with $g\in SU(2,1)$ and $k\in SU(2)\times U(1)$.  The
four scalar fields can take arbitrary real values.

It is convenient to define the Hermitian matrix
\beq\label{kdef}
\cosm = \cV\cV^\dagger
\eeq
that transforms as $\cosm\to g\cosm g^\dagger$ under the action of $g\in
SU(2,1)$. Explicitly, this matrix reads
\beq
\mc{K}=\left( \begin{array}{ccc}
e^{-2\phi}+|\lambda|^2+e^{2\phi}|\ga|^2 & i\bar\lambda+e^{2\phi} \bar\lambda\ga & e^{2\phi} \ga \\
-i{\lambda}+e^{2\phi} {\lambda}\bar{\ga} & 1+e^{2\phi} |\lambda|^2 & e^{2\phi}{\lambda}\\
e^{2\phi}\bar{\ga} & e^{2\phi} \bar\lambda & e^{2\phi} \\
\end{array} \right),
\label{generalizedmetric}
\eeq
where, for later convenience, we defined the complex variables
\beq\label{eisenvars}
\lambda:=\chi+\tilde\chi+i(\tilde\chi-\chi), \qquad \ga:= 2\psi+\f{i}{2}|\lambda|^2.
\eeq
From $\mc{K}$ one can define the metric on the symmetric space via
\beq
ds^2 = -\frac18\text{tr} \left(d\cosm \,d(\cosm^{-1})\right)
= \frac18\text{tr}\left( \cV^{-1}d\cV + (\cV^{-1}d\cV)^\dagger\right)^2\,.
\eeq
Working this out for the coset element (\ref{cosetel}) one finds the following
$SU(2,1)$ invariant metric
\beq\label{metricscalars}
ds^2 = d\phi^2 + e^{2\phi}(d\chi^2+d\tilde\chi^2)
  + e^{4\phi}(d\psi+\chi d\tilde\chi-\tilde\chi d\chi)^2\,.
\eeq
Comparing (\ref{metricscalars}) to (\ref{metricZ}) leads to the identification
\beqa\label{scalarsZ}
z_1 &=& 2\psi + i \left(e^{-2\phi}+\frac12 |z_2|^2\right) = 2\psi+i\left(e^{-2\phi}+\chi^2+\tilde\chi^2\right)\,,\nn\\
z_2 &=& \chi + \tilde\chi + i(\tilde\chi-\chi)\,.
\eqa
Note that $z_1=\ga +ie^{-2\phi}$ and $z_2=\lambda$. In Section
\ref{Section:Eisenstein} it will prove to be convenient to use the complex
variable $\ga$ rather than $z_1$.

We see that the condition $\cF(\cZ)>0$ is satisfied in the parametrization
(\ref{scalarsZ}) of $\cZ$ since $\cF(\cZ)=e^{-2\phi}$. The advantage of using
the complex hyperbolic upper half plane rather than the scalar coset in the
form (\ref{cosetel}) is that the action of $SU(2,1)$ is simpler to
evaluate. The action of $g\in SU(2,1)$ on $\cV$ requires a compensating
transformation $k\in SU(2)\times U(1)$ to restore the triangular gauge chosen
in (\ref{cosetel}) whenever $g$ is itself not of triangular form. Finding this
compensating transformation in general can be involved. Since it has been
implicitly carried out in the non-linear action (\ref{su21act}) one does not
require the precise form of the compensator. We note that setting $z_2=0$
gives back the real upper half plane but the parametrization (\ref{scalarsZ})
appears to have additional factors of two compared to the usual expressions at
first sight. However, these factors have a physical significance as will be
discussed further below. Besides being a quaternion-K\"ahler manifold,
$SU(2,1)/(SU(2)\times U(1))$ also has the stronger property of being K\"ahler,
as is clear from (\ref{kahlerpot}).

In the variables $\cZ=(z_1,z_2)$ given by (\ref{scalarsZ}), the matrix $\cosm$
of (\ref{kdef}) takes the simple form
\beq\label{cosetmatrix}
\cosm  = \tcosm+\eta \,,
\eeq
where $\eta$ is the defining matrix of $SU(2,1)$ given in
(\ref{invariantmetric}) and
\beq\label{ktildematrix}
\tcosm =  e^{2\phi}\left(\begin{array}{ccc}
|z_1|^2 & z_1\bar{z}_2 & z_1\\
\bar{z}_1 z_2 & |z_2|^2&z_2 \\
\bar{z}_1& \bar{z}_2 & 1
\end{array}\right)\,,
\eeq
bearing in mind that $e^{2\phi}=\cF(\cZ)^{-1}$.

\subsection{Coset Transformations and Subgroups of $SU(2,1)$}
\label{subsec:CosetTransf}

We now study the effect of some particular elements of $SU(2,1)$ on complex hyperbolic two-space. The specific transformations we investigate are the ones with an immediate physical interpretation.

\subsubsection*{Heisenberg Translations}
 Let $ N  $ denote the exponential of the nilpotent algebra of positive step operators $\mf{n}_+$. We define the following elements of $ N  $
\beq\label{heisgen}
T_1 = \left(\begin{array}{ccc}
1 & -1+i & i \\
0 & 1 & 1-i\\
0 &0 & 1\end{array}\right)\,,
\quad
\tilde{T}_1  = \left(\begin{array}{ccc}
1 & 1+i & i \\
0 & 1 & 1+i\\
0 &0 & 1\end{array}\right)\,,
\quad
T_2 = \left(\begin{array}{ccc}
1 & 0 & 1 \\
0 & 1 & 0\\
0 &0 & 1\end{array}\right)\,.
\eeq
These are such that $T_1=\exp(X_1)$ etc. Any element  $n\in N  $ then can be written as
\beqa\label{GeneralHeisenberg}
n &=& (T_1)^a (\tilde{T}_1)^b (T_2)^{c+2ab} = e^{a X_1 + b\tilde{X}_1 +c X_2} \nn\\
&=& \left(\begin{array}{ccc}
1 & a(-1+i) +b(1+i) & c +i (a^2+b^2)\\
0 &1 & a(1-i) +b(1+i) \\
0 &0 &1\end{array}\right)
\eqa
for $a,b,c\in\mbb{R}$. The effect of this transformation on $\cZ=(z_1,z_2)$ is
\beqa\label{HeisenbergAction}
z_1 & \longmapsto& z_1 + \big[a(-1+i)+b(1+i)\big] z_2 +c+i(a^2+b^2)\,,\nn\\
z_2 & \longmapsto& z_2 + a(1-i)+b(1+i)\,,
\eqa
or in terms of the four scalars fields of (\ref{cosetel})
\beqa\label{heisshiftsscalars}
\phi & \longmapsto& \phi\,,\nn\\
\chi & \longmapsto& \chi+a\,,\nn\\
\tilde\chi & \longmapsto& \tilde\chi +b\,,\nn\\
\psi & \longmapsto&\psi +\frac12c -a\tilde\chi +b \chi\,.
\eqa
The appearance of the shift parameters $a$ and $b$ in the transformation of
$\psi$ is due to the non-abelian structure of $\mf{n}_+$ given by the
Heisenberg algebra (\ref{HeisenbergAlgebra}). This effect is also evident in
the first line of the expression (\ref{GeneralHeisenberg}) for the general
element of $ N  $. From the point of view of the coset, the Heisenberg
translations do not require any compensating transformation as they preserve
the Borel gauge.

\subsubsection*{Rotations}
Rotations are generated by the compact Cartan element $J$ of
$\mf{su}(2,1)$ given in (\ref{cartans}). Let
\beq\label{rot}
R = \exp(\pi J/2) = \left(\begin{array}{ccc}
i & 0 & 0\\
0 & -1& 0\\
0 & 0 & i \end{array}\right),
\eeq
then the most general transformation of this type is given by $R^m$, for
$m\in [0,4)$, and acts on $\cZ=(z_1,z_2)$ via
\beq
z_1 \to  z_1\,,\quad
z_2 \to  e^{i\pi\sigma/2} z_2\,.
\eeq
In terms of the four scalar fields this transformation reads
\beqa
\phi & \longmapsto& \phi\,,\nn\\
\chi & \longmapsto& \cos(\pi\sigma/2)\chi-\sin(\pi\sigma/2)\tilde\chi\,,\nn\\
\tilde\chi & \longmapsto& \sin(\pi\sigma/2)\chi+\cos(\pi\sigma/2)\tilde\chi\,,\nn\\
\psi & \longmapsto&\psi\,
\eqa
and so rotates the two scalars $\chi$ and $\tilde\chi$ among each other while
leaving the other two invariant.  The compensating transformation to restore
the Borel gauge for the coset element (\ref{cosetel}) is $k=R^m$.

\subsubsection*{Involution}

 The last transformation of interest is the following involution
% \cite{FrancsicsLax}
\beq\label{inv}
S = \left(
\begin{array}{ccc}
0 & 0 & i \\
0 & -1 & 0 \\
-i & 0 & 0
\end{array}
\right),
\eeq
which acts on $\cZ=(z_1, z_2)$ according to
\beq
z_1  \mapsto  -\frac{1}{z_1}\,,\quad
z_2  \mapsto -i\frac{z_2}{z_1}\,,
\eeq
making it apparent that this is an inversion of $z_1$. For the real scalars
themselves we find the following transformation
\beqa
\phi & \longmapsto & -\frac{1}{2}\ln\left[
  \frac{e^{-2\phi}}{4\psi^2 +[e^{-2\phi}+ \chi^2 + \tchi^2]^2 }
\right]\,, \nn \\
\chi &  \longmapsto & \phantom{-}\frac{2\psi\tchi - (e^{-2\phi}+\chi^2 +
  \tchi^2 )\chi}{4\psi^2+ [e^{-2\phi}+\chi^2 + \tchi^2]^2}\,,\nn\\
\tchi &  \longmapsto & \phantom{-}\frac{2\psi\chi + (e^{-2\phi}+\chi^2 +
  \tchi^2 )\tchi}{4\psi^2+ [e^{-2\phi}+\chi^2 + \tchi^2]^2}\,, \nn \\
\psi &  \longmapsto & -\frac{\psi}{ 4\psi^2+[e^{-2\phi}+\chi^2 + \tchi^2]^2 }\,.
\eqa
It is straightforward to check that the required compensating transformation
in this case indeed belongs to the maximal compact subgroup $SU(2) \times
U(1)$.

We note that the involution (\ref{inv}) can also be written as
\beq\label{inv2}
S= e^{Y_{-2}} e^{X_2} e^{Y_{-2}} R\,,
\eeq
using the rotation matrix (\ref{rot}). The expression (\ref{inv2}) is almost
the expression of Kac~\cite{Kac} for Weyl reflections (see
Lemma~3.8 in \cite{Kac}). There is a difference in our expression for $S$ compared to \cite{Kac} since we are dealing
with a non-split real form. Firstly, one should take the generators
corresponding to the canonically normalised, split
$\mf{sl}(2,\mbb{R})\subset\mf{su}(2,1)$, that is, use the generators
$Y_{-2}$ and $-X_2$ of the graded decomposition (\ref{5grading}). Secondly,
one has to verify the action of the resulting transformation on the degree one
subspaces
in (\ref{5grading}) that are degenerate. Demanding that no additional
transformation on their basis is introduced requires the inclusion of $R$ in
(\ref{inv2}) since then $SX_1S^{-1}=Y_{-1}$ (and not $\tilde{Y}_{-1}$). We also
note that without the inclusion of $R$ the transformation (\ref{inv2}) would
be of order four and not two as required for a reflection.
The reduced root system $BC_1$ of $\mf{su}(2,1)$ is of (real) rank one and
therefore its Weyl group is generated by a single reflection, namely the one
displayed in (\ref{inv}). Therefore one should call the Weyl group
\beq
\cW (\mf{su}(2,1)) = \cW(BC_1) \cong \Zn_2\,.
\eeq
In the context of U-duality symmetries for torus compactifications it is often
the Weyl group that is retained as a minimal discrete symmetry group acting on
BPS states \cite{Elitzur:1997zn,Obers:1998fb}. In the present case this would
correspond to studying the action of the involution (\ref{inv}) on BPS
states.

\subsection{The Picard Modular Group}
\label{Section:PicardModularGroup}

We finally discuss the Picard modular group $SU(2, 1;\mbb{Z}[i])$. This group
can be defined as the intersection \cite{FrancsicsLax}
\beq
SU(2, 1;\mbb{Z}[i]) := SU(2,1)\cap SL(3,\mbb{Z}[i]),
\eeq
where $\mbb{Z}[i]$ denotes the Gaussian integers
\beq
\mbb{Z}[i]=\{ z \in \mbb{C} \,:\, \Re(z), \Im(z)\in \mbb{Z}\}.
\eeq
This definition implies that any element $g \in SU(2,1)$ which has only
Gaussian integer matrix entries belongs to $SU(2, 1;\mbb{Z}[i]) $. In view of
the discussion of $PU(2,1)\cong SU(2,1)$ the Picard modular group can also be
called $PU(2,1;\Zn[i])$.\footnote{The nomenclature ``Picard group'' is not unique, in fact our Picard group is a member of a family of similar groups $PSU(1,n+1;\mbb{Z}[i])$ of which the case $n=0$, corresponding to $PSL(2,\mbb{Z}[i])$ is also often called \emph{the} Picard group. In this chapter we will always mean $SU(2,1;\mbb{Z}[i])$ when speaking of the Picard group.}

Let us now examine the particular $SU(2,1)$-transformations of the previous
subsection to check whether they belong to the Picard group. The Heisenberg group $ N  \subset SU(2,1)$ contains a subgroup
$ N  (\mbb{Z}):= N  \cap \pg $. By inspection of
Eq. (\ref{GeneralHeisenberg}) we see that $ N  (\mbb{Z})$ must be of the
form
\beq
 N  (\mbb{Z})=\big\{ e^{a X_{1}+b\tilde{X}_{1}+cX_{2}}  \hs :\hs a,b,c
\in\mbb{Z} \big\}\,.
\eeq
In view of (\ref{GeneralHeisenberg}), a natural set of generators for
$ N  (\mbb{Z})$ is given by the three matrices in (\ref{heisgen}) $T_1$,
$\tilde{T}_1$ and $T_2$. The action of these discrete shifts are then as given in
(\ref{heisshiftsscalars}) with parameters $a,b,c\in\Zn$. The translations
(\ref{heisgen}) are of infinite order in the Picard modular group.

The rotation $R^m$ defined in (\ref{rot}) is only an element of $SU(2,1;\mbb{Z}[i])$ for the
discrete values of the exponent $m=0,1,2,3$, and $R$ is an element of
order $4$ in the Picard modular group. The action of $R$ on the scalar fields
is
\beq
R\,:\quad (\chi, \tilde{\chi}) \longmapsto (-\tilde\chi, \chi).
\label{emdual}
\eeq
Physically speaking, this corresponds to electric-magnetic duality, which is
expected to be preserved in the quantum theory \cite{BeckerBecker}.

Finally, we will examine the involution $S$ in Eq. (\ref{inv}). Clearly, the
involution is an element of the Picard modular group. The involution $S$ is of
order $2$ in the Picard modular group. As already noted above, the involution (\ref{inv}) corresponds to the Weyl
reflection of the restricted root system $BC_1$ of the non-split real form
$\mf{su}(2,1)$. The Weyl reflection is associated with the (long) root
$2\alpha$. We can also give an interpretation to the rotation $R$. This is a
transformation that rotates within the degenerate, two-dimensional $\alpha$
root space, spanned by the generators $X_1$ and $\tilde{X}_1$.

The Picard modular group acts discontinuously on the complex hyperbolic
space
$\cxh^2$. A fundamental domain for its action has been given in~\cite{FrancsicsLax}. Recently, the following has also be proven:

\vspace{5mm}
\noindent \rule[0.1in]{15.5cm}{0.3mm} \\
%\vspace{.2cm}
\noindent {\bf Theorem \cite{Falbel:2009cf}:}
\noindent {\it The Picard modular group $SU(2,1;\mbb{Z}[i])$ is
generated by the translations ${T}_1$ and $T_2$, together with the
rotation $R$ and the involution $S$.}

\vspace{.2cm}

\noindent \rule[0.1in]{15.5cm}{0.3mm}

\noindent Since the two translations $T_1$ and $\tilde{T}_1$ are related
through ``electric-magnetic duality'' by $\tilde{T}_1=RT_1R^{-1}$, one
may
equivalently choose either of the translations $T_1$ or $\tilde{T}_1$
associated with the $\alpha$ root space in the theorem. Since all three
translations $T_1$, $\tilde{T}_1$ and $T_2$ will turn out to have a
clear
physical interpretation we present the Picard modular group as generated
(non-minimally) by the following five elements:
 \beqa
 T_1=\left(\begin{array}{ccc}
1 & -1+i & i \\
 0& 1 & 1-i \\
0 &0 & 1†\\
 \end{array}\right), &
  \tilde{T}_1=\left(\begin{array}{ccc}
1 & 1+i & i \\
 0& 1 & 1+i \\
0 & 0& 1†\\
 \end{array}\right),  &
 T_2=\left(\begin{array}{ccc}
1 &  0& 1 \\
 0& 1 &0 \\
 0& 0& 1†\\
 \end{array}\right),
 \nn \\
 R=\left(\begin{array}{ccc}
i & 0& 0\\
0 & -1 &0 \\
0 &0 & i \\
 \end{array}\right), &
 S=\left(\begin{array}{ccc}
0 &0 & i \\
 0 & -1 & 0\\
  -i &0 & 0\\
  \end{array}\right).
  \eqa

In accordance with the $SL(2,\mbb{R})$ subgroup identified in (\ref{sl2sub})
we note that there is an $SL(2,\mbb{Z})\subset \pg $ that acts on
the slice $z_2=0$ of complex hyperbolic space as the usual modular group on
the remaining variable $z_1$.

\section{Eisenstein Series for the Picard Modular Group}
\label{Section:Eisenstein}

In this section we shall construct Eisenstein series for the Picard modular
group in the principal continuous series representation of $SU(2,1)$. We 
shall give three different constructions, which, despite being equivalent, mutually enlighten each other. 
In Section \ref{LatticeConstruction} we construct a manifestly $SU(2,1;\mbb{Z}[i])$-invariant function on $SU(2,1)/(SU(2)\times U(1))$ by summing over points in  
the three-dimensional Gaussian lattice $\mbb{Z}[i]^3$. This produces a non-holomorphic Eisenstein series $\mc{E}_s$, parametrized by $s$, which will be the central object of study in the remainder of this paper. In Section \ref{Section:PoincareSeries},  we use the isomorphism between the coset space $SU(2,1)/(SU(2)\times U(1))$ and the complex upper half plane $\mbb{CH}^2$ to construct a Poincar\'e series $\mc{P}_s$ on $\mbb{CH}^2$. This turns out to be identical to $\mc{E}_s$ up to an $s$-dependent Dedekind zeta function
factor. For completeness, in Section \ref{Section:pAdicConstruction} we give a third construction using standard 
adelic techniques, which illuminates the representation-theoretic nature of  $\mc{E}_s$.

\subsection{Lattice Construction and Quadratic Constraint}
\label{LatticeConstruction}
Our first method will be the one discussed in \cite{ObersPioline}. A
non-holomorphic Eisenstein series, of order $s$, on the quotient
\beq
\pg \bas SU(2,1)/ (SU(2)\times U(1)),
\eeq
can then be constructed in the following way
\beq
\mc{E}_s(\mc{K}):= \sum_{\vec{\om} \in \mbb{Z}[i]^3\atop \vec{\om}^{\dagger}\wedge
\vec{\om}=0} \Big[\vec{\om}^{\dagger}\cdot \mc{K} \cdot \vec{\om}\Big]^{-s},
\label{EisensteinPicard}
\eeq
where $\mc{K}=\mc{V}\mc{V}^{\dagger}$ is the ``generalized metric'' which was constructed explicitly in Eq. (\ref{generalizedmetric}), and $\vec{\om}^{\dagger}\wedge\vec{\om}=0$ is a quadratic constraint
on the Gaussian integers $\vec{\om}\in\mbb{Z}[i]^3$ which ensures that the
Eisenstein series is an eigenfunction of the Laplacian on the coset space
$SU(2,1)/(SU(2)\times U(1))$ \cite{ObersPioline}. We will discuss this constraint in detail below. Moreover, it is always understood that the
summation is restricted such that the vector $\vec{\om}=(0,0,0)$ is
excluded. Writing out Eq. (\ref{EisensteinPicard}) explicitly yields
\beq
\mc{E}_s(\phi, \lambda,
\ga)=\sum_{\vec{\om}\in\mbb{Z}[i]^3\atop \vec{\om}^{\dagger}\wedge
\vec{\om}=0}
e^{-2s\phi}\Big[|\bar\om_1+\bar\om_2\bar\lambda+\bar\om_3\bar{\ga}|^2+e^{-2\phi}|\bar\om_2-i\om_3{\lambda}|^2+e^{-4\phi}|\om_3|^2\Big]^{-s}.
\label{EisensteinPicard2}
\eeq
The variables $\lambda$ and $\gamma$ were defined as functions of $\cZ=(z_1,z_2)$ in (\ref{eisenvars}). To better understand the quadratic constraint
$\delta(\vec{\om}^{\dagger}\wedge\vec{\om})$, it is illuminating to utilize
the isomorphism between the coset space $SU(2,1)/(SU(2)\times U(1))$ and the complex hyperbolic space $\mbb{CH}^2$, as discussed in Section~\ref{sec:su21coset}. We recall from (\ref{ktildematrix}) that in terms of the variable $\mc{Z}=(z_1, z_2)\in \mbb{CH}^2$, the matrix $\mc{K}$ reads
\beq
\mc{K}=\tilde{\mc{K}}+\eta,
\eeq
where $\eta$ is the $SU(2,1)$-invariant metric, Eq. (\ref{invariantmetric}), and the matrix $\tilde{\mc{K}}$ is given by
\beq
\tcosm =  e^{2\phi}\left(\begin{array}{ccc}
|z_1|^2 & z_1\bar{z}_2 & z_1\\
\bar{z}_1 z_2 & |z_2|^2&z_2 \\
\bar{z}_1& \bar{z}_2 & 1
\end{array}\right).
\eeq
In this new parametrization, the Eisenstein series becomes
\beqa
 \mc{E}_s(\mc{Z})&=& \sum_{\vec{\om} \in \mbb{Z}[i]^3\atop \vec{\om}^{\dagger}\wedge
 \vec{\om}}
 \Big[\vec{\om}^{\dagger}\cdot \tilde{\mc{K}} \cdot
 \vec{\om}+\vec{\om}^{\dagger}\cdot \eta \cdot \vec{\om}\Big]^{-s}
 \nn \\
 &=& \sum_{\vec{\om} \in \mbb{Z}[i]^3\atop \vec{\om}^{\dagger}\wedge
 \vec{\om}} \Big[e^{2\phi}|\om_1+\om_2 z_2+\om_3 {z}_1|^2+\vec{\om}^{\dagger}\cdot \eta \cdot \vec{\om}\Big]^{-s}.
 \label{EisensteinPicard3}
 \eqa
This decomposition of the Eisenstein series can be understood as follows
\cite{ObersPioline}. By virtue of Eq. (\ref{cosetel}) the coset
representative $\mc{V}\in SU(2,1)/(SU(2)\times U(1))$ is constructed in the fundamental
representation $\mc{R}$ of $SU(2,1)$. This implies that the generalized metric
$\mc{K}=\mc{V}\mc{V}^{\dagger}$ transforms in the symmetric tensor product
$\mc{R}\otimes_s\mc{R}$. The contraction of $\mc{K}$ with the lattice vectors
$\vec{\om}$ is only an eigenfunction of the Laplacian on $SU(2,1)/(SU(2)\times U(1))$ if this
contraction is irreducible. The split of the summand in
Eq. (\ref{EisensteinPicard3}) precisely shows that this is not the case;
$\vec{\om}^{\dagger}\cdot \mc{K}\cdot \vec{\om}$ is reducible and its
irreducible components are the two terms in Eq. (\ref{EisensteinPicard3}). The
second term is the trivial (moduli-independent) component of the tensor
product, and must be projected out in order to obtain an eigenfunction of the
Laplacian. The Laplacian will be discussed in detail in Section \ref{Section:PoincareSeries} below. By the reasoning above we find that the quadratic constraint
$\delta(\vec{\om}^{\dagger}\wedge \vec{\om})$ must be
\beq
\vec{\om}^{\dagger}\wedge \vec{\om}:= \vec{\om}^{\dagger}\cdot \eta \cdot \vec{\om}=|\om_2|^2-2\Im(\om_1\bar{\om}_3)=0.
\label{constraint}
\eeq
The Eisenstein series then finally becomes
\beq
 \mc{E}_s(\mc{Z})=\sum_{\vec{\om}\in\mbb{Z}[i]^3\atop \vec{\om}^{\dagger}\cdot \eta \cdot \vec{\om}}e^{-2s\phi}|\om_1+\om_2 {z}_2+\om_3 {z}_1|^{-2s}.
 \label{EisensteinPicard4}
 \eeq
Setting 
\beq
\om_1=m_1+im_2, \qquad \om_2=n_1+in_2, \qquad \om_3=p_1+ip_2,
\eeq 
this may be rewritten as a sum over six integers $m_i,n_i,p_i$, not all vanishing, subject to the 
constraint 
\beq
\vec{\om}^{\dagger}\cdot \eta\cdot \vec{\om}
=n_1^2+n_2^2+2m_1p_2-2m_2p_1=0 .
\label{App:Constraint}
\eeq
The Eisenstein series defined in (\ref{EisensteinPicard}) converges absolutely for $\Re(s)>2$. 

It will also be convenient to have the equivalent form of the Eisenstein
 series in the original variables $\phi, \lambda$ and $\ga$. This is of course
 just Eq. (\ref{EisensteinPicard2}) with the explicit constraint inserted. Imposing the (real) quadratic constraint $|\om_2|^2-2\Im(\om_1\bar{\om}_3)=0$
eliminates one of the summation variables, so that we are effectively summing
over only five integers.

\subsection{Poincar\'e Series on the Complex Upper Half Plane}
\label{Section:PoincareSeries}
As we have seen in Chapter \ref{Chapter:ConstructingAutomorphicForms}, a standard way of constructing non-holomorphic Eisenstein series on a symmetric space $G/K$ is in terms of \emph{Poincar\'e series}. For example, for the case of $SL(2, \mbb{R})/SO(2)$, parametrized by a complex coordinate $\tau$, we have seen in Section \ref{Section:SL(2)Eisenstein} that such a Poincar\'e series is obtained by summing the function $\Im(\ga\cdot \tau)$ over the orbit $\ga\in \Gamma_{\infty}\bas SL(2, \mbb{Z})$, where $\Gamma_{\infty}$ is generated by $T : \tau \mapsto \tau+1$. This indeed produces a non-holomorphic Eisenstein series on the double quotient $SL(2, \mbb{Z})\bas SL(2, \mbb{R})/SO(2)$. 

Here we would like to generalize this construction to a Poincar\'e series on the complex upper half plane $\mbb{CH}^2$, parametrized by the variable $\mc{Z}=(z_1, z_2)$. The generalization of $\Im(\tau)$ is then given by the $N(\mbb{Z})$-invariant function $\mc{F}(\mc{Z})$, constructed in (\ref{KaehlerArgument}) \cite{Zhang}.\footnote{I am grateful to Genkai Zhang for helpful discussions on this construction.} The invariance of $\mc{F}(\mc{Z})$ under $N(\mbb{Z})$ can be checked by direct substitution of the Heisenberg translations in Eq. (\ref{HeisenbergAction}). As we have seen in Section \ref{Section:SU(2,1)}, the Picard modular group $\pg $ acts by
fractional transformations on $\mc{Z}\in \mbb{CH}^2$ such that the function $\mc{F}(\mc{Z})$
transforms as
\beq
\mc{F}(\ga\cdot \mc{Z})=\f{\mc{F}(\mc{Z})}{|C\mc{Z}+D|^2}, \qquad \ga=\left(\begin{array}{cc}
A & B \\
C & D \\
\end{array}\right)\in \pg .
\eeq
A Poincar\'e series for the Picard group may now be constructed as follows
\beqa
\mc{P}_s(\mc{Z}) := \sum_{\ga\in N(\mbb{Z})\bas \pg }\mc{F}(\ga\cdot \mc{Z})^{s}
= \sum_{\ga\in N(\mbb{Z})\bas \pg }\Big(\f{\mc{F}(\mc{Z})}{|C\mc{Z}+D|^2}\Big)^s\,.
\eqa
Taking $C\equiv (\om_3, \om_2)\in \mbb{Z}[i]^2$ and $D\equiv \om_1\in \mbb{Z}[i]$, and recalling $\mc{F}(\mc{Z})=e^{-2\phi}$, then reproduces the same form of the Eisenstein series as in Eq. (\ref{EisensteinPicard4}), i.e.
\beq
\mc{P}_s(\mc{Z})=\sum_{\ga\in N(\mbb{Z})\bas \pg }e^{-2s\phi}|\om_1+\om_2 z_2+\om_3 z_1|^{-2s}.
\label{PGPoincare}
\eeq
Defining $\vec{\om}=\vec{\om}^{\prime} \be$ with $\be= \gcd(\om_1,\om_2, \om_3)\in\mbb{Z}[i]$ and inserting this into (\ref{EisensteinPicard}) we then have the relation
\beq\label{EisenPoin}
\mc{E}_s(\phi, \lambda, \ga) = 4\zeta_{\mbb{Q}(i)}(s) \mc{P}_s(\mc{Z}),
\eeq
where $\zeta_{\mbb{Q}(i)}(s)$ is the Dedekind zeta function for the quadratic extension $\mbb{Q}(i)$ of the rational numbers, and the overall factor of $4$ originates from the four units in $\mbb{Z}[i]$. This will be discussed in more detail in Section \ref{Section:FirstConstantTerm} (see Eq. (\ref{dedezeta})).

Let us now also discuss the Laplacian condition on $\mc{P}_s$. The Laplacian on the coset space $\mbb{CH}^2=SU(2,1)/(SU(2)\times U(1))$ is most conveniently written in terms of the real variables $\{y=e^{-2\phi}, \chi, \tilde\chi, \psi\}$ and reads explicitly~\cite{FrancsicsLax3}
\beq
\Delta_{\mbb{CH}^2} =\frac{1}{4} y (\partial_\chi^2 + \partial_{\tilde\chi}^2)
+ \frac14 (y^2 + y(\chi^2+{\tilde\chi}^2))\partial_\psi^2
+ \frac12 y (\tilde\chi \partial_\chi - \chi \partial_{\tilde\chi})\partial_\psi
+ y^2 \partial_y^2 - y\partial_y.
\eeq
We furthermore recall that in terms of these variables the height function $\mc{F}(\mc{Z})$ takes a particularly simple form
\beq
\mc{F}(\mc{Z})=y.
\label{heightfunctiony}
\eeq
Hence, the only part of the Laplacian which is non-vanishing when acting on $\mc{F}(\mc{Z})$ is the ``radial part'' $y^2\pa_y^2-y\pa_y$.\footnote{Note the extra linear term compared to the ``radial part'' of the Laplacian on $SL(2,\mbb{R})/SO(2)$. This is what gives rise to the difference in the eigenvalues.} We then find $\Delta_{\mbb{CH}^2} \mc{F}(\mc{Z})^s=s(s-2)\mc{F}(\mc{Z})$, and consequently
\beq
\Delta_{\mbb{CH}^2} \mc{P}_s(\mc{Z})=s(s-2)\mc{P}_s(\mc{Z}).
\eeq
Moreover, since the Eisenstein series $\mc{E}_s$ in Eq. (\ref{EisensteinPicard2}) has the same summand as $\mc{P}_s$ when the constraint $\vec{\om}^{\dagger}\cdot \eta \cdot \vec{\om}=0$ is imposed, we may deduce that $\mc{E}_s$ is indeed an eigenfunction,
\beq
\Delta_{\mbb{CH}^2}\mc{E}_s(\phi, \lambda, \ga)=s(s-2)\mc{E}_s(\phi, \lambda, \ga)\,.
\label{EigenvalueEquation}
\eeq

Since $SU(2,1)$ admits two Casimir operators of degree 2 and 3, and since 
$\Delta_{\mbb{CH}^2}$ represents the action of the quadratic Casimir on
the space of (square-integrable) functions on $SU(2,1)/(SU(2)\times U(1))$, 
one may ask whether $ \mc{E}_s$ is also an eigenvector of an invariant
differential operator of degree 3. It turns out however, as already noticed in
\cite{PiolineGunaydin}, that the principal representation of the cubic Casimir 
in the space of functions on $SU(2,1)/(SU(2)\times U(1))$ vanishes identically.
In terms of the parametrization of the  Casimir eigenvalues by the
complex variables $(p,q)$ used in  \cite{Bars:1989bb,PiolineGunaydin},
the Eisenstein series $ \mc{E}_s$ is attached to the principal spherical representation 
with $p=q=s-2$ (see Appendix \ref{Section:pAdicConstruction} for some details on the principal series of $SU(2,1)$).

Let us also comment on the functional dimension of the representation associated to $ \mc{E}_s$.
The summation ranges over six (real) integers coordinating the lattice $\mbb{Z}[i]^3 \simeq\mbb{Z}^6$. Since both the summand and the constraint are homogeneous in $\vec{\omega}$ one can factor out an overall common Gaussian integer.
Among the remaining four real integers the (real) quadratic constraint $|\om_2|^2-2\Im(\om_1\bar{\om}_3)=0$
eliminates one of the summation variables, leaving effectively a sum
over 3 integers only. This is consistent with the functional dimension 3 of the
principal continuous series representation of $SU(2,1)$ and the number of expected different instanton contributions.

\subsection{Spherical Vector and $p$-Adic Eisenstein Series}
\label{Section:pAdicConstruction}
As we have discussed in detail in Chapter \ref{Chapter:ConstructingAutomorphicForms}, there exists a more general method for constructing automorphic forms on a coset
space $G/K$, as developed in
\cite{Kazhdan:2001nx,Kazhdan,Pioline:2003bk}. In this subsection, we show that this method may be used to construct the Eisenstein series
$\mc{E}_s(\phi, \lambda, \ga)$ for the Picard modular group. This
alternative approach also sheds light on the relation between the quadratic
constraint $\vec{\om}^{\dagger}\wedge \vec{\om}=0$ and the representation
theoretic structure of the Eisenstein series. Recall from Section \ref{Section:SphericalVector} that an automorphic form $\Psi$ can be rewritten in the following way
\beq
\Psi(\mc{V})=\sum_{\vec{x}\in \mbb{Q}^n} \Big[\prod_{p<\infty} f_p(\vec{x})\Big]\rho(\mc{V})\cdot f_{K}(\vec{x}),
\label{pAdicAutomorphicForm}
\eeq
where $\vec{x}$ is a vector of rational numbers in $\mbb{Q}^n$, and the
product is over all prime numbers $p$. 

To reproduce the Picard Eisenstein series \eqref{EisensteinPicard} by this method, 
we consider the principal continuous series representation of $SU(2,1)$,
induced from the Heisenberg parabolic $P$ whose Lie algebra 
consists of the non-positive grade part of the 5-grading (\ref{5grading}):
\beq
\mf{p}=\mf{g}_{-2}\oplus \mf{g}_{-1}\oplus \mf{g}_0 \subset \mf{su}(2,1).
\eeq
The parabolic group $P$ thus corresponds to the subgroup of lower-triangular matrices,
\beq
P = \left\{ \left(\begin{array}{ccc}
t_1 &  & \\
 *  & t_2 & \\
 *  &  *  & t_3 \\
\end{array} \right) \in SU(2,1)\hs :\hs  ~ t_1t_2t_3=1 \right\}.
\eeq
The coset space $P\bas SU(2,1)$ is isomorphic to
the Heisenberg group ${N}$,  and can be parameterized as follows:
\beq\label{HeisenbergGroup}
n=e^{xX_{1}+\tilde{x}\tilde X_{1}+2yX_{2}}=\left(\begin{array}{ccc}
1 & i\bar C_2 & C_1 \\
  & 1 & {C}_2\\
  &   & 1 \\
\end{array} \right):= \left(\begin{array}{c}
\vec{r}_1  \\
\vec{r}_2 \\
\vec{r}_3 \\
\end{array}\right) \in N,
\eeq
where
\beq
\label{C12}
C_1:=2y+\f{i}{2}|C_2|^2\,,\qquad C_2:= x+\tilde{x}+i(\tilde x-x)
\eeq
satisfy the quadratic relation
\beq
 |C_2|^2-2\Im (C_1) = 0\,,
\label{QuadraticRelation}
\eeq
and the last equality in (\ref{HeisenbergGroup}) defines the row vectors $\vec r_i$ of the Heisenberg group element. 

The coset space $N=P\bas SU(2,1)$ admits an action of $g\in SU(2,1)$ by multiplication from the right,
followed by a compensating action by $p(g)\in P$ from the left so as to restore the upper triangular gauge \eqref{HeisenbergGroup}. The principal continuous series representation consists
of functions $f(x,\tilde{x},y)$ on $N$ 
transforming by the character $\chi_s(p(g))$ under the action of $g$, where
\beq
\chi_s(p):=t_1^{-2s}, \qquad p=\left(\begin{array}{ccc}
t_1 &  & \\
 *  & t_2 & \\
 *  &  *  & t_3 \\
\end{array} \right)\in P.
\eeq
The spherical vector $f_K$ can be obtained straightforwardly as follows \cite{PiolineGunaydin}: 
while the compensating left-action of $P$ on the second and third rows, $\vec{r}_2$ and $\vec{r}_3$, of $n$ is quite complicated,  the action on the first row $\vec{r}_1$ is very simple: $p\in P$ simply modifies $\vec{r}_1$ by an overall factor of $t_1$. Moreover, the action of $k\in SU(2)\times U(1)$ leaves invariant the (complex) norms of the rows $\vec{r}_i$. The spherical vector $f_K$ 
can therefore be obtained by raising the norm of the first row $\vec{r}_1$ of $n$ to the appropriate power of $s$ \cite{PiolineGunaydin}\footnote{See also \cite{AutomorphicMembrane} for a similar construction in the context of $SL(3, \mbb{R})$.}:
\beq
f_{K}(x, \tilde{x}, y):=|\vec{r}_1|^{-2s}=\big(1+|C_1|^2+|C_2|^2\big)^{-s}=\Big(1+2(x^2+\tilde{x}^2)+4y^2+(x^2+\tilde{x}^2)^2\Big)^{-s}.
\nn \\
\eeq
This object is indeed invariant under $SU(2)\times U(1)$, since the right action of $k$ on $n$ is a ``rotation'' that preserves the norm, while the compensating left action of $p$ merely modifies $f_{K}$ by an overall factor $t_1^{2s}$, which in turn is canceled against the character $\chi_s(p)=t_1^{-2s}$ which is present since $f_{K}$ is in the principal series. 

The next step is to compute the action of $\rho(\mc{V})$ on $f_{K}$. Following the prescription above, this can be done by first computing $n\cdot \mc{V}=p_0 \cdot n^{\prime}$, with
\beq
p_0=\left(\begin{array}{ccc}
e^{-\phi} & & \\
& 1 & \\
& & e^{\phi}\\
\end{array}\right)\in P, \qquad n^{\prime}=\left(\begin{array}{ccc}
1 & ie^{\phi}(\bar\lambda+\bar{C}_2) & e^{2\phi}(\ga+i\bar{C}_2\lambda+C_1) \\
 & 1 & e^{\phi}(\lambda+C_2)\\
 &  &  1\\
 \end{array}\right)\in P\bas SU(2,1).
 \eeq
Applying this to the spherical vector $f_K(x, \tilde x, y)=f_K(n)$ yields
\beq
\rho(\mc{V})\cdot f_K(n)=f_K(n\mc{V})=f_K(p_0 n^{\prime})=\chi_s(p_0)f_K(n^{\prime})=e^{2s\phi}|{\vec{r}_1}^{\hs \prime}|^{-2s},
\eeq
which may be written explicitly in the form
\beq
\rho(\mc{V})\cdot f_{K}(C_1, C_2)=e^{-2s\phi}\Big(|\bar{C_1}-i{C_2}\bar\lambda+\bar\ga|^2+e^{-2\phi}|{C_2}+{\lambda}|^2+e^{-4\phi}\Big)^{-s}.
\eeq
The $p$-adic spherical vector $f_p(C_1,C_2)$ can now be found by
replacing the Euclidean norm $|\cdot |$
appearing in the real spherical vector $f_{K}$ by the $p$-adic counterpart 
(see e.g. \cite{BrekkeFreund}):
\beq
|z|_p^{\mbb{Q}(i)}:=\sqrt{\big|z\bar{z}\big|_p}, \qquad z\in \mbb{Q}(i),
\eeq
where the right hand side is evaluated using the standard $p$-adic norm $|\cdot |_p$ since $z\bar{z}\in \mbb{Q}$. The $p$-adic spherical vector is then given by
\beq
f_p(C_1,C_2):=\Big[\big|v_1\big|^{\mbb{Q}(i)}_p\Big]^{-2s}=\mathrm{max}\Big(1,\sqrt{\big|C_1\bar{C}_1\big|_p}, \sqrt{\big|C_2\bar{C}_2\big|_p}\Big)^{-2s}.
\eeq

\subsection{Product over Primes}

The automorphic form $\Psi(\mc{V})$ in this representation now reads
\beq\label{ratprod}
\Psi(\mc{V})=\sum^{\qquad \prime}_{(x,\tilde{x},y)\in\mbb{Q}^3}\bigg[\prod_{p<\infty}\mathrm{max}\Big(1,\sqrt{\big|C_1\bar{C}_1\big|_p}, \sqrt{\big|C_2\bar{C}_2\big|_p}\Big)^{-2s}\bigg]\rho(\mc{V})\cdot f_{K}(x, \tilde x, y).
\eeq
We can also write the summation over the complex rational variables $(C_1,C_2)\in \mbb{Q}(i)^2$ instead of the rational variables $(x, \tilde{x}, y)\in \mbb{Q}^3$, if we incorporate the constraint from Eq. (\ref{QuadraticRelation}) as follows
\beq
\Psi(\mc{V})=\sum^{\qquad \prime}_{(C_1,C_2)\in\mbb{Q}(i)^2\atop |C_2|^2-2\Im(C_1)=0}
\rho(\mc{V})\cdot  \Big[\prod_{p<\infty}f_p(C_1,C_2)\Big]f_{K}(C_1, C_2).
\label{padicEisensteinSeries}
\eeq
Next we must evaluate the infinite product over prime numbers. To this end we split the rational variables $C_1$ and $C_2$ in the following way:
\beq
{C_1}=\f{\om_1}{\om_3}\,,\quad {C_2}= \f{i \bar\om_2}{\bar\om_3}\,,
\label{variablerelation}
\eeq
with $\om_j\in \mbb{Z}[i]$, for $j=1,2,3$, $\gcd(\omega_1,\omega_2,\omega_3)=1$.\footnote{We note that the greatest common divisor in $\mbb{Z}[i]$ is defined up to Gaussian units which are a subgroup of order 4 in the Gaussian integers $\mbb{Z}[i]$.} In the case $\gcd(|\omega_1|^2,|\omega_2|^2,|\omega_3|^2)=1$, we can explicitly evaluate the infinite product over primes in (\ref{ratprod}) as
\beq
\prod_{p<\infty}\mathrm{max}\Big(1,\sqrt{\Big|\f{\om_1\bar{\om}_1}{\om_3\bar{\om}_3}\Big|_p},\sqrt{\Big|\f{\om_2\bar{\om}_2}{\om_3\bar{\om}_3}\Big|_p}\Big)^{-2s}=|\om_3|^{-2s}.
\label{primeproduct}
\eeq
This relation is not valid if there are Gaussian prime factors in the $\omega_i$ related by complex conjugation. This subtlety is reflected in the multiplicative properties of the abelian instanton measure discussed near (\ref{eqn:abelianmult}) and should be related to the Hecke algebra of $\pg$~\cite{Kudla}.  Multiplying the constraint \eqref{QuadraticRelation} further by a factor of $|\om_3|^2$ one obtains
\beq
|\om_3|^2\Big(|C_2|^2-2\Im(C_1)\Big)=  |\om_2|^2-2\Im(\om_1\bar{\om}_3)=\vec{\om}^{\dagger}\cdot \eta \cdot \vec{\om}=0.
\label{NewConstraint}
\eeq
Combining Eqs. (\ref{padicEisensteinSeries}), (\ref{primeproduct}) and (\ref{NewConstraint})
and adding  the contribution at $C_1=C_2=\infty$ (i.e. $\omega_3=0$) 
then yields the final form of $\Psi(\mc{V})$:
\beq
\Psi(\mc{V})=\sum^{\qquad \prime}_{\vec{\om}\in \mbb{Z}[i]^3,\  \gcd(\om_1,\om_2,\om_3)=1 \atop   |\om_2|^2-2\Im(\om_1\bar{\om}_3)=0}e^{-2s\phi}\Big[|\bar\om_1+\bar\om_2\bar\lambda+\bar\om_3\bar{\ga}|^2+e^{-2\phi}|\bar\om_2-i\bar\om_3{\lambda}|^2+e^{-4\phi}|\om_3|^2\Big]^{-s},
\label{padicEisensteinSeries2}
\eeq
which we recognize as the Eisenstein series $\mc{P}_s(\phi, \lambda, \ga)$ constructed in Section \ref{Section:PoincareSeries}.

\section{Fourier Exapnsion of $\mc{E}_s(\phi, \chi, \tilde\chi, \psi)$}
\label{Section:FourierExpansion}
In this section we compute the Fourier expansion of the Eisenstein series 
\eqref{EisensteinPicard}. We begin by recalling the general decomposition with respect to the action of the Heisenberg subgroup $N\subset SU(2,1)$.

\subsection{General Structure of the Non-abelian Fourier Expansion}
\label{section:generalFourierexpansion}

The main complication of the Fourier expansion stems from the non-abelian nature of the nilpotent group $ N \subset SU(2,1)$.  $N$ is isomorphic to a three-dimensional Heisenberg group, where the center $Z=[N,N]$ is parametrized by $\psi$. The Fourier expansion therefore splits into an 
abelian part and a non-abelian part. The abelian term corresponds to an expansion with respect to the abelianized group $N/Z$, while the non-abelian terms represent the expansion with respect to the center $Z$. This general structure of the Fourier expansion of automorphic forms for the Picard modular group is discussed in detail by Ishikawa \cite{Ishikawa}, to which we refer the interested reader. A similar discussion may also be found in the mathematics \cite{Vinogradov,Proskurin} and physics \cite{AutomorphicNS5} literature for the case of automorphic forms on $SL(3, \mbb{R})/SO(3)$. 

We have seen in Section \ref{Section:SU(2,1)} that the action of an arbitrary Heisenberg shift $\mc{U}_{a,b;c}\in N(\mbb{Z})=N\cap SU(2,1;\mbb{Z}[i])$ on $\chi, \tilde{\chi}$ and $\psi$ is given by:
 \beqa
\mc{U}_{a,b;c} &: & \chi\hs \longmapsto\hs  \chi + a,
 \nn \\
  & & \tilde\chi \hs \longmapsto\hs \tilde\chi +b,
  \nn \\
  & & \psi \hs \longmapsto \psi+ \f{1}{2}c-a\tilde\chi+b\chi
  \label{DiscreteHeisenbergAction}
  \eqa
for $a,b,c\in\mbb{Z}$. Since the Eisenstein series (\ref{EisensteinPicard}) is in particular invariant under $N(\mbb{Z})$ we can organize the Fourier expansion by diagonalizing different subgroups of the non-abelian Heisenberg group $N(\mbb{Z})$.

Explicitly, we  write the general form of the Fourier expansion as
 \beq
 \mc{E}_s(\phi, \chi, \tilde\chi, \psi)=\mc{E}_s^{(\text{const})}(\phi)+\mc{E}^{(\text{A})}_s(\phi, \chi, \tilde\chi)+\mc{E}_s^{(\text{NA})}(\phi, \chi, \tilde\chi, \psi),
 \eeq
 where $\mc{E}_s^{(\text{const})}(\phi)$ is the constant term and 
 \beqa
 \mc{E}^{(\text{A})}_s(\phi, \chi, \tilde\chi)&=& \sum^{\qquad \prime}_{(\ell_1,\ell_2)\in\mbb{Z}^2} \mf{C}_{\ell_1,\ell_2}^{\text{(A)}}(\phi;s) e^{-2\pi i (\ell_1\chi +\ell_2\tilde\chi)},
 \nn \\
 \mc{E}_s^{(\text{NA})}(\phi, \chi, \tilde\chi, \psi)&=&\phantom{kk} \sum^{\qquad \prime}_{k\in \mbb{Z}}\mf{C}_{k}^{\text{(NA)}}(\phi, \chi, \tilde\chi;s) e^{-4\pi i k \psi}
 \label{NonabelianFourierExpansionPicard}
 \eqa 
are called the abelian and non-abelian terms, respectively. Following \cite{Ishikawa,AutomorphicNS5}, we proceed to extract an additional phase factor in the non-abelian term which accounts for the shifts of $\psi$ along the non-central directions. This yields the following structure of the non-abelian term
\beq
  \mc{E}_s^{(\text{NA})}(\phi, \chi, \tilde\chi, \psi)=\sum^{\qquad \prime}_{k\in \mbb{Z}}\sum_{\ell=0}^{4|k|-1}\sum_{n\in\mbb{Z}+\f{\ell}{4|k|}} \mf{C}_{k, \ell}^{\text{(NA)}}(\phi, \tilde{\chi}-n;s) e^{8\pi i k n \chi-4\pi i k(\psi+\chi\tilde\chi)}.
  \label{NewNonabelianTerm}
  \eeq
The abelian term is manifestly invariant under shifts of the form $\mc{U}_{a,b;0}\in N(\mbb{Z})/Z$. For the non-abelian term, invariance under 
\beqa
\mc{U}_{1,0;0} &:& \chi\hs  \longmapsto\hs \chi +1
\nn \\
&:& \psi\hs \longmapsto\hs  \psi-\tilde\chi
\eqa
 is manifest since $4kn\in \mbb{Z}$. On the other hand, the transformation
 \beqa
\mc{U}_{0,1;0} &:& \tilde\chi\hs  \longmapsto\hs \tilde\chi +1
\nn \\
&:& \psi\hs \longmapsto\hs  \psi+\chi
\eqa
requires a compensating shift $n\mapsto n+1$ on the summation, under which the variation of the total phase cancels. Note also the restricted dependence on $\tilde\chi$ in the Fourier coefficient; upon shifting $\tilde\chi \mapsto \tilde\chi+1$ and compensating $n\mapsto n+1$ the coefficient is indeed invariant. Finally, invariance under $\mc{U}_{0,0;1}$ is manifest since this gives an overall phase $e^{-4\pi i k/2}=1$. 

To derive (\ref{NewNonabelianTerm}), note first that the non-abelian coefficient $\mf{C}_k^{(\text{NA})}$ in (\ref{NonabelianFourierExpansionPicard}) is invariant under shifts of $\psi$ since it does not depend on this variable. However, under shifts of $(\chi, \tilde\chi)$ it transforms with an overall phase:
\beq
\mf{C}_k^{(\text{NA})}(\phi,\chi+a, \tilde\chi+b;s)=\mf{C}_k^{(\text{NA})}(\phi,\chi, \tilde\chi;s)e^{-4\pi i k(a\tilde\chi- b\chi)},
\eeq
by virtue of (\ref{DiscreteHeisenbergAction}). Now define
\beq
\tilde{\mf{C}}_k^{(\text{NA})}(\phi, \chi, \tilde\chi;s) := \mf{C}_k^{(\text{NA})}(\phi, \chi, \tilde\chi;s) e^{4\pi i k \chi \tilde\chi},
\eeq
which transforms as follows under shifts of $\tilde\chi$:
\beq
\tilde{\mf{C}}_k^{(\text{NA})}(\phi, \chi, \tilde\chi+b;s)= \tilde{\mf{C}}_k^{(\text{NA})}(\phi, \chi, \tilde\chi;s) e^{8\pi i kb\chi},
\label{chitildeshift}
\eeq
while it is periodic under shifts of $\chi$ with period 1. This implies that $\tilde{\mf{C}}_k^{(\text{NA})}(\phi, \chi, \tilde\chi;s)$ has a Fourier expansion of the form
\beq
\tilde{\mf{C}}_k^{(\text{NA})}(\phi, \chi, \tilde\chi;s) = \sum_{m\in\mbb{Z}} \tilde{\mf{C}}_{k,m}^{(\text{NA})}(\phi, \tilde\chi;s) e^{2\pi i m\chi},
\label{subfourier}
\eeq
and the non-abelian term $\mc{E}_s^{(\text{NA})}(\phi, \chi, \tilde\chi, \psi)$ may thus be written as
\beqa
\mc{E}_s^{(\text{NA})}(\phi, \chi, \tilde\chi, \psi)&=& \sum_{k\neq 0}\tilde{\mf{C}}_k^{(\text{NA})}(\phi, \chi, \tilde\chi) e^{-4\pi i k(\psi+\chi\tilde\chi)}
\nn \\
&=& \sum_{k\neq 0} \sum_{m\in \mbb{Z}} \tilde{\mf{C}}_{k,m}^{(\text{NA})}(\phi, \tilde\chi;s) e^{2\pi i m\chi -4\pi i k(\psi+\chi\tilde\chi)}.
\eqa
To determine how the new coefficient $\tilde{\mf{C}}_{k,m}^{(\text{NA})}(\phi, \tilde\chi;s)$ depends on $m\in \mbb{Z}$, we note that (\ref{chitildeshift}) and (\ref{subfourier}) imply the following relation
\beq
\tilde{\mf{C}}_k^{(\text{NA})}(\phi, \chi, \tilde\chi+b;s) = \sum_{m\in\mbb{Z}} \tilde{\mf{C}}_{k,m}^{(\text{NA})}(\phi, \tilde\chi+b;s) e^{2\pi i m\chi}= \sum_{m\in\mbb{Z}} \tilde{\mf{C}}_{k,m}^{(\text{NA})}(\phi, \tilde\chi;s) e^{2\pi i (m+4kb)\chi}.
\label{NAderivation}
\eeq
Now shifting $m\mapsto  m-4kb$ in the summation in the last term, Eq. (\ref{NAderivation}) yields an equation for the coefficients
\beq
\tilde{\mf{C}}_{k,m}^{(\text{NA})}(\phi, \tilde\chi+b;s)=\tilde{\mf{C}}_{k,m-4kb}^{(\text{NA})}(\phi, \tilde\chi;s).
\eeq
Further defining the new summation variable
\beq
m:= 4nk, \qquad n\in \mbb{Z}+\f{\ell}{4|k|},
\eeq
and imposing the periodicity $\mf{C}_k^{(\text{NA})}(\phi, \chi, \tilde\chi+1;s)=\mf{C}_k^{(\text{NA})}(\phi, \chi, \tilde\chi;s)$ we find
\beq
\mf{C}_k^{(\text{NA})}(\phi, \chi, \tilde\chi;s)= \sum_{k\neq 0} \sum_{\ell=0}^{4|k|-1} \sum_{n\in \mbb{Z}+\ell/(4|k|)} \tilde{\mf{C}}_{k, n-1}(\phi, \tilde\chi;s) e^{8\pi i k(n-1)-4\pi i k (\psi+\chi\tilde\chi)}.
\label{NAderivation2}
\eeq 
From this we deduce that an integer shift of $\tilde\chi$ can be compensated by an opposite shift of the new summation variable $n$, and hence the dependence of the Fourier coefficients in (\ref{NAderivation2}) on $\tilde\chi$ and $n$ is restricted to the combination $\tilde\chi-n$. Therefore we define
\beq
\tilde{\mf{C}}_{k, n}(\phi, \tilde\chi;s):={\mf{C}}_{k}(\phi, \tilde\chi-n;s).
\eeq 
Inserting this into (\ref{NAderivation2}) then yields (\ref{NewNonabelianTerm}) as advertised. 

Note that in writing the non-abelian term (\ref{NewNonabelianTerm}) we have made an explicit choice of {polarization}, in the sense that we have manifestly diagonalized the action of Heisenberg shifts of the restricted form $\mc{U}_{a,0;c}$. We could have chosen the opposite polarization in which we instead diagonalize the action of $\mc{U}_{0,b;c}$. In this case, the non-abelian term reads
\beq
  \mc{E}_s^{(\text{NA})}(\phi, \chi, \tilde\chi, \psi)=\sum^{\qquad \prime}_{k\in\mbb{Z}}\sum_{\ell^{\prime}=0}^{4|k|-1}\sum_{\tilde{n} \in\mbb{Z}+\f{\ell^{\prime}}{4|k|}} \tilde{\mf{C}}_{k, \ell^{\prime}}^{\text{(NA)}}(\phi, {\chi}-\tilde n;s) e^{-8\pi i k \tilde{n} \tilde\chi-4\pi i k(\psi-\chi\tilde\chi)}.
  \label{NewNonabelianTermSecondPolarization}
  \eeq
The Fourier coefficients $\mf{C}_{k, \ell}^{\text{(NA)}}$ and $\tilde{\mf{C}}_{k, \ell^{\prime}}^{\text{(NA)}}$  in the two different polarizations are related via the following Fourier transform \cite{AutomorphicNS5}\footnote{Note that there are certain differences between this expression and the one given in Eq. 3.40 of \cite{AutomorphicNS5}. This is due to the fact that here we consider the 3-dimensional Heisenberg group $N(\mbb{Z})$ associated with $SU(2,1;\mbb{Z}[i])$, while in \cite{AutomorphicNS5} it was the Heisenberg subgroup of $SL(3,\mbb{Z})$ which was relevant. The difference lies in the normalization of the central generator. }:
\beq 
\tilde{\mf{C}}^{(\text{NA})}_{k, \ell'}(\chi)=\sum_{\ell=0}^{4|k|-1} e^{\pi i \f{\ell\ell'}{2|k|}} \int d\tilde\chi\ \mf{C}^{(\text{NA})}_{k,\ell}(\tilde\chi)\ e^{-8\pi i k\chi\tilde\chi},
\label{FT}
\eeq
while the abelian coefficients are invariant under Fourier transform. To derive\footnote{The following manipulations were shown to me by Boris Pioline.} (\ref{FT}), we begin by shifting the summation variable $n$ in (\ref{NewNonabelianTerm}) according to $n\mapsto n+\ell/ (4|k|)$, which yields
\beq
  \mc{E}_s^{(\text{NA})}(\phi, \chi, \tilde\chi, \psi)=\sum^{\qquad \prime}_{k\in \mbb{Z}}\sum_{\ell=0}^{4|k|-1}\sum_{n\in\mbb{Z}} \mf{C}_{k, \ell}^{\text{(NA)}}(\phi, \tilde{\chi}-n-\f{\ell}{4|k|};s) e^{8\pi i k \big(n+\f{\ell}{4|k|}\big) \chi-4\pi i k(\psi+\chi\tilde\chi)},
  \label{FTderivation1}
  \eeq
where we suppressed the $\phi$- and $s$-dependence in the Fourier coefficients in order to avoid cluttering. We now compute the Fourier transform of the summand in (\ref{FTderivation1}) with respect to the new variable $n\in\mbb{Z}$:
\beqa
\mc{FT}(k,\ell,m;\tilde\chi)&:=&\int dn\ \mf{C}_{k, \ell}^{\text{(NA)}} \big(\tilde{\chi}-n-\f{\ell}{4|k|}\big)\ e^{8\pi i k\big(n+\f{\ell}{4|k|}\big)\chi +2\pi i mn}
\nn \\
&=& \int d\xi\ \mf{C}_{k, \ell}^{\text{(NA)}}(\xi)\ e^{8\pi i k(\tilde\chi-\xi)\chi +2\pi i m\big(\tilde\chi-\xi-\f{\ell}{4|k|}\big)},
\label{FTderivation2}
\eqa
where in the second line we have performed the change of variables $\xi=\tilde\chi-n-\ell/(4|k|)$. Further defining 
\beq
\hat{\mf{C}}_{k,\ell}(\chi):= \int d\xi \ e^{-8\pi i k\xi \chi}\ \mf{C}_{k,\ell}(\xi),
\label{subFT}
\eeq
we may rewrite (\ref{FTderivation2}) as 
\beq
\mc{FT}(k,\ell,m;\tilde\chi)= \hat{\mf{C}}_{k, \ell}\big(\chi+\f{m}{4|k|}\big) \ e^{8\pi i k\chi\tilde\chi+2\pi i m \big(\tilde\chi-\f{\ell}{4|k|}\big)}.
\eeq
Inserting this result into (\ref{FTderivation1}) we find that the Fourier transformed non-abelian term reads
\beq
 \mc{E}_s^{(\text{NA})}(\phi, \chi, \tilde\chi, \psi)=\sum^{\qquad \prime}_{k\in \mbb{Z}}\sum_{m\in\mbb{Z}}\left[\sum_{\ell=0}^{4|k|-1} \hat{\mf{C}}_{k, \ell}\big(\chi+\f{m}{4|k|}\big) \ e^{2\pi i m \big(\tilde\chi-\f{\ell}{4|k|}\big)}\right] \ e^{-4\pi i k(\psi-\chi\tilde\chi)}.
 \eeq
We now want to identify this expression with the dual polarization (\ref{NewNonabelianTermSecondPolarization}). To this end, we make an additional change of summation variable: $m=-4|k|\tilde{m}-\ell'$, which yields
\beq
 \mc{E}_s^{(\text{NA})}(\phi, \chi, \tilde\chi, \psi)=\sum^{\qquad \prime}_{k\in \mbb{Z}}\sum_{\tilde{m}\in\mbb{Z}}\left[\sum_{\ell=0}^{4|k|-1} \hat{\mf{C}}_{k,\ell}\big(\chi-\tilde{m}-\f{\ell'}{4|k|}\big) e^{\pi i \f{\ell\ell'}{2|k|}}\right] e^{-8\pi i |k| \big(\tilde{m}+\f{\ell'}{4|k|}\big)\tilde\chi-4\pi i k(\psi-\chi\tilde\chi)}.
 \eeq
Finally, we define $\tilde{m}:= \tilde{n}-\ell'/(4|k|)$, with $\tilde{n}\in \mbb{Z}+\ell'/(4|k|)$, and write
\beq
 \mc{E}_s^{(\text{NA})}(\phi, \chi, \tilde\chi, \psi)=\sum^{\qquad \prime}_{k\in \mbb{Z}}\sum_{\ell'=0}^{4|k|-1} \sum_{\tilde{n}\in \mbb{Z}+\ell'/(4|k|)} \left[\sum_{\ell=0}^{4|k|-1} \hat{\mf{C}}_{k, \ell}(\chi-\tilde{n}) \ e^{\pi i \f{\ell\ell'}{2|k|}}\right] \ e^{-8\pi i k \tilde{n} \tilde\chi-4\pi i k(\psi-\chi\tilde\chi)}.
 \label{FTderivation3}
 \eeq
Comparing (\ref{FTderivation3}) with (\ref{NewNonabelianTermSecondPolarization}) leads, via (\ref{subFT}) and shifting $\chi\longmapsto \chi+\tilde{n}$, to the relation (\ref{FT}), as desired. In the sequel we work for definiteness with the first polarization defined by \ref{NewNonabelianTerm}). 

Besides invariance under the Heisenberg group we can also use invariance under the electric-magnetic duality transformation $R: (\chi,\tilde\chi)\mapsto(-\tilde\chi,\chi)$ of (\ref{emdual}). On the abelian term this implies that the coefficient $\mf{C}_{\ell_1,\ell_2}^{\text{(A)}}$ is invariant under $\pi/2$ rotations of $(\ell_1,\ell_2)$. On the non-abelian term (\ref{NewNonabelianTerm}) application of $R$ leads to 
\beq
\mc{E}_s^{(\text{NA})}(\phi, \chi, \tilde\chi, \psi)=\sum^{\qquad \prime}_{k\in \mbb{Z}}\sum_{\ell=0}^{4|k|-1}\sum_{n\in\mbb{Z}+\f{\ell}{4|k|}} \mf{C}_{k, \ell}^{\text{(NA)}}(\phi,\chi-n;s) e^{-8\pi i k n \tilde\chi-4\pi i k(\psi-\chi\tilde\chi)}
\eeq
and hence we have $\mf{C}_{k, \ell}^{\text{(NA)}}=\tilde{\mf{C}}_{k, \ell}^{\text{(NA)}}$, relating the two choices of polarization as to be expected from electric-magnetic duality. Applying $R$ again leads to relations among the coefficients $\mf{C}_{k, \ell}^{\text{(NA)}}$ and $\mf{C}_{k, \ell'}^{\text{(NA)}}$ for different $\ell$ and $\ell'$.

Finally, we can use the Laplacian condition on the Eisenstein series $\mc{E}_s$ (see Eq. (\ref{EigenvalueEquation})) to further constrain the Fourier coefficients $\mf{C}_{\ell_1, \ell_2}^{\text{(A)}}$ and $\mf{C}_{k, \ell}^{\text{(NA)}}$ and determine their functional dependence on the moduli. In all cases, we require normalizability of the solution, which physically means a well-behaved `weak-coupling' limit $e^\phi\to 0$. Plugging in the abelian term $\mc{E}_s^{(\text{A})}$ into the eigenvalue equation (\ref{EigenvalueEquation}) yields an equation for the $\phi$-dependence of the coefficients which is solved by a modified Bessel function. More precisely, we find that the abelian term in the expansion takes the form
\beq
\mc{E}_s^{(\text{A})}(\phi, \chi, \tilde\chi, \psi)=e^{-2\phi}\sum_{(\ell_1, \ell_2)\in\mbb{Z}^2}^{\quad \prime} C^{\text{(A)}}_{\ell_1, \ell_2}(s) K_{2s-2}\Big(2\pi e^{-\phi} \sqrt{\ell_1^2+\ell_2^2}\Big) e^{-2\pi i (\ell_1\chi+\ell_2\tilde\chi)},
\label{Generalabelian}
\eeq
where the remaining coefficients $C^{\text{(A)}}_{\ell_1, \ell_2}(s)$ are now independent of $\phi$ and encode the arithmetic information of the group $SU(2,1;\mbb{Z}[i])$. The precise form of these numerical coefficients will be computed in Section \ref{abelianCoefficients} below. 

Turning to the non-abelian term (\ref{NewNonabelianTerm}), the Laplacian condition on the coefficient separates into a harmonic oscillator equation in the variable $x=\tilde\chi-n$, with solution given by a Hermite polynomial $H$, as well as a hypergeometric equation in the variable $y=e^{-2\phi}$ whose solution can be written in terms of a Whittaker function $W$. The separation of variables induces a sum over the eigenvalues of the harmonic oscillator, leading to the following structure for the non-abelian term:
\beqa
\mc{E}_s^{(\text{NA})}(\phi, \chi, \tilde\chi, \psi)&=&e^{-\phi} \sum^{\qquad \prime}_{k\in\mbb{Z}} \sum_{\ell=0}^{4|k|-1}\sum_{n\in\mbb{Z}+\f{\ell}{4|k|}}\sum_{r=0}^{\infty} 
C^{\text{(NA)}}_{r,k, \ell}(s) \hs |k|^{1/2-s} e^{-4\pi |k| (\tilde\chi-n)^2}
\nn \\
& & \times \,H_{r}\Big(\sqrt{8\pi |k|} (\tilde\chi-n)\Big) W_{-r-\f{1}{2}, s-1} \Big(4\pi |k| e^{-2\phi}\Big) e^{8\pi i k n \chi-4\pi i k(\psi+\chi\tilde\chi)},
\nn \\
\label{GeneralNonabelianTerm}
\eqa
where the numerical coefficients $C^{\text{(NA)}}_{r,k, \ell}(s)$ will be further discussed in Section \ref{NonabelianCoefficients}. 

We shall now proceed to compute the explicit form of the Fourier expansion; that is, determine the constant term $\mc{E}_s^{(\text{const})}$ as well as the abelian and non-abelian numerical Fourier coefficients $C^{\text{(A)}}_{\ell_1, \ell_2}(s)$ and $C^{\text{(NA)}}_{r,k, \ell}(s)$.

\subsection{First Constant Term}
\label{Section:FirstConstantTerm}
The constant term\footnote{The terminology ``constant term'' originates from holomorphic Eisenstein series where these terms are truly constant and independent of the scalar fields. For non-holomorphic Eisenstein series, as the one studied here, the constant terms retain a dependence on the fields corresponding to Cartan generators.} is defined generally as
\beq
\mc{E}_s^{(\text{const})}(\phi)=\int_{0}^{1}d\chi \int_0^{1}d\tilde\chi \int_0^{1/2}d\psi \ \mc{E}_s(\phi, \chi, \tilde\chi, \psi),
\label{constanttermformula}
\eeq
where the integral over the coordinate $\psi$ (physically, the NS-axion modulus) runs from $0$ to $1/2$ because of the extra factor of 2 in front of $\psi$ in our parametrization of $N$ in Eq.~(\ref{cosetel}). Since the Cartan subgroup $A$ appearing in the Iwasawa decomposition of $SU(2,1)$ is one-dimensional, the constant term only depends on the dilatonic scalar $\phi$. Moreover, recall from the discussion in Section \ref{subsec:CosetTransf} that the Weyl group of $\mf{su}(2,1)$ is the Weyl group of the restricted root system $BC_1$, which is isomorphic with $\mbb{Z}_2$. Hence, the constant term $\mc{E}_s^{(\text{const})}(\phi)$ consists of two contributions, $\mc{E}_s^{(0)}$ and $\mc{E}_s^{(1)}$, which are permuted by $\mbb{Z}_2$~\cite{Langlands}.\footnote{I am grateful to Pierre Vanhove for helpful discussions on the constant terms.} 

The powers of $e^{\phi}$ in $\mc{E}_s^{(\text{const})}(\phi)$ may be determined by the Laplacian condition on $\mc{E}_s$. In Section \ref{Section:PoincareSeries} we have seen that the Eisenstein series is an eigenfunction of the Laplacian $\Delta_{\mbb{CH}^2}$ with eigenvalue $s(s-2)$. This implies that all the constant terms must individually be eigenfunctions of $\Delta_{\mbb{CH}^2}$ with the same eigenvalue. It turns out that there is a unique solution to this, and we find that $\mc{E}_s^{(0)}$ must be of the form
\beq
\mc{E}_s^{(\text{const})}(\phi)=\mc{E}_s^{(0)}+\mc{E}_s^{(1)}=A(s)e^{-2s\phi}+B(s) e^{-2(2-s)\phi}.
\label{ConstantTerm}
\eeq
Below we will compute the coefficients $A(s)$ and $B(s)$. The first constant term $\mc{E}_s^{(0)}$ corresponds to the leading order term in an expansion at the cusp $e^{\phi}\rightarrow 0$, which physically corresponds to the regime of weak coupling. 

Our strategy for performing the Fourier expansion is as follows: we first consider the
term $\omega_3=0$, which by virtue of the constraint (\ref{App:Constraint}) also requires
 $\omega_2=0$. The remaining sum over $\omega_1\neq 0$ yields the first constant term $\mc{E}_s^{(0)}$. We then consider the case $\omega_3\neq 0$ and solve the constraint (\ref{App:Constraint}) explicitly using the Euclidean algorithm which reduces the remaining sum to one over three integers. On these integers we will perform Poisson resummations to uncover the second constant 
 term, as well the abelian and non-abelian Fourier coefficients.  

Accordingly we start by extracting the $\om_3=0$ (implying $\om_2=0$) part of the sum in the Eisenstein series, $\mc{E}_s^{(0)}$, leaving a remainder $\mc{A}^{(s)}$
\beq\label{firstconstant}
\mc{E}_s(\phi, \lambda, \ga)=\mc{E}_s^{(0)}+\mc{A}^{(s)}.
\eeq
The first term $\mc{E}_s^{(0)}$ is the leading order contribution in the limit $e^{\phi}\rightarrow 0$ and corresponds to a sum over $\omega_1=m_1+i m_2$
\beq
\mc{E}_s^{(0)}=e^{-2s\phi}\sum^{\qquad \prime}_{(m_1,m_2)\in\mbb{Z}^2} \f{1}{(m_1^2+m_2^2)^s}= 4 \zeta_{\mbb{Q}(i)}(s) e^{-2s\phi}\, ,
\label{LeadingTerm}
\eeq
where $\zeta_{\mbb{Q}(i)}(s)$ is the Dedekind zeta function over the Gaussian integers
\beq\label{dedezeta}
\zeta_{\mbb{Q}(i)}(s)= \frac14 \sum_{ \omega\in \mbb{Z}[i]}^{\quad \prime} |\omega|^{-2s} = \frac14 \sum^{\qquad \prime}_{(m,n)\in\mbb{Z}^2} \f{1}{(m^2+n^2)^s}.
\eeq
The factor of $4$ is related to the units of the Gaussian integers.

The Dedekind zeta function $\zeta_{\mbb{Q}(i)}(s)$ satisfies a functional equation
which is most conveniently written in terms of the ``completed Dedekind zeta function"
\beq\label{completedDedekind}
\zeta_{\mbb{Q}(i)*} (s) := \pi^{-s} \Gamma(s)\zeta_{\mbb{Q}(i)}(s)\,,
\eeq
in terms of which one has
\beq
\zeta_{\mbb{Q}(i)*} (1-s) = \zeta_{\mbb{Q}(i)*} (s)\,.
\label{functionalrelationcompletedDedekind}
\eeq
It is known that the Dedekind function over a quadratic number field can be written as a Dirichlet L-function times the standard Riemann zeta function. In our case this reads (see, e.g., \cite{Cartier} for a proof)
\beq
\label{zeQ}
\zeta_{\mbb{Q}(i)}(s)=\be(s)\zeta(s),
\eeq
where the standard Riemann zeta function is defined as
\beq\label{zetaEuler}
\zeta(s) := \sum_{n=1}^{\infty} n^{-s} = \prod_{p\,\text{prime}} \frac{1}{1-p^{-s}}\quad\quad \text{for $\Re(s)>1$}
\eeq
and $\be(s)$ is the Dirichlet beta function,
\beq
\be(s):=\sum_{n=0}^{\infty}(-1)^n (2n+1)^{-s}\quad\quad\text{for $\Re(s)>0$}.
\eeq
We also note that $\beta(s)$ has an Euler product representation of the form 
\beq\label{betaEuler}
\beta(s)= \prod_{p\, :\, p=1\, \text{mod} \,4} \frac{1}{1-p^{-s}}
\prod_{p\, :\, p=3\, \text{mod} \, 4} \frac{1}{1+p^{-s}},
\eeq
which together with the Euler product form of the Riemann zeta function $\zeta(s)$ above will be useful later. The functional relation for $\beta(s)$ is again best stated using its completion
\beq\label{completedBeta}
\beta_*(s) := \left(\frac{\pi}{4}\right)^{-\frac{s+1}{2}} \Gamma\left(\frac{s+1}{2}\right)\, \beta(s),
\eeq
for which the functional relation takes the simple form
\beq
\beta_*(s) = \beta_*(1-s).
\label{functionalrelationcompletedbeta}
\eeq

In conclusion, we have found that the first coefficient $A(s)$ in (\ref{ConstantTerm}) is given by the Dedekind function $\zeta_{\mbb{Q}(i)}(s)$ and that it is related to the term $\omega_3 =0$ in the sum over the Gaussian integers. We will now proceed to evaluate the terms with $\omega_3\neq 0$, contained in $\mc{A}^{(s)}$ of (\ref{firstconstant}). We emphasize that the term with $\omega_3=0$ and $\omega_2\neq 0$ vanishes identically because of the quadratic constraint (\ref{App:Constraint}). Thus, $\mc{A}^{(s)}$ only contains terms for which $\om_3\neq 0$.

\subsection{Solution of Constraint and Poisson Resummation}

To solve the constraint (\ref{App:Constraint}) we shall make use of the Euclidean algorithm, which implies that for integers $p_1$ and $p_2$ the equation
\beq
\label{bezout}
q_1p_2-q_2p_1=d
\eeq
has integer solutions for $q_1$ and $q_2$ if and only if $d$ divides $\gcd(p_1,p_2)$. The most general solution is the sum of a particular solution $(q_1,q_2)$ plus an integer times $(p_1,p_2)$.
More precisely, in the case of our constraint (\ref{App:Constraint}) we find that for $\omega_3=p_1+ip_2\neq 0$ there are solutions in $\Zn[i]^3$ if and only if 
\beq
\f{|\om_2|^2}{2d}\in\mbb{Z}\ , \quad\quad \text{ where $d=\gcd(p_1,p_2)$}
\label{problem}
\eeq
and the most general solution for $\omega_1=m_1+i m_2$ is then
\beqa
{} m_1&=&-\f{|\om_2|^2}{2d}q_1+m\f{p_1}{d},
\nn \\
{} m_2 &=& -\f{|\om_2|^2}{2d} q_2 +m\f{p_2}{d}.
\label{SolutionConstraint}
\eqa
Here, $q_1$ and $q_2$ is any particular solution of  $q_1p_2-q_2p_1=d$ and $m\in\Zn$ is an {\em unconstrained} integer. Therefore, we can rewrite the constrained sum as
\beq
\sum_{\omega_3\neq 0} \sum_{\omega_2\in \mbb{Z}[i]\atop   2d\,\vline\, |\omega_2|^2} \sum_{m\in\Zn}\big(\cdots \big)
\eeq
where in the summand, $\omega_1=m_1+i m_2$ has to be replaced by the expression from (\ref{SolutionConstraint}).

Let us implement this procedure on our Eisenstein series. 
After solving the constraint, the first term in the bracket of (\ref{EisensteinPicard}) becomes
\beqa
|\om_1+\om_2\lambda+\om_3 \ga|^2&=&\f{|\om_3|^2}{d^2}\bigg[\Big(m-\f{|\om_2|^2}{2|\om_3|^2}(q_1p_1+q_2p_2)+\tilde\ell_1\chi+\tilde\ell_2\tilde\chi+2d\psi\Big)^2
\nn \\
& &\phantom{++} +\f{1}{16d^2}\Big((\tilde\ell_1+2d\tilde\chi)^2+(\tilde\ell_2-2d\chi)^2\Big)^2\bigg],
\label{ConstraintSolved}
\eqa
where we defined
\beqa
{}\tilde{\ell}_1&:=& \f{d}{|\om_3|^2}\Big[(p_1-p_2)n_1+(p_1+p_2)n_2\Big],
\nn \\
{}\tilde\ell_2&:=& \f{d}{|\om_3|^2}\Big[ (p_1+p_2)n_1-(p_1-p_2)n_2\Big].
\eqa
Extracting an overall factor of $|\om_3|^2/d^2$ the total summand may be written as follows
\beqa
\f{d^2}{|\om_3|^2}\om^{\dagger}\cdot \mc{K}\cdot \om &=& \bigg[m-\f{|\om_2|^2}{2|\om_3|^2}(q_1p_1+q_2p_2)+\tilde\ell_1\chi+\tilde\ell_2\tilde\chi+2d\psi\bigg]^2
\nn \\
& & +\f{e^{-4\phi}}{d^2}\bigg[ d^2+\f{e^{2\phi}}{4}\Big((\tilde\ell_1+2d\tilde\chi)^2+(\tilde\ell_2-2d\chi)^2\Big)\bigg]^2.
\eqa
The integer $m$ is now unconstrained and is amenable to Poisson resummation using the standard formula
\beq
\sum_{m\in\mbb{Z}}e^{-\pi x (m+a)^2+2\pi i m b}=\f{1}{\sqrt{x}}\sum_{\tm\in\mbb{Z}}e^{-\f{\pi}{x}(\tm+b)^2-2\pi i (\tm +b)a}.
\eeq
Using an integral representation for the summand in the remainder $\mc{A}^{(s)}$ defined in (\ref{firstconstant}),
\beq \Big[\vec{\om}^{\dagger}\cdot \mc{K} \cdot \vec{\om}\Big]^{-s} =\f{\pi^s}{\Gamma(s)} \int \f{dt}{t^{s+1}} e^{-\f{\pi}{t} \vec{\om}^{\dagger}\cdot \mc{K} \cdot \vec{\om}},
\eeq
and performing a Poisson resummation on $m$, we obtain
\beqa
{}\mc{A}^{(s)}&=&\f{\pi^s}{\Gamma(s)}e^{-2s\phi}\sum_{\tm \in\mbb{Z}}
\sum^{\qquad \prime}_{(p_1,p_2)\in\mbb{Z}^2} \sum_{(n_1,n_2)\in\mbb{Z}^2\atop  2d | n_1^2+n_2^2}\f{d}{|\om_3|} e^{-2\pi i \tm \big(-\f{|\om_2|^2}{2|\om_3|^2}(q_1p_1+q_2p_2)+\tilde\ell_1\chi+\tilde\ell_2\tilde\chi+2d\psi\big)}
\nn \\
{}& &\phantom{+}\times  \int_0^{\infty}\f{dt}{t^{s+1/2}} e^{-\pi t \f{d^2}{|\om_3|^2} \tm^2-\f{\pi}{t} \f{|\om_3|^2}{d^4}e^{-4\phi} \Big[ d^2+\f{e^{2\phi}}{4}\big((\tilde\ell_1+2d\tilde\chi)^2+(\tilde\ell_2-2d\chi)^2\big)\Big]^2},
\label{C2}
\eqa
where we have indicated explicitly the constraint from (\ref{problem}) that $2d$ must divide $|\om_2|^2$.

We now split off the non-abelian contribution with $\tm\neq 0$:
\beq
\mc{A}^{(s)} =\mc{D}^{(s)}+\mc{E}_s^{(\text{NA})}, 
\label{abelianNonabelianSplit}
\eeq
where $\mc{E}_s^{(\text{NA})}$ denotes the non-abelian term with $\tm\neq 0$, to be considered later. From $\mc{D}^{(s)}$ we will be able to extract the second constant term $\mc{E}_s^{(1)}$ as well as the abelian Fourier coefficients $\mf{C}^{\text{(A)}}_{\ell_1,\ell_2}(s)$. Explicitly we have
\beq
 \mc{D}^{(s)}=\f{\pi^s}{\Gamma(s)}e^{-2s\phi}\sum^{\qquad \prime}_{(p_1,p_2)\in\mbb{Z}^2} \sum_{(n_1,n_2)\in\mbb{Z}^2\atop  2d | n_1^2+n_2^2}\f{d}{|\om_3|} \int_0^{\infty}\f{dt}{t^{s+1/2}} e^{-\f{\pi}{t} \f{e^{-4\phi}|\om_3|^2}{d^4} \Big[ d^2+\f{e^{2\phi}}{4}\big((\tilde\ell_1+2d\tilde\chi)^2+(\tilde\ell_2-2d\chi)^2\big)\Big]^2}.
 \eeq
To get rid of the square in the exponent, we shall perform the integration over $t$ and then choose a new integral representation of the summand. The current form of the exponent will be convenient for the evaluation of the non-abelian terms in Section \ref{NonabelianCoefficients}, but for our present purposes we shall rewrite it in the following way
 \beq
 \f{e^{-4\phi}|\om_3|^2}{d^4}\bigg[d^2+\f{e^{2\phi}}{4}\big((\tilde\ell_1+2d\tilde\chi)^2+(\tilde\ell_2-2d\chi)^2\big)\bigg]^2=\f{1}{4|\om_3|^2}\Big[ |\mc{Y}|^2+2e^{-2\phi} |\om_3|^2\Big]^2,
 \eeq
where we defined the new variable $\mc{Y}=\mc{Y}_1+i\mc{Y}_2$, with
\beqa
\mc{Y}_1 &:=& n_1+(p_1-p_2) \tilde\chi -(p_1+p_2)\chi,
\nn \\
\mc{Y}_2 &:=& n_2 +(p_1+p_2) \tilde\chi+(p_1-p_2) \chi .
\eqa
Evaluating the integral over $t$ then yields
 \beq
\mc{D}^{(s)}=\f{2^{2s-1}\sqrt{\pi}\Gamma(s-1/2)}{\Gamma(s)} e^{-2s\phi} \sum^{\qquad \prime}_{(p_1,p_2)\in\mbb{Z}^2}\sum_{(n_1,n_2)\in\mbb{Z}^2\atop  2d | n_1^2+n_2^2} \f{d}{|\om_3|^{2-2s}}\bigg\{\Big[ |\mc{Y}|^2+2e^{-2\phi} |\om_3|^2\Big]\bigg\}^{1-2s}.
  \eeq
  After replacing the term within brackets by its integral representation we obtain
  \beq
    \mc{D}^{(s)}=\f{(2\pi)^{2s-1}\sqrt{\pi}\Gamma(s-1/2)}{\Gamma(s)\Gamma(2s-1)} e^{-2s\phi} \sum^{\qquad \prime}_{(p_1,p_2)\in\mbb{Z}^2}\sum_{(n_1,n_2)\in\mbb{Z}^2\atop  2d | n_1^2+n_2^2} \f{d}{|\om_3|^{2-2s}}\int_0^{\infty} \f{dt}{t^{2s}} e^{-\f{\pi}{t} \big[ |\mc{Y}|^2+2e^{-2\phi} |\om_3|^2\big]}.
    \label{D}
    \eeq
 Since all values of $n_1$ and $n_2$ are almost degenerate we shall perform a further Poisson resummation on these variables. Here we must take into account the remaining constraint that $2d$ divides $n_1^2+n_2^2$. The set of solutions to this constraint can be written as
\beq
n_1 = n_1^0 + \delta n_1 \ ,\qquad n_2 = n_2^0 + \delta n_2,
\label{n1n2solution}
\eeq
where $(\delta n_1,\delta n_2)$ runs over the lattice $L$
\beq
L =\left\{(\delta n_1,\delta n_2)\right\} =\left\{d ( k_1 + k_2 , k_1 - k_2 )\,:\, (k_1,k_2)\in \mbb{Z}^2 \right\}
\label{LatticeL}
\eeq
and $(n_1^0,n_2^0)$ runs over all solutions
of the quadratic equation $n_1^2+n_2^2=0 \mod 2d$ in a fundamental domain. We take this fundamental domain to be $0\leq n_1^0<d$ and $0\leq n_2^0<2d$, and it has area $2d^2$. The set of such solutions in the fundamental domains will be written as
\beq\label{fundsolutions}
\mc{F}(d) := \left\{ n_1^0 + i  n_2^0\,:\,n_1^2+n_2^2=0 \mod 2d, \ 0\leq n_1^0<d\,,\ 0\leq n_2^0<2d\right\}\,.
\eeq
The cardinality of this set gives the number of solutions
\beq
N(d) := \sharp\mc{F}(d).
\label{MultiplicativeCharacter}
\eeq
This number series is multiplicative but not completely multiplicative \cite{SloaneA086933} (see {\bf Paper VIII} for more details).

After inserting (\ref{n1n2solution}) into (\ref{D}) and performing a Poisson resummation on $\delta n_1$ and $\delta n_2$, we obtain
\beq
\begin{split}
\mc{D}^{(s)}=\f{(2\pi)^{2s-1}\sqrt{\pi}\Gamma(s-1/2)}{2\Gamma(s)\Gamma(2s-1)} e^{-2s\phi} 
\sum^{\qquad \prime}_{(p_1,p_2)\in\mbb{Z}^2}\sum_{\tilde{\omega}_2\in L^*}
\sum_{f\in \mc{F}(d)}
\f{1}{d |\om_3|^{2-2s}}  \\
    \times e^{2\pi i \Re(\tilde{\omega}_2 f)}\,  
    \int_0^{\infty} \f{dt}{t^{2s-1}} 
     e^{-\pi t (\tn_1^2+\tn_2^2)-\f{2\pi}{t} e^{-2\phi} |\om_3|^2 + 2\pi i (\ell_1 \chi+\ell_2 \tilde\chi)},
\end{split}
\label{Ds}
\eeq
where $L^*$ is the lattice dual to $L$, 
\beq\label{duallattice}
L^* = \left\{\tilde{\omega}_2 = \tilde{n}_1+i\tilde{n}_2 =\frac{1}{2d} ( \tilde k_1 +\tilde  k_2 , \tilde k_1 - \tilde k_2 )\,:\,(\tilde k_1,\tilde k_2)\in \mbb{Z}^2 \right\}\,,
\eeq
the set $\mc{F}(d)$ denotes the elements in the fundamental domain contributing to (\ref{fundsolutions}) for $d=\gcd(p_1,p_2)$, and we defined the new charges
\beqa
\ell_1 &:=& \tn_1 (p_1+p_2)-\tn_2 (p_1-p_2)\,, 
\nn \\
\ell_2 &:=& \tn_1 (p_2-p_1) -\tn_2 (p_1+p_2)\,.
\label{abeliancharges}
\eqa     

\subsection{Second Constant Term}
\label{Section:SecondConstantTerm}
We may now extract the second constant term from the $\ell_1=\ell_2=0$ part of the sum, and accordingly we split $\mc{D}^{(s)}$ as
\beq
\mc{D}^{(s)}=\mc{E}_s^{(1)}+\mc{E}_s^{(\text{A})},
\eeq
where $\mc{E}_s^{(\text{A})}$ is the abelian term in the Fourier expansion to be considered in the next subsection. The $\ell_1=\ell_2=0$ part arises from the $\tilde{\omega}_2=0$ term which reads
\beq
\mc{E}_s^{(1)}=\f{(2\pi)^{2s-1}\Gamma(s-1/2)}{2\sqrt{\pi}\Gamma(s)\Gamma(2s-1)} e^{-2s\phi} \sum^{\qquad \prime}_{(p_1,p_2)\in\mbb{Z}^2}\sum_{f\in\mc{F}(d)}  \f{1}{d |\om_3|^{2-2s}}\int_0^{\infty} \f{dt}{t^{2s-1}} e^{-\f{2\pi}{t} e^{-2\phi} |\om_3|^2}.
\eeq
The sum over $f\in\mc{F}(d)$ is now identified with the multiplicative function $N(d)$ in (\ref{MultiplicativeCharacter}) and the integral can be explicitly evaluated with the result
\beq
\mc{E}_s^{(1)}= \f{\pi^{3/2}\Gamma(s-1/2)\Gamma(2s-2)}{\Gamma(s)\Gamma(2s-1)} e^{-2(2-s)\phi} \sum^{\qquad \prime}_{(p_1,p_2)\in\mbb{Z}^2}N(d)\frac{1}{d |\om_3|^{2s-2}}.
\eeq
The sum can now be recast in terms of Riemann and Dedekind zeta functions as follows. Extract the greatest common divisor of $p_1$ and $p_2$, defining $p_1= d p_1^{\prime}$ and $p_2= d p_2^{\prime}$, with $d=\gcd(p_1, p_2)$ and $\gcd(p_1^{\prime}, p_2^{\prime})=1$. This yields a sum over $d$ and coprime $(p_1',p_2')$
\beq
\sum^{\qquad \prime}_{(p_1,p_2)\in\mbb{Z}^2} N(d) d^{-1} |p|^{2-2s}=
\left( \sum_{d>0}  N(d) d^{1-2s} \right) \, 
\left( \sum_{(p_1^{\prime}, p_2^{\prime})=1} \f{1}{(p_1^{\prime2}+p_2^{\prime 2})^{s-1}} \right).
\label{secondconstantterm}
\eeq
The second sum may be rewritten as a ratio of Riemann and Dedekind zeta functions as follows (see Section \ref{Section:FirstConstantTerm})
\beq
 \sum_{(p_1^{\prime}, p_2^{\prime})=1} \f{1}{(p_1^{\prime2}+p_2^{\prime 2})^{s-1}} =\f{4\zeta_{\mbb{Q}(i)}(s-1)}{\zeta(2s-2)}.
 \eeq
Let us now consider the first sum on the right hand side of (\ref{secondconstantterm}) which involves the combinatorial function $N(d)$ defined in (\ref{MultiplicativeCharacter}) (see also~\cite{SloaneA086933}). Given $N(d)$ we may construct a Dirichlet series
\beq\label{dirichletseries}
L(N, s): = \sum_{d=1}^{\infty} N(d) d^{-s}
\eeq 
that converges for $\Re(s)>2$. We shall evaluate this series using its Euler product presentation. To this end we note that the multiplicative series $N(d)$ exhibits the following properties~\cite{SloaneA086933}
\beq
N(2^m)=2^m\ ,\qquad N(p^m) = 
\left\{ \begin{matrix} 
(m(p-1)+p) p^{m-1}\ , & p=1\, \text{mod}\, 4 \\
p^{2\, \lfloor m/2 \rfloor}\ , &\, \, p=3\, \text{mod} \, 4 .
\end{matrix}  \right.
\label{MultiplicativityN}
\eeq
Therefore, the Dirichlet series (\ref{dirichletseries}) has an Euler product representation given by (see Appendix B of {\bf Paper VIII} for the derivation)
\beq
L(N, s)=\frac{1}{1-2^{1-s}} \prod_{p\, : \, p=1\, \text{mod} \, 4} \frac{1-p^{-s}}{(1-p^{1-s})^2}
\prod_{p\, :\, p=3\, \text{mod} \, 4} \frac{1+p^{-s}}{(1-p^{1-s})(1+p^{1-s})},
\label{EulerProductN(d)}
\eeq
where the product runs over all primes $p> 2$. 
Comparing to the Euler product presentations of the Dirichlet beta function (\ref{betaEuler}) and the Riemann zeta function (\ref{zetaEuler}), we deduce that the Dirichlet series $L(N,s)$ can be expressed explicitly as
\beq
L(N,s) = \frac{\beta(s-1) \zeta(s-1)}{\beta(s)}.
\eeq
Putting everything together we then find the following expression for the constant term 
\beq
\mc{E}_s^{(1)}= 4\f{\pi^{3/2}\Gamma(s-1/2)\Gamma(2s-2)}{\Gamma(s)\Gamma(2s-1)} \f{L(N,2s-1)}{\zeta(2s-2)} \zeta_{\mbb{Q}(i)}(s-1) e^{-2(2-s)\phi} .
\eeq
Referring back to the completed Dedekind zeta function (\ref{completedDedekind}) and Dirichlet beta function (\ref{completedBeta}) we define a completed ``Picard Zeta function'' by
\beq
\label{defxi}
\mathfrak{Z}(s) := \zeta_{\mbb{Q}(i)*}(s) \beta_*(2s-1)\ ,
\eeq 
in term of which the two constant terms can be neatly summarized by
\beq\label{constantTermstogether}
\mc{E}_s^{(\text{const})}=\mc{E}_s^{(0)}+ \mc{E}_s^{(1)} = 4\zeta_{\mbb{Q}(i)}(s) \left\{ e^{-2s\phi} + \frac{\mathfrak{Z}(2-s)}{\mathfrak{Z}(s)} e^{-2(2-s)\phi}\right\}.
\eeq
Eq. (\ref{constantTermstogether}) can be viewed as an extension of Langlands's constant term formula \cite{Langlands} for Eisenstein series associated to special linear groups to the case of the unitary group $SU(2,1)$. The completed Picard zeta function $\mathfrak{Z}(s)$ 
plays the same role as the  completed Riemann zeta function $\xi(s)=\pi^{-s/2}\Gamma(s/2)\zeta(s)$ 
in Langlands' formula. 

\subsection{Abelian Fourier Coefficients}
\label{abelianCoefficients}
We now turn to the abelian Fourier coefficients, corresponding to the terms $(\tn_1,\tn_2)\neq 0$
in \eqref{Ds}. The integral over $t$ leads to a modified Bessel function,
\beqa
\mc{E}_s^{(\text{A})}&=&\f{2\pi^{2s-1/2}\Gamma(s-1/2)}{\Gamma(s)\Gamma(2s-1)} 
e^{-2\phi} \sum^{\qquad \prime}_{(p_1,p_2)\in\mbb{Z}^2} 
\sum^{\qquad \prime}_{(\tilde{k}_1,\tilde{k}_2)\in\mbb{Z}^2}
\sum_{f\in\mc{F}(d)}
\frac1{d^{2s-1}} \, |u|^{2s-2}\nn\\
&&\quad\times
e^{\f{\pi i}{d}\Re\left[ u f (1-i)\right]}\,
 \, K_{2s-2}\Big(2\pi e^{-\phi}|\Lambda|\Big)e^{2\pi i (\ell_1\chi+\ell_2\tilde\chi)}\,,
\eqa
where we have introduced the following additional notation
\beq
u=\tilde{k}_1+i\tilde{k}_2\,,\quad \Lambda = \ell_2-i \ell_1
\eeq
for
\beq
\ell_1 =\f{1}{d} (\tilde{k}_1p_2+\tilde{k}_2p_1)\,,\quad
\ell_2 = \f{1}{d} (\tilde{k}_2p_2-\tilde{k}_1p_1).
\label{abeliancharges2}
\eeq
These charges are manifestly integral since $d$ divides $p_1$ and $p_2$. This last relation can also be written as
\beq
\Lambda=\frac{u\om_3}{d} = u \omega_3',
\eeq
where $\omega_3'=\omega_3/d$ is a Gaussian number with coprime real and imaginary part.
To extract the abelian Fourier coefficients $\mf{C}^{\text{(A)}}_{\ell_1, \ell_2}(\phi)$ we therefore replace the sum over $\omega_3$ and $u$ by a sum over $d$, $\Lambda$ and $\omega_3'$ where the primitive Gaussian integer $\omega_3'$ has to be a Gaussian divisor of $\Lambda$, to wit
\beqa\label{abterm}
\mc{E}_s^{(\text{A})}&=&C^{\text{(A)}}_s e^{-2\phi}  \sum^{\qquad \prime}_{\Lambda\in\mbb{Z}[i]}\left\{\sum_{\om_3' | \Lambda} \left|\f{\Lambda}{\omega_3'}\right|^{2s-2}
\left(\sum_{d>0}\f{1}{d^{2s-1}} \sum_{f\in\mc{F}(d)} e^{\f{\pi i}{d} \Re\big[\f{\Lambda}{\om_3'}f (1-i)\big]}\right)\right\}  \nn\\
& & \phantom{++++}\times K_{2s-2}\Big(2\pi e^{-\phi} |\Lambda|\Big) e^{ 2\pi i (\ell_1 \chi+\ell_2\tilde\chi)}\,,
\eqa
where the coefficient is given by
\beq\label{coeffab}
C^{\text{(A)}}_s=\f{2\pi^{2s-1/2}\Gamma(s-1/2)}{\Gamma(s)\Gamma(2s-1)} 
= \frac{8\zeta_{\mbb{Q}(i)}(s)\beta(2s-1)}{\mathfrak{Z}(s)}\,.
\eeq
To make contact with the general discussion of Section \ref{section:generalFourierexpansion}, we rewrite this result as a sum over the real variables $\ell_1$ and $\ell_2$:
\beq
\mc{E}_s^{(\text{A})}=2\zeta_{\mbb{Q}(i)}(s) \frac{e^{-2\phi}}{\mathfrak{Z}(s)}\sum^{\qquad \prime}_{(\ell_1, \ell_2)\in\mbb{Z}^2}\mu_s(\ell_1, \ell_2) \big[\ell_1^2+\ell_2^2\big]^{s-1} K_{2s-2}\Big(2\pi e^{-\phi}\sqrt{\ell_1^2+\ell_2^2}\Big) e^{2\pi i (\ell_1\chi+\ell_2\tilde\chi)},
\label{FinalabelianTerm}
\eeq
where we defined the summation measure
\beq
\label{abinstmes}
\mu_s(\ell_1, \ell_2):= 4\beta(2s-1) \sum_{\om_3' | \Lambda} |\omega_3'|^{2-2s} \left(\sum_{d>0} d^{1-2s} \sum_{f\in\mc{F}(d)} e^{\f{\pi i}{d} \Re\big[\f{\Lambda}{\om_3'}f(1-i)\big]}\right),
\eeq
containing the sum over primitive Gaussian divisors of $\Lambda=\ell_2-i\ell_1$. The sum over $d$ in the parenthesis may be carried out for fixed $\Lambda$ and $\omega_3'$ to give the Gaussian divisor function (see Appendix B of {\bf Paper VIII} for the derivation)
\beq
\sum_{d>0} d^{1-2s} \sum_{f\in\mc{F}(d)} e^{\f{\pi i}{d} \Re\big[\f{\Lambda}{\om_3'}f (1-i)\big]} = \frac{1}{4\beta(2s-1)}
\sum_{z|\frac{\Lambda}{\omega_3'}} |z|^{4-4s}\,,
\label{equality}
\eeq
whence the instanton measure (\ref{abinstmes}) simplifies to
\beq\label{FinalabelianMeasure}
\mu_s(\ell_1,\ell_2) = \sum_{\omega_3'|\Lambda} |\omega_3'|^{2-2s} 
\sum_{z|\frac{\Lambda}{\omega_3'}} |z|^{4-4s}\,.
\eeq
Thus, the abelian summation measure (\ref{FinalabelianMeasure}) involves both a sum over primitive divisors of $\Lambda$ and a sum over all divisors of $\Lambda/\omega_3^{\prime}$. By comparing (\ref{FinalabelianTerm}) to (\ref{Generalabelian}) we may now extract the numerical abelian Fourier coefficients:
\beq
C^{(\text{A})}_{\ell_1, \ell_2}(s)= \frac{2\zeta_{\mbb{Q}(i)}(s)}{\mathfrak{Z}(s)}\mu_s(\ell_1, \ell_2) \big[\ell_1^2+\ell_2^2\big]^{s-1}.
\label{NumericalabelianFourierCoefficients}
\eeq
We note that the abelian instanton measure (\ref{FinalabelianMeasure}) is not multiplicative over the Gaussian primes: The relation
\beq\label{eqn:abelianmult}
\mu_s(\Lambda_1) \mu_s(\Lambda_2) = \mu_s(\Lambda_1\Lambda_2)
\eeq
holds if and only if there are no common prime factors in the Gaussian integers $\Lambda_1$ and $\Lambda_2$ {\em and} furthermore no prime factors that are related by complex conjugation (as arising from split primes, see Appendix B of {\bf Paper VIII}).

\subsection{Non-Abelian Fourier Coefficients}
\label{NonabelianCoefficients}
Finally we consider 
the non-abelian term $\mc{E}_s^{(\text{NA})}$ in (\ref{abelianNonabelianSplit}). This term reads
\beqa
{}\mc{E}_s^{(\text{NA})}&=&\f{\pi^s}{\Gamma(s)}e^{-2s\phi}\sum^{\qquad \prime}_{\tm \in\mbb{Z} }\sum^{\qquad \prime}_{(p_1,p_2)\in\mbb{Z}^2}\sum_{(n_1,n_2)\in\mbb{Z}^2\atop  2d | n_1^2+n_2^2} \f{d}{|\om_3|} e^{-2\pi i \tm \big(-\f{|\om_2|^2}{2|\om_3|^2}(q_1p_1+q_2p_2)+\tilde\ell_1\chi+\tilde\ell_2\tilde\chi+2d\psi\big)}
\nn \\
{}& &\phantom{+}\times  \int_0^{\infty}\f{dt}{t^{s+1/2}} e^{-\pi t \f{d^2}{|\om_3|^2} \tm^2-\f{\pi}{t} \f{|\om_3|^2}{d^4}e^{-4\phi} \Big[ d^2+\f{e^{2\phi}}{4}\big((\tilde\ell_1+2d\tilde\chi)^2+(\tilde\ell_2-2d\chi)^2\big)\Big]^2}.
\label{NonabelianTerm}
\eqa
The integral is of Bessel type and yields
\beqa
\mc{E}_s^{(\text{NA})}&=& \f{2\pi^s}{\Gamma(s)}e^{-2s\phi} \sum^{\qquad \prime}_{\tm \in\mbb{Z} } \sum^{\qquad \prime}_{(p_1,p_2)\in\mbb{Z}^2}\sum_{(n_1,n_2)\in\mbb{Z}^2\atop  2d | n_1^2+n_2^2} \bigg[\f{d}{|\om_3|}\bigg]^{s+1/2} \bigg[\f{|\tm|^2}{\Re(S_{\ell_1,\ell_2,k})}\bigg]^{s-1/2}
\nn \\
& & \phantom{+++}\times K_{s-1/2}\Big(2\pi \Re(S_{\ell_1, \ell_2, k})\Big)e^{-2\pi i \Im(S_{\ell_1, \ell_2, k})}e^{- \f{\pi i}{2kd} (\ell_1^2+\ell_2^2)(q_1p_1+q_2p_2)} ,
\label{NonabelianTermBessel}
\eqa
where the real and imaginary parts of $S_{\ell_1,\ell_2,k}$ are given by
\beqa
\Re(S_{\ell_1,\ell_2,k})&=& |k| e^{-2\phi} +\f{1}{4|k|}\Big[ (\ell_1+2k\tilde\chi)^2+(\ell_2-2k\chi)^2\Big],
\nn \\
\Im(S_{\ell_1,\ell_2,k})&=&\ell_1\chi+\ell_2\tilde\chi+2k\psi,
\label{defSl12k}
\eqa
and we also defined\footnote{The non-abelian charges ${\ell}_i$ defined in \eqref{defl122} should
not be confused with the abelian charges ${\ell}_i$ in \eqref{abeliancharges}.}
\beqa
{}k&:=& \tm d,
\nn \\
{}\ell_1&:=& \tm\tilde{\ell}_1=\f{k}{|\om_3|^2}\Big[(p_1-p_2)n_1+(p_1+p_2)n_2\Big],
\nn \\
{}\ell_2&:=& \tm\tilde\ell_2=\f{k}{|\om_3|^2}\Big[ (p_1+p_2)n_1-(p_1-p_2)n_2\Big].
\label{defl122}
\eqa
In Gaussian notation, $\Lambda=\ell_2+i\ell_1$, the last two relations amount to
\beq
\Lambda=\f{(1+i)k\om_2}{\om_3}.
\eeq
Comparing the expression (\ref{NonabelianTermBessel}) with the general form of the non-abelian term (\ref{GeneralNonabelianTerm}), we see that the former involves a sum of nearly Gaussian wave-functions peaked around $(\ell_2,-\ell_1)/(2k)$ in the $(\chi, \tilde\chi)$ plane, while the latter 
is written in terms of a basis of
Landau-type wave functions which are eigenmodes of $\pa_\chi$ and $\pa_{\psi+\chi\tilde\chi}$,
with quantized charges $4kn$ and $k$. 
To extract the non-abelian Fourier coefficients $C^{(\text{NA})}_{r,k, \ell}(s)$ we must therefore transform (\ref{NonabelianTermBessel}) into the correct basis. This can be achieved via Fourier transform 
along the variable $\chi$ (or $\tilde\chi$ in the other polarization). 

To perform the Fourier transform we go back to the integral representation in (\ref{NonabelianTerm}). The integrand is quartic in $\chi$ and therefore not immediately amenable for Fourier transform. To remedy this we make the following change of integration variables:
\beq
t \hs \longrightarrow \hs \f{t|\om_3|^2 A}{k^2},
\eeq
where
\beq
A(y, \chi, \tilde\chi)=k \Big[y  + \Big(\tilde\chi+\f{\ell_1}{2k}\Big)^2+\Big(\chi-\f{\ell_2}{2k}\Big)^2\Big],
\eeq
and we recall that $y=e^{-2\phi}$. Implementing this in (\ref{NonabelianTerm}) we obtain 
\beqa
\mc{E}_s^{(\text{NA})}&=&\f{\pi^s}{\Gamma(s)} y^s\sum^{\qquad \prime}_{\tm \in\mbb{Z}}\sum^{\qquad \prime}_{(p_1,p_2)\in\mbb{Z}^2}\sum_{(n_1,n_2)\in\mbb{Z}^2\atop  2d | n_1^2+n_2^2}\f{d|k|^{2s-1}}{|\om_3|^{2s}}  
\nn \\
& & \times e^{-2\pi i \tm \big(-\f{|\om_2|^2}{2|\om_3|^2}(q_1p_1+q_2p_2)+\tilde\ell_1\chi+\tilde\ell_2\tilde\chi+2d\psi\big)}  \int_0^{\infty} \f{dt}{t^{s+1/2}} A^{1/2-s} e^{-\pi \big(t+\f{1}{t}\big) A},
\eqa
where the exponent is now quadratic in both $\chi$ and $\tilde\chi$. Using an integral representation for the factor $A^{1/2-s}$, we may rewrite this expression as follows
\beqa
\mc{E}_s^{(\text{NA})}&=&\f{\pi^{2s-1/2}}{\Gamma(s)\Gamma(s-1/2)} y^s\sum^{\qquad \prime}_{\tm \in\mbb{Z}}\sum^{\qquad \prime}_{(p_1,p_2)\in\mbb{Z}^2}\sum_{(n_1,n_2)\in\mbb{Z}^2\atop  2d | n_1^2+n_2^2}\f{d |k|^{2s-1}}{|\om_3|^{2s}} e^{ \f{\pi i \tm|\om_2|^2}{2|\om_3|^2}(q_1p_1+q_2p_2)}
\nn \\
& & \times e^{-4\pi i k\psi-2\pi i \ell_2\tilde\chi} \int_0^{\infty} \f{dt du}{t^{s+1/2}u^{3/2-s}} e^{-\pi k \big(u+t+\f{1}{t}\big) \big[y+\big(\tilde\chi+\f{\ell_1}{2k}\big)^2\big]} f(y,\chi,\tilde\chi; t,u),
\nn \\
\label{NonabelianTerm2}
\eqa
where all the $\chi$-dependence is contained in the function 
\beq
f(y, \chi ;t,u)=e^{-\pi k \big(u+t+\f{1}{t}\big) \big(\chi-\f{\ell_2}{2k}\big)^2-2\pi i \ell_1\chi}.
\label{functionf}
\eeq
The Fourier transform over $\chi$ is now implemented by substituting
\beq
 f(y, \chi ;t,u)=e^{-4\pi i k\chi\tilde\chi}\int dn \hat{f}(y, n, \tilde\chi ;t,u) e^{8\pi i kn \chi},
 \eeq
 with 
 \beqa
 \hat{f}(y, n, \tilde\chi ;t,u) &=& 4|k| \int d\xi \ e^{-8\pi i kn \xi+4\pi i k \xi \tilde\chi} f(y, \xi, \tilde\chi;t,u)
 \nn \\
 &=& 4|k|\ e^{-\f{\pi}{k} \f{(\ell_1-2k\tilde\chi+8\pi kn)^2}{u+t+1/t}}.
 \eqa
After Fourier transform, the non-abelian term thus becomes 
\beqa
\mc{E}_s^{(\text{NA})}&=&\f{4\pi^{2s-1/2}}{\Gamma(s)\Gamma(s-1/2)} y^s\sum^{\qquad \prime}_{\tm \in\mbb{Z}}\sum^{\qquad \prime}_{(p_1,p_2)\in\mbb{Z}^2}\sum_{(n_1,n_2)\in\mbb{Z}^2\atop  2d | n_1^2+n_2^2}\f{d |k|^{2s}}{|\om_3|^{2s}} e^{ \f{\pi i \tm|\om_2|^2}{2|\om_3|^2}(q_1p_1+q_2p_2)}
\nn \\
{} & & \times \int \f{dt du}{t^{s+1/2}u^{3/2-s}}  e^{-\pi k \big(u+t+\f{1}{t}\big) \big[y+\big(\tilde\chi+\f{\ell_1}{2k}\big)^2\big]} 
\nn \\
& & \times \int dn\ e^{-\f{\pi i}{k}\ell_2(\ell_1-4kn)}e^{-\f{\pi}{k} \f{(\ell_1-2k\tilde\chi+8\pi kn)^2}{u+t+1/t}}\ e^{8\pi i kn \chi-4\pi i k(\psi+\chi\tilde\chi)}.
\label{NonabelianTerm3}
\eqa

Let us now comment on the structure of Eq. (\ref{NonabelianTerm3}). After Fourier transforming we see that the non-abelian term indeed corresponds to an expansion in terms of the invariant wavefunctions on the twisted torus as in Eq. (\ref{GeneralNonabelianTerm}). However, we have not been able to further manipulate Eq. (\ref{NonabelianTerm3}) into the form displayed in (\ref{GeneralNonabelianTerm}) and therefore we cannot extract the numerical Fourier coefficients $C^{(\text{NA})}_{k, \ell}(s)$ in as compact a form as the abelian coefficients (\ref{NumericalabelianFourierCoefficients}). Nevertheless, as a consistency check we shall show that the leading order exponential behaviour of (\ref{NonabelianTerm3}) near the cusp $y\rightarrow \infty$ coincides with that of Eq. (\ref{GeneralNonabelianTerm}). To this end we may take the saddle point approximation for the integrals over $t$ and $u$ in (\ref{NonabelianTerm3}) for which the saddle points are located at $t=1$ and $u=0$. We thus find that the leading exponential dependence of (\ref{NonabelianTerm3}) at the saddle point is given by $e^{-S}$ with 
\beq
\Re(S)=2\pi |k|\Big[y+\Big(\tilde\chi+\f{\ell_1}{2|k|}\Big)^2\Big]+\f{\pi}{2|k|}\big(\ell_1-2|k|\tilde\chi+4|k|n\big)^2.
\eeq
Rearranging terms, this can be written as
\beq
\Re(S)=2\pi |k| y+4\pi |k|\big(\tilde\chi-n\big)^2+4\pi |k|\Big(n+\f{\ell_1}{2|k|}\Big)^2.
\label{saddlepointresult}
\eeq
Using the asymptotic behaviour of the Whittaker function $W_{k,m}(x)\sim e^{-x/2}$ one may indeed verify that the first two terms in (\ref{saddlepointresult}) exactly coincide with the leading behaviour of the general expression (\ref{GeneralNonabelianTerm}) in the limit $y\rightarrow\infty$. We further expect that the summation over $\ell_1$ and $\ell_2$ (or, more precisely, over $\om_2$ and $\om_3$) will restrict the integral over $n$ such that it localizes on the points in $\mbb{Z}+\ell/(4|k|)$ as is expected from the general expression (\ref{GeneralNonabelianTerm}). We stress that the result (\ref{saddlepointresult}) is valid in the polarization (\ref{NewNonabelianTerm}) we have chosen. There is an analogous result for the other polarization. 

Besides the representation of the non-abelian coefficients in the form (\ref{NonabelianTerm3}) one could also try to extract the coefficients by other means. One possibility would be to manipulate the expression (\ref{NonabelianTermBessel}) by expanding out the Bessel function and binomially expanding the resulting power series in $A(y,\chi,\tilde\chi)$ to make contact with the power series expansions of the Hermite polynomials and Whittaker functions of (\ref{GeneralNonabelianTerm}). Alternatively, one could try to compute the coefficients by going to a suitably chosen point in moduli space (e.g. the cusp $y=\infty$) or by using $\pg$ symmetry or Hecke operators to relate the non-abelian coefficients to the abelian ones. 

\subsection{Functional Relation}

The expression (\ref{constantTermstogether}) for the constant terms of the Eisentein series
$\mc{E}_s$ is suggestive of  a functional relation most conveniently written in terms of the Poincar\'e series \eqref{PGPoincare} and the 
Picard Zeta function \eqref{defxi}, 
\beq
\label{functional}
\mathfrak{Z}(s) \mc{P}_s = \mathfrak{Z}(2-s) \mc{P}_{2-s}\ .
\eeq
Indeed, it is easily checked that both the
constant terms (\ref{constantTermstogether})  and the abelian Fourier coefficients 
(\ref{FinalabelianTerm}), (\ref{FinalabelianMeasure}) satisfy this relation, taking into
account the symmetry of the modified Bessel function $K_{2s-2}(x)=K_{2-2s}(x)$.
Unfortunately, due the unwieldy form of the non-abelian terms we are unable to 
present a full proof of  \eqref{functional}, which would constitute 
an analog of the familiar functional relation for Eisenstein series associated to special linear groups \cite{Langlands}. 

\subsection{Summary of Fourier Expansion}

For the readers convenience we here give the complete result of the Fourier expansion:
\beq
\begin{split}
f(\phi, \chi, \tilde\chi, \psi)&=A(s)\, e^{-s\phi} + B(s)\, e^{-2(2-s)\phi} \\
&+ e^{-2\phi}\sideset{}{'}\sum_{(\ell_1, \ell_2)\in\mbb{Z}^2}
C^{(\text{A})}_{\ell_1, \ell_2}(s) \, 
K_{2s-2}\Big(2\pi e^{-\phi}\sqrt{\ell_1^2+\ell_2^2}\Big) e^{-2\pi i (\ell_1\chi+\ell_2\tilde\chi)} \\
&+ e^{-\phi} \sideset{}{'}\sum_{k\in\mbb{Z}} \sum_{\ell=0}^{4|k|-1}\sum_{n\in\mbb{Z}+\f{\ell}{4|k|}}\sum_{r=0}^{\infty} 
C^{\text{(NA)}}_{r,k, \ell}(s) \hs |k|^{1/2-s} e^{-4\pi |k| (\tilde\chi-n)^2} \\
&  \quad \times \,H_{r}\Big(\sqrt{8\pi |k|} (\tilde\chi-n)\Big) W_{-r-\f{1}{2}, s-1} \Big(4\pi |k| e^{-2\phi}\Big) e^{8\pi i k n \chi-4\pi i k(\psi+\chi\tilde\chi)},
\end{split}
\label{FirstFourier}
\eeq
where the abelian Fourier coefficients are given by:
\beq
C^{(\text{A})}_{\ell_1, \ell_2}(s)= \frac{2\zeta_{\mbb{Q}(i)}(s)}{\mathfrak{Z}(s)}\mu_s(\ell_1, \ell_2) \big[\ell_1^2+\ell_2^2\big]^{s-1}.
\label{NumericalabelianFourierCoefficients2}
\eeq

\section{Instanton Corrections to the Universal Hypermultiplet}
\label{Section:InstantonCorrections}
In this section we discuss in detail the physical interpretation of the Eisenstein series $\mc{E}_s(\phi, \lambda, \ga)$ and its Fourier expansion. To this end we begin by recalling what is known about quantum corrections to hypermultiplet moduli spaces in type II Calabi-Yau compactifications, with particular emphasis on recent developments involving twistor theory. We then discuss the restriction to rigid Calabi-Yau compactifications, and show that the Eisenstein series $\mc{E}_s(\phi, \lambda, \ga)$ encodes information about the non-perturbative corrections to the moduli space $\mc{M}_{{}_{\mathrm{UH}}}$ of the universal hypermultiplet.  
%
%We conclude the section by discussing certain aspects of the $c$-map, and speculate on the relation between the quantum corrected geometry of $\mc{M}_{{}_{\mathrm{UH}}}$ and the spectrum of BPS states in pure $D=4$, $ \mc{N} =2$ supergravity.

\subsection{Quantum Corrected Hypermultiplet Moduli Spaces}
\label{Section:QuantumCorrectedModuliSpaces}

Perturbative corrections to hypermultiplet moduli spaces are well understood, and it has been established that the metric on $\mc{M}_{{}_{\mathrm{H}}}$ receives only tree-level and one-loop corrections, but no perturbative corrections beyond one loop \cite{Strominger,AntoniadisMinasian1,AntoniadisMinasian2,Gunther:1998sc,Anguelova:2004sj,RoblesLlana:2006ez}. For the universal hypermultiplet this was rigorously proven in \cite{Anguelova:2004sj}. The general form of the perturbative corrections can be inferred from compactifications of higher derivative couplings in ten dimensions \cite{AntoniadisMinasian1,AntoniadisMinasian2}. Note, however, that due to field redefinition ambiguities for higher derivative terms, the precise coefficients must be inferred from explicit string theory calculations in $D=4$ \cite{AntoniadisMinasian2}.\footnote{I thank Pierre Vanhove for emphasizing this.} In the string frame, the tree-level correction is of the form $\zeta(3) \chi_{{}_X} g_s^{-2}$, where $\chi_{{}_X}$ is the Euler number of the Calabi-Yau threefold $X$. while the one-loop correction to the metric on $\mc{M}_{{}_{\mathrm{H}}}$ is of the form $\zeta(2)\chi_{{}_X}$. The complete perturbatively corrected metric can be found in \cite{RoblesLlana:2006ez, Alexandrov:2008nk}.

Since the work of Becker, Becker and Strominger  \cite{BeckerBeckerStrominger} it is also known that $\mc{M}_{{}_{\mathrm{H}}}$ should receive non-perturbative corrections due to D2-brane and NS5-brane instantons. The contributions to $\mc{M}_{{}_{\mathrm{H}}}$ from these effects have, however, remained very difficult to determine, mainly due to the fact that quaternion-K\"ahlergeometry is much more complicated than special K\"ahler geometry, and because techniques for instanton calculus in string theory are not at all well developed.

Nevertheless, in a series of recent papers \cite{VandorenAlexandrov,Vandoren1,Vandoren2,Alexandrov:2008ds,Alexandrov:2008nk,Alexandrov:2008gh}, part of the non-perturbative corrections to $\mc{M}_{{}_{\mathrm{H}}}$ have been gradually understood. The key idea, following advances in the mathematics \cite{Salamon,Swann,LeBrun} and physics literature \cite{deWitVandoren1,deWitVandoren2,deWitSaueressig} on quaternion-K\"ahlerspaces, is that linear deformations of the hypermultiplet moduli space $\mc{M}_{{}_{\mathrm{H}}}$ can be lifted to its \emph{twistor space} $\mc{Z}_{{}_{\mathrm{H}}}$, which is a $\mbb{C}P^1$ bundle over $\mc{M}_{{}_{\mathrm{H}}}$ which admits a canonical complex structure $\mc{J}$ and a complex contact structure $\mc{C}$. In contrast to $\mc{M}_{{}_{\mathrm{H}}}$, the $4h_{2,1}+6$-dimensional twistor space $\mc{Z}_{{}_{\mathrm{H}}}$ is K\"ahler-Einstein, and quantum corrections to $\mc{M}_{{}_{\mathrm{H}}}$ can thereby, in principle, be encoded in the K\"ahler potential of its twistor space $\mc{Z}_{{}_{\mathrm{H}}}$. Practically, however, the authors of \cite{Alexandrov:2008gh} emphasize that a more convenient way of describing corrections to $\mc{M}_{{}_{\mathrm{H}}}$ is by analyzing deformations of the transition functions which glue together local patches on $\mc{Z}_{{}_{\mathrm{H}}}$.  This can be done since the twistor space is identified with a complex $2h_{2,1}+3$-dimensional contact manifold (see, e.g., \cite{Etnyre} for an introduction to contact geometry), which, by Darboux's theorem, can be completely described by generic local complex coordinates $\big(\xi_{[i]}^{\Lambda}, \tilde{\xi}^{[i]}_{\Lambda}, \al_{[i]}\big)$, $\Lambda=1, \dots, h_{2,1}+1$, in each patch $U_i\subset \mc{Z}_{{}_{\mathrm{H}}} $, and contact transformations $S_{ij}$, relating the coordinates $\big(\xi_{[i]}^{\Lambda}, \tilde{\xi}^{[i]}_{\Lambda}, \al_{[i]}\big)$ to the coordinates $\big(\xi_{[j]}^{\Lambda}, \tilde{\xi}^{[j]}_{\Lambda}, \al_{[j]}\big)$ on the overlap $U_i\cap U_j$. The metric on $\mc{M}_{{}_{\mathrm{H}}}$ can be recovered from the K\"ahler-Einstein metric 
on $\mc{Z}_{\mc{M}_{{}_{\mathrm{H}}}}$ using
\beq
ds^2_{\mc{Z}_{\mc{M}_{{}_{\mathrm{H}}}}} = \frac14 \left( e^{-2 K_{\mc{Z}_{\mc{M}_{{}_{\mathrm{H}}}}}} | \mc{C} |^2 + \nu\,
ds^2_{\mc{M}_{{}_{\mathrm{H}}}} \right) ,
\eeq
where $\nu$ is a numerical constant related to the curvature of the base manifold $\mc{M}_{{}_{\mathrm{H}}}$. 

Locally, in the patch $U_i$, the contact structure $\mc{C}$ is captured by the contact one-form $\mc{C}^{[i]}$ which may be expressed in Darboux form as:
\beq
\label{DarbouxX}
\mc{C}^{[i]} = d\al_{[i]} + \xi_{[i]}^{\Lambda}\, d\tilde{\xi}^{[i]}_{\Lambda}
:= 2 \, e^{\Phi_{[i]}}\, \f{Dz}{z}\ .
\eeq
The second equality defines the ``contact potential'' $\Phi_{[i]}$ in the patch $U_i$, 
a complex function on $\mc{Z}_{\mc{M}}$ holomorphic along the
fiber. The contact potential in the patch $U_i$ is related to the 
K\"ahler potential $K_{\mc{Z}_{\mc{M}}}$ in the same patch via
\beq
\mc{K}^{[i]}_{\mc{Z}_{\mc{M}}}=\log \f{1+z\bar{z}}{|z|}+\Re\big[\Phi_{[i]}(x^{\mu}, z)\big].
\label{generaltwistorpotential}
\eeq

Globally, the complex contact structure on $\mc{Z}_{\mc{M}}$ is determined by the 
complex contact transformations $S^{[ij]}$ between the Darboux coordinate system 
on the overlap $U_i \cap U_j$. These can be described  e.g. 
by providing holomorphic generating functions 
$S_{ij}\big(\xi_{[i]}^{\Lambda}, \tilde{\xi}^{[j]}_{\Lambda}, \al_{[j]}\big)$, subject to 
compatibility conditions on triple overlaps $U_i \cap U_j \cap U_k$, equivalence under 
local contact transformations on $U_i$ and $U_j$, and reality constraints. 
The quaternion-K\"ahler metric on $\mc{M}_{{}_{\mathrm{H}}}$ can then be extracted from these holomorphic data,
by determining the contact twistor lines, i.e. expressing  
$\big(\xi_{\Lambda}, \tilde{\xi}^{\Lambda}, \alpha,\Phi \big)$ in some patch $U$
in terms of the coordinates $x^{\mu}\in \mc{M}_{{}_{\mathrm{H}}}$ on the base manifold and the complex coordinate $z\in \mbb{C}P^1$ on the fiber. Plugging the solution into \eqref{DarbouxX}
allows to extract the $SU(2)$ connection $p_\pm, p_3$, the quaternionic 2-forms and 
finally the metric on $\mc{M}_{{}_{\mathrm{H}}}$. 
More details on this construction can be found in  \cite{Alexandrov:2008nk,Alexandrov:2008gh}.

%The geometry of the hypermultiplet moduli space can then be extracted by determining the \emph{contact twistor lines}, i.e. expressing the generic coordinates $\big(\xi_{\Lambda}, \tilde{\xi}^{\Lambda}, \alpha\big)$, in some patch $U$, in terms of the coordinates $x^{\mu}\in \mc{M}_{{}_{\mathrm{H}}}$ on the base manifold, and the complex coordinate $z\in \mbb{C}P^1$ on the fiber. Deformations of the contact transformations $S_{ij}$ then determine the corrections to the contact twistor lines, from which the corrected geometry of $\mc{M}_{{}_{\mathrm{H}}}$ may be extracted in terms of the \emph{contact potential} $ e^{ {\Phi_{[i]}(x^{\mu}, z)}}$, which determines the K\"ahler potential on $\mc{Z}_{{}_{\mathrm{H}}}$ through
%\beq
%\mc{K}^{[i]}_{\mc{Z}_{{}_{\mathrm{H}}}}=\log \f{1+|z|^2}{|z|}+\Re\big[\Phi_{[i]}(x^{\mu}, z)\big].
%\label{generaltwistorpotential}
%\eeq
% For details on this construction we refer the reader to \cite{Alexandrov:2008nk,Alexandrov:2008gh}. 

\subsubsection*{{  $SL(2,\mbb{Z})$}-Invariance and D-Brane Instantons}

This elaborate scheme has been successfully carried out for all instanton corrections arising from Euclidean D2-branes wrapping both $A$- and $B$-cycles in the Calabi-Yau threefold $X$ \cite{Vandoren1,Vandoren2,Alexandrov:2008gh}. One of the crucial features of this analysis was to utilize the twistor techniques discussed above, while at the same time imposing $SL(2,\mbb{Z})$-invariance of the effective action \cite{Vandoren1}. This $SL(2,\mbb{Z})$-invariance descends from the familiar $SL(2, \mbb{Z})$-invariance of type IIB string theory, and carries over to the IIA side via mirror symmetry in four dimensions \cite{Vandoren2}. In type IIB, the vector multiplet moduli space $\mc{M}_{{}_{\mathrm{VM}}}^{{}_{\mathrm{IIB}}}$ is classically exact to all orders in $\alpha^{\prime}$, as well as in $g_s$. On the other hand, the hypermultiplet moduli space $\mc{M}_{{}_{\mathrm{H}}}^{{}_{\mathrm{IIB}}}$ receives perturbative $\alpha^{\prime}$- and $g_s$-corrections, as well as non-perturbative worldsheet instanton and brane instanton corrections. The hypermultiplet moduli in type IIB encode the K\"ahler structure of the Calabi-Yau threefold, and therefore receives corrections from Euclidean D$p$-branes  ($p=-1, 1, 3, 5$) wrapping even cycles in $X$, as well as Euclidean NS5-branes wrapping the entire Calabi-Yau.

We have learned in Chapter \ref{Chapter:Aspects} that the effects of D$(-1)$-instantons in ten dimensions are automatically captured by demanding invariance under $SL(2,\mbb{Z})$ \cite{GreenGutperle}. By assuming that $SL(2,\mbb{Z})$ remains unbroken upon compactification on $X$, the authors of \cite{Vandoren1} showed that the D(-1)-instanton corrections to the metric on the hypermultiplet moduli space $\mc{M}_{{}_{\mathrm{H}}}^{{}_{\mathrm{IIB}}}$ are also captured by the same Eisenstein series $\mc{E}_{3/2}^{SL(2, \mbb{Z})}(\tau, \bar{\tau})$, which in this context is interpreted as a contribution to the contact potential $e^{\Phi(x^{\mu}, z)}$ on the twistor space $\mc{Z}_{\mc{M}_{{}_{\mathrm{H}}}^{{}_{\mathrm{IIB}}}}$ \cite{Alexandrov:2008gh}.\footnote{To be more specific, references \cite{Vandoren1,Vandoren2} were analyzing this problem within the framework of the so-called \emph{hyper-K\"ahler cone} (or \emph{Swann bundle}) $\mc{S}$, which is a $\mbb{C}^2/\mbb{Z}_2$ bundle over the quaternion-K\"ahlerspace $\mc{M}_{{}_{\mathrm{H}}}$. The space $\mc{S}$ is then a hyperk\"ahler manifold of real dimension $4h_{2,1}+8$. Deformations of $\mc{M}_{{}_{\mathrm{H}}}$ can in this approach be described in terms of the twistor space $\mc{Z}_{\mc{S}}=\mbb{C}P^1\times \mc{S}$, a trivial $\mbb{C}P^1$ bundle over $\mc{S}$. This construction is geared towards the so-called \emph{projective superspace} description of the hypermultiplet moduli space, which relies on the existence of $h_{2,1}+2$ commuting isometries (in the IIA picture). This is indeed the case in the presence of only D2-brane instantons on the IIA side, or D(-1), D1 and F1-instantons in type IIB. The Eisenstein series $\mc{E}_{3/2}^{SL(2, \mbb{Z})}(\tau, \bar{\tau})$, encoding the D(-1)-instantons, then gives a contribution to the hyperk\"ahler potential $\chi$ on $\mc{S}$ \cite{Vandoren1}. This formalism fails, however, when $B$-type D2-brane instantons as well as NS5-brane instantons in IIA (or D3, D5, and NS5-brane instantons in IIB) are included. For such generic instanton configurations no isometries remain and therefore the projective superspace approach breaks down. However, the general method of \cite{Alexandrov:2008gh}, based on the twistor space $\mc{Z}_{{}_{\mathrm{H}}}$ of $\mc{M}_{{}_{\mathrm{H}}}$ itself, still applies.} Moreover, it is known that the (D1, F1)-system transforms as a doublet under $SL(2,\mbb{R})$, and hence by imposing $SL(2,\mbb{Z})$-invariance in four dimensions also these instanton corrections to $\mc{M}_{{}_{\mathrm{H}}}^{{}_{\mathrm{IIB}}}$ were determined \cite{Vandoren1}. However, $SL(2,\mbb{Z})$-invariance is not sufficient to determine the contributions to $\mc{M}_{{}_{\mathrm{H}}}^{{}_{\mathrm{IIB}}}$ from  Euclidean D3, D5- and NS5-brane instantons.

%\subsubsection*{Mirror Symmetry and D2-Brane Instantons in Type IIA}
In \cite{Vandoren2}, these results were translated into the type IIA picture using mirror symmetry. After the mirror map, the contact potential $e^{\Phi(x^{\mu}, z)}$ carries information about the corrected geometry of $\mc{M}_{{}_{\mathrm{H}}}:= \mc{M}_{{}_{\mathrm{H}}}^{{}_{\mathrm{IIA}}}$ due to Euclidean D2-branes wrapping $A$-cycles in $X$. However, $SL(2,\mbb{Z})$-invariance is insufficient also in the IIA picture for summing up all possible non-perturbative corrections, and indeed misses the contributions from $B$-type D2-brane instantons, and NS5-brane instantons. The insight of the recent paper \cite{Alexandrov:2008gh}, was that the contributions from D2-branes wrapping $B$-cycles may nevertheless be obtained by imposing \emph{electric-magnetic duality} (or, more generally, \emph{symplectic invariance}) between $A$- and $B$-cycles. At large volume, weak-coupling, this led to the following form of the contact potential $e^{\Phi(x^{\mu}, z)}$ encoding all D2-brane instanton corrections to the hypermultiplet moduli space $\mc{M}_{{}_{\mathrm{H}}}$ in type IIA on a Calabi-Yau threefold $X$:
\beq
\label{phipertD2}
e^\Phi = \frac{\tau_2^2 V_{\mc{X}}}{2} +\f{\chi_E}{192 \pi} + e^{\Phi_{(\text{D2})}}+\cdots 
\eeq
where $\tau_2=1/g_s$ is the ten-dimensional string coupling, $\chi_{E}$ is the Euler
number of $\mc{X}$ and $V_{\mc{X}}$ is the volume
of  $\mc{X}$ in string units. The complete perturbatively corrected metric 
corresponding to the contact potential \eqref{phipertD2} can be found in \cite{RoblesLlana:2006ez, Alexandrov:2008nk}. We note that in the corresponding expression in the type IIB hypermultiplet sector there are additional contributions arising from $\alpha^{\prime}$-corrections and worldsheet instantons. However, due to the fact that the metric on the complex structure moduli space of $\mc{X}$ is insensitive to $\alpha^{\prime}$-effects, these corrections are absent in the type IIA expression (\ref{phipertD2}). The corrections to the contact potential which are non-perturbative in $g_s$ are due to 
D2-brane and NS5-brane instantons \cite{BeckerBeckerStrominger}. Using S-duality and mirror symmetry, the form of the D2-brane instanton corrections was 
obtained in a series of works \cite{Alexandrov:2008nk,Alexandrov:2008gh,Vandoren1,Vandoren2,Alexandrov:2008ds,VandorenAlexandrov}. 
To first order away from the one-loop corrected metric, their contribution to the contact potential
reads
\beq
\label{phiD}
e^{\Phi_{(\text D2 )}} = \frac{1}{4\pi^2} \sum_\gamma n_\gamma 
\sum_{m>0} \frac{e^{-\phi} |Z_\gamma|}{m}\, K_1\left( 8\pi m  e^{-\phi} |Z_\gamma|\right)
e^{2\pi i m \int_{\ga} C_{(3)}}
\eeq
where $\gamma$ runs over the homology classes in $H^{3}(\mc{X},\mbb{Z})$, $Z_\gamma$ is the central charge associated to the cycle $\gamma$, $e^{\phi}$
is the 4D string coupling such that $e^{-2\phi} = \tau_2^2 V_{\mc{X}} / 2$, $C_{(3)}$ is the Ramond-Ramond 3-form and $n_\gamma$ is a
numerical factor which counts the number of BPS states in the homology class $\gamma$. 
The ellipsis in (\ref{phipertD2}) corresponds to NS5-brane contributions which have been discussed in \cite{VandorenAlexandrov,AutomorphicNS5}, 
but remain largely mysterious in general.

\subsubsection*{NS5-Brane Instantons Corrections}

The NS5-brane contributions were unfortunately still out of reach in the analysis of \cite{Vandoren2,Alexandrov:2008gh}, due to the fact that deformations of the transition functions $S_{ij}$ on the twistor space $\mc{Z}_{{}_{\mathrm{H}}}$ become considerably more complicated if the NS5-brane contribution is taken into account.
Nevertheless, a ``roadmap'' for how to incorporate NS5-brane effects was proposed in \cite{Vandoren2}. The main idea is to use mirror symmetry to transform the moduli space $\mc{M}_{{}_{\mathrm{H}}}$ back to the IIB side, after electric-magnetic duality between $A$- and $B$-cycles has been imposed. In type IIB this will then automatically include the previously unknown effects of Euclidean D3- and D5-branes. The NS5-branes should then in principle emerge after once more imposing $SL(2,\mbb{Z})$-invariance, since the (D5, NS5)-system transforms as a doublet under $SL(2,\mbb{R})$. The final step would be to use mirror symmetry to map the result back to the IIA side, where now the NS5-brane effects are included. Although  quite appealing, this program is technically very challenging due to the complicated action of $SL(2,\mbb{Z})$ on the moduli after mapping the results back to the IIB-side \cite{Alexandrov:2008gh}. In the next subsection, we give a proposal for how to circumvent this problem in the special case of type IIA compacifications on a rigid Calabi-Yau threefold $\mc{X}$.

\subsection{Instanton Corrections to the Universal Hypermultiplet}
\label{Section:InstantonCorrectionsUH}
As discussed in detail in Section \ref{Section:UniversalSector}, rigid Calabi-Yau compactifications give rise to pure $ \mc{N} =2$ supergravity in four dimensions (ignoring the vector multiplets), coupled to the universal hypermultiplet, whose classical moduli space $\mc{M}_{{}_{\mathrm{UH}}}$ coincides with the coset space $SU(2,1)/(SU(2)\times U(1))$, parametrized by the four-dimensional dilaton $e^{\phi}$, the Ramond-Ramond axions $\chi$ and $\tilde{\chi}$, and the axionic scalar $\psi$, dual to the NS 2-form. This may be viewed as a ``toy example'' of the more general discussion of Section \ref{Section:QuantumCorrectedModuliSpaces}, but one which still contains the essential features: namely D2-brane instantons as well as NS5-brane instantons \cite{BeckerBecker}. This follows from the fact that, even though there is no complex structure, Euclidean D2-branes may still wrap the universal holomorphic and anti-holomorphic 3-cycles $\mc{A}$ and $\mc{B}$, and Euclidean NS5-branes may wrap the entire rigid Calabi-Yau manifold $\mc{X}$. Perturbative and non-perturbative corrections to $\mc{M}_{{}_{\mathrm{UH}}}$ have been discussed extensively in the literature \cite{Strominger,AntoniadisMinasian2,BeckerBecker,Anguelova:2004sj,VandorenAlexandrov}, but no proposal has been put forward for the exact quantum corrected geometry including the effects of NS5-branes. 

\subsubsection*{On the Twistor Space of the Tree-Level Universal Hypermultiplet}
We now illustrate the twistorial construction in the case of the tree-level universal hypermultiplet moduli space. The twistor space $\mc{Z}_{\mc{M}_{\mathrm{UH}}}$ of the classical moduli space $\mc{M}_{\mathrm{UH}}$ can be nicely described group-theoretically as follows. 
Viewing the $\mbb{C}P^1$  twistor fiber as $S^2=SU(2)/U(1)$, the
fibration of $SU(2)/U(1)$ over $\mc{M}_{\mathrm{UH}}$ is such that 
the $SU(2)$ cancels: \cite{PiolineGunaydin,Wolf}:
\beq
\mc{Z}_{\mc{M}_{\mathrm{UH}}}= \f{SU(2)}{U(1)}\ltimes \f{SU(2,1)}{SU(2)\times U(1)}=\f{SU(2,1)}{U(1)\times U(1)}.
\label{TwistorSpace}
\eeq
The twistor space $\mc{Z}_{\mc{M}_{\mathrm{UH}}}$ is a complex 3-dimensional contact manifold, with local coordinates $(\xi, \tilde{\xi}, \al)$. These coordinates parametrize the complexified Heisenberg group $ N _{\mbb{C}}$, or, equivalently, coordinates on the complex coset space $P_{\mbb{C}}\bas SL(3,\mbb{C})$, where $P_{\mbb{C}}$ is the complexification of the parabolic subgroup $P\subset SU(2,1)$ discussed in Appendix~\ref{Section:pAdicConstruction} and $SL(3,\mbb{C})$ is the complexification of $SU(2,1)$. In terms of the coordinates $(\xi, \xit, \al)$ on $P_{\mbb{C}}\bas SL(3,\mbb{C})$ the K\"ahler potential of $\mc{Z}_{\mc{M}_{\mathrm{UH}}}$ takes the following form \cite{PiolineGunaydin}
\beq
K_{\mc{Z}_{\mc{M}_{\mathrm{UH}}}}=\f{1}{2}\log\left[\Big(\big(\xi-\xib\big)^2+\big(\xit-\xitb\big)^2\Big)^2+4\Big(\al-\bar{\al}+\xib\xit-\xi\xitb\Big)^2\right].
\label{TwistorPotentialGeneric}
\eeq
As mentioned above, the contact twistor lines for the unperturbed twistor space correspond to the change of variables that relate the coordinates $(\xi, \tilde{\xi}, \al)$ on $\mc{Z}_{\mc{M}_{\mathrm{UH}}}$ to the coordinates $x^{\mu} =\{ e^{\phi}, \chi, \tilde{\chi}, \psi\}$ on the base $\mc{M}_{\mathrm{UH}}$ and the coordinate $z$ on the fiber $\mbb{C}P^1=SU(2)/U(1)$. These twistor lines were obtained in \cite{PiolineGunaydin}. In our notations they read (away from the north pole $z=0$ and south pole $z=\infty$)
\beqa
\xi&=& -\sqrt{2} \chi+\f{1}{\sqrt{2}}e^{-\phi}\big(z-z^{-1}\big),
\nn \\
\tilde{\xi}&=&-\sqrt{2} \tilde{\chi}-\f{i}{\sqrt{2}}e^{-\phi}\big(z+z^{-1}\big),
\nn \\
\al&=& \phantom{-}2\psi -  e^{-\phi}\Big[z(\tilde\chi+i {\chi})-z^{-1}(\tilde\chi-i {\chi})\Big].
\label{twistorlines}
\eqa

Plugging these into (\eqref{DarbouxX}) and (\ref{TwistorPotentialGeneric}) we find that the K\"ahler potential for the twistor space of the classical universal hypermultiplet, in the coordinates $x^{\mu}\in\mc{M}_{\mathrm{UH}}$ and $z\in\mbb{C}P^1$, reads
\beq
K_{\mc{Z}_{\mc{M}_{\mathrm{UH}}}}=\log \f{1+|z|^2}{|z|}-2\phi,
\eeq
which indeed agrees with the general form of the K\"ahler potential in Eq. (\ref{generaltwistorpotential}) upon identifying the classical contact potential as follows
\beq
e^{\Phi_{\mathrm{classical}}(x^{\mu}, z)}=e^{-2\phi},
\label{classicalcontactpotential}
\eeq
%Plugging these into \eqref{DarbouxX} and (\ref{TwistorPotentialGeneric}), 
%we find that the contact potential in this patch is simply
%\beq
%e^{\Phi(x^{\mu}, z)}=e^{-2\phi}\ ,
%\label{classicalcontactpotential}
%\eeq
which is in particular independent of $z$. We further note that under an action of $SU(2,1)$, the contact potential and contact one-form transform as
\beq
\label{transPhi}
e^{\Phi}\hs \longmapsto \hs |C+D\mc{Z}|^{-2}\ e^{\Phi},\qquad
\mc{C} \longmapsto \hs (C+D\mc{Z})^{2}\,  \mc{C} \ ,
\eeq
which ensure that the K\"ahler potential $K_{\mc{Z}_{\mc{M}_{\text{UH}}}}$ transforms by a K\"ahler transformation, and that $SU(2,1)$ acts isometrically on both $\mc{Z}_{\mc{M}_{\text{UH}}}$
and $\mc{M}_{\text{UH}}$ itself.

\subsubsection*{On the Physical Relevance of the Picard Modular Group}
As discussed extensively in Chapters \ref{Chapter:Introduction} and \ref{Chapter:Aspects}, it is generally expected that the full quantum effective action in string compactifications should be invariant under a global discrete symmetry group $G(\mbb{Z})$ (see, e.g., \cite{HullTownsend,GreenGutperle,Obers:1998fb,ObersPioline,PiolineKiritsis}). However, for Calabi-Yau threefold compactifications little is known about the structure of $G(\mbb{Z})$. For the case of rigid Calabi-Yau compactifications the situation is different. In this case we know that the classical effective action exhibits a global $SU(2,1)$-symmetry, which is broken to a discrete subgroup $G(\mbb{Z})\subset SU(2,1)$ by quantum corrections \cite{BeckerBecker}. The problem is then reduced to finding which discrete subgroup $G(\mbb{Z})$ is preserved. As stressed in Section \ref{cmap}, we shall focus on rigid Calabi-Yau manifolds with period matrix given by $\tau=i$, i.e. with complex multiplication by the Gaussian integers $\mbb{Z}[i]$. On physical grounds, the subgroup $G(\mbb{Z})$ should then contain the following action on the moduli:
\begin{itemize}
\item  A discrete Heisenberg group $ N (\mbb{Z})$, acting by discrete (Peccei-Quinn) shift symmetries on the axions $\chi, \tilde{\chi}$ and $\psi$~\cite{BeckerBecker}:
\beqa\label{heisshiftsscalars2}
\chi & \longmapsto& \chi+a\,,\nn\\
\tilde\chi & \longmapsto& \tilde\chi +b\,,\nn\\
\psi & \longmapsto&\psi +\frac12c -a\tilde\chi +b \chi,
\eqa
where $a, b, c\in \mbb{Z}$, while leaving the dilaton $\phi$ invariant. 
In the type IIA setting, the breaking of the continuous shifts of $\chi$ and $\tilde{\chi}$ are due to D2-brane instantons, while the breaking of the shift of $\psi$ is due to the combined effects of D2- and NS5-brane instantons. The factor $1/2$ appearing in front of $c$ is in agreement with the quantization condition on the NS5-brane instantons derived in \cite{BeckerBecker}. The Heisenberg symmetry 
\eqref{heisshiftsscalars2} should hold more generally for any value of the period matrix $\tau$.

\item The ``electric-magnetic duality''   $R$ which interchanges the R-R scalars $\chi$ and $\tilde{\chi}$~\cite{BeckerBecker}:
\beq
R\hs :\hs (\chi, \tilde{\chi}) \hs \longmapsto \hs (-\tilde{\chi}, \chi).
\label{emduality}
\eeq
This symmetry is only expected to hold for rigid Calabi-Yau compactifications with $\tau=i$.
In that  case it amounts to a change of symplectic basis for $H_3(\mc{X}, \mbb{Z})$.

\item We further assume that a discrete subgroup $SL(2,\mbb{Z})$ of the four-dimensional 
$S$-duality (or, on the type IIB side, Ehlers symmetry), acting 
in the standard non-linear way on the complex parameter $\chi+i e^{-\phi}$
on the slice $\tilde\chi=\psi=0$, is left unbroken by quantum corrections.  
As in earlier endeavours \cite{Lambert:2006ny,Aspects,PiolineYann,ClaudiaAxel,Bao:2007fx},
it is difficult to justify this assumption rigorously, but the fact, demonstrated herein,
that it leads to physically sensible results can be taken as support for this assumption.\footnote{On non-rigid Calabi-Yau manifolds one can use mirror symmetry to obtain S-duality transformations in IIA from the $SL(2,\mbb{Z})$ symmetry of type IIB, as has been advocated in literature for example in~\cite{Gunther:1998sc,deWit:1996ix,Bohm:1999uk,Manschot:2009ia}.}
\end{itemize}
Based on these assumptions, it follows that, when $\tau=i$, $G(\mbb{Z})$ must be the {\em Picard modular group}\footnote{The nomenclature ``Picard group'' is not unique, in fact our Picard group is a member of a family of similar groups $PSU(1,n+1;\mbb{Z}[i])$ of which the case $n=0$, corresponding to $PSL(2,\mbb{Z}[i])$ is also often called {the} Picard group. In this paper we will always mean $SU(2,1;\mbb{Z}[i])$ when speaking of the Picard group.}
$SU(2,1;\mbb{Z}[i])$, defined as the intersection (see e.g. \cite{Falbel:2009cf,FrancsicsLax})
\beq
\label{defpic}
SU(2, 1;\mbb{Z}[i]) := SU(2,1)\cap SL(3,\mbb{Z}[i]).
\eeq
Indeed, the symmetries 1,2,3 above reproduce the list of generators of the Picard modular group
obtained in  \cite{Falbel:2009cf}.  We find it remarkable that adjoining electric-magnetic duality and S-duality to the physically well-established Peccei-Quinn shift symmetries  
generates an interesting discrete subgroup of $SU(2,1)$.

We would now like to test this proposal by analyzing whether the expected instanton corrections to the universal hypermultiplet can be determined by imposing invariance under $\pg $. More specifically, we want to understand if the Eisenstein series $\mc{E}_s(\phi, \lambda, \ga)$, constructed in Section \ref{Section:Eisenstein}, may be related to a generalized version of the contact potential $e^{\Phi(x^{\mu}, z)}$ given in \cite{Alexandrov:2008gh}, but here restricted to the special case of rigid Calabi-Yau compactifications. To this end we shall in the following section investigate in detail the physical interpretation if the Fourier expansion of $\mc{E}_s(\phi, \lambda, \ga)$, obtained in Section \ref{Section:FourierExpansion}, to determine if it encodes the expected instanton contributions, in the spirit of \cite{GreenGutperle,Vandoren1,Vandoren2,Alexandrov:2008gh}.

\subsubsection*{On the Contact Potential and the Picard Eisenstein Series}
\label{Section:TwistorialInterpretation}

We now restrict to the case of type IIA string theory compactified on a rigid Calabi-Yau threefold, 
and propose that the Picard Eisenstein Series $\mc{E}_{s}(\phi, \chi,\tilde\chi,\psi)$, 
for a suitable value of the parameter $s$, controls the exact, quantum corrected metric 
on the universal hypermultiplet moduli space. As in\cite{AutomorphicNS5} (see in particular Section~3.1),
we shall restrict our attention to the contact potential $\Phi(x^\mu,z)$ 
on a certain holomorphic section\footnote{In the presence of NS5-brane corrections, the contact potential
is no longer constant along the fiber. The quaternion-K\"ahler metric on $\mc{M}=\mc{M}_{\rm UH}$
can nevertheless be described, in many different ways, in terms of a single real function $h(x^\mu)$ 
on $\mc{M}$
subject to a non-linear partial differential equation \cite{VandorenAlexandrov,Przanowski1}. 
The latter can be identified with the K\"ahler potential $K_{\mc{Z}_\mc{M}}$ on any holomorphic section $z(x^\mu)$ of $\mc{Z}_\mc{M}$ \cite{Alexandrov:2009vj}. Our proposal refers to 
a specific choice of holomorphic section, which we are not able to specify at this stage. We stress that this technical point plays no role at the level of our present analysis.}
 $z(x^\mu)$ of the twistor space $\mc{Z}_{\mc{M}_{\mathrm{UH}}}$. We also  choose variables such that the action of $SU(2,1)$
on $\mc{M}_{\rm UH}$ is the tree-level action (though it is no longer isometric in general), 
and look for a completion of $\Phi(x^\mu,z(x^\mu))$ which  reproduces the 
expected perturbative contributions.
Determining the specific holomorphic section $z(x^\mu)$ and the exact twistor lines and hypermultiplet metric are  important open problems which lie outside the scope of this work.
%We first observe that for  compactifications on a rigid Calabi-Yau manifold, the volume $V$ is fixed, and the $(\alpha')^3$ correction proportional to  $\chi_{E} \zeta(3)$ is indistinguishable from the tree-level term in \eqref{phipertD2}. Indeed, both correspond to a term proportional to $i (X^0)^2$ in the prepotential, and so can be bundled together by an appropriate choice of normalization.

Matching the powers of the dilaton, we then propose that, on the holomorphic section $z(x^\mu)$ introduced above, the contact potential for the quantum corrected metric on $\mc{M}_{\rm UH}$ is given by\footnote{Due to the different power of $e^\phi$, the contact potential 
$\Phi_{\text{exact}}$ appears to transform differently from its tree-level 
counterpart \eqref{transPhi}; this is not a fatal flaw however, since the locus $z(x^\mu)$ 
is in general not 
fixed by the action of the Picard modular group.}
 \beq
e^{\Phi_{\text{exact}}(x^{\mu},z(x^\mu))}
= \kappa \, e^{\phi} \mc{E}_{3/2}(\phi, \chi, \tilde\chi, \psi)\ ,
\label{Conjecture}
\eeq
where $\mc{E}_s$ is the Picard Eisenstein series \eqref{EisensteinPicard}, and 
$\kappa$ is an adjustable numerical constant. Using the Fourier expansion
\eqref{FirstFourier}, we see that \eqref{Conjecture} predicts 
\beq
e^{\Phi_{\text{exact}}}=4\zeta_{\mbb{Q}(i)}(3/2)  \kappa 
\left( e^{-2\phi} + \frac{\mathfrak{Z}(1/2)}{\mathfrak{Z}(3/2)} \right) + 
e^{\Phi_{(\text{A})}} +
e^{\Phi_{(\text{NA})}}\ ,
\eeq
where the last two terms correspond to the abelian and non-abelian parts of the Fourier expansion, respectively. The two constant terms have the same dependence on the dilaton $e^{\phi}$ as the two perturbative contributions in \eqref{phipertD2}. We thus want to identify the second constant term at $s=3/2$ with the one-loop coefficient $\chi_E /192 \pi$ in (\ref{phipertD2}). Here we run into a problem since for $s=3/2$ we find
\beq
\frac{\mathfrak{Z}(1/2)}{\mathfrak{Z}(3/2)} 
 \approx - 2.32607\ ,
 \label{OneLoopMess}
\eeq
implying that matching with the physical one-loop term requires $\chi_E\sim -1403.05$, 
a negative, non-integer number. This contradicts the fact that $\chi_E=2h_{1,1}\in 2\mbb{N}$ 
for a  rigid Calabi-Yau threefold.  Hence, the value of the one-loop coefficient predicted by the second constant term in the Eisenstein series is not physically viable. While this invalidates the proposal that the principal Eisenstein series $\mc{E}_s$ describes the exact universal hypermultiplet metric,
it does not necessarily ruin the idea that the Picard modular group controls that metric. We proceed with our current proposal however, as the
form of the non-abelian Fourier expansion is largely independent of the details of the automorphic form under consideration. In particular, we show next that the form of the abelian and non-abelian contributions to the Fourier expansion of $\mc{E}_{3/2}(\phi, \chi, \tilde\chi, \psi)$ agrees
with the expected form of D2-brane and NS5-brane instanton contributions, respectively. 
\vspace{.3cm}

\noindent{\it D2-Brane Instantons}
\vspace{.2cm}

\noindent The abelian contribution \eqref{FinalabelianTerm} at $s=3/2$ becomes 
\beq
e^{\Phi_{(\text{A})}} = \frac{2\kappa\, \zeta_{\mbb{Q}(i)}(3/2)\,e^{-\phi}}{\mathfrak{Z}(3/2)}\sum^{\qquad \prime}_{(\ell_1, \ell_2)\in\mbb{Z}^2}\mu_{3/2}(\ell_1, \ell_2) \big[\ell_1^2+\ell_2^2\big]^{1/2} K_{1}\Big(2\pi e^{-\phi}\sqrt{\ell_1^2+\ell_2^2}\Big) e^{-2\pi i (\ell_1\chi+\ell_2\tilde\chi)},
\label{FinalabelianTerm32}
\eeq
where the summation measure $\mu_{3/2}(\ell_1, \ell_2)$ is given in \eqref{FinalabelianMeasure}.  
%This result agrees with the general form \eqref{phiD} of D2-brane contributions, 
%upon identifying $Z_\gamma = \ell_1+i \ell_2$ and $m (\gamma, \zeta)=-\ell_1\chi-\ell_2\tilde\chi$.
In the weak-coupling limit $e^{\phi}\rightarrow 0$ we may use the asymptotic expansion of the modified Bessel function at large~$x$,
\beq
K_t(x) \sim \sqrt{\f{\pi}{2x}} e^{-x}\sum_{n\geq 0} \frac{\Gamma\left(t+n+\frac12\right)}{\Gamma(n+1)\Gamma\left(t-n+\frac12\right)}(2x)^{-n}\ ,
\eeq
to approximate 
\beq 
e^{\Phi_{(\text{A})}}\sim \frac{\kappa \, \zeta_{\mbb{Q}(i)}(3/2)}{\mf{Z}(3/2)}e^{-\phi/2}\sum_{(\ell_1, \ell_2)\in\mbb{Z}^2}^{\quad \prime} \mu_{3/2}(\ell_1, \ell_2)(\ell_1^2+\ell_2^2)^{1/4} e^{-2\pi S_{\ell_1, \ell_2}}\Big[1+\mc{O}(e^\phi)\Big].
\label{D2Term}
\eeq
We thus find that $e^{\Phi_{(\text{A})}}$ exhibits exponentially suppressed corrections in the limit $e^\phi\rightarrow 0$, weighted by the instanton action
\beq
S_{\ell_1, \ell_2}= e^{-\phi} \sqrt{\ell_1^2+\ell_2^2} + i (\ell_1\chi+\ell_2\tilde\chi)\ .
\label{D2action}
\eeq 
This is recognized as the action for Euclidean D2-branes wrapping special Lagrangian 3-cycles in the homology class $\ell_1\mc{A}+\ell_2\mc{B}\in H_3(\mc{X}, \mbb{Z})$, where $(\mc{A},\mc{B})$ provides an integral symplectic basis of $H_3(\mc{X}, \mbb{Z})$. To see this, we note that generally the instanton action for D2-branes wrapping a special Lagrangian submanifold in the homology class $\ga\in H_3(X,\mbb{Z})$ inside a Calabi-Yau threefold $\mc{X}$ is given by 
\beq
S_{\ga}=\f{1}{g_s} \Big| \int_{\ga} \Omega\Big| + i\int_{\ga} C_{(3)},
\label{generalD2action}
\eeq
where ${g}_s$ is the ten-dimensional string coupling, $\Omega\in H_{3,0}(X)$ is the holomorphic 3-form and $C_{(3)}\in H^{3}(X,\mbb{R})/H^{3}(X,\mbb{Z})$ is the RR 3-form. The real  part of the action can further be written in terms of the central charge $Z_{\ga}=e^{K/2} \int_{\ga}\Omega$ as  $\Re(S_{\ga})= e^{-K/2} |Z_{\ga}| / g_s$, where $K=-\log \int_X \Omega\wedge \bar{\Omega} $ is the K\"ahler potential of the complex structure moduli space. Noting that $K=-\log V_{\mc{X}}$ we then find 
\beq
S_{\ga}= e^{-\phi} |Z_{\ga}| + i\int_{\ga} C_{(3)},
\eeq
where we defined the four-dimensional dilaton by $e^{\phi} := V_{\mc{X}}^{-1/2} g_s$. Restricting to a rigid Calabi-Yau threefold $\mc{X}$, we recall from Section \ref{cmap} that the prepotential is $F=\tau X /2$ with $\tau$ being the period ``matrix'' $\int_{\mc{B}} \Omega /\int_{\mc{A}}\Omega$. In this case the D2-brane wraps a 3-cycle in the homology class $\ga=\ell_1\mc{A}+\ell_2\mc{B}\in H_3(\mc{X}, \mbb{Z})$, which gives $Z_{\ga}=\ell_1+\tau\ell_2$, so the instanton action reduces to
\beq
S_{\ell_1,\ell_2}(\tau)= e^{-\phi} |\ell_1+\tau \ell_2| + i \int_{\ell_1\mc{A}+\ell_2\mc{B}}C_{(3)}.
\eeq
Further setting $\tau=i$, which is the relevant value for our analysis, and using Eq. (\ref{RRperiods}) for the periods of the Ramond-Ramond 3-form $C_{(3)}$, this action indeed coincides with the instanton action (\ref{D2action}) predicted from $SU(2,1;\mbb{Z}[i])$-invariance. Thus, we may conclude that the abelian term (\ref{FinalabelianTerm32}) in the Fourier expansion agrees with the general form of D-instanton corrections in \eqref{phiD} upon restricting to a rigid Calabi-Yau threefold which admits complex multiplication by $\mbb{Z}[i]$. 

The infinite series within the brackets in (\ref{D2Term}) should, in the spirit of \cite{GreenGutperle}, arise from perturbative contributions around  the instanton background. The summation measure is given by 
specifying \eqref{FinalabelianMeasure} to $s=3/2$,
\beq
\mu_{3/2}(\ell_1,\ell_2) = \sum_{\omega_3'|\Lambda} |\omega_3'|^{-1} \sum_{z|\frac{\Lambda}{\omega_3'}}|z|^{-2}\ ,
\label{D2measure}
\eeq
where we recall that $\Lambda=\ell_2-i\ell_1$ is a complex combination of the electric and magnetic charges $(\ell_1, \ell_2)$. The instanton measure $\mu_{3/2}(\ell_1, \ell_2)$ should count the degeneracy of Euclidean D2-branes in the homology class $\ell_1\mc{A}+\ell_2\mc{B}\in H_3(\mc{X}, \mbb{Z})$. For 
D2-instantons with $A$-type charge only,  i.e. $\ell_2=0$, and such that $\ell:=\ell_1$ 
is a product of inert primes  (those of the form $p=4n+3$, see Appendix B of {\bf Paper VIII}), the first sum collapses to $\omega_3'=1$ and the instanton measure reduces to 
\beq\label{D2AMeasure}
\mu_{3/2}(\ell, 0) = \sum_{z|\ell} |z|^{-2}\ .
\eeq
This reproduces the instanton measure found on the basis of $SL(2,\mbb{Z})$ invariance 
in \cite{Alexandrov:2008gh,Vandoren1,Vandoren2,PiolineVandoren},  which 
by analogy with \cite{GreenGutperle,Kostov:1998pg,Moore:1998et} 
should count ways of splitting a marginal bound state into smaller constituents.  However, it is possible
that $\ell$  be prime over the integers but
factorizable over the Gaussian integers, e.g. $2=-i(1+i)^2$ or $5=(2+i)(2-i)$, in which case
the measure \eqref{D2measure} involves additional contributions compared to \cite{Alexandrov:2008gh,Vandoren1,Vandoren2,PiolineVandoren}. 
This novel feature of compactifications on rigid Calabi-Yau manifolds is an interesting
prediction of $SU(2,1;\mbb{Z}[i])$-invariance which deserves further investigation.
\vspace{.3cm}

\noindent {\it NS5-Brane Instantons}

\vspace{.2cm}

\noindent As mentioned above, the non-abelian term $e^{\Phi_{(\text{NA})}}$
may be interpreted as NS5-brane instanton contributions.
 Although we have not been able to extract the coefficients  $C^{\text{(NA)}}_{r,k, \ell}(s)$
 in the non-abelian Fourier expansion \eqref{GeneralNonabelianTerm}, 
 we can still extract the instanton action by taking the semiclassical limit. This corresponds to the asymptotic behaviour of (\ref{GeneralNonabelianTerm}) in the limit $y\rightarrow\infty$, or, equivalently, to the saddle point approximation of the $t$-integral in (\ref{NonabelianTerm3}) as analyzed in Section \ref{NonabelianCoefficients}. Expanding the Whittaker function around $x=\infty$ yields
\beqa
W_{k,m}(x) &\sim&e^{-x/2} x^k \sum_{n\geq 0}\frac{\Gamma\left(m-k+n+\frac12\right)\Gamma\left(m+k+\frac12\right)}{\Gamma(n+1)\Gamma\left(m-k+\frac12\right)\Gamma\left(m+k-n+\frac12\right)} x^{-n}
\nn\\
&\sim& 
e^{-x/2} x^{k} \Big[1+\mc{O}\big(1/x\big)\Big].
\eqa
Implementing this in (\ref{GeneralNonabelianTerm}) and extracting the leading $r=0$ term, we deduce that the leading order contribution to $e^{\Phi_{(\text{NA})}}$ is given by
\beq
e^{\Phi_{(\text{NA})}}\sim e^{\phi} \sum^{\qquad \prime}_{k\in\mbb{Z}} \sum_{\ell=0}^{4|k|-1} \sum_{n\in\mbb{Z}+\f{\ell}{4|k|}} C_{r,k, \ell} |k|^{-s} e^{-2\pi S_{k, q}} \Big[1+\mc{O}\big(e^{2\phi}\big)\Big],
\label{NS5D2contribution}
\eeq
where we have defined
\beq S_{k, q}=|k| e^{-2\phi}+2 |k|\big(\tilde\chi-n\big)^2-iq \chi+2ik(\psi+\chi\tilde\chi).
\label{D2NS5action}
\eeq
This reproduces the Euclidean action of $k$ NS5-branes bound to 
$q:= 4nk$ D2-branes. Note that even in the absence of D2-brane instanton contributions, $q=0$, the real part of the action receives a contribution from the background Ramond-Ramond flux $\tilde\chi$,
as found  previously in \cite{VandorenAlexandrov}. For vanishing $\tilde\chi$, this reduces to the  pure NS5-brane instanton action of \cite{BeckerBeckerStrominger}:
\beq
S_{k}= |k| e^{-2\phi}+2ik\psi.
\eeq
It should be emphasized that the result (\ref{NS5D2contribution}) displays the contribution from $A$-type D2-brane instantons only. The $B$-type D2-branes could be exposed by choosing the alternative polarization displayed in (\ref{NewNonabelianTermSecondPolarization}), but then the $A$-type D2-brane effects are not visible. This is in contrast to the situation in \cite{AutomorphicNS5}, where the 
appearance of an extra summation in the non-abelian term made it possible to expose the 
D$(-1)$, D5 and NS5-brane effects simultaneously.\footnote{We note that the presence of an extra ``theta-angle'' in the NS5-brane instanton action of \cite{AutomorphicNS5}, compared to our result (\ref{D2NS5action}), is related to the fact that the spherical vector $f_K$ in the principal series of $SL(3,\mbb{R})$ displays a cubic phase factor \cite{AutomorphicMembrane} which is absent in the corresponding spherical vector for $SU(2,1)$ \cite{PiolineGunaydin}.} 

\vspace{.3cm}

\noindent {\bf Comments:}

\begin{itemize}

\item 
It is interesting to note that  the asymptotic expansion of the Whittaker function predicts an 
infinite series of perturbative corrections around the NS5-brane instanton background. This is in marked contrast to the case of type IIA Euclidean NS5-branes wrapping $K3\times T^2$, where
the perturbative corrections around the instanton background truncate at one loop \cite{Obers:2001sw}.

\item The instanton measure (\ref{D2AMeasure}) is similar to the dilogarithm sum
found in \cite{Alexandrov:2008gh}, with additional refinements
when the charges include non-inert prime factors. It would be interesting 
to compare the instanton summation measure with the generalized Donaldson-Thomas 
invariants of rigid Calabi-Yau manifolds \cite{KS1,KS2}.

\item While the sign of the one-loop term invalidates our proposal that the Eisenstein series \eqref{EisensteinPicard2} governs the exact metric on the hypermultiplet moduli space, and so 
forbids us to expect a detailed agreement between our summation measure and 
the  generalized Donaldson-Thomas invariants,  we do not think that
it ruins the basic postulate that the Picard modular group $SU(2,1;\mbb{Z}[i])$ should
act isometrically on the exact universal hypermultiplet moduli space. Rather, we take it
as an incentive to construct a more sophisticated automorphic form which would produce
the correct one-loop term, as well as produce a non-trivial dependence on the coordinate $z$ on the twistor fiber $\mbb{C}P^1$, which is generally expected when all isometries are broken. In fact, since the twistor space
is known to be described by holomorphic contact transformations, it is natural to expect
that automorphic forms attached to the quaternionic discrete series of $SU(2,1)$ should be
relevant \cite{GNPW}. Indeed, these forms can be lifted to sections of a certain complex line bundle 
on the twistor space $\mc{Z}_{\mc{M}_{\text{UH}}}=SU(2,1)/(U(1)\times U(1))$ \cite{PiolineGunaydin,GrossWallach}. 
It is challenging to construct such automorphic forms explicitly, and adapt the analysis in \cite{Alexandrov:2009qq} to produce a manifestly
$SU(2,1,\mbb{Z}[i])$-invariant description of the twistor space.\footnote{Some preliminary steps in this direction were taken recently in \cite{Alexandrov:2009vj}.} We anticipate that the resulting
instanton corrections will be qualitatively similar to the ones considered here,
although the summation measure will certainly be quite different.

\item In this work we have concentrated exclusively on rigid Calabi-Yau threefolds whose
intermediate Jacobian $J(\mc{X})$ admits complex multiplication by $\mbb{Z}[i]$. 
It is interesting to ask how our construction may generalize to other values 
of the period matrix $\tau$. When $J(\mc{X})$ admits complex multiplication by the ring of integers $\mc{O}_d$ in the imaginary quadratic number field $\mbb{Q}(i\sqrt{d}), \, d> 0$, it is natural to conjecture
that the relevant arithmetic subgroup of $SU(2,1)$ would be $SU(2,1;\mc{O}_d)$. For example, choosing $\tau=(1+i\sqrt{3})/2:= \omega$  should correspond to the ``Picard-Eisenstein" modular group $SU(2,1;\mbb{Z}[\omega])$, where $\mbb{Z}[\omega]$ are the Eisenstein integers, corresponding to the ring of integers $\mc{O}_3=\mbb{Z}[\om]$ in $\mbb{Q}(i\sqrt{3})$~\cite{Holzapfel}. In contrast to the $\tau=i$ case, it is interesting to note that $SU(2,1;\mbb{Z}[\omega])$ does not contain the ``rotation'' generator $R$ in (\ref{emduality}) \cite{FalbelParker}. Indeed, one does not generally
expect the full electric-magnetic duality group to be a quantum symmetry, but rather its
subgroup generated by monodromies in the moduli space of complex structures (which
is non-existent in the case of rigid CY threefolds).

\end{itemize}

\chapter{Compactification and the Automorphic Lift}
The purpose of this final chapter is to analyze the fate the continuous ``U-duality'' groups $G_3(\mbb{R})$, appearing in toroidal compactifications of supergravity theories when higher derivative interactions are turned on. Recently, several authors
\cite{LambertWest1,LambertWest2,Aspects,ClaudiaAxel,PiolineYann}
have considered this problem from various perspectives, with the consensus that
toroidal compactifications of quadratic and higher order corrections
give rise to terms in $D=3$ which are not $G_3$-invariant.\footnote{One
exception being ref. \cite{ClaudiaAxel} in which the authors
considered quadratic curvature corrections to pure gravity in four
dimensions. In that special case, the most general correction
can be related, through suitable field redefinitions, to the Gauss-Bonnet term which is topological in four dimensions and does not contribute to the dynamics. Hence, the $SL(2,
\mbb{R})$-symmetry of the compactified Lagrangian is trivially preserved.} We propose that the result of the compactification should then be viewed as the leading term in a large-volume expansion of some $G_3(\mbb{Z})$-invariant expression, where $G_3(\mbb{Z})\subset G_3(\mbb{R})$ is the underlying discrete duality group which is expected to be preserved in the full quantum theory \cite{HullTownsend}. This completion under $G_3(\mbb{Z})$ is performed through the introduction of $G_3(\mbb{Z})$-invariant automorphic forms, and is generally referred to as the \emph{automorphic lift}. The analysis suggests that generalized versions of Eisenstein series which transform in some representation of the maximal compact subgroup $K(G_3)$ are required. 

This chapter is based on {\bf Paper IV}, written in collaboration with Ling Bao, Johan Bielecki, Martin Cederwall and Bengt E. W. Nilsson.

\section{Automorphic Lift - General Philosophy}

Keeping in mind the discussion from Chapter  \ref{Chapter:Aspects} regarding the $\mc{R}^4$-correction in type IIB supergravity, a natural interpretation of the compactified higher-derivative Lagrangian presents itself: the result of the compactification should correspond to the leading contribution in a large volume expansion of some manifestly $G_3(\mbb{Z})$-invariant higher-derivative effective action. Inspired by the philosophy of Chapter \ref{Chapter:Aspects}, we would therefore like to ``complete'' the Lagrangian in $D=3$ using a certain $G_3(\mbb{Z})$-invariant automorphic form, whose Fourier expansion reproduces the result of the compactification to leading order. Let us now consider this appealing idea in some more detail. 

In the in the analysis of Chapter \ref{Chapter:Aspects} the completion to an S-duality
invariant $\mc{R}^4$-term was achieved through the use of an Eisenstein series which was completely
$SL(2, \mbb{Z})$-invariant. More generally, one might find terms in
the effective action whose non-perturbative completion requires
automorphic forms transforming under the maximal compact subgroup
$K(G_3)$. For example, this was found to be the case in
\cite{GreenGutperleKwon}, where interaction terms of sixteen
fermions were analyzed. These terms transform under the maximal
compact subgroup $U(1)\subset SL(2, \mbb{R})$ and so the U-duality
invariant completion requires in this case an automorphic form which
transforms with a $U(1)$ weight that compensates for the
transformation of the fermionic term, and thus renders the effective
action invariant.

The need for automorphic forms which transform under the maximal
compact subgroup $K(G_3)$ was also emphasized in
\cite{LambertWest2}, based on the observation that the dilaton
exponents in compactified higher curvature corrections correspond to
weights of the global symmetry group $G_3$, implying that these
terms transform non-trivially in some representation of
$K(G_3)$. An explicit realisation of these arguments was
found in \cite{PiolineYann} for the case of compactification on
$S^1$ of the four-dimensional coupled Einstein-Liouville system,
supplemented by a four-derivative curvature correction. The
resulting effective action was shown to explicitly break the Ehlers
$SL(2, \mbb{R})$-symmetry; however, an $SL(2,
\mbb{Z})_{\text{global}}\times U(1)_{\text{local}}$-invariant
effective action was obtained by ``lifting'' the scalar coefficients
to automorphic forms transforming with compensating $U(1)$ weights.
The non-perturbative completion implied by this lifting is in this
case attributed to gravitational Taub-NUT instantons
\cite{PiolineYann}.

Similar conclusions were drawn in \cite{Aspects}, in which
compactifications of derivative corrections of second, third and
fourth powers of the Riemann tensor were analyzed. Again, it was
concluded that the $G_3$-symmetry is explicitly broken by the
correction terms. It was argued, in accordance with the type IIB
analysis discussed above, that the result  of the compactification
-- being inherently perturbative in nature -- should be considered
as the large volume expansion of a $G_3(\mbb{Z})$-invariant
effective action. It was shown on general grounds that any term
resulting from such a compactification can always be lifted to a
U-duality invariant expression through the use of automorphic forms
transforming in some representation of $K(G_3)$.

In chapter we extend some aspects of the analysis of
\cite{Aspects}. In \cite{Aspects} only parts of the compactification
of the Riemann tensor squared, $\hat{R}_{ABCD}\hat{R}^{ABCD}$, were
presented. The terms which were analyzed were sufficient to show
that the continuous symmetry was broken, and to argue for the
necessity of introducing transforming automorphic forms to restore
the U-duality symmetry $G_3(\mbb{Z})$. Moreover, the overall
volume factor of the internal torus was  neglected in the analysis.

Here we shall restrict our attention to corrections quadratic in the Riemann tensor
in order for a complete compactification to be a feasible task. More
precisely, we shall focus on a four-derivative correction to the
Einstein-Hilbert action in the form of the Gauss-Bonnet term
$\hat{R}_{ABCD}\hat{R}^{ABCD}-4\hat{R}_{AB}\hat{R}^{AB}+\hat{R}^2$.
Modulo field equations, this is the only independent invariant
quadratic in the Riemann tensor. We extend the investigations of
\cite{Aspects} by giving the complete compactification on $T^n$ of
the Gauss-Bonnet term from $D$ dimensions to $D-n$ dimensions. In
the special case of compactifications to $D-n=3$ dimensions the
resulting expression simplifies, making it amenable for a more
careful analysis. In particular, one of the main points of this
chapter is to study the full structure of the dilaton exponents, with
the purpose of determining the $\mf{sl}(n+1,
\mbb{R})$-representation structure associated with quadratic
curvature corrections.%
\section{Compactification of the Gauss-Bonnet Term}

\label{section:Compactification}

In this section we outline the derivation of the toroidal
compactification of the Gauss-Bonnet term from $D$ dimensions to
$D-n$ dimensions. The full result for the compactification to arbitrary
dimensions is given in {\bf Paper IV}. Here we focus on the special case of $D-n=3$, which is
the most relevant case for the questions we pursue.

\subsection{The General Procedure}

The Gauss-Bonnet Lagrangian density is quadratic in the Riemann
tensor and takes the explicit form
\beq
\mc{L}_{GB}=\hat{e}\big[\hat{R}_{ABCD}\hat{R}^{ABCD} -
4\hat{R}_{AB}\hat{R}^{AB}+\hat{R}^2\big].
\eqnlab{GaussBonnet}
\eeq
The compactification of the $D$-dimensional Riemann tensor
${\hat{R}^{A}}_{BCD}$ on an $n$-torus, $T^n$, is done in three steps:
first we perform a Weyl-rescaling of the internal vielbein, followed
by a splitting of the external and internal indices, and finally we
perform yet another rescaling of the vielbein. In the following we
shall always assume that the torsion vanishes.

%\subsubsection*{Conventions and Reduction Ansatz}

Our index conventions are as follows. $M,N,\dots $ denote $D$
dimensional curved indices, and $A, B, \dots$ denote $D$ dimensional
flat indices. Upon compactification we split the indices according
to $M=(\mu,m)$, where $\mu, \nu, \dots$ and $m, n, \dots$ are curved
external and internal indices, respectively. Similarly, the flat
indices split into external and internal parts according to
$A=(\alpha,a)$.

Our reduction Ansatz for the vielbein is
\begin{equation}
\hvb{M}{A} = e^{\varphi} \tvb{M}{A} = e^{\varphi} \left(
\begin{array}{cc} \vb{\mu}{\alpha} & \mathcal{A}_{\mu}^m
\tvb{m}{a} \\
0 & \tvb{m}{a}\\
\end{array} \right),
\end{equation}
where the internal vielbein $\tvb{m}{a}$ is an element of the
isometry group $GL(n, \mbb{R})$ of the $n$-torus. Later on we shall
parametrise $\tvb{m}{a}$ in various ways. With this Ansatz,
the line element becomes
\beq
ds^2_D = e^{2\varphi} \Big\{ ds^2_{D-n} +
\big[(dx^m+\mathcal{A}_{(1)}^m)\tvb{m}{a}\big]^2 \Big\}.
\eeq

%\subsubsection*{Weyl-Rescaling}

In order to obtain a Lagrangian in Einstein frame after dimensional
reduction, we perform a Weyl-rescaling of the $D$-dimensional
vielbein,
\beq
{\hat{e}_M}^{\ph{M}A}\hs \longrightarrow \hs
{\tilde{e}_M}^{\ph{M}A}=e^{-\varphi}{\hat{e}_M}^{\ph{M}A}.
\eeq

Note that all $D$-dimensional objects before rescaling are denoted
$\hat{X}$, the Weyl-rescaled objects are denoted $\tilde{X}$, while
the $d=(D-n)$-dimensional objects are written without any
diacritics. After the Weyl-rescaling the Gauss-Bonnet Lagrangian,
including the volume measure $\hat{e} = e^{D\varphi}\tilde{e}$, can be conveniently organized in terms of equations of motion and
total derivatives. This is achieved using integration by parts,
where $\tilde{\nabla}_{(A}\tilde{\partial}_{B)}\varphi$ does not
appear explicitly. The resulting Lagrangian is (see the appendix of {\bf Paper IV} for the derivation):
\beqa
{}Ê\mathcal{L}_{\mathrm{GB}} &=& \tilde{e} e^{(D-4)\varphi} \bigg\{
\tilde{R}_{\mathrm{GB}}^2 - (D-3)(D-4)\big[
2(D-2)(\tilde{\partial}\varphi)^2\tilde{\square}\varphi +
(D-2)(D-3)(\tilde{\partial}\varphi)^4 \nn \\
{}Ê& & + 4 \big(\tilde{R}_{AB} - \frac{1}{2}\eta_{AB}\tilde{R}\big)
(\tilde{\partial}^A\varphi)(\tilde{\partial}^B\varphi)\big]\big\}
\nn \\
{}Ê& & + 2(D-3)\tilde{e}\tilde{\nabla}_A\big\{
e^{(D-4)\varphi}\big[(D-2)^2(\tilde{\partial}
\varphi)^2\tilde{\partial}^A\varphi + 2(D-2)(\tilde{\square}
\varphi)\tilde{\partial}^A\varphi
\nn \\
{}Ê& & - (D-2)\tilde{\partial}^A(\tilde{\partial}
\varphi)^2 + 4\big(\tilde{R}^{AB} -
\frac{1}{2}\eta^{AB}\tilde{R}\big)
\tilde{\partial}_B\varphi]\big]\bigg\},
\eqnlab{eqn:GBweyl1}
\eqa
where $\tilde{R}_{GB}^{2}$ represents the rescaled Gauss-Bonnet combination. In $D=4$ the Lagrangian is only altered by a total derivative, while
in $D=3$ the Lagrangian it is merely rescaled by a factor of
$e^{-\varphi}$. The total derivative terms here will remain total
derivatives even after the compactification. Along with the volume
factor the Weyl-rescaling will determine the overall exponential
dilaton factor, which shall play an important role in the analysis
that follows.

%\subsubsection*{Compactification}

\subsection{Tree-Level Scalar Coset Symmetries}

The internal vielbein $\hat{e}_m^{\phantom{m}a}$ can be used to
construct the internal metric $\hat{g}_{mn} =
\hat{e}_m^{\phantom{m}a}\hat{e}_n^{\phantom{m}b}\delta_{ab}$, which
is manifestly invariant under local $SO(n)$ rotations in the reduced
directions. Thus we are free to fix a gauge for the internal
vielbein using the $SO(n)$-invariance. After compactification the
volume measure becomes $\tilde{e}=e\tvbint$, where $e$ is the
determinant of the spacetime vielbein and $\tvbint$ is the
determinant of the internal vielbein. Defining the Weyl-rescaling
coefficient as $e^{-(D-2)\varphi} \equiv \tvbint$ ensures that the
reduced Lagrangian is in Einstein frame.

The $GL(n, \mbb{R})$ group element $\tvb{m}{a}$ can now be
parameterized in several ways, and we will discuss the two most
natural choices here. The first choice is included for completeness, while it is the second choice which we shall subsequently employ in the compactification of the Gauss-Bonnet term.

\subsubsection*{First Parametrisation - Making the Symmetry Manifest}

First, there is the possibility of separating out the determinant of
the internal vielbein according to $\tvb{m}{a} = (\tvbint)^{1/n}
\varepsilon_m^{\phantom{m}a} =
e^{-\frac{(D-2)}{n}\varphi}\varepsilon_m^{\phantom{m}a}$, where
$\varepsilon_m^{\phantom{m}a}$ is an element of $SL(n, \mbb{R})$ in
any preferred gauge. The line element takes the form
\beq
ds^2_D = e^{2\varphi} \left\{ ds^2_{D-n} +
e^{-2\frac{(D-2)}{n}\varphi} \left[(dx^m +
\mathcal{A}_{(1)}^m)\varepsilon_m^{\phantom{m}a}\right]^2 \right\}.
\eeq
This Ansatz is nice for investigating the symmetry properties of the
reduced Lagrangian because the $GL(n, \mbb{R})$-symmetry of the
internal torus is manifestly built into the formalism. More
precisely, the reduction of the tree-level Einstein-Hilbert
Lagrangian, $\hat{e}\hat{R}$, to $d=D-n$ dimensions becomes,
\beq
\mathcal{L}^{[d]}_{\mathrm{EH}} = e\left[ R -
\frac{1}{4}e^{-2\frac{(D-2)}{n}\xi\rho}
F_{c\alpha\beta}F^{c\alpha\beta} - \frac{1}{2}(\partial\rho)^2 -
\mathrm{tr}(P_\alpha P^\alpha) - 2\xi\square\rho \right],
\eqnlab{ReducedEH1}
\eeq
where $F^c_{\phantom{c}\alpha\beta} \equiv
\varepsilon_m^{\phantom{m}a} F^m_{\phantom{m}\alpha\beta}$ and
\beq
P_\alpha^{\phantom{\alpha}bc} \equiv \varepsilon^{m(b}
\partial_\alpha \varepsilon_m^{\phantom{m}c)} =
\tP_\alpha^{\phantom{\alpha} bc} +
\frac{(D-2)}{n}\xi\partial_\alpha\rho\delta^{bc}.
\eeq
Notice that $P_\alpha^{\phantom{\alpha} bc}$ is $\mf{sl}(n,
\mbb{R})$ valued and hence fulfills $\tr(P_\alpha) = 0$. To obtain
\Eqnref{ReducedEH1} we also performed a scaling $\varphi = \xi\rho$
with $\xi = \sqrt{\frac{n}{2(D-2)(D-n-2)}}$, so as to ensure that
the scalar field $\rho$ appears canonically normalized in the
Lagrangian.

The $SL(n, \mbb{R})$-symmetry is manifest in this Lagrangian because
the term $\mathrm{tr}(P_\alpha P^\alpha)$ is constructed using the
invariant Killing form on $\mf{sl}(n, \mbb{R})$. By dualising the
two-form field strength $F_{(2)}$, the symmetry is enhanced to
$SL(n+1, \mbb{R})$. With a slight abuse of terminology we call this
the (classical) ``U-duality'' group. Since we are only investigating
the pure gravity sector, this is of course only a subgroup of the
full continuous U-duality group.

It was already shown in \cite{Aspects}, that the tree-level symmetry
$SL(n+1, \mbb{R})$ is not realized in the compactified Gauss-Bonnet
Lagrangian. It was argued, however, that the quantum symmetry
$SL(n+1, \mbb{Z})$ could be reinstated by ``lifting'' the result of
the compactification through the use of automorphic forms. In the present analysis we take the same point of view, but since we now have access
to the complete expression of the compactified Gauss-Bonnet
Lagrangian we can here extend the analysis of \cite{Aspects} in some
aspects. In order to do this we shall make use of a different
parametrisation than the one displayed above, which illuminates the
structure of the dilaton exponents in the Lagrangian. The dilaton
exponents reveals the weight structure of the global symmetry group
and so can give information regarding which representation of the
U-duality group we are dealing with. Because we here have access to
a complete expression this analysis is more exhaustive than the one
presented in \cite{LambertWest1,LambertWest2}.

\subsubsection*{Second Parametrisation - Revealing the Root Structure}

The second natural choice of the internal vielbein is to
parameterize it in triangular form by using dimension by dimension
compactification \cite{Dualisation1,Dualisation2}. Instead
of extracting only the determinant of the vielbein, one dilaton
scalar is pulled out for each compactified dimension according to
$\tvb{m}{a} = e^{-\frac{1}{2}\fphi{a}}\ivb{m}{a}$, where $\vec{\phi}
= (\phi_1,\dots,\phi_n)$ and
\beq
\vec{f}_a = 2(\alpha_1,\dots,\alpha_{a-1},(D-n-2+a)\alpha_a,
\underbrace{0,\dots,0}_{n - a}),
\eqnlab{vectorf}
\eeq
with
\beq
\alpha_{a} = \frac{1}{\sqrt{2(D-n-2+a)(D-n-3+a)}}.
\eeq
The internal vielbein is now the Borel representative of the coset
$GL(n, \mbb{R})/SO(n)$, with the diagonal degrees of freedom
$e^{-\frac{1}{2}\fphi{a}}$ corresponding to the Cartan generators
and the upper triangular degrees of freedom
\beq
\ivb{m}{a} = [(1 - \mathcal{A}_{(0)})^{-1}]_m^{\phantom{m}a} = [1 +
\mathcal{A}_{(0)} + (\mathcal{A}_{(0)})^2 +
\dots]_m^{\phantom{m}a}
\eqnlab{uma}
\eeq
corresponding to the positive root generators. The form of
\Eqnref{uma} follows naturally from a step by step compactification,
where the scalar potentials $(\mathcal{A}_{(0)})^i_j$, arising from
the compactification of the graviphotons, are nonzero only when
$i>j$. The sum of the vectors $\vec{f}_a$ can be shown to be
\begin{equation}
\sum_{a=1}^n \vec{f}_a = \frac{D-2}{3}\vec{g},
\eqnlab{sumoff}
\end{equation}
$\vec{g} \equiv 6(\alpha_1,\alpha_2\dots,\alpha_n)$. In addition,
$\vec{g}$ and $\vec{f}_a$ obey
\beqa
{}Ê\vec{g}\cdot\vec{g} &=& \frac{18n}{(D-2)(D-n-2)},
\nn \\
{}Ê\vec{g}\cdot\vec{f}_a &= &\frac{6}{D-n-2},
\nn \\
{}Ê\vec{f}_a \cdot\vec{f}_b &= &2\delta_{ab} + \frac{2}{D-n-2},
\eqnlab{exponentrelations}
\eqa
and
\beq
\sum_{a=1}^n (\vec{f}_a\cdot \vec{x})(\vec{f}_a\cdot \vec{y}) = 2
(\vec{x}\cdot\vec{y}) + \frac{D-2}{9}(\vec{g}\cdot
\vec{x})(\vec{g}\cdot\vec{y}).
\eeq
These scalar products can naturally be used to define the Cartan
matrix, once a set of simple root vectors are found. The line
element becomes
\beq
ds^2_D = e^{\frac{1}{3}\gphi} \Big\{ ds^2_{D-n} + \sum_{a=1}^n
e^{-\fphi{a}} \left[(dx^m+\mathcal{A}_{(1)}^m)\ivb{m}{a}\right]^2
\Big\},
\eqnlab{SecondAnsatz}
\eeq
yielding the corresponding Einstein-Hilbert Lagrangian in $d$
dimensions
\beq
\mathcal{L}_{\mathrm{EH}}^{[d]} = e\bigg[ R -
\frac{1}{2}(\partial\vec{\phi})^2 - \frac{1}{4}\sum_{a=1}^n
e^{-\fphi{a}}F_{a\beta\gamma}F^{a\beta\gamma} -
\frac{1}{2}\sum_{\substack{b,c=1 \\ b<c}}^n e^{(\vec{f}_b -
\vec{f}_c)\cdot\vec{\phi}}G_{\alpha bc} G^{\alpha bc} -
\frac{1}{3}\vec{g}\cdot\square\vec{\phi} \bigg],
\eeq
with $F^c_{\phantom{c}\alpha\beta} \equiv \ivb{m}{a}
F^m_{\phantom{m}\alpha\beta}$ and
\beq
G_\alpha^{\phantom{\alpha}bc} = u^{mb}\partial_\alpha\ivb{m}{c} =
e^{-\frac{1}{2}(\vec{f}_b -
\vec{f}_c)\cdot\vec{\phi}}\left[(\tP_\alpha^{\phantom{\alpha} bc} +
\frac{1}{2}\vec{f}_b\cdot\partial_\alpha \vec{\phi}\delta^{bc}) +
Q_\alpha^{\phantom{\alpha}bc}\right].
\eeq
Here, no Einstein's summation rule is assumed for the flat internal
indices.

We shall refer to the various exponents of the form $e^{\vec{x}\cdot
\vp}$ ($\vec{x}$ being some vector in $\mbb{R}^{n}$) collectively as
``dilaton exponents''. If relevant, this also includes the
contribution from the overall volume factor.

All the results
obtained in this parametrisation can be converted to the first
parametrisation simply by using the following identifications
\beqa
{}Ê\frac{1}{3}(\gphi) &=& 2\xi\rho, \nn \\
{}Ê\fphi{a} &=& 2\frac{(D-2)}{n}\xi\rho, \hspace{0.5 cm} \forall a,
\nn \\
{}Ê\vscalar{\phi}{\phi} &=& \rho^2.
\eqa
Notice also that our compactification procedure breaks down at $D-n
= 2$, in which case the scalar products in
\Eqnref{exponentrelations} become ill-defined.

Even though proving the symmetry contained in the Lagrangian is
somewhat more cumbersome compared to the first choice of
parametrisation, since all the group actions have to be carried out
adjointly in a formal manner, the second choice comes to its power
when dealing with the exceptional symmetry groups of the
supergravities for which no matrix representations exist. This
parametrisation is particularly suitable for reading off the root
vectors of the underlying symmetry algebra; they appear as
exponential factors in front of each term in the Lagrangian.
Identifying a complete set of root vectors in this way gives a
necessary but not sufficient constraint on the underlying symmetry.

\subsection{The Gauss-Bonnet Lagrangian Reduced to Three Dimensions}

When reducing to $D-n=3$ dimensions, we can dualise the two-form
field strength $\tF_{\phantom{a}\alpha\beta}^{a}\equiv
\tilde{e}_m^{\phantom{m}a}F^m_{\phantom{m}\alpha\beta}$ of the
graviphoton $\mc{A}_{(1)}$ into the one-form $\tH_{a \alpha}$. More
explicitly, we employ the standard dualisation so that
\beq
\delta_{ab}\tF^b_{\phantom{b}\alpha\beta} =
\epsilon_{\alpha\beta\gamma}\tilde{e}^m_{\phantom{m}a}
\partial^\gamma \chi_m \equiv \epsilon_{\alpha\beta\gamma}
\tH_a^{\phantom{a}\gamma}.
\eeq
When we go to Einstein frame, the appearance of the inverse vielbein
$\tilde{e}^m_{\phantom{m}a}$ in the definition of the one-form
$\tH_{a\alpha}$ implies there is a sign flip in the Lagrangian after
dualisation. The dualisation presented here follows from the
tree-level Lagrangian, but in general receives higher order
$\al^{\prime}$-corrections. However, these lead to terms of higher
derivative order than quartic  and so can be neglected in the
present analysis \cite{LambertWest1,Aspects}.

The final result for the compactification is written
in such way that the only explicit derivative terms appearing are
divergences, total derivatives and first derivatives on the dilatons
$\varphi$. The
end result reads explicitly
\beqa
{}Ê\mathcal{L}_{\mathrm{GB}}^{[3]}&=& \sqrt{|g|} e^{-2\varphi}
\Big\{ -\frac{1}{4}\tHHHH + \frac{1}{4}\tH^2 \tH^2 - 4\tH^2
(\partial\varphi)^2 + 2\tHPPHa  \nn \\
{}Ê& & - 2\tHPPHb + 4\tHPHa \partial_\beta \varphi - 6\tHPHb
\partial_\beta \varphi \nn \\
{}Ê& & + 2\tr(\tP_\alpha\tP_\beta \tP^\alpha\tP^\beta) +
2\tr(\tP_\alpha\tP_\beta) \tr(\tP^\alpha\tP^\beta) - (\tP^2)^2 +
8\tr(\tP_\alpha\tP_\beta \tP^\beta) \partial^\alpha\varphi \nn \\
{}Ê& &  - 4(D-2)\tr(\tP_\alpha
\tP_\beta)\partial^\alpha\varphi
\partial^\beta\varphi + 2(D+2)\tP^2 (\partial\varphi)^2 +
(D-2)(D-4)(\partial\varphi)^2(\partial\varphi)^2 \Big\},
\nn \\
\eqnlab{GB3}
\eqa
where $\tH^2 \equiv \tH_{a\beta}\tH^{a\beta}$ and $\tP^2 \equiv
\tP_{\alpha bc}\tP^{\alpha bc}$. Note that contributions from the
boundary terms and terms proportional to the equations of motion
have been ignored. The one-form $\tilde{P}_{\al}$ is the
Maurer-Cartan form associated with the internal vielbein
$\tilde{e}_{m}^{\ph{a}a}$, and so takes values in the Lie algebra
$\mf{gl}(n, \mbb{R})=\mf{sl}(n, \mbb{R})\oplus \mbb{R}$. Here, the
abelian summand $\mbb{R}$ corresponds to the ``trace-part'' of
$\tilde{P}_{\al}$. Explicitly, we have
$\text{tr}(\tilde{P}_{\al})=-(D-2)\pa_{\al}\varphi$. We shall
discuss various properties of $\tilde{P}_{\al}$ in more detail
below.

Finally, we note that the three-dimensional Gauss-Bonnet term is
absent from the reduced Lagrangian because it vanishes identically
in three dimensions:
\beq
R_{\al\be\ga\de}R^{\al\be\ga\de} - 4R_{\al\be}R^{\al\be}+R^2=0,
\qquad (\al, \be, \ga, \de = 1, 2, 3).
\eeq
The remainder of this chapter is devoted to a detailed analysis of the
symmetry properties of \Eqnref{GB3}.

\section{Algebraic Structure of the Compactified Gauss-Bonnet Term}

\label{section:AlgebraicStructure}

We have seen that the Ansatz presented in \Eqnref{SecondAnsatz} is
particularly suitable for identifying the roots of the relevant
symmetry algebra from the dilaton exponents associated with the
diagonal components of the internal vielbein. Through this analysis
one may deduce that for the lowest order effective action, the terms
in the action are organized according to the adjoint representation
of $\mf{sl}(n+1, \mbb{R})$, for which the weights are the roots. The
aim of this section is to extend the analysis to the Gauss-Bonnet
Lagrangian. By general arguments \cite{LambertWest1,LambertWest2},
it has been shown that the exponents no longer correspond to roots
of the symmetry algebra but rather they now lie on the weight
lattice. Here, however, we have access to the complete compactified
Lagrangian and we may therefore present an explicit counting of the
weights in the dilaton exponents and identify the relevant
$\mf{sl}(n+1, \mbb{R})$-representation.

An exhaustive analysis of the $\mf{sl}(4, \mbb{R})$-representation
structure of the Gauss-Bonnet term compactified from $6$ to $3$
dimensions on $T^3$ is performed. We do this in two alternative
ways.

First, we neglect the contribution from the overall dilaton factor
$e^{-2\varphi}$ in the representation structure. This is consistent
before dualisation because this factor is $SL(3,
\mbb{R})$-invariant. However, after dualisation this is no longer
true and one must understand what role this factor plays in the
algebraic structure. If one continues to neglect this factor then
all the weights fit into the ${\bf 84}$-representation of
$\mf{sl}(4, \mbb{R})$ with Dynkin labels $[2, 0, 2]$.

On the other hand, including the overall exponential dilaton factor
in the weight structure induces a shift on the weights which
``destroys'' the ${\bf 84}$ of $\mf{sl}(4, \mbb{R})$. Instead one
finds that the highest weight is now associated with the ${\bf
36}$-representation of $\mf{sl}(4, \mbb{R})$, with Dynkin labels
$[2, 0, 1]$. However, this representation is not ``big enough'' to
incorporate all the weights in the Lagrangian. It turns out that
there are additional weights outside of the ${\bf 36}$ that fit into
a ${\bf 27}$ of $\mf{sl}(3, \mbb{R})$. Unfortunately, we are unable
to determine which $\mf{sl}(4,\mbb{R})$-representation this belongs
to, since the associated highest weight does not seem to be
represented in the Lagrangian.

These results indicate that the general analysis performed in
\cite{LambertWest1} is only partly correct, or, rather, that the
interpretation given there might be incorrect. The highest weight of
the representation $[2, 0, 1]$ does indeed appear in the reduced
Lagrangian but it does not incorporate the full representation
structure of the compactified quadratic curvature correction.

\subsection{Kaluza-Klein Reduction and $\mf{sl}(n,
\mbb{R})$-Representations}

We shall begin by rewriting the reduction Ansatz in a way which has
a more firm Lie algebraic interpretation. Recall from
\Eqnref{SecondAnsatz} that the standard Kaluza-Klein Ansatz for the
metric is
\beq ds_D^{2}=e^{\f{1}{3}\vec{g}\cdot
\vec{\phi}}ds_{d}^{2}+e^{\f{1}{3}\gp}\sum_{i=1}^{n}e^{-\fip}
\left[\left(dx^{m} + \mc{A}^{m}_{(1)}\right){u_m}^{a}\right]^2.
\eqnlab{KaluzaKleinAnsatz}
\eeq
The exponents in this Ansatz are linear forms on the space of
dilatons. Let $\ve_i, \hs i=1, \dots, n$, constitute an
$n$-dimensional orthogonal basis of $\mbb{R}^{n}$, \beq
\ve_i\cdot\ve_j=\delta_{ij}. \eeq Since there is a non-degenerate
metric on the space of dilatons (the Cartan subalgebra
$\mf{h}\subset \mf{sl}(n+1, \mbb{R})$) we can use this to identify
this space with its dual space of linear forms. Thus, we may express
all exponents in the orthogonal basis $\ve_i$ and the vectors
$\vf_i, \vg$ and $\vp$ may then be written as

\beqa
{}Ê\vf_i& =&
\sqrt{2}\ve_i+\al\vg,
\nn \\
{}Ê\vg&=&\be \sum_{i=1}^{n}\ve_i,
\nn \\
{}Ê\vp& =& \phi_i\ve_i,
\eqnlab{dilatonvectors}
\eqa
where the constants $\al$ and $\be$ are defined as
\beqa
{}\al&=&\f{1}{3n}\bigg(D-2-\sqrt{(D-n-2)(D-2)}\bigg),
\nn \\
{}Ê  \be& =&\sqrt{\f{18}{(D-n-2)(D-2)}}.
\eqa
Note here that the constant $\al$ is not the same as the $\al_a$ of
\Eqnref{vectorf}.

The combinations
\beq
\vf_i-\vf_j = \sqrt{2}\ve_i-\sqrt{2}\ve_j
\eeq
span an $(n-1)$-dimensional lattice which can be identified with the
root lattice of $A_{n-1}=\mf{sl}(n, \mbb{R})$. For compactification
of the pure Einstein-Hilbert action to three dimensions, the dilaton
exponents precisely organize into the complete set of positive roots
of $\mf{sl}(n, \mbb{R})$, revealing that it is the adjoint
representation which is the relevant one for the U-duality
symmetries of the lowest order (two-derivative) action. After
dualisation of the Kaluza-Klein one forms $\mc{A}_{(1)}$ the
symmetry is lifted to the full adjoint representation of
$\mf{sl}(n+1, \mbb{R})$.

When we compactify higher derivative corrections to the
Einstein-Hilbert action it is natural to expect that other
representations of $\mf{sl}(n, \mbb{R})$ and $\mf{sl}(n+1, \mbb{R})$
become relevant. In order to pursue this question for the
Gauss-Bonnet Lagrangian, we shall need some features of the
representation theory of $A_n=\mf{sl}(n+1, \mbb{R})$.

\subsubsection*{Representation Theory of $A_n=\mf{sl}(n+1, \mbb{R})$}

For the infinite class of simple Lie algebras $A_n$, it is possible
to choose an embedding of the weight space $\mf{h}^{\star}$ in
$\mbb{R}^{n+1}$ such that $\mf{h}^{\star}$ is isomorphic to the
subspace of $\mbb{R}^{n+1}$ which is orthogonal to the vector
$\sum_{i=1}^{n+1}\ve_i$. We can use this fact
to construct an embedding of the $(n-1)$-dimensional weight space of
$A_{n-1}=\mf{sl}(n, \mbb{R})$ into the $n$-dimensional weight space of $A_n=\mf{sl}(n+1, \mbb{R})$, in terms of the $n$ basis vectors $\ve_i$ of $\mbb{R}^{n}$. 

To this end we define the new vectors
\beqa
{}Ê \vo_i &=&
\vf_i-(\al+\f{\sqrt{2}}{n\be})\vg
\nn \\
{}Ê& =& \sqrt{2}\ve_i-\f{\sqrt{2}}{n}\sum_{j=1}^{n}\ve_j,
\eqa
which have the property that
\beq
\vo_i\cdot \vg
=\sqrt{2}\be-\sqrt{2}\be=0.
\eqnlab{zeroscalarproduct}
\eeq
This implies that the vectors $\vo_i$ form a (non-orthogonal) basis
of the $(n-1)$-dimensional subspace $U\subset \mbb{R}^{n}$,
orthogonal to $\vg$. The space $U$ is then isomorphic to the weight
space $\mf{h}^{\star}$ of $A_{n-1}=\mf{sl}(n, \mbb{R})$. Since there
are $n$ vectors $\vo_i$, this basis is overcomplete. However, it is
easy to see that not all $\vo_i$ are independent, but are subject to
the relation
\beq
\sum_{i=1}^{n}\vo_i=0.
\eqnlab{tracelessness}
\eeq
A basis of simple roots of $\mf{h}^{\star}$ can now be written in
three alternative ways
\beq
\vec{\al}_i=\vf_i-\vf_{i+1}=\vo_i-\vo_{i+1}=\sqrt{2}(\ve_i-\ve_{i+1}),
\qquad (i=1, \dots, n-1).
\eeq

What is the algebraic interpretation of the vectors $\vo_i$? It
turns out that they may be identified with the weights of the
$n$-dimensional fundamental representation of $\mf{sl}(n, \mbb{R})$.
The condition $\sum_{i=1}^{n}\vo_i=0$ then reflects the fact that
the generators of the fundamental representation are traceless.

In addition, we can use the weights of the fundamental
representation to construct the fundamental weights
$\vec{\Lambda_i}$, defined by
\beq
\vec{\al}_i\cdot \vec{\Lambda}_j=2\delta_{ij}.
\eqnlab{fundamentalweightsdefinition}
\eeq
One finds
\beq
\vL_i=\sum_{j=1}^{i}\vo_j, \qquad (i=1, \dots, n-1),
\eqnlab{weightrelation}
\eeq
which can be seen to satisfy \Eqnref{fundamentalweightsdefinition}.

The relation, \Eqnref{weightrelation}, between the fundamental
weights $\vL_i$ and the weights of the fundamental representation
$\vo_i$ can be inverted to
\beq
\vo_i = \vL_i-\vL_{i-1}, \qquad (i=1, \dots, n-1).
\eeq
In addition, the $n$:th weight is
\beq \vo_n=-\vL_{n-1},
\eqnlab{nthWeight}
\eeq
corresponding to the lowest weight of the fundamental
representation.

We may now rewrite the Kaluza-Klein Ansatz in a way such that the
weights $\vo_i$ appear explicitly in the metric\footnote{A similar
construction was given in \cite{Arjan}.}
\beq
ds_D^{2} = e^{\f{1}{3}\vec{g}\cdot
\vec{\phi}}ds_{d}^{2}+e^{\ga\gp}\sum_{i=1}^{n}e^{-\vo_i\cdot
\vp}\Big[\big(dx^{m}+\mc{A}^{m}_{(1)}\big){u_m}^{a}\Big]^2,
\eqnlab{KaluzaKleinAnsatz2}
\eeq
with
\beq
\ga = \f{1}{3} - \al-\f{\sqrt{2}}{n\be}.
\eeq

\subsection{The Algebraic Structure of Gauss-Bonnet in Three
Dimensions}

We are interested in the dilaton exponents in the scalar part of the
three-dimensional Lagrangian. For the Einstein-Hilbert action we
know that these are of the forms
\beq
Ê\vf_a-\vf_b \quad (b>a), \qquad \text{and}\qquad \vf_a.
\eeq
The first set of exponents $\vf_a-\vf_b$ correspond to the positive
roots of $\mf{sl}(n, \mbb{R})$ and the second set $\vf_a$, which
contributes to the scalar sector after dualisation, extends the
algebraic structure to include all positive roots of $\mf{sl}(n+1,
\mbb{R})$. The highest weight $\vl^{\text{hw}}_{\text{ad}, n}$ of
the adjoint representation of $A_n=\mf{sl}(n+1, \mbb{R})$ can be
expressed in terms of the fundamental weights as
\beq
\vl^{\text{hw}}_{\text{ad}, n}=\vL_{1}+\vL_{n},
\eeq
corresponding to the Dynkin labels $$[1, 0, \dots, 0, 1].$$ We see
that before dualisation the highest weight of the adjoint
representation of $\mf{sl}(n, \mbb{R})$ arises in the dilaton
exponents in the form
$\vf_1-\vf_n=\vo_1-\vo_n=\vL_1+\vL_{n-1}=\vl^{\text{hw}}_{\text{ad},
n-1}$.

We proceed now to analyze the various dilaton exponents arising from
the Gauss-Bonnet term after compactification to three dimensions.
These can be extracted from each term in the Lagrangian \Eqnref{GB3}
by factoring out the diagonal components of the internal vielbein
according to $\tvb{m}{a} = e^{-\frac{1}{2}\fphi{a}}\ivb{m}{a}$. For
example, before dualisation we have the manifestly $SL(n,
\mbb{R})$-invariant term $\text{tr}(\tilde{P}_{\al}\tilde{P}_{\be}
\tilde{P}^{\al}\tilde{P}^{\be})$. Expanding this gives (among
others) the following types of terms
\beqa
{}Ê\text{tr}\big(\tilde{P}_{\al}\tilde{P}_{\be}
\tilde{P}^{\al}\tilde{P}^{\be}\big)& \sim &\ph{+}
\sum_{\tiny\begin{array}{l}c< a, b \\ d< a, b
\end{array}} e^{-2(\vf_c+\vf_d-\vf_a-\vf_b)\cdot \vp} G_{\al c
a}{G^{\al c}}_b{G_{\be d}}^{a}G^{\be d b} + \cdots
\nn \\
{}Ê& & + \sum_{\tiny\begin{array}{l}a>c>b\\
a>d>b \end{array}} e^{-2(\vf_b-\vf_a)\cdot \vp} G_{\al c
a}{{G^{\al}}_b}^{c}G^{\be d a}{{G_{\be}}^{b}}_d +\cdots.
\eqnlab{IndexExtraction}
\eqa
After dualisation, we need to take into account also terms
containing $\tilde{H}^{\al}_{\ph{a} a}$. We have then, for example,
the term
\beq
\tilde{H}^4 \hs \sim \hs \sum_{a, b}e^{-2(\vf_a+\vf_b)\cdot \vp} H^4.
\eeq
Many different terms in the Lagrangian might in this way give rise
to the same dilaton exponents. As can be seen from
\Eqnref{IndexExtraction}, the internal index contractions yield
constraints on the various exponents. We list below all the
``independent'' exponents, i.e., those which are the least
constrained. All other exponents follow as special cases of these.
Before dualisation we then find the following exponents:
\beqa
{}Ê\vf_a-\vf_b &\quad & (b>a),
\nn \\
{}Ê\vf_c+\vf_d-\vf_a-\vf_b &\quad & (c<a,\hs c<b, \hs d<a, \hs d<b),
\nn \\
{}Ê\vf_a+\vf_b-\vf_c-\vf_d & \quad & (b<c, \hs a<d),
\eqnlab{GBexponents1}
\eqa
and after dualisation we also get a contribution from
\beqa
{} \vf_a, & &
\nn \\
{}Ê\vf_a+\vf_b, &Ê&
\nn \\
{}Ê\vf_a+\vf_b-\vf_c, &\quad & (b<c).
\eqnlab{GBexponents2}
\eqa
Let us investigate the general weight structure of the dilaton
exponents before dualisation. The highest weight arises from the
terms of the form $\vf_c+\vf_d-\vf_a-\vf_b$ when $c=d=1$ and
$a=b=n$, i.e., for the dilaton vector $2\vf_1-2\vf_n$. This can be
written in terms of the fundamental weights as follows
\beq
2\vf_1-2\vf_n=2\vo_1-2\vo_n=2\vL_1+2\vL_{n-1},
\eeq
which is the highest weight of the $[2, 0, \dots, 0,
2]$-representation of $\mf{sl}(n, \mbb{R})$.

\subsection{Special Case: Compactification from $D=6$ on $T^{3}$}

\label{section:SpecialCaseNoVolumeFactor}

In order to determine if this is indeed the correct representation
for the Gauss-Bonnet term, we shall now restrict to the case of
$n=3$, i.e., compactification from $D=6$ on $T^{3}$. We do this so
that a complete counting of the weights in the Lagrangian is a
tractable task. Before dualisation we then expect to find the
representation ${\bf 27}$ of $\mf{sl}(3, \mbb{R})$, with Dynkin
labels $[2,2]$. We wish to investigate if, after dualisation, this
representation lifts to the representation ${\bf 84}$ of $\mf{sl}(4,
\mbb{R})$, with Dynkin labels $[2, 0, 2]$.

It is important to realize that of course the Lagrangian will not
display the complete set of weights in these representations, but
only the \emph{positive} weights, i.e., the ones that can be
obtained by summing positive roots only. Let us begin by analyzing
the weight structure before dualisation. From \Eqnref{GBexponents1}
we find the weights
\beqa
{}Ê& & \vf_1-\vf_2, \qquad \vf_2-\vf_3, \qquad \vf_1-\vf_3,
\nn \\
{}Ê& & 2(\vf_1-\vf_2), \qquad 2(\vf_2-\vf_3), \qquad 2(\vf_1-\vf_3),
\nn \\
{}Ê& & 2\vf_1-\vf_2-\vf_3, \qquad \vf_1+\vf_2-2\vf_3.
\eqa
The first three may be identified with the positive roots of
$\mf{sl}(3, \mbb{R})$, $\val_1=\vf_1-\vf_2,\hs \val_2=\vf_2-\vf_3$
and $\val_{\theta}=\vf_1-\vf_3$. The second line then corresponds to
$2\val_2,\hs 2\val_2$ and $2\val_{\theta}$. The remaining weights
are
\beqa
{}Ê\vf_1+\vf_2-2\vf_3&=&\val_1+2\val_2,
\nn \\
{}Ê 2\vf_1-\vf_2-\vf_3&=&2\val_1+\val_2.
\eqa
These weights are precisely the eight positive weights of the {\bf
27} representation of $\mf{sl}(3, \mbb{R})$.

We now wish to see if this representation lifts to the corresponding
representation of $\mf{sl}(4, \mbb{R})$, upon inclusion of the
weights in \Eqnref{GBexponents2}. As mentioned above, the
representation of $\mf{sl}(4, \mbb{R})$ with Dynkin labels $[2, 0,
2]$ is 84-dimensional. It is illuminating to first decompose this in
terms of representations of $\mf{sl}(3, \mbb{R})$,
\beq
{\bf 84}={\bf 27}\oplus {\bf 15}\oplus {\bf \bar{15}}\oplus {\bf
6}\oplus {\bf \bar{6}}\oplus {\bf 8}\oplus {\bf 3}\oplus {\bf
\bar{3}}\oplus {\bf 1},
\eqnlab{84inSL3}
\eeq
or, in terms of Dynkin labels,
\beq
[2, 0, 2]=[2,2]+[2,1]+[1,2]+[2,0]+[0,2]+[1,1]+[1,0]+[0,1]+[0,0].
\eeq
We may view this decomposition as a \emph{level decomposition} of
the representation ${\bf 84}$, with the level $\ell$ being
represented by the number of times the third simple root $\val_3$
appears in each representation. From this point of view, and as we
shall see in more detail shortly, the representations ${\bf 27},
{\bf 8}$ and ${\bf 1}$ reside at $\ell=0$, the representations ${\bf
15}$ and ${\bf 3}$ at $\ell=1$, and the representation ${\bf 6}$ at
$\ell=2$. The ``barred'' representations then reside at the
associated negative levels. We know that we can only expect to find
the strictly positive weights in these representations. Let us
therefore begin to count these.

\begin{figure}
\begin{center}
\includegraphics[width=150mm]{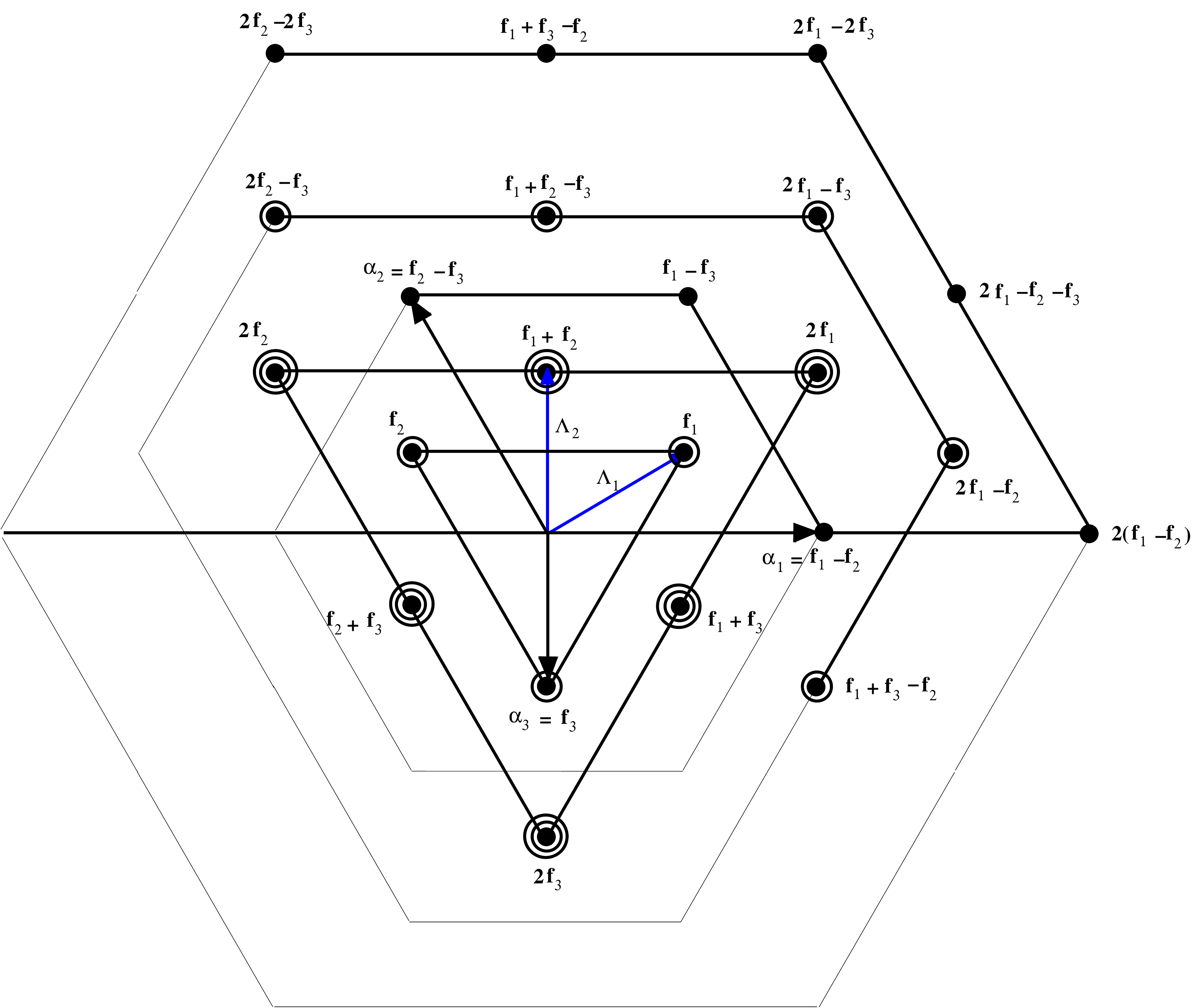}
\caption{Graphical presentation of the representation structure of
the compactified Gauss-Bonnet term. The black nodes arise from
distinct dilaton exponents in the three-dimensional Lagrangian. The
figure displays the level decomposition of the ${\bf
84}$-representation of $\mf{sl}(4, \mbb{R})$ into representations of
$\mf{sl}(3, \mbb{R})$. Only positive levels are displayed. The black
nodes correspond to positive weights of ${\bf 84}$ of $\mf{sl}(4,
\mbb{R})$. Nodes with no rings represent the positive weights of the
level zero representation ${\bf 27}$, nodes with one ring represent
the positive weights of the level one representations ${\bf 15}$ and
${\bf 3}$, while nodes with two rings represent the positive weights
of the level two representation ${\bf 6}$. The shaded lines complete
the representations with non-positive weights which are not
displayed explicitly. }
\label{figure:84ofSL4}
\end{center}
\end{figure}

Firstly, we may neglect all representations at negative levels since
these do not contain any positive weights. However, not all weights
for $\ell\geq 0$ are positive. If we had decomposed the adjoint
representation of $\mf{sl}(4, \mbb{R})$ this problem would not have
been present since all roots are either positive or negative, and
hence all weights at positive level are positive and vice versa. In
our case this is not true because for representations larger than
the adjoint many weights are neither positive nor negative. It is
furthermore important to realize that after dualisation it is the
positive weights of $\mf{sl}(4, \mbb{R})$ that we will obtain and
not of $\mf{sl}(3, \mbb{R})$. As can be seen in Figure
\ref{figure:84ofSL4} the decomposition indeed includes weights which
are negative weights of $\mf{sl}(3, \mbb{R})$ but nevertheless
positive weights of $\mf{sl}(4, \mbb{R})$. An explicit counting
reveals the following number of positive weights at each level (not
counting weight multiplicities):
\beqa
{}Ê\ell=0 &:&  8,
\nn \\
{}Ê\ell=1 &:& 8,
\nn \\
{}Ê\ell=2 &:& 6.
\eqa
The eight weights at level zero are of course the positive weights
of the ${\bf 27}$ representation of $\mf{sl}(3, \mbb{R})$ that we
had before dualisation. In order to verify that we find all positive
weights of ${\bf 84}$ we must now check explicitly that after
dualisation we get $8+6$ additional positive weights. The total
number of distinct weights of $\mf{sl}(4, \mbb{R})$ that should
appear in the Lagrangian after compactification and dualisation is
thus 22.

The lifting from $\mf{sl}(3, \mbb{R})$ to $\mf{sl}(4, \mbb{R})$ is
done by adding the third simple root $\val_3\equiv \vf_3$, from
\Eqnref{GBexponents2}. The complete set of new weights arising from
\Eqnref{GBexponents2} is then
\beqa
{}Ê\ell=1 &:&  \vf_1=\val_1+\val_2+\val_3, \qquad \qquad \quad
\hs\vf_2=\val_2+\val_3,
\nn \\
{}Ê& & 2\vf_1-\vf_2=2\val_1+\val_2+\val_3, \qquad
2\vf_2-\vf_3=2\val_2+\val_3,
\nn \\
{}Ê& & 2\vf_1-\vf_3=2\val_1+2\val_2+\val_3 \qquad
\vf_1+\vf_2-\vf_3=\val_1+2\val_2+\val_3,
\nn \\
{}Ê& & \vf_1+\vf_3-\vf_2=\val_1+\val_3,\qquad \quad \vf_3=\val_3,
\nn \\
\nn \\
{}Ê\ell=2 &:& 2\vf_1=2\val_1+2\val_2+2\val_3, \qquad \quad \hs
2\vf_2=2\val_2+2\val_3,
\nn \\
{}Ê& & 2\vf_3=2\val_3, \qquad \qquad \qquad \quad
\qquad\vf_1+\vf_2=\val_1+2\val_2+2\val_3,
\nn \\
{}Ê& & \vf_1+\vf_3=\val_1+\val_2+2\val_3, \qquad \quad
\vf_2+\vf_2=\val_2+2\val_3.
\eqa
In Table \ref{table:Reps} we indicate which representations these
weights belong to and in Figure \ref{figure:84ofSL4} we give a
graphical presentation of the level decomposition. These results
show that the Gauss-Bonnet term in $D=6$ compactified on $T^3$ to
three dimensions indeed gives rise to all strictly positive weights
of the ${\bf 84}$-representation of $\mf{sl}(4, \mbb{R})$.

\begin{table}
\begin{center}
\begin{tabular}{|m{15mm}|m{15mm}|m{80mm}|}
\hline
& &  \\
Reps & $\ell$ & Positive Weights of $\mf{sl}(4, \mbb{R})$ \\
& &  \\
\hline
& & \\
${\bf 3}$ & 1Ê& $\val_3, \hs \val_2+\val_3, \hs
\val_1+\val_2+\val_3$ \\
& & \\
\hline
& & \\
${\bf 15}$ & 1 & $2\val_2+\val_3, \hs \val_1+2\val_2+\val_3, \hs
2\val_1+2\val_2+\val_3$, \\
& & \\
& & $2\val_1+\val_2+\val_3, \hs \val_1+\val_3$ \\
& & \\
\hline
& & \\
$ {\bf 6}$ & 2 & $2\val_2+\val_3, \hs \val_1+2\val_2+\val_3, \hs
2\val_1+2\val_2+2\val_3$, \\
 & & \\
 & & $\val_1+\val_2+2\val_3, \hs 2\val_3, \hs \val_2+2\val_3$ \\
 & & \\
\hline
\end{tabular}
\caption{Positive weights at levels one and two.}
\label{table:Reps}
\end{center}
\end{table}

\subsubsection*{Weight Multiplicities}

We have shown that the six-dimensional Gauss-Bonnet term
compactified to three dimensions gives rise to all positive weights
of the ${\bf 84}$-representation of $\mf{sl}(4, \mbb{R})$. However,
we have not yet addressed the issue of weight multiplicities. It is
not clear how to approach this problem. Naively, one might argue that
if $k$ distinct terms in the Lagrangian are multiplied by the same
dilaton exponential, corresponding to some weight $\vec{\lambda}$,
then this weight has multiplicity $k$. Unfortunately, this type of
counting does not seem to work.

Consider, for instance, the representations at $\ell=1$. Both
representations ${\bf 15}$ and ${\bf 3}$ contain the weights $\vf_1,
\hs \vf_2$ and $\vf_3$. In ${\bf 15}$ these have all multiplicity 2,
while in ${\bf 3}$ they have multiplicity 1. Thus, in total these
weights have multiplicity 3 as weights of $\mf{sl}(3, \mbb{R})$.
Now, investigation of the dilaton exponents in \Eqnref{GBexponents2}
reveals that these weights arise from the two types of exponents
$\vf_a$ and $\vf_a+\vf_b-\vf_c$, $(b<c)$. The first type gives rise
to all three weights $\vf_1, \hs \vf_2$ and $\vf_3$, while the
second type only contributes $\vf_2=\vf_3+\vf_2-\vf_3$ and
$\vf_1=\vf_3+\vf_1-\vf_3$. Thus, by the arguments above the weights
$\vf_1$ and $\vf_2$ has multiplicity 2 in the Lagrangian, and
$\vf_3$ has multiplicity 1. We therefore deduce that for all these
weights there appears to be a mismatch in the multiplicity.

We suggest that the correct way to interpret this discrepancy in the weight multiplicities is as an indicative of the need to introduce transforming automorphic forms in order to restore the $SL(4, \mbb{Z})$-invariance. This will be discussed more closely in Section
\ref{section:resolution}.

\subsubsection*{Including the Dilaton Prefactor}

\label{section:SpecialCaseWithVolumeFactor}

We will now revisit the analysis from Section
\ref{section:SpecialCaseNoVolumeFactor}, but here we include the
contribution from the overall exponential factor $e^{-2\varphi}$ in
the Lagrangian \Eqnref{GB3}. This factor arises as follows. The
determinant of the $D$-dimensional vielbein is given by
$\hat{e}=e^{D\varphi}\tilde{e}$, because of the Weyl-rescaling.
Moreover, upon compactification the determinant of the rescaled
vielbein splits according to $\tilde{e}=e \tilde{e}_{\text{int}}$,
where $e$ represents the external vielbein and
$\tilde{e}_{\text{int}}$ the internal vielbein. By the definition of
the Weyl-rescaling we have
$\tilde{e}_{\text{int}}=e^{-(D-2)\varphi}$. This represents the
volume of the $n$-torus, upon which we perform the reduction. Thus,
the overall scaling contribution from the measure is
$e^{D\varphi}e^{-(D-2)\varphi}=e^{2\varphi}$. In addition, we have a
factor of $e^{-4\varphi}$ from the Gauss-Bonnet term. This gives a
total overall dilaton prefactor of $e^{-2\varphi}$, which, after
inserting $\varphi=\f{1}{6}\vp\cdot \vg$, becomes
$e^{-\f{1}{3}\vp\cdot \vg}$.

The importance of the volume factor for compactified higher
derivative terms was emphasized in \cite{LambertWest1}, using the
argument that after dualisation this factor is no longer invariant
under the extended symmetry group $SL(n+1, \mbb{R})$ and so must be
included in the weight structure. We shall see that the inclusion of
this factor drastically modifies the previous structure and destroys
the representation ${\bf 84}$ of $\mf{sl}(4, \mbb{R})$.

In order to perform this analysis, it is useful to first
rewrite the simple roots and fundamental weights in a way which
makes a comparison with \cite{LambertWest1} possible. We define
arbitrary $3$-vectors in $\mbb{R}^3$ as follows
\beq
\hat{\vec{v}}=v_1\vL_1+v_2\vL_2+v_g\vg = \big(\vec{v},
v_g\big)=\big(v_1, v_2, v_g\big),
\eeq
where $\vL_1$ and $\vL_2$ are the fundamental weights of $\mf{sl}(3,
\mbb{R})$ and $\vg$ is the basis vector taking us from the weight
space $\mbb{R}^2$ of $\mf{sl}(3, \mbb{R})$ to the weight space
$\mbb{R}^3$ of $\mf{sl}(4, \mbb{R})$. Note that
\beq
\vL_1\cdot \vg=\vL_2\cdot \vg=0,
\eeq
by virtue of \Eqnref{zeroscalarproduct} and \Eqnref{weightrelation},
which implies
\beq
\hat{\vec{v}}\cdot \hat{\vec{u}}=\vec{v}\cdot \vec{u}+v_gv_g\vg\cdot
\vg.
\eeq

The scalar products may all be deduced from the orthonormal basis
$\ve_i$ of $\mbb{R}^3$. Restricting to $D=6$ and $n=3$ gives
\beq
\vf_a=\sqrt{2}\ve_a+\f{2}{9}\vg,
\eeq
and thus
\beq
\vo_a=\vf_a-\f{4}{9}\vg.
\eeq
The relevant scalar products become
\beqa
{}Ê\vg\cdot \vg &=& \f{54}{4},
\nn \\
{}Ê\vg\cdot \vf_a&=&6,
\nn \\
{}Ê\vf_a\cdot \vf_b&=& 2\delta_{ab}+2,
\nn \\
{}Ê\vo_a\cdot \vo_b&=& 2\delta_{ab}-\f{2}{3}.
\eqa
The simple roots of $\mf{sl}(3, \mbb{R})$ may now be written as
\beqa
{}Ê\hat{\val}_1&=& \big(\val_1, 0\big)=\big(2, -1, 0\big),
\nn \\
{}Ê\hat{\val}_2 &=& \big(\val_2, 0\big)=\big(-1, 2, 0\big),
\eqa
and the third simple root becomes
\beq
\hat{\val}_3= \vf_3=\vo_3+\f{4}{9}\vg=-\vL_2+\f{4}{9}\vg=\big(0, -1,
\f{4}{9}\big).
\eeq
In addition, the associated fundamental weights $\hat{\vL}_i, \hs
i=1, 2, 3, $ of $\mf{sl}(4, \mbb{R})$, defined by
\beq
\hat{\val}_i\cdot \hat{\vL}_j=2\delta_{ij},
\eeq
become
\beq \hat{\vL}_1=\big(1, 0, \f{1}{9}\big),\quad \hat{\vL}_2 = \big(0,
1, \f{2}{9}\big),\quadÊ\hat{\vL}_3 = \big(0,0, \f{1}{3}\big).
\eeq
Let us check that these indeed correspond to the fundamental weights
of $\mf{sl}(4, \mbb{R})$, by computing the highest weight
$2\hat{\vL}_1+2\hat{\vL}_3$ explicitly,
\beqa
{}Ê2\hat{\vL}_1+2\hat{\vL}_3&=&2\vL_1+\f{2}{9}\vg+\f{2}{3}\vg
\nn \\
{}Ê&= &2(\vo_1+\f{4}{9}\vg)
\nn \\
{}Ê& =&2\vf_1
\nn \\
{}Ê&=&2\hat{\val}_1+2\hat{\val}_2+2\hat{\val}_3.
\eqa
This result is consistent with being the highest weight of the ${\bf
84}$ representation of $\mf{sl}(4, \mbb{R})$ as can be seen in
Figure \ref{figure:84ofSL4}.

 Let us now include the dilaton prefactor in the analysis. In
terms of $\mf{sl}(4, \mbb{R})$-vectors the volume factor can be
identified with a negative shift in $\hat{\vL}_3$, i.e.,
\beq
e^{-\f{1}{3}\vg\cdot \vp}=e^{-\hat{\vL}_3\cdot \vp}.
\eeq
As already mentioned above, this factor is irrelevant before
dualisation because $\vg\cdot \vp$ is invariant under $SL(3,
\mbb{R})$. Thus, before dualisation the manifest $SL(3,
\mbb{R})$-symmetry of the compactified Gauss-Bonnet term is
associated with the ${\bf 27}$-representation of $\mf{sl}(3,
\mbb{R})$.

After dualisation, all the dilaton exponents in
\Eqnref{GBexponents1} and \Eqnref{GBexponents2} become shifted by a
factor of $-\hat{\vL}_3$. In particular, the new highest weight is
\beq
\big(2\hat{\vL}_1+2\hat{\vL}_3\big)-\hat{\vL}_3=2\hat{\vL}_1 +
\hat{\vL}_3,
\eeq
corresponding to the ${\bf 36}$ representation of $\mf{sl}(4,
\mbb{R})$, with Dynkin labels $[2, 0, 1]$. This is consistent with
the general result of \cite{LambertWest1} that a generic curvature
correction to pure Einstein gravity of order $l/2$ should be
associated with an $\mf{sl}(n+1, \mbb{R})$-representation with
highest weight $\f{l}{2}\hat{\vL}_1+\hat{\vL}_n$.

However, this is not the full story. A more careful examination in
fact reveals that the ${\bf 36}$ representation cannot incorporate
all the dilaton exponents appearing in the Lagrangian, in contrast
to the ${\bf 84}$-representation of Figure \ref{figure:84ofSL4}. To
see this, let us decompose ${\bf 36}$ in terms of representations of
$\mf{sl}(3, \mbb{R})$. The result is:
\beqa
{}Ê{\bf 36} &=& {\bf 15}\oplus {\bf 8}\oplus {\bf 6}\oplus {\bf
3}\oplus {\bf \bar{3}}\oplus {\bf 1},
\nn \\
{}Ê[2, 0, 1]&=& [2,1] + [1,1] + [2,0] + [1,0] + [0,1] + [0,0].
\eqnlab{36inSL3}
\eqa
Comparing this with \Eqnref{84inSL3}, we see that the
representations ${\bf 27}, {\bf \bar{15}}$ and ${\bf \bar{6}}$ are
no longer present. For the latter two this is not a problem since
they were never present in the previous analysis. What happens is
that the ${\bf 6}$ of ${\bf 84}$ gets shifted ``downwards'' and
becomes the ${\bf 6}$ of ${\bf 36}$. Similarly, the ${\bf 15}$ and
${\bf 3}$ of ${\bf 84}$ become the ${\bf 15}$ and ${\bf 3}$ of ${\bf
36}$. This takes into account all the shifted dilaton exponents
arising from the dualisation process. However, since there is not
enough ``room'' for the ${\bf 27}$ of $\mf{sl}(3, \mbb{R})$ in
\Eqnref{36inSL3}, some of the dilaton exponents (the ones
corresponding to $2\vf_2-2\vf_3, \hs \vf_1+\vf_3-\vf_2, \hs
2\vf_1-2\vf_3, 2\vf_1-\vf_2-\vf_3$ and $2\vf_1-2\vf_2$) arising from
pure $\tilde{P}$-terms, i.e., before dualisation, remain outside of
${\bf 36}$. In fact, due to the shift of $-\hat{\vL}_3$ these have
now become \emph{negative} weights of $\mf{sl}(4, \mbb{R})$, because
they are below the hyperplane defined by $\vg\cdot \vp=0$. Although
we know that these weights still correspond to positive weights of
the ${\bf 27}$ of $\mf{sl}(3, \mbb{R})$, we cannot determine which
representation of $\mf{sl}(4, \mbb{R})$ they belong to.

By a straightforward generalisation of this analysis to
compactifications of quadratic curvature corrections from arbitrary
dimensions $D$, we may conclude that the highest weight
$2\hat{\vL}_1+\hat{\vL}_n$, can never incorporate the dilaton
exponents associated with the $[2, 0,$ $ \dots, 0,
2]$-representation of $\mf{sl}(n, \mbb{R})$ before dualisation.

\section{An $SL(n+1, \mbb{Z})$-Invariant Quartic Effective Action?}
\label{section:resolution}

It is clear from the analysis in the previous section that the
overall dilaton factor $e^{-\hat{\vL}_3\cdot \vp}$ (or, more
generally, $e^{-\hat{\vL}_n\cdot \vp}$) complicates the
interpretation of the dilaton exponents in terms of $\mf{sl}(n+1,
\mbb{R})$-representations. To analyze this question, let us discuss what kind of information is encoded in the weight structure. Apart
from the overall dilaton factor, the reduction of any higher
derivative term $\sim \mc{R}^p$ will give rise to terms with
$\Pc^{2p}$ (and terms with more derivatives and fewer $\Pc$'s),
where $\Pc$ represents any of the ``building blocks'' $P$, $H$ and
$\partial\phi$ (we suppress all 3-dimensional indices). The
appearance of weights of $\mf{sl}(n+1, \mbb{R})$ (without the
uniform shift from the overall dilaton factor) reflects the fact
that we use fields which are components of the symmetric part of the
left-invariant Maurer--Cartan form $\Pc$ of $\mf{sl}(n+1, \mbb{R})$.
Moreover, the dilaton factor contains information about the number
of such fields. A term $\mc{R}^{l/2}$ will generically give weights
in the weight space of the representation $[l/2,0,\ldots,0,l/2]$ of
$\mf{sl}(n+1, \mbb{R})$, and fill out the positive part of this
weight space.\footnote{We note that the representation structure
encountered here is of the same type as for the lattice of BPS
charges in string theory on $T^{n}$ \cite{ObersPioline}.} This much
is clear from the observation that the overall dilaton factor really
is ``overall''.

The presence of the overall dilaton factor shifts this weight space
uniformly in a negative direction. This shift happens to be by a
vector in the weight lattice of $\mf{sl}(n+1, \mbb{R})$ for any
value of $p$. However, we emphasize that the dilaton exponents still
lie in the weight space of the representation
$[l/2,0,\ldots,0,l/2]$, albeit shifted ``downwards''. From this
point of view, the weight space of the representation with the
shifted highest weight of $[l/2,0,\ldots,0,l/2]$ as highest weight
-- for example, the representation $[2, 0, 1]$ in the case discussed
above -- does not contain all the weights that appear in the reduced
Lagrangian, and is thus not relevant.\\

\subsection{Completion Under $SL(n+1, \mbb{Z})$}

Consider now the fact that it is really the discrete ``U-duality''
group $SL(n+1, \mbb{Z}) \subset SL(n+1, \mbb{R})$ which is expected
to be a symmetry of the complete effective action. Therefore, the
compactified action should be seen as a remnant of the full
U-duality invariant action, arising from a ``large volume
expansion'' of certain automorphic forms.

Schematically, a generic, quartic, scalar term in the action
%(again, restricting to $D=6, n=3$ for convenience)
after compactification of
the Gauss-Bonnet term is of the form
\beq
%\int d^3 x \sqrt{g^{(3)}} e^{-\hat{\vL}_3\cdot \vp} k(\vp)
%F^{[R]}_{IJ}(X),
\int d^3 x \sqrt{|g|} e^{-\hat{\vL}_n\cdot \vp}F(\Pc),
\eeq
where
%$F^{[R]}_{IJ}(X)$
$F(\Pc)$ is a quartic polynomial
%in the scalars $X$ of the theory,
%and $k(\vp)$ is some function of the ``dilatons'' $\vp$.
in the components of the Maurer--Cartan form mentioned above. $F$
will be invariant under $SO(n)$ by construction, but generically not
under $SO(n+1)$.
%This term is not an $SO(4)$ scalar, since $F^{[R]}_{IJ}(X)$ is
%constructed from a nonlinear realisation of the $\mf{sl}(4,
%\mbb{R})$-representation $[2, 0, 2]$, found in Section
%\ref{section:SpecialCaseNoVolumeFactor}. Therefore $F^{[R]}_{IJ}(X)$
%transforms in an $\mf{so}(4)$-representation $R$ which depends on $[2,
%0, 2]$ of $\mf{sl}(4, \mbb{R})$.

To obtain an action which is a scalar under $SO(n+1)$ we must first
``lift'' the result of the compactification to a globally $SL(n+1,
\mbb{Z})$-invariant expression. This can be done by replacing
%the coefficient $k(\vp)$
$e^{-\hat{\vL}_n\cdot \vp}F(\Pc)$
by a suitable automorphic form contracted with four $\Pc$'s:
\beq
\Psi_{I_1\ldots I_8}(X)\Pc^{I_1I_2}\Pc^{I_3I_4}\Pc^{I_5I_6}\Pc^{I_7I_8},
\eqnlab{completion}
\eeq
where the $I$'s are vector indices of $SO(n+1)$. Here, $\Psi(X)$ is
an automorphic form transforming in some representation of
$SO(n+1)$, and is constructed as an Eisenstein series, following,
e.g., refs. \cite{LambertWest2,Aspects}.
%, which transforms in the same
%representation (or, rather, the conjugate representation $\bar{R}$) of
%$SO(4)$ as the polynomial $F^{[R]}_{IJ}(X)$. In addition,
%$\Psi^{IJ}_{[\bar{R}]}(X)$ must reproduce the coefficient $k(\vp)$ as
%its first order term in a ``large volume expansion'',
%\beq
%\Psi^{IJ}_{[\bar{R}]}(X) \sim k(\vp) \quad \text{as}\quad
%\vp\longrightarrow \infty.
%\eeq
We must demand that when the large volume limit, $\hat{\vL}_n\cdot
\vp\rightarrow-\infty$, is imposed, the leading behaviour is
\beq
\Psi_{I_1\ldots
I_8}(X)\Pc^{I_1I_2}\Pc^{I_3I_4}\Pc^{I_5I_6}\Pc^{I_7I_8}\quad
\longrightarrow \quad e^{-\hat{\vL}_n\cdot \vp}F(\Pc).
\eqnlab{largevolumelimit}
\eeq
This limit was taken explicitly in \cite{LambertWest2,Aspects}. This
gives conditions on which irreducible $SO(n+1)$ representations the
automorphic forms transform under (from the tensor structure), as
well as a single condition on the ``weights'' of the automorphic
forms (from the matching of the overall dilaton factor). Automorphic
forms exist for continuous values of the weight (unlike holomorphic
Eisenstein series) above some minimal value derived from convergence
of the Eisenstein series. It was proven in \cite{Aspects} that any
$SO(n)$-covariant tensor structure can be reproduced as the large
volume limit of some automorphic form, and that the weight dictated
by the overall dilaton factor is consistent with the convergence
criterion.

Under the assumption that these arguments are valid, we may conclude
that the representation theoretic structure of the dilaton exponents
in the polynomial $F$ should be analyzed without inclusion of the
volume factor $e^{-\hat{\vL}_n\cdot \vp}$, and hence, for the
Gauss-Bonnet term ($l=4$), it is the $[2, 0, \dots, 0,
2]$-representation which is the relevant one (in the sense above,
that we are dealing with products of four Maurer--Cartan forms), and
\emph{not} the representation $[2, 0, \dots, 0, 1]$ which was
advocated in \cite{LambertWest1}. Another indication for why the
representation with highest weight $2\hat{\vL}_1+\hat{\vL}_n$ cannot
be the relevant one is that it is not contained in the tensor
product of the adjoint representation $[1, 0, \dots, 0, 1]$ of
$\mf{sl}(n+1, \mbb{R})$ with itself.

The present point of view also suggest a possible explanation for the discrepancy of the weight multiplicities observed in the previous section. In the complete $SL(n+1, \mbb{Z})$-invariant four-derivative effective action the multiplicities of the weights in the $[2, 0, \dots, 0, 2]$-representation necessarily match because the action is constructed directly from the $\mf{sl}(n+1, \mbb{R})$-valued building block $\mc{P}$. When taking the large volume limit, \Eqnref{largevolumelimit}, a lot of information is lost (see, e.g., \cite{Aspects}) and it is therefore natural that the result of the compactification does not display the correct weight multiplicities. Thus, it is only after taking the non-perturbative completion, \Eqnref{completion}, that we can expect to reproduce correctly the weight multiplicities of the representation $[2, 0, \dots, 0, 2]$.

%%%%%%%%%%%%%%%%%%%%%%%%%%%%%%%%%%%%%%%%%%%%
%
% Appendix
%
%%%%%%%%%%%%%%%%%%%%%%%%%%%%%%%%%%%%%%%%%%%%

\appendix
\chapter{Details on the $E_{10}/$Massive IIA Correspondence}
\label{Appendix:MassiveIIA}
\section{Details For Massive IIA Supergravity} 
\label{Appendix:mIIAdetails}

In this appendix we give all the relevant details of massive IIA
supergravity which are required to establish the correspondence with the $E_{10}$-sigma model. The complete Lagrangian and supersymmetry variations were already given in Section 2, and will not be repeated here. Here we complement this with information regarding our conventions, as well as explicit expressions for the bosonic and fermionic equations of motion, and the Bianchi identities. Moreover, we discuss in detail the truncations that we need to impose on the supergravity side to ensure the matching with the geodesic sigma model. Finally, we also discuss how our conventions for massless type IIA supergravity matches with a reduction from eleven-dimensional supergravity. This appendix is an extract from {\bf Paper VI}.

\subsection{Conventions} 
\label{app:conventions}

We use the signature $(-+\ldots +)$ for space-time. The indices $\mu=(t,m)$
are $(1+9)$-dimensional curved indices, whereas $\alpha=(0,a)$ are the
corresponding flat indices.  Partial derivatives with flat indices are defined
via conversion with the inverse vielbein: $\partial_\alpha = e_\alpha{}^\mu
\partial_\mu$. The Lorentz covariant derivative $D_\al$ acts on (co-)vectors via
$D_\al 
V_\beta= \partial_\al V_\beta + \omega_{\al\,\beta}{}^\gamma V_\gamma$ in terms of
the Lorentz connection, which is defined in turn as
\beq 
\omega_{\al\,\beta\gamma} = \frac12\left(\Omega_{\al\beta\,\gamma} 
  + \Omega_{\gamma\al\,\beta} -\Omega_{\beta\gamma\,\al}\right)
\eeq
in terms of the anholonomy of the orthonormal frame $e_\mu{}^\al$ defined by 
$\Omega_{\mu\nu}{}^\al= 2\partial_{[\mu}e_{\nu]}{}^\al$.

Our fermions are Majorana--Weyl spinors of $SO(1,9)$ of real dimension
$16$. As the theory is type II non-chiral we can combine two spinors into a
$32$-dimensional Majorana representation on which the $\Gamma$-matrices of
$SO(1,10)$ act. These are the eleven real $32\times 32$ matrices $(\Gamma^0,
\Gamma^a, \Gamma^{10})$ which are symmetric except for $\Gamma^0$ which is
antisymmetric. We choose the representation such that $\Gamma^{10}$ is block
diagonal and projects on the two $16$ component spinors of opposite
chirality. $\Gamma^0$ is the charge conjugation matrix such that our
conventions are identical to those of~\cite{Kleinschmidt:2004dy}. 

A useful identity for our $\Gamma$-matrices is 
\beq 
\Gamma^{a_1\ldots a_k} = \frac{(-1)^{(k+1)(k+2)/2}}{(9-k)!}\epsilon^{a_1\ldots
  a_k b_1\ldots b_{9-k}} \Gamma_{b_1\ldots b_{9-k}} \Gamma^0\Gamma^{10}\,. 
\eeq
The various $\epsilon$ tensors we use are such that 
\beq 
\eps^{0\,1\ldots 10} = +1\,,\quad 
\eps^{0\,1\ldots 9} = +1\,,\quad 
\eps^{1\,\ldots 9} = +1\,.
\eeq

\subsection{Bianchi Identities} 

For the comparison with $E_{10}$ it is useful to write all supergravity
equations in a non-coordinate orthonormal frame described by the vielbein
$e_\mu{}^\al$ and use only Lorentz covariant objects. In these flat 
indices the Bianchi identities following from (\ref{curvs}) are 
\begin{subequations}\label{Bianchis} 
\beqa
\label{SugraBianchi1} 3D_{[\al_1} F_{\al_2\al_3]} &=& m F_{\al_1\al_2\al_3} \,,\\ 
4D_{[\al_1} F_{\al_2\al_3\al_4]} &=& 0 \,,\\
5D_{[\al_1} F_{\al_2\al_3\al_4\al_5]} &=&  
  10 F_{[\al_1\al_2} F_{\al_3\al_4\al_5]} \,. 
\eqa
\end{subequations}

\subsection{Bosonic Equations of Motion}
\label{SugraBeom}

The form equations of motion can be rewritten in $10$-dimensional flat indices 
as 
\begin{subequations}\label{eoms} 
\beqa
\label{eom1}D_\alpha(e^{3\phi/2}F^{\alpha\beta}) 
   &=& -\frac1{3!}e^{\phi/2} F^{\alpha_1\alpha_2\alpha_3\beta}F_{\alpha_1\alpha_2\alpha_3} \,,\\
\label{eom2}D_\alpha(e^{-\phi}F^{\alpha\beta_1\beta_2})  
   &=& m e^{3\phi/2} F^{\beta_1\beta_2} +\frac1{2!}e^{\phi/2}
   F^{\beta_1\beta_2\alpha_1\alpha_2} F_{\alpha_1\alpha_2} \nn\\ 
    &&    - \frac1{1152}F_{\alpha_1\ldots \alpha_4} F_{\alpha_5\ldots \alpha_8}\eps^{\alpha_1\ldots \alpha_8\beta_1\beta_2}\,,\\ 
\label{eom3}D_\alpha(e^{\phi/2}F^{\alpha\beta_1\ldots \beta_3})
   &=& \frac1{144} F_{\alpha_1\ldots \alpha_4}F_{\alpha_5\ldots
     \alpha_7}\eps^{\alpha_1\ldots \alpha_7\beta_1\ldots \beta_3} ,
\eqa
\end{subequations} 
while the dilaton and gravity equations are 
\begin{subequations}\label{lev0}
\beqa
D^\alpha \partial_\alpha \phi &=& \frac38e^{3\phi/2} |F_{(2)}|^2 
   - \frac1{12} e^{-\phi} |F_{(3)}|^2
   + \frac1{96} e^{\phi/2} |F_{(4)}|^2 
   + \frac54 m^2 e^{5\phi/2} \,,\label{dilsugra}\\
R_{\alpha\beta} &=& \frac12 \partial_\alpha\phi \partial_\beta\phi  
   +\frac{m^2}{16} \eta_{\alpha\beta}e^{5\phi/2} 
   +\frac12e^{3\phi/2}F_{\alpha\gamma}F_\beta{}^\gamma 
       -\frac1{32}\eta_{\alpha\beta} e^{3\phi/2}F_{\gamma_1\gamma_2}F^{\gamma_1\gamma_2} \nn\\
&&\quad  +\frac14 e^{-\phi}
F_{\alpha\gamma_1\gamma_2}F_\beta{}^{\gamma_1\gamma_2}   
       -\frac1{48} \eta_{\alpha\beta} e^{-\phi}
       F_{\gamma_1\gamma_2\gamma_3}F^{\gamma_1\gamma_2\gamma_3}\label{Einsteinsugra}\\ 
&&\quad  +\frac1{12} e^{\phi/2} F_{\alpha\gamma_1\gamma_2\gamma_3}F_\beta{}^{\gamma_1\gamma_2\gamma_3}  
       -\frac1{128} \eta_{\alpha\beta} e^{\phi/2}
       F_{\gamma_1\gamma_2\gamma_3\gamma_4}F^{\gamma_1\gamma_2\gamma_3\gamma_4}.\nn 
\eqa
\end{subequations} 

\subsection{Fermionic Equations of Motion} 
\label{app:sugraf}

Besides the bosonic equations one also deduces the fermionic equations of 
motion from the Lagrangian (\ref{lag}) which we write out in flat indices. For
the dilatino this gives 
\beqa 
\label{eomlam} 
\G^\al D_\al \lt &-&
\f{5}{32}e^{3\phi/4}F_{\al_1\al_2}\G^{\al_1\al_2}\G_{10}\lt 
  +\f3{16}e^{3\phi/4}F_{\al_1\al_2}\G^{\beta}\G^{\al_1\al_2}\G_{10}\pt_\beta\nn\\ 
&+&\f1{24}e^{-\phi/2}F_{\al_1\cdots\al_3}\G^\beta\G^{\al_1\cdots\al_3}\G_{10}\pt_\beta\nn\\ 
&+&\f1{128}e^{\phi/4}F_{\al_1\cdots \al_4}\G^{\al_1\cdots \al_4}\lt
  - \f1{192} e^{\phi/4}F_{\al_1\cdots \al_4}\G^\beta\G^{\al_1\cdots \al_4}\pt_\beta\nn\\
&-&\f1{2}\partial_\al\phi\G^\beta\G^\al\pt_\beta\nn\\ 
&-&\f{21}{16}me^{5\phi/4}\lt-\f5{8}me^{5\phi/4}\G^\al\pt_\al = 0, 
\eqa
The  gravitino equation obtained directly from the variation of (\ref{lag}) is
of the form 
\beq
\label{psisimple}
\mathcal E^\alpha = \G^{\alpha\beta\gamma} D_\beta\pt_{\gamma}+R^{\alpha}=0.
\eeq
After multiplication with two gamma matrices it can be rewritten as
\beq
\label{psisimple+}
E_{\alpha}=\G^{\beta}(D_\alpha\pt_{\beta}-D_\beta\pt_{\al})+L_{\alpha}=0,
\eeq
where
\beq
L_{\alpha}=\f18(\G_{\alpha\beta}R^{\beta}-7R_{\alpha}),
\eeq
Although $\mathcal E^\alpha=0$ is equivalent to $E^\alpha=0$, the spatial
components $E^a$ and $\mathcal E^a$ are only equivalent when the
supersymmetry constraint $\mathcal E^0$ is also taken into account. It turns out
that the dynamical equation that corresponds directly to the $K(E_{10})$ Dirac
equation is $E^a=0$, not too surprisingly since in this form one obtains
directly a Dirac equation for $\psi_a$. The $SO(1,9)$ covariant equation
$E_{\alpha}=0$ reads explicitly as follows:  
\beqa \label{eompsi} 
\G^\beta\left( D_\al\pt_\beta -D_\beta\pt_\al\right)  
  &+&\f{21}{64} e^{3\phi/4} F_{\al\beta}\G^\beta \G_{10}\lt  
  -\f{3}{128} e^{3\phi/4} F_{\beta_1\beta_2}\G_{\al}{}^{\beta_1\beta_2}
  \G_{10}\lt\nn\\ 
&+& \f1{64}e^{3\phi/4} F_{\beta_1\beta_2} \G_{\al}{}^{\beta_1\beta_2\gamma}\G_{10}\pt^\gamma
  -\f1{32} e^{3\phi/4} F_{\beta_1\beta_2}
  \G_{\al}{}^{\beta_1}\G_{10}\pt^{\beta_2}\nn\\ 
&-&\f7{32} e^{3\phi/4} F_{\al\beta}\G^{\beta\gamma}\G_{10}\pt_\gamma
  -\f7{64} e^{3\phi/4} F_{\beta_1\beta_2}\G^{\beta_1\beta_2}\G_{10}\pt_\al
  \nn\\
&+& \f7{32} e^{3\phi/4} F_{\al\beta}\G_{10}\pt^\beta\nn\\ 
&+&\f1{96} e^{-\phi/2} F_{\beta_1\beta_2\beta_3}
\G_{\al}{}^{\beta_1\beta_2\beta_3}\G_{10}\lt 
 -\f3{32} e^{-\phi/2} F_{\al\beta_1\beta_2} \G^{\beta_1\beta_2}\G_{10}\lt\nn\\ 
&+& \f1{96} e^{-\phi/2} F_{\beta_1\beta_2\beta_3} \G_{\al}{}^{\beta_1\beta_2\beta_3\gamma}\G_{10}\pt^\gamma
  +\f1{32} e^{-\phi/2} F_{\beta_1\beta_2\beta_3}
  \G^{\beta_1\beta_2\beta_3}\G_{10}\pt_\al\nn\\ 
&-& \f1{32} e^{-\phi/2} F_{\beta_1\beta_2\beta_3}
\G_{\al}{}^{\beta_1\beta_2}\G_{10}\pt^{\beta_3} 
  -\f3{32} e^{-\phi/2} F_{\al\beta_1\beta_2} \G^{\beta_1\beta_2\gamma}\G_{10}\pt_\gamma\nn\\
&+& \f3{16} e^{-\phi/2} F_{\al\beta_1\beta_2}
\G^{\beta_1}\G_{10}\pt^{\beta_2}\nn\\ 
&-&\f1{512} e^{\phi/4} F_{\beta_1\ldots \beta_4} \G_{\al}{}^{\beta_1\ldots \beta_4}\lt
  +\f5{384} e^{\phi/4} F_{\al\beta_1\ldots \beta_3} \G^{\beta_1\ldots
    \beta_3}\lt\nn\\ 
&-& \f1{256} e^{\phi/4} F_{\beta_1\ldots \beta_4} \G_{\al}{}^{\beta_1\ldots\beta_4\gamma}\pt^\gamma
  +\f5{768} e^{\phi/4} F_{\beta_1\ldots \beta_4} \G^{\beta_1\ldots
    \beta_4}\pt_\al\nn\\ 
&+& \f1{64} e^{\phi/4} F_{\beta_1\ldots \beta_4} \G_{\al}{}^{\beta_1\ldots \beta_3}\pt^{\beta_4}
  +\f5{192} e^{\phi/4} F_{\al\beta_1\ldots \beta_3} \G^{\beta_1\ldots
    \beta_3\gamma}\pt_\gamma\nn\\ 
&-& \f5{64} e^{\phi/4} F_{\al\beta_1\ldots \beta_3}\G^{\beta_1\beta_2}\pt^{\beta_3} +\f12\partial_\al \phi \lt\nn\\
&+& \f1{32}e^{5\phi/4}m \G_{\al\beta}\pt^\beta +\f9{32}e^{5\phi/4}m \pt_\al
+\f5{64}e^{5\phi/4} m \G_\al \lt=0 \,. 
\eqa

\subsection{Truncation on the Supergravity Side}
\label{truncation} 

As explained in Section \ref{EOMandTruncations} of the main text, the correspondence between the dynamics of the $E_{10}$-invariant sigma model and the dynamics of type IIA supergravity only works if a certain truncation is applied. 

As dictated from the BKL analysis, the following truncations must be imposed on the equations of motion and Bianchi identities:
\begin{subequations}\label{truncbos}\beqa 
\partial_a (N \partial_0 \phi) = \partial_a ( N \omega_{0\,ab}) = \partial_a (N\omega_{a\,b 0})  
= \partial_a(N\partial_b\phi) = \partial_a (N\omega_{b\,cd}) =0 
\eqa 
and 
\beqa \label{truncsugra}
\partial_a\left(N e^{3\phi/4} F_{0b}\right)  
= \partial_a\left(N e^{-\phi/2} F_{0b_1b_2}\right) 
= \partial_a\left(N e^{\phi/4} F_{0b_1b_2b_3}\right)&=&0\,,\nn\\ 
\partial_a\left(N e^{3\phi/4} F_{b_1b_2}\right) 
= \partial_a\left(N e^{-\phi/2} F_{b_1b_2b_3}\right)
= \partial_a\left(N e^{\phi/4} F_{b_1b_2b_3b_4}\right) &=&0\,,\nn\\
\partial_a \left(N e^{5\phi/4}m\right) &=&0\,. 
\eqa 
Furthermore, as already indicated in (\ref{spintrace}), the spatial trace of the spin connection has to be set to zero
\beq 
\omega_{b\,ba} = 0\,.
\eeq
\end{subequations} 
Equations (\ref{truncbos}) exhaust all truncations of the bosonic variables. As explained in the text, with these truncations the bosonic geodesic equations agree with the supergravity up to one term in the Einstein equation coming from the contribution to the Ricci tensor $R_{ab}$ which is proportional to $\Omega_{ac\,d}\Omega_{bd\,c}$. We emphasize that there are no mismatches associated with the mass parameter $m$.
 
For the fermionic variables one also needs to apply appropriate truncations of spatial gradients. These turn out to be
\beq \label{fermtrunc}
N^{-1} \partial_a (N\lt) = N^{-1/2} \partial_a ( N^{1/2} \pt_b ) = 0\,.
\eeq
With this choice of truncation the Dirac equation of the coset match exactly
the fermionic equations of motion of supergravity if in addition the
supersymmetric gauge 
\beq 
\pt_0-\G_0\G^a\pt_a = 0 
\eeq 
of (\ref{susygauge}) is adopted.

\subsection{Reduction from $D=11$} 
 Our conventions for massless type IIA supergravity are consistent with the
 reduction of eleven-dimensional supergravity through the reduction ansatz
 and field redefinitions given in this section. This construction of massless
 type IIA was first carried out
 in~\cite{Giani:1984wc,Campbell:1984zc,Huq:1983im}. In this section only, we
 will denote the $D=11$ gravitino by $\Psi_M$.
 
 The supersymmetry variation of the gravitino in eleven-dimensional
supergravity is  
\beq 
\delta_{\varepsilon^{(11)}}\Psi_M^{(11)}=D_M^{(11)}\varepsilon^{(11)}+\f{1}{288}\Big({\Gamma_M}^{N_1\cdots N_4}-8\delta_M^{[N_1}\Gamma^{N_2N_3N_4]}\Big)\varepsilon^{(11)}F^{(11)}_{N_1\cdots N_4}.
\eeq
We reduce this expression along $x^{10}$ with the following ansatz for the eleven-dimensional vielbein:
\beq 
{E_M}^{A}=\left(\begin{array}{cc}
e^{-\f{1}{12}\phi}{e_\mu}^{\al} & e^{\f{2}{3}\phi}A_\mu\\ 
0 & e^{\f{2}{3}\phi}\\ 
\end{array}\right). 
\eeq
The four-form field strength is reduced as follows in curved indices
\beqa 
{}F^{(10)}_{\mu\nu\rho}&\equiv & F^{(11)}_{\mu\nu\rho\tten}, 
\nn \\
{}F^{(10)}_{\mu\nu\rho\sigma}&\equiv &
F^{(11)}_{\mu\nu\rho\sigma}+4A_{[\mu}F^{(10)}_{\nu\rho\sigma]}, 
\eqa
where $\tten$ denotes a curved index. The eleven-dimensional gravitino
$\Psi_M^{(11)}$ splits into the ten-dimensional gravitino $\pt_\mu$ and the
dilatino $\lt$, according to the following field redefinitions 
\beqa \label{kkferm}
{}\Psi^{(11)}_\mu &=& e^{-\f{1}{24}\phi}\big(\pt_\mu-\f{1}{12}\Gamma_\mu\lt\big)+\f{2}{3}e^{\f{1}{24}\phi}\Gamma_{\tten}A_\mu\lt,
\nn \\ 
{}\Gamma^{\tten}\Psi^{(11)}_{\tten}&=&\f{2}{3}e^{\f{1}{24}\phi}\lt. 
\eqa 
We also rescale the supersymmetry parameter according to
\beq 
\vet\equiv e^{\f{1}{24}\phi}\varepsilon^{(11)}. 
\eeq
 
\section{Details on the IIA Level Decomposition of $\mf{e}_{10}$ and $\mf{k}(\mf{e}_{10})$}
\label{Appendix:LevelDecomp} 
 
In this appendix we give all the details of the level deomposition of $\mf{e}_{10}$ with respect to $\mf{sl}(9, \mbb{R})$, up to level $(\ell_1, \ell_2)=(4,1)$. In particular we give all the relevant $\mf{e}_{10}$ commutators which are needed to compute the explicit expressions of the bosonic and fermionic equations of motion in Appendix \ref{app:EOMe10}. Moreover, we give details on the spinor and and vector-spinor representations of $\mf{k}(\mf{e}_{10})$. 

\begin{table} 
\centering 
\begin{tabular}{|c|c|c|}
\hline
$(\ell_1,\ell_2)$&$\mf{sl}(9, \mbb{R})$ Dynkin labels&$\mf{e}_{10}$ root $\alpha$ of lowest weight \\
\hline\hline
$(0,0)$&$[1,0,0,0,0,0,0,1]\oplus [0,0,0,0,0,0,0,0]$&$(-1,-1,-1,-1,-1,-1,-1,-1,0,0)$\\
$(0,0)$&$[0,0,0,0,0,0,0,0]$&$(0,0,0,0,0,0,0,0,0,0)$\\
$(0,1)$&$[0,0,0,0,0,0,0,1]$&$(0,0,0,0,0,0,0,0,1,0)$\\
$(1,0)$&$[0,0,0,0,0,0,1,0]$&$(0,0,0,0,0,0,0,0,0,1)$\\
$(1,1)$&$[0,0,0,0,0,1,0,0]$&$(0,0,0,0,0,0,1,1,1,1)$\\
$(2,1)$&$[0,0,0,1,0,0,0,0]$&$(0,0,0,0,1,2,3,2,1,2)$\\
$(2,2)$&$[0,0,1,0,0,0,0,0]$&$(0,0,0,1,2,3,4,3,2,2)$\\
$(3,1)$&$[0,1,0,0,0,0,0,0]$&$(0,0,1,2,3,4,5,3,1,3)$\\
$(3,2)$&$[1,0,0,0,0,0,0,0]$&$(0,1,2,3,4,5,6,4,2,3)$\\
$(3,2)$&$[0,1,0,0,0,0,0,1]$&$(0,0,1,2,3,4,5,3,2,3)$\\
$(4,1)$&$[0,0,0,0,0,0,0,0]$&$(1,2,3,4,5,6,7,4,1,4)$\\
\hline
$(3,3)$&$[1,0,0,0,0,0,0,1]$&$(0,1,2,3,4,5,6,4,3,3)$\\
\hline
\end{tabular}
\caption{\label{longerdec}\sl $\mf{sl}(9, \mbb{R})$ level decomposition of
  $\mf{e}_{10}$ with root vectors. All shown levels are complete. The very
  last entry is that of a mixed symmetry generator not studied for the
  dictionary of Table~\ref{dicoeom}. Its possible relation to trombone
  gauging is discussed in {\bf Paper VI}.} 
\end{table}

\subsection{Commutation Relations for Fields Appearing in the Dictionary}
\label{app:commutators}

At level $(\ell_1, \ell_2)=(0,0)$ there is a copy of $\mf{gl}(9, \mbb{R})=\mf{sl}(9, \mbb{R})\oplus \mbb{R}$, as well as a scalar generator associated with the dilaton. Their relations are ($a,b=1,\ldots, 9$)
\begin{align}
\lb K^a{}_b , K^c{}_d \rb &= \delta^c_b K^a{}_d - \delta^a_d K^c{}_b\,,\quad&
     \langle K^a{}_b| K^c{}_d \rangle &= \delta^a_d\delta^c_b 
       - \delta^a_b\delta^c_d \,,&\nn\\
\lb T, K^a{}_b \rb &= 0 \,,\quad &
    \langle T | T \rangle &=\frac12 \,,&\quad \langle T|K^a{}_b\rangle =0\,.
\end{align}
Here, $\langle\cdot|\cdot\rangle$ is the invariant bilinear form. We define also the trace $K=\sum_{a=1}^9 K^a{}_a$. For completeness
\beqa 
K &=& 8 h_1 + 16 h_2 + 24 h_3 + 32 h_4 + 40 h_5 + 48 h_6 
      + 56 h_7 +37 h_8 + 18 h_9 + 27 h_{10}\,,\nn\\
T &=& \frac12 h_1 + h_2 +\frac32 h_3 + 2 h_4 + \frac52 h_5 + 3 h_6
      +\frac72 h_7 +\frac{25}{12}h_8 + \frac23 h_9 + \frac{23}{12}h_{10}\,.
\eqa

All objects transform as $\mf{gl}(9, \mbb{R})$ tensors in the obvious way.
The $T$ commutator relations are
\begin{align}
\lb T, E^{a_1} \rb &= \frac34 E^{a_1} \,,& 
   \lb T, E^{a_1a_2} \rb &= -\frac12  E^{a_1a_2} \,, \nn\\
\lb T, E^{a_1a_2a_3} \rb &= \frac14 E^{a_1a_2a_3} \,,&
   \lb T, E^{a_1\ldots a_5} \rb &= -\frac14  E^{a_1\ldots a_5} \,, \nn\\
\lb T, E^{a_1\ldots a_6} \rb &= \frac12 E^{a_1\ldots a_6} \,,&
   \lb T, E^{a_1\ldots a_7} \rb &= -\frac34  E^{a_1\ldots a_7} \,, \\
\lb T, E^{a_1\ldots a_9} \rb &= -\frac54 E^{a_1\ldots a_9} \,,&\nn\\
\lb T, E^{a_0|a_1\ldots a_7} \rb &= 0 \,,&
   \lb T, E^{a_1\ldots a_8} \rb &= 0 \,.\nn 
\end{align}
The positive level generators are generated by the simple (fundamental)
generators on levels $(0,1)$ and $(1,0)$ by
\begin{align}\label{poslevs}
\lb E^{a_1}, E^{a_2} \rb &= 0 \,,&
  \lb E^{a_1a_2}, E^{a_3a_4} \rb &= 0 \,,&\nn\\
\lb E^{a_1a_2}, E^{a_3} \rb &= E^{a_1a_2a_3} \,,&
  \lb E^{a_1a_2}, E^{a_3\ldots a_5} \rb &= E^{a_1\ldots a_5} \,,& \nn\\
\lb E^{a_1a_2}, E^{a_3\ldots a_7} \rb &= E^{a_1\ldots a_7} \,,&
  \lb E^{a_1a_2}, E^{a_3\ldots a_9} \rb &= E^{a_1\ldots a_9} \,,&\\
\lb E^{a_1}, E^{a_2\ldots a_6} \rb &= E^{a_1\ldots a_6} \,,& 
  \lb E^{a_0} , E^{a_1\ldots a_7} \rb &= E^{a_0|a_1\ldots a_7} + \frac32
    E^{a_0a_1\ldots a_7} \,.\nn
\end{align}
These defining relations imply for example
\beqa 
\lb
E^{a_1a_2a_3}, E^{a_4a_5a_6}\rb &=& -E^{a_1\ldots a_6} \,,\nn\\
\lb E^{a_1a_2}, E^{a_3\ldots a_8} \rb &=& 
    -2 E^{[a_1|a_2]a_3\ldots a_8} + E^{a_1\ldots a_8}\,,\nn\\
\lb E^{a_1a_2a_3}, E^{a_4\ldots a_8} \rb &=& 
    -3 E^{[a_1|a_2a_3]a_4\ldots a_8} -\frac12 E^{a_1\ldots a_8}\,,\nn\\
\lb E^{a_1\ldots a_5}, E^{a_6a_7a_8} \rb &=&
    5 E^{[a_1|a_2\ldots a_5]a_6a_7a_8} -\frac12 E^{a_1\ldots a_8}\,,\\
\lb E^{a_1\ldots a_6}, E^{a_7a_8} \rb &=&
    -6 E^{[a_1|a_2\ldots a_6]a_7a_8} - E^{a_1\ldots a_8}\,,\nn\\
\lb E^{a_1\ldots a_7}, E^{a_8} \rb &=&
    -7 E^{[a_1|a_2\ldots a_7]a_8} +\frac32 E^{a_1\ldots a_8}\,.\nn
\eqa
The Young symmetry on the dual graviton implies $E^{a_0|a_1\ldots a_7} =
7E^{[a_1|a_2\ldots a_7]a_0}$. The two irreducible representations on $(3,2)$
are projected onto via
\beqa 
E^{a_1\ldots a_8} &=& \frac23 \lb E^{[a_1}, E^{a_2\ldots a_8]} \rb \,,\quad
E^{a_0|a_1\ldots a_7} = \frac78\left(\lb E^{a_0}, E^{a_1\ldots a_7} \rb
       + \lb E^{[a_1}, E^{a_2\ldots a_7]a_0}\rb \right) \,,\nn\\
E^{a_0|a_1\ldots a_7} &=& -\frac78\left(\lb E^{a_0[a_1}, E^{a_2\ldots a_7]} \rb
       + \lb E^{[a_1a_2}, E^{a_3\ldots a_7]a_0}\rb \right) \,.
\eqa 
The definitions (\ref{poslevs}) are such that the normalisations are
\beqa 
\langle E^{a_1\ldots a_p} | F_{b_1\ldots b_p}\rangle
   &=& p!\,\delta^{a_1\ldots  a_p}_{b_1\ldots b_p} \,
              \quad\quad\quad\Rightarrow\quad 
   \langle E^{1\,\ldots\,p} |F_{1\,\ldots\,p} \rangle =1 \quad
   (p\ne8)\,,\nn\\ 
\langle E^{a_1\ldots a_8} | F_{b_1\ldots b_8}\rangle
  &=& \frac12\cdot 8! \delta^{a_1\ldots  a_8}_{b_1\ldots b_8} \,
              \quad\quad\Rightarrow\quad 
   \langle E^{1\,\ldots\,8} |F_{1\,\ldots\,8} \rangle =\frac12\,,\nn\\
\langle E^{a_0|a_1\ldots a_7} | F_{b_0|b_1\ldots b_7}\rangle
  &=& \frac78\cdot 7! \left(\delta^{a_0}_{b_0}
       \delta^{a_1\ldots a_7}_{b_1\ldots b_7} 
    + \delta^{[a_1}_{b_0} \delta^{a_2\ldots a_7]a_0}_{b_1\,\,\ldots\ldots\,\,
      b_7}\right)\,. 
\eqa
The additional factor of $2$ in the normalisation of the $8$-form is chosen
such that all structure constants remain rational.

Defining the transposed generators via $F\equiv E^T$ as usual gives the
following commutations relation between the form generators and
their transposes:
\beqa \label{ef0}
\lb E^a, F_b \rb &=& -\frac18 \delta^a_b K + K^a{}_b 
   +\frac32 \delta^a_b T \,,\nn\\
\lb E^{a_1a_2}, F_{b_1b_2} \rb &=& -\frac12 \delta^{a_1a_2}_{b_1b_2} K 
    +4 \delta^{[a_1}_{[b_1} K^{a_2]}_{\,\,\,\,\, b_2]}   
    -2 \delta^{a_1a_2}_{b_1b_2} T \,,\nn\\
\lb E^{a_1a_2a_3}, F_{b_1b_2b_3} \rb &=& 
    -\frac38\cdot 3!\, \delta^{a_1a_2a_3}_{b_1b_2b_3} K 
    +3\cdot 3!\, \delta^{[a_1a_2}_{[b_1b_2} K^{a_3]}_{\,\,\,\,\, b_3]}   
    +3 \delta^{a_1a_2a_3}_{b_1b_2b_3} T \,,\nn\\
\lb E^{a_1\ldots a_5}, F_{b_1\ldots b_5} \rb &=& 
    -\frac58\cdot 5!\, \delta^{a_1\ldots a_5}_{b_1\ldots b_5} K 
    +5\cdot 5!\, \delta^{[a_1\ldots a_4}_{[b_1\ldots b_4} 
             K^{a_5]}_{\,\,\,\,\, b_5]}   
    -\frac12\cdot 5!\, \delta^{a_1\ldots a_5}_{b_1\ldots b_5} T \,,\nn\\
\lb E^{a_1\ldots a_6}, F_{b_1\ldots b_6} \rb &=& 
    -\frac34\cdot 6!\, \delta^{a_1\ldots a_6}_{b_1\ldots b_6} K 
    +6\cdot 6!\, \delta^{[a_1\ldots a_5}_{[b_1\ldots b_5} 
             K^{a_6]}_{\,\,\,\,\, b_6]}   
    +6!\, \delta^{a_1\ldots a_6}_{b_1\ldots b_6} T \,,\\
\lb E^{a_1\ldots a_7}, F_{b_1\ldots b_7} \rb &=& 
    -\frac78\cdot 7!\, \delta^{a_1\ldots a_7}_{b_1\ldots b_7} K 
    +7\cdot 7!\, \delta^{[a_1\ldots a_6}_{[b_1\ldots b_6} 
             K^{a_7]}_{\,\,\,\,\, b_7]}   
    -\frac32\cdot 7!\, \delta^{a_1\ldots a_7}_{b_1\ldots b_7} T \,,\nn\\
\lb E^{a_1\ldots a_9}, F_{b_1\ldots b_9} \rb &=& 
    -\frac98\cdot 9!\, \delta^{a_1\ldots a_9}_{b_1\ldots b_9} K 
    +9\cdot 9!\, \delta^{[a_1\ldots a_8}_{[b_1\ldots b_8} 
             K^{a_9]}_{\,\,\,\,\, b_9]}   
    -\frac52\cdot 9!\, \delta^{a_1\ldots a_9}_{b_1\ldots b_9} T \,,\nn\\
\lb E^{a_1\ldots a_8}, F_{b_1\ldots b_8} \rb &=&
   -\frac12\cdot 8!\, \delta^{a_1\ldots a_8}_{b_1\ldots b_8} K 
     + 4\cdot 8!\, \delta^{[a_1\ldots a_7}_{[b_1\ldots b_7} 
             K^{a_8]}_{\,\,\,\,\, b_8]} \,.\nn
\eqa
The commutator of the dual graviton generator $E^{a_0|a_1\ldots a_7}$ can be
most conveniently written using a dummy tensor $X_{a_0|a_1\ldots a_7}$ as 
\beq \label{efdg}
\lb  F_{b_0|b_1\ldots b_7} ,X_{a_0|a_1\ldots a_7}E^{a_0|a_1\ldots a_7}\rb = 
  7!\left(X_{b_0|b_1\ldots b_7} K - X_{c|b_1\ldots b_7} K^c{}_{b_0} 
   -7 X_{b_0|c[b_1\ldots b_6}K^c_{\,\,\,b_7]} \right).
\eeq

The generators of different rank commute in the following non-trivial way:
\begin{align}\label{ef}
\lb E^a , F_{b_1b_2b_3} \rb &= 3 \delta^a_{[b_1} F_{b_2b_3]} \,,&
\lb E^{a_1a_2}, F_{b_1b_2b_3} \rb &= 
    -6 \delta^{a_1a_2}_{[b_1b_2} F_{b_3]}\,,&\nn\\
\lb E^{a_1a_2} ,F_{b_1\ldots b_5} \rb &=
   -20 \delta^{a_1a_2}_{[b_1b_2}  F_{b_3b_4b_5]} \,,&
\lb E^{a_1a_2a_3} ,F_{b_1\ldots b_5} \rb &=
   60 \delta^{a_1a_2a_3}_{[b_1b_2b_3}  F_{b_4b_5]} \,,& \nn\\
\lb E^a, F_{b_1\ldots b_6} \rb &=
   -6 \delta^a_{[b_1} F_{b_2\ldots b_6]} \,,&
\lb E^{a_1a_2a_3}, F_{b_1\ldots b_6} \rb &= 
   120 \delta^{a_1a_2a_3}_{[b_1b_2b_3} F_{b_4b_5b_6]} \,,&\nn\\
\lb E^{a_1\ldots a_5}, F_{b_1\ldots b_6} \rb &=
   -6!\, \delta^{a_1\ldots a_5}_{[b_1\ldots b_5} F_{b_6]} \,,&
\lb E^{a_1a_2}, F_{b_1\ldots b_7} \rb &=
   -7\cdot 6 \delta^{a_1a_2}_{[b_1b_2} F_{b_3\ldots b_7]} \,,&\\
\lb E^{a_1\ldots a_5}, F_{b_1\ldots b_7} \rb &=
   \frac12\cdot 7!\, \delta^{a_1\ldots a_5}_{[b_1\ldots
     b_5}F_{b_6b_7]}\,,&
\lb E^{a_1a_2}, F_{b_1\ldots b_9} \rb &=
   -9\cdot 8 \delta^{a_1a_2}_{[b_1b_2}F_{b_3\ldots b_9]}\,,&\nn\\
\lb E^{a_1\ldots a_7}, F_{b_1\ldots b_9} \rb &= 
   \frac12\cdot 9!\,\delta^{a_1\ldots a_7}_{[b_1\ldots b_7}F_{b_8b_9]}\,.&\nn
\end{align}
Anticipating the geodesic equation we know that (\ref{ef}) describes all the
couplings between the different forms occurring in the matter equations, so
that for example the $9$-form (=mass term) occurs only in the Bianchi identity
for $F_{(2)}$ and in the eom of $F_{(3)}$, consistent with (\ref{Bianchis})
and (\ref{eoms}). The dilaton and Einstein equation are described by the
couplings of equations (\ref{ef0}) and (\ref{efdg}).

The commutators with the dual dilaton are
\begin{align}
\lb E^a, F_{b_1\ldots b_8} \rb &=
   -6 \delta^a_{[b_1} F_{b_2\ldots b_8]} \,,&
    \lb E^{a_1a_2}, F_{b_1\ldots b_8} \rb &= 
    - 4\cdot 7 \delta^{a_1a_2}_{[b_1b_2}F_{b_3\ldots b_8]} \,,&\nn\\
\lb E^{a_1a_2a_3}, F_{b_1\ldots b_8} \rb &=
    2\cdot 7\cdot 6 \delta^{a_1a_2a_3}_{[b_1b_2b_3}F_{b_4\ldots b_8]} \,,&
  \lb E^{a_1\ldots a_5}, F_{b_1\ldots b_8} \rb &=
    2\cdot 7\cdot 5!\, \delta^{a_1\ldots a_5}_{[b_1\ldots b_5}F_{b_6b_7b_8]} \,,&\nn\\
\lb E^{a_1\ldots a_6}, F_{b_1\ldots b_8} \rb &=
    2\cdot 7!\, \delta^{a_1\ldots a_6}_{[b_1\ldots b_6}F_{b_7b_8]} \,,&
  \lb E^{a_1\ldots a_7}, F_{b_1\ldots b_8} \rb &=
    -6\cdot 7!\, \delta^{a_1\ldots a_7}_{[b_1\ldots b_7}F_{b_8]} \,,&
\end{align}
whereas for the dual graviton one finds
\beqa 
\lb E^a, F_{b_0|b_1\ldots b_7} \rb &=& 
   -\frac78\left(\delta^a_{b_0} F_{b_1\ldots b_7} + \delta^a_{[b_1}F_{b_2\ldots b_7]b_0}\right) \,, \nn\\
\lb E^{a_1a_2}, F_{b_0|b_1\ldots b_7} \rb &=& 
   \frac{21}2\left(\delta^{a_1a_2}_{b_0[b_1} F_{b_2\ldots b_7]} + \delta^{a_1a_2}_{[b_1b_2}F_{b_3\ldots b_7]b_0}\right) \,, \nn\\
\lb E^{a_1a_2a_3}, F_{b_0|b_1\ldots b_7} \rb &=& 
  \f{45\cdot 7}{4}\left(\delta^{a_1a_2a_3}_{b_0[b_1b_2} F_{b_3\ldots b_7]} + \delta^{a_1a_2a_3}_{[b_1b_2b_3}F_{b_4\ldots b_7]b_0}\right) \,, \nn\\
\lb E^{a_1\ldots a_5}, F_{b_0|b_1\ldots b_7} \rb &=& 
  -\frac{5 \cdot 7!}{16}\left(\delta^{a_1\ldots a_5}_{b_0[b_1\ldots b_4} F_{b_5\ldots b_7]} + \delta^{a_1\ldots a_5}_{[b_1\ldots b_5}F_{b_6b_7]b_0}\right) \,, \nn\\
\lb E^{a_1\ldots a_6}, F_{b_0|b_1\ldots b_7} \rb &=& 
  \frac{3 \cdot 7!}{4}\left(\delta^{a_1\ldots a_6}_{b_0[b_1\ldots b_5} F_{b_6b_7]} + \delta^{a_1\ldots a_6}_{[b_1\ldots b_6}F_{b_7]b_0}\right) \,, \nn\\
  \lb E^{a_1\ldots a_7}, F_{b_0|b_1\ldots b_7} \rb &=& 
  \frac{7 \cdot 7!}{8}\left(\delta^{a_1\ldots a_7}_{b_0[b_1\ldots b_6} F_{b_7]} + \delta^{a_1\ldots a_7}_{b_1\ldots b_7}F_{b_0}\right) \,.
\eqa

\subsection{Spinor Representations of $\mf{k}(\mf{e}_{10})$ }
\label{app:SpinorReps}

For the $E_{10}$ model to incorporate all low energy limits of M-theory in a single model, the fermionic representations used in (\ref{e10fermlag}) should not depend on the particular supergravity one wishes to study. Rather the unfaithful ${\bf 320}$ and ${\bf 32}$ representations of $\mf{k}(\mf{e}_{10})$ should be decomposed under a suitable subalgebra. Here, this subalgebra is $\mf{so}(9)\subset \mf{gl}(9,{\mathbb{R}})$ and this appendix provides the details of the action of the $\mf{k}(\mf{e}_{10})$ generators in this basis. When doing the following calculations we found the computer package GAMMA \cite{GAMMA} useful.\footnote{We are grateful to Ulf Gran for generous help in modifying GAMMA to suit our needs.}

\subsubsection{Dirac Spinor}
\label{app:DiracSpinor}

The result of writing the $K(E_{10})$ action on the Dirac spinor is (we recall the notation $M^{a_1a_2}$ for the level $(0,0)$ generator of $K(E_{10})$ from (\ref{00gens}))
\begin{align}
{}M^{a_1a_2}\cdot \epsilon &= \f{1}{2}\Gamma^{a_1a_2}\epsilon ,& 
  \qquad J_{(0,1)}^{a}\cdot \epsilon &= \f{1}{2}\Gamma_{10}\Gamma^{a}\epsilon,&\nn \\
{} J_{(1,0)}^{a_1a_2}\cdot \epsilon &= \f{1}{2}\Gamma_{10}\Gamma^{a_1a_2}\eps,&
  \qquad  J_{(1,1)}^{a_1a_2a_3}\cdot \eps &=\f{1}{2}\Gamma^{a_1a_2a_3}\eps,&\nn \\
{} J_{(2,1)}^{a_1\cdots a_5}\cdot \eps&=\f{1}{2}\Gamma_{10}\Gamma^{a_1\cdots a_5}\epsilon,&
  \qquad  J_{(2,2)}^{a_1\cdots a_6}\cdot \eps&= -\f{1}{2}\Gamma^{a_1\cdots a_6}\eps,&\nn \\
{} J_{(3,1)}^{a_1\cdots a_7}\cdot \eps&= \f{1}{2}\Gamma^{a_1\cdots a_7}\eps,&
  \qquad J_{(3,2)}^{a_0|a_1\cdots a_7}\cdot \eps&=\f{7}{2}\Gamma_{10}\delta_{a_0}^{[a_1}\Gamma^{a_2\cdots a_7]}\eps,&\nn \\
{} J_{(3,2)}^{a_1\cdots a_8}\cdot \eps&=0 ,&
  \qquad J_{(4,1)}^{a_1\cdots a_9}\cdot \eps&=\f{1}{2}\Gamma_{10}\Gamma^{a_1\cdots a_9}\eps,&
\end{align}
where the generators above levels $(0,1)$ and $(1,0)$ are defined through the lower levels as follows 
\beqa 
{}J_{(1,1)}^{a_1a_2a_3}\cdot\epsilon &:=& \lb J_{(1,0)}^{[a_1a_2}, J_{(0,1)}^{a_3]}\rb\cdot\epsilon ,
\nn \\
{}J_{(2,1)}^{a_1\cdots a_5}\cdot\epsilon &:= & \lb J_{(1,0)}^{[a_1a_2}, J_{(1,1)}^{a_3a_4a_5]}\rb\cdot\epsilon ,
\nn\\
{} J_{(2,2)}^{a_1\cdots a_6}\cdot\epsilon &:= & \lb J_{(1,1)}^{[a_1}, J_{(1,1)}^{a_2\ldots a_6]}\rb\cdot\epsilon ,
\nn \\
J_{(3,1)}^{a_1\cdots a_7}\cdot\epsilon &:= & \lb J_{(1,0)}^{[a_1a_2}, J_{(2,1)}^{a_3\cdots a_7]}\rb\cdot\epsilon ,
\nn \\
J_{(3,2)}^{a_0|a_1\cdots a_7}\cdot\epsilon &:= &-\f{7}{8}\Big(\lb J_{(0,1)}^{a_0}, J_{(3,1)}^{a_1\cdots a_7}\rb+\lb J_{(0,1)}^{[a_1}, J_{(3,1)}^{a_2\cdots a_7]a_0}\rb\Big)\cdot\epsilon ,
\nn \\
J_{(3,2)}^{a_1\cdots a_8}\cdot\epsilon &:= & \lb J_{(1,0)}^{[a_1a_2}, J_{(2,2)}^{a_3\cdots a_8]}\rb\cdot\epsilon ,
\nn \\
J_{(4,1)}^{a_1\cdots a_9}\cdot\epsilon &:= &\lb J_{(1,0)}^{[a_1a_2}, J_{(3,1)}^{a_3\cdots a_9]}\rb \cdot\epsilon .
\eqa
We stress that this kind of construction is guaranteed to yield a consistent unfaithful representation of all of $\mf{k}(\mf{e}_{10})$ given that a few simple consistency conditions between the lowest level fundamental generators are satisfied~\cite{Damour:2006xu}. That these conditions are satisfied here can be checked easily directly, but it also follows from the branching of the transformation rules given in~\cite{deBuyl:2005zy,Damour:2005zs}.

\subsection*{Vector-Spinor}
\label{app:VectorSpinor}

By reduction of the transformation rules of~\cite{Damour:2005zs,deBuyl:2005mt} one obtains that
the fundamental $\mf{k}(\mf{e}_{10})$-generators act on the vector spinor representation as follows on the $\Psi_{10}$ component 
\beqa 
J_{(0,1)}^{a}\cdot \Psi_{10} &=& \f12 \G_{10}\G^a\Psi_{10}+\Psi^a\nn\\
J_{(1,0)}^{a_1a_2}\cdot \Psi_{10} &=& \f16 \G_{10}\G^{a_1a_2}\Psi_{10}+\f43 \G^{[a_1}\Psi^{a_2]}\nn.
\eqa
On the $\Psi_a$ ($a=1,\ldots,9$) component they act as 
\beqa 
J_{(0,1)}^{a}\cdot \Psi_{b}&=&\f12 \G_{10}\G^a\Psi_{b}-\delta^a_b\Psi_{10}\nn\\
J_{(1,0)}^{a_1a_2}\cdot \Psi_{b} &=& \f12\G_{10}\G^{a_1a_2}\Psi_b -\f43\G_{10}\delta^{[a_1}_b\Psi^{a_2]}+\f23\G_{10}\G_b^{\ [a_1}\Psi^{a_2]}\nn\\
&&+\f43\delta_b^{[a_1}\G^{a_2]}\Psi_{10}-\f13\G_b^{\ a_1a_2}\nn\Psi_{10}.
\eqa
The other levels action is then computed to be on $\Psi_{10}$ as
\beqa 
M^{a_1a_2}\cdot  \Psi_{10} &=&\phantom{-} \f12\G^{a_1a_2}\Psi_{10}\nn\\
J_{(1,1)}^{a_1\cdots a_3}\cdot \Psi_{10} &=& \phantom{-}  \f12\G^{a_1\cdots a_3}\Psi_{10}-\G_{10}\G^{[a_1a_2}\Psi^{a_3]}\nn\\
J_{(2,1)}^{a_1\cdots a_5}\cdot \Psi_{10} &=& - \f16 \G_{10}\G^{a_1\cdots a_5}\Psi_{10}-\f53\G^{[a_1\cdots a_4}\Psi^{a_5]}\nn\\
J_{(2,2)}^{a_1\cdots a_6}\cdot \Psi_{10} &=& -\f12\G^{a_1\cdots a_6}\Psi_{10}-4\G_{10}\G^{[a_1\cdots a_5}\Psi^{a_6]}\nn\\
J_{(3,1)}^{a_1\cdots a_7}\cdot \Psi_{10} &=& -\f32\G^{a_1\cdots a_7}\Psi_{10}+7\G_{10}\G^{[a_1\cdots a_6}\Psi^{a_7]}\nn\\
J_{(3,2)}^{a_1\cdots a_8}\cdot \Psi_{10} &=& - \f43 \G_{10}\G^{a_1\cdots a_8}\Psi_{10}-\f{32}3\G^{[a_1\cdots a_7}\Psi^{a_8]}\nn\\
J_{(3,2)}^{a_0|a_1\cdots a_7}\cdot \Psi_{10}&=& - \f{7}{2}\Gamma_{10}\delta_{a_0}^{[a_1}\Gamma^{a_2\cdots a_7]}\Psi_{10}\nn\\
J_{(4,1)}^{a_1\cdots a_9}\cdot \Psi_{10} &=& \phantom{-}  \f92 \G_{10}\G^{a_1\cdots a_9}\Psi_{10}-15\G^{[a_1\cdots a_8}\Psi^{a_9]}.
\label{onpsi10}
\eqa

On $\Psi_a$ we obtain similarly
\beqa 
M^{a_1a_2}\cdot  \Psi_{b} &=& \f12\G^{a_1a_2}\Psi_{b}+2\delta_b^{[a_1}\Psi^{a_2]}\nn\\
J_{(1,1)}^{a_1\cdots a_3}\cdot \Psi_{b} &=& \f12\G^{a_1\cdots a_3}\Psi_{b}+4\delta_b^{[a_1}\G^{a_2}\Psi^{a_3]}-\G_b{}^{[a_1a_2}\Psi^{a_3]}\nn\\
J_{(2,1)}^{a_1\cdots a_5}\cdot \Psi_{b} &=&  \f12 \G_{10}\G^{a_1\cdots a_5}\Psi_{b}+\f{20}{3}\Gamma_{10}\delta_b^{[a_1}\Gamma^{a_2\cdots a_4}\Psi^{a_5]}-\f{10}3\Gamma_{10}{\Gamma_b}^{[a_1\cdots a_4}\Psi^{a_5]}
\nn \\
& & +\f53\delta_b^{[a_1}\Gamma^{a_2\cdots a_5]}\Psi_{10}-\f{2}{3}{\Gamma_b}^{a_1\cdots a_5}\Psi_{10}\nn\\
J_{(2,2)}^{a_1\cdots a_6}\cdot \Psi_{b} &=& -\f12\G^{a_1\cdots a_6}\Psi_{b}+10\delta_b^{[a_1}\Gamma^{a_2\cdots a_5}\Psi^{a_6]}-4{\Gamma_b}^{[a_1\cdots a_5}\Psi^{a_6]}\nn\\
J_{(3,1)}^{a_1\cdots a_7}\cdot \Psi_{b} &=& \f12\G^{a_1\cdots a_7}\Psi_{b}-7{\Gamma_b}^{[a_1\cdots a_6}\Psi^{a_7]} -2\Gamma_{10}{\Gamma_b}^{a_1\cdots a_7}\Psi_{10}\nn\\
J_{(3,2)}^{a_1\cdots a_8}\cdot \Psi_{b} &=&-\f43\G_{10}\G_b^{\ [a_1\cdots a_7}\Psi^{a_8]}-\f{28}3\G_{10}\delta_b^{[a_1}\G^{a_2\cdots a_7}\Psi^{a_8]} \nn\\
&&-\f43\Gamma_b^{\ a_1\cdots a_8}\Psi_{10}+\f43\delta_b^{[a_1}\G^{a_2\cdots a_8]}\Psi_{10}\nn\\
J_{(3,2)}^{a_0|a_1\cdots a_7}\cdot \Psi_{b}&=&\f{7}{2}\Gamma_{10}\delta^{[a_1}_{a_0}\Gamma^{a_2\cdots a_7]}\Psi_{b}-\f{21}{2}\Gamma_{10}\delta_b^{[a_1}\Gamma_{a_0}^{\ \,a_2\cdots a_6}\Psi^{a_7]}-\f{49}{4}\Gamma_{10}\delta_b^{a_0}\Gamma^{[a_1\cdots a_6}\Psi^{a_7]}
\nn \\
 & & +42 \Gamma_{10}\delta_{a_0}^{[a_1}{\Gamma_b}^{a_2\cdots a_6}\Psi^{a_7]}-7\delta_{a_0}^{[a_1}{\Gamma_b}^{a_2\cdots a_7]}\Psi_{10}
 \nn \\
 & & +\f{7}{4}\Big(\Gamma_{10}{\Gamma_{b}}^{a_0[a_1\cdots a_6}\Psi^{a_7]}-\Gamma_{10}\delta_b^{[a_1}\Gamma^{a_2\cdots a_7]}\Psi^{a_0}
 \nn \\
 & &\phantom{7}+\Gamma_{10}{\Gamma_b}^{a_1\cdots a_7}\Psi^{a_0} +\delta_b^{[a_1}\Gamma^{a_2\cdots a_7] a_0}\Psi_{10} +\delta_b^{a_0}\Gamma^{a_1\cdots a_7}\Psi_{10}\Big)
  \nn\\
J_{(4,1)}^{a_1\cdots a_9}\cdot \Psi_{b} &=&\f{1}{2}\Gamma_{10}\Gamma^{a_1\cdots a_9}\Psi_b -12\Gamma_{10}{\Gamma_{b}}^{[a_1\cdots a_8}\Psi^{a_9]}
\nn \\
& & -24\Gamma_{10}\delta_b^{[a_1}\Gamma^{a_2\cdots a_8}\Psi^{a_9]}
-9\delta_b^{[a_1}\Gamma^{a_2\cdots a_9]}\Psi_{10}.
\label{onpsib}
\eqa
We note that the mixed symmetry generator at level $(3,2)$ indeed satisfies 
\beq
J_{(3,2)}^{[a_0|a_1\cdots a_7]}\cdot \Psi_{b}=0
\eeq
as desired.

\section{Equations of Motion of the $E_{10}/K(E_{10})$ Coset Model}
\label{app:EOMe10}

Using all the explicit commutators and representations of the Appendix \ref{Appendix:LevelDecomp}, we can now write out the bosonic and fermionic equations of motion in their full glory.

\subsection{Bosonic Equations of Motion}
\label{Appendix:EOMBosonic}

The level zero equations of motion read 
\begin{subequations}\label{levelzeroeom}
\beqa
\p^2_t\phi&=&e^{3\phi/2}P_aP_a-2 e^{-\phi}P_{a_1a_2}P_{a_1a_2}+\f{1}{3}e^{\phi/2}P_{a_1a_2a_3}P_{a_1a_2a_3}\nn\\
&&-\frac2{5!}e^{-\phi/2}P_{a_1\ldots a_5}P_{a_1\ldots a_5}+\frac4{6!}e^{\phi}P_{a_1\ldots a_6}P_{a_1\ldots a_6}\nn\\
&&-\frac6{7!}e^{-3\phi/2}P_{a_1\ldots a_7}P_{a_1\ldots a_7}-\frac{10}{9!}e^{-5\phi/2}P_{a_1\ldots a_9}P_{a_1\ldots a_9}\label{dilatoneom}\\
\cD p_{ab} &=& -\frac1{4}e^{3\phi/2}\delta_{ab}P_cP_c+2e^{3\phi/2}P_aP_b \nn \\
& & -\f{1}{4}e^{-\phi}\delta_{ab}P_{cd}P_{cd}+2e^{-\phi}P_{ca}P_{cb}
\nn \\
& & -\f1{8}e^{\phi/2}\delta_{ab}P_{c_1c_2c_3}P_{c_1c_2c_3}+e^{\phi/2}P_{ac_1c_2}P_{bc_1c_2}
\nn \\
& & -\f1{4\cdot 4!}e^{-\phi/2}\delta_{ab}P_{c_1\cdots c_5}P_{c_1\cdots c_5}+\f{2}{4!}e^{-\phi/2}P_{ac_1\cdots c_4}P_{bc_1\cdots c_4}
\nn \\
& & -\f{3}{2\cdot 6!}e^{\phi}\delta_{ab}P_{c_1\cdots c_6}P_{c_1\cdots c_6}+\f{2}{5!}e^{\phi}P_{ac_1\cdots c_5}P_{bc_1\cdots c_5}
\nn \\
& & -\f1{4\cdot 6!}e^{-3\phi/2}\delta_{ab}P_{c_1\cdots c_7}P_{c_1\cdots c_7}+\f{2}{6!}e^{-3\phi/2}P_{ac_1\cdots c_6}P_{bc_1\cdots c_6}
\nn \\
& & -\f1{8!}\delta_{ab}P_{c_1\cdots c_8}P_{c_1\cdots c_8}+\f1{7!}P_{ac_1\cdots c_7}P_{bc_1\cdots c_7}
\nn \\
& & +\f1{4\cdot 8!}\big(-\delta_{ab}P_{c_0|c_1\cdots c_7}P_{c_0|c_1\cdots c_7}+P_{a|c_1\cdots c_7}P_{b|c_1\cdots c_7}
\nn \\
& & +7P_{c_0|c_1\cdots c_6 a}P_{c_0|c_1\cdots c_6 b}\big)
\nn \\
& & -\f1{4\cdot 8!}e^{-5\phi/2}\delta_{ab}P_{c_1\cdots c_9}+\f{2}{8!}e^{-5\phi/2}P_{ac_1\cdots c_8}P_{bc_1\cdots c_8}\,.\label{Einstein}
\eqa
\end{subequations}
{\allowdisplaybreaks For the higher level fields the equations of motion become
\begin{subequations}\label{higherleveom}
\beqa
\cD (e^{3\phi/2} P_a) &=&
    -e^{\phi/2} P_{ac_1c_2} P_{c_1c_2}   + \frac2{5!} e^\phi P_{ac_1\ldots c_5}P_{c_1\ldots c_5}     +\frac{12}{8!} P_{ac_1\ldots c_7}P_{c_1\ldots c_7} \nn\\
 &&       -\frac{7}{4\cdot 8!}(P_{c_1|ac_2\ldots c_7}P_{c_1\ldots c_7}+P_{a|c_1\ldots c_7}P_{c_1\ldots c_7})\,,\label{boseom1}\\
\cD (e^{-\phi} P_{a_1a_2}) &=& 
    2 e^{\phi/2}  P_{a_1a_2c}P_c +\frac13 e^{-\phi/2} P_{a_1a_2c_1c_2c_3}P_{c_1c_2c_3}\nn\\
 &&  +\frac2{5!}e^{-3\phi/2} P_{a_1a_2c_1\ldots c_5}P_{c_1\ldots c_5} 
     +\frac2{7!} e^{-5\phi/2}P_{a_1a_2c_1\ldots c_7}P_{c_1\ldots c_7} \nn\\
 && +\frac1{6!}P_{a_1a_2c_1\ldots c_6}P_{c_1\ldots c_6} \nn\\
 && + \frac{3}{8\cdot 6!} (P_{c_1|a_1a_2c_2\ldots c_6}P_{c_1\ldots c_6}+P_{a_1|a_2c_1\ldots c_6}P_{c_1\ldots c_6})\,, \label{boseom2} \\
\cD (e^{\phi/2}P_{a_1a_2a_3}) &=&
    -e^{-\phi/2} P_{a_1a_2a_3c_1c_2}P_{c_1c_2} 
        -\frac13 e^\phi P_{a_1a_2a_3c_1c_2c_3}P_{c_1c_2c_3}\nn\\
        && -\frac1{2\cdot 5!} P_{a_1a_2a_3c_1\ldots c_5}P_{c_1\ldots c_5}\nn\\
 && +\frac{1}{256}(P_{c_1|a_1a_2a_3c_2\ldots c_5}P_{c_1\ldots c_5}+P_{a_1|a_2a_3c_1\ldots c_5}P_{c_1\ldots c_5}) \,,\label{boseom3}\\
\cD (e^{-\phi/2} P_{a_1\ldots a_5}) &=&
    2 e^\phi P_{a_1\ldots a_5c}P_c-e^{-3\phi/2}P_{a_1\ldots a_4 c_1c_2}P_{c_1c_2}\nn\\
    && -\frac1{12}P_{a_1\ldots a_5c_1c_2c_3}P_{c_1c_2c_3}\nn\\
    && +\frac{1}{28}(P_{c_1|a_1\ldots a_5c_2c_3}P_{c_1\ldots c_3}+P_{a_1|a_2\ldots a_5c_1\ldots c_3}P_{c_1\ldots c_3}) \,,\label{Bianchi1}\\
\cD (e^\phi P_{a_1\ldots a_6}) &=&
   -\frac12 P_{a_1\ldots a_6c_1c_2} P_{c_1c_2}\nn\\
   &&-\frac{3}{16}(P_{c_1|a_1\ldots a_6c_2}P_{c_1c_2}+P_{a_1|a_2\ldots a_6c_1c_2}P_{c_1 c_2})\,,\label{Bianchi2} \\
\cD (e^{-3\phi/2} P_{a_1\ldots a_7}) &=&
  - e^{-5\phi/2}P_{a_1\ldots a_7c_1c_2} P_{c_1c_2}+\frac32P_{a_1\ldots a_7 c}P_{c} \nn\\
  &&-\frac7{32}(P_{c|a_1\ldots a_7}P_{c}+P_{a_1|a_2\ldots a_7c}P_{c})  \,,\\
\cD (e^{-5\phi/2} P_{a_1\ldots a_9}) &=& 0 \,,\label{Bianchi3}\\
\cD P_{a_1\ldots a_8} &=& 0 \,,\\
\cD P_{a_0|a_1\ldots a_7} &=& 0 \,.
\eqa
\end{subequations}}

\subsection{Fermionic Equations of Motion}
\label{app:FermionicSigmaModelEOM}

The fermionic sector of the $E_{10}$-invariant Lagrangian involves a
Dirac-type kinetic term for the 320-dimensional vector-spinor representation
$\Psi$ of $\mf{k}(\mf{e}_{10})$ which was given in (\ref{e10fermlag}). The
resulting Dirac equation can be evaluated for both the $\Psi_a$ and the
$\Psi_{10}$ components as in (\ref{GravitinoDilatinoEOM})  using the
expressions for the $K(E_{10})$ action which were derived in
Appendix~\ref{app:SpinorReps}. 

The result for the $\Psi_{10}$ component up to level $(4,1)$ is {\allowdisplaybreaks
\beqa 
\label{eompsi10}
0&=&\pa_t \Psi_{10}-\f{1}{4}q_{a_1a_2} \G^{a_1a_2}\Psi_{10}\nn\\
&-&\f12e^{3\phi/4}P_{a}\G_{10}\G^a\Psi_{10}-e^{3\phi/4}P_{a}\Psi^a\nn\\
&-&\f{1}{12}e^{-\phi/2}P_{a_1a_2}\G_{10}\G^{a_1a_2}\Psi_{10}
-\f23e^{-\phi/2}P_{a_1a_2} \G^{a_1}\Psi^{a_2}\nn\\
&-&\f{1}{12}e^{\phi/4}P_{a_1a_2a_3}\G^{a_1\cdots a_3}\Psi_{10}+\f16e^{\phi/4}P_{a_1a_2a_3}\G_{10}\G^{a_1a_2}\Psi^{a_3}\nn\\
&+& \f{1}{6!}e^{-\phi/4}P_{a_1\cdots a_5} \G_{10}\G^{a_1\cdots a_5}\Psi_{10}+\f1{3\cdot4!}e^{-\phi/4}P_{a_1\cdots a_5} \G^{a_1\cdots a_4}\Psi^{a_5}\nn\\
&+&\f{1}{2\cdot6!}e^{\phi/2}P_{a_1\cdots a_6}\G^{a_1\cdots a_6}\Psi_{10}+\f1{180}e^{\phi/2}P_{a_1\cdots a_6}\G_{10}\G^{a_1\cdots a_5}\Psi^{a_6}\nn\\
&+&\f{3}{2\cdot7!}e^{-3\phi/4}P_{a_1\cdots a_7}\G^{a_1\cdots a_7}\Psi_{10}
-\f1{6!}e^{-3\phi/4}P_{a_1\cdots a_7}\G_{10}\G^{a_1\cdots a_6}\Psi^{a_7}\nn\\
&+&\f{4}{3\cdot8!}P_{a_1\cdots a_8}\G_{10}\G^{a_1\cdots a_8}\Psi_{10}+\f{4}{3\cdot7!}P_{a_1\cdots a_8}\G^{a_1\cdots a_7}\Psi^{a_8} \nn \\
&+&\f{7}{2\cdot8!}P_{c|ca_1\cdots a_6}\Gamma_{10}\Gamma^{a_1\cdots a_6}\Psi_{10}\nn\\
&-&\f{1}{2\cdot8!}e^{-5\phi/4}P_{a_1\cdots a_9} \G_{10}\G^{a_1\cdots
  a_9}\Psi_{10}+\f{15}{9!}e^{-5\phi/4}P_{a_1\cdots a_9} \G^{a_1\cdots
  a_8}\Psi^{a_9} + \ldots\,,
\eqa
and is related to the dilatino equation of motion in the body of the article.}

For the gravitino component $\Psi_a$ one finds similarly{\allowdisplaybreaks{
\beqa 
\label{eomke10}
0 &=& \pa_t \Psi_a-\f{1}{4}q_{b_1b_2}\G^{b_1b_2}\Psi_{a}-q_{ab}\Psi_{b}
\nn\\
&-&\f12e^{3\phi/4}P_{b} \G_{10}\G^b\Psi_{a}+e^{3\phi/4}P_{a}\Psi_{10}
\nn \\
&-&\f{1}{4}e^{-\phi/2}P_{b_1b_2}\G_{10}\G^{b_1b_2}\Psi_a +\f23e^{-\phi/2}P_{ab}\G_{10}\Psi^{b}-\f13e^{-\phi/2}P_{b_1b_2}\G_{10}\G_a^{\ b_1}\Psi^{b_2}
\nn\\
&&-\f23e^{-\phi/2}P_{ab}\G^{b}\Psi_{10}
 +\f16e^{-\phi/2}P_{b_1b_1}\G_a^{\ b_1b_2}\Psi_{10}
\nn\\
&-&\f{1}{12}e^{\phi/4}P_{b_1b_2b_3}\G^{b_1\cdots b_3}\Psi_{a}
-\f23e^{\phi/4}P_{ab_1b_2}\G^{b_1}\Psi^{b_2}
+\f16e^{\phi/4}P_{b_1b_2b_3}\G_a{}^{b_1b_2}\Psi^{b_3}
\nn \\
&-& \f{1}{2\cdot5!}e^{-\phi/4}P_{b_1\cdots b_5} \G_{10}\G^{b_1\cdots b_5}\Psi_{a}
-\f{1}{18}e^{-\phi/4}P_{ab_1\cdots b_4} \Gamma_{10}\Gamma^{b_1\cdots
  b_3}\Psi^{b_4}\nn\\
&&+\f{1}{36}e^{-\phi/4}P_{b_1\cdots b_5} \Gamma_{10}{\Gamma_a}^{b_1\cdots
  b_4}\Psi^{b_5} 
 -\f1{3\cdot4!}e^{-\phi/4}P_{ab_1\cdots b_4}\Gamma^{b_1\cdots
   b_4}\Psi_{10}\nn\\ 
&&+\f{2}{3\cdot5!}e^{-\phi/4}P_{b_1\cdots b_5}{\Gamma_a}^{b_1\cdots b_5}\Psi_{10}\nn \\
&+&\f{1}{2\cdot6!}e^{\phi/2}P_{b_1\cdots b_6}\G^{b_1\cdots b_6}\Psi_{a}
-\f{1}{3\cdot 4!}e^{\phi/2}P_{ab_1\cdots b_5}\Gamma^{b_1\cdots b_4}\Psi^{b_5}
+\f4{6!}e^{\phi/2}P_{b_1\cdots b_6}{\Gamma_a}^{b_1\cdots b_5}\Psi^{b_6}
\nn \\
&-&\f{1}{2\cdot7!}e^{-3\phi/4}P_{b_1\cdots b_7}\G^{b_1\cdots b_7}\Psi_{a}
+\f1{6!}e^{-3\phi/4}P_{b_1\cdots b_7}{\Gamma_a}^{b_1\cdots b_6}\Psi^{b_7}\nn\\
&&+\f2{7!}e^{-3\phi/4}P_{b_1\cdots b_7}\Gamma_{10}{\Gamma_a}^{b_1\cdots b_7}\Psi_{10}
 \nn \\
&+&\f{1}{6\cdot7!}P_{b_1\cdots b_8}\G_{10}\G_a^{\ b_1\cdots b_7}\Psi^{b_8}
+\f1{6\cdot6!}P_{ab_1\cdots b_7}\G_{10}\G^{b_1\cdots b_6}\Psi^{b_7}
 \nn\\
&&+\f{1}{6\cdot7!}P_{b_1\cdots b_8}\Gamma_a^{\ b_1\cdots b_8}\Psi_{10}
-\f{1}{6\cdot7!}P_{ab_1\cdots b_7}\G^{b_1\cdots b_7}\Psi_{10}
\nn \\
&-&\f{7}{2\cdot8!}P_{c|cb_1\cdots b_6}\Gamma_{10}\Gamma^{b_1\cdots b_6}\Psi_{a}
+\f{21}{2\cdot8!}P_{b_0|ab_1\cdots b_6}\Gamma_{10}\Gamma^{b_0b_1\cdots
  b_5}\Psi^{b_6}\nn\\ 
&&+\f{49}{4\cdot8!}P_{a|b_1\cdots b_7}\Gamma_{10}\Gamma^{b_1\cdots b_6}\Psi^{b_7}
-\f{42}{8!} P_{c|cb_1\cdots b_6}\Gamma_{10}{\Gamma_a}^{b_1\cdots b_5}\Psi^{b_6}
 +\f7{8!}P_{c|cb_1\cdots b_6}{\Gamma_a}^{b_1\cdots b_6}\Psi_{10}
 \nn \\
 & & -\f{7}{4\cdot8!}P_{b_0|b_1\cdots
   c_7}\Gamma_{10}{\Gamma_{a}}^{b_0b_1\cdots b_6}\Psi^{b_7} 
+\f{7}{4\cdot 8!}P_{b_0|ab_1\cdots b_6}\Gamma_{10}\Gamma^{b_1\cdots
  b_6}\Psi^{b_0} 
 \nn \\
 & &-\f{7}{4\cdot 8!}P_{b_0|b_1\cdots b_7}\Gamma_{10}{\Gamma_a}^{b_1\cdots
   b_7}\Psi^{b_0} 
  -\f{7}{4\cdot 8!}P_{b_0|ab_1\cdots b_6}\Gamma^{b_0b_1\cdots
    b_6}\Psi_{10}\nn\\
&&   -\f{7}{4\cdot 8!}P_{a|b_1\cdots b_7}\Gamma^{b_1\cdots b_7}\Psi_{10}\nn
   \\ 
&-&\f{1}{2\cdot9!}e^{-5\phi/4}P_{b_1\cdots b_9}\Gamma_{10}\Gamma^{b_1\cdots b_9}\Psi_a
 +\f{12}{9!}e^{-5\phi/4}P_{b_1\cdots b_9}\Gamma_{10}{\Gamma_{a}}^{b_1\cdots b_8}\Psi^{b_9}
\nn \\
& & +\f{24}{9!}e^{-5\phi/4}P_{ab_1\cdots b_8}\Gamma_{10}\Gamma^{b_1\cdots b_7}\Psi^{b_8}
+\f1{8!}e^{-5\phi/4}P_{ab_1\cdots b_8}\Gamma^{b_1\cdots b_8}\Psi_{10}+
\ldots\,. 
\eqa}}

\subsection{Supersymmetry Variation}
\label{app:susyvarcoset}

In the same fashion, the supersymmetry variation given in
(\ref{SusyTransfSigmaModel})
can be written explicitly as 
\beqa 
\label{susyexpli}
\delta\Psi_t &=& \pa_t \epsilon -\f{1}{4}q_{a_1a_2}\Gamma^{a_1a_2}\epsilon
- \f{1}{2}e^{3\phi/4}P_{a}\Gamma_{10}\Gamma^{a}\epsilon
-\f{1}{4}e^{-\phi/2}P_{a_1a_2}\Gamma_{10}\Gamma^{a_1a_2}\eps\nn\\
&-&\f{1}{2\cdot3!}e^{\phi/4}P_{a_1a_2a_3}\Gamma^{a_1a_2a_3}\eps
- \f{1}{2\cdot5!}e^{-\phi/4}P_{a_1\cdots a_5}\Gamma_{10}\Gamma^{a_1\cdots a_5}\epsilon
+\f{1}{2\cdot6!}e^{\phi/2}P_{a_1\cdots a_6}\Gamma^{a_1\cdots a_6}\eps\nn \\
& -&\f{1}{2\cdot7!}e^{-3\phi/4}P_{a_1\cdots a_7}\Gamma^{a_1\cdots a_7}\eps
-\f{7}{2\cdot8!}P_{c|ca_1\cdots a_6}\Gamma_{10}\Gamma^{a_1\cdots a_6}\eps\nn\\
&-&\f{1}{2\cdot9!}e^{-5\phi/4}P_{a_1\cdots a_9}\Gamma_{10}\Gamma^{a_1\cdots a_9}\eps+ \cdots
\eqa

%%%%%%%%%%%%%%%%%%%%%%%%%%%%%%%%%%%%%%%%%%%%
%
% Bibliography
%
%%%%%%%%%%%%%%%%%%%%%%%%%%%%%%%%%%%%%%%%%%%%
%\pagestyle{empty}

%\cleardoublepage

\clearpage
\pagestyle{plain}
\def\href#1#2{#2}
\bibliographystyle{utphysmod2}
\addcontentsline{toc}{chapter}{\sffamily\bfseries Bibliography}
\bibliography{ThesisBibliography}

%\cleardoublepage
\end{document}